\documentclass[PhD]{iitddiss}
 \usepackage{t1enc}

\usepackage{enumitem}
\usepackage{cite}
\AtBeginDocument{}

\let\proof\relax 
\let\endproof\relax
\usepackage{amsthm}
\usepackage{amsfonts}
\usepackage{amssymb,bm}
\usepackage{mathrsfs}
\usepackage{mathtools}
\usepackage{amsmath}

\usepackage{accents}

\newlength{\dhatheight}
\newcommand{\doublehat}[1]{%
    \settoheight{\dhatheight}{\ensuremath{\hat{#1}}}%
    \addtolength{\dhatheight}{-0.25ex}%
    \hat{\vphantom{\rule{1pt}{\dhatheight}}%
    \smash{\hat{#1}}}}

\newtheorem{theorem}{Theorem}[chapter]
\newtheorem{definition}{Definition}[chapter]
\newtheorem{lemma}{Lemma}[chapter]
\newtheorem{proposition}{Proposition}[chapter]
\newtheorem{corollary}{Corollary}[chapter]

\newtheorem{claim}{Claim}[chapter]

\newtheorem{remark}{Remark}[chapter]
\newenvironment{claimproof}{\proof}{\endproof}

\usepackage{comment}

\usepackage[switch]{lineno}
\usepackage[named]{algo}
\usepackage[noend]{algpseudocode}
\usepackage{algorithm}

\usepackage{fancyhdr}
\usepackage{enumitem}
\usepackage{setspace}
\usepackage{subfiles}

\begin{document}
\begin{titlepage}
	\thispagestyle{empty}
	\centering
        \addtolength{\topmargin}{2em}
	{\Large \bf INVERSE STOCHASTIC FILTERING\\ FOR INVERSE COGNITION IN COUNTER-ADVERSARIAL SYSTEMS}
	
	\vspace{5 cm}
	\centering{\Large \bf HIMALI SINGH}\\
	\vspace {2cm}
	
	\begin{figure}[h]                                                                                          {\centering\resizebox*{2.25 in}{2.25 in} {\includegraphics{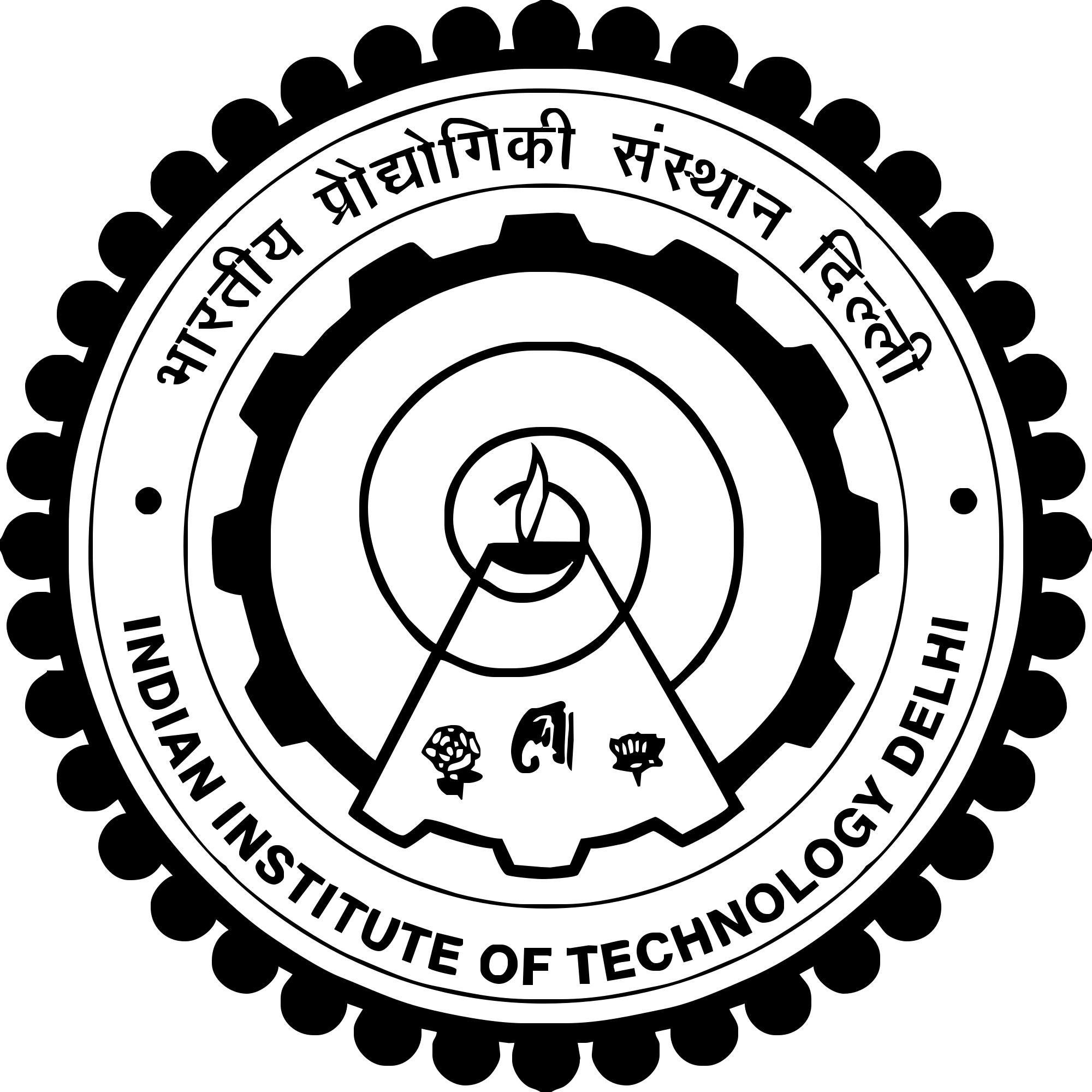}}\par}
	\end{figure}
	\vspace{6em}
	{\Large \centering \bf DEPARTMENT OF ELECTRICAL ENGINEERING}\\\centering{\Large \bf INDIAN INSTITUTE OF TECHNOLOGY DELHI} \\
	\centering  {\Large \bf FEBRUARY 2025}
\end{titlepage}
\begin{titlepage}
	\thispagestyle{empty}
	\addtolength{\topmargin}{2em}
	\centering
	{\Large \bf Inverse Stochastic Filtering for Inverse Cognition\\ in counter-adversarial systems}
	\\
	\vspace{1.0 cm}
	{\bf \centering by}
	\\
	\vspace{.3 cm}
	\centering {\Large \bf HIMALI SINGH}\\
	{\centering \bf DEPARTMENT OF ELECTRICAL ENGINEERING}\\
	\vspace {1.0 cm}
	
	\centering{Submitted} \\
	\centering{\it in fulfillment of the requirements of the degree of Doctor of Philosophy}\\
	\centering{to the}
	\vspace{1.0 cm}
	
	\begin{figure}[h]
		{\centering\resizebox*{2.25 in}{2.25 in} {\includegraphics{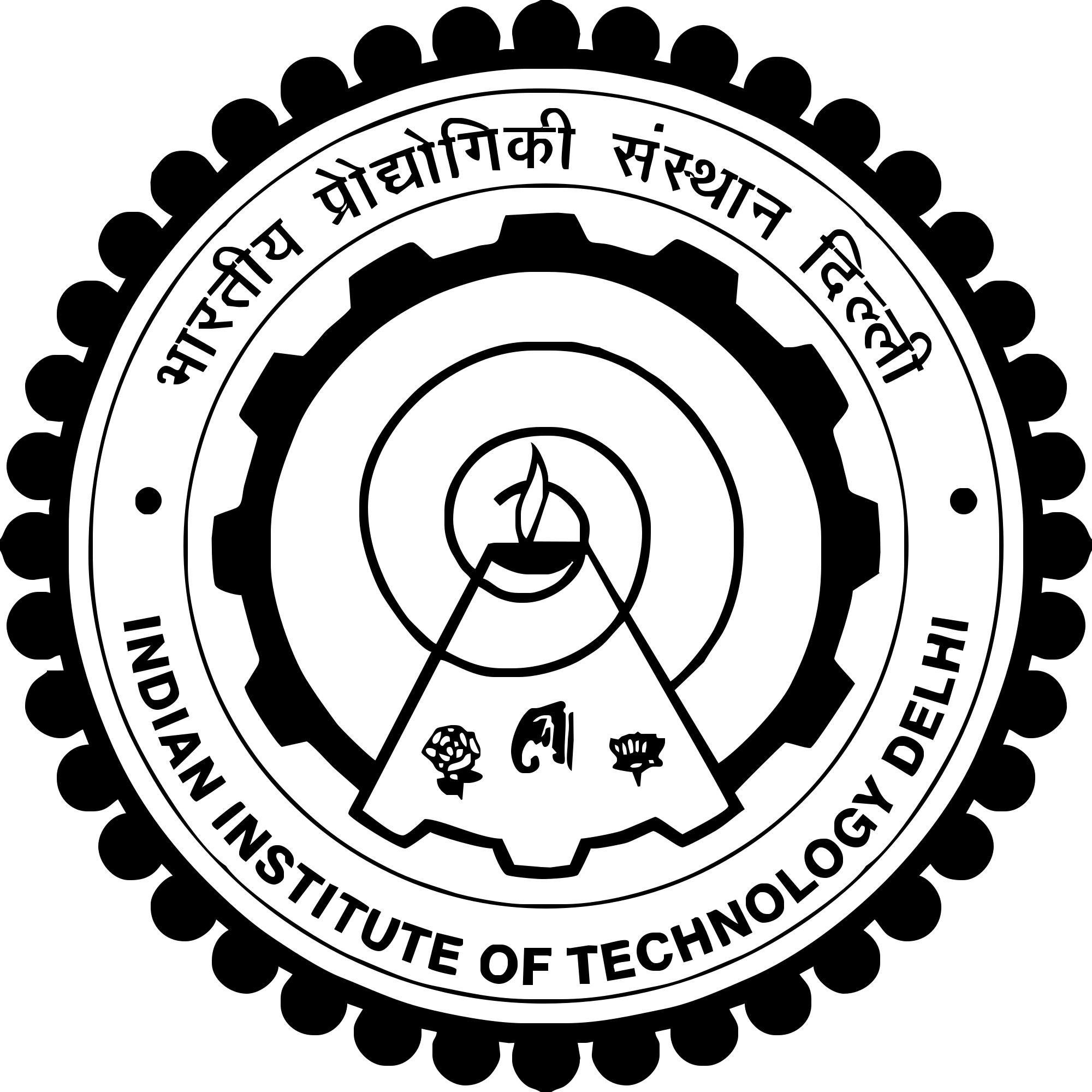}}\par}
	\end{figure}
	\vspace{6em}
	\centering{\large \bf INDIAN INSTITUTE OF TECHNOLOGY DELHI} \\
	\centering  {\large \bf FEBRUARY 2025}
\end{titlepage}

\pagenumbering{roman}



\abstract

\noindent
Recent advances in cognitive and counter-adversarial systems with the ability to infer their adversary's beliefs have garnered significant research interest in inverse filtering from a Bayesian perspective. Such counter-adversarial systems have potential applications where an autonomous system interacts with its environment, learns relevant information about it, and then adapts itself to achieve its goals optimally. An important example is the \textit{inverse cognition} problem in cognitive radar-target dynamics. A cognitive radar adapts its waveform, beam pattern, and other parameters in real-time to accurately localize its target. An intelligent target can employ these inverse filtering results to evade detection and even deceive the adversarial radar. In this setting, a cognitive adversary or `attacker' tracks its target of interest using a stochastic framework such as a Kalman filter (KF). The target or `defender' then employs another \textit{inverse} stochastic filter to infer the forward filter's estimates of the defender's state computed by the attacker. For instance, recently, \textit{inverse KF} (I-KF) has been developed to predict the attacker's future steps by estimating its (forward) KF's estimates. Existing works on inverse stochastic filtering and inverse cognition consider either linear Gaussian or finite state-space models, which limits their applicability in many practical engineering problems. Hence, in this dissertation, we study inverse filtering in the general non-linear non-Gaussian system framework. However, the non-linear Bayesian filtering is an intractable problem that necessitates sub-optimal approximation techniques. In our work, we develop inverse non-linear filters using different sub-optimal filtering approaches. We further study the theoretical performance guarantees for the developed filters. Finally, several numerical experiments demonstrate the estimation performance of our methods using the recursive Cram\'{e}r-Rao lower bound and non-credibility index (NCI) as benchmarks.

In the inverse non-linear filtering context, we first investigate the extended KF (EKF), which uses Taylor series expansion to extend the standard KF to non-linear dynamics. To this end, we develop inverse EKF (I-EKF) for non-linear systems with and without unknown exogenous inputs. We then provide theoretical stability guarantees for I-EKF in the exponential-mean-squared-boundedness sense using both bounded non-linearity and unknown matrix approaches. In particular, the inverse filter is stable under mild system-level conditions, provided that the forward filter is stable. Additionally, we prove that I-EKF is a conservative estimator, i.e., I-EKF's estimated error covariance upper bounds its true value. In the process, we also obtain I-KF for linear systems with unknown inputs and provide sufficient conditions for its asymptotic stability. Our I-EKF can be further modified using the maximum correntropy criterion (MCC) to tackle non-Gaussian noises. However, the standard EKF technique has several disadvantages. The EKF, and hence I-EKF, performs poorly when the dynamic system is highly non-linear, is sensitive to initialization, and may even fail completely. These drawbacks of EKF have led to the development of several advanced variants to enhance estimation accuracy, stability, and practical feasibility. Hence, we generalize our I-EKF formulation to highly non-linear models and develop inverse second-order EKF (I-SOEKF), inverse Gaussian-sum EKF (I-GS-EKF), and inverse dithered EKF (I-DEKF). These advanced I-EKF variants overcome some of I-EKF's limitations and provide improved performance, depending on the system. Finally, we derive I-SOEKF's theoretical stability guarantees using the bounded non-linearity approach. Our numerical experiments further suggest that an advanced inverse filter may provide better performance at the cost of computational efforts, even when assuming an incorrect forward filter employed by the attacker. 

Despite its simplicity, EKF suffers from linearization errors in highly non-linear applications, which can be efficiently handled with derivative-free sigma-point KFs (SPKFs). In contrast to EKF, SPKFs use function evaluation at a deterministic set of sigma points to approximate the statistics of a random variable under non-linear transformation. In light of this, we explore the unscented transform method or, equivalently, statistical linearization technique and develop inverse unscented KF (I-UKF). However, practical systems often involve continuous-time state evolution or complex-valued states and observations. Hence, we generalize our I-UKF theory to obtain continuous-discrete and complex I-UKFs for inverse filtering applications. We also investigate enhanced SPKFs based on efficient numerical integration techniques and develop inverse cubature KF (I-CKF), inverse quadrature KF (I-QKF), and inverse cubature-quadrature KF (I-CQKF). Finally, we derive sufficient conditions for the stochastic stability of the aforementioned inverse SPKFs using the unknown matrix approach. In the process, we also obtain hitherto unreported general stability results for forward UKF and CKF. Inverse SPKFs, like I-EKFs, have been shown to be conservative estimators.

Our inverse EKFs and SPKFs adopt the inverse filtering framework of I-KF and assume that both the attacker and defender agents have perfect knowledge of the system model. However, in practice, agents employing stochastic filters often lack prior system model information. In this context, we address the unknown system model case by developing reproducing kernel Hilbert space (RKHS)-based EKF and UKF to learn system parameters and estimate the state simultaneously. Our RKHS-based filters can be employed by both the attacker and defender to estimate the defender's state and the attacker's state estimate, respectively. While both RKHS-based filters adopt an online approximate Expectation Maximization (EM) algorithm to learn the system parameters, the expectations under non-linear transformations in RKHS-EKF and RKHS-UKF are computed using Taylor series-based linearization and the unscented transform, respectively. Our numerical experiments further demonstrate that RKHS-UKF outperforms RKHS-EKF in terms of estimation accuracy. Finally, we provide theoretical stability guarantees for the developed RKHS-based filters using the unknown matrix approach. In our RKHS-based KFs, the recursive Bayesian state estimation framework is extended to unknown systems by coupling an additional EM-based parameter update step with the EKF/UKF recursions. As a result, we are able to examine their stochastic stability, which is not possible with other kernel-based methods reported in the literature.  

The sub-optimal non-linear filtering approaches discussed so far assume additive Gaussian system models and/or locally approximate the non-linear dynamics at the state estimates. Consequently, these methods are difficult to implement in many practical scenarios. Hence, we also consider the global but computationally expensive sequential Monte Carlo (SMC) filtering approach to develop inverse filters for general non-linear and non-Gaussian systems. In this context, we first develop an inverse particle filter (I-PF) that assumes perfect system model information. Unlike EKF and SPKFs, which assume a Gaussian posterior density, the PF framework uses MC methods to approximate arbitrary posterior distributions. However, the particles in PFs are not statistically independent, and standard convergence results for MC methods based on central limit theorems are not applicable. In this work, we study the convergence of the developed I-PF in the $L^{4}$-sense and show that under mild system-level conditions, our I-PF's estimates converge to the optimal inverse filter's estimates. Despite the fact that the Gaussianity assumption restricts practical applications, it frequently presents a trade-off between accuracy and ease of implementation. Hence, we next explore the application of SMC techniques to Gaussian systems and develop inverse Gaussian PF (I-GPF) and inverse ensemble KF (I-EnKF). GPFs employ sequential importance sampling (SIS) to approximate the mean and covariance of a Gaussian posterior density. The EnKF, on the other hand, is an MC approximation of the standard KF but has been successfully applied to a wide range of extremely high-dimensional and non-linear problems. Our I-GPF and I-EnKF can further efficiently handle non-Gaussian noises with suitable modifications. Finally, we introduce the differentiable I-PF, differentiable I-EnKF, and RKHS-EnKF methods, leveraging SMC techniques to tackle situations where the defender lacks knowledge of the system information.

\pagebreak

\begin{singlespace}
\tableofcontents
\listoffigures
\addcontentsline{toc}{chapter}{List of Figures}
\listoftables
\addcontentsline{toc}{chapter}{List of Tables}

\abbreviations
\noindent
\begin{tabbing}
xxxxxxxxxxxxxxxxxxxxxxxxxxx \= xxxxxxxxxxxxxxxxxxxxxxxxxxxxxxxxxxxxxxxxxxxxxxxxxxxxxxxx \kill
ADF   \> Assumed density filtering\\
ALD  \> Approximate linear dependence\\
CKF/ I-CKF  \> Cubature Kalman filter/ Inverse CKF\\
CQKF/ I-CQKF   \> Cubature quadrature Kalman filter/ Inverse CQKF\\
DEKF/ I-DEKF   \> Dithered Extended Kalman Filter/ Inverse DEKF\\
DF   \> Direct feed-through\\
DPF \> Differentiable Particle filter\\
EKF/ I-EKF   \> Extended Kalman filter/ Inverse EKF\\
EM   \> Expectation Maximization\\
EnKF/ I-EnKF  \> Ensemble Kalman filter/ Inverse EnKF\\
FM  \> Frequency modulation\\
GM/ GS \> Gaussian Mixture/ Gaussian sum\\
GPF/ I-GPF  \> Gaussian Particle filter/ Inverse GPF\\
GS-EKF/ I-GS-EKF   \> Gaussian-sum EKF/ Inverse GS-EKF\\
GSPF/ I-GSPF   \> Gaussian-sum PF/ Inverse GSPF\\
HMMs  \> Hidden Markov Models \\
KF/ I-KF  \> Kalman filter/ Inverse KF\\
KKF  \> Kernel Kalman filter\\
KRLS/ Ex-KRLS  \> Kernel recursive least squares/ Extended KRLS\\
MC/ MCMC   \> Monte Carlo/ Markov chain MC\\
MCC \> Maximum Correntropy criterion\\
ML \> Machine learning\\
MSE/ MMSE \> Mean-squared error/ Minimum MSE\\
NCI   \> Non-credibility index\\
NN/ RNN \> Neural Network/ Recurrent NN\\
PF/ I-PF  \> Particle filter/ Inverse PF\\
P-MCMC \> Particle MCMC\\
QKF/ I-QKF   \> Quadrature Kalman filter/ Inverse QKF\\
RCRLB   \> Recursive Cramer-Rao lower bound\\
RKHS   \> Reproducing kernel Hilbert space\\
RKHS-EKF/UKF/EnKF  \> RKHS-based EKF/UKF/EnKF\\
RL/ IRL \> Reinforcement learning/ Inverse RL\\
RMSE/ AMSE   \> Root mean-squared error/ Time-averaged RMSE\\
SIS   \> Sequential Importance sampling\\
SLT   \> Statistical Linearization technique\\
SMC   \> Sequential Monte Carlo\\
SOEKF/ I-SOEKF   \> Second-order EKF/ Inverse SOEKF\\
SPKF   \> Sigma-point Kalman filter\\
UKF/ I-UKF   \> Unscented Kalman filter/ Inverse UKF\\
\end{tabbing}

\chapter*{\centering Notations}
\addcontentsline{toc}{chapter}{Notations}
\noindent
\begin{tabbing}
xxxxxxxxxxxxxxxxxx \= xxxxxxxxxxxxxxxxxxxxxxxxxxxxxxxxxx \kill
$\textbf{Parameter/}$ \> $\textbf{Definition}$\\
$\textbf{Function}$ \>\\
\\
$k$ \> Discrete-time index\\
$\mathbf{x}_{k}$   \> Defender's true state ($n_{x}$-dimensional vector) at $k$-th time instant\\
$\mathbf{u}_{k}$   \> Unknown exogenous control input ($n_{u}$-dimensional vector) at $k$-th time instant\\
$\mathbf{y}_{k}$   \> Attacker's observation ($n_{y}$-dimensional vector) of $\mathbf{x}_{k}$ at $k$-th time instant\\
$\hat{\mathbf{x}}_{k}$   \> Attacker's estimate of $\mathbf{x}_{k}$ given observations $\{\mathbf{y}_{j}\}_{1\leq j\leq k}$ computed via\\
\> a forward filter\\
$\hat{\mathbf{u}}_{k}$ \> Attacker's estimate of the unknown input via forward filter\\
$\bm{\Sigma}_{k}$   \> Forward filter's estimate of error covariance for estimate $\hat{\mathbf{x}}_{k}$\\
$\mathbf{a}_{k}$   \> Defender's observation ($n_{a}$-dimensional vector) of $\hat{\mathbf{x}}_{k}$ at $k$-th time instant\\
$\doublehat{\mathbf{x}}_{k}$   \> Defender's estimate of $\hat{\mathbf{x}}_{k}$ given observations $\{\mathbf{a}_{j}\}_{1\leq j\leq k}$ and true states\\
$\doublehat{\mathbf{u}}_{k}$ \> Defender's estimate of $\hat{\mathbf{u}}_{k}$ via inverse filter\\
\> $\{\mathbf{x}_{j}\}_{0\leq j\leq k}$ computed via an inverse filter\\
$\overline{\bm{\Sigma}}_{k}$   \> Inverse filter's estimate of error covariance for estimate $\doublehat{\mathbf{x}}_{k}$\\
$[\mathbf{a}]_{i}/\;[\mathbf{A}]_{i,j}$ \> The $i$-th component of vector $\mathbf{a}$/ $(i,j)$-th component of matrix $\mathbf{A}$\\
$[\mathbf{A}]_{(i_{1}:i_{2},j_{1}:j_{2})}$ \> The sub-matrix of $\mathbf{A}$ consisting of rows $i_{1}$ to $i_{2}$ and columns $j_{1}$ to $j_{2}$\\
$[\mathbf{A}]_{(i,:)}/\;[\mathbf{A}]_{(:,j)}$ \> The $i$-th row/ $j$-th column of matrix $\mathbf{A}$\\
$(\cdot)^{T/H}$   \> Transpose/ Hermitian operation\\
$\|\cdot\|_{2}/\;\|\cdot\|_{\mathbf{A}}$ \> $l_{2}$ norm/ norm with respect to matrix $\mathbf{A}$ of a vector, i.e., $\mathbf{x}^{T}\mathbf{A}\mathbf{x}$\\
$\|\mathbf{x}\|_{\infty}$ \> $l_{\infty}$ norm of vector $\mathbf{x}$\\
$\textrm{Tr}/\textrm{det}/\textrm{rank}(\mathbf{A})$\> Trace/ Determinant/Rank of matrix $\mathbf{A}$\\
$\textrm{diag}(a_{1},a_{2},\hdots,a_{l})$\> A `$l\times l$' diagonal matrix with diagonal elements $a_{1},a_{2},\hdots,a_{l}$\\
$\|\mathbf{A}\|/\;\|\mathbf{A}\|_{\infty}$ \> Spectral/ maximum row sum norm of matrix $\mathbf{A}$\\
$\mathbf{A}\preceq\mathbf{B}$ \> The matrix $\mathbf{B}-\mathbf{A}$ is a positive semidefinite (p.s.d.) matrix\\
$\sqrt{\mathbf{A}}$ \> Square-root of matrix $\mathbf{A}$ such that $\mathbf{A}=\sqrt{\mathbf{A}}\sqrt{\mathbf{A}}^{T}$\\
$\nabla f/\;\frac{\partial f}{\partial\mathbf{x}}$ \> The $\mathbb{R}^{m \times n}$ Jacobian matrix for a function $f:\mathbb{R}^{n}\rightarrow\mathbb{R}^{m}$ or\\
\> $\mathbb{R}^{n \times 1}$ gradient vector for $f:\mathbb{R}^{n}\rightarrow\mathbb{R}$\\
$\nabla^{2}f/\;\frac{\partial^{2} f}{\partial\mathbf{x}^{2}}$ \> The $\mathbb{R}^{n\times n}$ Hessian matrix for a function $f:\mathbb{R}^{n}\rightarrow\mathbb{R}$.\\
$\|f\|_{\infty}$ \> The supremum norm of the real-valued function $f(\cdot)$\\
$\delta(x-x_{0})$ \> The Dirac-delta function in variable $x$ centered at $x_{0}$\\
$\mathbf{I}_{n}/\;\mathbf{0}_{n\times m}$ \> $n\times n$ identity matrix/ $n\times m$ all zero matrix\\
$\mathbf{I}/\;\mathbf{0}$ \> Identity matrix/ all zero matrix of appropriate size\\
i.i.d. \> Independent and identically distributed\\
p.d./ p.s.d. \> Positive definite/ positive semi-definite\\
$\mathbb{E}[\cdot]$ \> Expectation operation\\
$\mathcal{N}(\bm{\mu},\mathbf{Q})/\;\mathcal{N}(\mathbf{x};\bm{\mu},\mathbf{Q})$ \> Gaussian distribution with mean $\bm{\mu}$ and covariance matrix $\mathbf{Q}$\\
$\mathcal{U}[u_l,u_u]$ \> The uniform distribution over interval $[u_l,u_u]$\\
$\textrm{Cov}(\mathbf{x})/\;\textrm{Cov}(\mathbf{x},\mathbf{y})$ \> Covariance/ cross-covariance matrix for random variables $\mathbf{x}$ and $\mathbf{y}$
\end{tabbing}

\end{singlespace}

\newpage
\pagenumbering{arabic}


\chapter{Introduction}
\label{chap:intro}
Inference and control form an integral part of many dynamic systems in various engineering applications, including navigation\cite{simon2006optimal}, guidance \cite{bar2004estimation}, and radar target tracking \cite{bell2015cognitive}. Often, these applications, such as communications, sensing, and robotics, involve intelligent and autonomous \textit{cognitive} agents that sense the environment, learn relevant information about it, and then adapt their actions in real-time to achieve optimal performance. For instance, a cognitive radar \cite{mishra2020toward} adapts its transmit waveform and receive processing to improve target detection \cite{mishra2023next} and tracking \cite{bell2015cognitive,sharaga2015optimal}. However, in many applications, the target is also intelligent and might want to assist or desist the estimation by intelligently taking control actions, for which it requires an estimate of the estimate made by the radar (or any other sensor). For example, the estimate made by an anti-aircraft gun using a radar can be inferred by a fighter airplane and used to perform control maneuvers to avoid being hit. In this context, \textit{inverse cognition} has been recently proposed for a \textit{defender} agent to detect its cognitive adversarial agent and infer the information adversary has learned about the defender \cite{krishnamurthy2019how,krishnamurthy2020identifying}. This aids in designing counter-adversarial systems to assist or desist the adversary\cite{mattila2020hmm,krishnamurthy2019how}. Similar examples abound in interactive learning\cite{krishnamurthy2019how}, fault diagnosis, cyber-physical security\cite{mattila2017inverse}, and inverse reinforcement learning\cite{ng2000algorithms}. In this dissertation, we focus on inverse stochastic filtering for these inverse cognition applications.

Stating in the most general terms, the inverse cognition problem involves two agents: a `\textit{defender}' (e.g., an intelligent target) and an `\textit{attacker}' (e.g., a sensor or radar) equipped with a Bayesian tracker (forward filter). The attacker infers an estimate of the defender's kinematic state and cognitively adapts its actions based on this estimate. The defender observes the attacker's actions with the goal of predicting its future actions in a Bayesian sense. To this end, the defender requires an estimate of the attacker's inference. In this context, \textit{inverse Bayesian filtering}\cite{krishnamurthy2019how} has been proposed to infer the attacker's inference at the defender's end. In (forward) Bayesian filtering, we obtain a posterior distribution of the underlying state given noisy observations of the state. A
common example is the Kalman filter (KF), which recursively estimates the state in linear systems with Gaussian process and measurement noises and
is optimal in the minimum mean-squared error (MMSE) sense. The inverse filtering problem, on the other hand, is concerned with estimating this posterior distribution of the forward Bayesian filter given the noisy measurements of the attacker's actions. An example of such a system is the recently introduced inverse KF (I-KF)\cite{krishnamurthy2019how} for linear Gaussian systems. In this work, we develop inverse filters to handle general non-linear and non-Gaussian attacker-defender dynamics. Given the intractability of non-linear Bayesian filtering, we explore various sub-optimal approximation methods for inverse non-linear filtering. Along with developing these algorithms, we examine their theoretical performance guarantees and provide numerical demonstrations of their estimation accuracy across several different systems.

In the following, we first review the existing literature on inverse filtering and inverse cognition in Section~\ref{sec:inverse cognition}. Section~\ref{sec:non-linear filtering} then briefly overviews some of the widely used non-linear Bayesian filtering techniques, which are subsequently considered to develop inverse filters in our work. In Section~\ref{sec:contributions}, we list the key contributions of our research while Section~\ref{sec:organization} briefly outlines the organization of the thesis.

\section{Inverse filtering and Inverse Cognition}\label{sec:inverse cognition}
In many engineering applications, it is desired to infer the parameters of a filtering system by observing its output. Conventionally, this inverse filtering was limited to non-dynamic systems for applications such as system identification, fault detection, image deblurring, and signal deconvolution \cite{idier2013bayesian,gustafsson2007statistical,wipf2014deconvolution}. However, recent cognitive and counter-adversarial systems applications \cite{haykin2006cognitive,bell2015cognitive,mattila2020hmm} have motivated the design of inverse stochastic filters. These types of inverse problems may be traced to \cite{kalman1964linear} that aimed to find the cost criterion for a given control policy. An analogous formulation also appears in inverse reinforcement learning (IRL)\cite{ng2000algorithms,choi2011irl}. In particular, IRL passively learns the associated reward function by observing the optimal behavior of an expert. In contrast, the inverse cognition agent actively probes its adversarial agent and can thus be regarded as a generalization of IRL. In fact, \cite{krishnamurthy2020identifying} proposed optimally adapting the target's states to probe the radar and minimize its detection probability.

In counter-adversarial applications, it is imperative for the defender agent to first detect its attacker's cognitive behavior. To this end, \cite{krishnamurthy2020identifying} considered the radar-target interaction where the radar cognitively adapts its waveform and beam allocation. Stochastic revealed preferences-based algorithms were developed to ascertain if the adversarial radar's actions are consistent with optimizing a utility function and, if so, estimate that utility function. Autonomous system identification has been further considered in \cite{krishnamurthy2021adversarial,bookman2022autonomous}. On the other hand, \cite{krishnamurthy2021adversarial,kang2023smart} dealt with smart interference design to force the radar to change its transmit waveform using the inverse filtering and cognitive system identification results. In \cite{krishnamurthy2019how}, optimal probe signals are designed to estimate the adversary's sensor characteristics.

After detecting the cognitive attacker, it is essential for the defender to employ an inverse filter to estimate the attacker's inference if it wants to guard against the latter's future actions. Inverse filtering for Hidden Markov models (HMMs) was considered in \cite{mattila2017inverse} to estimate the attacker's observation sequence and observation likelihood given a sequence of posterior state distributions from an HMM with known dynamics. The developed inverse filtering algorithm was then applied for real-time fault diagnosis of an automatic sleep-tracking system. On the other hand, \cite{mattila2020hmm} addressed inverse filtering for HMM by proposing a convex optimization-based solution for reconstructing the state transition kernel and attacker's observation likelihood and observation sequences given measurements of posterior distributions from a Bayesian filter. A sequential stochastic decision framework was considered for the attacker in \cite{mattila2019estimating}. The attacker optimized its local utility and announced its decision to the world. By observing these actions, the inverse filtering agent aimed to estimate the attacker's private belief, i.e., its posterior distribution about the state.

Inverse filtering algorithms were developed for single output linear Gaussian state-space models in \cite{mattila2018inverse} to reconstruct the sensor characteristics and observations given the system model and noisy measurements of the attacker's posterior. Further, in \cite{mattila2020did}, an optimal smoother was designed to estimate all the attacker's past beliefs. In \cite{krishnamurthy2019calibrate} and further in \cite{krishnamurthy2019how}, inverse Kalman filter (I-KF) was proposed to infer the forward KF's tracked estimates. The learning rate of the proposed I-KF was studied in \cite{krishnamurthy2019how} by considering a sequential localization game between the attacker and defender. Given the defender's measurements of the attacker's actions, the maximum likelihood estimate of the attacker's gain matrix was also obtained for the linear Gaussian case. Note that, historically, the Wiener filter - a special case of KF when the process is stationary - has long been used for frequency-domain inverse filtering for deblurring in image processing \cite{biemond1990iterative}.

These prior works model inverse cognition as a linear Gaussian or finite state-space system. In practice, many engineering problems involve non-linear processes\cite{haykin2004kalman,simon2006optimal}. While inverse non-linear filters have been studied for adaptive systems in some previous works \cite{broomhead1996nonlinear,shen2001robust}, the inverse of non-linear stochastic filters remains unexamined so far.

\section{Non-linear filtering}\label{sec:non-linear filtering}
Counter-adversarial applications involve \textit{inference} by both the defender and attacker, wherein they estimate the posterior distributions of an underlying state that cannot be directly observed but is inferred through an observation process conditioned on that state. Posterior distributions provide not only \textit{point estimates}, i.e., single-number estimates such as the mean of the posterior distributions, but also quantify their uncertainty. In sequential state estimation, states are inferred from a sequence of observations, with posteriors updated recursively. When state-space models are available, Bayesian filtering is a probabilistic technique for sequential state estimation that has been extensively applied to visual tracking in computer vision \cite{zhang2017multi,dai2019visual}, localization in robotics \cite{ullah2019localization,fox1999monte}, and other signal processing problems \cite{yousefi2019efficient,liu2019audio}.

The non-linear Bayesian filters can be broadly categorized into local and global approximation approaches. The local filters, also known as Gaussian filters, assume a Gaussian posterior density and approximate the non-linear system model and the posterior statistics locally at the state estimates\cite{li2017approximate}. These Gaussian filtering approaches are further connected to the general approximate Bayesian inference approaches in machine learning (ML). In particular, Gaussian filters are a special case of assumed density filtering (ADF)\cite{maybeck1982stochastic} or online Bayesian learning\cite{opper1999bayesian}, which sequentially computes the approximate posterior distribution of the underlying state. Expectation propagation is a further extension of ADF where new observations are also used to refine the previous approximations iteratively \cite{minka2013expectation}. Contrarily, the global approximations based on sequential Monte Carlo (SMC) methods start with a set of sample points that capture the initial state distribution and then propagate them through the actual non-linear dynamics. An ensemble of these samples then provides the approximate posterior. As a result, these methods are easily applicable to non-Gaussian dynamics and have been successfully applied to mobile robot localization \cite{fox1999monte}, simultaneous localization and mapping (SLAM) \cite{montemerlo2003fastslam}, and planning in partially observable environments \cite{somani2013despot}. In this study, we examine both Gaussian and SMC-based methods to address inverse filtering in general non-linear, non-Gaussian systems.

\subsection{Extended KF and its variants}\label{subsec:EKF literature}
Although KF and its continuous-time variant Kalman-Bucy filter \cite{kalman1961new} are highly effective in many practical applications, they are optimal for only linear and Gaussian models. In non-linear cases, a linearized KF is used, wherein the states of a linear system represent the deviations from a nominal trajectory of a non-linear system. The KF estimates the deviations from the nominal trajectory and obtains an estimate of the states of the non-linear system. The linearized KF is extended to directly estimate the states of a non-linear system in the extended KF (EKF) \cite{schmidt1966application}. In EKF, the linearization is locally at the state estimates through Taylor series expansion. This is very similar to the Volterra series filters  \cite{zaknich2005principles} that are non-linear counterparts of adaptive linear filters. Besides the traditional state estimation applications, EKF has also been considered in learning applications like dual and joint estimation of state and parameters\cite{haykin2004kalman}, parameter optimization for fuzzy logic systems\cite{khanesar2011extended}, and training neural networks\cite{simon2002training,wang2011convergence}.

In spite of its ease of implementation, the standard EKF has its limitations. Unlike KF, EKF and its (filter) gain equations are not decoupled; hence, offline computations are infeasible. The EKF performs poorly when the dynamic system is significantly non-linear\cite{li2017approximate}. It is also very sensitive to initialization/ modeling errors and may even completely fail \cite{tenney1977tracking}. Recently, KF's initialization has been studied rigorously in \cite{zhao2020trial}, while \cite{salvoldi2018process} deals with process noise covariance design. However, they do not consider the non-linear EKF case. These EKF drawbacks have led to the development of several advanced variants, each aiming toward improving either the estimation accuracy (higher-order EKFs \cite{jazwinski2007stochastic,wang2019second}, Gaussian-sum EKF \cite{tam1977gaussian}), convergence (iterated EKF \cite{wishner1969comparison}), stability (dithered EKF \cite{weiss1980improved}) or practical feasibility (hybrid EKF\cite{simon2006optimal}). The literature also reports grid-based EKF \cite{bucy1971digital} that is a precursor of the particle filter. These filters reduce the linearization errors that are inherent to the EKF and may provide an improved estimation at the cost of higher complexity and computations. A comparison of non-linear KFs is available in \cite{schwartz1968computational,netto1978optimal}.

\subsection{Sigma-point KFs}\label{subsec:SPKF literature}
Despite being a widely used non-linear filter, EKF suffers from linearization errors in highly non-linear applications\cite{daum2005nonlinear}. In many practical applications, the computation of the Jacobian matrices required for EKF is also non-trivial. These drawbacks can be efficiently tackled through derivative-free sigma-point KFs (SPKFs)\cite{bhaumik2019nonlinear}, which are based on a weighted sum of function evaluations at a finite number of deterministic \textit{sigma-points}. These points aim to approximate the posterior probability density of a random variable under non-linear transformation. While EKF is applicable only to differentiable functions, SPKFs handle discontinuities. For example, the unscented KF (UKF) \cite{julier2004unscented} draws points using the unscented transform and delivers estimates that are exact in mean for monomials up to third-degree; the covariance computation is exact only for linear functions. The basic intuition of unscented transform is that it is easier to approximate a probability distribution than it is to approximate an arbitrary non-linear function \cite{julier2004unscented}. EKF, on the other hand, considers a linear approximation for the non-linear functions. The unscented transform is further equivalent to statistically linearizing the non-linear functions using specific regression points, and UKF can be viewed as a special case of linear regression KF\cite{lefebvre2002comment}. To this end, unlike EKF, the UKF takes into account the increased uncertainty due to linearization errors, providing enhanced performance in many applications. Square-root UKF\cite{van2001square} and Gaussian-sum UKF\cite{kottakki2014state} have also been proposed for, respectively, enhanced numerical stability and non-Gaussian models. 

The SPKF performance is further improved by employing better numerical integration techniques to calculate the recursive integrals in Bayesian estimation. For example, cubature KF (CKF) \cite{arasaratnam2009cubature} and quadrature KF (QKF) \cite{ito2000gaussian,arasaratnam2007qkf} numerically approximate the multidimensional integral based on, respectively, cubature and Gauss-Hermite quadrature rules. The cubature-quadrature KF (CQKF)\cite{bhaumik2013cubature} uses the cubature and quadrature rules together while central difference KF\cite{ito2000gaussian} considers polynomial interpolation methods with central difference approximation of the derivatives. Analogous to EKFs, several advanced SPKFs have also been developed for improving either estimation accuracy (higher-degree CKFs\cite{jia2013high}) or numerical stability (square-root CKF\cite{arasaratnam2009cubature}). On the other hand, Gaussian-sum QKF (GS-QKF)\cite{kottakki2014state,arasaratnam2007qkf} handles non-Gaussian system models by representing the posterior density as a weighted sum of finite Gaussians. However, GS filters are computationally expensive, and hence, adaptive GS filters have been proposed\cite{bhaumik2019nonlinear}, wherein the Gaussian components are adapted or reduced online for efficient representation of the posterior density.

\subsection{Sequential Monte-Carlo filters}\label{subsec:SMC literature}
KF and its non-linear extensions, such as EKF and UKF, being Gaussian filters, assume Gaussian state posteriors and are only applicable to systems with Gaussian noises \cite{li2017approximate}. While they provide closed-form analytic solutions for posteriors, their appropriateness varies by application, often failing with highly non-linear models or multi-modal posteriors \cite{kotecha2003gaussian,cappe2007overview}. Techniques like Gaussian mixture (GM) \cite{alspach1972nonlinear} or grid-based \cite{kramer1988recursive} filters, proposed to mitigate these limitations, become computationally expensive in high-dimensional systems. The curse of dimensionality also affects deterministic numerical integration methods like CKF and QKF, making them difficult to implement for high-dimensional states, with the rate of convergence of the approximation error decreasing as the state dimension increases \cite{crisan2002survey}. On the other hand, practical engineering systems are frequently non-linear and non-Gaussian, which restricts the use of Gaussian approximations. Non-linear filtering methods that do not rely on this assumption use sequential importance sampling (SIS) and resampling \cite{gordon1993novel}, leading to SMC or particle filtering (PF) methods \cite{cappe2007overview}. SMC methods employ random sampling to achieve asymptotically exact integral computation, though with higher computational complexity. PFs \cite{ristic2003beyond,arulampalam2002tutorial} are the most general and widely used SMC-based filters. Additionally, quasi-MC methods \cite{guo2006quasi,avron2016quasi} are deterministic alternatives to MC methods, using regularly distributed points rather than random ones to approximate the posterior.

Following \cite{gordon1993novel}'s bootstrap PF formulation, several variants of PFs have been developed for enhanced performance. Auxiliary PFs \cite{pitt1999filtering,branchini2021optimized} direct particles to higher-density regions of the posterior distribution, whereas unscented PFs \cite{van2000unscented_pf,rui2001better} include the current observation into the proposal distribution using UKF. In \cite{murphy2001rao}, Rao-Blackwellised PF is proposed for systems wherein the state can be partitioned such that the posterior distribution of one part is tractably conditioned on the other. \cite{kurle2020deep} further generalizes Rao-Blackwellised PFs to switching linear Gaussian systems. On the other hand, high-dimensional systems can be efficiently handled by multiple PF \cite{djuric2007multiple} and particle flow PFs \cite{bunch2016approximations,li2017particle}. In \cite{vermaak2003maintaining}, multiple targets are tracked using mixture PF models. The PF framework has also been used to implement Bernoulli filters \cite{ristic2013tutorial} (for randomly switching systems), possibility PFs \cite{ristic2019robust} (for mismatched models), and probability hypothesis density (PHD) filters \cite{mahler2003multitarget} (for high-dimensional multi-object Bayesian inference).

Despite the fact that the Gaussianity assumption limits practical applications, it frequently presents a trade-off between accuracy and ease of implementation. Hence, SMC approaches have also been used in Gaussian systems. Gaussian PF (GPF) \cite{kotecha2003gaussian} assumes a Gaussian posterior and approximates its mean and covariance using the PF framework. In particular, the posterior mean and covariance estimates are computed from a set of randomly generated particles with associated weights based on SIS. GPF can be further viewed as an extension of the conventional Gaussian filters using the MC integration and Bayesian update rules \cite{wu2005comments_gpf}. In \cite{kotecha2003gaussian_sum}, Gaussian-sum PFs (GSPFs) are developed for non-Gaussian systems. On the other hand, ensemble KF (EnKF) \cite{evensen2003ensemble,katzfuss2016understandingEnKf} is an MC extension of standard KF. EnKF approximates the state distribution by storing, propagating, and updating an ensemble of state vectors \cite{evensen2003ensemble}. While the standard KF works with the entire state distribution explicitly, EnKF can be viewed as a form of dimension reduction wherein a small ensemble is propagated instead of the joint distribution with full covariance matrix \cite{katzfuss2016understandingEnKf}. In particular, EnKF, when combined with localization or covariance tapering techniques  \cite{furrer2007estimation,houtekamer2001sequential,furrer2006covariance}, leads to computationally tractable filters for very high-dimensional systems. Mixture KF \cite{chen2000mixture}, and MC-KF \cite{song2000monte} are two other examples that use latent variable representation to extend the KF framework to conditional linear Gaussian systems and discrete models, respectively. In \cite{stordal2011bridging}, a hybrid PF-EnKF is also proposed that combines local KF-type update with PF's weighting/resampling.

\subsection{Unknown system dynamics}\label{subsec:unknown system literature}
Classical Bayesian filters assume known state-space models, but in many real-world applications, the state evolution and observation models contain unknown parameters that need to be estimated. The uncertainty in system parameters or noise statistics may result in large estimation errors or even filter divergence \cite{vila2021robust,ge2016performance}. Recent studies \cite{vila2021robust,huang2017novel,zhu2021adaptive,yi2021robust,mohamed2011robust} proposed various adaptive filters to reduce the sensitivity of KF to system uncertainties. However, the general non-linear filtering problem is not considered. Among prior works, for non-linear filtering with unknown system models, \cite{liu2009extended} developed the extended kernel recursive least squares (Ex-KRLS) method by transforming the input space to a kernel-based feature space. The KRLS\cite{engel2004kernel} was coupled with EKF to learn the unknown non-linear measurement model in \cite{zhu2011extended,singh2022rkhs}, but with a known linear state-transition. In \cite{zhu2013learning}, the KF algorithm was reformulated for non-linear state transition functions in reproducing kernel Hilbert space (RKHS) using a conditional embedding operator, but linear observations were assumed. For linear Gaussian state-space models, an iterative expectation maximization (EM)-based parameter learning was proposed in \cite{wen2012data}.

PFs also rely heavily on accurate system dynamics and effective proposal densities. The simplest method for parameter estimation is augmenting the state with the unknown parameters, but this often results in poor parameter space exploration due to the pseudo-dynamic nature of the parameters \cite{ionides2006inference}. A widely used PF-based Bayesian approach to parameter estimation is Particle Markov chain Monte Carlo (P-MCMC) \cite{andrieu2010particle,lindsten2014particle}, which combines SMC with MCMC sampling. P-MCMC is a likelihood-based batch-processing method using PF to obtain an unbiased estimate of data log-likelihood. Alternatively, SMC$^2$ estimates both parameters and latent states simultaneously with two layers of PF \cite{chopin2013smc2}, while \cite{singh2023particle} uses Fisher's identity to approximate the log-likelihood gradient with PF. Variational inference \cite{bishop2006pattern}, another approach, employs parametric distributions to approximate the state posterior and jointly optimizes them with model parameters using a lower bound on the log-likelihood. Variational SMC methods \cite{maddison2017filtering,naesseth2018variational} use PF to construct this lower bound.

While the aforementioned methods work well when the system dynamics has a state-space representation, sometimes building probabilistic models itself is infeasible. For instance, in robot localization with an onboard camera, the observation model is a probabilistic distribution over all possible camera images conditioned on a continuous robot state and given environment. Such an enormous observation space can not be represented using state-space models. Hence, ML techniques such as neural networks (NNs) and gradient descent have been suggested for learning both the model and PF's proposal distribution, resulting in differentiable PFs (DPFs) \cite{karkus2018particle,wen2021end,chen2021differentiable,corenflos2021differentiable}. Incorporating NN into Bayesian filters, particularly PFs, provides flexibility in model learning. \cite{haarnoja2016backprop} further proposed differentiable KF with Gaussian belief and end-to-end learnable measurement model whereas \cite{chen2022autodifferentiable} developed differentiable EnKF. \cite{karl2016deep,watter2015embed} learn the latent state space model from raw images for the long-term sequence prediction. In \cite{okuma2004boosted}, mixture PF is integrated with Adaboost to learn proposal distributions for automatic multiple object tracking. In an IRL scenario \cite{samejima2003estimating}, PF is employed to estimate the internal parameters of an RL agent given a sequence of observable variables.

\section{Motivation and Contributions}\label{sec:contributions}
As remarked in Section~\ref{sec:inverse cognition}, the prior works on inverse filtering only consider finite state-space or linear Gaussian system models, which limits their practical applications. Furthermore, these prior works assume perfect system model information on both the attacker's and defender's side. The defender also assumes a known forward filter employed by the attacker. However, in practice, the defender may not have information about the attacker's forward filter and the strategy employed by the attacker to adapt its actions. Similarly, the attacker may also lack complete system information, including the defender's state evolution process or the noise distributions. In such situations, the performance and applicability of inverse filters are limited.

Addressing the aforementioned limitations, we first develop inverse non-linear filters for scenarios with perfect system information, utilizing both EKF and SPKF-based Gaussian filtering techniques. In the process, we also develop inverse filters for systems with unknown exogenous inputs. For systems with unknown parameters, we develop RKHS-based EKF and UKF. We then explore SMC-based approaches to handle inverse filtering in a general non-linear, non-Gaussian framework and also generalize these methods for cases without any prior system model information. Besides formulating the inverse filtering algorithms, we also rigorously analyze their theoretical performance guarantees and demonstrate their estimation performance through numerical simulations using the recursive Cram\'{e}r-Rao lower bound (RCRLB) \cite{tichavsky1998posterior} and non-credibility index (NCI) \cite{li2001practical} as performance metrics. In particular, our theoretical analyses show that the forward filter’s stability is sufficient to guarantee the same for the inverse filter under mild conditions imposed on the system. However, it should be noted that a non-linear filter's performance also depends on the system itself. In practice, selecting the most appropriate filter for a given application typically involves a trade-off between estimation accuracy and computational efforts \cite{li2017approximate}. The same also holds true for inverse non-linear filters.

The key contributions of this dissertation are summarized as follows.

\begin{enumerate}
    \item \textbf{Inverse EKF:} Based on the Taylor series linearization, we introduce and develop the inverse EKF (I-EKF) to infer the attacker's estimate in non-linear systems with additive Gaussian noises. The attacker is assumed to employ EKF as the forward Bayesian tracker to infer the defender's state. We then derive sufficient conditions for the stochastic stability of the proposed I-EKF using the bounded non-linearity\cite{reif1999stochastic} and unknown matrix\cite{xiong2006performance_ukf} techniques. In the process, we also obtain novel stability results for forward EKF based on the unknown matrix approach. We further show that I-EKF is a conservative estimator, i.e., its estimated error covariance upper bounds the true one. Our developed I-EKF can also be suitably modified based on the maximum correntropy criterion (MCC) to handle non-Gaussian noises.
    
    Note that the I-EKF is different from the \textit{inversion of EKF} \cite{zhengyu2021iterated}, which may not take the same form as EKF, is employed on the attacker's side, and is unrelated to our inverse cognition problem. Similarly, the non-linear extended information filter (EIF) proposed in \cite{mutambara1999information} used the inverse of the covariance matrix and was compared with KF for estimation of the same states. Our I-EKF has a different formulation focused on estimating the inference of an attacker that uses an EKF to estimate the defender's state.
    \item \textbf{Inverses of EKF variants:} Among various advanced EKF variants for highly non-linear systems discussed in Section~\ref{subsec:EKF literature}, we focus on the inverses of second-order EKF (SOEKF), Gaussian-sum EKF (GS-EKF), and dithered EKF (DEKF). We develop inverse SOEKF (I-SOEKF) and provide sufficient conditions for its stability. In the process, we obtain novel stability conditions for forward SOEKF as well and derive the one-step prediction formulation of SOEKF, which is analytically more useful while deriving the stability conditions of the filter. The presence of second-order terms poses additional challenges in deriving the theoretical stability guarantees for both forward and inverse SOEKF. On the other hand, GS-EKF comprises of several individual EKFs \cite{tam1977gaussian,alspach1972nonlinear}. Here, the \textit{a posteriori} density function of the state given the observations is approximated by a sum of Gaussian density functions, each of which has its own separate EKF. GS-EKF and hence, its inverse, not only provides enhanced estimation accuracy but is also a popular method to handle non-Gaussianity in system models. In situations where the estimation error is small, the \textit{a posteriori} density is approximated adequately by one Gaussian density, and the Gaussian sum filter reduces to the EKF; the same is true for their inverses. Finally, we also examine the inverse of DEKF \cite{weiss1980improved}, which has been shown to perform better when non-linearities are cone-bounded. This filter introduces dither signals prior to the non-linearities to tighten the cone-bounds and hence, improve stability properties.
    \item \textbf{I-KFs and I-EKFs with unknown inputs:} In the inverse cognition scenario, the target may introduce additional motion or jamming that is known to the target but not to the adversarial cognitive sensor. In this context, we consider a more general non-linear system model with unknown exogenous inputs that affect the state transition and observations but are not known to the attacker. We develop I-EKFs with unknown inputs for both with and without \textit{direct feed-through} cases. In the process, we also obtain I-KFs-with-unknown-inputs that were not examined in the I-KF developed in \cite{krishnamurthy2019how} and provide sufficient conditions for their asymptotic stability. For systems with unknown inputs, the attacker's state estimate depends on its estimate of the unknown input. As a result, the attacker's forward filters vary with system models. Depending on the state transitions for the inverse filter, we consider different augmented states so that the unknown input estimation is performed jointly with state estimation.
    \item \textbf{Inverse SPKFs:} To address the limitations of I-EKF's linearization, we first consider the unscented transform and develop inverse UKF (I-UKF). Our I-UKF estimates an attacker's inference, who also deploys a forward UKF. The inverse filter is formulated using augmented states to take into account the non-additive process noise terms in the forward UKF’s state estimate evolution. We further provide the stability conditions for I-UKF based on the unknown matrix approach. In the process, we also obtain hitherto unreported general stability results for forward UKF. We then prove that I-UKF's recursive estimates are also conservative. We again remark that the I-UKF is different from the \textit{inversion of UKF} \cite{zhengyu2021iterated}, which estimates the input based on the output. Clearly, such an inversion of UKF may not take the same mathematical form because the UKF is employed on the attacker's side. Hence, this formulation is unrelated to our inverse cognition problem. Furthermore, our proposed I-UKF does not follow trivially from I-KF\cite{krishnamurthy2019how} or I-EKF; the differences are highlighted in Remark~\ref{remark:I-UKF with I-KF and I-EKF}.
    
    We then extend our developed I-UKF theory to propose inverse filters based on different efficient numerical integration techniques. In particular, we develop inverse CKF (I-CKF), inverse QKF (I-QKF), and inverse CQKF (I-CQKF) based on the cubature and quadrature rules. Our theoretical analyses also provide improved stability results, hitherto unreported in the literature, for the forward CKF. In practice, the systems often involve continuous-time state evolution or are complex-valued, requiring suitable continuous-discrete or complex filters. Hence, we also generalize our I-UKF theory to obtain continuous-discrete and complex I-UKFs for inverse filtering applications. Similar to I-EKF, we also discuss MCC-based modification of I-UKF for non-Gaussian systems.
    \item \textbf{RKHS-EKF and RKHS-UKF:} We examine the scenario where both the state-transition and observation models are unknown to the agent using a stochastic filter in a non-linear system. In this context, we propose RKHS-based EKF and UKF, which may be employed by both the attacker and defender, respectively, to estimate the defender's state and the attacker's state estimate. In particular, the unknown non-linear functions are represented using kernel function approximation. The EKF/UKF provides required state estimates, while an online approximate expectation maximization (EM) algorithm is used to learn the system parameters. The presence of non-linear transformations in expectation computations makes the filter derivation non-trivial. RKHS-EKF and RKHS-UKF, respectively, approximate these expectations using Taylor series linearization and unscented transform. Because of the unscented transform, RKHS-UKF avoids Jacobian computations and is observed to outperform RKHS-EKF in terms of estimation accuracy in our numerical experiments. Using the unknown matrix approach, we further provide the stability conditions for the developed RKHS-based filters.
    \item \textbf{Inverse PF:} Gaussian inverse filters such as I-EKF and I-UKF are not applicable to general non-Gaussian systems. To address this, we develop inverse PF (I-PF) assuming perfect system model information, including a general but known forward filter at the defender's end. At each time instant, our I-PF considers the joint conditional distribution of the attacker's current state estimate and observation, given the defender's knowledge of its own true states and observations of the attacker's actions up to the current instant. Our I-PF seeks to empirically approximate the optimal inverse filter's (joint) posterior and samples the particles from the optimal importance sampling density. This is in contrast to the typical PF, where the optimal density is often unavailable. The known forward filter assumption allows sampling from the optimal density in I-PF, further discussed in Remark~\ref{remark:IPF optimal density}.
    
    In PFs, the particles interact and are not statistically independent, rendering classical convergence results for MC methods, which rely on central limit theorems under i.i.d. assumptions, inapplicable. Despite this, it is essential to study PFs' convergence to the true posterior as they approximate optimal filters. In this work, we examine the convergence of our proposed I-PF in the $L^{4}$-sense. Specifically, we demonstrate that our I-PF's estimates converge to the optimal inverse filter's estimates for bounded observation densities, given that the estimated function grows at a slower rate than the defender's observation density. Moreover, convergence in the $L^{4}$-sense implies almost sure convergence of our I-PF to the optimal inverse filter's posterior.

    \item \textbf{Inverse GPF and EnKF:} We explore SMC approaches for Gaussian approximations in the inverse filtering context and develop inverse GPF (I-GPF) and inverse EnKF (I-EnKF). Note that EnKF implicitly assumes linear Gaussian state-space models. In fact, for linear Gaussian systems, EnKF's estimates converge in probability to KF's estimates as the ensemble size grows \cite{butala2008asymptotic}. Hence, in our work, we develop I-EnKF considering non-linear systems with additive Gaussian noises. On the other hand, I-GPF and I-PF consider general probabilistic system dynamics. While I-GPF, like I-PF, also considers a general forward filter, I-EnKF follows the inverse filtering framework of I-KF \cite{krishnamurthy2019how}, I-EKF, and I-UKF. In particular, I-EnKF assumes a forward EnKF employed by the attacker, while I-KF, I-EKF, and I-UKF, respectively, assume forward KF, EKF, and UKF.
    
    Under certain conditions, GPF and, hence, I-GPF can be applied to non-Gaussian systems as well. Similarly, EnKF shows remarkable robustness to deviations from the Gaussianity assumption and has been successfully applied to many extremely high-dimensional, non-linear, and non-Gaussian problems \cite{katzfuss2016understandingEnKf}. To efficiently handle non-Gaussian and highly non-linear systems, maximum correntropy and maximum likelihood EnKFs have also been proposed in \cite{tao2023maximum} and \cite{zupanski2005maximum}, respectively. Hence, we also generalize our I-GPF and I-EnKF formulations to non-Gaussian systems. To this end, we consider Gaussian mixture (GM) \cite{kotecha2003gaussian_sum} and MCC \cite{izanloo2016kalman} to modify I-GPF and I-EnKF and obtain inverse GSPF (I-GSPF) and MCC-modified I-EnKF, respectively. Exploring SMC methods for unknown system dynamics, we propose differentiable I-PF, differentiable I-EnKF, and RKHS-based EnKF (RKHS-EnKF). Recall that DPF and differentiable EnKF are differentiable implementations of PF and EnKF, respectively, that exploit NNs to learn the system dynamics. On the other hand, RKHS-EnKF considers non-linear state-space dynamics with the state-transition and observation models unknown to the agent employing the stochastic filter.
\end{enumerate}

Notably, our inverse filters, except RKHS-based and differentiable filters, assume a known attacker's forward filter. If the defender is not certain about the attacker's forward filter, the filters capable of handling unknown system dynamics are helpful in estimating the attacker's state estimate. However, when the defender employs an inverse filter assuming a forward filter that is not the same as the attacker's true forward filter, our numerical experiments show that the defender is still able to estimate the attacker's state estimate with reasonable accuracy. Furthermore, we observe that assuming a sophisticated forward filter and employing a similar inverse filter may even provide better performance at the cost of computational efforts.

\section{Organization of the thesis}\label{sec:organization}
The thesis is organized in eight chapters. This chapter, Chapter~\ref{chap:intro}, introduces and motivates the inverse stochastic filtering problem addressed in our work, along with the relevant literature survey and the key contributions of the thesis. The rest of the thesis is organized as follows.\\
\textbf{Chapter~\ref{chap:system model and optimal inverse filter}} provides the system model for the inverse filtering problem. Additionally, the optimal inverse Bayesian filter recursions are also derived in Chapter~\ref{chap:system model and optimal inverse filter}.\\
\textbf{Chapter~\ref{chap:inverse EKFs}} first presents the I-EKF for systems with perfect prior model information, along with I-EKF's theoretical stability and conservativeness guarantees. Further, inverses of different EKF variants, namely, SOEKF, GS-EKF, and DEKF, are also developed in Chapter~\ref{chap:inverse EKFs} for highly non-linear systems. In this chapter, we also derive the stochastic stability guarantees for forward and inverse SOEKF. Finally, we corroborate our results with numerical experiments, including the mismatched forward and inverse filters case.\\
\textbf{Chapter~\ref{chap:unknown inputs}} considers systems wherein the state evolution and observations are affected by the presence of unknown exogenous inputs. To this end, we develop I-KFs and I-EKFs for both with and without direct feed-through cases for linear and non-linear systems, respectively. In this chapter, we also include the asymptotic stability conditions for I-KF-without-unknown-input and demonstrate the inverse filters' estimation accuracy through different numerical experiments.\\
\textbf{Chapter~\ref{chap:inverse SPKFs}} develops different inverse SPKFs, namely, I-UKF, I-CKF, I-QKF, and I-CQKF. Besides the filter formulations, this chapter also provides the rigorous theoretical performance analyses of the developed inverse filters. Finally, we demonstrate the filters' estimation performance through extensive numerical experiments, including mismatched forward and inverse filters cases and comparisons with the I-EKF developed in Section~\ref{sec:IEKF}.\\
In \textbf{Chapter~\ref{chap:rkhs}}, we relax the perfect system information assumption and derive RKHS-based KFs. To this end, we first provide an online approximate version of the EM algorithm for parameter learning, which is then coupled with EKF and UKF recursions to yield RKHS-EKF and RKHS-UKF, respectively. This chapter also derives the stochastic stability conditions for these filters. Different numerical experiments further validate our methods and compare their performance to forward and inverse filters with prior system model information.\\
\textbf{Chapter~\ref{chap:smc}} explores SMC-based techniques to develop inverse filters for general non-linear non-Gaussian systems. The SMC-based inverse filters presented in this chapter are I-PF, I-GPF, and I-EnKF and their relevant generalizations to non-Gaussian noises and unknown system cases. The developed I-PF is also proved to converge to the optimal inverse filter under mild system-level conditions. Finally, extensive numerical experiments are considered to demonstrate these filters' estimation performance and time complexity, including comparisons with the Gaussian inverse filters developed in Chapters~\ref{chap:inverse EKFs} and \ref{chap:inverse SPKFs}.\\
\textbf{Chapter~\ref{chap:conclusions}} concludes the work presented in this thesis and discusses the scope for future work.

Throughout this dissertation, we reserve boldface lowercase and uppercase letters for vectors (column vectors) and matrices, respectively, and $\lbrace a_{i}\rbrace_{i_{1}\leq i\leq i_{2}}$ denotes a set of elements indexed by an integer $i$.  For simplicity, we consider Cholesky decomposition to obtain the factorization of matrix $\mathbf{A}$ as $\mathbf{A}=\sqrt{\mathbf{A}}\sqrt{\mathbf{A}}^{T}$. Other robust but computationally expensive factorization methods include singular value and eigenvector decomposition\cite{arasaratnam2007qkf,bierman2006factorization}. Also, we use shorthand notation $\mathbb{P}(\mathbf{X}\in d\mathbf{x})$ to refer to the probability of random variable $\mathbf{X}\in [\mathbf{x},\mathbf{x}+d\mathbf{x}]$, where $d\mathbf{x}$ is an infinitesimal interval length.

\chapter{System model and Optimal inverse filter}
\label{chap:system model and optimal inverse filter}
In the following, we first describe the system model for the inverse filtering problem. We begin by presenting a general probabilistic framework for the attacker-defender dynamics in Section~\ref{sec:system model}. Within this framework, additive Gaussian noise systems are considered as a special case, discussed in Section~\ref{subsec:Gaussian system} and subsequently adopted in the KF-based inverse non-linear filters. Furthermore, before exploring the approximate filtering approaches in subsequent chapters, we derive the optimal Bayesian recursions for the inverse filtering problem in Section~\ref{sec:optimal filter}.

\section{System model}\label{sec:system model}
Consider the `$n_{x}$'-dimensional stochastic process $\mathbf{X}=\{\mathbf{X}_{k}\}_{k\geq 0}$ as the defender's state evolution process. The defender perfectly knows its state $\mathbf{x}_{k}\in\mathbb{R}^{n_{x}\times 1}$ for all $k\geq 0$. The state process $\mathbf{X}$ is a Markov process with initial state $\mathbf{X}_{0}\sim\pi^{x}_{0}(d\mathbf{x}_{0})$ and evolves as
\par\noindent\small
\begin{align}
    \mathbb{P}(\mathbf{X}_{k+1}\in d\mathbf{x}_{k+1}|\mathbf{X}_{k}=\mathbf{x}_{k})=\mathcal{K}(\mathbf{x}_{k+1}|\mathbf{x}_{k})d\mathbf{x}_{k+1},\label{eqn:state x}
\end{align}
\normalsize
where $\mathcal{K}(\cdot)$ denotes the transition kernel density (with respect to a Lebesgue measure). The attacker observes the defender's state as a `$n_{y}$'-dimensional observation process $\mathbf{Y}=\{\mathbf{Y}_{k}\}_{k\geq 1}$. The observations  $\mathbf{Y}$ are conditionally independent given $\mathbf{X}$ with
\par\noindent\small
\begin{align}
    \mathbb{P}(\mathbf{Y}_{k}\in d\mathbf{y}_{k}|\mathbf{X}_{k}=\mathbf{x}_{k})=\rho(\mathbf{y}_{k}|\mathbf{x}_{k})d\mathbf{y}_{k},\label{eqn:observation y}
\end{align}
\normalsize
where $\rho(\cdot)$ is the attacker's conditional observation density and the $k$-th observation $\mathbf{y}_{k}\in\mathbb{R}^{n_{y}\times 1}$.

The attacker computes an estimate $\hat{\mathbf{x}}_{k}$ of the defender's state $\mathbf{x}_{k}$ given the available observations $\{\mathbf{y}_{j}\}_{1\leq j\leq k}$ using the forward filter. Consider $\hat{\mathbf{X}}=\{\hat{\mathbf{X}}_{k}\}_{k\geq 0}$ as the attacker's state estimation process. The forward filter recursively computes the current estimate $\hat{\mathbf{x}}_{k}$ from the previous estimate $\hat{\mathbf{x}}_{k-1}$ and current observation $\mathbf{y}_{k}$ in a deterministic manner as
\par\noindent\small
\begin{align}
   \hat{\mathbf{x}}_{k}=T(\hat{\mathbf{x}}_{k-1},\mathbf{y}_{k}).\label{eqn:filter T}
\end{align}
\normalsize
For instance, $T(\cdot)$ represents the standard EKF/ UKF recursive update if the attacker employs a forward EKF/ UKF to compute state estimate $\hat{\mathbf{x}}_{k}$. Note that $T(\cdot)$ can be a time-dependent function for many forward filters. In the case of EKF/ UKF, the mapping $T(\cdot)$ at the $k$-th time instant depends on the covariance matrix estimate computed at the previous $(k-1)$-th time instant. For the sake of brevity, we simply denote the forward filter recursion as in \eqref{eqn:filter T} but implement the appropriate function for the given time instant. The attacker then uses the estimate $\hat{\mathbf{x}}_{k}$ to administer an action which the defender observes as a `$n_{a}$'-dimensional observation process $\mathbf{A}=\{\mathbf{A}_{k}\}_{k\geq 1}$. Given $\hat{\mathbf{X}}$, the observations $\mathbf{A}$ are conditionally independent and
\par
\begin{figure}
  \centering
  \includegraphics[width = 0.7\columnwidth]{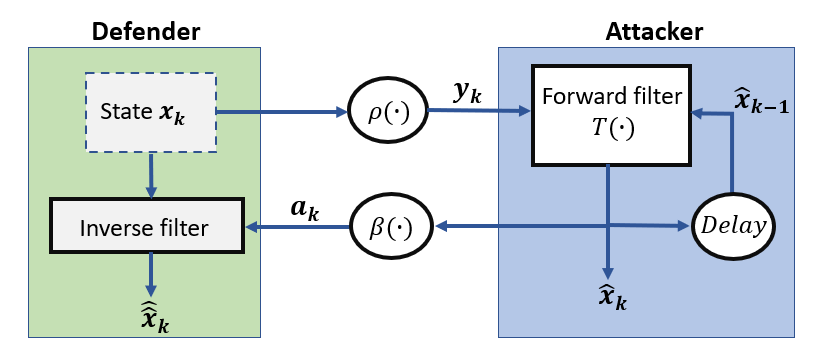}
  \caption{Graphical representation of the defender-attacker interaction.}
 \label{fig:system model}
\end{figure}

\noindent\small
\begin{align}
    \mathbb{P}(\mathbf{A}_{k}\in d\mathbf{a}_{k}|\hat{\mathbf{X}}_{k}=\hat{\mathbf{x}}_{k})=\beta(\mathbf{a}_{k}|\hat{\mathbf{x}}_{k})d\mathbf{a}_{k},\label{eqn:observation a}
\end{align}
\normalsize
where $\beta(\cdot)$ is the defender's conditional observation density and the $k$-th observation $\mathbf{a}_{k}\in\mathbb{R}^{n_{a}\times 1}$. Finally, the defender uses $\{\mathbf{a}_{j},\mathbf{x}_{j}\}_{1\leq j\leq k}$ to compute the estimate $\doublehat{\mathbf{x}}_{k}\in\mathbb{R}^{n_{x}\times 1}$ of $\hat{\mathbf{x}}_{k}$ in the inverse filter. Fig.~\ref{fig:system model} graphically illustrates the system dynamics for inverse filtering.

It should be noted that the attacker and defender are entirely different agents employing independent sensors to observe each other. Furthermore, in the inverse filtering problem, the attacker is unaware that the defender is observing the former. The attacker also does not seek to hide its actions from the defender, which is a more challenging problem recently addressed in \cite{lourencco2020protect,lourencco2021hidden}. If the attacker also guards itself against the defender, the attacker-defender interaction then requires an inverse-inverse reinforcement learning-based representation of the problem, which is not the focus of this dissertation and has been addressed in other recent works \cite{pattanayak2022inverse,pattanayak2022meta}.

In Chapters~\ref{chap:inverse EKFs}-\ref{chap:inverse SPKFs} and \ref{chap:smc}, we assume that both defender and attacker have perfect knowledge of the system model, i.e., densities $\mathcal{K}(\cdot)$, $\rho(\cdot)$ and $\beta(\cdot)$. Additionally, the defender assumes a known forward filter $T(\cdot)$ employed by the attacker. This inverse cognition framework is similar to the one investigated in \cite{krishnamurthy2019how,mattila2020hmm}. However, in our numerical experiments, we also analyze the mismatched forward and inverse filter cases. In particular, we observe that the developed inverse filters provide reasonably accurate estimates even when assuming an incorrect forward filter. In some cases, a sophisticated inverse filter may even provide better estimates. We address the unknown system dynamics case in Chapter~\ref{chap:rkhs} wherein RKHS-based EKF and UKF are developed to estimate the state and system parameters jointly. Additionally, Chapter~\ref{chap:smc} explores methods to generalize SMC-based inverse filters for the unknown systems case. For simplicity, the presence of known exogenous inputs is ignored in the state evolution \eqref{eqn:state x}, and observations \eqref{eqn:observation y} and \eqref{eqn:observation a}. However, extending our inverse filters for these modifications in the system model is trivial. Throughout the dissertation, we focus on discrete-time and real-valued models but generalize our methods to handle continuous-time state evolution and complex systems in Section~\ref{sec:IUKF}.

\subsection{Systems with additive Gaussian noises}\label{subsec:Gaussian system}
Consider the special case of additive system noises such that the state evolution and observations are modeled as
\par\noindent\small
\begin{align}
    &\mathbf{x}_{k+1}=f(\mathbf{x}_{k})+\mathbf{w}_{k},\label{eqn:non-linear state x}
\end{align}
\begin{align}
    &\mathbf{y}_{k}=h(\mathbf{x}_{k})+\mathbf{v}_{k},\label{eqn:non-linear observation y}\\
    &\mathbf{a}_{k}=g(\hat{\mathbf{x}}_{k})+\bm{\epsilon}_{k}.\label{eqn:non-linear observation a}
\end{align}
\normalsize
Here, $f(\cdot)$, $h(\cdot)$ and $g(\cdot)$ represent a general non-linear system dynamics with function $g(\cdot)$ representing the combined effect of the attacker's action strategy and the defender's observation. For example, inspired by linear quadratic Gaussian (LQG) control problems, I-KF\cite{krishnamurthy2019how} selects $\mathbf{a}_{k}$ based on a linear relationship with $\hat{\mathbf{x}}_{k}$, adjusted by its estimated error covariance matrix $\bm{\Sigma}_{k}$. However, \eqref{eqn:non-linear observation a} considers a general non-linear observation function. The noise processes $\{\mathbf{w}_{k}\}_{k\geq 0}$, $\{\mathbf{v}_{k}\}_{k\geq 1}$ and $\{\bm{\epsilon}_{k}\}_{k\geq 1}$ are mutually independent and i.i.d. across time. In practical systems, $\mathbf{w}_{k}$ typically represents the modeling uncertainties in state transition \eqref{eqn:non-linear state x} while $\mathbf{v}_{k}$ and $\bm{\epsilon}_{k}$ are measurement noises of two different agents/ sensors. Hence, these noises are mutually independent. If the probability density functions of $\mathbf{w}_{k}$, $\mathbf{v}_{k}$ and $\bm{\epsilon}_{k}$ are denoted by $p_{w}(\cdot)$, $p_{v}(\cdot)$ and $p_{\epsilon}(\cdot)$, respectively, then $\mathcal{K}(\mathbf{x}_{k+1}|\mathbf{x}_{k})=p_{w}(\mathbf{x}_{k+1}-f(\mathbf{x}_{k}))$, $\rho(\mathbf{y}_{k}|\mathbf{x}_{k})=p_{v}(\mathbf{y}_{k}-h(\mathbf{x}_{k}))$, and $\beta(\mathbf{a}_{k}|\hat{\mathbf{x}}_{k})=p_{\epsilon}(\mathbf{a}_{k}-g(\hat{\mathbf{x}}_{k}))$. Under the Gaussian noise assumption, $\mathbf{w}_{k}$, $\mathbf{v}_{k}$ and $\bm{\epsilon}_{k}$ are distributed as $\mathcal{N}(\mathbf{0},\mathbf{Q})$, $\mathcal{N}(\mathbf{0},\mathbf{R})$ and $\mathcal{N}(\mathbf{0},\bm{\Sigma}_{\epsilon})$, respectively, where $\mathbf{Q} \in \mathbb{R}^{n_{x}\times n_{x}}$, $\mathbf{R}\in \mathbb{R}^{n_{y}\times n_{y}}$ and $\bm{\Sigma}_{\epsilon} \in \mathbb{R}^{n_{a}\times n_{a}}$ are the noise covariance matrices. For simplicity, these covariances are assumed to be time-varying only when analyzing the filters' theoretical performance guarantees.

\textbf{Systems with unknown inputs:} In the case of systems with unknown inputs, the state evolution and attacker's observations also depend on an exogenous control input $\mathbf{u}_k \in \mathbb{R}^{n_{u} \times 1}$ such that \eqref{eqn:non-linear state x} and \eqref{eqn:non-linear observation y}, respectively, become
\par\noindent\small
\begin{align}
    &\mathbf{x}_{k+1}=f(\mathbf{x}_{k},\mathbf{u}_{k})+\mathbf{w}_{k},\label{eqn:non-linear state x with unknown input}\\
    &\mathbf{y}_{k}=h(\mathbf{x}_{k},\mathbf{u}_{k})+\mathbf{v}_{k},\label{eqn:non-linear observation y with unknown input}
\end{align}
\normalsize
The input $\mathbf{u}_{k}$ is known to the defender but not to the attacker. This is a \textit{direct feed-through} (DF) model, wherein $\mathbf{y}_{k}$ depends on the unknown input. Without DF, the defender's state evolves as in \eqref{eqn:non-linear state x with unknown input}  but attacker's observations are as in \eqref{eqn:non-linear observation y}. We show in Chapter~\ref{chap:unknown inputs} that the presence or absence of the unknown input in the attacker's observation model leads to different solution approaches towards the forward and inverse filters.

When the system dynamics are linear, then the non-linear functions $f(\cdot, \cdot)$, $h(\cdot,\cdot)$, and $g(\cdot)$ are replaced by the matrix pairs $\{\mathbf{F, B}\}$, $\{\mathbf{H, D}\}$, and the matrix $\mathbf{G}$, respectively, as
\par\noindent\small
\begin{align}
&\mathbf{x}_{k+1}=\mathbf{Fx}_{k}+\mathbf{Bu}_{k}+\mathbf{w}_{k},\label{eqn: linear x with input}\\
&\mathbf{y}_{k}=\mathbf{Hx}_{k}+\mathbf{Du}_{k}+\mathbf{v}_{k},\label{eqn: linear y withdf}\\
&\mathbf{a}_{k}=\mathbf{G}\hat{\mathbf{x}}_{k}+\bm{\epsilon}_{k},\label{eqn: linear a}
\end{align}
\normalsize
where $\mathbf{F}\in \mathbb{R}^{n_{x}\times n_{x}}$, $\mathbf{B} \in \mathbb{R}^{n_{x}\times n_{u}}$, $\mathbf{H} \in \mathbb{R}^{n_{y}\times n_{x}}$, $\mathbf{D} \in \mathbb{R}^{n_{y}\times n_{u}}$ and $\mathbf{G} \in \mathbb{R}^{n_a \times n_{x}}$. When the unknown input is absent, either $\mathbf{B}=\mathbf{0}_{n_{x} \times n_{u}}$ or $\mathbf{D}=\mathbf{0}_{n_{y} \times n_{u}}$ or both vanish.

\section{Optimal inverse filter}\label{sec:optimal filter}
Consider the (joint) conditional distribution of $(\hat{\mathbf{x}}_{k},\mathbf{y}_{k})$ given the defender's knowledge of its own true states and observations of the attacker's actions at the $k$-th time instant. As in forward Bayesian filter, the optimal inverse filter computes this conditional distribution recursively using the time and measurement updates. Define the conditional distributions $\pi_{k|k-1}(d\hat{\mathbf{x}}_{k},d\mathbf{y}_{k})$ and $\pi_{k|k}(d\hat{\mathbf{x}}_{k},d\mathbf{y}_{k})$ as
\par\noindent\small
\begin{align}
&\pi_{k|k-1}(d\hat{\mathbf{x}}_{k},d\mathbf{y}_{k})\doteq p(\hat{\mathbf{x}}_{k},\mathbf{y}_{k}|\mathbf{x}_{0:k},\mathbf{a}_{1:k-1})d\hat{\mathbf{x}}_{k}d\mathbf{y}_{k},\label{eqn:pi predict}\\
&\pi_{k|k}(d\hat{\mathbf{x}}_{k},d\mathbf{y}_{k})\doteq p(\hat{\mathbf{x}}_{k},\mathbf{y}_{k}|\mathbf{x}_{0:k},\mathbf{a}_{1:k})d\hat{\mathbf{x}}_{k}d\mathbf{y}_{k}.\label{eqn:pi update}
\end{align}
\normalsize
In the time-update step, we obtain $\pi_{k|k-1}$ from $\pi_{k-1|k-1}$ considering the true states $\mathbf{x}_{0:k}$ but observations $\mathbf{a}_{1:k-1}$ excluding the current observation $\mathbf{a}_{k}$. Finally, $\pi_{k|k-1}$ is updated using the current observation $\mathbf{a}_{k}$ in the measurement update step to obtain $\pi_{k|k}$. The defender's MMSE estimate $\doublehat{\mathbf{x}}_{k}$ is then $\doublehat{\mathbf{x}}_{k}=\iint\hat{\mathbf{x}}_{k}\pi_{k|k}(d\hat{\mathbf{x}}_{k},d\mathbf{y}_{k})$.

First, we consider the optimal filter's time update. We have
\par\noindent\small
\begin{align}
&p(\hat{\mathbf{x}}_{k},\mathbf{y}_{k}|\mathbf{x}_{0:k},\mathbf{a}_{1:k-1})=\int p(\hat{\mathbf{x}}_{k},\mathbf{y}_{k}|\mathbf{x}_{0:k},\mathbf{a}_{1:k-1},\hat{\mathbf{x}}_{k-1})p(\hat{\mathbf{x}}_{k-1}|\mathbf{x}_{0:k},\mathbf{a}_{1:k-1})d\hat{\mathbf{x}}_{k-1}.\label{eqn:time update start}
\end{align}
\normalsize
However, $p(\hat{\mathbf{x}}_{k},\mathbf{y}_{k}|\mathbf{x}_{0:k},\mathbf{a}_{1:k-1},\hat{\mathbf{x}}_{k-1})=p(\hat{\mathbf{x}}_{k}|\mathbf{y}_{k},\mathbf{x}_{0:k},\mathbf{a}_{1:k-1},\hat{\mathbf{x}}_{k-1})p(\mathbf{y}_{k}|\mathbf{x}_{0:k},\mathbf{a}_{1:k-1},\hat{\mathbf{x}}_{k-1})=\\p(\hat{\mathbf{x}}_{k}|\mathbf{y}_{k},\hat{\mathbf{x}}_{k-1})p(\mathbf{y}_{k}|\mathbf{x}_{k}),$ because observation $\mathbf{y}_{k}$ is conditionally independent given state $\mathbf{x}_{k}$ with $p(\mathbf{y}_{k}|\mathbf{x}_{k})$ given by \eqref{eqn:observation y}. Also, the estimate $\hat{\mathbf{x}}_{k}$ is independent of $\{\mathbf{x}_{0:k},\mathbf{a}_{1:k-1}\}$ given $\{\mathbf{y}_{k},\hat{\mathbf{x}}_{k-1}\}$. In particular, $\hat{\mathbf{x}}_{k}$ is a deterministic function of $\mathbf{y}_{k}$ and $\hat{\mathbf{x}}_{k-1}$ such that $p(\hat{\mathbf{x}}_{k}|\mathbf{y}_{k},\hat{\mathbf{x}}_{k-1})=\delta(\hat{\mathbf{x}}_{k}-T(\hat{\mathbf{x}}_{k-1},\mathbf{y}_{k}))$. Hence, \eqref{eqn:time update start} yields
\par\noindent\small
\begin{align}
&p(\hat{\mathbf{x}}_{k},\mathbf{y}_{k}|\mathbf{x}_{0:k},\mathbf{a}_{1:k-1})=\int\delta(\hat{\mathbf{x}}_{k}-T(\hat{\mathbf{x}}_{k-1},\mathbf{y}_{k}))\rho(\mathbf{y}_{k}|\mathbf{x}_{k})p(\hat{\mathbf{x}}_{k-1}|\mathbf{x}_{0:k},\mathbf{a}_{1:k-1})d\hat{\mathbf{x}}_{k-1}.\label{eqn:time update inter}
\end{align}
\normalsize
Now, $p(\hat{\mathbf{x}}_{k-1},\mathbf{y}_{k-1}|\mathbf{x}_{0:k},\mathbf{a}_{1:k-1})=p(\hat{\mathbf{x}}_{k-1},\mathbf{y}_{k-1}|\mathbf{x}_{0:k-1},\mathbf{a}_{1:k-1})$ because $\{\hat{\mathbf{x}}_{k-1},\mathbf{y}_{k-1}\}$ do not depend on future state $\mathbf{x}_{k}$ given $\mathbf{x}_{0:k-1}$ and $\mathbf{a}_{1:k-1}$. Hence, using this and \eqref{eqn:pi update}, we have the marginal distribution $\mathbb{P}(\hat{\mathbf{x}}_{k-1}\in d\hat{\mathbf{x}}_{k-1}|\mathbf{x}_{0:k},\mathbf{a}_{1:k-1})=\int\pi_{k-1|k-1}(d\hat{\mathbf{x}}_{k-1},d\mathbf{y}_{k-1})$. Substituting in \eqref{eqn:time update inter}, the optimal time update becomes
\par\noindent\small
\begin{align}
&p(\hat{\mathbf{x}}_{k},\mathbf{y}_{k}|\mathbf{x}_{0:k},\mathbf{a}_{1:k-1})=\iint\delta(\hat{\mathbf{x}}_{k}-T(\hat{\mathbf{x}}_{k-1},\mathbf{y}_{k}))\rho(\mathbf{y}_{k}|\mathbf{x}_{k})\pi_{k-1|k-1}(d\hat{\mathbf{x}}_{k-1},d\mathbf{y}_{k-1}).\label{eqn:time update}
\end{align}
\normalsize

Now, consider the measurement update with current observation $\mathbf{a}_{k}$. The joint conditional distribution $p(\hat{\mathbf{x}}_{k},\mathbf{y}_{k},\mathbf{a}_{k}|\mathbf{x}_{0:k},\mathbf{a}_{1:k-1})=\beta(\mathbf{a}_{k}|\hat{\mathbf{x}}_{k})\pi_{k|k-1}(d\hat{\mathbf{x}}_{k},d\mathbf{y}_{k})$ because observation $\mathbf{a}_{k}$ is conditionally independent of everything else given $\hat{\mathbf{x}}_{k}$ with $p(\mathbf{a}_{k}|\hat{\mathbf{x}}_{k})$ given by \eqref{eqn:observation a}. Hence, using Bayes' theorem, we have
\par\noindent\small
\begin{align}
\pi_{k|k}(d\hat{\mathbf{x}}_{k},d\mathbf{y}_{k})=\frac{\beta(\mathbf{a}_{k}|\hat{\mathbf{x}}_{k})\pi_{k|k-1}(d\hat{\mathbf{x}}_{k},d\mathbf{y}_{k})}{\iint\beta(\mathbf{a}_{k}|\hat{\mathbf{x}}_{k})\pi_{k|k-1}(d\hat{\mathbf{x}}_{k},d\mathbf{y}_{k})},\label{eqn:measurement update}
\end{align}
\normalsize
which is the optimal measurement update.

In the case of SMC methods, the defender estimates a general function $\phi(\hat{\mathbf{x}},\mathbf{y})$ as\\ $\mathbb{E}[\phi(\hat{\mathbf{x}},\mathbf{y})|\mathbf{x}_{0:k},\mathbf{a}_{1:k}]$ using the inverse filter's posterior distribution. For instance, considering $\phi(\hat{\mathbf{x}},\mathbf{y})=\hat{\mathbf{x}}$ yields the defender's MMSE estimate $\doublehat{\mathbf{x}}_{k}$ of $\hat{\mathbf{x}}_{k}$. For the sake of I-PF's convergence analysis in Section~\ref{subsec:IPF convergence}, 
we now introduce some simplified notations for the integrals involved in the optimal filter recursions. Given a measure $\nu$, a function $\phi$ and a Markov transition kernel $\mathcal{K}$, we define
\par\noindent\small
\begin{align*}
\langle\nu,\phi\rangle\doteq\int\phi(x)\nu(dx),\;\;\;\mathcal{K}\phi(x)\doteq\int\phi(z)\mathcal{K}(dz|x).
\end{align*}
\normalsize
With this notation, the optimal filter recursions \eqref{eqn:time update} and \eqref{eqn:measurement update} can be expressed as
\par\noindent\small
\begin{align}
\langle\pi_{k|k-1},\phi\rangle&=\langle\pi_{k-1|k-1},\delta_{T}\rho\phi\rangle,\label{eqn:optimal filter time update}\\
\langle\pi_{k|k},\phi\rangle&=\frac{\langle\pi_{k|k-1},\beta\phi\rangle}{\langle\pi_{k|k-1},\beta\rangle},\label{eqn:optimal filter measurement update}
\end{align}
\normalsize
where $\delta_{T}$ denotes function $\delta(\hat{\mathbf{x}}_{k}-T(\hat{\mathbf{x}}_{k-1},\mathbf{y}_{k}))$. For brevity, we drop the time parameter $k$ in our notation $\delta_{T}$, but while implementing, the given time-instant is taken into consideration. Note that the optimal inverse filter exists only if $\langle\pi_{k|k-1},\beta\rangle>0$.

\chapter{Inverse EKF and its variants}
\label{chap:inverse EKFs}
In this chapter, we first introduce the I-EKF for non-linear systems with additive Gaussian noises in Section~\ref{sec:IEKF}. The sufficient conditions for the stability and conservativeness of I-EKF are derived in Section~\ref{subsec:IEKF stability}, considering both bounded non-linearity and unknown matrix approaches for stability analyses. In Section~\ref{sec:I-SOEKF}, we develop the I-SOEKF and provide its stability guarantees based on the bounded non-linearity approach. Similarly, Section~\ref{sec:I-GSEKF and I-DEKF} presents inverse GS-EKF (I-GS-EKF) and inverse DEKF (I-DEKF). Finally, we compare the estimation accuracy of these filters numerically in Section~\ref{sec:IEKF numericals} before concluding in Section~\ref{sec:IEKF conclusions}. Throughout the chapter, we consider the defender's state transition as in \eqref{eqn:non-linear state x} while the attacker's and defender's observations are given by \eqref{eqn:non-linear observation y} and \eqref{eqn:non-linear observation a}, respectively. The defender also assumes that the attacker employs an appropriate forward filter to estimate the defender's state.

\section{Inverse EKF}\label{sec:IEKF}
\subsection{Filter formulation}\label{subsec:IEKF formulation}
\textit{Forward filter:} For I-EKF formulation, we assume that the attacker employs a forward EKF to compute estimate $\hat{\mathbf{x}}_{k}$ of defender's state $\mathbf{x}_{k}$. Define the Jacobians as $\mathbf{F}_{k} \doteq \nabla_{\mathbf{x}}f(\mathbf{x})\vert_{\mathbf{x}=\hat{\mathbf{x}}_{k}}$ and 
$\mathbf{H}_{k+1} \doteq \nabla_{\mathbf{x}}h(\mathbf{x})\vert_{\mathbf{x}=\hat{\mathbf{x}}_{k+1|k}}$. The forward EKF's recursive state estimation procedure first obtains the prediction $\hat{\mathbf{x}}_{k+1|k}$ of the current state using the previous state estimate, with $\bm{\Sigma}_{k+1|k}$ as the associated prediction error covariance matrix estimate. In the measurement update step, the state estimate $\hat{\mathbf{x}}_{k+1}$ and the associated error covariance matrix estimate $\bm{\Sigma}_{k+1}$ are updated using current observation $\mathbf{y}_{k+1}$, predicted observation $\hat{\mathbf{y}}_{k+1|k}$, and the Kalman gain matrix $\mathbf{K}_{k+1}$. The attacker's forward EKF recursions are\cite{anderson2012optimal}
\par\noindent\small
\begin{align}
&\textit{Time update:}\;\;\;\hat{\mathbf{x}}_{k+1|k}=f(\hat{\mathbf{x}}_{k}),\label{eqn:forward ekf predict}\\
&\bm{\Sigma}_{k+1|k}=\mathbf{F}_{k}\bm{\Sigma}_{k}\mathbf{F}_{k}^{T}+\mathbf{Q}.\label{eqn:forward ekf covariance predict}\\
&\textit{Measurement update:}\;\;\;\hat{\mathbf{y}}_{k+1|k}=h(\hat{\mathbf{x}}_{k+1|k}),\nonumber\\
&\bm{\Sigma}^{y}_{k+1}=\mathbf{H}_{k+1}\bm{\Sigma}_{k+1|k}\mathbf{H}_{k+1}^{T}+\mathbf{R},\;\;\;\mathbf{K}_{k+1}=\bm{\Sigma}_{k+1|k}\mathbf{H}_{k+1}^{T}(\bm{\Sigma}^{y}_{k+1})^{-1},\nonumber\\
&\hat{\mathbf{x}}_{k+1}=\hat{\mathbf{x}}_{k+1|k}+\mathbf{K}_{k+1}(\mathbf{y}_{k+1}-\hat{\mathbf{y}}_{k+1|k}),\label{eqn:forward ekf update}\\
&\bm{\Sigma}_{k+1}=\bm{\Sigma}_{k+1|k}-\bm{\Sigma}_{k+1|k}\mathbf{H}_{k+1}^{T}(\bm{\Sigma}^{y}_{k+1})^{-1}\mathbf{H}_{k+1}\bm{\Sigma}_{k+1|k}.\nonumber
\end{align}
\normalsize

\textit{Inverse filter:} I-EKF aims to find an estimate $\doublehat{\mathbf{x}}_{k}$ of the state estimate $\hat{\mathbf{x}}_{k}$ given the actual states $\mathbf{x}_{1:k}$ and observations $\mathbf{a}_{1:k}$ given by \eqref{eqn:non-linear observation a}. Note that even when assuming perfect system model information and a known forward filter, the estimate $\hat{\mathbf{x}}_{k}$ is unknown to the defender. In particular, the forward EKF computes $\hat{\mathbf{x}}_{k}$ given attacker's noisy observations $\{\mathbf{y}_{j}\}_{1\leq j\leq k}$ up to $k$-th time instant, which are not available to the defender. Hence, the defender computes an estimate $\doublehat{\mathbf{x}}_{k}$ of $\hat{\mathbf{x}}_{k}$ given $\{\mathbf{a}_{j},\mathbf{x}_{j}\}_{1\leq j\leq k}$. The state transition equation for the I-EKF is obtained by substituting \eqref{eqn:non-linear observation y} and \eqref{eqn:forward ekf predict} in \eqref{eqn:forward ekf update} and represented by the time-varying non-linear function $\widetilde{f}_{k}(\cdot)$ as
\par\noindent\small
\begin{align}
&\hat{\mathbf{x}}_{k+1}=\widetilde{f}_{k}(\hat{\mathbf{x}}_{k},\mathbf{x}_{k+1},\mathbf{v}_{k+1})=f(\hat{\mathbf{x}}_{k})+\mathbf{K}_{k+1}(h(\mathbf{x}_{k+1})+\mathbf{v}_{k+1}-h(f(\hat{\mathbf{x}}_{k}))).\label{eqn:iekf state transition}
\end{align}
\normalsize
In this state transition, the true state $\mathbf{x}_{k+1}$ is perfectly known to the defender and henceforth treated as a known exogenous input. Further, $\mathbf{v}_{k+1}$ is the process noise involved which is non-additive because $\mathbf{K}_{k+1}$ depends on the Jacobians evaluated at the previous estimate $\hat{\mathbf{x}}_{k}$. Then, the I-EKF is formulated based on EKF with non-additive noise and known exogenous input formulation. I-EKF's recursions to compute the estimate $\doublehat{\mathbf{x}}_{k}$ and the associated error covariance matrix estimate $\overline{\bm{\Sigma}}_{k}$ are
\par\noindent\small
\begin{align}
&\textit{Time update:}\;\;\;\doublehat{\mathbf{x}}_{k+1|k}=\widetilde{f}_{k}(\doublehat{\mathbf{x}}_{k},\mathbf{x}_{k+1},\mathbf{0}),\;\;\;\overline{\bm{\Sigma}}_{k+1|k}=\widetilde{\mathbf{F}}_{k}\overline{\bm{\Sigma}}_{k}\widetilde{\mathbf{F}}_{k}^{T}+\overline{\mathbf{Q}}_{k}.\nonumber\\
&\textit{Measurement update:}\;\;\;\hat{\mathbf{a}}_{k+1|k}=g(\doublehat{\mathbf{x}}_{k+1|k}),\nonumber\\
&\overline{\bm{\Sigma}}^{a}_{k+1}=\mathbf{G}_{k+1}\overline{\bm{\Sigma}}_{k+1|k}\mathbf{G}_{k+1}^{T}+\bm{\Sigma}_{\epsilon},\;\;\;\overline{\mathbf{K}}_{k+1}=\overline{\bm{\Sigma}}_{k+1|k}\mathbf{G}_{k+1}^{T}(\overline{\bm{\Sigma}}^{a}_{k+1})^{-1},\nonumber\\
&\doublehat{\mathbf{x}}_{k+1}=\doublehat{\mathbf{x}}_{k+1|k}+\overline{\mathbf{K}}_{k+1}(\mathbf{a}_{k+1}-\hat{\mathbf{a}}_{k+1|k}),\label{eqn:IEKF state update}\\
&\overline{\bm{\Sigma}}_{k+1}=\overline{\bm{\Sigma}}_{k+1|k}-\overline{\bm{\Sigma}}_{k+1|k}\mathbf{G}_{k+1}^{T}(\overline{\bm{\Sigma}}^{a}_{k+1})^{-1}\mathbf{G}_{k+1}\overline{\bm{\Sigma}}_{k+1|k},\label{eqn:IEKF covariance update}
\end{align}
\normalsize
where $\widetilde{\mathbf{F}}_{k}\doteq\nabla_{\mathbf{x}}\widetilde{f}_{k}(\mathbf{x},\mathbf{x}_{k+1},\mathbf{0})\vert_{\mathbf{x}=\doublehat{\mathbf{x}}_{k}}=\mathbf{F}_{k}-\mathbf{K}_{k+1}\mathbf{H}_{k+1}\mathbf{F}_{k}$ (with Jacobians $\mathbf{F}_{k}$ evaluated at $\doublehat{\mathbf{x}}_{k}$ and $\mathbf{H}_{k+1}$ evaluated at $f(\doublehat{\mathbf{x}}_{k})$), $\overline{\mathbf{Q}}_{k}=\widetilde{\mathbf{V}}_{k}\mathbf{R}\widetilde{\mathbf{V}}_{k}^{T}$, $\widetilde{\mathbf{V}}_{k}\doteq\nabla_{v}\widetilde{f}_{k}(\doublehat{\mathbf{x}}_{k},\mathbf{x}_{k+1},\mathbf{v})\vert_{\mathbf{v}=\mathbf{0}}=\mathbf{K}_{k+1}$ and $\mathbf{G}_{k+1}\doteq\nabla_{\mathbf{x}}g(\mathbf{x})\vert_{\mathbf{x}=\doublehat{\mathbf{x}}_{k+1|k}}$.

In the I-EKF formulation, $\mathbf{K}_{k+1}$ is treated as a time-varying parameter of the state transition function $\widetilde{f}_{k}(\cdot)$ and not as a function of the state estimate $\hat{\mathbf{x}}_{k}$. It is approximated by evaluating its value using I-EKF's estimate $\doublehat{\mathbf{x}}_{k}$ recursively in the same manner as the forward EKF evaluates at its own estimate $\hat{\mathbf{x}}_{k}$ (using current covariance matrix $\bm{\Sigma}_{k}$, the Jacobians $\mathbf{F}_{k}$ and $\mathbf{H}_{k}$ evaluated at the current estimate $\hat{\mathbf{x}}_{k}$ and noise covariances $\mathbf{Q}$ and $\mathbf{R}$). On the contrary, in I-KF, the forward Kalman gain $\mathbf{K}_{k+1}$ is deterministic and fully determined by the model parameters given the initial covariance estimate $\bm{\Sigma}_{0}$\cite{krishnamurthy2019how}. It is computed offline independent of the current I-KF's estimate. Note that the proposed I-EKFs' recursions are obtained from that of a standard EKF, but with the state transition equation representing the evolution of the corresponding forward filter's state estimate. Hence, I-EKF has similar computational complexity as a standard EKF, i.e., $\mathcal{O}(n_{x}^3)$, where $n_{x}$ is the state dimension\cite{daum2005nonlinear}. I-EKF could be applied in various non-linear target tracking applications, where EKF is a popular forward filter\cite{ristic2003beyond}.

We considered additive noise in our system model. If the noise is non-additive in the system model, both the forward and inverse EKF can be modified suitably by changing the corresponding covariance matrices. For instance, let the noise $\mathbf{w}_{k}$ be non-additive in the state evolution such that \eqref{eqn:non-linear state x} becomes $\mathbf{x}_{k+1}=f(\mathbf{x}_{k},\mathbf{w}_{k})$. In the forward EKF, the prediction step \eqref{eqn:forward ekf predict} is modified to $\hat{\mathbf{x}}_{k+1|k}=f(\hat{\mathbf{x}}_{k},\mathbf{0})$ while \eqref{eqn:forward ekf covariance predict} to $\bm{\Sigma}_{k+1|k}=\mathbf{F}_{k}\bm{\Sigma}_{k}\mathbf{F}_{k}^{T}+\mathbf{Q}_{k}$, where $\mathbf{F}_{k}\doteq\nabla_{\mathbf{x}}f(\mathbf{x},\mathbf{0})\vert_{\mathbf{x}=\hat{\mathbf{x}}_{k}}$ and $\mathbf{Q}_{k}=\mathbf{W}_{k}\mathbf{Q}\mathbf{W}_{k}^{T}$ with $\mathbf{W}_{k}\doteq\nabla_{\mathbf{w}}f(\hat{\mathbf{x}}_{k},\mathbf{w})\vert_{\mathbf{w}=\mathbf{0}}$. While formulating the I-EKF, these modifications need to be taken into account in the state transition equation of the inverse filter.

\begin{remark}[I-EKF for non-Gaussian noise]\label{remark:ekf non Gaussian}
Our system model \eqref{eqn:non-linear state x}-\eqref{eqn:non-linear observation a} considered the Gaussian process and measurement noises. To tackle the non-Gaussianity of the noise, GS-EKF \cite{anderson2012optimal} and its inverse developed in Section~\ref{subsec:I-GSEKF} may be considered. Alternatively, one may employ the maximum correntropy criterion (MCC)-modified KFs\cite{cinar2012hidden,izanloo2016kalman,yang2019map,liu2017maximum,wang2017maximum}. For instance, the forward MCC-EKF in \cite{yang2019map} introduces a scalar ratio $d_{k+1}=G_{\sigma}(\|\mathbf{y}_{k+1}-h(\hat{\mathbf{x}}_{k+1|k})\|_{R^{-1}})/G_{\sigma}(\|\hat{\mathbf{x}}_{k+1|k}-f(\hat{\mathbf{x}}_{k})\|_{\bm{\Sigma}_{k+1|k}^{-1}})$, where $G_{\sigma}(\cdot)$ is the Gaussian kernel. The forward gain matrix $\mathbf{K}_{k+1}$ then becomes $\mathbf{K}_{k+1}=\bm{\Sigma}_{k+1|k}\mathbf{H}^{T}_{k+1}(\mathbf{H}_{k+1}\bm{\Sigma}_{k+1|k}\mathbf{H}^{T}_{k+1}+d_{k+1}^{-1}\mathbf{R})^{-1}$. The state prediction and update steps are the same as in forward EKF. While formulating the inverse filter, these modifications need to be taken into account in the inverse filter's state-transition equation. Also, I-EKF's gain matrix $\overline{\mathbf{K}}_{k+1}$ is similarly modified using $\overline{d}_{k+1}$ which is the counterpart of $d_{k+1}$ for the inverse filter's dynamics.
\end{remark}

\textit{One-step prediction formulation:} The two-step prediction-update formulation (as discussed for forward EKF and I-EKF so far) infers an estimate of the current state. However, often, for stability analyses, the one-step prediction formulation is analytically more useful. In this formulation, the estimate $\hat{\mathbf{x}}_{k}$ is the one-step prediction estimate, i.e., an estimate of state $\mathbf{x}_{k}$ at $k$-th instant given the observations $\lbrace\mathbf{y}_{j}\rbrace_{1\leq j\leq k-1}$ up to time instant $k-1$ with $\bm{\Sigma}_{k}$ as the corresponding prediction error covariance matrix. The forward one-step prediction EKF formulation\cite{reif1999stochastic} for the same system but with $\mathbf{F}_{k}\doteq\nabla_{\mathbf{x}}f(\mathbf{x})\vert_{\mathbf{x}=\hat{\mathbf{x}}_{k}}$ and $\mathbf{H}_{k}\doteq\nabla_{\mathbf{x}}h(\mathbf{x})\vert_{\mathbf{x}=\hat{\mathbf{x}}_{k}}$ is
\par\noindent\small
\begin{align}
&\mathbf{K}_{k}=\mathbf{F}_{k}\bm{\Sigma}_{k}\mathbf{H}_{k}^{T}(\mathbf{H}_{k}\bm{\Sigma}_{k}\mathbf{H}_{k}^{T}+\mathbf{R})^{-1},\label{eqn: one step forward ekf gain}\\
&\hat{\mathbf{x}}_{k+1}=f(\hat{\mathbf{x}}_{k})+\mathbf{K}_{k}(\mathbf{y}_{k}-h(\hat{\mathbf{x}}_{k})),\label{eqn: one step forward ekf update}\\
&\bm{\Sigma}_{k+1}=\mathbf{F}_{k}\bm{\Sigma}_{k}\mathbf{F}_{k}^{T}+\mathbf{Q}-\mathbf{K}_{k}(\mathbf{H}_{k}\bm{\Sigma}_{k}\mathbf{H}_{k}^{T}+\mathbf{R})\mathbf{K}_{k}^{T}.\label{eqn: one step forward ekf covariance}
\end{align}
\normalsize

From \eqref{eqn:non-linear observation y} and \eqref{eqn: one step forward ekf update}, the state transition equation for one-step formulation of I-EKF is $\hat{\mathbf{x}}_{k+1}=\widetilde{f}_{k}(\hat{\mathbf{x}}_{k},\mathbf{x}_{k},\mathbf{v}_{k})\doteq f(\hat{\mathbf{x}}_{k})-\mathbf{K}_{k}h(\hat{\mathbf{x}}_{k})+\mathbf{K}_{k}h(\mathbf{x}_{k})+\mathbf{K}_{k}\mathbf{v}_{k}$. With this state transition, the I-EKF's one-step prediction formulation follows directly from EKF's one-step prediction formulation treating $\mathbf{a}_{k}$ as the observation with the Jacobians with respect to state estimate $\widetilde{\mathbf{F}}^{x}_{k}=\nabla_{\mathbf{x}}\widetilde{f}_{k}(\mathbf{x},\mathbf{x}_{k},\mathbf{0})\vert_{\mathbf{x}=\doublehat{\mathbf{x}}_{k}}=\mathbf{F}_{k}-\mathbf{K}_{k}\mathbf{H}_{k}$ and $\mathbf{G}_{k}=\nabla_{\mathbf{x}}g(\mathbf{x})\vert_{\mathbf{x}=\doublehat{\mathbf{x}}_{k}}$, and the process noise covariance matrix $\overline{\mathbf{Q}}_{k}=\mathbf{K}_{k}\mathbf{R}\mathbf{K}_{k}^{T}$.

\subsection{Performance analyses}\label{subsec:IEKF stability}
The treatment of linear filters includes filter stability and model error sensitivity. But, in general, stability and convergence results for non-linear KFs, and more so for their inverses, are difficult to obtain. Some convergence results were mentioned in \cite{krener2003convergenceOfEKF} for continuous-time non-linear Kalman filtering. In the case of EKF, sufficient conditions for stability of non-linear systems with linear output map were described in \cite{la1995conditionsforEKFforfreq}. Recently, the stability of deterministic EKF was studied based on contraction theory in \cite{bonnabel2014contraction}. The asymptotic convergence of EKF for a special class of systems, where EKF is applied for joint state and parameter estimation of linear stochastic systems, was studied in \cite{ljung1979asymptotic,ursin1980asymptotic}. If the non-linearities have known bounds, then the Riccati equation is slightly modified to guarantee stability for the continuous-time EKF \cite{reif1998ekf}.

To derive sufficient conditions for stochastic stability of non-linear filters, one of the common approaches is to introduce unknown instrumental matrices to account for the linearization errors \cite{xiong2006performance_ukf}. It does not assume any bound on the estimation error, but its sufficient conditions for stability, especially the bounds assumed on the unknown matrices, are difficult to verify for practical systems. Besides providing sufficient conditions for error boundedness, this approach also rigorously justifies the enlarging of the noise covariance matrices to stabilize the filter \cite{wu2007comments}. Alternatively, \cite{reif1999stochastic} considers the one-step prediction formulation of the filter and provides sufficient conditions under which the state prediction error is \textit{exponentially bounded in mean-squared} sense. This approach is based on bounded non-linearities, which has been earlier employed for proving stochastic stability of discrete-time \cite{reif1999stochastic} and continuous-time \cite{reif2000continuousEKFstability} EKFs.

In the following, we consider both the unknown matrix and bounded non-linearity techniques to analyze I-EKF's stability. Since the I-EKF's error dynamics depends on the forward filter's recursive updates, the derivations of these theoretical guarantees are not straightforward. In the process, we also derive the forward EKF stability conditions using the unknown matrix approach; note that the same was obtained using the bounded non-linearity method in \cite{reif1999stochastic}. Finally, we show that the I-EKF's estimates are conservative. Furthermore, we consider the general case of time-varying process and measurement noise covariances $\mathbf{Q}_{k}$, $\mathbf{R}_{k}$ and $\overline{\mathbf{R}}_{k}$ instead of $\mathbf{Q}$, $\mathbf{R}$ and $\bm{\Sigma}_{\epsilon}$ respectively.  Note that, analogous to the stability literature of KF, the following theorems (including those in subsequent chapters) provide sufficient but not necessary conditions for stability. Hence, these conditions are not verified for the example systems in the numerical experiments throughout the thesis. Furthermore, EKF (and other non-linear KF), are local approximation approaches \cite{li2017approximate} such that, unlike KF, the gain computation and covariance update steps are coupled with the state updates. Therefore, the asymptotic convergence of the filter is not defined for EKF and I-EKF. Instead, we show that the I-EKF provides conservative estimates in Theorem~\ref{thm:iekf consistency}.

We restate some definitions and a useful Lemma from \cite{reif1999stochastic}.
\begin{definition}[Exponential mean-squared boundedness \cite{reif1999stochastic}]\label{defn:exponential boundedness} A stochastic process $\{\bm{\zeta}_{k} \}_{k \geq 0}$ is defined to be exponentially bounded in mean-squared sense if there are real numbers $\eta,\nu>0$ and $0<\lambda<1$ such that $\mathbb{E}\left[\|\bm{\zeta}_{k}\|_{2}^{2}\right]\leq \eta\mathbb{E}\left[\|\bm{\zeta}_{0}\|_{2}^{2}\right]\lambda^{k}+\nu$ holds for every $k\geq 0$.
\end{definition}
\begin{definition}[Boundedness with probability one \cite{reif1999stochastic}] A stochastic process $\{\bm{\zeta}_{k} \}_{k \geq 0}$  is defined to be bounded with probability one if $\sup_{k\geq 0}\|\bm{\zeta}_{k}\|_{2} < \infty$ holds with probability one.  
\end{definition}
\begin{lemma}[Boundedness of stochastic process {\cite[Lemma 2.1]{reif1999stochastic}}]
\label{lemma:exponential boundedness}
Consider a function $V_{k}(\bm{\zeta}_{k})$ of the stochastic process $\bm{\zeta}_{k}$ and real numbers $v_{\textrm{min}}$, $v_{\textrm{max}}$, $\mu>0$, and $0<\lambda\leq 1$ such that $v_{\textrm{min}}\|\bm{\zeta}_{k}\|_{2}^{2}\leq V_{k}(\bm{\zeta}_{k})\leq v_{\textrm{max}}\|\bm{\zeta}_{k}\|_{2}^{2}$ and $\mathbb{E}\left[ V_{k+1}(\bm{\zeta}_{k+1})|\bm{\zeta}_{k}\right]-V_{k}(\bm{\zeta}_{k})\leq\mu-\lambda V_{k}(\bm{\zeta}_{k})$ for all $k\geq 0$. Then, the stochastic process $\{\bm{\zeta}_{k}\}_{k \geq 0}$ is exponentially bounded in mean-squared sense, i.e.,
\par\noindent\small
\begin{align*}
\mathbb{E}\left[\|\bm{\zeta}_{k}\|_{2}^{2}\right]\leq\frac{v_{\textrm{max}}}{v_{\textrm{min}}}\mathbb{E}\left[\|\bm{\zeta}_{0}\|_{2}^{2}\right](1-\lambda)^{k}+\frac{\mu}{v_{\textrm{min}}}\sum_{i=1}^{k-1}(1-\lambda)^{i},
\end{align*}
\normalsize
for every $k\geq 0$. Further, $\{\bm{\zeta}_{k}\}_{k \geq 0}$ is also bounded with probability one.
\end{lemma}
\begin{remark}[Bounded estimation error]\label{remark:reif result remark}
In the context of bounded non-linearity stability analysis, \cite[Sec. III]{reif1999stochastic} showed that, while the two-step prediction and update recursion (described in previous sections) and one-step formulation of (forward) filters may differ in their performance and transient behavior, they have similar convergence properties. However, the conditions of Lemma \ref{lemma:exponential boundedness} were proved to hold when the error remained within suitable bounds; the guarantees fail if the error exceeds this bound at any instant. However, it was numerically shown \cite[Sec. V]{reif1999stochastic} that the bound on the error was only of theoretical interest and, in practice, the filter remained stable for much larger estimation errors.
\end{remark}

\subsubsection{1) I-EKF's stability: Unknown matrix approach}\label{subsubsec:ekf stable unknown}
The I-EKF's estimation error can be shown to be exponentially bounded if the forward EKF is stable and the system satisfies some additional assumptions. Particularly, the I-EKF's error dynamics can be shown to satisfy the stability conditions of a standard EKF. Therefore, we first examine the stochastic stability of forward EKF.

\textit{Forward EKF stability:} Denote the forward EKF's state prediction, state estimation and measurement prediction errors by $\widetilde{\mathbf{x}}_{k+1|k}\doteq\mathbf{x}_{k+1}-\hat{\mathbf{x}}_{k+1|k}$, $\widetilde{\mathbf{x}}_{k}\doteq\mathbf{x}_{k}-\hat{\mathbf{x}}_{k}$ and $\widetilde{\mathbf{y}}_{k}\doteq\mathbf{y}_{k}-\hat{\mathbf{y}}_{k}$, with $\hat{\mathbf{y}}_{k}=h(\hat{\mathbf{x}}_{k|k-1})$, respectively. Using \eqref{eqn:non-linear state x}, \eqref{eqn:forward ekf predict} and the Taylor series expansion of $f(\cdot)$ at $\hat{\mathbf{x}}_{k}$, we get
\par\noindent\small
\begin{align*}
    \widetilde{\mathbf{x}}_{k+1|k}=\mathbf{F}_{k}(\mathbf{x}_{k}-\hat{\mathbf{x}}_{k})+\mathbf{w}_{k}+\mathcal{O}(\|\mathbf{x}_{k}-\hat{\mathbf{x}}_{k}\|_{2}^{2})\approx\mathbf{F}_{k}\widetilde{\mathbf{x}}_{k}+\mathbf{w}_{k}.
\end{align*}
\normalsize
To account for the residuals and obtain an exact equality, we introduce an unknown instrumental diagonal matrix $\mathbf{U}^{x}_{k} \in\mathbb{R}^{n_{x}\times n_{x}}$\cite{xiong2006performance_ukf,li2012stochastic_ukf} as
\par\noindent\small
\begin{align}
    \widetilde{\mathbf{x}}_{k+1|k}=\mathbf{U}^{x}_{k}\mathbf{F}_{k}\widetilde{\mathbf{x}}_{k}+\mathbf{w}_{k}\label{eqn:forward EKF predict error with alpha}.
\end{align}
\normalsize
However, using \eqref{eqn:forward ekf update}, we have $\widetilde{\mathbf{x}}_{k}=\widetilde{\mathbf{x}}_{k|k-1}-\mathbf{K}_{k}\widetilde{\mathbf{y}}_{k}$, which when substituted in \eqref{eqn:forward EKF predict error with alpha} yields $\widetilde{\mathbf{x}}_{k+1|k}=\mathbf{U}^{x}_{k}\mathbf{F}_{k}\widetilde{\mathbf{x}}_{k|k-1}-\mathbf{U}^{x}_{k}\mathbf{F}_{k}\mathbf{K}_{k}\widetilde{\mathbf{y}}_{k}+\mathbf{w}_{k}$. Similarly, using Taylor series expansion of $h(\cdot)$ at $\hat{\mathbf{x}}_{k+1|k}$ in \eqref{eqn:non-linear observation y} and introducing an unknown diagonal matrix $\mathbf{U}^{y}_{k+1}\in\mathbb{R}^{n_{y}\times n_{y}}$ gives $\widetilde{\mathbf{y}}_{k+1}=\mathbf{U}^{y}_{k+1}\mathbf{H}_{k+1}\widetilde{\mathbf{x}}_{k+1|k}+\mathbf{v}_{k+1}$. The prediction error dynamics of the forward EKF becomes
\par\noindent\small
\begin{align}
    \widetilde{\mathbf{x}}_{k+1|k}=\mathbf{U}^{x}_{k}\mathbf{F}_{k}(\mathbf{I}-\mathbf{K}_{k}\mathbf{U}^{y}_{k}\mathbf{H}_{k})\widetilde{\mathbf{x}}_{k|k-1}-\mathbf{U}^{x}_{k}\mathbf{F}_{k}\mathbf{K}_{k}\mathbf{v}_{k}+\mathbf{w}_{k}.\label{eqn:forward EKF prediction error dynamics}
\end{align}
\normalsize

Denote the true prediction covariance by  $\mathbf{P}_{k+1|k}=\mathbb{E}\left[\widetilde{\mathbf{x}}_{k+1|k}\widetilde{\mathbf{x}}_{k+1|k}^{T}\right]$. Define $\delta\mathbf{P}_{k+1|k}$ as the difference of estimated prediction covariance $\bm{\Sigma}_{k+1|k}$ and the true prediction covariance $\mathbf{P}_{k+1|k}$ while $\Delta\mathbf{P}_{k+1|k}$ as the error in the approximation of the expectation\\
\small
$\mathbb{E}\left[\mathbf{U}^{x}_{k}\mathbf{F}_{k}(\mathbf{I}-\mathbf{K}_{k}\mathbf{U}^{y}_{k}\mathbf{H}_{k})\widetilde{\mathbf{x}}_{k|k-1}\widetilde{\mathbf{x}}_{k|k-1}^{T}(\mathbf{I}-\mathbf{K}_{k}\mathbf{U}^{y}_{k}\mathbf{H}_{k})^{T}\mathbf{F}_{k}^{T}\mathbf{U}^{x}_{k}\right]$
\normalsize
by\\
\small
$\mathbf{U}^{x}_{k}\mathbf{F}_{k}(\mathbf{I}-\mathbf{K}_{k}\mathbf{U}^{y}_{k}\mathbf{H}_{k})\bm{\Sigma}_{k|k-1}(\mathbf{I}-\mathbf{K}_{k}\mathbf{U}^{y}_{k}\mathbf{H}_{k})^{T}\mathbf{F}_{k}^{T}\mathbf{U}^{x}_{k}$.
\normalsize
Denoting $\hat{\mathbf{Q}}_{k}=\mathbf{Q}_{k}+\mathbf{U}^{x}_{k}\mathbf{F}_{k}\mathbf{K}_{k}\mathbf{R}_{k}\mathbf{K}_{k}^{T}\mathbf{F}_{k}^{T}\mathbf{U}^{x}_{k}+\delta\mathbf{P}_{k+1|k}+\Delta\mathbf{P}_{k+1|k}$ and following similar steps as in \cite{xiong2006performance_ukf, li2012stochastic_ukf}, we have
\par\noindent\small
\begin{align}
    &\bm{\Sigma}_{k+1|k}=\mathbf{U}^{x}_{k}\mathbf{F}_{k}(\mathbf{I}-\mathbf{K}_{k}\mathbf{U}^{y}_{k}\mathbf{H}_{k})\bm{\Sigma}_{k|k-1}(\mathbf{I}-\mathbf{K}_{k}\mathbf{U}^{y}_{k}\mathbf{H}_{k})^{T}\mathbf{F}_{k}^{T}\mathbf{U}^{x}_{k}+\hat{\mathbf{Q}}_{k}.\label{eqn:forward ekf stable sig}
\end{align}
\normalsize

Similarly, denoting the true measurement prediction covariance and true cross-covariance by $\mathbf{P}^{yy}_{k+1}$ and $\mathbf{P}^{xy}_{k+1}$, respectively, we obtain
\par\noindent\small
\begin{align}
    \bm{\Sigma}^{y}_{k+1}&=\mathbf{U}^{y}_{k+1}\mathbf{H}_{k+1}\bm{\Sigma}_{k+1|k}\mathbf{H}_{k+1}^{T}\mathbf{U}^{y}_{k+1}+\hat{\mathbf{R}}_{k+1},\label{eqn:forward ekf stable sig y}\\
    \bm{\Sigma}^{xy}_{k+1}&=\begin{cases}\bm{\Sigma}_{k+1|k}\mathbf{U}^{xy}_{k+1}\mathbf{H}_{k+1}^{T}\mathbf{U}^{y}_{k+1}, & n_{x}\geq n_{y}\\
    \bm{\Sigma}_{k+1|k}\mathbf{H}_{k+1}^{T}\mathbf{U}^{y}_{k+1}\mathbf{U}^{xy}_{k+1}, & n_{x}<n_{y}\end{cases},\label{eqn:forward ekf stable sig xy}
\end{align}
\normalsize
where $\hat{\mathbf{R}}_{k+1}=\mathbf{R}_{k+1}+\Delta\mathbf{P}^{yy}_{k+1}+\delta\mathbf{P}^{yy}_{k+1}$ and $\mathbf{U}^{xy}_{k+1}$ is an unknown instrumental matrix introduced to account for errors in the estimated cross-covariance $\bm{\Sigma}_{k+1}^{xy}$\cite{xiong2007authorreply}.

The following Theorem~\ref{theorem:Forward ekf stable unknown matrix} provides stability conditions for the forward EKF using the unknown matrices $\mathbf{U}^{x}_{k}$, $\mathbf{U}^{y}_{k}$ and $\mathbf{U}^{xy}_{k}$.
\begin{theorem}[Stochastic stability of forward EKF]
\label{theorem:Forward ekf stable unknown matrix}
Consider the non-linear stochastic system in \eqref{eqn:non-linear state x} and \eqref{eqn:non-linear observation y}, and the two-step prediction-update forward EKF formulation. Let the following assumptions hold true:\\
\textbf{C3.a)} There exist positive real numbers $\overline{f}$, $\overline{h}$, $\overline{\alpha}$, $\overline{\beta}$, $\overline{\gamma}$, $\underline{\sigma}$, $\overline{\sigma}$, $\overline{q}$, $\overline{r}$, $\hat{q}$ and $\hat{r}$ such that the following bounds are fulfilled for all $k\geq 0$.
    \par\noindent\small
    \begin{align*}
        &\|\mathbf{F}_{k}\|\leq\overline{f},\;\;\|\mathbf{H}_{k}\|\leq\overline{h},\;\;\|\mathbf{U}^{x}_{k}\|\leq\overline{\alpha},\;\;\|\mathbf{U}^{y}_{k}\|\leq\overline{\beta},\;\;\|\mathbf{U}^{xy}_{k}\|\leq\overline{\gamma},\;\;\mathbf{Q}_{k}\preceq\overline{q}\mathbf{I},\;\;    \mathbf{R}_{k}\preceq\overline{r}\mathbf{I},\;\;\hat{q}\mathbf{I}\preceq\hat{\mathbf{Q}}_{k},\\
        &\hat{r}\mathbf{I}\preceq\hat{\mathbf{R}}_{k},\;\;
        \underline{\sigma}\mathbf{I}\preceq\bm{\Sigma}_{k|k-1}\preceq\overline{\sigma}\mathbf{I}.
    \end{align*}
\textbf{C3.b)} $\mathbf{U}^{x}_{k}$ and $\mathbf{F}_{k}$ are non-singular for every $k\geq 0$.\\
\textbf{C3.c)} The constants satisfy the inequality $\overline{\sigma}\overline{\gamma}\overline{h}^{2}\overline{\beta}^{2}<\hat{r}$.\\
Then, the prediction error $\widetilde{\mathbf{x}}_{k|k-1}$ and the estimation error $\widetilde{\mathbf{x}}_{k}$ of the forward EKF are exponentially bounded in mean-squared sense and bounded with probability one.
\end{theorem}
\begin{proof}
See Appendix~\ref{App-thm-Forward ekf stable unknown matrix}.
\end{proof}

\textit{Inverse EKF stability:} For a stable forward EKF, we prove the stochastic stability of the I-EKF as an extension of Theorem \ref{theorem:Forward ekf stable unknown matrix}. Similar to the forward EKF, we introduce unknown matrices $\overline{\mathbf{U}}^{x}_{k}$ and $\overline{\mathbf{U}}^{a}_{k}$ to account for the errors in the linearization of functions $\widetilde{f}_{k}(\cdot)$ and $g(\cdot)$, respectively, and $\overline{\mathbf{U}}^{xa}_{k}$ for the errors in cross-covariance matrix estimation. Similarly, denote $\hat{\overline{\mathbf{Q}}}_{k}$ and $\hat{\overline{\mathbf{R}}}_{k}$ as the counterparts of $\hat{\mathbf{Q}}_{k}$ and $\hat{\mathbf{R}}_{k}$, respectively, in the I-EKF dynamics. The following Theorem~\ref{theorem: inverse EKF stable unknown matrix} states the stability criteria for I-EKF. Note that, when compared  to Theorem~\ref{theorem:Forward ekf stable unknown matrix}, the following result requires an additional condition $\underline{r}\mathbf{I}\preceq\mathbf{R}_{k}$ for all $k\geq 0$ for some $\underline{r}>0$.
\begin{theorem}[Stochastic stability of I-EKF]
\label{theorem: inverse EKF stable unknown matrix}
Consider the attacker's forward EKF that is stable as per Theorem \ref{theorem:Forward ekf stable unknown matrix}. Additionally, assume that the following hold true for all $k\geq 0$.
\par\noindent\small
\begin{align*}
    &\underline{r}\mathbf{I}\preceq\mathbf{R}_{k},\;\;\|\mathbf{G}_{k}\|\leq\overline{g},\;\;\|\overline{\mathbf{U}}^{a}_{k}\|\leq\overline{c},\;\;
    \|\overline{\mathbf{U}}^{xa}_{k}\|\leq\overline{d},\;\;\overline{\mathbf{R}}_{k}\preceq\overline{\epsilon}\mathbf{I},\;\;\hat{c}\mathbf{I}\preceq\hat{\overline{\mathbf{Q}}}_{k},\;\;   \hat{d}\mathbf{I}\preceq\hat{\overline{\mathbf{R}}}_{k},\;\;
    \underline{p}\mathbf{I}\preceq\overline{\bm{\Sigma}}_{k|k-1}\preceq\overline{p}\mathbf{I},
\end{align*}
\normalsize
for some real positive constants $\underline{r}, \overline{g}, \overline{c}, \overline{d}, \overline{\epsilon}, \hat{c}, \hat{d}, \underline{p}, \overline{p}$. Then, the state estimation error of I-EKF is exponentially bounded in mean-squared sense and bounded with probability one provided that the constants satisfy the inequality
  $\overline{p}\overline{d}\overline{g}^{2}\overline{c}^{2}<{\hat{d}}$. 
\end{theorem}
\begin{proof}
See Appendix~\ref{App-thm-inverse EKF stable unknown matrix}.
\end{proof}

Note that Theorem \ref{theorem:Forward ekf stable unknown matrix} requires both $\hat{\mathbf{Q}}_{k}$ and $\hat{\mathbf{R}}_{k}$ to be positive definite (p.d.). In general, the difference matrices $\Delta\mathbf{P}_{k+1|k}$, $\delta\mathbf{P}_{k+1|k}$, $\Delta\mathbf{P}^{yy}_{k+1}$ and $\delta\mathbf{P}^{yy}_{k+1}$ may not be p.d. One could enhance the stability of EKF by enlarging the noise covariance matrices by adding sufficiently large $\Delta\mathbf{Q}_{k}$ and $\Delta\mathbf{R}_{k}$ to $\mathbf{Q}_{k}$ and $\mathbf{R}_{k}$, respectively \cite{xiong2006performance_ukf,xiong2007authorreply}. The same argument also holds true for I-EKF noise covariance matrices.

\subsubsection{2) I-EKF's stability: Bounded non-linearity method}\label{subsubsec:ekf stable Reif}
\textit{Forward filter:} Consider the forward EKF's one step prediction formulation \eqref{eqn: one step forward ekf gain}-\eqref{eqn: one step forward ekf covariance}. Using Taylor series expansion around the estimate $\hat{\mathbf{x}}_{k}$, we have
\par\noindent\small
\begin{align*}
&f(\mathbf{x}_{k})-f(\hat{\mathbf{x}}_{k})=\mathbf{F}_{k}(\mathbf{x}_{k}-\hat{\mathbf{x}}_{k})+\phi(\mathbf{x}_{k},\hat{\mathbf{x}}_{k}),\\
&h(\mathbf{x}_{k})-h(\hat{\mathbf{x}}_{k})=\mathbf{H}_{k}(\mathbf{x}_{k}-\hat{\mathbf{x}}_{k})+\chi(\mathbf{x}_{k},\hat{\mathbf{x}}_{k}),
\end{align*}
\normalsize
where $\phi(\cdot)$ and $\chi(\cdot)$ are suitable non-linear functions to account for the higher-order terms of the expansions. Denoting the estimation error by $\mathbf{e}_{k}\doteq\mathbf{x}_{k}-\hat{\mathbf{x}}_{k}$, the error dynamics of the forward filter is
\par\noindent\small
\begin{align}
\mathbf{e}_{k+1}=(\mathbf{F}_{k}-\mathbf{K}_{k}\mathbf{H}_{k})\mathbf{e}_{k}+\mathbf{r}_{k}+\mathbf{s}_{k},\label{eqn: forward ekf error}
\end{align}
\normalsize
where $\mathbf{r}_{k}=\phi(\mathbf{x}_{k},\hat{\mathbf{x}}_{k})-\mathbf{K}_{k}\chi(\mathbf{x}_{k},\hat{\mathbf{x}}_{k})$ and $\mathbf{s}_{k}=\mathbf{w}_{k}-\mathbf{K}_{k}\mathbf{v}_{k}$. The following Theorem~\ref{theorem: ekf stable Reif} (reproduced from \cite{reif1999stochastic}) provides sufficient conditions for forward EKF's stochastic stability.
\begin{theorem}[Exponential boundedness of forward EKF's error \cite{reif1999stochastic}]\label{theorem: ekf stable Reif} Consider a non-linear stochastic system defined by \eqref{eqn:non-linear state x} and \eqref{eqn:non-linear observation y}, and the one-step prediction formulation of forward EKF \eqref{eqn: one step forward ekf gain}-\eqref{eqn: one step forward ekf covariance}. Let the following assumptions hold true.\\
\textbf{C3.d)} There exist positive real numbers $\overline{f}$,$\overline{h}$,$\underline{\sigma}$,$\overline{\sigma}$,$\underline{q}$,$\underline{r}$, $\delta$ such that the following bounds are fulfilled for all $k\geq 0$.
\par\noindent\small
\begin{align*}
&\underline{\sigma}\mathbf{I}\preceq\bm{\Sigma}_{k}\preceq\overline{\sigma}\mathbf{I},\;\;\underline{q}\mathbf{I}\preceq \mathbf{Q}_{k}\preceq\delta\mathbf{I},\;\;\underline{r}\mathbf{I}\preceq \mathbf{R}_{k}\preceq\delta\mathbf{I},\;\;\|\mathbf{F}_{k}\|\leq\overline{f},\;\;\|\mathbf{H}_{k}\|\leq\overline{h}.
\end{align*}
\normalsize
\textbf{C3.e)} $\mathbf{F}_{k}$ is non singular for every $k\geq 0$.\\
\textbf{C3.f)} There exist positive real numbers $\kappa_{\phi}$, $\epsilon_{\phi}$, $\kappa_{\chi}$, $\epsilon_{\chi}$ such that the non-linear functions $\phi(\cdot)$ and $\chi(\cdot)$ satisfy
\par\noindent\small
\begin{align*}
&\|\phi(\mathbf{x},\hat{\mathbf{x}})\|_{2}\leq \kappa_{\phi}\|\mathbf{x}-\hat{\mathbf{x}}\|_{2}^{2}\;\;\;\text{for}\;\;\;\|\mathbf{x}-\hat{\mathbf{x}}\|_{2}\leq\epsilon_{\phi},\\
&\|\chi(\mathbf{x},\hat{\mathbf{x}})\|_{2}\leq \kappa_{\chi}\|\mathbf{x}-\hat{\mathbf{x}}\|_{2}^{2}\;\;\;\text{for}\;\;\;\|\mathbf{x}-\hat{\mathbf{x}}\|_{2}\leq\epsilon_{\chi}.
\end{align*}
\normalsize
Then, the estimation error given by \eqref{eqn: forward ekf error} is exponentially bounded in mean-squared sense provided that the estimation error is bounded by suitable constant $\epsilon>0$. 
\end{theorem}
Theorem~\ref{theorem: ekf stable Reif} guarantees that the estimation error remains exponentially bounded in mean-squared sense as long as the error is within suitable $\epsilon$ bounds. Further, the mean drift $\mathbb{E}[V_{k+1}(\mathbf{e}_{k+1})|\mathbf{e}_{k}]\\-V_{k}(\mathbf{e}_{k})$ for a suitably defined $V_{k}(\cdot)$ (for application of Lemma \ref{lemma:exponential boundedness}) is negative when $\widetilde{\epsilon}\leq\|\mathbf{e}_{k}\|_{2}\leq\epsilon$, which drives the system towards zero error in an expected sense. However, with some finite probability, the estimation error at some time-steps may be outside the $\epsilon$ bound. In this case, we cannot guarantee with probability one that the error will be within $\epsilon$ bound again at some future time-steps. As mentioned in Remark~\ref{remark:reif result remark}, bounded non-linearity approach may not provide theoretical guarantees for the filter to be stable for all time-steps but, practically, the filter remains stable even if the estimation error is outside the $\epsilon$ bound provided that the assumed bounds on the system model are satisfied.

\textit{Inverse filter:} For the inverse filter observations \eqref{eqn:non-linear observation a}, the Taylor series expansion of $g(\cdot)$ at estimate $\doublehat{\mathbf{x}}_{k}$ of I-EKF's one step prediction formulation, considering suitable non-linear function $\overline{\chi}(\cdot)$ is
\par\noindent\small
\begin{align*}
&g(\hat{\mathbf{x}}_{k})-g(\doublehat{\mathbf{x}}_{k})=\mathbf{G}_{k}(\hat{\mathbf{x}}_{k}-\doublehat{\mathbf{x}}_{k})+\overline{\chi}(\hat{\mathbf{x}}_{k},\doublehat{\mathbf{x}}_{k}).
\end{align*}
\normalsize
Finally, the error dynamics of the inverse filter, with the estimation error denoted by $\overline{\mathbf{e}}_{k}\doteq\hat{\mathbf{x}}_{k}-\doublehat{\mathbf{x}}_{k}$ and the inverse filter's Kalman gain and estimation error covariance matrix by $\overline{\mathbf{K}}_{k}$ and $\overline{\bm{\Sigma}}_{k}$, respectively, is
\par\noindent\small
\begin{align}
\overline{\mathbf{e}}_{k+1}=(\widetilde{\mathbf{F}}^{x}_{k}-\overline{\mathbf{K}}_{k}\mathbf{G}_{k})\overline{\mathbf{e}}_{k}+\overline{\mathbf{r}}_{k}+\overline{\mathbf{s}}_{k},\label{eqn: inverse ekf error}
\end{align}
\normalsize
where $\overline{\mathbf{r}}_{k}=\overline{\phi}_{k}(\hat{x}_{k},\doublehat{\mathbf{x}}_{k})-\overline{\mathbf{K}}_{k}\overline{\chi}(\hat{\mathbf{x}}_{k},\doublehat{\mathbf{x}}_{k})$ and $\overline{\mathbf{s}}_{k}=\mathbf{K}_{k}\mathbf{v}_{k}-\overline{\mathbf{K}}_{k}\bm{\epsilon}_{k}$ with $\overline{\phi}_{k}(\hat{\mathbf{x}}_{k},\doublehat{\mathbf{x}}_{k})=\phi(\hat{\mathbf{x}}_{k},\doublehat{\mathbf{x}}_{k})-\mathbf{K}_{k}\chi(\hat{\mathbf{x}}_{k},\doublehat{\mathbf{x}}_{k})$.

The following Theorem~\ref{theorem: inverse ekf stable Reif} guarantees the stability of I-EKF. Note the additional assumption of $\mathbf{H}_{k}$ to be full column rank for all $k\geq 0$, which implies $n_{y}\geq n_{x}$.
\begin{theorem}[Exponential boundedness of I-EKF's error]\label{theorem: inverse ekf stable Reif} 
Consider the attacker's forward one-step prediction EKF that is stable as per Theorem \ref{theorem: ekf stable Reif}. Additionally, assume that the following hold true.\\
\textbf{C3.g)} There exist positive real numbers $\overline{g}$, \underline{$m$}, $\overline{m}$, $\underline{\epsilon}$, $\overline{\delta}$ such that the following bounds are fulfilled for all $k\geq 0$.
\par\noindent\small
\begin{align*}
&\|\mathbf{G}_{k}\|\le\overline{g},\;\;\underline{m}\mathbf{I}\preceq\overline{\bm{\Sigma}}_{k}\preceq\overline{m}\mathbf{I},\;\;\underline{\epsilon}\mathbf{I}\preceq \overline{\mathbf{R}}_{k}\preceq\overline{\delta}\mathbf{I}.
\end{align*}
\normalsize
\textbf{C3.h)} $\mathbf{H}_{k}$ is full column rank for every $k\geq 0$.\\
\textbf{C3.i)} There exist positive real numbers $\kappa_{\bar{\chi}}$ and $\epsilon_{\bar{\chi}}$ such that the non-linear function $\overline{\chi}(\cdot)$ satisfies
\par\noindent\small
\begin{align*}
\|\overline{\chi}(\hat{\mathbf{x}},\doublehat{\mathbf{x}})\|_{2}\leq \kappa_{\bar{\chi}}\|\hat{\mathbf{x}}-\doublehat{\mathbf{x}}\|_{2}^{2}\;\;\;\text{for}\;\;\;\|\hat{\mathbf{x}}-\doublehat{\mathbf{x}}\|_{2}\leq\epsilon_{\bar{\chi}}.
\end{align*}
\normalsize
Then, the estimation error for I-EKF given by \eqref{eqn: inverse ekf error} is exponentially bounded in mean-squared sense provided that the estimation error is bounded by suitable constant $\overline{\epsilon}>0$.
\end{theorem}
\begin{proof}
See Appendix~\ref{App-thm-inverse ekf stable Reif}.
\end{proof}
\begin{remark}[Observability of the system]\label{remark:observability}
The inequality $\underline{m}\mathbf{I}\preceq\overline{\bm{\Sigma}}_{k}\preceq\overline{m}\mathbf{I}$ assumed in Theorem~\ref{theorem: inverse ekf stable Reif} is closely related to the observability of the non-linear inverse filtering model. In particular, \cite{reif1999stochastic} showed that the condition $\underline{\sigma}\mathbf{I}\preceq\bm{\Sigma}_{k}\preceq\overline{\sigma}\mathbf{I}$ assumed for forward EKF's stability in Theorem~\ref{theorem: ekf stable Reif} is satisfied if the non-linear observability rank condition holds, i.e., the non-linear observability matrix $\mathbf{A}_{k}=\begin{bsmallmatrix}\nabla h(\mathbf{x}_{k})\\\nabla h(\mathbf{x}_{k+1})\nabla f(\mathbf{x}_{k})\\\vdots\\\nabla h(\mathbf{x}_{k+n_{x}-1})\nabla f(\mathbf{x}_{k+n_{x}-2})\hdots\nabla f(\mathbf{x}_{k})\end{bsmallmatrix}$ has full rank $n_{x}$ at $\mathbf{x}_{k}$. Note that the inequality $\underline{m}\mathbf{I}\preceq\overline{\bm{\Sigma}}_{k}\preceq\overline{m}\mathbf{I}$ is a weaker assumption than the observability. The same argument holds for the inequality $\underline{m}\mathbf{I}\preceq\overline{\bm{\Sigma}}_{k}\preceq\overline{m}\mathbf{I}$ for I-EKF's stability.
\end{remark}

\subsubsection{3) I-EKF's conservativeness}\label{subsec:iekf consistency}
Recall the definition of a conservative estimate pair $(\hat{\mathbf{x}},\bm{\Sigma})$. Note that \cite{battistelli2014kullback} defines the same as a consistent estimator.
\begin{definition}[Conservative estimate\cite{battistelli2014kullback}]\label{def:consistency}
Consider an unbiased estimate $\hat{\mathbf{x}}$ of random variable $\mathbf{x}$ and its error covariance estimate $\bm{\Sigma}$. The pair ($\hat{\mathbf{x}}$,$\bm{\Sigma}$) are said to be conservative if $\mathbb{E}[(\mathbf{x}-\hat{\mathbf{x}})(\mathbf{x}-\hat{\mathbf{x}})^{T}]\preceq\bm{\Sigma}$, i.e., the estimated covariance $\bm{\Sigma}$ upper bounds the true error covariance. 
\end{definition}
To prove that I-EKF is a conservative estimator, we consider the statistical linearization technique (SLT)\cite{arasaratnam2007qkf}. Linearize the state transition \eqref{eqn:iekf state transition} and observation \eqref{eqn:non-linear observation a}, respectively, at $[\hat{\mathbf{x}}_{k}^{T},\mathbf{v}_{k+1}^{T}]^{T}$ and $\hat{\mathbf{x}}_{k}$ as
\par\noindent\small
\begin{align}
    &\hat{\mathbf{x}}_{k+1}=\mathbf{U}^{xv}_{k}\overline{\mathbf{F}}^{x}_{k}\hat{\mathbf{x}}_{k}+\mathbf{U}^{xv}_{k}\overline{\mathbf{F}}^{v}_{k}\mathbf{v}_{k+1},\label{eqn:SLT state transition}\\
    &\mathbf{a}_{k}=\mathbf{U}^{a}_{k}\overline{\mathbf{G}}_{k}\hat{\mathbf{x}}_{k}+\bm{\epsilon}_{k},\label{eqn:SLT observation}
\end{align}
\normalsize
where $\overline{\mathbf{F}}_{k}=[\overline{\mathbf{F}}^{x}_{k},\overline{\mathbf{F}}^{v}_{k}]$ and $\overline{\mathbf{G}}_{k}$ are the respective linear pseudo transition matrices. Also, $\mathbf{U}^{xv}_{k}$ and $\mathbf{U}^{a}_{k}$ are unknown diagonal matrices introduced to account for the approximation errors in SLT. Note that these unknown matrices are different from the ones introduced earlier in the unknown matrix stability approach to account for the higher-order terms in the Taylor approximation.
\begin{theorem}[I-EKF's conservative estimates]
    \label{thm:iekf consistency}
    Consider an I-EKF initialized with a conservative initial estimate pair $(\doublehat{\mathbf{x}}_{0},\overline{\bm{\Sigma}}_{0})$. Then for any $k\geq 1$, the estimate $(\doublehat{\mathbf{x}}_{k},\overline{\bm{\Sigma}}_{k})$ computed recursively by the I-EKF are also conservative such that $\mathbb{E}[(\hat{\mathbf{x}}_{k}-\doublehat{\mathbf{x}}_{k})(\hat{\mathbf{x}}_{k}-\doublehat{\mathbf{x}}_{k})^{T}]\preceq\overline{\bm{\Sigma}}_{k}$, where $\hat{\mathbf{x}}_{k}$ is the forward EKF's state estimate.
\end{theorem}
\begin{proof}
    See Appendix~\ref{App-thm-inverse EKF consistency}.
\end{proof}

\section{Inverse SOEKF}\label{sec:I-SOEKF}
\subsection{Filter formulation}\label{subsec:I-SOEKF formulation}
While the EKF formulations are limited to only first-order Taylor series expansion, the forward and inverse filters of SOEKF also include second-order terms. For I-SOEKF formulation, we assume that the attacker employs a forward SOEKF to estimate the defender's state. Again, we define the Jacobians as $\mathbf{F}_{k}\doteq\nabla_{\mathbf{x}}f(\mathbf{x})\vert_{\mathbf{x}=\hat{\mathbf{x}}_{k}}$ and 
$\mathbf{H}_{k+1}\doteq\nabla_{\mathbf{x}}h(\mathbf{x})\vert_{\mathbf{x}=\hat{\mathbf{x}}_{k+1|k}}$.

\textit{Forward filter:} Denoting the $i$-th Euclidean basis vectors in $\mathbb{R}^{n_{x}\times 1}$ and $\mathbb{R}^{n_{y}\times 1}$ by $\widetilde{\mathbf{a}}_{i}$ and $\mathbf{b}_{i}$, respectively, the forward SOEKF recursions are \cite{bar2004estimation}
\par\noindent\small
\begin{align}
&\textit{Time update:}\;\;\hat{\mathbf{x}}_{k+1|k}=f(\hat{\mathbf{x}}_{k})+\frac{1}{2}\sum_{i=1}^{n_{x}}\widetilde{\mathbf{a}}_{i}\textrm{Tr}\left(\nabla^{2}\left[f(\hat{\mathbf{x}}_{k})\right]_{i}\bm{\Sigma}_{k}\right),\label{eqn: forward SOEKF x prediction}\\
&\bm{\Sigma}_{k+1|k}=\mathbf{F}_{k}\bm{\Sigma}_{k}\mathbf{F}_{k}^{T}+\frac{1}{2}\sum_{i=1}^{n_{x}}\sum_{j=1}^{n_{x}}\widetilde{\mathbf{a}}_{i}\widetilde{\mathbf{a}}_{j}^{T}\textrm{Tr}\left(\nabla^{2}\left[f(\hat{\mathbf{x}}_{k})\right]_{i}\bm{\Sigma}_{k}\nabla^{2}\left[f(\hat{\mathbf{x}}_{k})\right]_{j}\bm{\Sigma}_{k}\right)+\mathbf{Q},\nonumber\\
&\textit{Measurement update:}\;\;\hat{\mathbf{y}}_{k+1|k}=h(\hat{\mathbf{x}}_{k+1|k})+\frac{1}{2}\sum_{i=1}^{n_{y}}\mathbf{b}_{i}\textrm{Tr}\left(\nabla^{2}\left[h(\hat{\mathbf{x}}_{k+1|k})\right]_{i}\bm{\Sigma}_{k+1|k}\right),\label{eqn: forward SOEKF y prediction}\\
&\bm{\Sigma}^{y}_{k+1}=\mathbf{H}_{k+1}\bm{\Sigma}_{k+1|k}\mathbf{H}_{k+1}^{T}+\frac{1}{2}\sum_{i=1}^{n_{y}}\sum_{j=1}^{n_{y}}\mathbf{b}_{i}\mathbf{b}_{j}^{T}\textrm{Tr}\left(\nabla^{2}\left[h(\hat{\mathbf{x}}_{k+1|k})\right]_{i}\bm{\Sigma}_{k+1|k}\nabla^{2}\left[h(\hat{\mathbf{x}}_{k+1|k})\right]_{j}\bm{\Sigma}_{k+1|k}\right)+\mathbf{R},\nonumber\\
&\mathbf{K}_{k+1}=\bm{\Sigma}_{k+1|k}\mathbf{H}_{k+1}^{T}(\bm{\Sigma}^{y}_{k+1})^{-1},\nonumber
\end{align}
\begin{align}
&\textit{Measurement update (contd.):}\;\;\hat{\mathbf{x}}_{k+1}=\hat{\mathbf{x}}_{k+1|k}+\mathbf{K}_{k+1}(\mathbf{y}_{k+1}-\hat{\mathbf{y}}_{k+1|k}),\label{eqn: forward SOEKF x update}\\
&\bm{\Sigma}_{k+1}=\bm{\Sigma}_{k+1|k}-\bm{\Sigma}_{k+1|k}\mathbf{H}_{k+1}^{T}(\bm{\Sigma}^{y}_{k+1})^{-1}\mathbf{H}_{k+1}\bm{\Sigma}_{k+1|k},\nonumber
\end{align}
\normalsize
where $\mathbf{K}_{k+1}$ is SOEKF's gain matrix. Unlike EKF, the SOEKF includes Hessian terms in the prediction equations.

\textit{Inverse filter:} Define $\widetilde{f}(\hat{\mathbf{x}},\bm{\Sigma})\doteq f(\hat{\mathbf{x}})+\frac{1}{2}\sum_{i=1}^{n_{x}}\widetilde{\mathbf{a}}_{i}\textrm{Tr}\left(\nabla^{2}\left[f(\hat{\mathbf{x}})\right]_{i}\bm{\Sigma}\right)$ and $\widetilde{h}(\hat{\mathbf{x}},\bm{\Sigma})\doteq h(\hat{\mathbf{x}})+\frac{1}{2}\sum_{i=1}^{n_{y}}\mathbf{b}_{i}\textrm{Tr}\left(\nabla^{2}\left[h(\hat{\mathbf{x}})\right]_{i}\bm{\Sigma}\right)$. Using \eqref{eqn:non-linear observation y} and \eqref{eqn: forward SOEKF x prediction}-\eqref{eqn: forward SOEKF x update}, the state transition equation for the I-SOEKF is
\par\noindent\small
\begin{align*}
\hat{\mathbf{x}}_{k+1}&=\overline{f}_{k}(\hat{\mathbf{x}}_{k},\mathbf{x}_{k+1},\mathbf{v}_{k+1})=\widetilde{f}(\hat{\mathbf{x}}_{k},\bm{\Sigma}_{k})-\mathbf{K}_{k+1}\widetilde{h}(\widetilde{f}(\hat{\mathbf{x}}_{k},\bm{\Sigma_{k}}),\bm{\Sigma}_{k+1|k})+\mathbf{K}_{k+1}h(\mathbf{x}_{k+1})+\mathbf{K}_{k+1}\mathbf{v}_{k+1}.
\end{align*}
\normalsize

The I-SOEKF recursions now follow directly from forward SOEKF's recursions treating $\mathbf{a}_{k}$ from \eqref{eqn:non-linear observation a} as the observation. Denote the Jacobians with respect to the state estimate $\hat{\mathbf{x}}_{k}$ as $\overline{\mathbf{F}}_{k}\doteq\nabla_{\mathbf{x}}\overline{f}_{k}(\mathbf{x},\mathbf{x}_{k+1},\mathbf{0})|_{\mathbf{x}=\doublehat{\mathbf{x}}_{k}}$ and $\mathbf{G}_{k+1}\doteq\nabla_{\mathbf{x}}g(\mathbf{x})|_{\mathbf{x}=\doublehat{\mathbf{x}}_{k+1|k}}$, while the Hessians are denoted as $\nabla^{2}\left[\overline{f}_{k}(\doublehat{\mathbf{x}}_{k})\right]_{i}\doteq\nabla_{x}^{2}\left[\overline{f}_{k}(\mathbf{x},\mathbf{x}_{k+1},\mathbf{0})\right]_{i}|_{\mathbf{x}=\doublehat{\mathbf{x}}_{k}}$ and $\nabla^{2}[g(\doublehat{x}_{k+1|k})]_{i}=\nabla^{2}_{\mathbf{x}}[g(\mathbf{x})]_{i}|_{\mathbf{x}=\doublehat{\mathbf{x}}_{k+1|k}}$. The process noise covariance matrix is $\overline{\mathbf{Q}}_{k}=\overline{\mathbf{V}}_{k}\mathbf{R}\overline{\mathbf{V}}_{k}^{T}$ with $\overline{\mathbf{V}}_{k}\doteq\nabla_{v}\overline{f}_{k}(\doublehat{\mathbf{x}}_{k},\mathbf{x}_{k+1},\mathbf{v})|_{\mathbf{v}=\mathbf{0}}$. The I-SOEKF's state and observation predictions are
\par\noindent\small
\begin{align*}
&\doublehat{\mathbf{x}}_{k+1|k}=\overline{f}_{k}(\doublehat{\mathbf{x}}_{k},\mathbf{x}_{k+1},\mathbf{0})+\frac{1}{2}\sum_{i=1}^{n_{x}}\widetilde{\mathbf{a}}_{i}\textrm{Tr}\left(\nabla^{2}\left[\overline{f}_{k}(\doublehat{\mathbf{x}}_{k})\right]_{i}\overline{\bm{\Sigma}}_{k}\right),\\
&\overline{\bm{\Sigma}}_{k+1|k}=\overline{\mathbf{F}}_{k}\overline{\bm{\Sigma}}_{k}\overline{\mathbf{F}}_{k}^{T}+\frac{1}{2}\sum_{i=1}^{n_{x}}\sum_{j=1}^{n_{x}}\widetilde{\mathbf{a}}_{i}\widetilde{\mathbf{a}}_{j}^{T}\textrm{Tr}\left(\nabla^{2}\left[\overline{f}_{k}(\doublehat{\mathbf{x}}_{k})\right]_{i}\overline{\bm{\Sigma}}_{k}\nabla^{2}\left[\overline{f}_{k}(\doublehat{\mathbf{x}}_{k})\right]_{j}\overline{\bm{\Sigma}}_{k}\right)+\overline{\mathbf{Q}}_{k},\\
&\hat{\mathbf{a}}_{k+1|k}=g(\doublehat{\mathbf{x}}_{k+1|k})+\frac{1}{2}\sum_{i=1}^{n_{a}}\mathbf{d}_{i}\textrm{Tr}\left(\nabla^{2}\left[g(\doublehat{\mathbf{x}}_{k+1|k})\right]_{i}\overline{\bm{\Sigma}}_{k+1|k}\right),\\
&\overline{\bm{\Sigma}}^{a}_{k+1}=\mathbf{G}_{k+1}\overline{\bm{\Sigma}}_{k+1|k}\mathbf{G}_{k+1}^{T}+\frac{1}{2}\sum_{i=1}^{n_{a}}\sum_{j=1}^{n_{a}}\mathbf{d}_{i}\mathbf{d}_{j}^{T}\textrm{Tr}\left(\nabla^{2}\left[g(\doublehat{\mathbf{x}}_{k+1|k})\right]_{i}\overline{\bm{\Sigma}}_{k+1|k}\nabla^{2}[g(\doublehat{\mathbf{x}}_{k+1|k})]_{j}\overline{\bm{\Sigma}}_{k+1|k}\right)+\bm{\Sigma}_{\epsilon},
\end{align*}
\normalsize
where $\mathbf{d}_{i}$ denotes the $i$-th Euclidean basis vector in $\mathbb{R}^{n_{a}\times 1}$. After these prediction steps, the update steps to compute estimate $\doublehat{\mathbf{x}}_{k+1}$ and $\overline{\bm{\Sigma}}_{k+1}$ are same as \eqref{eqn:IEKF state update} and \eqref{eqn:IEKF covariance update}, respectively.

Here, besides the gain matrix $\mathbf{K}_{k+1}$, the Hessian terms are also treated as time-varying parameters of the state transition and approximated by evaluating them using the previous I-SOEKF's estimate $\doublehat{\mathbf{x}}_{k}$. When the second-order terms are neglected, the forward SOEKF and I-SOEKF reduce to, respectively, forward EKF and I-EKF. Similar to I-EKF, I-SOEKF also follows from general SOEKF recursions and hence, has similar computational complexity, i.e., $\mathcal{O}(n_{x}^{5})$ where $n_{x}$ is the state dimension. Note that a derivative-free efficient implementation \cite{roth2011efficientSOEKF} of SOEKF yields a $\mathcal{O}(n_{x}^{4})$ complexity.

\textit{One-step prediction formulation:} The one-step prediction formulation of SOEKF for the system model \eqref{eqn:non-linear state x}-\eqref{eqn:non-linear observation y} can be derived following the derivation of two-step recursion formulation of SOEKF as outlined in \cite{bar2004estimation}. Denoting the Jacobians as $\mathbf{F}_{k}\doteq\nabla_{\mathbf{x}}f(\mathbf{x})\vert_{\mathbf{x}=\hat{\mathbf{x}}_{k}}$ and $\mathbf{H}_{k}\doteq\nabla_{\mathbf{x}}h(\mathbf{x})\vert_{\mathbf{x}=\hat{\mathbf{x}}_{k}}$, the one-step prediction SOEKF recursions are
\par\noindent\small
\begin{align}
&\widetilde{\mathbf{x}}_{k}=f(\hat{\mathbf{x}}_{k})+\frac{1}{2}\sum_{i=1}^{n_{x}}\widetilde{\mathbf{a}}_{i}\textrm{Tr}\left(\nabla^{2}\left[f(\hat{\mathbf{x}}_{k})\right]_{i}\bm{\Sigma}_{k}\right),\label{eqn: one step SOEKF first update}
\end{align}
\begin{align}
&\widetilde{\bm{\Sigma}}_{k}=\mathbf{F}_{k}\bm{\Sigma}_{k}\mathbf{F}_{k}^{T}+\frac{1}{2}\sum_{i=1}^{n_{x}}\sum_{j=1}^{n_{x}}\widetilde{\mathbf{a}}_{i}\widetilde{\mathbf{a}}_{j}^{T}\textrm{Tr}\left(\nabla^{2}\left[f(\hat{\mathbf{x}}_{k})\right]_{i}\bm{\Sigma}_{k}\nabla^{2}\left[f(\hat{\mathbf{x}}_{k})\right]_{j}\bm{\Sigma}_{k}\right)+\mathbf{Q},\label{eqn: one step SOEKF sig tilde}\\
&\bm{\Sigma}^{y}_{k}=\mathbf{H}_{k}\bm{\Sigma}_{k}\mathbf{H}_{k}^{T}+\frac{1}{2}\sum_{i=1}^{n_{y}}\sum_{j=1}^{n_{y}}\mathbf{b}_{i}\mathbf{b}_{j}^{T}\textrm{Tr}\left(\nabla^{2}\left[h(\hat{\mathbf{x}}_{k})\right]_{i}\bm{\Sigma}_{k}\nabla^{2}\left[h(\hat{\mathbf{x}}_{k})\right]_{j}\bm{\Sigma}_{k}\right)+\mathbf{R},\\
&\mathbf{M}_{k}=\frac{1}{2}\sum_{i=1}^{n_{x}}\sum_{j=1}^{n_{y}}\widetilde{\mathbf{a}}_{i}\mathbf{b}_{j}^{T}\textrm{Tr}\left(\nabla^{2}\left[f(\hat{\mathbf{x}}_{k})\right]_{i}\bm{\Sigma}_{k}\nabla^{2}\left[h(\hat{\mathbf{x}}_{k})\right]_{j}\bm{\Sigma}_{k}\right),\\
&\mathbf{K}_{k}=(\mathbf{F}_{k}\bm{\Sigma}_{k}\mathbf{H}_{k}^{T}+\mathbf{M}_{k})(\bm{\Sigma}^{y}_{k})^{-1},\label{eqn: one step SOEKF Kk}\\
&\hat{\mathbf{x}}_{k+1}=\widetilde{\mathbf{x}}_{k}+\mathbf{K}_{k}\left(\mathbf{y}_{k}-h(\hat{\mathbf{x}}_{k})-\frac{1}{2}\sum_{i=1}^{n_{y}}\mathbf{b}_{i}\textrm{Tr}\left(\nabla^{2}\left[h(\hat{\mathbf{x}}_{k})\right]_{i}\bm{\Sigma}_{k}\right)\right),\label{eqn: one step SOEKF final update}\\
&\bm{\Sigma}_{k+1}=\widetilde{\bm{\Sigma}}_{k}-\mathbf{K}_{k}\bm{\Sigma}^{y}_{k}\mathbf{K}_{k}^{T}.\label{eqn: one step SOEKF sig update}
\end{align}
\normalsize

From \eqref{eqn:non-linear observation y}, \eqref{eqn: one step SOEKF first update} and \eqref{eqn: one step SOEKF final update}, the state transition equation for one-step formulation of I-SOEKF is $\hat{\mathbf{x}}_{k+1}\doteq\overline{f}_{k}(\hat{\mathbf{x}}_{k},\mathbf{x}_{k},\mathbf{v}_{k})$ where
\par\noindent\small
\begin{align}
    \overline{f}_{k}(\hat{\mathbf{x}}_{k},\mathbf{x}_{k},\mathbf{v}_{k})&=f(\hat{\mathbf{x}}_{k})-\mathbf{K}_{k}h(\hat{\mathbf{x}}_{k})+\frac{1}{2}\sum_{i=1}^{n_{x}}\widetilde{\mathbf{a}}_{i}\textrm{Tr}\left(\nabla^{2}\left[f(\hat{\mathbf{x}}_{k})\right]_{i}\bm{\Sigma}_{k}\right)-\frac{1}{2}\mathbf{K}_{k}\sum_{i=1}^{n_{y}}\mathbf{b}_{i}\textrm{Tr}\left(\nabla^{2}\left[h(\hat{\mathbf{x}}_{k})\right]_{i}\bm{\Sigma}_{k}\right)\nonumber\\
    &+\mathbf{K}_{k}h(\mathbf{x}_{k})+\mathbf{K}_{k}\mathbf{v}_{k}.\label{eqn: one step I-SOEKF state transition}
\end{align}
\normalsize
Considering the aforementioned state transition and observation \eqref{eqn:non-linear observation a}, the inverse filter follows, \textit{mutatis mutandis}, from SOEKF's one-step prediction formulation. The Jacobians with respect to the state estimate are $\overline{\mathbf{F}}_{k}\doteq\nabla_{\mathbf{x}}\overline{f}_{k}(\mathbf{x},\mathbf{x}_{k},\mathbf{0})\vert_{\mathbf{x}=\doublehat{\mathbf{x}}_{k}}=\mathbf{F}_{k}-\mathbf{K}_{k}\mathbf{H}_{k}$ and $\mathbf{G}_{k}\doteq\nabla_{\mathbf{x}}g(\mathbf{x})\vert_{\mathbf{x}=\doublehat{\mathbf{x}}_{k}}$, with the process noise covariance matrix $\overline{\mathbf{Q}}_{k}=\mathbf{K}_{k}\mathbf{R}_{k}\mathbf{K}_{k}^{T}$.

\subsection{Stability guarantees}\label{subsec:I-SOEKF stability}
Similar to I-EKF's stability analysis in Section~\ref{subsec:IEKF stability}, we first obtain stability conditions of forward SOEKF in the exponentially-bounded-mean-squared sense and then, extend those results to I-SOEKF's stability. To this end, we adopt the bounded non-linearity approach \cite{reif1999stochastic} and consider the one-step prediction formulation of forward and inverse SOEKF derived in the previous section. As mentioned in Remark~\ref{remark:reif result remark}, the convergence behaviors of both one- and two-step formulations are similar.

\textit{Forward SOEKF stability:} Consider the one-step SOEKF's formulation \eqref{eqn: one step SOEKF first update}-\eqref{eqn: one step SOEKF sig update}. Considering second-order terms as well, the Taylor series expansion of functions $f(\cdot)$ and $h(\cdot)$ at $\hat{\mathbf{x}}_{k}$ are
 \par\noindent\small
\begin{align*}
&f(\mathbf{x}_{k})-f(\hat{\mathbf{x}}_{k})=\mathbf{F}_{k}(\mathbf{x}_{k}-\hat{\mathbf{x}}_{k})+\frac{1}{2}\sum_{i=1}^{n_{x}}\widetilde{\mathbf{a}}_{i}(\mathbf{x}_{k}-\hat{\mathbf{x}}_{k})^{T}\nabla^{2}\left[f(\hat{\mathbf{x}}_{k})\right]_{i}(\mathbf{x}_{k}-\hat{\mathbf{x}}_{k})+\phi(\mathbf{x}_{k},\hat{\mathbf{x}}_{k}),\\
&h(\mathbf{x}_{k})-h(\hat{\mathbf{x}}_{k})=\mathbf{H}_{k}(\mathbf{x}_{k}-\hat{\mathbf{x}}_{k})+\frac{1}{2}\sum_{i=1}^{n_{y}}\mathbf{b}_{i}(\mathbf{x}_{k}-\hat{\mathbf{x}}_{k})^{T}\nabla^{2}\left[h(\hat{\mathbf{x}}_{k})\right]_{i}(\mathbf{x}_{k}-\hat{\mathbf{x}}_{k})+\chi(\mathbf{x}_{k},\hat{\mathbf{x}}_{k}),
\end{align*}
\normalsize
where $\phi(\cdot)$ and $\chi(\cdot)$ are suitable non-linear functions to account for third and higher-order terms in the expansions. Using these expansions, the error dynamics of the forward filter with $\mathbf{e}_{k}\doteq\mathbf{x}_{k}-\hat{\mathbf{x}}_{k}$ is
\par\noindent\small
\begin{align}
\mathbf{e}_{k+1}=(\mathbf{F}_{k}-\mathbf{K}_{k}\mathbf{H}_{k})\mathbf{e}_{k}+\mathbf{r}_{k}+\mathbf{q}_{k}+\mathbf{s}_{k},\label{eqn: forward SOEKF error}
\end{align}
\normalsize
where $\mathbf{r}_{k}=\phi(\mathbf{x}_{k},\hat{\mathbf{x}}_{k})-\mathbf{K}_{k}\chi(\mathbf{x}_{k},\hat{\mathbf{x}}_{k})$, $\mathbf{q}_{k}=\frac{1}{2}\sum_{i=1}^{n_{x}}\widetilde{\mathbf{a}}_{i}\mathbf{e}_{k}^{T}\nabla^{2}\left[f(\hat{\mathbf{x}}_{k})\right]_{i}\mathbf{e}_{k}-\frac{1}{2}\sum_{i=1}^{n_{x}}\widetilde{\mathbf{a}}_{i}\textrm{Tr}\left(\nabla^{2}\left[f(\hat{\mathbf{x}}_{k})\right]_{i}\bm{\Sigma}_{k}\right)-\frac{1}{2}\mathbf{K}_{k}\sum_{i=1}^{n_{y}}\mathbf{b}_{i}\mathbf{e}_{k}^{T}\nabla^{2}\left[h(\hat{\mathbf{x}}_{k})\right]_{i}\mathbf{e}_{k}+\frac{1}{2}\mathbf{K}_{k}\sum_{i=1}^{n_{y}}\mathbf{b}_{i}\textrm{Tr}\left(\nabla^{2}\left[h(\hat{\mathbf{x}}_{k})\right]_{i}\bm{\Sigma}_{k}\right)$ and $\mathbf{s}_{k}=\mathbf{w}_{k}-\mathbf{K}_{k}\mathbf{v}_{k}$. The following Theorem~\ref{theorem: forward SOEKF stability} provides sufficient conditions for the stochastic stability of forward SOEKF.
\begin{theorem}[Exponential boundedness of forward SOEKF's error]\label{theorem: forward SOEKF stability}
Consider the non-linear stochastic system defined by \eqref{eqn:non-linear state x} and \eqref{eqn:non-linear observation y}, and SOEKF's one-step prediction formulation \eqref{eqn: one step SOEKF first update}-\eqref{eqn: one step SOEKF sig update}. Let the following assumptions hold true.\\
\textbf{C3.j)} There exist positive real numbers $\overline{f}$, $\overline{h}$, $\underline{\sigma}$, $\overline{\sigma}$, $\underline{q}$, $\underline{r}$, $\overline{a}$, $\overline{b}$, $\delta$ and real numbers $\underline{a}$, $\underline{b}$ (not necessarily positive) such that the following bounds are satisfied for all $k\geq 0$.
    \par\noindent\small
    \begin{align*}
    &\underline{\sigma}\mathbf{I}\preceq\bm{\Sigma}_{k}\preceq\overline{\sigma}\mathbf{I},\;\;\|\mathbf{F}_{k}\|\leq\overline{f},\;\;\|\mathbf{H}_{k}\|\leq\overline{h},\;\;\underline{r}\mathbf{I}\preceq \mathbf{R}_{k}\preceq\delta\mathbf{I},\;\;\underline{q}\mathbf{I}\preceq\mathbf{Q}_{k}\preceq\delta\mathbf{I},\\
    &\underline{a}\mathbf{I}\preceq\nabla^{2}\left[f(\hat{\mathbf{x}}_{k})\right]_{i}\preceq\overline{a}\mathbf{I}\;\;\forall i\in\lbrace1,2,\hdots,n_{x}\rbrace,\;\;
    \underline{b}\mathbf{I}\preceq\nabla^{2}\left[h(\hat{\mathbf{x}}_{k})\right]_{j}\preceq\overline{b}\mathbf{I}\;\;\forall j\in\lbrace1,2,\hdots,n_{y}\rbrace.
    \end{align*}
    \normalsize
\textbf{C3.k)} $\mathbf{F}_{k}$ is non-singular and $\mathbf{F}_{k}^{-1}$ satisfies the bound $\|\mathbf{F}_{k}^{-1}\|\leq\widetilde{f}$ for all $k\geq 0$ for some positive real number $\widetilde{f}$.\\
\textbf{C3.l)} There exist positive real numbers $\kappa_{\phi}$, $\epsilon_{\phi}$, $\kappa_{\chi}$, $\epsilon_{\chi}$ such that the non-linear functions $\phi(\cdot)$ and $\chi(\cdot)$ satisfy
    \par\noindent\small
    \begin{align*}
        &\|\phi(\mathbf{x},\hat{\mathbf{x}})\|_{2}\leq\kappa_{\phi}\|\mathbf{x}-\hat{\mathbf{x}}\|_{2}^{3}\hspace{0.5cm}for\hspace{0.5cm}\|\mathbf{x}-\hat{\mathbf{x}}\|_{2}\leq\epsilon_{\phi},\\
        &\|\chi(\mathbf{x},\hat{\mathbf{x}})\|_{2}\leq\kappa_{\chi}\|\mathbf{x}-\hat{\mathbf{x}}\|_{2}^{3}\hspace{0.5cm}for\hspace{0.5cm}\|\mathbf{x}-\hat{\mathbf{x}}\|_{2}\leq\epsilon_{\chi}.
    \end{align*}
    \normalsize
Then the estimation error given by \eqref{eqn: forward SOEKF error} is exponentially bounded in mean-squared sense if the estimation error is within $\epsilon$ bound for a suitable constant $\epsilon>0$,
\par\noindent\small
\begin{align}
&\widetilde{f}<\frac{2\underline{r}}{\overline{h}\overline{a}\overline{b}\overline{\sigma}^{2}n\sqrt{np}},\label{eqn: SOEKF stable constraint on inverse norm}\\
&\underline{q}> c,\;\;\;\textrm{and}\label{eqn: SOEKF stable constraint on q}\\
&\delta=\frac{1}{\kappa_{\textrm{noise}}}\left(\frac{\alpha\widetilde{\epsilon}^{2}}{2\overline{\sigma}}-c_{\textrm{sec}}\right),\label{eqn: SOEKF stable constraint on delta}
\end{align}
\normalsize
for some $\widetilde{\epsilon}<\epsilon$, where $c$, $\alpha$, $\kappa_{\textrm{noise}}$ and $c_{\textrm{sec}}$ are constants that depend on the bounds assumed on the system.
\end{theorem}
\begin{proof}
See Appendix~\ref{App-thm-forward SOEKF stability}.
\end{proof}

It follows from the proof of Lemma~\ref{lemma: SOEKF stable alpha term} (see Appendix~\ref{App-thm-forward SOEKF stability}) that the constant $c$ depends on $\delta$. The conditions \eqref{eqn: SOEKF stable constraint on q} and \eqref{eqn: SOEKF stable constraint on delta} suggest that $\delta$ should be chosen appropriately so that both the conditions could be satisfied simultaneously. However, the exact bounds on $\delta$ depend on other bounds assumed on the system matrices in Theorem~\ref{theorem: forward SOEKF stability}. Furthermore, as with the bounded non-linearity approach for EKF, these bounds may be conservative and, consequently, the estimation error may remain bounded outside this range \cite[Sec. V]{reif1999stochastic}.

\textit{Inverse SOEKF stability:} Considering a suitable non-linear function $\overline{\chi}(\cdot)$, the Taylor series expansion of $g(\cdot)$ at estimate $\doublehat{\mathbf{x}}_{k}$ of I-SOEKF's one-step prediction formulation is
\par\noindent\small
\begin{align*}
&g(\hat{\mathbf{x}}_{k})-g(\doublehat{\mathbf{x}}_{k})=\mathbf{G}_{k}(\hat{\mathbf{x}}_{k}-\doublehat{\mathbf{x}}_{k})+\frac{1}{2}\sum_{i=1}^{n_{a}}\mathbf{d}_{i}(\hat{\mathbf{x}}_{k}-\doublehat{\mathbf{x}}_{k})^{T}\nabla^{2}\left[g(\doublehat{\mathbf{x}}_{k})\right]_{i}(\hat{\mathbf{x}}_{k}-\doublehat{\mathbf{x}}_{k})+\overline{\chi}(\hat{\mathbf{x}}_{k},\doublehat{\mathbf{x}}_{k}),
\end{align*}
\normalsize
where $\mathbf{d}_{i}$ is the $i$-th Euclidean basis vector in $\mathbb{R}^{n_{a}\times 1}$. Finally, the error dynamics of the inverse filter with estimation error denoted by $\overline{\mathbf{e}}_{k}\doteq\hat{\mathbf{x}}_{k}-\doublehat{\mathbf{x}}_{k}$ and the inverse filter's Kalman gain and estimation error covariance matrix by $\overline{\mathbf{K}}_{k}$ and $\overline{\bm{\Sigma}}_{k}$, respectively, is
\par\noindent\small
\begin{align}
    \overline{\mathbf{e}}_{k+1}=(\overline{\mathbf{F}}_{k}-\overline{\mathbf{K}}_{k}\mathbf{G}_{k})\overline{\mathbf{e}}_{k}+\overline{\mathbf{r}}_{k}+\overline{\mathbf{q}}_{k}+\overline{\mathbf{s}}_{k},\label{eqn: inverse SOEKF error}
\end{align}
\normalsize
where $\overline{\mathbf{r}}_{k}=\overline{\phi}_{k}(\hat{\mathbf{x}}_{k},\doublehat{\mathbf{x}}_{k})-\overline{\mathbf{K}}_{k}\overline{\chi}(\hat{\mathbf{x}}_{k},\doublehat{\mathbf{x}}_{k})$, $\overline{\mathbf{s}}_{k}=\mathbf{K}_{k}\mathbf{v}_{k}-\overline{\mathbf{K}}_{k}\bm{\epsilon}_{k}$, and $\overline{\mathbf{q}}_{k}=\frac{1}{2}\sum_{i=1}^{n_{x}}\widetilde{\mathbf{a}}_{i}\overline{\mathbf{e}}_{k}^{T}\nabla^{2}\left[\overline{f}_{k}(\doublehat{\mathbf{x}}_{k})\right]_{i}\overline{\mathbf{e}}_{k}-\frac{1}{2}\sum_{i=1}^{n_{x}}\widetilde{\mathbf{a}}_{i}\textrm{Tr}\left(\nabla^{2}\left[\overline{f}_{k}(\doublehat{\mathbf{x}}_{k})\right]_{i}\overline{\bm{\Sigma}}_{k}\right)-\frac{1}{2}\overline{\mathbf{K}}_{k}\sum_{i=1}^{n_{a}}\mathbf{d}_{i}\overline{\mathbf{e}}_{k}^{T}\nabla^{2}\left[g(\doublehat{\mathbf{x}}_{k})\right]_{i}\overline{\mathbf{e}}_{k}+\frac{1}{2}\overline{\mathbf{K}}_{k}\sum_{i=1}^{n_{a}}\mathbf{d}_{i}\textrm{Tr}\left(\nabla^{2}\left[g(\doublehat{\mathbf{x}}_{k})\right]_{i}\overline{\bm{\Sigma}}_{k}\right)$ with $\frac{1}{2}\sum_{i=1}^{n_{x}}\widetilde{\mathbf{a}}_{i}\overline{\mathbf{e}}_{k}^{T}\nabla^{2}\left[\overline{f}_{k}(\doublehat{\mathbf{x}}_{k})\right]_{i}\overline{\mathbf{e}}_{k}=\frac{1}{2}\sum_{i=1}^{n_{x}}\widetilde{\mathbf{a}}_{i}\overline{\mathbf{e}}_{k}^{T}\nabla^{2}\left[f(\doublehat{\mathbf{x}}_{k})\right]_{i}\overline{\mathbf{e}}_{k}-\frac{1}{2}\mathbf{K}_{k}\sum_{i=1}^{n_{y}}\mathbf{b}_{i}\overline{\mathbf{e}}_{k}^{T}\nabla^{2}\left[h(\doublehat{\mathbf{x}}_{k})\right]_{i}\overline{\mathbf{e}}_{k}$ and $\overline{\phi}_{k}(\hat{\mathbf{x}}_{k},\doublehat{\mathbf{x}}_{k})=\phi(\hat{\mathbf{x}}_{k},\doublehat{\mathbf{x}}_{k})-\mathbf{K}_{k}\chi(\hat{\mathbf{x}}_{k},\doublehat{\mathbf{x}}_{k})$. Here, the error in approximations of the terms $\frac{1}{2}\sum_{i=1}^{n_{x}}\widetilde{\mathbf{a}}_{i}\textrm{Tr}\left(\nabla^{2}\left[f(\hat{\mathbf{x}}_{k})\right]_{i}\bm{\Sigma}_{k}\right)$ and $\frac{1}{2}\mathbf{K}_{k}\sum_{i=1}^{n_{y}}\mathbf{b}_{i}\textrm{Tr}\left(\nabla^{2}\left[h(\hat{\mathbf{x}}_{k})\right]_{i}\bm{\Sigma}_{k}\right)$ by I-SOEKF are neglected. Also, using the bounds assumed in Theorem \ref{theorem: forward SOEKF stability}, these approximation errors are bounded by positive constants. The following Theorem~\ref{theorem: inverse SOEKF stability} states the conditions for stability of I-SOEKF.

\begin{theorem}[Exponential boundedness of I-SOEKF's error]
\label{theorem: inverse SOEKF stability}
Consider the attacker's forward SOEKF's one-step prediction formulation that is stable as per Theorem \ref{theorem: forward SOEKF stability}. Additionally, assume that the following hold true.\\
\textbf{C3.m)} There exist positive real numbers $\overline{g}$, $\underline{m}$, $\overline{m}$, $\underline{\epsilon}$, $\overline{c}$, $\overline{\delta}$ and a real number $\underline{c}$ (not necessarily positive) such that the following bounds are fulfilled for all $k\geq 0$.
    \par\noindent\small
    \begin{align*}
&\|\mathbf{G}_{k}\|\leq\overline{g},\;\;\underline{m}\mathbf{I}\preceq\overline{\bm{\Sigma}}_{k}\preceq\overline{m}\mathbf{I},\;\;\underline{\epsilon}\mathbf{I}\preceq\overline{\mathbf{R}}_{k}\preceq\overline{\delta}\mathbf{I},\;\;\underline{c}\mathbf{I}\preceq\nabla^{2}\left[g(\doublehat{\mathbf{x}}_{k})\right]_{i}\preceq\overline{c}\mathbf{I}\;\;\forall i\in\lbrace 1,2,\hdots,n_{a}\rbrace.
    \end{align*}
    \normalsize
\textbf{C3.n)} $\mathbf{H}_{k}\bm{\Sigma}_{k}\mathbf{F}_{k}^{T}+\mathbf{M}_{k}^{T}$ is full column rank matrix for every $k\geq 0$.\\
\textbf{C3.o)} There exist positive real numbers $\kappa_{\bar{\chi}}$, $\epsilon_{\bar{\chi}}$ such that the non-linear function $\overline{\chi}(\cdot)$ satisfies
    \par\noindent\small
    \begin{align*}
        \|\overline{\chi}(\hat{\mathbf{x}},\doublehat{\mathbf{x}})\|_{2}\leq\kappa_{\bar{\chi}}\|\hat{\mathbf{x}}-\doublehat{\mathbf{x}}\|_{2}^{3}\;\;\textrm{for}\;\;\|\hat{\mathbf{x}}-\doublehat{\mathbf{x}}\|_{2}\leq\epsilon_{\bar{\chi}}.
    \end{align*}
    \normalsize
Then, the estimation error of I-SOEKF given by \eqref{eqn: inverse SOEKF error} is exponentially bounded in mean-squared sense if the estimation error is bounded by a suitable constant $\overline{\epsilon}>0$ and the bound constants also satisfy the equivalent conditions of \eqref{eqn: SOEKF stable constraint on inverse norm}, \eqref{eqn: SOEKF stable constraint on q}, and \eqref{eqn: SOEKF stable constraint on delta} for the inverse filter dynamics.
\end{theorem}
\begin{proof}
See Appendix~\ref{App-thm-inverse SOEKF stability}.
\end{proof}

As discussed in Remark~\ref{remark:observability}, the inequality $\underline{m}\mathbf{I}\preceq\overline{\bm{\Sigma}}_{k}\preceq\overline{m}\mathbf{I}$ assumed in Theorem~\ref{theorem: inverse SOEKF stability} is closely related to the theoretical observability of the non-linear inverse filtering model. Further, the inequality $\underline{m}\mathbf{I}\preceq\overline{\bm{\Sigma}}_{k}\preceq\overline{m}\mathbf{I}$ is a weaker assumption than observability.

\section{Inverse Gaussian sum and dithered EKFs}\label{sec:I-GSEKF and I-DEKF}
\subsection{Inverse GS-EKF}\label{subsec:I-GSEKF}
The GS-EKF \cite{anderson2012optimal} assumes the posterior distribution of the state estimate to be a weighted sum of Gaussian densities, which are updated recursively based on the current observation. This multiple-model filtering technique is also widely used to handle non-Gaussian process and measurement noises in the system model\cite{bilik2010mmse}. Hence, the I-GS-EKF developed in the following is also applicable to non-Gaussian inverse filtering problems.

\textit{Forward filter:} In the forward GS-EKF (employed by the attacker), we consider $l$ Gaussians denoting the $i$-th Gaussian probability density function, with mean $\overline{\mathbf{x}}_{i,k}$ and covariance $\bm{\Sigma}_{i,k}$ at $k$-th time-step, as $\gamma(\mathbf{x}-\overline{\mathbf{x}}_{i},\bm{\Sigma}_{i})=(2\pi)^{-n_{x}/2}\textrm{det}(\bm{\Sigma}_{i})^{-1/2}\exp{\left\{-\frac{1}{2}(\mathbf{x}-\overline{\mathbf{x}}_{i})^{T}\bm{\Sigma}_{i}^{-1}(\mathbf{x}-\overline{\mathbf{x}}_{i}) \right\}}$. Given observations $\mathbf{Y}^{k}=\lbrace\mathbf{y}_{j}\rbrace_{1\leq j\leq k}$ up to $k$-th time instant, the posterior distribution of state $\mathbf{x}_{k}$, is approximated as $p(\mathbf{x}_{k}|\mathbf{Y}^{k})=\sum_{i=1}^{l}c_{i,k}\gamma(\mathbf{x}_{k}-\overline{\mathbf{x}}_{i,k},\bm{\Sigma}_{i,k})$, where $c_{i,k}$ is $i$-th Gaussian's weight at $k$-th time step. Linearizing the functions as $\mathbf{F}_{i,k}\doteq\nabla_{\mathbf{x}}f(\mathbf{x})|_{\mathbf{x}=\overline{\mathbf{x}}_{i,k}}$ and $\mathbf{H}_{i,k+1}\doteq\nabla_{\mathbf{x}}h(\mathbf{x})|_{\mathbf{x}=\overline{\mathbf{x}}_{i,k+1|k}}$ for the $i$-th Gaussian, the means $\lbrace\overline{\mathbf{x}}_{i,k}\rbrace_{1\leq i\leq l}$ and covariance matrices $\lbrace\bm{\Sigma}_{i,k}\rbrace_{1\leq i\leq l}$ are updated based on the current observation $\mathbf{y}_{k+1}$ using independent EKF recursions for each Gaussian. Finally, the weights $\lbrace c_{i,k}\rbrace_{1\leq i\leq l}$ are updated as\cite{anderson2012optimal}:
\par\noindent\small
\begin{align*}
c_{i,k+1}=\frac{c_{i,k}\gamma(\mathbf{y}_{k+1}-h(\overline{\mathbf{x}}_{i,k+1|k}),\bm{\Sigma}^{y}_{i,k+1})}{\sum_{i'=1}^{l}c_{i',k}\gamma(\mathbf{y}_{k+1}-h(\overline{\mathbf{x}}_{i',k+1|k}),\bm{\Sigma}^{y}_{i',k+1})},  
\end{align*}
\normalsize
where $\overline{\mathbf{x}}_{i,k+1|k}$ is the $i$-th Gaussian's predicted mean and $\bm{\Sigma}^{y}_{i,k+1}=\mathbf{H}_{i,k+1}\bm{\Sigma}_{i,k+1|k}\mathbf{H}_{i,k+1}^{T}+\mathbf{R}$ with $\bm{\Sigma}_{i,k+1|k}$ as the prediction covariance matrix of $\overline{\mathbf{x}}_{i,k+1|k}$. With these updated Gaussians, the point-estimate $\hat{\mathbf{x}}_{k}$ and the associated error covariance matrix $\bm{\Sigma}_{k}$ are $\hat{\mathbf{x}}_{k+1}=\sum_{i=1}^{l}c_{i,k+1}\overline{\mathbf{x}}_{i,k+1}$ and $\bm{\Sigma}_{k+1}=\sum_{i=1}^{l}c_{i,k+1}\left(\bm{\Sigma}_{i,k+1}+(\hat{\mathbf{x}}_{k+1}-\overline{\mathbf{x}}_{i,k+1})(\hat{\mathbf{x}}_{k+1}-\overline{\mathbf{x}}_{i,k+1})^{T}\right)$.

\textit{Inverse filter:} Consider an augmented state vector $\mathbf{z}_{k}=\lbrace\lbrace\overline{\mathbf{x}}_{i,k}\rbrace_{1\leq i\leq l},\lbrace c_{i,k}\rbrace_{1\leq i\leq l}\rbrace$ (means and weights of forward GS-EKF). Then, substituting for observation $\mathbf{y}_{k+1}$ from \eqref{eqn:non-linear observation y} in the forward filter's updates (similar to I-EKF/SOEKF) and denoting the $i$-th EKF's gain matrix as $\mathbf{K}_{i,k+1}=\bm{\Sigma}_{i,k+1|k}\mathbf{H}_{i,k+1}^{T}(\bm{\Sigma}^{y}_{i,k+1})^{-1}$ yields the state transition equations for I-GS-EKF as
\par\noindent\small
\begin{align}
&\overline{\mathbf{x}}_{i,k+1}=f(\overline{\mathbf{x}}_{i,k})+\mathbf{K}_{i,k+1}\left(h(\mathbf{x}_{k+1})+\mathbf{v}_{k+1}-h(f(\overline{\mathbf{x}}_{i,k}))\right),\nonumber\\
&c_{i,k+1}=\frac{c_{i,k}\gamma(h(\mathbf{x}_{k+1})+\mathbf{v}_{k+1}-h(f(\overline{\mathbf{x}}_{i,k})),\bm{\Sigma}^{y}_{i,k+1})}{\sum_{i'=1}^{l}c_{i',k}\gamma(h(\mathbf{x}_{k+1})+\mathbf{v}_{k+1}-h(f(\overline{\mathbf{x}}_{i',k})),\bm{\Sigma}^{y}_{i',k+1})}.\label{eqn: inverse GS-EKF weight transition}
\end{align}
\normalsize
Treating $\lbrace\mathbf{K}_{i,k+1},\bm{\Sigma}^{y}_{i,k+1}\rbrace_{1\leq i\leq l}$ as time-varying parameters of the state transition equations, which are approximated in a similar way as we approximated the gain matrix for I-EKF/SOEKF, the overall state transition in terms of the augmented state is $\mathbf{z}_{k+1}=\overline{f}_{k}(\mathbf{z}_{k},\mathbf{x}_{k+1},\mathbf{v}_{k+1})$. Similarly, the observation $\mathbf{a}_{k}$ as a function of augmented state $\mathbf{z}_{k}$ is $\mathbf{a}_{k}=\overline{g}(\mathbf{z}_{k})+\bm{\epsilon}_{k}=g\left(\sum_{i=1}^{l}c_{i,k}\overline{\mathbf{x}}_{i,k}\right)+\bm{\epsilon}_{k}$.

The I-GS-EKF approximates the posterior distribution of the augmented state as a sum of $\overline{l}$ Gaussians with its recursions again following directly from forward GS-EKF's recursions treating $\mathbf{a}_{k}$ as the observation. However, the inverse filter estimates an $l(n_{x}+1)$-dimensional augmented state $\mathbf{z}_{k}$ with the Jacobians with respect to the state denoted as $\overline{\mathbf{F}}_{j,k}\doteq\nabla_{\mathbf{z}}\overline{f}_{k}(\mathbf{z},\mathbf{x}_{k+1},\mathbf{0})|_{\mathbf{z}=\overline{\mathbf{z}}_{j,k}}$ and $\mathbf{G}_{j,k+1}\doteq\nabla_{\mathbf{z}}\overline{g}(\mathbf{z})|_{\mathbf{z}=\overline{\mathbf{z}}_{j,k+1|k}}$, and the process noise covariance matrix as $\overline{\mathbf{Q}}_{k}=\overline{\mathbf{V}}_{j,k}\mathbf{R}\overline{\mathbf{V}}_{j,k}^{T}$ with $\overline{\mathbf{V}}_{j,k}\doteq\nabla_{\mathbf{v}}\overline{f}_{k}(\overline{\mathbf{z}}_{j,k},\mathbf{x}_{k+1},\mathbf{v})|_{\mathbf{v}=\mathbf{0}}$ for the $j$-th inverse filter's Gaussian updates. The point estimate $\hat{\mathbf{z}}_{k}$ consists of the estimates $\lbrace\hat{\overline{\mathbf{x}}}_{i,k},\hat{c}_{i,k}\rbrace_{1\leq i\leq l}$ of the forward GS-EKF's means and weights such that the point estimate $\doublehat{\mathbf{x}}_{k}$ of the forward filter's estimate $\hat{\mathbf{x}}_{k}$ is $\doublehat{\mathbf{x}}_{k}=\sum_{i=1}^{l}\hat{c}_{i,k}\hat{\overline{\mathbf{x}}}_{i,k}$.

When the forward filter considers only one Gaussian ($l=1$), the forward GS-EKF reduces to forward EKF with the only weight $c_{1,k}=1$ for all $k$. Hence, this weight need not be considered in the augmented state and $\mathbf{z}_{k}$ reduces to $\overline{\mathbf{x}}_{1,k}$ which is the estimate $\hat{\mathbf{x}}_{k}$ itself. Similarly, I-GS-EKF also reduces to I-EKF if only one Gaussian is considered ($\overline{l}=1$).
\begin{remark}[I-GS-EKF's computational complexity]\label{remark: GSEKF complexity}
Since GS-EKF consists of several independent EKF, its computational load is larger than an EKF considering only one Gaussian. Furthermore, I-GS-EKF considers an augmented state of dimension $l(n_{x}+1)$, which is greater than that of the forward GS-EKF. Hence, in general, I-GS-EKF is computationally more complex than the forward GS-EKF and the I-EKF. However, our numerical experiments in Section~\ref{subsec:FM with EKF SOEKF GSEKF} suggest that I-GS-EKF can provide reasonable accuracy even when considering a smaller number of Gaussians than that in forward GS-EKF, i.e., when $\overline{l}<l$. A survey of several methods to reduce the computational complexity of these KF extensions is available in \cite{raitoharju2019computational}.
\end{remark}

\subsection{Inverse DEKF}\label{subsec:I-DEKF}
Consider the attacker employing DEKF \cite{weiss1980improved} as its forward filter. In DEKF, the output non-linearities are modified using dither signals so as to tighten the cone-bounds. Dithering tightens this cone such that the non-linearities are smoothened but it may also degrade the near-optimal performance of the EKF after the initial transient phase of estimation. Therefore, dithering is introduced only during the initial transient phase with the aim to improve the filter's transient performance and avoid divergence. Denote the dither amplitude which controls the tightness of the cone-bounds by $d$ and its amplitude probability density function by $p(a)$. The observation function $h(x)$ is dithered as $h^{*}(x)=\int_{-d}^{d}h(a+x)p(a)da$. If $d=d_{0}e^{-k/\tau}$, where $d_{0}$ and $\tau$ are constants and `$k$' denotes the time index, then $h^{*}(x)\to h(x)$ exponentially as $k\to\infty$ during the transient phase.

The forward DEKF follows from conventional forward EKF described in Section~\ref{subsec:IEKF formulation} by replacing $h(\cdot)$ with $h^{*}(\cdot)$ as the observation function of $\mathbf{y}_{k}$ during the initial transient phase and hence, the I-DEKF also follows from I-EKF of Section~\ref{subsec:IEKF formulation}. The dither of the attacker's filter is assumed to be known to us. Otherwise, the I-DEKF may also proceed with the unmodified observation function. We show in Section~\ref{subsec:coordinate DEKF} that these two formulations, labeled I-DEKF-1 and I-DEKF-2, respectively, generally vary in their estimation performances especially during the transient phase where the modified observation function is considered. A DEKF differs from a standard EKF only because of the modified observation function and hence, has the same $\mathcal{O}(n_{x}^{3})$ computational complexity as EKF. The same argument holds for I-DEKF, which also follows from standard I-EKF recursions.

\section{Numerical experiments}\label{sec:IEKF numericals}
We illustrate the efficacy of the proposed inverse filters by comparing the estimation error with recursive Cram\'{e}r-Rao lower bound (RCRLB) for different example systems. The CRLB provides a lower bound on mean-squared error (MSE) and is widely used to assess the performance of an estimator. For the discrete-time non-linear filtering, we employ the RCRLB as $\mathbb{E}\left[(\mathbf{x}_{k}-\hat{\mathbf{x}}_{k})(\mathbf{x}_{k}-\hat{\mathbf{x}}_{k})^{T}\right]\succeq\mathbf{J}_{k}^{-1}$ where $\mathbf{J}_{k}=\mathbb{E}\left[-\frac{\partial^{2}\ln{p(Y^{k},X^{k})}}{\partial\mathbf{x}_{k}^{2}}\right]$ is the Fisher information matrix\cite{tichavsky1998posterior}. Here, $X^{k}=\lbrace\mathbf{x}_{0},\mathbf{x}_{1},\hdots,\mathbf{x}_{k}\rbrace$ is the state vector series while $Y^{k}=\lbrace\mathbf{y}_{1},\mathbf{y}_{2},\hdots,\mathbf{y}_{k}\rbrace$ are the noisy observations. Also, $p(Y^{k},X^{k})$ is the joint probability density of pair $(Y^{k},X^{k})$ and $\hat{\mathbf{x}}_{k}$ (a function of $Y^{k}$) is an estimate of $\mathbf{x}_{k}$ with $\frac{\partial^{2}(\cdot)}{\partial\mathbf{x}^{2}}$ denoting the Hessian with second order partial derivatives. The information matrix $\mathbf{J}_{k}$ can be computed recursively as \cite{tichavsky1998posterior}
\par\noindent\small
\begin{align}
    \mathbf{J}_{k}&=\mathbf{D}_{k}^{22}-\mathbf{D}_{k}^{21}(\mathbf{J}_{k-1}+\mathbf{D}_{k}^{11})^{-1}\mathbf{D}_{k}^{12},\label{eqn: general Jk recursions}
\end{align}
\normalsize
where $\mathbf{D}_{k}^{11}=\mathbb{E}\left[-\frac{\partial^{2}\ln{p(\mathbf{x}_{k}\vert\mathbf{x}_{k-1})}}{\partial\mathbf{x}_{k-1}^{2}}\right]$, $\mathbf{D}_{k}^{12}=\mathbb{E}\left[-\frac{\partial^{2}\ln{p(\mathbf{x}_{k}\vert\mathbf{x}_{k-1})}}{\partial\mathbf{x}_{k}\partial\mathbf{x}_{k-1}}\right]=(\mathbf{D}_{k}^{21})^{T}$, and $\mathbf{D}_{k}^{22}=\mathbb{E}\left[-\frac{\partial^{2}\ln{p(\mathbf{x}_{k}\vert\mathbf{x}_{k-1})}}{\partial\mathbf{x}_{k}^{2}}\right]\\+\mathbb{E}\left[-\frac{\partial^{2}\ln{p(\mathbf{y}_{k}\vert\mathbf{x}_{k})}}{\partial\mathbf{x}_{k}^{2}}\right]$. For the non-linear system given by \eqref{eqn:non-linear state x} and \eqref{eqn:non-linear observation y}, the forward information matrices $\lbrace\mathbf{J}_{k}\rbrace$ recursions reduces to \cite{xiong2006performance_ukf}
\par\noindent\small
\begin{align}
    &\mathbf{J}_{k+1}=\mathbf{Q}_{k}^{-1}+\mathbf{H}_{k+1}^{T}\mathbf{R}_{k+1}^{-1}\mathbf{H}_{k+1}-\mathbf{Q}_{k}^{-1}\mathbf{F}_{k}(\mathbf{J}_{k}+\mathbf{F}_{k}^{T}\mathbf{Q}_{k}^{-1}\mathbf{F}_{k})^{-1}\mathbf{F}_{k}^{T}\mathbf{Q}_{k}^{-1},\label{eqn: additive Jk recursions}
\end{align}
\normalsize
where $\mathbf{F}_{k}=\nabla_{\mathbf{x}}f(\mathbf{x})\vert_{\mathbf{x}=\mathbf{x}_{k}}$ and $\mathbf{H}_{k}=\nabla_{\mathbf{x}}h(\mathbf{x})\vert_{\mathbf{x}=\mathbf{x}_{k}}$. Note that, for the information matrices recursion, the Jacobians $\mathbf{F}_{k}$ and $\mathbf{H}_{k}$ are evaluated at the true state $\mathbf{x}_{k}$ while for forward filters, these are evaluated at the estimates of the state. These recursions can be trivially extended to compute the information matrix $\overline{\mathbf{J}}_{k}$ for inverse filter's estimate $\doublehat{\mathbf{x}}_{k}$. Some recent studies on cognitive radar target tracking instead consider posterior CRLB \cite{bell2015cognitive} as a metric to tune tracking filters.

Throughout all experiments, the initial information matrices $\mathbf{J}_{0}$ and $\overline{\mathbf{J}}_{0}$ were set to $\bm{\Sigma}_{0}^{-1}$ and $\overline{\bm{\Sigma}}_{0}^{-1}$, respectively, unless mentioned otherwise. Note that these initial estimates only affect the RCRLB in the transient phase. The steady-state RCRLB is independent of the initialization. Furthermore, the forward and inverse filters are compared only to highlight the relative estimation accuracy. Unless stated otherwise, the forward and inverse filters' time-averaged root MSE (AMSE) at $k$-th time step are, respectively, $r_{k}$ and $\overline{r}_{k}$, as follows.
\par\noindent\small
\begin{align*}
    &r_{k}=\sqrt{\frac{1}{k}\sum_{k'=1}^{k}\left(\frac{1}{M}\sum_{m=1}^{M}\|\mathbf{x}_{k'}-\hat{\mathbf{x}}_{k'}\|_{2}^{2}\right)},\;\;\;\overline{r}_{k}=\sqrt{\frac{1}{k}\sum_{k'=1}^{k}\left(\frac{1}{M}\sum_{m=1}^{M}\|\hat{\mathbf{x}}_{k'}-\doublehat{\mathbf{x}}_{k'}\|_{2}^{2}\right)},
\end{align*}
\normalsize
where $M$ is the number of independent Monte-Carlo runs.

In the following, we also consider the inverse filters' performance under incorrect forward filter assumption such that the misspecified CRB (MCRB)\cite{richmond2015parameter,fortunati2017performance} may provide further insights. However, the MCRB, as it exists in the literature, cannot be directly applied to the mismatched forward and inverse filter case. Firstly, MCRB's existence can be guaranteed only under certain regularity conditions, like the non-singularity of a generalized Fisher information matrix and the existence of a unique pseudo-true parameter vector minimizing the Kullback-Leibler divergence (KLD) between the true and assumed densities\cite{richmond2015parameter}. These regularity conditions cannot be trivially verified for our inverse filtering problem because of the non-linear transformation of the Gaussian noise terms. Furthermore, discrete-time filtering problems require a recursively computed lower bound as the observations $\mathbf{Y}^{k}=\{\mathbf{y}_{1},\mathbf{y}_{2},\hdots,\mathbf{y}_{k}\}$ and the states $\mathbf{X}^{k}=\{\mathbf{x}_{0},\mathbf{x}_{1},\hdots,\mathbf{x}_{k}\}$ upto $k$-th time step cannot be processed jointly. To the best of our knowledge, such recursive bounds for the misspecified non-linear filtering problems have not been proposed so far. Hence, in our experiments, we only consider RCRLB assuming a perfectly specified model.
\begin{figure}
  \centering
  \includegraphics[width = 1.0\columnwidth]{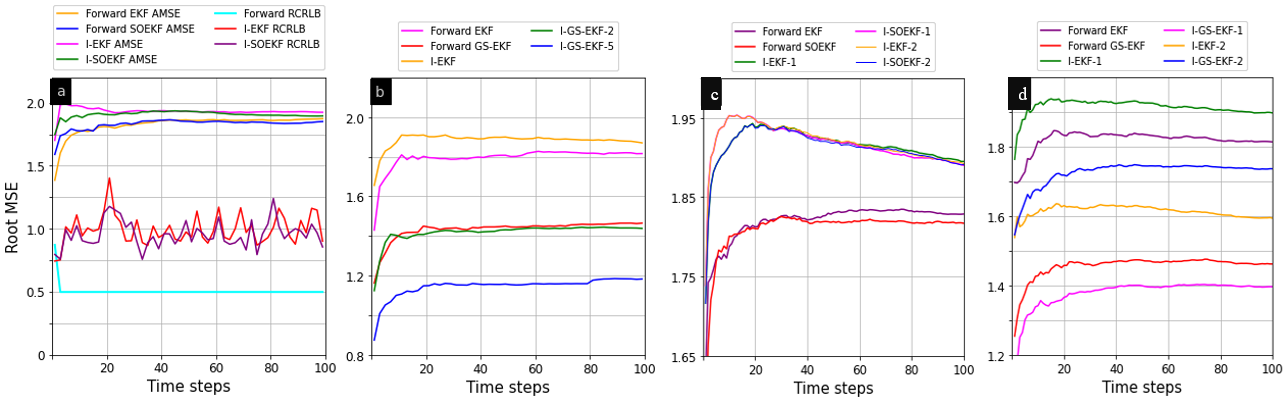}
  \caption{AMSE and RCRLB for forward and inverse filters: (a) EKF and SOEKF; (b) EKF and GS-EKF (I-GS-EKF-2 and I-GS-EKF-5, respectively, for $\overline{l}=2$ and $5$); mismatched (c) I-EKF and I-SOEKF; and (d)I-EKF and I-GS-EKF. The assumptions of different inverse filters are detailed in Table~\ref{tbl:ekf mismatch describe}.}
 \label{fig:FM ekf soekf gsekf}
\end{figure}

\subsection{FM demodulation with I-EKF, I-SOEKF and I-GS-EKF}\label{subsec:FM with EKF SOEKF GSEKF}
Consider the discrete-time non-linear system model of FM demodulator as \cite[Sec. 8.2]{anderson2012optimal}
\par\noindent\small
\begin{align*}
&\mathbf{x}_{k+1}\doteq\begin{bmatrix}\lambda_{k+1}\\\theta_{k+1}\end{bmatrix}=\begin{bmatrix}\exp{(-T/\beta)}&0\\-\beta \exp{(-T/\beta)}-1&1\end{bmatrix}\begin{bmatrix}\lambda_{k}\\\theta_{k}\end{bmatrix}+\begin{bmatrix}1\\-\beta\end{bmatrix}w_{k},\\
&\mathbf{y}_{k}=\sqrt{2}\begin{bmatrix}\sin{\theta_{k}}\\\cos{\theta_{k}}\end{bmatrix}+\mathbf{v}_{k},\;\;
a_{k}=\hat{\lambda}_{k}^{2}+\epsilon_{k},
\end{align*}
\normalsize
with $w_{k}\sim\mathcal{N}(0,0.01)$, $\mathbf{v}_{k}\sim\mathcal{N}(\mathbf{0},\mathbf{I}_{2})$, $\epsilon_{k}\sim\mathcal{N}(0,5)$, $T=2\pi/16$ and $\beta=100$. Here, the observation function $g(\cdot)$ for the inverse filter is quadratic. Also, $\hat{\lambda}_{k}$ is the forward EKF's estimate of $\lambda_{k}$.

The initial state $\mathbf{x}_{0}\doteq[\lambda_{0},\theta_{0}]^{T}$ and its estimates for all forward and inverse filters including mean estimates for GS-EKF were set randomly with $\lambda_{0}\sim\mathcal{N}(0,1)$ and $\theta_{0}\sim\mathcal{U}[-\pi,\pi]$. The initial covariances were set to $\bm{\Sigma}_{0}=10\mathbf{I}_{2}$ and $\overline{\bm{\Sigma}}_{0}=5\mathbf{I}_{2}$ for forward and inverse EKF and SOEKF. In the case of GS-EKF, we considered $l=5$ Gaussians for the forward filter with the initial covariances and weights set to $10\mathbf{I}_{2}$ and $1/5$, respectively. For I-GS-EKF, an augmented state $\mathbf{z}_{k}$ of $\lbrace\overline{\mathbf{x}}_{i,k},c_{i,k}\rbrace_{1\leq i\leq 5}$ was considered resulting in a $15$-dimensional state vector with the initial weight estimates set to $1/5$ and the initial covariance estimates $\lbrace\overline{\bm{\Sigma}}_{j,0}\rbrace_{1\leq j\leq\overline{l}}$ as $5\mathbf{I}_{15}$. The phase term of the state $\theta$ and its estimates $\hat{\theta}$ and $\doublehat{\theta}$ (for both prediction and measurement updates) were considered to be modulo $2\pi$ \cite{anderson2012optimal}. Note that the process covariance $\mathbf{Q}$ is a singular matrix. For numerical stability and to facilitate computation of $\mathbf{Q}^{-1}$ for evaluating information matrices $\mathbf{J}_{k}$, we used an enlarged covariance matrix by adding $10^{-10}\mathbf{I}_{2}$ to $\mathbf{Q}$ in the forward filters. Similarly, we added $10^{-10}\mathbf{I}_{2}$ to $\overline{\mathbf{Q}}_{k}$ in the inverse filter because $\overline{\mathbf{Q}}_{k}$ is time-varying and may be ill-conditioned. The initial $\overline{\mathbf{J}}_{0}$ was taken close to the inverse of the steady state estimation covariance matrix of the forward filter. The initial $\overline{\mathbf{J}}_{0}$ only affects the RCRLB calculated for the initial few time steps. The RCRLB after these initial time-steps (around 20 for the considered system) shows the same behaviour irrespective of the initial $\overline{\mathbf{J}}_{0}$. Additionally, the Gaussian noise term $\mathbf{v}_{k+1}$ in I-GS-EKF's state transition \eqref{eqn: inverse GS-EKF weight transition} is transformed through a non-linear function $\gamma(\cdot,\cdot)$ such that \eqref{eqn: additive Jk recursions} is not applicable. The RCRLB, in this case, is derived using the general $\mathbf{J}_{k}$ recursions given by \eqref{eqn: general Jk recursions}, which we omit here.
    \begin{table}
    \caption{Summary of forward-inverse filters in Fig.~\ref{fig:FM ekf soekf gsekf}(c)-(d)}
    \label{tbl:ekf mismatch describe}
    \centering
    \begin{tabular}{p{1.0cm}p{3.5cm}p{4.5cm}p{3.0cm}}
    \hline\noalign{\smallskip}
    Fig. & True forward filter & Assumed forward filter & Inverse filter\\
    \noalign{\smallskip}
    \hline
    \noalign{\smallskip}
    \ref{fig:FM ekf soekf gsekf}c & EKF & EKF & I-EKF-1\\
    \ref{fig:FM ekf soekf gsekf}c & SOEKF & EKF & I-EKF-2\\
    \ref{fig:FM ekf soekf gsekf}c & SOEKF & SOEKF & I-SOEKF-1\\
    \ref{fig:FM ekf soekf gsekf}c & EKF & SOEKF & I-SOEKF-2\\
    \ref{fig:FM ekf soekf gsekf}d & EKF & EKF & I-EKF-1\\
    \ref{fig:FM ekf soekf gsekf}d & GS-EKF & EKF & I-EKF-2\\
    \ref{fig:FM ekf soekf gsekf}d & GS-EKF & GS-EKF & I-GS-EKF-1\\
    \ref{fig:FM ekf soekf gsekf}d & EKF & GS-EKF & I-GS-EKF-2\\
    \noalign{\smallskip}
    \hline\noalign{\smallskip}
    \end{tabular}
    \end{table}
 
Fig. \ref{fig:FM ekf soekf gsekf}a-b shows the AMSE and RCRLB for both forward and inverse EKF, SOEKF and GS-EKF, averaged over $500$ runs. The RCRLB value for state estimation is $\sqrt{\textrm{Tr}(\mathbf{J}^{-1})}$ with $\mathbf{J}$ denoting the associated information matrix. Fig. \ref{fig:FM ekf soekf gsekf}a-b shows that the forward GS-EKF with $l=5$ performs better than both forward EKF and forward SOEKF. Considering second-order terms of the Taylor series expansion, in addition to the first-order terms, does not improve the estimation performance for this system as observed in Fig. \ref{fig:FM ekf soekf gsekf}a. When including second-order terms in the forward SOEKF yields smaller gains, then this is also reflected in the RCRLB of the inverse filters, i.e., I-EKF and I-SOEKF have similar lower bound values. Both I-EKF and I-SOEKF converge to the same steady-state estimation error values, which are also higher than that of the corresponding forward filters. Being suboptimal filters, the forward as well as inverse EKF and SOEKF do not achieve the RCRLB on the estimation error. However, the difference between AMSE and RCRLB for the inverse filters is less than that for the forward filters. We conclude that I-EKF and I-SOEKF are more efficient here.

For GS-EKF in Fig. \ref{fig:FM ekf soekf gsekf}b, the estimation error of the inverse filter is the same as that of the forward filter when $\overline{l}=2$ but improves significantly when $\overline{l}=5$. Note that this improvement in performance comes at the expense of increased computational complexity because the inverse filter estimates an augmented state of dimension `$l(n_{x}+1)$', which is larger than the forward filter's state dimension `$n_{x}$'. The I-EKF/SOEKF assume initial covariance $\bm{\Sigma}_{0}$ as $5\mathbf{I}_{2}$ (the true $\bm{\Sigma}_{0}$ of forward filters are $10\mathbf{I}_{2}$) and a random initial state for these recursions. In spite of this difference in the initial estimates, I-EKF/SOEKF's error performance is comparable to that of the forward EKF/SOEKF. Interestingly, despite similar differences in the initial estimates, I-GS-EKF with $\overline{l}=5$ outperforms the forward GS-EKF.

In Fig.~\ref{fig:FM ekf soekf gsekf}c-d, we consider the case when the defender assumes a different forward filter from the attacker's actual filter. In Fig.~\ref{fig:FM ekf soekf gsekf}c, the I-GS-EKF with $\overline{l}=5$ is considered. The attacker's true forward filter and that assumed by the defender for different inverse filters of Fig.~\ref{fig:FM ekf soekf gsekf}c-d are provided in Table~\ref{tbl:ekf mismatch describe}. The forward as well as inverse filters (with perfect information of forward filter) are also included for comparison. From Fig.~\ref{fig:FM ekf soekf gsekf}c-d, we observe that the I-SOEKF and I-EKF have similar performance regardless of the true forward filter employed by the attacker because both the forward EKF and SOEKF also have same estimation accuracy. On the other hand, forward GS-EKF performs better than EKF for the considered system. Hence, I-EKF-2 based on observations from forward GS-EKF has lower estimation error than I-EKF-1 (with EKF as true forward filter) even though our assumption about forward filter is not true. However, I-EKF-2's estimation error is still more than I-GS-EKF-1 which knows the true forward filter. Furthermore, I-GS-EKF assuming forward GS-EKF even though the true forward filter is EKF (I-GS-EKF-2 case) also has lower estimation error than I-EKF-1, which uses perfect forward filter information. Hence, we conclude that assuming a more sophisticated forward GS-EKF and using I-GS-EKF provides higher estimation accuracy than I-EKF at the expense of computational efforts, regardless of the forward filter employed by the attacker.
\begin{figure}
  \centering
  \includegraphics[width = 0.3\columnwidth]{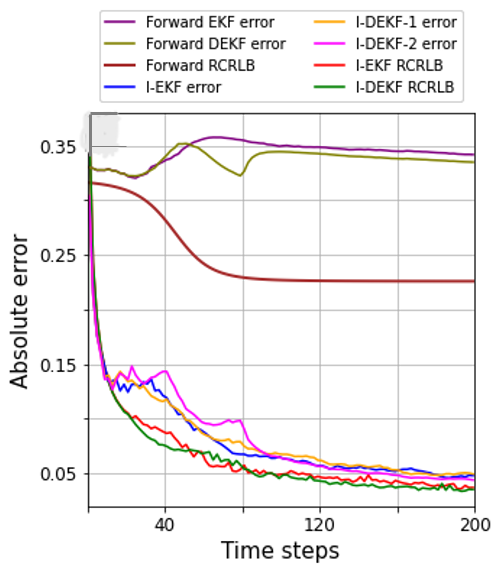}
  \caption{Absolute error and RCRLB for forward and inverse EKF as well as DEKF. I-DEKF-1 is the I-DEKF without considering the modified dithered function in the inverse filter formulation, while I-DEKF-2 considers the modified function.}
 \label{fig:coordinate dekf}
\end{figure}

\subsection{Coordinate estimation with I-DEKF}\label{subsec:coordinate DEKF}
We consider the application of coordinate estimation of a stationary target from bearing observations taken by a moving sensor \cite{weiss1980improved}. The actual coordinates of the stationary target are $(X,Y)$ and that of the sensor at $k$-th time instant are $(p^{x}_{k},p^{y}_{k})$. The constant velocity of sensor is $s$. The forward and inverse EKF as well as DEKF were implemented in a modified coordinate basis with the state estimate $\mathbf{x}_{k}=[p^{x}_{k}/Y, s/Y, s, X/Y]^{T}$ and system model
\par\noindent\small
\begin{align*}
    &\mathbf{x}_{k+1}=\begin{bmatrix}1&\Delta t&0&0\\0&1&0&0\\0&0&1&0\\0&0&0&1\end{bmatrix}\mathbf{x}_{k}+\begin{bmatrix}0\\\Delta t/Y\\\Delta t\\0\end{bmatrix}w_{k},\\
    &y_{k}=\arctan([\mathbf{x}_{k}]_{4}-[\mathbf{x}_{k}]_{1})+v_{k},\;\;
    a_{k}=([\hat{\mathbf{x}}_{k}]_{4})^{2}+\epsilon_{k},
\end{align*}
\normalsize
where $w_{k}\sim\mathcal{N}(0,0.1^{2})$, $v_{k}\sim\mathcal{N}(0,2^{2})$, $\epsilon_{k}\sim\mathcal{N}(0,1.5^{2})$ and $\Delta t=20$ s. The initial covariance estimates were set to $\bm{\Sigma}_{0}=\textrm{diag}(4.44\times 10^{-7}, 0.5\times 10^{-6},1,0.1)$ and $\overline{\bm{\Sigma}}_{0}=\textrm{diag}(10^{-6},6\times 10^{-7},5,0.5)$, respectively, for the forward and inverse filters. Again, we add $10^{-10}\mathbf{I}_{4}$ to both $\mathbf{Q}$ and $\overline{\mathbf{Q}}_{k}$ for numerical stability. The initial state estimate for inverse filters were $[0,0.002,200,2]^{T}$. All other parameters of the system, including the dither and the estimates, were identical to those in \cite{weiss1980improved}.

With 200 time steps, the modified observation function $h^{*}(\cdot)$ in the forward DEKF replaced $h(\cdot)$ up to 80 time steps. While implementing I-DEKF, two cases are considered: (a) the inverse filter has knowledge of the modified observation function $h^{*}(\cdot)$ in the forward filter during the transient phase (I-DEKF-2 case); and (b) the inverse filter considers the unmodified observation function $h(\cdot)$ in its formulation neglecting the modification by the forward filter (I-DEKF-1 case). Fig. \ref{fig:coordinate dekf} shows the absolute error and RCRLB, averaged over 400 runs, for estimation of $X/Y$ whose estimate at the $k$-th time instant are given by $[\hat{\mathbf{x}}_{k}]_{4}$ and $[\doublehat{\mathbf{x}}_{k}]_{4}$ of the forward and inverse filters. The RCRLB at $k$-th time instant is $\sqrt{\left[\mathbf{J}_{k}^{-1}\right]_{4,4}}$. The inverse filters' estimation errors were significantly lower than that of the forward filters. While I-DEKF-1 and I-DEKF-2 differ in their transient performance, they converge to the same steady-state error as I-EKF.

\section{Concluding remarks}\label{sec:IEKF conclusions}
In this chapter, we investigated the non-linear inverse filtering problem for additive Gaussian systems under the assumption of perfect system information. We developed I-EKF based on Taylor series linearization and extended the theory to highly non-linear systems that employ inverses of SOEKF, GS-EKF, and DEKF. These variants may provide improved estimation performance or stability depending on the system. The stochastic stability of a forward filter, given certain additional system assumptions, is also sufficient for the stability of the inverse filter. For I-EKF, we examined its stability using two different approaches, each with unique advantages. We provided sufficient conditions for the exponential mean-squared boundedness of I-SOEKF's estimation error and obtained stability guarantees for the forward SOEKF in the process. The effectiveness of the proposed inverse filters was demonstrated through numerical examples using RCRLB as a performance measure. Our experiments with mismatched forward filters suggested that the inverse filter assuming a similar advanced forward filter may provide better state estimates but at a high computational cost. In certain systems, the inverse filter may even outperform the forward filter in terms of efficiency.

\chapter{Inverse filters for systems with unknown inputs}
\label{chap:unknown inputs}
In this chapter, we develop I-EKFs and I-KFs for systems with unknown inputs in Section~\ref{sec:IEKF with unknown} and \ref{sec:IKF with unknown}, respectively. Unknown inputs refer to exogenous excitations to the system that affect the state transition and observations but are not known to the agent employing the stochastic filter. Examples of systems with unknown inputs include fault detection with unknown excitations \cite{yang2007adaptive} and missile-target interception with unknown target acceleration \cite{pan2010applying}. The inverse cognition in these applications would then resort to the I-EKFs outlined in the following section. Recall from Section~\ref{sec:contributions} that for systems with unknown inputs, the attacker's state estimate also depends on its estimate of the unknown input such that the attacker's forward filter varies with system models. Furthermore, the I-KF-with-unknown-input for linear systems was also not examined in \cite{krishnamurthy2019how} wherein I-KF (without unknown input) was proposed. In this chapter, we also present conditions for asymptotic stability of I-KF-with-unknown-input. Finally, in Section~\ref{sec:unknown input numericals}, we corroborate our results with numerical experiments before concluding in Section~\ref{sec:unknown input conclusions}.

One of the earliest approaches to treat the unknown input was to model the inputs as a stochastic process with known evolution dynamics and jointly estimate the state and inputs. Relaxing the known input dynamics assumption, \cite{kitanidis1987unbiased,gillijns2007unknownkf,gillijns2007kfb,zhang2022boundedness} developed and analyzed unbiased minimum variance linear filters with unknown inputs. Recently, \cite{marco2022regularized,kong2021kalman} have also considered non-persistent and norm-constrained unknown input estimation in linear systems. Various EKF variants to handle unknown inputs in non-linear systems have also been proposed\cite{yang2007adaptive,pan2010applying,xiao2018adaptive,meyer2020unknown,kim2020simultaneous}. In the following, we consider a more general EKF with unknown inputs based on a weighted least squared error criterion in case of both without \cite{pan2010applying} and with \cite{yang2007adaptive} \emph{direct feed-through} (DF). We do not make any other assumptions about the inputs. Note that the EKF linearizes the model based on the nominal values of the state vector and control input. It is similar to the iterated least squares (ILS) method except that the former is for dynamical systems and the latter is not \cite{mendel1995lessons}.

\begin{remark}[Forward filters with and without DF]\label{remark:with and without diff}
Note that the optimal forward filters with and without DF are conceptually different. In the latter case, while the observation $\mathbf{y}_{k}$ in \eqref{eqn:non-linear observation y} is unaffected by the unknown input $\mathbf{u}_{k}$, it is still dependent on $\mathbf{u}_{k-1}$ through $\mathbf{x}_k$; this induces a one-step delay in the attacker's estimate of $\mathbf{u}_k$. On the other hand, with DF, i.e., when observation $\mathbf{y}_{k}$ is given by \eqref{eqn:non-linear observation y with unknown input}, there is no such delay in estimating $\mathbf{u}_k$.
\end{remark}

We now show that this difference results in different inverse filters for these two cases. We define $\hat{\mathbf{u}}_{k}$ to be the estimate of $\mathbf{u}_k$ as computed in the attacker's forward filter, while $\doublehat{\mathbf{u}}_{k}$ is an estimate of $\hat{\mathbf{u}}_{k}$ as computed by the defender's inverse filter.

\section{Inverse EKFs with unknown input}\label{sec:IEKF with unknown}
\subsection{I-EKF-without-DF}\label{subsec:IEKF without DF}
Consider the non-linear system without DF given by \eqref{eqn:non-linear state x with unknown input} and \eqref{eqn:non-linear observation y}. We linearize the model functions as $\mathbf{F}_{k}\doteq\nabla_{\mathbf{x}}f(\mathbf{x},\hat{\mathbf{u}}_{k-1})|_{\mathbf{x}=\hat{\mathbf{x}}_{k}}$, $\mathbf{B}_{k} \doteq\nabla_{\mathbf{u}}f(\hat{\mathbf{x}}_{k},\mathbf{u})|_{\mathbf{u}=\hat{\mathbf{u}}_{k-1}}$ and $\mathbf{H}_{k+1}\doteq\nabla_{\mathbf{x}}h(\mathbf{x})|_{\mathbf{x}=\hat{\mathbf{x}}_{k+1|k}}$.

\textit{Forward filter:} Similar to forward EKF described in Section~\ref{subsec:IEKF formulation}, the forward filter consists of time and measurement update steps. In the time update, we predict the state $\mathbf{x}_{k+1}$ as $\hat{\mathbf{x}}_{k+1|k}$ using the previous state and input estimates, with $\bm{\Sigma}^{x}_{k+1|k}$ as the associated state prediction error covariance matrix. Then, the state and input gain matrices $\mathbf{K}^{x}_{k+1}$ and $\mathbf{K}^{u}_{k}$, respectively, are computed along with the input estimation (with delay) covariance matrix $\bm{\Sigma}^{u}_{k}$. In the measurement update, the state estimate $\hat{\mathbf{x}}_{k+1}$, input estimate $\hat{\mathbf{u}}_{k}$, and covariance matrix $\bm{\Sigma}^{x}_{k+1}$ are updated using current observation $\mathbf{y}_{k+1}$, and gain matrices $\mathbf{K}^{x}_{k+1}$ and $\mathbf{K}^{u}_{k}$. Note that the current observation $\mathbf{y}_{k+1}$ provides an estimate $\hat{\mathbf{u}}_{k}$ of the input $\mathbf{u}_{k}$ at the previous time step. The attacker's forward EKF's recursions are\cite{pan2010applying}:
\par\noindent\small
\begin{align}
&\textit{Time update:}\;\;\hat{\mathbf{x}}_{k+1|k}=f(\hat{\mathbf{x}}_{k},\hat{\mathbf{u}}_{k-1}),\label{eqn: ekfwithoutdf predict}\\
&\bm{\Sigma}^{x}_{k+1|k}=\mathbf{F}_{k}\bm{\Sigma}^{x}_{k}\mathbf{F}_{k}^{T}+\mathbf{Q},\nonumber\\
&\textit{Gain computation:}\;\;\mathbf{K}^{x}_{k+1}=\bm{\Sigma}^{x}_{k+1|k}\mathbf{H}_{k+1}^{T}\left(\mathbf{H}_{k+1}\bm{\Sigma}^{x}_{k+1|k}\mathbf{H}_{k+1}^{T}+\mathbf{R}\right)^{-1},\nonumber\\
&\bm{\Sigma}^{u}_{k}=\left(\mathbf{B}_{k}^{T}\mathbf{H}_{k+1}^{T}\mathbf{R}^{-1}(\mathbf{I}_{n_{y}\times n_{y}}-\mathbf{H}_{k+1}\mathbf{K}^{x}_{k+1})\mathbf{H}_{k+1}\mathbf{B}_{k}\right)^{-1},\nonumber\\
&\mathbf{K}^{u}_{k}=\bm{\Sigma}^{u}_{k}\mathbf{B}_{k}^{T}\mathbf{H}_{k+1}^{T}\mathbf{R}^{-1}(\mathbf{I}_{n_{y}\times n_{y}}-\mathbf{H}_{k+1}\mathbf{K}^{x}_{k+1}),\nonumber\\
&\textit{Measurement update:}\;\hat{\mathbf{x}}_{k+1}=\hat{\mathbf{x}}_{k+1|k}+\mathbf{K}^{x}_{k+1}(\mathbf{y}_{k+1}-h(\hat{\mathbf{x}}_{k+1|k})),\label{eqn: ekfwithoutdf update x}\\
&\hat{\mathbf{u}}_{k}=\mathbf{K}^{u}_{k}(\mathbf{y}_{k+1}-h(\hat{\mathbf{x}}_{k+1|k})+\mathbf{H}_{k+1}\mathbf{B}_{k}\hat{\mathbf{u}}_{k-1}),\label{eqn: ekfwithoutdf update u}\\
&\bm{\Sigma}^{x}_{k+1}=(\mathbf{I}_{n_{x}\times n_{x}}-\mathbf{K}^{x}_{k+1}\mathbf{H}_{k+1})\left(\bm{\Sigma}^{x}_{k+1|k}+\mathbf{B}_{k}\bm{\Sigma}^{u}_{k}\mathbf{B}_{k}^{T}(\mathbf{I}_{n_{x}\times n_{x}}-\mathbf{K}^{x}_{k+1}\mathbf{H}_{k+1})^{T}\right).\nonumber
\end{align}
\normalsize
The forward filter exists if $\textrm{rank}(\bm{\Sigma}^{u}_{k})=n_{u}$, for all $k\geq 0$, and $n_{y} \geq n_{u}$ \cite{pan2010applying}, where $n_{u}$ and $n_{y}$ are the dimensions of unknown input $\mathbf{u}_{k}$ and observation $\mathbf{y}_{k}$, respectively. We provide a detailed derivation of the forward EKF-without-DF recursions (omitted in \cite{pan2010applying}) in Appendix~\ref{App-thm-forward-EKF-without-DF}.

\textit{Inverse filter:} Consider an augmented state vector $\mathbf{z}_{k}=\begin{bmatrix}
\hat{\mathbf{x}}_{k}^{T} & \hat{\mathbf{u}}_{k-2}^{T}
\end{bmatrix}^{T}$. The defender's observation $\mathbf{a}_{k}$ in \eqref{eqn:non-linear observation a} is the first observation that contains the information about unknown input estimate $\hat{\mathbf{u}}_{k-2}$, because of the delay in forward filter input estimate. Hence, the delayed estimate $\hat{\mathbf{u}}_{k-2}$ is considered in the augmented state $\mathbf{z}_{k}$. Define $\widetilde{\phi}_{k}(\hat{\mathbf{x}}_{k},\hat{\mathbf{u}}_{k-1},\mathbf{x}_{k+1},\mathbf{v}_{k+1})=f(\hat{\mathbf{x}}_{k},\hat{\mathbf{u}}_{k-1})-\mathbf{K}^{x}_{k+1}h(f(\hat{\mathbf{x}}_{k},\hat{\mathbf{u}}_{k-1}))+\mathbf{K}^{x}_{k+1}h(\mathbf{x}_{k+1})+\mathbf{K}^{x}_{k+1}\mathbf{v}_{k+1}$.
From \eqref{eqn:non-linear observation y} and \eqref{eqn: ekfwithoutdf predict}-\eqref{eqn: ekfwithoutdf update u}, state transition equations of augmented state vector are $\hat{\mathbf{x}}_{k+1}=\widetilde{f}_{k}(\hat{\mathbf{x}}_{k},\hat{\mathbf{u}}_{k-2},\hat{\mathbf{x}}_{k-1},\mathbf{x}_{k},\mathbf{x}_{k+1},\mathbf{v}_{k},\mathbf{v}_{k+1})$ and 
$\hat{\mathbf{u}}_{k-1}=\widetilde{h}_{k}(\hat{\mathbf{u}}_{k-2},\hat{\mathbf{x}}_{k-1},\mathbf{x}_{k},\mathbf{v}_{k})$, where
\par\noindent\small
\begin{align}
&\widetilde{h}_{k}(\hat{\mathbf{u}}_{k-2},\hat{\mathbf{x}}_{k-1},\mathbf{x}_{k},\mathbf{v}_{k})=\mathbf{K}^{u}_{k-1}(\mathbf{H}_{k}\mathbf{B}_{k-1}\hat{\mathbf{u}}_{k-2}-h(f(\hat{\mathbf{x}}_{k-1},\hat{\mathbf{u}}_{k-2}))+h(\mathbf{x}_{k})+\mathbf{v}_{k}),\label{eqn: state transition ekf without df input}\\
&\widetilde{f}_{k}(\hat{\mathbf{x}}_{k},\hat{\mathbf{u}}_{k-2},\hat{\mathbf{x}}_{k-1},\mathbf{x}_{k},\mathbf{x}_{k+1},\mathbf{v}_{k},\mathbf{v}_{k+1})=\widetilde{\phi}_{k}(\hat{\mathbf{x}}_{k},\widetilde{h}_{k}(\hat{\mathbf{u}}_{k-2},\hat{\mathbf{x}}_{k-1},\mathbf{x}_{k},\mathbf{v}_{k}),\mathbf{x}_{k+1},\mathbf{v}_{k+1}).\label{eqn: state transition ekf without df state}
\end{align}
\normalsize
In these state transition equations, the actual states $\mathbf{x}_{k}$ and $\mathbf{x}_{k+1}$ are perfectly known to the defender and henceforth treated as known exogenous inputs. Unlike the forward filter, the process noise terms $\mathbf{v}_{k}$ and $\mathbf{v}_{k+1}$ are non-additive because the filter gains $\mathbf{K}^{x}_{k+1}$ and $\mathbf{K}^{u}_{k-1}$ depend on the previous estimates (through the Jacobians).

Denote $\hat{\mathbf{z}}_{k+1} \doteq \begin{bmatrix}
\doublehat{\mathbf{x}}_{k+1}^{T} & \doublehat{\mathbf{u}}_{k-1}^{T}
\end{bmatrix}^{T}$. The state transition of the augmented state $\mathbf{z}_{k+1}$ depends on the estimate $\hat{\mathbf{x}}_{k-1}$ which the defender approximates by its previous estimate $\doublehat{\mathbf{x}}_{k-1}$. With this approximation, $\hat{\mathbf{x}}_{k-1}$ is treated as another known exogenous input for the inverse filter while the augmented process noise vector is $\begin{bmatrix}
\mathbf{v}_{k}^{T} & \mathbf{v}_{k+1}^{T}
\end{bmatrix}^{T}$. Define the Jacobians $\widetilde{\mathbf{F}}^{z}_{k}\doteq \begin{bmatrix}
\nabla_{\doublehat{\mathbf{x}}_{k}}\widetilde{f}_{k} & \nabla_{\doublehat{\mathbf{u}}_{k-2}}\widetilde{f}_{k}\\
\mathbf{0}_{n_{u}\times n_{x}} & \nabla_{\doublehat{\mathbf{u}}_{k-2}}\widetilde{h}_{k}
\end{bmatrix}$, and $\mathbf{G}_{k+1}\doteq\begin{bmatrix}
\nabla_{\doublehat{\mathbf{x}}_{k+1|k}}g & \mathbf{0}_{n_{a}\times n_{u}}\end{bmatrix}$ with respect to the augmented state; Jacobian $\widetilde{\mathbf{F}}^{v}_{k}\doteq\begin{bmatrix}
\nabla_{\mathbf{v}_{k}}\widetilde{f}_{k} & \nabla_{\mathbf{v}_{k+1}}\widetilde{f}_{k}\\
\nabla_{\mathbf{v}_{k}}\widetilde{h}_{k} & \mathbf{0}_{n_{u}\times n_{y}}
\end{bmatrix}$ with respect to the augmented process noise vector; and $\overline{\mathbf{Q}}_{k}=\widetilde{\mathbf{F}}^{v}_{k}\begin{bmatrix}
\mathbf{R} & \mathbf{0}_{n_{y}\times n_{y}}\\ \mathbf{0}_{n_{y}\times n_{y}} & \mathbf{R}
\end{bmatrix}(\widetilde{\mathbf{F}}^{v}_{k})^{T}$. Then, the I-EKF-without-DF's recursions yield the estimate $\hat{\mathbf{z}}_{k}$ of the augmented state and the associated covariance matrix $\overline{\bm{\Sigma}}_{k}$ as:
\par\noindent\small
\begin{align}
&\textit{Time update:}\;\doublehat{\mathbf{x}}_{k+1|k}=\widetilde{f}_{k}(\doublehat{\mathbf{x}}_{k},\doublehat{\mathbf{u}}_{k-2},\doublehat{\mathbf{x}}_{k-1},\mathbf{x}_{k},\mathbf{x}_{k+1},\mathbf{0}_{n_{y}\times 1},\mathbf{0}_{n_{y}\times 1}),\nonumber\\
&\doublehat{\mathbf{u}}_{k-1|k}=\widetilde{h}_{k}(\doublehat{\mathbf{u}}_{k-2},\doublehat{\mathbf{x}}_{k-1},\mathbf{x}_{k},\mathbf{0}_{n_{y}\times 1}),\;\;\;\hat{\mathbf{z}}_{k+1|k}=\begin{bmatrix}
\doublehat{\mathbf{x}}_{k+1|k}^{T} & \doublehat{\mathbf{u}}_{k-1|k}^{T}
\end{bmatrix}^{T},\nonumber\\
&\overline{\bm{\Sigma}}_{k+1|k}=\widetilde{\mathbf{F}}^{z}_{k}\overline{\bm{\Sigma}}_{k}(\widetilde{\mathbf{F}}^{z}_{k})^{T}+\overline{\mathbf{Q}}_{k},\label{eqn: I-EKF without DF covariance predict}\\
&\textit{Measurement update:}\;\overline{\bm{\Sigma}}^{a}_{k+1}=\mathbf{G}_{k+1}\overline{\bm{\Sigma}}_{k+1|k}\mathbf{G}_{k+1}^{T}+\bm{\Sigma}_{\epsilon},\label{eqn: I-EKF without DF S compute}\\
&\hat{\mathbf{z}}_{k+1}=\hat{\mathbf{z}}_{k+1|k}+\overline{\bm{\Sigma}}_{k+1|k}\mathbf{G}_{k+1}^{T}(\overline{\bm{\Sigma}}^{a}_{k+1})^{-1}\left(\mathbf{a}_{k+1}-g(\doublehat{\mathbf{x}}_{k+1|k})\right),\label{eqn: I-EKF without DF a predict}\\
&\overline{\bm{\Sigma}}_{k+1}=\overline{\bm{\Sigma}}_{k+1|k}-\overline{\bm{\Sigma}}_{k+1|k}\mathbf{G}_{k+1}^{T}(\overline{\bm{\Sigma}}^{a}_{k+1})^{-1}\mathbf{G}_{k+1}\overline{\bm{\Sigma}}_{k+1|k}.\label{eqn: I-EKF without DF covariance update}
\end{align}
\normalsize
\begin{figure}
  \centering
  \includegraphics[width = 0.7\columnwidth]{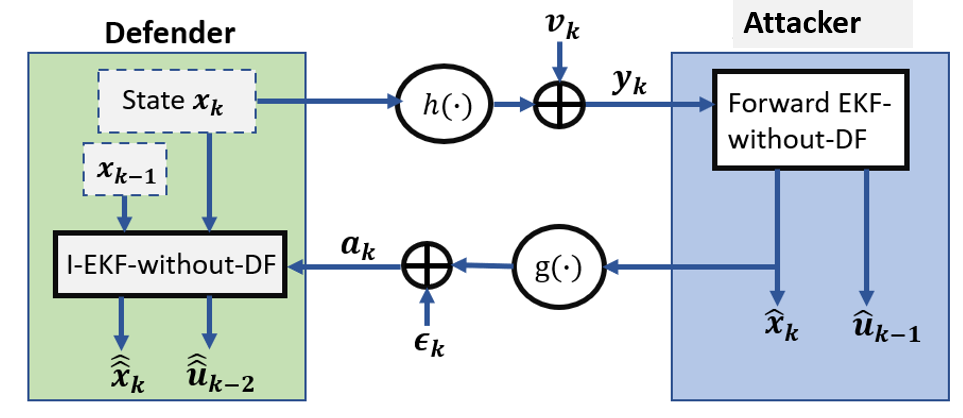}
  \caption{Graphical representation of I-EKF-without-DF recursion at $k$-th time step.}
 \label{fig:I-EKF-without-DF}
\end{figure}

Fig.~\ref{fig:I-EKF-without-DF} provides a schematic diagram for the I-EKF-without-DF updates. The inverse filter recursions take the same form as that of the standard EKF but with modified system matrices. In particular, the former employs an augmented state such that the Jacobian of the state transition function with respect to the state is computed as $\widetilde{\mathbf{F}}^{z}_{k}$ while for the latter, it is simply $\mathbf{F}_{k}\doteq\nabla_{\hat{\mathbf{x}}_{k}} f(\mathbf{x})$. Further, unlike standard KF or EKF, the noise terms $\mathbf{v}_{k}$ and $\mathbf{v}_{k+1}$ in \eqref{eqn: state transition ekf without df input} and \eqref{eqn: state transition ekf without df state} are non-additive such that an approximate process noise covariance matrix $\overline{\mathbf{Q}}_{k}$ is computed using linearization $\widetilde{\mathbf{F}}^{v}_{k}$ of the state transition function with respect to these noise terms. Similar to I-EKF-without-unknown-input of Section~\ref{subsec:IEKF formulation}, the forward filter gains $\mathbf{K}^{x}_{k+1}$ and $\mathbf{K}^{u}_{k-1}$ are treated as time-varying parameters of the state transition equation and approximated by evaluating their values at the inverse filter's estimates ($\doublehat{\mathbf{x}}_{k}$ and $\doublehat{\mathbf{u}}_{k-1}$) in the similar manner as the forward filter evaluates them using its own estimates. 

\subsection{I-EKF-with-DF}\label{subsec:IEKF with DF}
Consider the non-linear system with DF given by \eqref{eqn:non-linear state x with unknown input} and \eqref{eqn:non-linear observation y with unknown input}. We linearize the functions as $\mathbf{F}_{k}\doteq\nabla_{\mathbf{x}}f(\mathbf{x},\hat{\mathbf{u}}_{k})|_{\mathbf{x}=\hat{\mathbf{x}}_{k}}$, $\mathbf{H}_{k+1}\doteq\nabla_{\mathbf{x}}h(\mathbf{x},\hat{\mathbf{u}}_{k})|_{\mathbf{x}=\hat{\mathbf{x}}_{k+1|k}}$ and $\mathbf{D}_{k}\doteq\nabla_{\mathbf{u}}h(\hat{\mathbf{x}}_{k+1|k},\mathbf{u})|_{\mathbf{u}=\hat{\mathbf{u}}_{k}}$.

\textit{Forward filter:} Denote the state and input estimation covariance and gain matrices identical to Section~\ref{subsec:IEKF without DF}. Here, the current observation $\mathbf{y}_{k+1}$ depends on the current unknown input $\mathbf{u}_{k+1}$ such that the forward filter infers $\hat{\mathbf{u}}_{k+1}$ without any delay. For input estimation covariance without delay, we use $\bm{\Sigma}^{u}_{k+1}$. Then, the forward EKF-with-DF's recursions are \cite{yang2007adaptive}
\par\noindent\small
\begin{align}
&\textit{Time update:}\;\;\hat{\mathbf{x}}_{k+1|k}=f(\hat{\mathbf{x}}_{k},\hat{\mathbf{u}}_{k}),\;\bm{\Sigma}^{x}_{k+1|k}=\mathbf{F}_{k}\bm{\Sigma}^{x}_{k}\mathbf{F}_{k}^{T}+\mathbf{Q},\label{eqn: ekfwithdf predict}\\
&\textit{Gain computation:}\;\;\mathbf{K}^{x}_{k+1}=\bm{\Sigma}^{x}_{k+1|k}\mathbf{H}_{k+1}^{T}(\mathbf{H}_{k+1}\bm{\Sigma}^{x}_{k+1|k}\mathbf{H}_{k+1}^{T}+\mathbf{R})^{-1},\nonumber\\
&\bm{\Sigma}^{u}_{k+1}=\left(\mathbf{D}_{k}^{T}\mathbf{R}^{-1}(\mathbf{I}_{n_{y}\times n_{y}}-\mathbf{H}_{k+1}\mathbf{K}^{x}_{k+1})\mathbf{D}_{k}\right)^{-1},\nonumber\\
&\mathbf{K}^{u}_{k+1}=\bm{\Sigma}^{u}_{k+1}\mathbf{D}_{k}^{T}\mathbf{R}^{-1}(\mathbf{I}_{n_{y}\times n_{y}}-\mathbf{H}_{k+1}\mathbf{K}^{x}_{k+1}),\nonumber\\
&\textit{Measurement update:}\;\;\hat{\mathbf{u}}_{k+1}=\mathbf{K}^{u}_{k+1}\left(\mathbf{y}_{k+1}-h(\hat{\mathbf{x}}_{k+1|k},\hat{\mathbf{u}}_{k})+\mathbf{D}_{k}\hat{\mathbf{u}}_{k}\right),\label{eqn: ekfwithdf update u}\\
&\hat{\mathbf{x}}_{k+1}=\hat{\mathbf{x}}_{k+1|k}+\mathbf{K}^{x}_{k+1}\left(\mathbf{y}_{k+1}-h(\hat{\mathbf{x}}_{k+1|k},\hat{\mathbf{u}}_{k})-\mathbf{D}_{k}(\hat{\mathbf{u}}_{k+1}-\hat{\mathbf{u}}_{k})\right),\label{eqn: ekfwithdf update x}\\
&\bm{\Sigma}^{x}_{k+1}=\bm{\Sigma}^{x}_{k+1|k}(\mathbf{I}_{n_{x}\times n_{x}}+\mathbf{K}^{x}_{k+1}\mathbf{D}_{k}\bm{\Sigma}^{u}_{k+1}\mathbf{D}_{k}^{T}\mathbf{R}^{-1}\mathbf{H}_{k+1})(\mathbf{I}_{n_{x}\times n_{x}}-\mathbf{K}^{x}_{k+1}\mathbf{H}_{k+1}).\nonumber
\end{align}
\normalsize
The forward filter exists if $\textrm{rank}(\mathbf{D}_{k})=n_{u}$ for all $k\geq 0$, which implies $n_{y}\geq n_{u}$\cite{yang2007adaptive}.

\textit{Inverse filter:} Consider an augmented state vector $\mathbf{z}_{k}=\begin{bmatrix}
\hat{\mathbf{x}}_{k}^{T} & \hat{\mathbf{u}}_{k}^{T}
\end{bmatrix}^{T}$  (note the absence of delay in the input estimate). Define $\widetilde{\phi}_{k}(\hat{\mathbf{x}}_{k},\hat{\mathbf{u}}_{k},\hat{\mathbf{u}}_{k+1},\mathbf{x}_{k+1},\mathbf{u}_{k+1},\mathbf{v}_{k+1})=f(\hat{\mathbf{x}}_{k},\hat{\mathbf{u}}_{k})-\mathbf{K}^{x}_{k+1}h(f(\hat{\mathbf{x}}_{k},\hat{\mathbf{u}}_{k}),\hat{\mathbf{u}}_{k})-\mathbf{K}^{x}_{k+1}\mathbf{D}_{k}(\hat{\mathbf{u}}_{k+1}-\hat{\mathbf{u}}_{k})+\mathbf{K}^{x}_{k+1}h(\mathbf{x}_{k+1},\mathbf{u}_{k+1})+\mathbf{K}^{x}_{k+1}\mathbf{v}_{k+1}$.
From \eqref{eqn:non-linear observation y with unknown input} and \eqref{eqn: ekfwithdf predict}-\eqref{eqn: ekfwithdf update x}, state transitions for inverse filter are $\hat{\mathbf{x}}_{k+1}=\widetilde{f}_{k}(\hat{\mathbf{x}}_{k},\hat{\mathbf{u}}_{k},\mathbf{x}_{k+1},\mathbf{u}_{k+1},\mathbf{v}_{k+1})$ and $\hat{\mathbf{u}}_{k+1}=\widetilde{h}_{k}(\hat{\mathbf{x}}_{k},\hat{\mathbf{u}}_{k},\mathbf{x}_{k+1},\mathbf{u}_{k+1},\mathbf{v}_{k+1})$, 
where 
\par\noindent\small
\begin{align}
&\widetilde{h}_{k}(\hat{\mathbf{x}}_{k},\hat{\mathbf{u}}_{k},\mathbf{x}_{k+1},\mathbf{u}_{k+1},\mathbf{v}_{k+1})=\mathbf{K}^{u}_{k+1}(h(\mathbf{x}_{k+1},\mathbf{u}_{k+1})+\mathbf{v}_{k+1}-h(f(\hat{\mathbf{x}}_{k},\hat{\mathbf{u}}_{k}),\hat{\mathbf{u}}_{k})+\mathbf{D}_{k}\hat{\mathbf{u}}_{k})\nonumber\\
&\widetilde{f}_{k}(\hat{\mathbf{x}}_{k},\hat{\mathbf{u}}_{k},\mathbf{x}_{k+1},\mathbf{u}_{k+1},\mathbf{v}_{k+1})=\widetilde{\phi}_{k}(\hat{\mathbf{x}}_{k},\hat{\mathbf{u}}_{k},\widetilde{h}_{k}(\hat{\mathbf{x}}_{k},\hat{\mathbf{u}}_{k},\mathbf{x}_{k+1},\mathbf{u}_{k+1},\mathbf{v}_{k+1}),\mathbf{x}_{k+1},\mathbf{u}_{k+1},\mathbf{v}_{k+1}).\label{eqn:state transition ekf with df}
\end{align}
\normalsize

\begin{figure}
  \centering
  \includegraphics[width = 0.7\columnwidth]{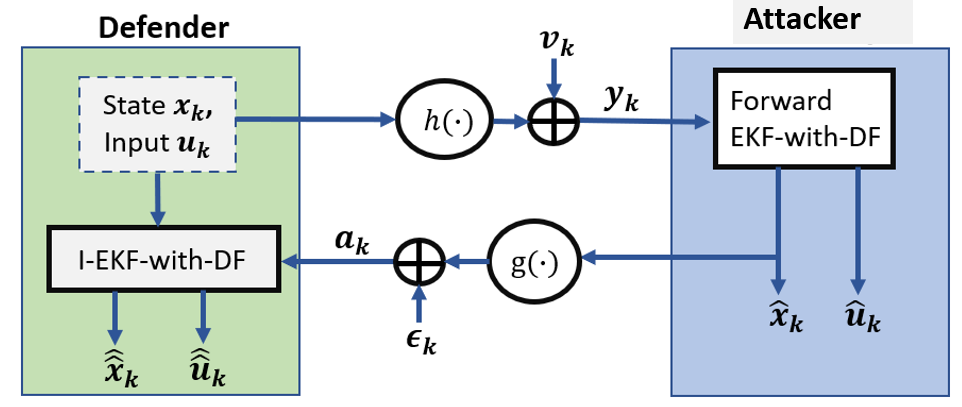}
  \caption{Graphical representation of I-EKF-with-DF recursion at $k$-th time step.}
 \label{fig:I-EKF-with-DF}
\end{figure}

Then, \textit{ceteris paribus}, following similar steps as in I-EKF-without-DF, the I-EKF-with-DF estimate $\hat{\mathbf{z}}_{k}=\begin{bmatrix}
\doublehat{\mathbf{x}}_{k}^{T} & \doublehat{\mathbf{u}}_{k}^{T}
\end{bmatrix}^{T}$ from observations \eqref{eqn:non-linear observation a} is computed recursively. The predicted augmented state is $\hat{\mathbf{z}}_{k+1|k}=\begin{bmatrix}
\doublehat{\mathbf{x}}_{k+1|k}^{T} & \doublehat{\mathbf{u}}_{k+1|k}^{T}
\end{bmatrix}^{T}$, where $\doublehat{\mathbf{x}}_{k+1|k}=\widetilde{f}_{k}(\doublehat{\mathbf{x}}_{k},\doublehat{\mathbf{u}}_{k},\mathbf{x}_{k+1},\mathbf{u}_{k+1},\mathbf{0}_{n_{y}\times 1})$ and $\doublehat{\mathbf{u}}_{k+1|k}=\widetilde{h}_{k}(\doublehat{\mathbf{x}}_{k},\doublehat{\mathbf{u}}_{k},\mathbf{x}_{k+1},\mathbf{u}_{k+1},\mathbf{0}_{n_{y}\times 1})$. Hereafter, the remaining steps are as in \eqref{eqn: I-EKF without DF covariance predict}-\eqref{eqn: I-EKF without DF covariance update}. For I-EKF-with-DF, the Jacobians with respect to the augmented state are $\widetilde{\mathbf{F}}^{z}_{k}\doteq\begin{bmatrix}
\nabla_{\doublehat{\mathbf{x}}_{k}}\widetilde{f}_{k} & \nabla_{\doublehat{\mathbf{u}}_{k}}\widetilde{f}_{k} \\
\nabla_{\doublehat{\mathbf{x}}_{k}}\widetilde{h}_{k} & \nabla_{\doublehat{\mathbf{u}}_{k}}\widetilde{h}_{k}
\end{bmatrix}$ and $\mathbf{G}_{k+1}\doteq\begin{bmatrix}
\nabla_{\doublehat{\mathbf{x}}_{k+1|k}}g & \mathbf{0}_{n_{a}\times n_{u}}\end{bmatrix}$; the Jacobian with respect to the process noise term is $\widetilde{\mathbf{F}}^{v}_{k}\doteq\begin{bmatrix}
\nabla_{\mathbf{v}_{k+1}}\widetilde{f}_{k}\\
\nabla_{\mathbf{v}_{k+1}}\widetilde{h}_{k}
\end{bmatrix}$; and $\overline{\mathbf{Q}}_{k}=\widetilde{\mathbf{F}}^{v}_{k}\mathbf{R}(\widetilde{\mathbf{F}}^{v}_{k})^{T}$. Fig.~\ref{fig:I-EKF-with-DF} shows these updates graphically. Note that unlike I-EKF-without-DF, I-EKF-with-DF requires the true input $\mathbf{u}_{k}$ information. Here, unlike I-EKF-without-DF, the inverse filter's prediction dispenses with any approximation of $\hat{\mathbf{x}}_{k-1}$. The absence of delay in input estimation also results in a simplified process noise term $\mathbf{v}_{k+1}$, in place of I-EKF-without-DF's augmented noise vector.

\begin{remark}[I-EKF-with-unknown-input's computational complexity]\label{remark:iekf unknown input computation complexity}
Note that the I-EKF-with-unknown-input recursions are obtained from that of a standard EKF but with the state transition equation representing the corresponding forward filter's state estimate evolution. Hence, these I-EKFs have similar computational complexity as a standard EKF, i.e., $\mathcal{O}(d^3)$ where $d$ is the dimension of the estimated state vector\cite{daum2005nonlinear}. However, I-EKF with and without DF would estimate the augmented states $\mathbf{z}_{k}=[\hat{\mathbf{x}}_{k}^{T},\hat{\mathbf{u}}_{k-2}^{T}]^{T}$ and $\mathbf{z}_{k}=[\hat{\mathbf{x}}_{k}^{T},\hat{\mathbf{u}}_{k}^{T}]^{T}$, respectively. Hence, the overall computational complexity of these filters depends on the dimension of state $\mathbf{x}_{k}$ as well as the dimension of unknown inputs $\mathbf{u}_{k}$ in the system.
\end{remark}

\section{Inverse KFs with unknown input}\label{sec:IKF with unknown}
For linear Gaussian state-space models, our methods developed in the previous section are useful in extending the I-KF mentioned in \cite{krishnamurthy2019how} to unknown input. Again, the forward KFs employed by the attacker with and without DF are conceptually different \cite{gillijns2007kfb} because of the delay involved in input estimation. The forward KFs with unknown input provide unbiased minimum variance state and input estimates.

\subsection{I-KF-without-DF}\label{subsec:IKF without DF}
Consider the linear system without DF given by \eqref{eqn: linear x with input} and \eqref{eqn: linear y withdf} with $\mathbf{D}=\mathbf{0}_{n_{y}\times n_{u}}$.

\textit{Forward filter:} Unlike EKF-without-DF, the forward KF-without-DF considers an intermediate state update step using the estimated unknown input before the final state updates. In this step, the unknown input is first estimated (with one-step delay) using the current observation $\mathbf{y}_{k+1}$ and input estimation gain matrix $\mathbf{M}_{k+1}$. In the update step, the current state estimate $\hat{\mathbf{x}}_{k+1}$ is computed by again considering the current observation $\mathbf{y}_{k+1}$ as\cite{gillijns2007unknownkf}
\par\noindent\small
\begin{align}
&\textit{Time update:}\;\;\hat{\mathbf{x}}_{k+1|k}=\mathbf{F}\hat{\mathbf{x}}_{k},\;\bm{\Sigma}_{k+1|k}=\mathbf{F}\bm{\Sigma}_{k}\mathbf{F}^{T}+\mathbf{Q},\label{eqn: kfwithoutdf predict}\\
&\textit{Unknown input estimation:}\;\bm{\Sigma}^{y}_{k+1}=\mathbf{H}\bm{\Sigma}_{k+1|k}\mathbf{H}^{T}+\mathbf{R},\\
&\mathbf{M}_{k+1}=(\mathbf{B}^{T}\mathbf{H}^{T}(\bm{\Sigma}^{y}_{k+1})^{-1}\mathbf{HB})^{-1}\mathbf{B}^{T}\mathbf{H}^{T}(\bm{\Sigma}^{y}_{k+1})^{-1},
\end{align}
\begin{align}
&\textit{Unknown input estimation (contd.):}\;\hat{\mathbf{u}}_{k}=\mathbf{M}_{k+1}(\mathbf{y}_{k+1}-\mathbf{H}\hat{\mathbf{x}}_{k+1|k}),\label{eqn: kfwithoutdf update u}\\
&\widetilde{\mathbf{x}}_{k+1|k+1}=\hat{\mathbf{x}}_{k+1|k}+\mathbf{B}\hat{\mathbf{u}}_{k},\label{eqn: kfwithoutdf update x with u}\\
&\widetilde{\bm{\Sigma}}_{k+1|k+1}=(\mathbf{I}_{n_{x}\times n_{x}}-\mathbf{BM}_{k+1}\mathbf{H})\bm{\Sigma}_{k+1|k}(\mathbf{I}_{n_{x}\times n_{x}}-\mathbf{BM}_{k+1}\mathbf{H})^{T}+\mathbf{BM}_{k+1}\mathbf{RM}_{k+1}^{T}\mathbf{B}^{T},\\
&\textit{Measurement update:}\;\;\mathbf{K}_{k+1}=\bm{\Sigma}_{k+1|k}\mathbf{H}^{T}(\bm{\Sigma}^{y}_{k+1})^{-1},\\
&\hat{\mathbf{x}}_{k+1}=\widetilde{\mathbf{x}}_{k+1|k+1}+\mathbf{K}_{k+1}(\mathbf{y}_{k+1}-\mathbf{H}\widetilde{\mathbf{x}}_{k+1|k+1}),\label{eqn: kfwithoutdf update x}\\
&\bm{\Sigma}_{k+1}=\widetilde{\bm{\Sigma}}_{k+1|k+1}-\mathbf{K}_{k+1}(\widetilde{\bm{\Sigma}}_{k+1|k+1}\mathbf{H}^{T}-\mathbf{BM}_{k+1}\mathbf{R})^{T}.\label{eqn: kfwithoutdf sigma update}
\end{align}
\normalsize
The forward filter exists if $\textrm{rank}(\mathbf{HB})=\textrm{rank}(\mathbf{B})=n_{u}$ which implies $n_{x}\geq n_{u}$ and $n_{y}\geq n_{u}$\cite{gillijns2007unknownkf}. Here, unlike I-EKFs, the gain matrices $\mathbf{K}_{k+1}$ and $\mathbf{M}_{k+1}$, are deterministic and completely determined by the model parameters and the initial covariance matrix similar to I-KF\cite{krishnamurthy2019how}.

\textit{Inverse filter:} Denote $\widetilde{\mathbf{F}}_{k}=(\mathbf{I}_{n_{x}\times n_{x}}-\mathbf{K}_{k+1}\mathbf{H})(\mathbf{I}_{n_{x}\times n_{x}}-\mathbf{BM}_{k+1}\mathbf{H})\mathbf{F}$ and $\mathbf{E}_{k}=\mathbf{BM}_{k+1}-\mathbf{K}_{k+1}\mathbf{HBM}_{k+1}+\mathbf{K}_{k+1}$. From \eqref{eqn: linear y withdf} with $\mathbf{D}=\mathbf{0}_{n_{y}\times n_{u}}$, and \eqref{eqn: kfwithoutdf predict}-\eqref{eqn: kfwithoutdf update x}, the state transition equation for I-KF-without-DF is
\par\noindent\small
\begin{align}
\label{eqn: state for kfwithoutdf}
\hat{\mathbf{x}}_{k+1}=\widetilde{\mathbf{F}}_{k}\hat{\mathbf{x}}_{k}+\mathbf{E}_{k}\mathbf{Hx}_{k+1}+\mathbf{E}_{k}\mathbf{v}_{k+1}.
\end{align}
\normalsize
Unlike the state transition \eqref{eqn: state transition ekf without df input} and \eqref{eqn: state transition ekf without df state} of I-EKF-without-DF, the state transition for I-KF-without-DF is not an explicit function of the forward filter input estimate and hence, an augmented state is not needed. The difference arises from the forward EKF-without-DF, where the current input estimate explicitly depends on the previous input estimates as observed in \eqref{eqn: ekfwithoutdf update u}, which is not the case in KF-without-DF. The I-KF-without-DF's recursions with observation \eqref{eqn: linear a} are:
\par\noindent\small
\begin{align}
&\textit{Time update:} \;\;
\doublehat{\mathbf{x}}_{k+1|k}=\widetilde{\mathbf{F}}_{k}\doublehat{\mathbf{x}}_{k}+\mathbf{E}_{k}\mathbf{Hx}_{k+1},\label{eqn: inverse kfwithoutdf state predict}\\
&\overline{\bm{\Sigma}}_{k+1|k}=\widetilde{\mathbf{F}}_{k}\overline{\bm{\Sigma}}_{k}\widetilde{\mathbf{F}}_{k}^{T}+\overline{\mathbf{Q}}_{k},\label{eqn: inverse kfwithoutdf covariance predict}\\
&\textit{Measurement update:}\;\;\overline{\bm{\Sigma}}^{a}_{k+1}=\mathbf{G}\overline{\bm{\Sigma}}_{k+1|k}\mathbf{G}^{T}+\bm{\Sigma}_{\epsilon},\label{eqn: inverse kfwithoutdf gain}\\
&\doublehat{\mathbf{x}}_{k+1}=\doublehat{\mathbf{x}}_{k+1|k}+\overline{\bm{\Sigma}}_{k+1|k}\mathbf{G}^{T}(\overline{\bm{\Sigma}}^{a}_{k+1})^{-1}(\mathbf{a}_{k+1}-\mathbf{G}\doublehat{\mathbf{x}}_{k+1|k}),\label{eqn: inverse kfwithoutdf state update}\\
&\overline{\bm{\Sigma}}_{k+1}=\overline{\bm{\Sigma}}_{k+1|k}-\overline{\bm{\Sigma}}_{k+1|k}\mathbf{G}^{T}(\overline{\bm{\Sigma}}^{a}_{k+1})^{-1}\mathbf{G}\overline{\bm{\Sigma}}_{k+1|k},\label{eqn: inverse kfwithoutdf covariance update}
\end{align}
\normalsize
where (inverse) process noise covariance matrix $\overline{\mathbf{Q}}_{k}=\mathbf{E}_{k}\mathbf{R}\mathbf{E}_{k}^{T}$.

\subsection{I-KF-with-DF}\label{subsec:IKF with DF}
Consider the linear system model with DF given by \eqref{eqn: linear x with input} and \eqref{eqn: linear y withdf}.

\textit{Forward filter:} Denote the state estimation covariance, input estimation (without delay) covariance, and cross-covariance of state and input estimates by $\bm{\Sigma}^{x}_{k}$, $\bm{\Sigma}^{u}_{k}$ and $\bm{\Sigma}^{xu}_{k}$, respectively. The forward KF-with-DF is  \cite{gillijns2007kfb}:
\par\noindent\small
\begin{align}
&\textit{Time update:}\;\hat{\mathbf{x}}_{k+1|k}=\mathbf{F}\hat{\mathbf{x}}_{k}+\mathbf{B}\hat{\mathbf{u}}_{k},\label{eqn: kfwithdf predict}\\
&\bm{\Sigma}^{x}_{k+1|k}=\begin{bmatrix}
\mathbf{F} & \mathbf{B}
\end{bmatrix}\begin{bmatrix}
\bm{\Sigma}^{x}_{k} & \bm{\Sigma}^{xu}_{k}\\
\bm{\Sigma}^{ux}_{k} & \bm{\Sigma}^{u}_{k}
\end{bmatrix}\begin{bmatrix}
\mathbf{F}^{T}\\
\mathbf{B}^{T}
\end{bmatrix}+\mathbf{Q},\nonumber
\end{align}
\begin{align}   
&\textit{Gain computation:}\;\bm{\Sigma}^{y}_{k+1}=\mathbf{H}\bm{\Sigma}^{x}_{k+1|k}\mathbf{H}^{T}+\mathbf{R},\nonumber\\
&\mathbf{M}_{k+1}=(\mathbf{D}^{T}(\bm{\Sigma}^{y}_{k+1})^{-1}\mathbf{D})^{-1}\mathbf{D}^{T}(\bm{\Sigma}^{y}_{k+1})^{-1},\;\;\;\mathbf{K}_{k+1}=\bm{\Sigma}^{x}_{k+1|k}\mathbf{H}^{T}(\bm{\Sigma}^{y}_{k+1})^{-1},\nonumber\\
&\textit{Measurment update:}\;\hat{\mathbf{u}}_{k+1}=\mathbf{M}_{k+1}(\mathbf{y}_{k+1}-\mathbf{H}\hat{\mathbf{x}}_{k+1|k}),\label{eqn: kfwithdf update u}\\
&\hat{\mathbf{x}}_{k+1}=\hat{\mathbf{x}}_{k+1|k}+\mathbf{K}_{k+1}(\mathbf{y}_{k+1}-\mathbf{H}\hat{\mathbf{x}}_{k+1|k}-\mathbf{D}\hat{\mathbf{u}}_{k+1}),\label{eqn: kfwithdf update x}\\
&\bm{\Sigma}^{u}_{k+1}=(\mathbf{D}^{T}(\bm{\Sigma}^{y}_{k+1})^{-1}\mathbf{D})^{-1},\nonumber\\
&\bm{\Sigma}^{x}_{k+1}=\bm{\Sigma}^{x}_{k+1|k}-\mathbf{K}_{k+1}(\bm{\Sigma}^{y}_{k+1}-\mathbf{D}\bm{\Sigma}^{u}_{k+1}\mathbf{D}^{T})\mathbf{K}_{k+1}^{T},\nonumber\\
&\bm{\Sigma}^{xu}_{k+1}=(\bm{\Sigma}^{ux}_{k+1})^{T}=-\mathbf{K}_{k+1}\mathbf{D}\bm{\Sigma}^{u}_{k+1}.\nonumber
\end{align}
\normalsize
The forward filter exists if $\textrm{rank}(\mathbf{D})=n_{u}$ which implies $n_{y}\geq n_{u}$.

\textit{Inverse filter:} Consider an augmented state vector $\mathbf{z}_{k}=\begin{bmatrix}
\hat{\mathbf{x}}_{k}^{T} & \hat{\mathbf{u}}_{k}^{T}
\end{bmatrix}^{T}$. Denote $\widetilde{\mathbf{F}}_{k}=(\mathbf{I}_{n_{x}\times n_{x}}-\mathbf{K}_{k+1}\mathbf{H}+\mathbf{K}_{k+1}\mathbf{DM}_{k+1}\mathbf{H})\mathbf{F}$, $\widetilde{\mathbf{B}}_{k}=(\mathbf{I}_{n_{x}\times n_{x}}-\mathbf{K}_{k+1}\mathbf{H}+\mathbf{K}_{k+1}\mathbf{DM}_{k+1}\mathbf{H})\mathbf{B}$, $\mathbf{E}_{k}=\mathbf{K}_{k+1}(\mathbf{I}_{n_{y}\times n_{y}}-\mathbf{DM}_{k+1})$, $\widetilde{\mathbf{H}}_{k}=-\mathbf{M}_{k+1}\mathbf{HF}$ and $\widetilde{\mathbf{D}}_{k}=-\mathbf{M}_{k+1}\mathbf{HB}$. From \eqref{eqn: linear y withdf}, and \eqref{eqn: kfwithdf predict}-
\eqref{eqn: kfwithdf update x}, the  state transition equations for I-KF-with-DF are
\par\noindent\small
\begin{align*}
&\hat{\mathbf{x}}_{k+1}=\widetilde{\mathbf{F}}_{k}\hat{\mathbf{x}}_{k}+\widetilde{\mathbf{B}}_{k}\hat{\mathbf{u}}_{k}+\mathbf{E}_{k}\mathbf{H}\mathbf{x}_{k+1}+\mathbf{E}_{k}\mathbf{D}\mathbf{u}_{k+1}+\mathbf{E}_{k}\mathbf{v}_{k+1},\\
&\hat{\mathbf{u}}_{k+1}=\widetilde{\mathbf{H}}_{k}\hat{\mathbf{x}}_{k}+\widetilde{\mathbf{D}}_{k}\hat{\mathbf{u}}_{k}+\mathbf{M}_{k+1}\mathbf{H}\mathbf{x}_{k+1}+\mathbf{M}_{k+1}\mathbf{D}\mathbf{u}_{k+1}+\mathbf{M}_{k+1}\mathbf{v}_{k+1}.
\end{align*}
\normalsize
Also, $\begin{bmatrix}
(\mathbf{E}_{k}\mathbf{v}_{k+1})^{T} & (\mathbf{M}_{k+1}\mathbf{v}_{k+1})^{T}
\end{bmatrix}^{T}$ is the augmented noise vector involved in this state transition with noise covariance matrix $\overline{\mathbf{Q}}_{k}=\begin{bmatrix}
\mathbf{E}_{k}\mathbf{R}\mathbf{E}_{k}^{T} & \mathbf{E}_{k}\mathbf{R}\mathbf{M}_{k+1}^{T}\\
\mathbf{M}_{k+1}\mathbf{R}\mathbf{E}_{k}^{T} & \mathbf{M}_{k+1}\mathbf{R}\mathbf{M}_{k+1}^{T}
\end{bmatrix}$. Then, \textit{ceteris paribus}, following similar steps as in I-KF-without-DF, the I-KF-with-DF computes the estimate $\hat{\mathbf{z}}_{k}=\begin{bmatrix}
\doublehat{\mathbf{x}}_{k}^{T} & \doublehat{\mathbf{u}}_{k}^{T}
\end{bmatrix}^{T}$ of the augmented state vector using the observation $\mathbf{a}_{k}$ given by \eqref{eqn: linear a}. The system matrices for the augmented state are $\widetilde{\mathbf{F}}^{z}_{k}=\begin{bmatrix}
\widetilde{\mathbf{F}}_{k} & \widetilde{\mathbf{B}}_{k}\\
\widetilde{\mathbf{H}}_{k} & \widetilde{\mathbf{D}}_{k}
\end{bmatrix}$ and $\overline{\mathbf{G}}=\begin{bmatrix}
\mathbf{G} & \mathbf{0}_{n_{a}\times n_{u}}
\end{bmatrix}$. The I-KF-with-DF predicts the augmented state as
\par\noindent\small
\begin{align*}
&\doublehat{\mathbf{x}}_{k+1|k}=\widetilde{\mathbf{F}}_{k}\doublehat{\mathbf{x}}_{k}+\widetilde{\mathbf{B}}_{k}\doublehat{\mathbf{u}}_{k}+\mathbf{E}_{k}\mathbf{Hx}_{k+1}+\mathbf{E}_{k}\mathbf{Du}_{k+1},\\
&\doublehat{\mathbf{u}}_{k+1|k}=\widetilde{\mathbf{H}}_{k}\doublehat{\mathbf{x}}_{k}+\widetilde{\mathbf{D}}_{k}\doublehat{\mathbf{u}}_{k}+\mathbf{M}_{k+1}\mathbf{Hx}_{k+1}+\mathbf{M}_{k+1}\mathbf{Du}_{k+1},\\
&\hat{\mathbf{z}}_{k+1|k}=\begin{bmatrix}
\doublehat{\mathbf{x}}_{k+1|k}^{T} & \doublehat{\mathbf{u}}_{k+1|k}^{T}
\end{bmatrix}^{T},\;
\overline{\bm{\Sigma}}_{k+1|k}=\widetilde{\mathbf{F}}^{z}_{k}\overline{\bm{\Sigma}}_{k}(\widetilde{\mathbf{F}}^{z}_{k})^{T}+\overline{\mathbf{Q}}_{k},
\end{align*}
\normalsize
followed by the update procedure \eqref{eqn: inverse kfwithoutdf gain}-\eqref{eqn: inverse kfwithoutdf covariance update} with $\mathbf{G}$ and $\doublehat{x}_{k+1}$ replaced by $\overline{\mathbf{G}}$ and $\hat{\mathbf{z}}_{k+1}$, respectively. Since the observation $\mathbf{y}_{k}$ explicitly depends on the unknown input $\mathbf{u}_{k}$ for a system with DF, I-KF-with-DF and I-EKF-with-DF require perfect knowledge of the current input $\mathbf{u}_{k}$ as a known exogenous input to obtain their state and input estimates, which is not the case in I-KF-without-DF and I-EKF-without-DF. Similar to Remark~\ref{remark:iekf unknown input computation complexity}, the I-KFs with unknown inputs are also derived from standard KF recursions and have similar computational complexity. However, I-KF-without-DF is not formulated using an augmented state and hence, is computationally less complex than I-KF-with-DF.

\begin{remark}[I-KFs and I-EKFs with unknown inputs]\label{remark:unknown input diff}
Note that the I-KFs with unknown inputs for linear system models are not special cases of I-EKFs with unknown inputs for non-linear system models. In I-KF-with-unknown-inputs, the attacker employs a forward KF, which provides unbiased minimum variance estimates of the state and the unknown inputs\cite{gillijns2007unknownkf,gillijns2007kfb}. On the other hand, in I-EKF-with-unknown-inputs, the forward filter's state and unknown inputs estimates are computed based on a weighted least squared error criterion\cite{pan2010applying,yang2007adaptive}. The different forward filters employed by the attacker result in different inverse filters for the defender to estimate the attacker's state estimate.
\end{remark}

\subsection{Stability analysis}\label{subsec:IKF unknown stability}
We derive the stability conditions for I-KF-without-DF of Section~\ref{subsec:IKF without DF} in which we rely on the asymptotic stability of the forward KF-without-DF as proved in \cite{fang2012on}. The procedure is similar for the stability of I-KF-with-DF and I-KF-without-unknown-input \cite{krishnamurthy2019how} and hence, we omit the details for these filters. Consider the forward KF-without-DF asymptotically stable under the sufficient conditions provided by \cite{fang2012on}. The following Theorem~\ref{theorem: inverse kf without DF} then states conditions for stability of the inverse filter.
 
\begin{theorem} [Stability of I-KF-without-DF]\label{theorem: inverse kf without DF}
Consider an asymptotically stable forward KF-without-DF \eqref{eqn: kfwithoutdf predict}-\eqref{eqn: kfwithoutdf sigma update} such that the gain matrices $\mathbf{M}_{k}$ and $\mathbf{K}_{k}$ asymptotically approach to limiting gain matrices $\overline{\mathbf{M}}$ and $\overline{\mathbf{K}}$, respectively. The measurement noise covariance matrix $\bm{\Sigma}_{\epsilon}$ is positive definite (p.d.). Denote the limiting matrices $\overline{\mathbf{F}}=(\mathbf{I}-\overline{\mathbf{K}}\mathbf{H})(\mathbf{I}-\mathbf{B}\overline{\mathbf{M}}\mathbf{H})\mathbf{F}$ and $\overline{\mathbf{Q}}=\overline{\mathbf{E}}\mathbf{R}\overline{\mathbf{E}}^{T}$, where $\overline{\mathbf{E}}=\mathbf{B}\overline{\mathbf{M}}-\overline{\mathbf{K}}\mathbf{HB}\overline{\mathbf{M}}+\overline{\mathbf{K}}$. Then, the I-KF-without-DF \eqref{eqn: inverse kfwithoutdf state predict}-\eqref{eqn: inverse kfwithoutdf covariance update} is asymptotically stable under the assumption that pair ($\overline{\mathbf{F}}$,$\mathbf{G}$) is observable and the pair ($\overline{\mathbf{F}}$,$\mathbf{C}$) is controllable for the system given by \eqref{eqn: linear a} and \eqref{eqn: state for kfwithoutdf}, where $\mathbf{C}$ is such that $\overline{\mathbf{Q}}=\mathbf{C}^{T}\mathbf{C}$. 
\end{theorem}
\begin{proof}
See Appendix~\ref{App-thm-kf-without-DF}.
\end{proof}

Note that, for I-KF-with-DF's stability, the stability conditions of basic KF need to hold for the augmented state considered in inverse filter formulation of Section~\ref{subsec:IKF with DF}. For forward KF-with-DF's stability conditions, we refer the reader to \cite{fang2012on}.

\section{Numerical experiments}\label{sec:unknown input numericals}
As in Section~\ref{sec:IEKF numericals}, we consider different example systems to compare the estimation accuracy of I-KFs and I-EKFs with unknown inputs with RCRLB as the performance metric. Also, throughout all experiments, the initial information matrices $\mathbf{J}_{0}$ and $\overline{\mathbf{J}}_{0}$ for RCRLB computations were set to $\bm{\Sigma}_{0}^{-1}$ and $\overline{\bm{\Sigma}}_{0}^{-1}$, respectively, unless mentioned otherwise.

\subsection{I-KFs with unknown inputs}\label{subsec:IKF experiment}
Consider a discrete-time linear system without DF\cite{hsieh2000robust},
\par\noindent\small
\begin{align*}
&\mathbf{x}_{k+1}=\begin{bmatrix}0.1 & 0.5 & 0.08\\ 0.6 & 0.01 & 0.04\\ 0.1 & 0.7 & 0.05\end{bmatrix}\mathbf{x}_{k}+\begin{bmatrix}0\\ 2\\ 1\end{bmatrix}u_{k}+\mathbf{w}_{k},\\
&\mathbf{y}_{k}=\begin{bmatrix}1 & 1 & 0\\ 0 & 1 & 1\end{bmatrix}\mathbf{x}_{k}+\mathbf{v}_{k},\;\;\;a_{k}=\begin{bmatrix}1 & 1 & 1\end{bmatrix}\hat{\mathbf{x}}_{k}+\epsilon_{k},
\end{align*}
\normalsize
with $\mathbf{w}_{k}\sim\mathcal{N}(\mathbf{0},\mathbf{I}_{3})$, $\mathbf{v}_{k}\sim\mathcal{N}(\mathbf{0},2\mathbf{I}_{2})$ and $\epsilon_{k}\sim\mathcal{N}(0,5)$. The unknown input $u_{k}$ was set to $50$ for $1\leq k \leq 50$ and $-50$ thereafter. The initial state was $\mathbf{x}_{0}=[1,1,1]^{T}$. For the forward filter, the initial state estimate was set to $[0,0,0]^{T}$ with $\bm{\Sigma}_{0}=\mathbf{I}_{3}$. For the inverse filter, the initial state estimate was set to $\mathbf{x}_{0}$ (known to the defender) itself with $\overline{\bm{\Sigma}}_{0}=5\mathbf{I}_{3}$.

For KF-with-DF, we modify the forward filter's observations as\cite{pan2011study}:
\par\noindent\small
\begin{align*}
\mathbf{y}_{k}=\begin{bmatrix}1 & 1 & 0\\ 0 & 1 & 1\end{bmatrix}\mathbf{x}_{k}+\begin{bmatrix}0\\1\end{bmatrix}u_{k}+\mathbf{v}_{k}.
\end{align*}
\normalsize
Here, the initial input estimate was set to $10$ with initial input estimate covariance $\bm{\Sigma}^{u}_{0}=10$ and initial cross-covariance $\bm{\Sigma}^{xu}_{0}=[0,0,0]^{T}$. The inverse filter's initial augmented state estimate $\mathbf{z}_{0}$ was set to $[1,1,1,50]^{T}$ with initial covariance $\overline{\bm{\Sigma}}_{0}=5\mathbf{I}_{4}$.

\begin{figure}
  \centering
  \includegraphics[width = 0.8\columnwidth]{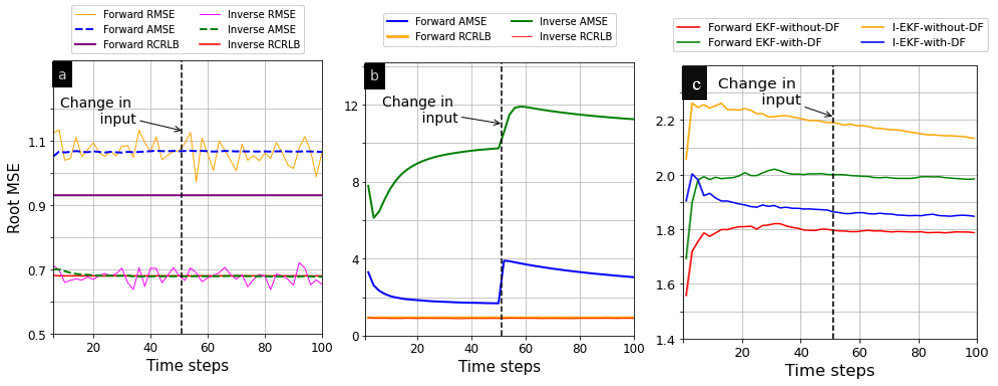}
  \caption{RMSE, AMSE, and RCRLB for forward and inverse filters (a) KF-without-DF; (b) KF-with-DF; and (c) AMSE for forward and inverse EKF with and without DF.}
 \label{fig:KF EKF unknown input}
\end{figure}

Fig. \ref{fig:KF EKF unknown input}a-b shows the AMSE and RCRLB for state estimation for both forward and inverse filters in the two cases, respectively, averaged over 200 runs. For KF-without-DF, we plot the RMSE for comparison here but omit it for later plots (including subsequent chapters) for clarity. Note that in Fig.~\ref{fig:KF EKF unknown input}a, the I-KF-without-DF's RMSE fluctuates about the RCRLB because of a finite number of sample paths; see also similar phenomena in \cite{xiong2006performance_ukf,djuric2008target,vsimandl2001filtering}. The RCRLB value for state estimation is $\sqrt{\textrm{Tr}(\mathbf{J}^{-1})}$ with $\mathbf{J}$ denoting the associated information matrix. Fig. \ref{fig:KF EKF unknown input}a-b shows that the effect of change in unknown input after 50 time-steps is negligible for KF-without-DF in both forward and inverse filters. However, for KF-with-DF, the sudden change in unknown input leads to an increase in state estimation error of the forward filter and, consequently, of the inverse filter. The estimation error of I-KF-without-DF is less than the corresponding forward filter while for KF-with-DF, the inverse filter has a higher estimation error than the forward filter. Only I-KF-without-DF efficiently achieves the RCRLB bound on the estimation error.

\subsection{I-EKFs with unknown inputs}\label{subsec:IEKF experiment}
For I-EKFs with unknown input, we modified the non-linear system model of Section~\ref{subsec:FM with EKF SOEKF GSEKF} to include an unknown input $u_{k}$ as
\par\noindent\small
\begin{align*}
\mathbf{x}_{k+1}=\begin{bmatrix}\exp{(-T/\beta)}&0\\-\beta \exp{(-T/\beta)}-1&1\end{bmatrix}\begin{bmatrix}\lambda_{k}\\\theta_{k}\end{bmatrix}+\begin{bmatrix}0.001\\1\end{bmatrix}u_{k}+\begin{bmatrix}1\\-\beta\end{bmatrix}w_{k},
\end{align*}
\normalsize
where $u_{k}$ was set to $\pi/4$ for $1\leq k \leq 50$ and $-\pi/4$ thereafter. The observation $\mathbf{y}_{k}$ of the forward EKF-without-DF was same as in Section~\ref{subsec:FM with EKF SOEKF GSEKF}. Consider a linear measurement $a_{k}$ for the inverse filter as $a_{k}=\hat{\lambda}_{k}+\epsilon_{k}$. For the forward filter, the initial input estimate was set to $0$ while the inverse filter initial augmented state estimate consisted of the true state $\mathbf{x}_{0}$ and true input $u_{0}$ (known to the defender) with initial covariance estimate $\overline{\bm{\Sigma}}_{0}=15\mathbf{I}_{3}$. All other system parameters and initial estimates are considered to be same as in Section~\ref{subsec:FM with EKF SOEKF GSEKF}. 

Similarly, for system with DF, we again considered the same non-linear system (without any unknown input in $\mathbf{x}_{k}$ state transition) but with a modified forward filter's observation $\mathbf{y}_{k}=\sqrt{2}\begin{bmatrix}\sin{(\theta_{k}+u_{k})}\\\cos{(\theta_{k}+u_{k})}\end{bmatrix}+\mathbf{v}_{k}$.
The input estimates $\hat{u}$ and $\doublehat{u}$ were also, as before, modulo $2\pi$. The Gaussian noise terms in the inverse filter state transitions (\eqref{eqn: state transition ekf without df state} and \eqref{eqn:state transition ekf with df}) are transformed through non-linear functions such that \eqref{eqn: additive Jk recursions} is not applicable. The RCRLB in this case is derived using the general $\mathbf{J}_{k}$ recursions given by \eqref{eqn: general Jk recursions}, which is omitted here. Fig. \ref{fig:KF EKF unknown input}c shows that for both EKF with and without DF, the change in unknown input after $50$ time-steps does not increase the estimation error (as for KF-with-DF in Fig. \ref{fig:KF EKF unknown input}b). The estimation error of I-EKF-without-DF (I-EKF-with-DF) is higher (lower) than that of the corresponding forward filter. Any change in unknown input affects the inverse filter's performance only when a significant change occurs in the forward filter's performance.

\section{Concluding remarks}\label{sec:unknown input conclusions}
In this chapter, we studied the inverse filtering problem for systems with unknown inputs, wherein the attacker's observations may or may not be influenced by the exogenous input known to the defender but not the attacker. To address these scenarios, we developed I-EKF and I-KF (each with and without DF) for non-linear and linear system dynamics, respectively. These I-KFs for systems with unknown inputs have also not been examined in the previous work \cite{krishnamurthy2019how} on I-KF (without unknown inputs). We then derived the asymptotic stability conditions for I-KF-without-DF by extending standard KF stability results.  Similar to I-EKFs, the stability of a forward KF-with-unknown-input ensures the stability of the corresponding I-KF-with-unknown-input, provided the system meets certain additional assumptions. Our numerical experiments suggested that the impact of the unknown input on the inverse filter's performance strongly depends on its impact on the forward filter.

\chapter{Inverse SPKFs}
\label{chap:inverse SPKFs}
In this chapter, we develop various inverse SPKFs and provide their theoretical performance guarantees. To this end, we first explore the unscented transform technique and introduce I-UKF in Section~\ref{sec:IUKF}, along with sufficient conditions to guarantee its stochastic stability (using unknown matrix approach) and conservative estimates. In the process, we also obtain hitherto unreported general stability results for forward UKF. We then generalize our I-UKF to obtain continuous-discrete, complex, and MCC-based I-UKFs for continuous-time state evolution, complex-valued systems, and non-Gaussian noise cases, respectively. In Section~\ref{sec:ICKF and IQKF}, we extend the I-UKF theory to derive inverses of different cubature and quadrature KFs, namely, I-CKF, I-QKF and I-CQKF; and provide enhanced stability results for the forward CKF. Finally, Section~\ref{sec:ISPKF numericals} validate the estimation performance of these inverse filters through extensive numerical experiments using RCRLB and non-credibility index (NCI) as the performance metrics, before concluding in Section~\ref{sec:ISPKF conclusions}.

\section{Inverse UKF}\label{sec:IUKF}
The UKF generates a set of `$2n_{x}+1$' sigma points deterministically from the previous state estimate, including the previous estimate itself as one of the sigma points. The sigma points are then propagated through the non-linear system model, and the state estimates are obtained as a weighted sum of these propagated points. In I-UKF, we assume that the attacker is employing a forward UKF to compute its estimate $\hat{\mathbf{x}}_{k}$ with known state transition \eqref{eqn:non-linear state x} and observation \eqref{eqn:non-linear observation y}. The I-UKF then infers the estimate $\doublehat{\mathbf{x}}_{k}$ of $\hat{\mathbf{x}}_{k}$ using observation \eqref{eqn:non-linear observation a}.

\subsection{Filter formulation}\label{subsec:IUKF formulation}
\textit{Forward filter:} Consider the scaling parameter $\kappa\in\mathbb{R}$ controlling the spread of the forward UKF's sigma points around the previous estimate. The sigma points $\{\widetilde{\mathbf{x}}_{i}\}_{0\leq i\leq 2n_{x}}$ are generated from state estimate $\hat{\mathbf{x}}$ and its error covariance matrix $\bm{\Sigma}$ as
\par\noindent\small
\begin{align}
  &\{\widetilde{\mathbf{x}}_{i}\}_{0\leq i\leq 2n_{x}}=S_{gen}(\hat{\mathbf{x}},\bm{\Sigma})=\begin{cases}
  \hat{\mathbf{x}},\;\;\;\;\;\;\;i=0,\\
 \hat{\mathbf{x}}+\left[\sqrt{(n_{x}+\kappa)\bm{\Sigma}}\right]_{(:,i)},\;i=1,2,\hdots,n_{x}\\
    \hat{\mathbf{x}}-\left[\sqrt{(n_{x}+\kappa)\bm{\Sigma}}\right]_{(:,i-n_{x})},\;i=n_{x}+1,n_{x}+2,\hdots,2n_{x}\end{cases},\label{eqn:sigma points generation}
\end{align}
\normalsize
with their weights $\omega_{i}=\begin{cases}\frac{\kappa}{n_{x}+\kappa} & i=0\\\frac{1}{2(n_{x}+\kappa)} & i=1,2,\hdots,2n_{x}\end{cases}$.

Denote the sigma points generated and propagated for the time update at $k$-th time instant by $\lbrace\mathbf{s}_{i,k}\rbrace$ and $\lbrace\mathbf{s}^{*}_{i,k+1|k}\rbrace$, respectively. Similarly, $\lbrace\mathbf{q}_{i,k+1|k}\rbrace$ and $\lbrace\mathbf{q}^{*}_{i,k+1|k}\rbrace$ are the sigma points, respectively, generated and propagated to predict observation $\mathbf{y}_{k+1}$ as $\hat{\mathbf{y}}_{k+1|k}$. The attacker's forward UKF recursions to compute state estimate $\hat{\mathbf{x}}_{k+1}$ and the associated error covariance matrix estimate $\bm{\Sigma}_{k+1}$ are \cite{simon2006optimal}
\par\noindent\small
\begin{align}
    &\textrm{Time update:}\;\;\lbrace\mathbf{s}_{i,k}\rbrace_{0\leq i\leq 2n_{x}}=S_{gen}(\hat{\mathbf{x}}_{k},\bm{\Sigma}_{k}),\label{eqn:forward ukf prediction sigma points}\\
    &\mathbf{s}^{*}_{i,k+1|k}=f(\mathbf{s}_{i,k})\;\;\;\forall i=0,1,\hdots,2n_{x},\nonumber\\
    &\hat{\mathbf{x}}_{k+1|k}=\sum_{i=0}^{2n_{x}}\omega_{i}\mathbf{s}^{*}_{i,k+1|k},\label{eqn:forward ukf x predict}\\
    &\bm{\Sigma}_{k+1|k}=\sum_{i=0}^{2n_{x}}\omega_{i}\mathbf{s}^{*}_{i,k+1|k}(\mathbf{s}^{*}_{i,k+1|k})^{T}-\hat{\mathbf{x}}_{k+1|k}\hat{\mathbf{x}}_{k+1|k}^{T}+\mathbf{Q},\nonumber\\
    &\textrm{Measurement update:}\;\;\lbrace\mathbf{q}_{i,k+1|k}\rbrace_{0\leq i\leq 2n_{x}}=S_{gen}(\hat{\mathbf{x}}_{k+1|k},\bm{\Sigma}_{k+1|k}),\label{eqn:forward UKF update sigma points}\\
    &\mathbf{q}^{*}_{i,k+1|k}=h(\mathbf{q}_{i,k+1|k})\;\;\;\forall i=0,1,\hdots,2n_{x},\nonumber\\
    &\hat{\mathbf{y}}_{k+1|k}=\sum_{i=0}^{2n_{x}}\omega_{i}\mathbf{q}^{*}_{i,k+1|k},\label{eqn:forward ukf y predict}\\
    &\bm{\Sigma}^{y}_{k+1}=\sum_{i=0}^{2n_{x}}\omega_{i}\mathbf{q}^{*}_{i,k+1|k}(\mathbf{q}^{*}_{i,k+1|k})^{T}-\hat{\mathbf{y}}_{k+1|k}\hat{\mathbf{y}}_{k+1|k}^{T}+\mathbf{R},\nonumber\\
    &\bm{\Sigma}^{xy}_{k+1}=\sum_{i=0}^{2n_{x}}\omega_{i}\mathbf{q}_{i,k+1|k}(\mathbf{q}^{*}_{i,k+1|k})^{T}-\hat{\mathbf{x}}_{k+1|k}\hat{\mathbf{y}}_{k+1|k}^{T}\nonumber,\\
    &\hat{\mathbf{x}}_{k+1}=\hat{\mathbf{x}}_{k+1|k}+\mathbf{K}_{k+1}(\mathbf{y}_{k+1}-\hat{\mathbf{y}}_{k+1|k}),\label{eqn:forward ukf x update}\\
    &\bm{\Sigma}_{k+1}=\bm{\Sigma}_{k+1|k}-\mathbf{K}_{k+1}\bm{\Sigma}^{y}_{k+1}\mathbf{K}_{k+1}^{T},\label{eqn:forward UKF sigma update}
\end{align}
\normalsize
where gain matrix $\mathbf{K}_{k+1}=\bm{\Sigma}^{xy}_{k+1}\left(\bm{\Sigma}^{y}_{k+1}\right)^{-1}$. The step to generate the second set of sigma points may be omitted, and $\lbrace\mathbf{q}^{*}_{i,k+1|k}\rbrace$ may be obtained by propagating $\lbrace\mathbf{s}^{*}_{i,k+1|k}\rbrace$ through the observation function $h(\cdot)$ to save computational efforts. However, this may degrade the performance of the classical UKF because the effect of process noise here is unaccounted for. On the other hand, pure-propagation UKF \cite{straka2014design} generates a modified sigma-point set with increased covariance only once without compromising the performance.

\textit{Inverse filter:} Under the known forward UKF assumption, substituting \eqref{eqn:non-linear observation y}, \eqref{eqn:forward ukf x predict}, and \eqref{eqn:forward ukf y predict} in \eqref{eqn:forward ukf x update}, yields the inverse filter's state transition as
\par\noindent\small
\begin{align}
    \hat{\mathbf{x}}_{k+1}&=\sum_{i=0}^{2n_{x}}\omega_{i}\left(\mathbf{s}^{*}_{i,k+1|k}-\mathbf{K}_{k+1}\mathbf{q}^{*}_{i,k+1|k}\right)+\mathbf{K}_{k+1}h(\mathbf{x}_{k+1})+\mathbf{K}_{k+1}\mathbf{v}_{k+1}.\label{eqn: IUKF state transition detail}
\end{align}
\normalsize
In this state transition, $\mathbf{x}_{k+1}$ is a known exogenous input, while $\mathbf{v}_{k+1}$ represents the process noise involved. Since the functions $f(\cdot)$ and $h(\cdot)$ are known, the propagated sigma points $\{\mathbf{s}^{*}_{i,k+1|k}\}$ and $\{\mathbf{q}^{*}_{i,k+1|k}\}$, and the gain matrix $\mathbf{K}_{k+1}$ are functions of the first set of sigma points $\{\mathbf{s}_{i,k}\}$. These sigma points, in turn, are obtained deterministically from the previous state estimate $\hat{\mathbf{x}}_{k}$ and covariance matrix $\bm{\Sigma}_{k}$ using \eqref{eqn:forward ukf prediction sigma points}. Hence, I-UKF's state transition, \emph{under the assumption that parameter $\kappa$ is known to the defender}, is
\par\noindent\small
\begin{align}
    \hat{\mathbf{x}}_{k+1}=\widetilde{f}(\hat{\mathbf{x}}_{k},\bm{\Sigma}_{k},\mathbf{x}_{k+1},\mathbf{v}_{k+1}).\label{eqn:inverse ukf state transition}
\end{align}
\normalsize
Note that the process noise $\mathbf{v}_{k+1}$ is non-additive because $\mathbf{K}_{k+1}$ depends on the previous estimates. Furthermore, $\bm{\Sigma}_{k}$ does not depend on the current forward filter's observation $\mathbf{y}_{k}$ but is evaluated recursively using the previous estimate. We approximate $\bm{\Sigma}_{k}$ as $\bm{\Sigma}_{k}^{*}$ by computing the covariance matrix using its own previous estimate, i.e. $\doublehat{\mathbf{x}}_{k}$, in the same recursive manner as the forward filter estimates at its estimate $\hat{\mathbf{x}}_{k}$ (using \eqref{eqn:forward UKF sigma update} after computing gain matrix $\mathbf{K}_{k+1}$ from the generated sigma points). Hence, the inverse filter treats $\bm{\Sigma}_{k}$ as another known exogenous input in the state transition \eqref{eqn:inverse ukf state transition}.

Since the state transition \eqref{eqn:inverse ukf state transition} involves non-additive noise term, we consider an augmented state vector $\mathbf{z}_{k}=[\hat{\mathbf{x}}_{k}^{T},\mathbf{v}_{k+1}^{T}]^{T}$ of dimension $n_{z}=n_{x}+n_{y}$ for I-UKF formulation such that the state transition \eqref{eqn:inverse ukf state transition} becomes $\hat{\mathbf{x}}_{k+1}=\widetilde{f}(\mathbf{z}_{k},\bm{\Sigma}_{k},\mathbf{x}_{k+1})$. Denote
\par\noindent\small
\begin{align}
    \hat{\mathbf{z}}_{k}=[\doublehat{\mathbf{x}}_{k}^{T},\mathbf{0}_{1\times n_{y}}]^{T},\;\;\overline{\bm{\Sigma}}^{z}_{k}=\begin{bsmallmatrix}\overline{\bm{\Sigma}}_{k}&\mathbf{0}_{n_{x}\times n_{y}}\\\mathbf{0}_{n_{y}\times n_{x}}&\mathbf{R}\end{bsmallmatrix}.\label{eqn:IUKF z hat and sigma z}
\end{align}
\normalsize
Considering the I-UKF's scaling parameter as $\overline{\kappa}\in\mathbb{R}$, the sigma points $\{\overline{\mathbf{s}}_{j,k}\}_{0\leq j\leq 2n_{z}}$ are generated from $\hat{\mathbf{z}}_{k}$ and $\overline{\bm{\Sigma}}^{z}_{k}$ similar to \eqref{eqn:sigma points generation} with weights $\overline{\omega}_{j}$. I-UKF then computes $\doublehat{\mathbf{x}}_{k}$ and its associated error covariance matrix $\overline{\bm{\Sigma}}_{k}$ recursively as
\par\noindent\small
\begin{align}
    &\textit{Time update:}\;\;\;\overline{s}^{*}_{j,k+1|k}=\widetilde{f}(\overline{\mathbf{s}}_{j,k},\bm{\Sigma}_{k}^{*},\mathbf{x}_{k+1})\;\;\;\forall j=0,1,\hdots,2n_{z},\label{eqn:IUKF f propagate}\\   
    &\doublehat{\mathbf{x}}_{k+1|k}=\sum_{j=0}^{2n_{z}}\overline{\omega}_{j}\overline{\mathbf{s}}^{*}_{j,k+1|k},\label{eqn:IUKF x predict}\\
    &\overline{\bm{\Sigma}}_{k+1|k}=\sum_{j=0}^{2n_{z}}\overline{\omega}_{j}\overline{\mathbf{s}}^{*}_{j,k+1|k}(\overline{\mathbf{s}}^{*}_{j,k+1|k})^{T}-\doublehat{\mathbf{x}}_{k+1|k}\doublehat{\mathbf{x}}_{k+1|k}^{T},\label{eqn:IUKF sig predict}\\
    &\textit{Measurement update:}\;\;\;\mathbf{a}^{*}_{j,k+1|k}=g(\overline{\mathbf{s}}^{*}_{j,k+1|k})\;\;\;\forall j=0,1,\hdots,2n_{z},\label{eqn:IUKF g propagate}\\
    &\hat{\mathbf{a}}_{k+1|k}=\sum_{j=0}^{2n_{z}}\overline{\omega}_{j}\mathbf{a}^{*}_{j,k+1|k},\label{eqn:IUKF a predict}\\
    &\overline{\bm{\Sigma}}^{a}_{k+1}=\sum_{j=0}^{2n_{z}}\overline{\omega}_{j}\mathbf{a}^{*}_{j,k+1|k}(\mathbf{a}^{*}_{j,k+1|k})^{T}-\hat{\mathbf{a}}_{k+1|k}\hat{\mathbf{a}}_{k+1|k}^{T}+\bm{\Sigma}_{\epsilon},\label{eqn:IUKF sig a}\\
    &\overline{\bm{\Sigma}}^{xa}_{k+1}=\sum_{j=0}^{2n_{z}}\overline{\omega}_{j}\overline{\mathbf{s}}^{*}_{j,k+1|k}(\mathbf{a}^{*}_{j,k+1|k})^{T}-\doublehat{\mathbf{x}}_{k+1|k}\hat{\mathbf{a}}_{k+1|k}^{T},\label{eqn:IUKF cross cov}\\
     &\overline{\mathbf{K}}_{k+1}=\overline{\bm{\Sigma}}^{xa}_{k+1}\left(\overline{\bm{\Sigma}}^{a}_{k+1}\right)^{-1},\label{eqn:IUKF gain}\\
    &\doublehat{x}_{k+1}=\doublehat{x}_{k+1|k}+\overline{\mathbf{K}}_{k+1}(\mathbf{a}_{k+1}-\hat{\mathbf{a}}_{k+1|k}),\label{eqn:IUKF state update}\\
    &\overline{\bm{\Sigma}}_{k+1}=\overline{\bm{\Sigma}}_{k+1|k}-\overline{\mathbf{K}}_{k+1}\overline{\bm{\Sigma}}^{a}_{k+1}\overline{\mathbf{K}}_{k+1}^{T}.\label{eqn:IUKF covariance update}
\end{align}
\normalsize

I-UKF's recursions follow from UKF's non-additive noise formulation\cite{wan2000unscented} with the sigma points generated in higher ($n_{z}=n_{x}+n_{y}$) dimensional state space than the $n_{x}$ dimensions in forward UKF. However, the latter requires a new set for the measurement update, while I-UKF generates these points only once. The defender's assumed observation \eqref{eqn:non-linear observation a} depends only on the state estimate $\hat{\mathbf{x}}_{k}$. However, in many applications, the attacker may also consider the estimated covariance $\bm{\Sigma}_{k}$ of $\hat{\mathbf{x}}_{k}$ in deciding its actions. In such cases, our proposed I-UKF can be trivially modified to include $\bm{\Sigma}_{k}$ in observations \eqref{eqn:non-linear observation a}. In particular, I-UKF's state transition \eqref{eqn:inverse ukf state transition} depends on forward UKF's $\bm{\Sigma}_{k}$ and hence, we compute the approximation $\bm{\Sigma}^{*}_{k}$. If observation \eqref{eqn:non-linear observation a} depends on $\bm{\Sigma}_{k}$, I-UKF's update step \eqref{eqn:IUKF g propagate} also uses $\bm{\Sigma}^{*}_{k}$. As mentioned earlier, I-UKF computes $\bm{\Sigma}^{*}_{k}$ using its estimate $\doublehat{\mathbf{x}}_{k}$, in the same manner as the forward UKF computes $\bm{\Sigma}_{k}$ from its estimate $\hat{\mathbf{x}}_{k}$. In particular, we obtain $\mathbf{K}_{k+1}$ (and intermediately, $\bm{\Sigma}_{k+1|k}$ and $\bm{\Sigma}^{y}_{k+1}$) while propagating sigma-point $\overline{\mathbf{s}}_{j,k}$ through state transition \eqref{eqn:inverse ukf state transition} (equivalently, \eqref{eqn: IUKF state transition detail}). Hence, for each $\overline{\mathbf{s}}_{j,k}$, we obtain an estimate $\widetilde{\bm{\Sigma}}_{j,k+1}$ of forward UKF's covariance update $\bm{\Sigma}_{k+1}$ using \eqref{eqn:forward UKF sigma update}. We choose the average of these $\{\widetilde{\bm{\Sigma}}_{j,k+1}\}_{0\leq j\leq 2n_{z}}$ as $\bm{\Sigma}^{*}_{k+1}$ for the next I-UKF recursion. Note that under the Gaussian posterior assumption, $\doublehat{\mathbf{x}}_{k}$ and $\bm{\Sigma}^{*}_{k}$, respectively, provide an estimate of forward filter's $\hat{\mathbf{x}}_{k}$ and $\bm{\Sigma}_{k}$, i.e., the mean and covariance of the assumed Gaussian density.

\begin{remark}[Unknown $\kappa$]\label{remark:unknown kappa}
In the I-UKF formulation, we assumed that the parameter $\kappa$ of the forward UKF is known. In general, assuming a $\kappa$ different from the attacker's actual $\kappa$ may increase estimation errors in I-UKF. However, our numerical experiments in Section~\ref{sec:ISPKF numericals} show that the I-UKF provides reasonable estimates even when assuming a different $\kappa$ from its true value. Further, the I-UKF's stability (Theorem~\ref{theorem:IUKF stability}) requires a stable forward UKF and is independent of the assumed $\kappa$. Note that the choice of forward UKF's $\kappa$ by the inverse filter is independent of its own control parameter $\overline{\kappa}$.
\end{remark}
\begin{remark}[Differences from I-KF and I-EKF]\label{remark:I-UKF with I-KF and I-EKF}
   Unlike I-KF \cite{krishnamurthy2019how} and I-EKF, the forward gain matrix $\mathbf{K}_{k+1}$ is not treated as a time-varying parameter of I-UKF's state transition \eqref{eqn: IUKF state transition detail}. In KF, the gain matrix is fully deterministic given the model parameters and the initial covariance estimate $\bm{\Sigma}_{0}$. In EKF, it depends on the linearized model functions at the state estimates. However, the UKF gain matrix is computed from the covariance matrix estimates obtained as a weighted average of the generated sigma points, which are explicit functions of the state estimates. This prevents I-UKF from treating $\mathbf{K}_{k+1}$ as a parameter of \eqref{eqn: IUKF state transition detail}. 
\end{remark}
\begin{remark}[I-UKF's computational complexity]\label{remark:UKF complexity}
    Note that I-UKF's recursions are obtained from a general UKF algorithm and, hence, have similar computational complexity. UKF recursions have an approximate complexity of $\mathcal{O}(d^{3})$ where $d$ is the dimension of the estimated state vector\cite{daum2005nonlinear}. However, I-UKF estimates an $n_{z}=n_{x}+n_{y}$ dimensional state such that the actual complexity depends on the dimensions of both the defender's state and the attacker's observation.
\end{remark}

\subsubsection{Generalizations of I-UKF}
\textbf{1) Non-Gaussian noises:} Similar to Remark~\ref{remark:ekf non Gaussian}, UKF and hence, I-UKF also assume Gaussian noises, but I-UKF can be suitably modified based on MCC for non-Gaussian system models. For instance, forward MCC-UKF proposed in \cite{wang2017maximum} introduces a scalar $l_{k+1}=G_{\sigma}(\|\mathbf{y}_{k+1}-\hat{\mathbf{y}}_{k+1|k}\|_{\widetilde{\mathbf{R}}_{k+1}^{-1}})$ with $G_{\sigma}(\cdot)$ as the Gaussian kernel and computes the gain matrix as $\mathbf{K}_{k+1}=\bm{\Sigma}_{k+1|k}l_{k+1}\widetilde{\mathbf{H}}_{k+1}^{T}(\widetilde{\mathbf{R}}_{k+1}+\widetilde{\mathbf{H}}_{k+1}\bm{\Sigma}_{k+1|k}l_{k+1}\widetilde{\mathbf{H}}_{k+1}^{T})^{-1}$. Here, $\widetilde{\mathbf{H}}_{k+1}$ and $\widetilde{\mathbf{R}}_{k+1}$, respectively, are the pseudo-measurement matrix and modified covariance matrix. The estimated error covariance then becomes $\bm{\Sigma}_{k+1}=(\mathbf{I}-\mathbf{K}_{k+1}\widetilde{\mathbf{H}}_{k+1})\bm{\Sigma}_{k+1|k}(\mathbf{I}-\mathbf{K}_{k+1}\widetilde{\mathbf{H}}_{k+1})^{T}+\mathbf{K}_{k+1}\widetilde{\mathbf{R}}_{k+1}\mathbf{K}_{k+1}^{T}$ while all other state prediction and update steps remain same as in forward UKF. These modifications need to be taken into account in the inverse filter's state transition equation while formulating the inverse filter. I-UKF's gain matrix $\overline{\mathbf{K}}_{k+1}$ and covariance estimate $\overline{\bm{\Sigma}}_{k+1}$ are also similarly modified using scalar $\overline{l}_{k+1}$ which is the counterpart of $l_{k+1}$ for the inverse filter's dynamics.\\

\noindent\textbf{2) Continuous-time state evolution:} The state-evolution \eqref{eqn:non-linear state x} and observations \eqref{eqn:non-linear observation y} represent discrete-time processes. In many practical applications like radar tracking \cite{li2003survey,kulikov2015accurate} and chemical systems' estimation \cite{kulikov2019numerical}, the state evolution is inherently a continuous-time process while the observations are obtained at discrete-time instants. In the forward filtering case, state estimation for such systems is efficiently handled through the continuous-discrete Kalman-Bucy filter \cite{kalman1961new} and its non-linear extensions \cite{jazwinski2007stochastic,kulikov2015accurate,sarkka2007unscented}. On the contrary, if the defender observes the attacker's actions (as $\mathbf{a}_{k}$) and estimates $\hat{\mathbf{x}}_{k}$ at the same discrete-time instants only, the inverse filtering problem remains a discrete-time problem even when the state evolution is modeled as a continuous-time process. Our I-UKF can handle the continuous-time state evolution case with some trivial modifications. In particular, the forward continuous-discrete UKF's time update numerically integrates a pair of differential equations \cite[eq.~(34)]{sarkka2007unscented} to compute $\hat{\mathbf{x}}_{k+1|k}$ and $\bm{\Sigma}_{k+1|k}$ using estimates $\hat{\mathbf{x}}_{k}$ and $\bm{\Sigma}_{k}$ as the initial conditions. The measurement update steps are the same as in the discrete-time forward UKF. These differences need to be taken into account while formulating the inverse filter.

Denote the forward continuous-discrete UKF's time update (solutions of differential equations) as $\hat{\mathbf{x}}_{k+1|k}=\chi_{1}(\hat{\mathbf{x}}_{k})$ and $\bm{\Sigma}_{k+1|k}=\chi_{2}(\hat{\mathbf{x}}_{k},\bm{\Sigma}_{k})$. I-UKF's state transition \eqref{eqn: IUKF state transition detail} becomes
\par\noindent\small
\begin{align}
        \hat{\mathbf{x}}_{k+1}&=\chi_{1}(\hat{\mathbf{x}}_{1})-\sum_{i=0}^{2n_{x}}\omega_{i}\mathbf{K}_{k+1}\mathbf{q}^{*}_{i,k+1|k}+\mathbf{K}_{k+1}h(\mathbf{x}_{k+1})+\mathbf{K}_{k+1}\mathbf{v}_{k+1}.\label{eqn:state transition continuous discrete}
\end{align}
\normalsize
Here, the propagated points $\{\mathbf{q}^{*}_{i,k+1|k}\}$ are again obtained deterministically from the predicted state $\hat{\mathbf{x}}_{k+1|k}$ and covariance estimate $\bm{\Sigma}_{k+1|k}$, which in turn are functions of $\hat{\mathbf{x}}_{k}$ and $\bm{\Sigma}_{k}$ via solutions $\chi_{1}(\cdot)$ and $\chi_{2}(\cdot)$. Hence, \eqref{eqn:state transition continuous discrete} simply becomes $\hat{\mathbf{x}}_{k+1}=\widetilde{f}(\hat{\mathbf{x}}_{k},\bm{\Sigma}_{k},\mathbf{x}_{k+1},\mathbf{v}_{k+1})$ with $\widetilde{f}(\cdot)$ now denoting the modified state transition function.\\

\noindent\textbf{3) Complex-valued systems:} The system model described by equations \eqref{eqn:non-linear state x}-\eqref{eqn:non-linear observation a} pertains to scenarios involving real-valued states and observations. However, certain applications, such as frequency estimation and neural network training, necessitate the consideration of complex-valued systems, often employing complex KFs \cite{dini2011widely,dini2012class}. While the simplest complex KFs assume second-order circularity and only use covariance matrix information, the widely linear processing \cite{dini2011widely,dini2012class,zhang2022unscented} has recently emerged as a technique that considers the general non-circular case and leverages both covariance and pseudo-covariance matrices information. In the realm of complex-valued inverse filtering problems, our I-UKF can be suitably adapted to yield the widely linear complex counterparts.

Consider the defender's state $\mathbf{x}_{k}\in\mathbb{C}^{n_{x}\times 1}$ with $\bm{\Sigma}_{k}$ and $\bm{\Sigma}^{p}_{k}$ denoting its covariance and pseudo-covariance matrices, respectively. In \cite{dini2011widely}, the forward widely linear complex UKF defines an augmented state $\bm{\xi}_{k}\doteq[\mathbf{x}_{k}^{T},\mathbf{x}^{H}_{k}]^{T}$ and its covariance matrix $\bm{\Sigma}^{\xi}_{k}\doteq\begin{bsmallmatrix}
    \bm{\Sigma}_{k}&\bm{\Sigma}^{p}_{k}\\
    (\bm{\Sigma}^{p}_{k})^{H}&\bm{\Sigma}_{k}^{H}
\end{bsmallmatrix}$. The forward filter recursions to estimate $\bm{\xi}_{k}$ then follow from the standard UKF with $(\cdot)^{T}$ replaced by $(\cdot)^{H}$. Note that the sigma points are then generated using estimate $\hat{\bm{\xi}}_{k}$ (of $\bm{\xi}_{k}$) and $\bm{\Sigma}^{\xi}_{k}$ which includes pseudo-covariance $\bm{\Sigma}^{p}_{k}$. While formulating the inverse filter, we need to consider the forward filter's augmented state $\bm{\xi}_{k}$ and modify the state transition \eqref{eqn: IUKF state transition detail}. Finally, \textit{mutatis mutandis}, the general complex I-UKF's recursions stem from the augmented complex UKF of \cite{dini2011widely} adopting the I-UKF's modified state transition as the state evolution process and $\mathbf{a}_{k}$ as observations. On the other hand, the (forward) complex UKF proposed in \cite{zhang2022unscented} introduces modified sigma points and state updates using both innovation and its conjugate. These changes are similarly accommodated in the inverse filter's formulation to obtain an alternative complex I-UKF.

\subsection{Performance analyses}\label{subsec:IUKF stability}
We adopt the unknown matrix approach to derive I-UKF's stability conditions considering general time-varying noise covariances as in Section~\ref{subsec:IEKF stability}. The unknown matrix approach has been previously applied for UKF with linear observations\cite{xiong2006performance_ukf}, UKF with intermittent observations\cite{li2012stochastic_ukf} and consensus-based UKF\cite{li2015weighted}. Similar to Section~\ref{subsec:IEKF stability}, we first examine the stochastic stability of forward UKF before showing that the I-UKF's error dynamics satisfy the stability conditions of a generalized UKF. Finally, we prove that I-UKF's estimates are also conservative.

\subsubsection{1) I-UKF's stability}\label{subsubsec:IUKF stability}
\textit{Forward UKF's stability:} Consider the forward UKF with state-transition \eqref{eqn:non-linear state x} and observation \eqref{eqn:non-linear observation y}. Denote the state prediction, state estimation, and measurement prediction errors by $\widetilde{\mathbf{x}}_{k+1|k}\doteq\mathbf{x}_{k+1}-\hat{\mathbf{x}}_{k+1|k}$, $\widetilde{\mathbf{x}}_{k}\doteq\mathbf{x}_{k}-\hat{\mathbf{x}}_{k}$ and $\widetilde{\mathbf{y}}_{k+1}\doteq\mathbf{y}_{k+1}-\hat{\mathbf{y}}_{k+1|k}$, respectively. From \eqref{eqn:non-linear state x} and \eqref{eqn:forward ukf x predict}, we have $\widetilde{\mathbf{x}}_{k+1|k}=f(\mathbf{x}_{k})+\mathbf{w}_{k}-\sum_{i=0}^{2n_{x}}\omega_{i}\mathbf{s}^{*}_{i,k+1|k}$, which on substituting $\mathbf{s}^{*}_{i,k+1|k}=f(\mathbf{s}_{i,k})$ yields $\widetilde{\mathbf{x}}_{k+1|k}=f(\mathbf{x}_{k})+\mathbf{w}_{k}-\sum_{i=0}^{2n_{x}}\omega_{i}f(\mathbf{s}_{i,k})$.
Using the first-order Taylor series expansion of $f(\cdot)$ at $\hat{\mathbf{x}}_{k}$, we have $\widetilde{\mathbf{x}}_{k+1|k}\approx f(\hat{\mathbf{x}}_{k})+\mathbf{F}_{k}(\mathbf{x}_{k}-\hat{\mathbf{x}}_{k})+\mathbf{w}_{k}-\sum_{i=0}^{2n_{x}}\omega_{i}(f(\hat{\mathbf{x}}_{k})+\mathbf{F}_{k}(\mathbf{s}_{i,k}-\hat{\mathbf{x}}_{k}))$, where $\mathbf{F}_{k}\doteq\frac{\partial f(\mathbf{x})}{\partial\mathbf{x}}\vert_{\mathbf{x}=\hat{\mathbf{x}}_{k}}$. The sigma points $\lbrace\mathbf{s}_{i,k}\rbrace_{0\leq i\leq 2n_{x}}$ are chosen symmetrically about $\hat{\mathbf{x}}_{k}$. Substituting for $\mathbf{s}_{i,k}$ in terms of $\hat{\mathbf{x}}_{k}$ and $\bm{\Sigma}_{k}$ using \eqref{eqn:forward ukf prediction sigma points} simplifies the state prediction error to $\widetilde{\mathbf{x}}_{k+1|k}\approx\mathbf{F}_{k}\widetilde{\mathbf{x}}_{k}+\mathbf{w}_{k}$. Similar to \cite{xiong2006performance_ukf}, we introduce an unknown instrumental diagonal matrix $\mathbf{U}^{x}_{k}\in\mathbb{R}^{n_{x}\times n_{x}}$ to account for the linearization errors as
\par\noindent\small
\begin{align}
    \widetilde{\mathbf{x}}_{k+1|k}=\mathbf{U}^{x}_{k}\mathbf{F}_{k}\widetilde{\mathbf{x}}_{k}+\mathbf{w}_{k}.\label{eqn:linearized x}
\end{align}
\normalsize
Similarly, linearizing $h(\cdot)$ in \eqref{eqn:non-linear observation y} and introducing unknown diagonal matrix $\mathbf{U}^{y}_{k}\in\mathbb{R}^{n_{y}\times n_{y}}$ in \eqref{eqn:forward ukf y predict} yields
\par\noindent\small
\begin{align}
    \widetilde{\mathbf{y}}_{k+1}=\mathbf{U}^{y}_{k+1}\mathbf{H}_{k+1}\widetilde{\mathbf{x}}_{k+1|k}+\mathbf{v}_{k+1},\label{eqn:linearized y}
\end{align}
\normalsize
where $\mathbf{H}_{k+1}\doteq\frac{\partial h(\mathbf{x})}{\partial\mathbf{x}}\vert_{\mathbf{x}=\hat{\mathbf{x}}_{k+1|k}}$. Using \eqref{eqn:forward ukf x update}, we have $\widetilde{\mathbf{x}}_{k}=\widetilde{\mathbf{x}}_{k|k-1}-\mathbf{K}_{k}\widetilde{\mathbf{y}}_{k}$, which when substituted in \eqref{eqn:linearized x} with \eqref{eqn:linearized y} yields the forward UKF's prediction error dynamics as
\par\noindent\small
\begin{align}
    \widetilde{\mathbf{x}}_{k+1|k}=\mathbf{U}^{x}_{k}\mathbf{F}_{k}(\mathbf{I}-\mathbf{K}_{k}\mathbf{U}^{y}_{k}\mathbf{H}_{k})\widetilde{\mathbf{x}}_{k|k-1}-\mathbf{U}^{x}_{k}\mathbf{F}_{k}\mathbf{K}_{k}\mathbf{v}_{k}+\mathbf{w}_{k}.\label{eqn:forward ukf error dynamics}
\end{align}
\normalsize

Following similar procedure as in Section~\ref{subsec:IEKF stability} for deriving I-EKF's stability conditions based on the unknown matrix approach, we can obtain
\par\noindent\small
\begin{align*}
    \bm{\Sigma}_{k+1|k}&=\mathbf{U}^{x}_{k}\mathbf{F}_{k}(\mathbf{I}-\mathbf{K}_{k}\mathbf{U}^{y}_{k}\mathbf{H}_{k})\bm{\Sigma}_{k|k-1}(\mathbf{I}-\mathbf{K}_{k}\mathbf{U}^{y}_{k}\mathbf{H}_{k})^{T}\mathbf{F}_{k}^{T}\mathbf{U}^{x}_{k}+\hat{\mathbf{Q}}_{k},\\
    \bm{\Sigma}^{y}_{k+1}&=\mathbf{U}^{y}_{k+1}\mathbf{H}_{k+1}\bm{\Sigma}_{k+1|k}\mathbf{H}_{k+1}^{T}\mathbf{U}^{y}_{k+1}+\hat{\mathbf{R}}_{k+1},
\end{align*}
\begin{align*}
    \bm{\Sigma}^{xy}_{k+1}&=\begin{cases}\bm{\Sigma}_{k+1|k}\mathbf{U}^{xy}_{k+1}\mathbf{H}_{k+1}^{T}\mathbf{U}^{y}_{k+1}, & n_{x}\geq n_{y}\\
    \bm{\Sigma}_{k+1|k}\mathbf{H}_{k+1}^{T}\mathbf{U}^{y}_{k+1}\mathbf{U}^{xy}_{k+1}, & n_{x}<n_{y}\end{cases},
\end{align*}
\normalsize
where $\hat{\mathbf{Q}}_{k}=\mathbf{Q}_{k}+\mathbf{U}^{x}_{k}\mathbf{F}_{k}\mathbf{K}_{k}\mathbf{R}_{k}\mathbf{K}_{k}^{T}\mathbf{F}_{k}^{T}\mathbf{U}^{x}_{k}+\delta\mathbf{P}_{k+1|k}+\Delta\mathbf{P}_{k+1|k}$ and $\hat{\mathbf{R}}_{k+1}=\mathbf{R}_{k+1}+\Delta\mathbf{P}^{y}_{k+1}+\delta\mathbf{P}^{y}_{k+1}$ with  $\delta\mathbf{P}_{k+1|k}$ ($\delta\mathbf{P}^{y}_{k+1}$) and $\Delta\mathbf{P}_{k+1|k}$ ($\Delta\mathbf{P}^{y}_{k+1}$), respectively, accounting for the difference in true and estimated prediction (measurement prediction) covariances, and errors in the approximation of the expectation. Also, $\mathbf{U}^{xy}_{k+1}$ is the unknown matrix introduced to account for errors in the estimated cross-covariance $\bm{\Sigma}^{xy}_{k+1}$. The error dynamics \eqref{eqn:forward ukf error dynamics} and the various covariances have the same form as that for forward EKF in \eqref{eqn:forward EKF prediction error dynamics}-\eqref{eqn:forward ekf stable sig xy}. Hence, Theorem~\ref{theorem:Forward ekf stable unknown matrix} is applicable for forward UKF stability as well. This is formalized in Theorem \ref{theorem:forward ukf stability} below.
\begin{theorem}[Stochastic stability of forward UKF]\label{theorem:forward ukf stability}
Consider the forward UKF with the non-linear stochastic system given by \eqref{eqn:non-linear state x} and \eqref{eqn:non-linear observation y}. The forward UKF's estimation error $\widetilde{\mathbf{x}}_{k}$ is exponentially bounded in mean-squared sense and bounded with probability one if the following conditions hold true.\\
\textbf{C5.a)} There exist positive real numbers $\bar{f}$, $\bar{h}$, $\bar{\alpha}$, $\bar{\beta}$, $\bar{\gamma}$, $\underline{\sigma}$, $\bar{\sigma}$, $\bar{q}$, $\bar{r}$, $\hat{q}$ and $\hat{r}$ such that the following bounds are fulfilled for all $k\geq 0$.
\par\noindent\small
\begin{align*}    &\|\mathbf{F}_{k}\|\leq\bar{f},\;\;\;\|\mathbf{H}_{k}\|\leq\bar{h},\;\;\;\|\mathbf{U}^{x}_{k}\|\leq\bar{\alpha},\;\;\;\|\mathbf{U}^{y}_{k}\|\leq\bar{\beta},\;\;\;\|\mathbf{U}^{xy}_{k}\|\leq\bar{\gamma},\\        &\mathbf{Q}_{k}\preceq\bar{q}\mathbf{I},\;\;\;\mathbf{R}_{k}\preceq\bar{r}\mathbf{I},\;\;\;\hat{q}\mathbf{I}\preceq\hat{\mathbf{Q}}_{k},\;\;\;\hat{r}\mathbf{I}\preceq\hat{\mathbf{R}}_{k},\;\;\;\underline{\sigma}\mathbf{I}\preceq\bm{\Sigma}_{k|k-1}\preceq\bar{\sigma}\mathbf{I}.
\end{align*}
\normalsize
\textbf{C5.b)} $\mathbf{U}^{x}_{k}$ and $\mathbf{F}_{k}$ are non-singular for every $k\geq 0$.\\
\textbf{C5.c)} The constants satisfy the inequality $\bar{\sigma}\bar{\gamma}\bar{h}^{2}\bar{\beta}^{2}<\hat{r}$.
\end{theorem}

The UKF stability mentioned in \cite[Theorem~1]{xiong2006performance_ukf} for only linear measurements requires a lower bound on measurement noise covariance $\mathbf{R}_{k}$ and also, an upper bound on $\hat{\mathbf{Q}}_{k}$. But $\hat{\mathbf{Q}}_{k}$ is not upper-bounded in Theorem~\ref{theorem:forward ukf stability}. Our non-linear stability guarantee requires upper (lower) bounds on noise covariances $\mathbf{Q}_{k}$ ($\hat{\mathbf{Q}}_{k}$) and $\mathbf{R}_{k}$ ($\hat{\mathbf{R}}_{k}$). Both $\hat{\mathbf{Q}}_{k}$ and $\hat{\mathbf{R}}_{k}$ can be made positive definite to satisfy the lower bounds and enhance the filter's stability by enlarging the noise covariance matrices $\mathbf{Q}_{k}$ and $\mathbf{R}_{k}$, respectively\cite{xiong2006performance_ukf,xiong2007authorreply}. For practical systems where we have a reasonable estimate of the state $\mathbf{x}_{k}$ (due to the process's constraints), the bounds on unknown matrices $\mathbf{U}^{x}_{k}$ and $\mathbf{U}^{y}_{k}$ can be estimated using functions $f(\cdot)$ and $h(\cdot)$\cite{boutayeb1999strong}.

\textit{Inverse UKF's stability:} We now consider the I-UKF with state-transition \eqref{eqn:inverse ukf state transition} and observation \eqref{eqn:non-linear observation a}. To this end, we introduce unknown matrices $\overline{\mathbf{U}}^{x}_{k}$ and $\overline{\mathbf{U}}^{a}_{k}$ to account for the errors in the linearization of functions $\widetilde{f}(\cdot)$ and $g(\cdot)$, respectively, and $\overline{\mathbf{U}}^{xa}_{k}$ for the errors in cross-covariance matrix estimation. Also, $\hat{\overline{\mathbf{Q}}}_{k}$ and $\hat{\overline{\mathbf{R}}}_{k}$ denote the counterparts of $\hat{\mathbf{Q}}_{k}$ and $\hat{\mathbf{R}}_{k}$, respectively, in the I-UKF's error dynamics. Define $\widetilde{\mathbf{F}}_{k}\doteq\left.\frac{\partial\widetilde{f}(\mathbf{x},\bm{\Sigma}_{k},\mathbf{x}_{k+1},\mathbf{0})}{\partial\mathbf{x}}\right\vert_{\mathbf{x}=\doublehat{\mathbf{x}}_{k}}$ and $\mathbf{G}_{k}\doteq\left.\frac{\partial g(\mathbf{x})}{\partial\mathbf{x}}\right\vert_{\mathbf{x}=\doublehat{\mathbf{x}}_{k|k-1}}$. While approximating $\bm{\Sigma}_{k}$ by $\bm{\Sigma}^{*}_{k}$ in I-UKF, we ignore any possible errors. Assume that the forward gain $\mathbf{K}_{k+1}$ computed from $\doublehat{\mathbf{x}}_{k}$ is approximately same as that computed from $\hat{\mathbf{x}}_{k}$ in forward UKF. Additionally, these approximation errors can be bounded by positive constants because $\bm{\Sigma}_{k}$ and $\mathbf{K}_{k+1}$ can be proved to be bounded matrices under the I-UKF's stability assumptions. The bounds required on various matrices for forward UKF's stability are also satisfied when these matrices are evaluated by I-UKF at its own estimates, i.e., $\left\|\frac{\partial f(\mathbf{x})}{\partial\mathbf{x}}\right\|\leq \bar{f}$ and $\left\|\frac{\partial h(\mathbf{x})}{\partial\mathbf{x}}\right\|\leq \bar{h}$ where $\mathbf{x}$ can be any sigma point.

\begin{theorem}[Stochastic stability of I-UKF]\label{theorem:IUKF stability}
Consider the attacker's forward UKF that is stable as per Theorem~\ref{theorem:forward ukf stability}. The I-UKF's state estimation error is exponentially bounded in mean-squared sense and bounded with probability one if the following assumptions hold true.\\
 \textbf{C5.d)} There exist positive real numbers $\bar{g},\bar{c},\bar{d},\bar{\epsilon},\hat{c},\hat{d},\underline{p}$ and $\bar{p}$ such that the following bounds are fulfilled for all $k\geq 0$.
 \par\noindent\small
\begin{align*}
&\|\mathbf{G}_{k}\|\leq\bar{g},\;\;\|\overline{\mathbf{U}}^{a}_{k}\|\leq\bar{c},\;\;\|\overline{\mathbf{U}}^{xa}_{k}\|\leq\bar{d},\;\;\overline{\mathbf{R}}_{k}\preceq\bar{\epsilon}\mathbf{I},\;\;\hat{c}\mathbf{I}\preceq\hat{\overline{\mathbf{Q}}}_{k},\;\;\hat{d}\mathbf{I}\preceq\hat{\overline{\mathbf{R}}}_{k},\;\;\underline{p}\mathbf{I}\preceq\overline{\bm{\Sigma}}_{k|k-1}\preceq\bar{p}\mathbf{I}.
\end{align*}
\normalsize
\textbf{C5.e)} There exist a real constant $\underline{y}$ (not necessarily positive) such that $\bm{\Sigma}^{y}_{k}\succeq\underline{y}\mathbf{I}$ for all $k\geq 0$.\\
\textbf{C5.f)} The functions $f(\cdot)$ and $h(\cdot)$ have bounded outputs i.e. $\|f(\cdot)\|_{2}\leq\delta_{f}$ and $\|h(\cdot)\|_{2}\leq\delta_{h}$ for some real positive numbers $\delta_{f}$ and $\delta_{h}$.\\
\textbf{C5.g)} For all $k\geq 0$, $\widetilde{\mathbf{F}}_{k}$ is non-singular and its inverse satisfies $\|\widetilde{\mathbf{F}}^{-1}_{k}\|\leq\bar{a}$ for some positive real constant $\bar{a}$.\\
\textbf{C5.h)} The constants satisfy the inequality $\bar{p}\bar{d}\bar{g}^{2}\bar{c}^{2}<\hat{d}$.
\end{theorem}
\begin{proof}
See Appendix~\ref{App-thm-IUKF}.
\end{proof}

While there are no constraints on the constant $\underline{y}$ in \textbf{C5.e}, Theorem~\ref{theorem:IUKF stability} requires an additional lower bound on $\bm{\Sigma}^{y}_{k}$ which was not needed for forward UKF's stability. Also, $\underline{y}\neq 0$ because $(\bm{\Sigma}^{y}_{k})^{-1}$ exists for forward UKF to compute its gain. Conditions \textbf{C5.e} and \textbf{C5.f} are necessary to upper-bound the Jacobian $\widetilde{\mathbf{F}}_{k}$. The bounds $\|f(\cdot)\|_{2}\leq\delta_{f}$ and $\|h(\cdot)\|_{2}\leq\delta_{h}$ help to bound the magnitude of the propagated sigma points $\lbrace\mathbf{s}^{*}_{i,k+1|k}\rbrace$ and $\lbrace\mathbf{q}^{*}_{i,k+1|k}\rbrace$, respectively, generated from a state estimate, which in turn, upper bounds the various covariance estimates. Also, the computation of gain matrix $\mathbf{K}_{k}$ involves $(\bm{\Sigma}^{y}_{k})^{-1}$, which can be upper-bounded by assuming a lower bound on $\bm{\Sigma}^{y}_{k}$ as in \textbf{C5.e}. The additional conditions \textbf{C5.e-g} are also not necessary for I-EKF's stability in Theorem~\ref{theorem: inverse EKF stable unknown matrix}. 

\begin{remark}[Practical bounds on system functions]
    Intuitively, the bounds in \textbf{C5.f} represent the physical constraints on the state of the process being observed and its observations. For instance, in a radar's target localization problem, the target location is reasonably upper-bounded by the maximum unambiguous range and beam pattern (main lobe) of the radar. The noises $\mathbf{w}_{k}$ and $\mathbf{v}_{k}$ then represent, respectively, the modeling and measurement uncertainties that are assumed to be Gaussian to obtain simplified closed-form solutions\cite{ristic2003beyond}. Note that Theorem~\ref{theorem:IUKF stability} assumes the state transition function $f(\cdot)$ and the attacker's observation function $h(\cdot)$ to be bounded, but the defender's observation function $g(\cdot)$ is, in general, unbounded.
\end{remark}

\subsubsection{2) I-UKF's conservativeness}\label{subsubsec:consistency}
Using SLT\cite{arasaratnam2007qkf} to linearize \eqref{eqn:inverse ukf state transition} and \eqref{eqn:non-linear observation a} with respect to I-UKF's augmented state $\mathbf{z}_{k}$ and (forward) estimate $\hat{\mathbf{x}}_{k}$, respectively, we obtain
\par\noindent\small
\begin{align}       \hat{\mathbf{x}}_{k+1}&=\mathbf{U}^{z}_{k}\overline{\mathbf{F}}^{x}_{k}\hat{\mathbf{x}}_{k}+\mathbf{U}^{z}_{k}\overline{\mathbf{F}}^{v}_{k}\mathbf{v}_{k+1},\label{eqn:SLT state transition ukf}\\
    \mathbf{a}_{k}&=\mathbf{U}^{a}_{k}\overline{\mathbf{G}}_{k}\hat{\mathbf{x}}_{k}+\bm{\epsilon}_{k},\label{eqn:SLT observation ukf}
\end{align}
\normalsize
where $\overline{\mathbf{F}}_{k}=[\overline{\mathbf{F}}^{x}_{k},\overline{\mathbf{F}}^{v}_{k}]$ and $\overline{\mathbf{G}}_{k}$ are the respective linear pseudo transition matrices. Also, $\mathbf{U}^{z}_{k}$ and $\mathbf{U}^{a}_{k}$ are unknown diagonal matrices introduced to account for the approximation errors in SLT and are different from the ones introduced in Theorem~\ref{theorem:IUKF stability} for the higher-order terms in the Taylor approximation.
\begin{theorem}[I-UKF's conservative estimates]
    \label{thm:ukf consistency}
    Consider a conservative initial estimate pair $(\doublehat{\mathbf{x}}_{0},\overline{\bm{\Sigma}}_{0})$ for the I-UKF. Then for any $k\geq 1$, the I-UKF's recursive estimate pair $(\doublehat{\mathbf{x}}_{k},\overline{\bm{\Sigma}}_{k})$ is also conservative such that $\mathbb{E}[(\hat{\mathbf{x}}_{k}-\doublehat{\mathbf{x}}_{k})(\hat{\mathbf{x}}_{k}-\doublehat{\mathbf{x}}_{k})^{T}]\preceq\overline{\bm{\Sigma}}_{k}$, where $\hat{\mathbf{x}}_{k}$ is the forward UKF's state estimate.
\end{theorem}
\begin{proof}
    The result follows from the principle of mathematical induction and the symmetry of generated sigma-points, similar to I-EKF's proof in Appendix~\ref{App-thm-inverse EKF consistency}.
\end{proof}
It follows from Theorem~\ref{thm:ukf consistency} and Definition~\ref{def:consistency} that I-UKF is a conservative estimator if the initial estimate $\doublehat{\mathbf{x}}_{0}$ is unbiased and initial error covariance estimate $\overline{\bm{\Sigma}}_{0}$ is chosen sufficiently large. However, Theorem~\ref{thm:ukf consistency} holds only when the SLT-based linearized models \eqref{eqn:SLT state transition ukf} and \eqref{eqn:SLT observation ukf} well approximate the non-linear equations \eqref{eqn:inverse ukf state transition} and \eqref{eqn:observation a}, respectively. In particular, if the unknown matrices $\mathbf{U}^{z}_{k}$ and $\mathbf{U}^{a}_{k}$ are not sufficient to account for the approximation errors, the estimates $\doublehat{\mathbf{x}}_{k}$ will be in general biased and hence, not conservative.

\section{Inverse Cubature and Quadrature KFs}\label{sec:ICKF and IQKF}
The (forward) CKF generates a set of `$2n_{x}$' cubature points deterministically about the state estimate based on the third-degree spherical-radial cubature rule to numerically compute a standard Gaussian weighted non-linear integral \cite{arasaratnam2009cubature}. Similarly, the ($m$-point) QKF employs a $m$-point Gauss-Hermite quadrature rule to generate $m^{n_{x}}$ quadrature points\cite{ito2000gaussian}. In \cite{arasaratnam2007qkf}, QKF was reformulated using statistical linear regression, wherein the linearized function is more accurate in a statistical sense than EKF's first-order Taylor series approximation. The CQKF generalizes CKF by efficiently employing the cubature and quadrature integration rules together\cite{bhaumik2013cubature}. In particular, the $n_{x}$-dimensional recursive Bayesian integral is decomposed into a surface and line integral approximated using third-degree spherical cubature and one-dimensional Gauss-Laguerre quadrature rules, respectively.

\subsection{Inverse CKF}\label{subsec:ICKF}
Denote the $i$-th standard basis vector in $\mathbb{R}^{n_{x}\times 1}$ by $\mathbf{e}_{i}$. Define $\bm{\xi}_{i}$ as the $i$-th element of the $2n_{x}$-points set $\{\sqrt{n_{x}}\mathbf{e}_{1},\sqrt{n_{x}}\mathbf{e}_{2},\hdots,\sqrt{n_{x}}\mathbf{e}_{n_{x}},-\sqrt{n_{x}}\mathbf{e}_{1},\hdots,-\sqrt{n_{x}}\mathbf{e}_{n_{x}}\}$. The cubature points generated from a state estimate $\hat{\mathbf{x}}$ and its error covariance matrix $\bm{\Sigma}$ are $\widetilde{\mathbf{x}}_{i}=\hat{\mathbf{x}}+\sqrt{\bm{\Sigma}}\bm{\xi}_{i}$ for all $i=1,2,\hdots,2n_{x}$ with each weight $\omega_{i}=1/2n_{x}$. The third-degree cubature rule is exact for polynomials up to the third degree and computes the posterior mean accurately but the error covariance approximately. It has been reported in \cite{arasaratnam2009cubature} that higher-degree rules may not necessarily yield any improvement in the CKF performance. In I-CKF, we assume a forward CKF to estimate the defender's state.

Consider the state transition \eqref{eqn:non-linear state x} and observations \eqref{eqn:non-linear observation y} of the forward filter. Then, \textit{ceteris paribus}, using these cubature points in place of the sigma points in the forward UKF of Section~\ref{subsec:IUKF formulation}, we obtain the forward CKF. The corresponding time and measurement updates become
\par\noindent\small
\begin{align*}
    &\mathbf{s}_{i,k}=\hat{\mathbf{x}}_{k}+\sqrt{\bm{\Sigma}_{k}}\bm{\xi}_{i}\;\;\;\forall\;\; i=1,2,\hdots,2n_{x},\\
    &\mathbf{q}_{i,k+1|k}=\hat{\mathbf{x}}_{k+1|k}+\sqrt{\bm{\Sigma}_{k+1|k}}\bm{\xi}_{i}\;\;\;\forall\;\; i=1,2,\hdots,2n_{x},.
\end{align*}
\normalsize

Assuming perfect system model information, the I-CKF's state transition is
\par\noindent\small
\begin{align*}
    \hat{\mathbf{x}}_{k+1}&=\frac{1}{2n_{x}}\sum_{i=1}^{2n_{x}}\left(\mathbf{s}^{*}_{i,k+1|k}-\mathbf{K}_{k+1}\mathbf{q}^{*}_{i,k+1|k}\right)+\mathbf{K}_{k+1}h(\mathbf{x}_{k+1})+\mathbf{K}_{k+1}\mathbf{v}_{k+1}.
\end{align*}
\normalsize
In terms of the augmented state $\mathbf{z}_{k}=[\hat{\mathbf{x}}_{k}^{T},\mathbf{v}_{k+1}^{T}]^{T}$ of dimension $n_{z}=n_{x}+n_{y}$, the state transition is $\hat{\mathbf{x}}_{k+1}=\widetilde{f}(\mathbf{z}_{k},\bm{\Sigma}_{k},\mathbf{x}_{k+1})$. Approximate $\bm{\Sigma}_{k}$ by $\bm{\Sigma}^{*}_{k}$ (evaluated similarly as in I-UKF). Denote $\hat{\mathbf{z}}_{k}$ and $\overline{\bm{\Sigma}}^{z}_{k}$ as in \eqref{eqn:IUKF z hat and sigma z}. Also, $\overline{\bm{\xi}}_{j}$ denotes the $j$-th element of $2n_{z}$-points set $\{\sqrt{n_{z}}\overline{\mathbf{e}}_{1},\sqrt{n_{z}}\overline{\mathbf{e}}_{2},\hdots,\sqrt{n_{z}}\overline{\mathbf{e}}_{n_{z}},-\sqrt{n_{z}}\overline{\mathbf{e}}_{1},-\sqrt{n_{z}}\overline{\mathbf{e}}_{2},\hdots,-\sqrt{n_{z}}\overline{\mathbf{e}}_{n_{z}}\}$ where $\overline{\mathbf{e}}_{j}$ is the $j$-th standard basis vector in $\mathbb{R}^{n_{z}\times 1}$. The I-CKF's cubature points in the $n_{z}$-dimensional state space are
\par\noindent\small
\begin{align*}
    \overline{\mathbf{s}}_{j,k}=\hat{\mathbf{z}}_{k}+\sqrt{\overline{\bm{\Sigma}}^{z}_{k}}\overline{\bm{\xi}}_{j}\;\;\;\forall\;\; i=1,2,\hdots,2n_{z},
\end{align*}
\normalsize
with each weight $\overline{\omega}_{j}=1/2n_{z}$. The I-CKF's recursions then follow the time and measurement update procedure in \eqref{eqn:IUKF f propagate}-\eqref{eqn:IUKF covariance update}. The I-CKF recursions follow from the non-additive noise formulation of CKF \cite{wang2017augmentedckf} with a higher $(n_{x}+n_{y})$-dimensional cubature points. Again, unlike forward CKF, these cubature points are generated only once for the time update, taking into account the process noise statistics (covariance $\mathbf{R}$).

\begin{remark}[Differences with I-UKF]
Contrary to CKF, UKF generates a set of `$2n_{x}+1$' sigma points around the previous state estimate, including one at the previous estimate itself, with their spread and weights controlled by parameter $\kappa$. The I-UKF in Section~\ref{subsec:IUKF formulation} also assumes a known forward filter's $\kappa$. The I-CKF, however, does not require any such parameter information. Note that forward UKF with $\kappa$ set to $0$ results in CKF as the attacker's forward filter. In that case, I-UKF with I-UKF's control parameter $\overline{\kappa}=0$ reduces to I-CKF.
\end{remark}
\begin{remark}[I-CKF for non-Gaussian noises]
    As in the case of I-UKF, I-CKF can also be generalized to non-Gaussian systems. For instance, forward mixture correntropy-based CKF proposed in \cite{wang2020outlier} considers a time-varying noise covariance $\mathbf{R}_{k}$ instead of $\mathbf{R}$ and then iteratively modifies the gain matrix. In particular, \cite{wang2020outlier} introduces a diagonal matrix $\bm{\Lambda}_{k}$ with $i$-th diagonal element $[\bm{\Lambda}_{k}]_{i,i}=(\lambda/n_{y})(\alpha G_{\sigma_{1}}([\mathbf{e}_{k}]_{i})/\sigma_{1}^{2}+(1-\alpha)G_{\sigma_{2}}([\mathbf{e}_{k}]_{i})/\sigma_{2}^{2})$ where $\mathbf{e}_{k}=\mathbf{R}_{k}^{-1/2}(\mathbf{y}_{k}-h(\mathbf{x}_{k}))$; $\alpha$ and $\lambda$ are parameters of the mixture correntropy; and $G_{\sigma}(\cdot)$ denotes a Gaussian kernel with parameter $\sigma$. The gain matrix is then obtained by replacing $\mathbf{R}_{k}$ by $(\mathbf{R}_{k}^{1/2})^{T}\bm{\Lambda}_{k}^{-1}\mathbf{R}_{k}^{1/2}$ in the measurement update of the forward CKF algorithm while all other state prediction and update steps remain same. These modifications need to be taken into account in the I-CKF's state transition equation while formulating the inverse filter. I-CKF's gain matrix is also similarly modified by introducing diagonal matrix $\overline{\bm{\Lambda}}_{k}$ which is the counterpart of $\bm{\Lambda}_{k}$ for the inverse filter's dynamics.
\end{remark}

\subsubsection{Stability guarantees}

The theoretical performance guarantees for I-CKF can be derived following similar procedures as in Section~\ref{subsec:IUKF stability}. In fact, the unknown matrix approach has earlier been considered for forward CKF's stability in \cite{zarei2014convergence,wanasinghe2015stability}. However, \cite{zarei2014convergence} considered CKF with only linear observations and suggested modifications in CKF to enhance stability in the local asymptotic convergence sense. The stability conditions for CKF with non-linear measurements were derived in \cite{wanasinghe2015stability} using the exponential-mean-squared-boundedness sense. However, we provide improved stability results for the general forward CKF in the exponential-boundedness sense and obtain the same for the I-CKF. The following Proposition~\ref{proposition:forward CKF} and Theorem~\ref{thm:I-CKF} formalize the stability results for forward CKF and I-CKF, respectively, with the errors and matrices defined as in Section~\ref{subsec:IUKF stability}. Furthermore, the results of Theorem~\ref{thm:ukf consistency} can be trivially shown to be applicable to I-CKF's estimates considering the symmetry of cubature points and hence omitted here.

\begin{proposition}[Stochastic stability of forward CKF]
\label{proposition:forward CKF}
Consider the system \eqref{eqn:non-linear state x} and \eqref{eqn:non-linear observation y} with forward CKF. Forward CKF's estimation error $\widetilde{\mathbf{x}}_{k}$ is exponentially bounded in the mean-squared sense and bounded with probability one if the following holds true.\\
\textbf{C5.i)} There exist positive real numbers $\bar{f}$, $\bar{h}$, $\bar{\alpha}$, $\bar{\beta}$, $\bar{\gamma}$, $\underline{\sigma}$, $\bar{\sigma}$, $\bar{q}$, $\bar{r}$, $\hat{q}$ and $\hat{r}$ such that for all $k\geq 0$,
\par\noindent\small
\begin{align*}    &\|\mathbf{F}_{k}\|\leq\bar{f},\;\;\;\|\mathbf{H}_{k}\|\leq\bar{h},\;\;\;\|\mathbf{U}^{x}_{k}\|\leq\bar{\alpha},\;\;\;\|\mathbf{U}^{y}_{k}\|\leq\bar{\beta},\;\;\;\|\mathbf{U}^{xy}_{k}\|\leq\bar{\gamma},\\        &\mathbf{Q}_{k}\preceq\bar{q}\mathbf{I},\;\;\;\mathbf{R}_{k}\preceq\bar{r}\mathbf{I},\;\;\;\hat{q}\mathbf{I}\preceq\hat{\mathbf{Q}}_{k},\;\;\;\hat{r}\mathbf{I}\preceq\hat{\mathbf{R}}_{k},\;\;\;\underline{\sigma}\mathbf{I}\preceq\bm{\Sigma}_{k|k-1}\preceq\bar{\sigma}\mathbf{I}.
\end{align*}
\normalsize
\textbf{C5.j)} $\mathbf{U}^{x}_{k}$ and $\mathbf{F}_{k}$ are non-singular for every $k\geq 0$.\\
\textbf{C5.k)} The constants satisfy the inequality $\bar{\sigma}\bar{\gamma}\bar{h}^{2}\bar{\beta}^{2}<\hat{r}$.
\end{proposition}
\begin{proof}
The proof follows the stability conditions of (forward) CKF mentioned in \cite{wanasinghe2015stability}, which are the same as the bounds in \textbf{C5.i}. However, the proof in \cite{wanasinghe2015stability} uses invertibility of $\mathbf{U}^{x}_{k}\mathbf{F}_{k}(\mathbf{I}-\mathbf{K}_{k}\mathbf{H}_{k})$ for all $k\geq 1$, which may not be true in general. In Proposition~\ref{proposition:forward CKF}, similar to Appendix~\ref{App-thm-Forward ekf stable unknown matrix}, the inequality in \textbf{C5.k} guarantees $(\mathbf{I}-\mathbf{K}_{k}\mathbf{H}_{k})$ to be invertible for all $k\geq 1$. This, in turn, ensures invertibility of $\mathbf{U}^{x}_{k}\mathbf{F}_{k}(\mathbf{I}-\mathbf{K}_{k}\mathbf{H}_{k})$ under \textbf{C5.j}.
\end{proof}
\begin{theorem}[Stochastic stability of I-CKF]
\label{thm:I-CKF}
Assume a stable forward CKF as per Proposition~\ref{proposition:forward CKF}. The I-CKF's state estimation error is exponentially bounded in mean-squared sense and bounded with probability one if the following conditions hold true.\\
 \textbf{C5.l)} There exist positive real numbers $\bar{g},\bar{c},\bar{d},\bar{\epsilon},\hat{c},\hat{d},\underline{p}$ and $\bar{p}$ such that for all $k\geq 0$,
 \par\noindent\small
\begin{align*}      &\|\mathbf{G}_{k}\|\leq\bar{g},\;\;\|\overline{\mathbf{U}}^{a}_{k}\|\leq\bar{c},\;\;\|\overline{\mathbf{U}}^{xa}_{k}\|\leq\bar{d},\;\;\overline{\mathbf{R}}_{k}\preceq\bar{\epsilon}\mathbf{I},\;\;\hat{c}\mathbf{I}\preceq\hat{\overline{\mathbf{Q}}}_{k},\;\;\hat{d}\mathbf{I}\preceq\hat{\overline{\mathbf{R}}}_{k},\;\;\underline{p}\mathbf{I}\preceq\overline{\bm{\Sigma}}_{k|k-1}\preceq\bar{p}\mathbf{I}.
\end{align*}
\normalsize
\textbf{C5.m)} There exist a real constant $\underline{y}$ (not necessarily positive) such that $\bm{\Sigma}^{y}_{k}\succeq\underline{y}\mathbf{I}$ for all $k\geq 0$.\\
\textbf{C5.n)} The functions $f(\cdot)$ and $h(\cdot)$ are bounded as $\|f(\cdot)\|_{2}\leq\delta_{f}$ and $\|h(\cdot)\|_{2}\leq\delta_{h}$ for some real positive numbers $\delta_{f}$ and $\delta_{h}$.\\
\textbf{C5.o)} For all $k\geq 0$, $\widetilde{\mathbf{F}}_{k}$ is non-singular and satisfies $\|\widetilde{\mathbf{F}}^{-1}_{k}\|\leq\bar{a}$ for some real constant $\bar{a}>0$.\\
\textbf{C5.p)} The constants satisfy the inequality $\bar{p}\bar{d}\bar{g}^{2}\bar{c}^{2}<\hat{d}$.
\end{theorem}
\begin{proof}
    Under the assumptions of Theorem~\ref{thm:I-CKF}, I-CKF's error dynamics can be shown to satisfy the stability conditions \textbf{C5.i}-\textbf{C5.k} for a general CKF provided in Proposition~\ref{proposition:forward CKF}. Employing the procedure to bound the derivatives as in Lemma~\ref{lemma: jacobian bound} of Appendix~\ref{App-thm-IUKF}, it follows that the bounds \textbf{C5.m} and \textbf{C5.n} ensure that the Jacobian $\widetilde{\mathbf{F}}_{k}$ is upper-bounded by a constant $c_{f}>0$ for all $k\geq 0$ as $\|\widetilde{\mathbf{F}}_{k}\|\leq c_{f}$. Further, using the unknown matrices from forward CKF's error dynamics as in Appendix~\ref{subsec:proof of IUKF theorem}, it follows that $\overline{\mathbf{U}}^{x}_{k}=(\mathbf{I}-\mathbf{K}_{k+1}\mathbf{U}^{y}_{k+1}\mathbf{H}_{k+1})\mathbf{U}^{x}_{k}\mathbf{F}_{k}\widetilde{\mathbf{F}}_{k}^{-1}$. Hence, the non-singularity of $\widetilde{\mathbf{F}}_{k}$ (\textbf{C5.o}) leads to $\overline{\mathbf{U}}^{x}_{k}$ being invertible because $\mathbf{F}_{k}$, $\mathbf{U}^{x}_{k}$ and $(\mathbf{I}-\mathbf{K}_{k+1}\mathbf{U}^{y}_{k+1}\mathbf{H}_{k+1})$ are invertible under \textbf{C5.i}-\textbf{C5.k} of Proposition~\ref{proposition:forward CKF}. The bound in \textbf{C5.o} provides an upper-bound $\|\overline{\mathbf{U}}^{x}_{k}\|\leq\bar{\alpha}\bar{f}\bar{a}(1+\bar{k}\bar{\beta}\bar{h})$ where $\|\mathbf{K}_{k+1}\|\leq\bar{k}=\bar{\sigma}\bar{\gamma}\bar{h}\bar{\beta}/\hat{r}$ for forward CKF is obtained similarly as in Appendix~\ref{App-thm-Forward ekf stable unknown matrix}. All other conditions for CKF stability trivially hold true for the I-CKF's error dynamics under \textbf{C5.l}-\textbf{C5.p}.
\end{proof}

\subsection{Inverse QKF}\label{subsec:IQKF}
Define $\mathbf{M}$ as the `$m\times m$' symmetric tridiagonal matrix with zero diagonal elements such that $[\mathbf{M}]_{(i,i+1)}=\sqrt{i/2}$ for all $1\leq i\leq m$. For the one-dimensional case, the $i$-th quadrature point of the $m$-point quadrature rule is $\zeta_{i}=\sqrt{2}\beta_{i}$ where $\beta_{i}$ is the $i$-th eigenvalue of $\mathbf{M}$. The corresponding weight $\omega_{i}=[\bm{\nu_{i}}]_{1}^{2}$, where $[\bm{\nu}_{i}]_{1}$ is the first element of the $i$-th normalized eigenvector of $\mathbf{M}$. A $m$-point quadrature rule computes the mean exactly for polynomials of order less than or equal to $(2m-1)$, and the covariance is exact for polynomials of degree less than $m$ \cite{arasaratnam2007qkf}. In a $n_{x}$-dimensional state space, the $m$-point (per-axis) QKF considers $m^{n_{x}}$ quadrature points obtained by extending the scalar quadrature points as $\bm{\zeta}_{i}=[\zeta_{i_{1}}, \zeta_{i_{2}},\hdots,\zeta_{i_{n_{x}}}]^{T}$ and $\omega_{i}=\prod_{j=1}^{n_{x}}\omega_{i_{j}}$ where $\lbrace\zeta_{i_{j}},\omega_{i_{j}}\rbrace_{1\leq i_{j}\leq m}$ are the $m$ scalar quadrature points corresponding to the $j$-th dimension. In I-QKF, we assume a forward QKF to estimate the defender's state. 

Then, \textit{ceteris paribus}, considering these quadrature points in place of sigma points in the forward UKF, we obtain the $m$-point forward QKF with the time and measurement updates as follows
\par\noindent\small
\begin{align*}  
    \mathbf{s}_{i,k}&=\hat{\mathbf{x}}_{k}+\sqrt{\bm{\Sigma}_{k}}\bm{\zeta}_{i}\;\;\;\forall\;\; i=1,2,\hdots,m^{n_{x}},\\
    \mathbf{q}_{i,k+1|k}&=\hat{\mathbf{x}}_{k+1|k}+\sqrt{\bm{\Sigma}_{k+1|k}}\bm{\zeta}_{i}\;\;\;\forall\;\; i=1,2,\hdots,m^{n_{x}}.
\end{align*}
\normalsize
Assuming the parameter $m$ of the forward QKF to be known, the I-QKF's state transition is
\par\noindent\small
\begin{align*}
    \hat{\mathbf{x}}_{k+1}&=\sum_{i=1}^{m^{n_{x}}}\omega_{i}\left(\mathbf{s}^{*}_{i,k+1|k}-\mathbf{K}_{k+1}\mathbf{q}^{*}_{i,k+1|k}\right)+\mathbf{K}_{k+1}h(\mathbf{x}_{k+1})+\mathbf{K}_{k+1}\mathbf{v}_{k+1}.
\end{align*}
\normalsize
In terms of the augmented state $\mathbf{z}_{k}=[\hat{\mathbf{x}}_{k}^{T},\mathbf{v}_{k+1}^{T}]^{T}$, state transition is $\hat{\mathbf{x}}_{k+1}=\widetilde{f}(\mathbf{z}_{k},\bm{\Sigma}_{k},\mathbf{x}_{k+1})$. For the $\overline{m}$-point I-QKF, we denote the quadrature points in the $n_{z}$-dimensional state space by $\lbrace\overline{\bm{\zeta}}_{j},\overline{\omega}_{j}\rbrace_{1\leq j\leq \overline{m}^{n_{z}}}$ such that I-QKF generates a set of $\overline{m}^{n_{z}}$ quadrature points as
\par\noindent\small
\begin{align*}
    \overline{\mathbf{s}}_{j,k}=\hat{\mathbf{z}}_{k}+\sqrt{\overline{\bm{\Sigma}}^{z}_{k}}\overline{\bm{\zeta}}_{j},
\end{align*}
\normalsize
for all $j=1,2,\hdots,\overline{m}^{n_{z}}$, where $\hat{\mathbf{z}}_{k}$ and $\overline{\bm{\Sigma}}^{z}_{k}$ are as defined in \eqref{eqn:IUKF z hat and sigma z}. The I-QKF's recursions then follow the time and measurement update procedure in \eqref{eqn:IUKF f propagate}-\eqref{eqn:IUKF covariance update}. Note that the choice of I-QKF's $\overline{m}$ is independent of any assumption about the forward QKF's parameter $m$.

I-QKF recursions also similarly follow from the non-additive noise formulation of QKF\cite{arasaratnam2007qkf}. Analogous to I-UKF and I-CKF, I-QKF also generates only one set of quadrature points per recursion, but the relative increase in the state dimension from $n_{x}$ to $n_{z}$ is more significant in the latter. The QKF and, hence, I-QKF suffer from the curse of dimensionality because the number of quadrature points increases exponentially with the state-space dimension. On the contrary, the cubature/sigma points in CKF/UKF scale up linearly. However, the expensive computations required to estimate $\{\bm{\zeta}_{i},\omega_{i}\}_{1\leq i\leq m^{n_{x}}}$ ($\{\overline{\bm{\zeta}}_{j},\overline{\omega}_{j}\}_{1\leq j\leq\overline{m}^{n_{z}}}$) in forward QKF (I-QKF) are performed offline \cite{arasaratnam2007qkf}. Further, \cite{closas2012multiple,closas2015computational} suggest methods for QKF's complexity reduction. For the one-dimensional state space ($n_{x}=1$), the forward $3$-point QKF coincides with the forward UKF with $\kappa=2$ \cite{ito2000gaussian}. In this case, I-QKF with $\overline{m}=3$ also coincides with I-UKF with $\overline{\kappa}=2$.

\subsection{Inverse CQKF}\label{subsec:ICQKF}
Denote $\mathbf{e}_{i}$ as the $i$-th standard basis vector in $\mathbb{R}^{n_{x}\times 1}$ and define $2n_{x}$-points set $\{\mathbf{u}_{i'}\}_{1\leq i'\leq 2n_{x}}=\{\mathbf{e}_{1},\mathbf{e}_{2},\hdots,\mathbf{e}_{n_{x}},-\mathbf{e}_{1},\hdots,-\mathbf{e}_{n_{x}}\}$. The $m$-th order CQKF generates $2mn_{x}$ CQ points for $n_{x}$-dimensional state space. Algorithm~\ref{alg:CQ points} provides the procedure to compute the CQ points $\{\bm{\xi}_{i}\}_{1\leq i\leq 2mn_{x}}$ and their corresponding weights $\{\omega_{i}\}_{1\leq i\leq 2mn_{x}}$.
\begin{algorithm}
	\caption{Cubature-quadrature points generation}
	\label{alg:CQ points}
    \begin{algorithmic}[1]
    \Statex \textbf{Input:} $n_{x}$, $m$
    \Statex \textbf{Output:} $\{\bm{\xi}_{i},\omega_{i}\}_{1\leq i\leq 2mn_{x}}$
\State Compute the roots $\{\lambda_{j}\}_{1\leq j\leq m}$ of $m$-th order Chebyshev-Laguerre polynomial $L^{\beta}_{m}(\lambda)$ for $\beta=(n_{x}/2)-1$ given by
\begin{align*}
    L^{\beta}_{m}(\lambda)&=\lambda^{m}-\frac{m}{1!}(m+\beta)\lambda^{m-1}+\frac{m(m-1)}{2!}(m+\beta)(m+\beta-1)\lambda^{m-2}-\hdots,
\end{align*}
while $(\cdot)!$ denotes factorial.
\For{$i'\gets 1$ to $2n_{x}$}
\For{$j\gets 1$ to $m$}
        \State $\bm{\xi}_{m(i'-1)+j}=\sqrt{2\lambda_{j}}\mathbf{u}_{i'}$.
        \State $\omega_{m(i'-1)+j}=\frac{m!\Gamma(\beta+m+1)}{2n_{x}\Gamma(n_{x}/2)\lambda_{j}(\partial L^{\beta}_{m}/\partial\lambda_{j})^{2}}$ where $\Gamma(\cdot)$ is the Gamma function.
    \EndFor
\EndFor
\Statex \Return $\{\bm{\xi}_{i},\omega_{i}\}_{1\leq i\leq 2mn_{x}}$.
    \end{algorithmic}
\end{algorithm}

Then, \textit{ceteris paribus}, considering the CQ points in place of sigma points in the forward UKF, we obtain the forward CQKF with the following time and measurement updates:
\par\noindent\small
\begin{align*}  
    \mathbf{s}_{i,k}&=\hat{\mathbf{x}}_{k}+\sqrt{\bm{\Sigma}_{k}}\bm{\xi}_{i}\;\;\;\forall\;\; i=1,2,\hdots,2mn_{x},\\
    \mathbf{q}_{i,k+1|k}&=\hat{\mathbf{x}}_{k+1|k}+\sqrt{\bm{\Sigma}_{k+1|k}}\bm{\xi}_{i}\;\;\;\forall\;\; i=1,2,\hdots,2mn_{x}.
\end{align*}
\normalsize
Again, assuming the parameter $m$ of the forward CQKF to be known, the I-CQKF's state transition is
\par\noindent\small
\begin{align*}
    \hat{\mathbf{x}}_{k+1}&=\sum_{i=1}^{2mn_{x}}\omega_{i}\left(\mathbf{s}^{*}_{i,k+1|k}-\mathbf{K}_{k+1}\mathbf{q}^{*}_{i,k+1|k}\right)+\mathbf{K}_{k+1}h(\mathbf{x}_{k+1})+\mathbf{K}_{k+1}\mathbf{v}_{k+1},
\end{align*}
\normalsize
which in terms of the augmented state $\mathbf{z}_{k}=[\hat{\mathbf{x}}_{k}^{T},\mathbf{v}_{k+1}^{T}]^{T}$ becomes $\hat{\mathbf{x}}_{k+1}=\widetilde{f}(\mathbf{z}_{k},\bm{\Sigma}_{k},\mathbf{x}_{k+1})$. For the $\overline{m}$-point I-CQKF, we denote the CQ points in the $n_{z}$-dimensional state space by $\lbrace\overline{\bm{\xi}}_{j},\overline{\omega}_{j}\rbrace_{1\leq j\leq 2\overline{m}n_{z}}$ such that I-CQKF generates a set of $2\overline{m}n_{z}$ CQ points as
\par\noindent\small
\begin{align*}
    \overline{\mathbf{s}}_{j,k}=\hat{\mathbf{z}}_{k}+\sqrt{\overline{\bm{\Sigma}}^{z}_{k}}\overline{\bm{\xi}}_{j},
\end{align*}
\normalsize
for all $j=1,2,\hdots,2\overline{m}n_{z}$ with $\hat{\mathbf{z}}_{k}$ and $\overline{\bm{\Sigma}}^{z}_{k}$ as defined in \eqref{eqn:IUKF z hat and sigma z}. The I-CQKF's recursions then follow the time and measurement update procedure in \eqref{eqn:IUKF f propagate}-\eqref{eqn:IUKF covariance update}.

Note that an $m$-th order CQKF considers $2n_{x}m$ CQ points for an $n_{x}$-dimensional state-space. Hence, contrary to QKF, the number of CQ points increases linearly with the state-space dimension. However, computationally, CQKF is still more expensive than CKF. The same argument holds for I-CQKF. Again, the CQ points and their weights $\{\bm{\xi}_{i},\omega_{i}\}$ ($\{\overline{\bm{\xi}}_{i},\overline{\omega}_{i}\}$) in forward CQKF (I-CQKF) are generated offline following Algorithm~\ref{alg:CQ points}.

\section{Numerical experiments}\label{sec:ISPKF numericals}
We consider different example systems widely used to analyze UKF, CKF, QKF, and CQKF performances and also compare these filters' accuracy with I-EKF using RCRLB as the theoretical benchmark. Throughout all experiments, the initial information matrices $\mathbf{J}_{0}$ and $\overline{\mathbf{J}}_{0}$ (for RCRLB computations) were set to $\bm{\Sigma}_{0}^{-1}$ and $\overline{\bm{\Sigma}}_{0}^{-1}$, respectively. Besides achieving RCRLB, an estimator also needs to be credible, i.e., its estimated error covariance $\bm{\Sigma}$ is statistically close to the actual MSE matrix $\mathbf{P}$. In \cite{li2001practical}, averaged normalized estimation error squared (ANEES) and NCI have been proposed as credibility measures. However, NCI is preferable for comparing different estimators because it penalizes optimism and pessimism to the same degree. An optimistic (pessimistic) estimator's $\bm{\Sigma}$ is statistically smaller (larger) than $\mathbf{P}$. Define $NCI=(10/M)\sum_{m=1}^{M}\log_{10}(\epsilon_{m}/\epsilon^{*}_{m})$ where $\epsilon_{m}=\widetilde{\mathbf{x}}_{m}^{T}\bm{\Sigma}_{m}^{-1}\widetilde{\mathbf{x}}_{m}$ and $\epsilon^{*}_{m}=\widetilde{\mathbf{x}}_{m}^{T}\mathbf{P}_{m}^{-1}\widetilde{\mathbf{x}}_{m}$ with $M$ as the total number of Monte-Carlo independent runs. Here, $\widetilde{\mathbf{x}}_{m}=\mathbf{x}_{m}-\hat{\mathbf{x}}_{m}$ is the estimation error at $m$-th run for actual state $\mathbf{x}_{m}$ and its estimate $\hat{\mathbf{x}}_{m}$. A perfect NCI is $0$ while positive (negative) NCI represents optimism (pessimism).

In the following, we consider NCI as another performance metric for comparing I-UKF's performance with I-EKF in Sections~\ref{subsec:FM with UKF CKF QKF} and \ref{subsec:reentry UKF}. On the other hand, in Sections~\ref{subsec:tracking CKF} and \ref{subsec:lorenz QKF CQKF}, we compare the time complexity of different inverse SPKFs. For brevity, we do not include the NCI and time complexity comparisons in all experiments. Recall from Section~\ref{sec:contributions} that, in general, the estimation performance of different non-linear filtering approaches also depends on the system models. This principle also extends to the non-linear inverse filters. However, in contrast to the forward filter, the inverse filter utilizes its perfect knowledge of the defender's true state, in addition to observations $\{\mathbf{a}_{j}\}_{1\leq j\leq k}$. Also, the inverse filter's process noise is non-additive even when the noise terms $\mathbf{w}_{k}$, $\mathbf{v}_{k}$ and $\bm{\epsilon}_{k}$ are assumed to be additive and Gaussian.

\subsection{FM demodulation with I-UKF, I-CKF and I-QKF}\label{subsec:FM with UKF CKF QKF}
Consider the FM demodulator system of Section~\ref{subsec:FM with EKF SOEKF GSEKF}. For this system, EKF is observed to be more accurate than UKF in the forward filtering case. Here, we first compare I-UKF's and I-EKF's performance using RCRLB and NCI. We then compare I-CKF and I-QKF's performance for this system with I-EKF and I-UKF, including different mismatched forward filter cases. For forward and inverse UKF, $\kappa$ and $\overline{\kappa}$ both were set to $1$, but I-UKF assumed the forward UKF's $\kappa$ to be $2$. Similarly, we considered a $3$-point forward QKF and a $3$-point I-QKF, but I-QKF assumed forward QKF's $m$ to be $5$. Other parameters and initial estimates were as in Section~\ref{subsec:FM with EKF SOEKF GSEKF}.

\begin{figure}
  \centering
  \includegraphics[width = 0.7\columnwidth]{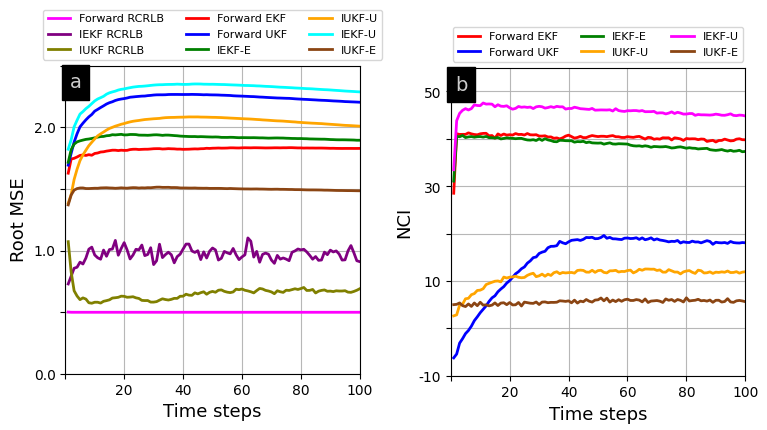}
  \caption{(a) Time-averaged RMSE and RCRLB, and (b) NCI for forward and inverse UKF for FM demodulator system.}
 \label{fig:fmdemod iukf}
\end{figure}

\textit{I-UKF and I-EKF:} Fig.~\ref{fig:fmdemod iukf} shows the AMSE, RCRLB, and NCI for state estimation for forward and inverse UKF and EKF averaged over 500 runs.  In the mismatched forward filter case, IUKF-E in Fig.~\ref{fig:fmdemod iukf}a denotes the I-UKF's estimation error which assumes a forward UKF when the true forward filter is EKF. The other notations in Fig.~\ref{fig:fmdemod iukf} and also, in further experiments are similarly defined. From Fig.~\ref{fig:fmdemod iukf}, we observe that the forward EKF has lower estimation error but higher NCI than forward UKF. Hence, with correct forward filter assumption, IUKF-U has a higher error than IEKF-E. However, IUKF-E outperforms I-EKF even with incorrect forward filter assumption. On the other hand, incorrect forward filter assumption degrades I-EKF's performance, i.e., IEKF-U has lower estimation accuracy and higher NCI than IEKF-E. While all filters considered here are optimistic, I-UKF is the most credible filter. I-UKF also outperforms forward UKF in terms of credibility because it uses additional true state $\mathbf{x}_{k}$ information. Note that even though I-UKF assumes forward UKF's $\kappa$ to be different from its true value, IUKF-U performs better than the forward UKF.

\begin{figure}
  \centering
  \includegraphics[width = 1.0\textwidth]{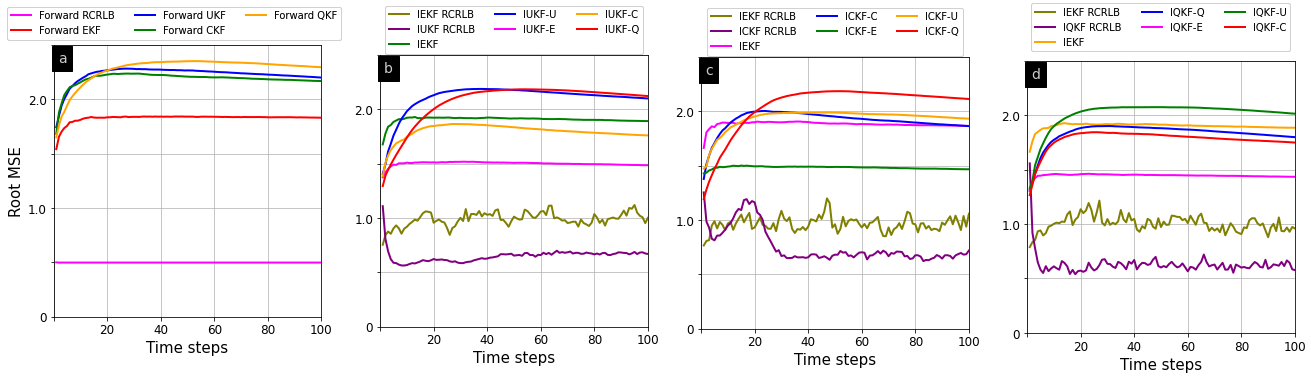}
  \caption{Time-averaged RMSE for (a) forward filters, (b) I-UKFs, (c) I-CKFs, and (d) I-QKFs for FM demodulator system.}
 \label{fig:fmdemod ckf qkf}
\end{figure}

\textit{I-CKF and I-QKF:} Fig.~\ref{fig:fmdemod ckf qkf}a-d shows the AMSE and RCRLB for state estimation for forward and inverse EKF, UKF, CKF, and QKF, including mismatched forward filter case. From Fig.~\ref{fig:fmdemod ckf qkf}a, we observe that the forward EKF has the lowest error while forward QKF performs worse than all other forward filters. Interestingly, forward UKF and CKF have similar accuracy, but IUKF-C has significantly lower errors than IUKF-U. Unlike I-UKF, in Fig.~\ref{fig:fmdemod ckf qkf}c, both ICKF-C, and I-EKF have similar performance with correct forward filter assumption. Contrarily, IQKF-Q in Fig.~\ref{fig:fmdemod ckf qkf}d performs slightly better than I-EKF, but with higher computational efforts. However, both ICKF-E and IQKF-E, respectively, in Fig.~\ref{fig:fmdemod ckf qkf}c and \ref{fig:fmdemod ckf qkf}d, again outperform I-EKF. Since forward EKF estimates its state most accurately, we observe that IUKF-E, ICKF-E, and IQKF-E have the lowest estimation error even without true forward filter information.

\subsection{Vehicle reentry with I-UKF}\label{subsec:reentry UKF}
Consider a radar tracking a vehicle's reentry using range and bearing measurements, widely used to illustrate UKF's performance \cite{julier2004unscented,sarkka2007unscented}. Here, we compare I-UKF's performance with I-EKF's under both correct and incorrect forward filter assumptions using RCRLB and NCI as performance metrics. We denote the position of the vehicle at $k$-th time instant as $[\mathbf{x}_{k}]_{1}$ and $[\mathbf{x}_{k}]_{2}$, its velocity as $[\mathbf{x}_{k}]_{3}$ and $[\mathbf{x}_{k}]_{4}$, and its constant aerodynamic parameter as $[\mathbf{x}_{k}]_{5}$. The vehicle's state continuous-time evolution follows $[\dot{\mathbf{x}}_{k}]_{1}=[\mathbf{x}_{k}]_{3}$, $[\dot{\mathbf{x}}_{k}]_{2}=[\mathbf{x}_{k}]_{4}$, $[\dot{\mathbf{x}}_{k}]_{3}=d_{k}[\mathbf{x}_{k}]_{3}+g_{k}[\mathbf{x}_{k}]_{1}+w_{1}$, $[\dot{\mathbf{x}}_{k}]_{4}=d_{k}[\mathbf{x}_{k}]_{4}+g_{k}[\mathbf{x}_{k}]_{2}+w_{2}$, $[\dot{\mathbf{x}}_{k}]_{5}=w_{3}$, where $[\dot{\mathbf{x}}_{k}]_{i}$ is the first-order partial derivative of $[\mathbf{x}_{k}]_{i}$ with respect to time, and $w_{1}, w_{2}$ and $w_{3}$ represent process noise. We consider the discretized version of this system with a time step of $0.1$ sec in our experiment. The quantities $d_{k}=\beta_{k}\exp{(\left(\rho_{0}-\rho_{k})/h_{0}\right)}V_{k}$ and $g_{k}=-Gm_{0}\rho_{k}^{-3}$ where $\beta_{k}=\beta_{0}\exp{([\mathbf{x}_{k}]_{5})}$, $V_{k}=\sqrt{[\mathbf{x}_{k}]_{3}^{2}+[\mathbf{x}_{k}]_{4}^{2}}$ and $\rho_{k}=\sqrt{[\mathbf{x}_{k}]_{1}^{2}+[\mathbf{x}_{k}]_{2}^{2}}$ with $\rho_{0}$, $h_{0}$, $G$, $m_{0}$ and $\beta_{0}$ as constants. The radar's range and bearing measurements are $[\mathbf{y}_{k}]_{1}=\sqrt{([\mathbf{x}_{k}]_{1}-\rho_{0})^{2}+[\mathbf{x}_{k}]_{2}^{2}}+v_{1}$, and $[\mathbf{y}_{k}]_{2}=\tan^{-1}{\left(\frac{[\mathbf{x}_{k}]_{2}}{[\mathbf{x}_{k}]_{1}-\rho_{0}}\right)}+v_{2}$, where $v_{1}$ and $v_{2}$ represent measurement noises \cite{julier2004unscented}. 

For the inverse filter, we consider a linear observation $\mathbf{a}_{k}=\left[[\hat{\mathbf{x}}_{k}]_{1},[\hat{\mathbf{x}}_{k}]_{2}\right]^{T}+\bm{\epsilon}_{k}$, where $\bm{\epsilon}_{k}\sim\mathcal{N}(\mathbf{0},3\mathbf{I}_{2})$. The initial state was $\mathbf{x}_{0}=[6500.4,349.14,-1.8093,-6.7967,0.6932]^{T}$. The initial state estimate $\Hat{\Hat{\mathbf{x}}}_{k}$ for I-UKF was set to actual $\mathbf{x}_{0}$ with initial covariance estimate $\overline{\bm{\Sigma}}_{0}=diag(10^{-5},10^{-5},10^{-5},10^{-5},1)$. For forward UKF, $\kappa$ was chosen as $2.5$ such that the weight for $0$-th sigma point at $\hat{\mathbf{x}}_{k}$ is $1/3$, and all other sigma points have equal weights. Similarly, $\overline{\kappa}$ of I-UKF was set to $3.5$. Here, we assumed that the forward UKF's $\kappa$ was perfectly known to I-UKF. All other system parameters and initial estimates were identical to \cite{julier2004unscented}.

\begin{figure}
  \centering
  \includegraphics[width = 0.7\columnwidth]{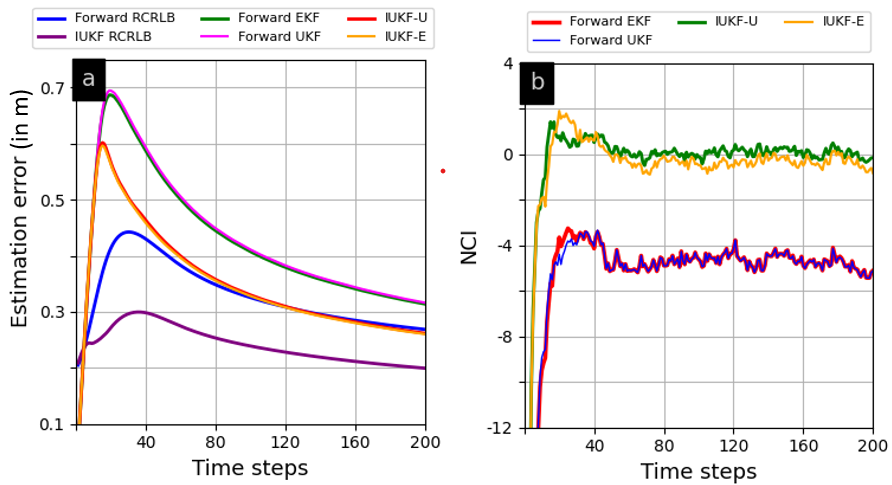}
  \caption{(a) Time-averaged estimation error and RCRLB, and (b) NCI for forward and inverse UKF for vehicle reentry system averaged over $100$ runs.}
 \label{fig:reentry iukf}
\end{figure}

Fig. \ref{fig:reentry iukf} shows the (root) time-averaged error in position estimation, its RCRLB (also, time-averaged) and NCI for forward and inverse UKF (IUKF-U), including forward EKF and mismatched IUKF-E case. Here, the RCRLB is computed as $\sqrt{[\mathbf{J}^{-1}]_{1,1}+[\mathbf{J}^{-1}]_{2,2}}$. The I-UKF's error are observed to be lower than that of forward UKF, as is the case with their corresponding RCRLBs. I-UKF's NCI is approximately $0$ (perfect NCI) while forward UKF and EKF are pessimistic. Further, incorrect forward filter assumption (IUKF-E case) does not affect the I-UKF's estimation because forward UKF and EKF have similar performances. For the vehicle re-entry example, I-EKF's error and NCI were similar to I-UKF and hence, omitted in Fig. \ref{fig:reentry iukf}.

\subsection{Target tracking with I-CKF}\label{subsec:tracking CKF}
Consider a target maneuvering at an unknown constant turn rate $\Omega$ in a horizontal plane with a fixed radar tracking its trajectory with range and bearing measurements using CKF\cite{arasaratnam2009cubature}. Denote the target's state at $k$-th time instant as $\mathbf{x}_{k}=[p^{x}_{k},v^{x}_{k},p^{y}_{k},v^{y}_{k},\Omega]^{T}$ where $p^{x}_{k}$ and $p^{y}_{k}$ are the positions, and $v^{x}_{k}$ and $v^{y}_{k}$ are the velocities in $x$ and $y$ directions, respectively. The non-linear system model is\cite{bar2004estimation} 
\par\noindent\small
\begin{align*}
    \mathbf{x}_{k+1}&=\begin{bsmallmatrix}
        1&\sin{(\Omega T)}/\Omega& 0 &-(1-\cos{(\Omega T)})/\Omega & 0\\
        0&\cos{(\Omega T)}& 0 &-\sin{(\Omega T)}& 0\\
        0&(1-\cos{(\Omega T)})/\Omega& 1 &\sin{(\Omega T)}/\Omega& 0\\
        0&\sin{(\Omega T)}& 0 &\cos{(\Omega T)}& 0\\
        0& 0& 0& 0& 1
    \end{bsmallmatrix}\mathbf{x}_{k}+\mathbf{w}_{k},\\
    \mathbf{y}_{k}&=\begin{bsmallmatrix}
        \sqrt{(p^{x}_{k})^{2}+(p^{y}_{k})^{2}}\\
        \tan^{-1}(p^{y}_{k}/p^{x}_{k})
    \end{bsmallmatrix}+\mathbf{v}_{k},\;\;\;\mathbf{a}_{k}=\begin{bmatrix}
        \sqrt{(\hat{p}^{x}_{k})^{2}+(\hat{p}^{y}_{k})^{2}}\\
        \tan^{-1}(\hat{p}^{y}_{k}/\hat{p}^{x}_{k})
    \end{bmatrix}+\bm{\epsilon}_{k},
\end{align*}
\normalsize
where we considered the inverse filter's observations similar to that of the forward filter. Note that we chose similar observation functions $h(\cdot)$ and $g(\cdot)$ only to compare the relative performance of the forward and inverse filters. Here, $\hat{p}^{x}_{k}$ and $\hat{p}^{y}_{k}$ are the forward filter's estimates of $p^{x}_{k}$ and $p^{y}_{k}$, respectively. The noise terms $\mathbf{w}_{k}\sim\mathcal{N}(\mathbf{0},\mathbf{Q})$, $\mathbf{v}_{k}\sim\mathcal{N}(\mathbf{0},\mathbf{R})$ and $\bm{\epsilon}_{k}\sim\mathcal{N}(\mathbf{0},\bm{\Sigma}_{\epsilon})$, with $\bm{\Sigma}_{\epsilon}=\mathbf{R}$. All other parameters, including the initial estimates and noise covariance matrices, were identical to \cite{arasaratnam2009cubature}. In particular, we set $\Omega=-3^{\circ}$ $\textrm{s}^{-1}$, $T=1$ $\textrm{s}$, $\mathbf{Q}=\textrm{diag}(q_{1}\mathbf{M},q_{1}\mathbf{M},q_{2}T)$ and $\mathbf{R}=\textrm{diag}(\sigma_{r}^{2},\sigma_{\theta}^{2})$, where $q_{1}=0.1$ $\textrm{m}^{2}\textrm{s}^{-3}$, $q_{2}=1.75\times 10^{-4}$ $\textrm{s}^{-3}$, $\sigma_{r}=10$ $\textrm{m}$, $\sigma_{\theta}=\sqrt{10}$ $\textrm{mrad}$ and $\mathbf{M}=\begin{bsmallmatrix}T^{3}/3 & T^{2}/2\\ T^{2}/2 & T\end{bsmallmatrix}$. Also, $\mathbf{x}_{0}=\doublehat{\mathbf{x}}_{0}=[1000$ $\textrm{m},300$ $\textrm{ms}^{-1},1000$ $\textrm{m},0$ $\textrm{ms}^{-1},-3^{\circ}$ $\textrm{s}^{-1}]^{T}$ while $\hat{\mathbf{x}}_{0}\sim\mathcal{N}(\mathbf{x}_{0},\bm{\Sigma}_{0})$ with $\bm{\Sigma}_{0}=\overline{\bm{\Sigma}}_{0}=\textrm{diag}(100$ $\textrm{m}^{2},10$ $\textrm{m}^{2}\textrm{s}^{-2},100$ $\textrm{m}^{2},10$ $\textrm{m}^{2}\textrm{s}^{-2},100$ $\textrm{mrad}^{2}\textrm{s}^{-2})$. We compared I-CKF's performance with I-UKF and set both forward UKF's $\kappa$ and I-UKF's $\overline{\kappa}$ to $1$.

\begin{figure}
  \centering
  \includegraphics[width = 1.0\textwidth]{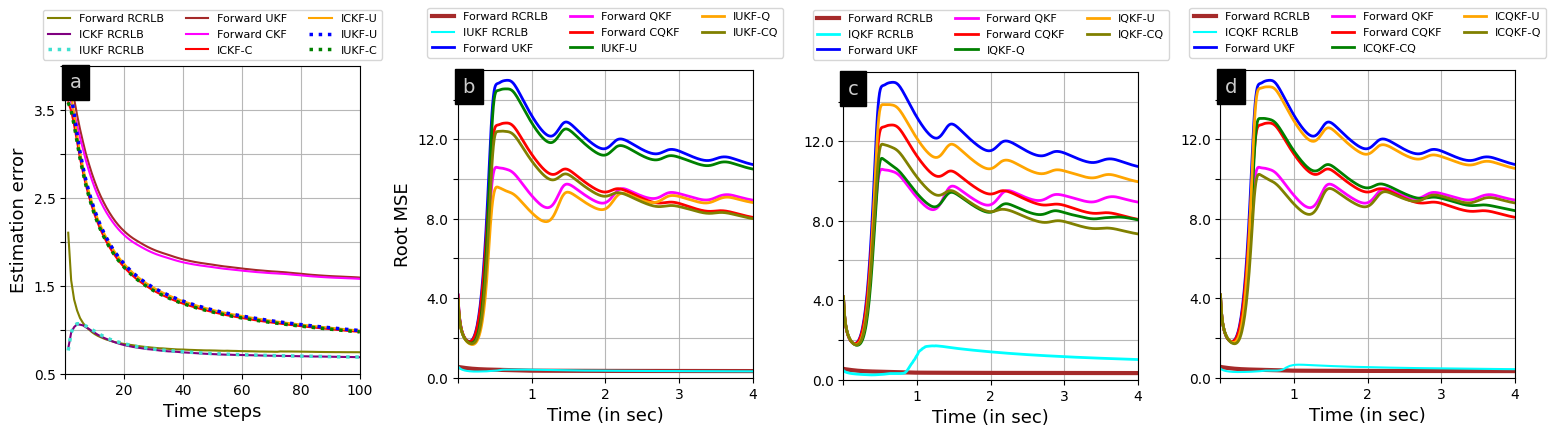}
  \caption{Time-averaged estimation error and RCRLB of forward and inverse (a) CKFs for target tracking system (averaged over $250$ runs); (b) UKFs, (c) QKFs, and (d) CQKFs for Lorenz system (averaged over $50$ runs).}
 \label{fig:tracking lorenz}
\end{figure}

Fig.~\ref{fig:tracking lorenz}a shows the time-averaged RMSE in velocity estimation and its RCRLB (also, time-averaged) for a system that employs forward and inverse CKF and UKF. The RCRLB is computed as $\sqrt{[\mathbf{J}^{-1}]_{2,2}+[\mathbf{J}^{-1}]_{4,4}}$, where $\mathbf{J}$ is the corresponding information matrix. We observe that forward CKF and UKF have similar estimation accuracy. Hence, both ICKF-C and ICKF-U yield similar estimation errors regardless of the actual forward filter. Although the forward and inverse filters have similar RCRLBs, the difference between the estimation error and RCRLB for I-CKF is less than that for the forward CKF. Hence, I-CKF outperforms forward CKF in terms of achieving the lower bound. For the considered system, I-UKF's error and RCRLB are similar to that of I-CKF.

Table~\ref{tbl:CKF target tracking} lists the total run time for $100$ time steps (in one Monte-Carlo run) of different filters for the target tracking system. Forward CKF and UKF have similar complexity, while all inverse filters require a longer run time than forward filters. Besides increased state dimension, the computational efforts of the inverse filters are also increased because of the computation of approximation $\bm{\Sigma}^{*}_{k}$ of the forward filter's $\bm{\Sigma}_{k}$ at each time-step. However, assuming a forward CKF and employing I-CKF is slightly less expensive than I-UKF. This reduction is observed because I-CKF processes $2n_{z}$ cubature points while I-UKF considers $2n_{z}+1$ sigma points.
    \begin{table}
    \caption{Run time (in seconds) of different filters for target tracking system.}
    \label{tbl:CKF target tracking}
    \centering
    \begin{tabular}{p{3.0cm}p{2.0cm}}
    \hline\noalign{\smallskip}   Filter & Time\\
    \noalign{\smallskip}
    \hline
    \noalign{\smallskip}
    Forward UKF & 0.0179\\
    Forward CKF & 0.0188\\
    IUKF-U & 0.1388\\
    IUKF-C & 0.1214\\
    ICKF-U & 0.0902\\
    ICKF-C & 0.1137\\
    \noalign{\smallskip}
    \hline\noalign{\smallskip}
    \end{tabular}
    \end{table}

\subsection{Lorenz system with I-QKF and I-CQKF}\label{subsec:lorenz QKF CQKF}
Consider the following $3$-dimensional system derived from the Lorenz stochastic differential system \cite{ito2000gaussian}: \par\noindent\small
\begin{align*}
&\mathbf{x}_{k+1}=\begin{bsmallmatrix}
    [\mathbf{x}_{k}]_{1}+\Delta tr_{1}(-[\mathbf{x}_{k}]_{1}+[\mathbf{x}_{k}]_{2})\\
    [\mathbf{x}_{k}]_{2}+\Delta t(r_{2}[\mathbf{x}_{k}]_{1}-[\mathbf{x}_{k}]_{2}-[\mathbf{x}_{k}]_{1}[\mathbf{x}_{k}]_{3})\\
    [\mathbf{x}_{k}]_{3}+\Delta t(-r_{3}[\mathbf{x}_{k}]_{3}+[\mathbf{x}_{k}]_{1}[\mathbf{x}_{k}]_{2})
\end{bsmallmatrix}+\begin{bsmallmatrix}
    0\\0\\0.5
\end{bsmallmatrix}w_{k},\\
& y_{k}=\Delta t\sqrt{([\mathbf{x}_{k}]_{1}-0.5)^{2}+[\mathbf{x}_{k}]_{2}^{2}+[\mathbf{x}_{k}]_{3}^{2}}+0.065v_{k},\\
& a_{k}=\Delta t\sqrt{[\hat{\mathbf{x}}_{k}]_{1}^{2}+([\hat{\mathbf{x}}_{k}]_{2}-0.5)^{2}+[\hat{\mathbf{x}}_{k}]_{3}^{2}}+0.1\epsilon_{k},
\end{align*}
\normalsize
where $w_{k}, v_{k}, \epsilon_{k}\sim\mathcal{N}(0,\Delta t)$ with parameters $\Delta t=0.01$, $r_{1}=10$, $r_{2}=28$ and $r_{3}=8/3$ such that the system has three unstable equilibria. The observations $\mathbf{a}_{k}$ are of the same form as $\mathbf{y}_{k}$. The attacker employs a $5$-point forward QKF and a second-order forward CQKF, while the defender considers a $3$-point I-QKF assuming the forward QKF's $m=3$ and a second-order I-CQKF. The inverse filters' accuracy is compared with I-UKF with $\overline{\kappa}=2$ and a forward UKF with $\kappa=1.5$. The initial state $\mathbf{x}_{0}$ and the inverse filters' estimate $\doublehat{\mathbf{x}}_{0}$ were all set to $[-0.2,-0.3,-0.5]^{T}$. The forward filters' estimate $\hat{\mathbf{x}}_{0}$ were chosen as $[1.35,-3,6]^{T}$. All the initial covariance estimate were set to $0.35\mathbf{I}$ with $\mathbf{J}_{0}=\bm{\Sigma}_{0}^{-1}$ and $\overline{\mathbf{J}}_{0}=\overline{\bm{\Sigma}}_{0}^{-1}$.
    \begin{table}
    \caption{Run time (in seconds) of different filters for Lorenz system.}
    \label{tbl:QKF lorenz}
    \centering
    \begin{tabular}{p{3.5cm}p{2.0cm}}
    \hline\noalign{\smallskip}
    Filter & Time\\
    \noalign{\smallskip}
    \hline
    \noalign{\smallskip}
    Forward UKF & 0.0541\\
    Forward QKF & 0.6026\\
    Forward CQKF & 0.0742\\
    IUKF-U & 0.4438\\
    IUKF-Q & 0.3904\\
    IUKF-CQ & 0.3921\\
    IQKF-U & 11.2705\\
    IQKF-Q & 11.3383\\
    IQKF-CQ & 11.2561\\
    ICQKF-U & 1.0726\\
    ICQKF-Q & 1.0782\\
    ICQKF-CQ & 1.106\\
    \noalign{\smallskip}
    \hline\noalign{\smallskip}
    \end{tabular}
    \end{table}

Fig.~\ref{fig:tracking lorenz}b-d shows the AMSE and RCRLB for state estimation for forward and inverse UKF, QKF, and CQKF, including the mismatched inverse filter cases. The RCRLB for state estimation is $\sqrt{\textrm{Tr}(\mathbf{J}^{-1})}$, where $\mathbf{J}$ is the corresponding information matrix. For the Lorenz system, forward QKF and CQKF estimate the state more accurately than forward UKF. Regardless of the forward filter assumption, I-UKF has a similar performance as the corresponding forward filter. For instance, IUKF-U and forward UKF have similar estimation errors. On the other hand, from Fig.~\ref{fig:tracking lorenz}c, we observe that I-QKF outperforms the forward filters in all cases, i.e., IQKF-Q, IQKF-U and IQKF-CQ have lower errors than forward QKF, UKF and CQKF, respectively. Contrarily, in Fig.~\ref{fig:tracking lorenz}d, ICQKFs closely follow the corresponding forward filters' errors. Interestingly, for the considered system, I-UKF's and I-CQKF's RCRLB is the same as that for the forward filters, which is slightly less than that for I-QKF. In spite of this, I-QKF has higher estimation accuracy than I-UKF, I-CQKF, and the forward filters, but with higher computational efforts.

Table~\ref{tbl:QKF lorenz} lists different filters' run time for the Lorenz system. Note that $\{\bm{\zeta}_{i},\omega_{i}\}$ ($\{\bm{\xi}_{i},\omega_{i}\}$) and $\{\overline{\bm{\zeta}}_{j},\overline{\omega}_{j}\}$ ($\{\overline{\bm{\xi}}_{i},\overline{\omega}_{i}\}$) in forward QKF (CQKF) and I-QKF (I-CQKF), respectively, are computed offline and not included in their run times. As expected, forward QKF is computationally more expensive than forward UKF, while forward CQKF is comparable to the latter. Again, inverse filters have longer run times than the corresponding forward filters. However, I-QKF's increase in run time is much more than that for I-UKF and I-CQKF. As noted earlier, QKF's complexity increases exponentially with the state dimension, which is more significant in I-QKF. Note that I-CQKF's run time is longer than I-UKF's but significantly lower than I-QKF's.

\section{Concluding remarks}\label{sec:ISPKF conclusions}
In this chapter, we developed derivative-free sigma-points-based inverse non-linear filters for cases with perfect system model information. Specifically, we utilized the unscented transform in I-UKF to generate deterministic sigma points, while I-CKF, I-QKF, and I-CQKF leveraged the cubature and quadrature integration rules for the same purpose. Our methods can also be adapted for scenarios involving non-Gaussian noises, continuous-time state evolution, and complex-valued systems. Theoretically, we guaranteed that if the attacker's forward UKF (CKF) is stable, then the I-UKF (I-CKF) is also stable under mild conditions. These developed filters are conservative, provided that the initial estimate pair is conservative. Although these inverse filters assume the attacker's forward filter to be known, our numerical results suggested that these filters provide reasonable estimates, even while incorrectly assuming the form of the forward filter. In fact, for the FM demodulator system, these filters outperformed the I-EKF even with correct forward filter information. I-UKF also surpassed forward UKF because of the perfect actual state information. Similarly, in some systems, I-CKF and I-QKF were observed to outperform I-UKF. However, I-QKF is computationally expensive, while I-CQKF provides reasonable estimates with lower computational costs.

\chapter{RKHS-based KFs for unknown system dynamics}
\label{chap:rkhs}
In Chapters~\ref{chap:inverse EKFs}-\ref{chap:inverse SPKFs}, we formulated the inverse filters assuming that both the attacker and defender had perfect information, including a known forward filter at the defender. However, in practice, the defender's state transition function $f(\cdot)$ of \eqref{eqn:non-linear state x} may not be known to the attacker, and similarly, the defender may not know the attacker's specific forward filter and action computation function $g(\cdot)$ from \eqref{eqn:non-linear observation a}. In this chapter, we introduce recursive RKHS-EKF and RKHS-UKF methods to simultaneously estimate the desired state and learn the unknown system model. We derive the stochastic stability conditions for these RKHS-based filters in Section~\ref{sec:rkhs stability} and compare their estimation efficiency with EKFs and UKFs (having perfect system model information) through various numerical experiments in Section~\ref{sec:rkhs numericals}. We conclude our findings in Section~\ref{sec:rkhs conclusions}. 

We consider a general stochastic filtering agent that lacks information about the underlying state evolution and observation models and adopt the iterative EM framework from \cite{wen2012data} (for linear Gaussian systems) for non-linear parameter learning in our RKHS-based EKF and UKF. These RKHS-based filters can be employed by both attacker and defender to infer, respectively, the defender's state (as a forward filter) and the attacker's state estimate (as an inverse filter). Similar to EKF, the RKHS-EKF's performance degrades due to the linearization of non-linear functions. To address this, our RKHS-UKF uses the unscented transform to approximate non-linear expectations instead of Taylor series linearization. Additionally, our RKHS-based KFs differ fundamentally from previously proposed kernel-based KFs (KKFs) for unknown non-linear time series prediction \cite{ralaivola2005time,zhu2013learning,dang2019kernel}; as discussed further Remark~\ref{remark:kernel KF}.

\section{Filter formulations}\label{sec:RKHS formulation}
Consider the state transition \eqref{eqn:non-linear state x} and observation \eqref{eqn:non-linear observation y} with the functions $f(\cdot)$ and $h(\cdot)$ as well as the noise covariances $\mathbf{Q}$ and $\mathbf{R}$ being unknown. We represent the unknown non-linear functions using an RKHS-based function approximation\cite{aronszajn1950theory}. From the representer theorem \cite{scholkopf2001generalized}, the optimal approximation of a function $s(\cdot):\mathbb{R}^{n}\to\mathbb{R}$ in the RKHS induced by a kernel $K(\cdot,\cdot):\mathbb{R}^{n}\times\mathbb{R}^{n}\to\mathbb{R}$  with respect to an arbitrary loss function takes the form $s(\cdot)\approx\sum_{i=1}^{M}a_{i}K(\widetilde{\mathbf{x}}_{i},\cdot)$
where $\{\widetilde{\mathbf{x}}_{i}\}_{1\leq i\leq M}$ are the $M$ input training samples or dictionary, and $\{a_{i}\}_{1\leq i\leq M}$ are the unknown coefficients to be learned. This kernel function approximation has been widely used for non-linear state-space modeling\cite{tobar2015unsupervised,ralaivola2003dynamical} and recursive least-squares algorithms with unknown non-linear functions\cite{engel2004kernel,liu2009extended,van2006sliding}. Often, a Gaussian kernel such as $K(\mathbf{x}_{i},\mathbf{x}_{j})=\exp{\left(-\frac{\|\mathbf{x}_{i}-\mathbf{x}_{j}\|^{2}_{2}}{\sigma^2}\right)}$, with kernel width $\sigma>0$ controlling the smoothness of the approximation, is commonly used. It is a universal kernel \cite{steinwart2001influence}, i.e., its induced RKHS is dense in the space of continuous functions. 

\textit{System models for uncertain dynamics:} Consider a kernel function $K(\cdot,\cdot)$ and a dictionary $\{\widetilde{\mathbf{x}}_{l}\}_{1\leq l\leq L}$ of size $L$. Denote $\bm{\Phi}(\mathbf{x})=\begin{bmatrix}K(\widetilde{\mathbf{x}}_{1},\mathbf{x})&\hdots&K(\widetilde{\mathbf{x}}_{L},\mathbf{x})\end{bmatrix}^{T}$. Using the kernel function approximation, state transition \eqref{eqn:non-linear state x} and observation \eqref{eqn:non-linear observation y} are approximated, respectively, as
\par\noindent\small
\begin{align}
    &\mathbf{x}_{k+1}=\mathbf{A}\bm{\Phi}(\mathbf{x}_{k})+\mathbf{w}_{k},\label{eqn:RKHS state transition approx}\\
    &\mathbf{y}_{k}=\mathbf{B}\bm{\Phi}(\mathbf{x}_{k})+\mathbf{v}_{k},\label{eqn:RKHS observation approx}
\end{align}
\normalsize
where the coefficient matrices $\mathbf{A}\in\mathbb{R}^{n_{x}\times L}$ and $\mathbf{B}\in\mathbb{R}^{n_{y}\times L}$ consists of the unknown mixing parameters to be learned. The dictionary $\{\widetilde{\mathbf{x}}_{l}\}_{1\leq l\leq L}$ can be formed using a sliding window\cite{van2006sliding} or approximate linear dependence (ALD)\cite{engel2004kernel} criterion. Besides the state estimate $\hat{\mathbf{x}}_{k}$, an RKHS-based filter needs to estimate the unknown parameters $\Theta=\{\mathbf{A},\mathbf{B},\mathbf{Q},\mathbf{R}\}$ based on the observations $\{\mathbf{y}_{i}\}_{1\leq i\leq k}$ upto $k$-th time step. Note that the system model \eqref{eqn:non-linear state x}-\eqref{eqn:non-linear observation y} and hence, system parameters $\Theta$ are not time-varying. For parameter learning, we derive an online approximate EM algorithm, which in turn is coupled with EKF/UKF recursions to obtain the state estimates. The EM algorithm\cite{hajek2015random} is widely used to compute maximum likelihood estimates in the presence of missing data. Since our RKHS-based filters learn their state transition based on the available observations themselves, we do not require any prior forward filter information to employ them as the defender's inverse filter. Our RKHS-EKF/UKF can also be trivially simplified if the agent has perfect prior information about the state evolution, observation function, and/or noise covariance matrices.

\subsection{Online approximate EM}\label{subsec:EM}
Given $\Theta$, the state estimate $\hat{\mathbf{x}}_{k}$ can be computed using EKF/UKF-based recursions. To estimate the unknown parameters $\Theta$, we consider the EM algorithm. Consider the states upto time $k$ as $\mathbf{X}^{k}=\{\mathbf{x}_{j}\}_{0\leq j\leq k}$ and the corresponding observations $\mathbf{Y}^{k}=\{\mathbf{y}_{j}\}_{1\leq j\leq k}$. The joint conditional probability density given the parameters $\Theta$ is
\par\noindent\small
\begin{align}
    p(\mathbf{X}^{k},\mathbf{Y}^{k}|\Theta)=p(\mathbf{x}_{0})\prod_{j=1}^{k}p(\mathbf{x}_{j}|\mathbf{x}_{j-1},\Theta)\prod_{j=1}^{k}p(\mathbf{y}_{j}|\mathbf{x}_{j},\Theta).\label{eqn:RKHS joint density}
\end{align}
\normalsize
Under the Gaussian noise assumption, the conditional probability densities are
\par\noindent\small
\begin{align}
    &p(\mathbf{x}_{j}|\mathbf{x}_{j-1},\Theta)=\gamma(\mathbf{x}_{j}-\mathbf{A}\bm{\Phi}(\mathbf{x}_{j-1}),\mathbf{Q}),\label{eqn:RKHS conditional density x}\\
    &p(\mathbf{y}_{j}|\mathbf{x}_{j},\Theta)=\gamma(\mathbf{y}_{j}-\mathbf{B}\bm{\Phi}(\mathbf{x}_{j}),\mathbf{R}),\label{eqn:RKHS conditional density y}
\end{align}
\normalsize
with the Gaussian probability density function $\gamma(\mathbf{x}-\bm{\mu},\bm{\Sigma})\doteq(2\pi)^{-d/2}\textrm{det}(\bm{\Sigma})^{-1/2}\\\times\exp{\left\{-\frac{1}{2}(\mathbf{x}-\bm{\mu})^{T}\bm{\Sigma}^{-1}(\mathbf{x}-\bm{\mu})\right\}}$ where $d$ is dimension of vector $\mathbf{x}$. Assume $\mathbf{x}_{0}\sim\mathcal{N}(\hat{\mathbf{x}}_{0},\bm{\Sigma}_{0})$. Note that this assumption is also needed to initialize the EKF/UKF recursions. Using this in \eqref{eqn:RKHS joint density} along with \eqref{eqn:RKHS conditional density x} and \eqref{eqn:RKHS conditional density y}, we have
\par\noindent\small
\begin{align}
    \log p(\mathbf{X}^{k},\mathbf{Y}^{k}|\Theta)&=-\frac{1}{2}\log |\bm{\Sigma}_{0}|-\frac{1}{2}(\mathbf{x}_{0}-\hat{\mathbf{x}}_{0})^{T}\bm{\Sigma}_{0}^{-1}(\mathbf{x}_{0}-\hat{\mathbf{x}}_{0})\nonumber\\
    &+\sum_{j=1}^{k}\left(-\frac{1}{2}\log |\mathbf{Q}|-\frac{1}{2}(\mathbf{x}_{j}-\mathbf{A}\bm{\Phi}(\mathbf{x}_{j-1}))^{T}\mathbf{Q}^{-1}(\mathbf{x}_{j}-\mathbf{A}\bm{\Phi}(\mathbf{x}_{j-1}))\right)\nonumber\\
    &+\sum_{j=1}^{k}\left(-\frac{1}{2}\log |\mathbf{R}|-\frac{1}{2}(\mathbf{y}_{j}-\mathbf{B}\bm{\Phi}(\mathbf{x}_{j}))^{T}\mathbf{R}^{-1}(\mathbf{y}_{j}-\mathbf{B}\bm{\Phi}(\mathbf{x}_{j}))\right)+\alpha,\label{eqn:RKHS log all terms}
\end{align}
\normalsize
where $\alpha$ denotes the constant terms which do not affect the maximization.

A basic EM algorithm to estimate $\Theta$ based on the observations $\mathbf{Y}^{k}$ consists of following two steps that are iterated a fixed number of times or until convergence: \textit{a) E-step:} Given the estimate $\hat{\Theta}$ of the unknown parameters $\Theta$, we compute $Q(\Theta,\hat{\Theta})=\mathbb{E}[\log p(\mathbf{X}^{k},\mathbf{Y}^{k}|\Theta)|\mathbf{Y}^{k},\hat{\Theta}]$, and \textit{b) M-step:} The updated estimate $\hat{\Theta}^{(new)}=\textrm{arg max}_{\Theta} Q(\Theta,\hat{\Theta})$. In our RKHS-based filters, we adopt an approximate online version of the EM algorithm such that all the available observations need not be processed at each time step. Denote $\hat{\Theta}_{k}=\{\hat{\mathbf{A}}_{k},\hat{\mathbf{B}}_{k},$ $\hat{\mathbf{Q}}_{k},\hat{\mathbf{R}}_{k}\}$ as the estimate of $\Theta$ given observations upto time $k$, i.e., $\mathbf{Y}^{k}$. For online EM, at $k$-th time step, the current estimate $\hat{\Theta}$ is $\hat{\Theta}_{k-1}$ and the updated estimate $\hat{\Theta}^{(\textrm{new})}$ is $\hat{\Theta}_{k}$. Further, as we will describe in the E-step, the required conditional expectations can be approximated using the estimates from EKF/UKF recursions itself. Hence, we first consider the M-step. Consider the $k$-th time step such that $k$ observations are available to us up to this instant.

\textit{Parameter estimates (M-step):} For simplicity, denote the conditional expectation operator $\mathbb{E}[\cdot|\mathbf{Y}^{k},\hat{\Theta}_{k-1}]$ given $k$ observations by $\mathbb{E}_{k}[\cdot]$. Now, from \eqref{eqn:RKHS log all terms}, maximizing $\mathbb{E}_{k}[\log p(\mathbf{X}^{k},\mathbf{Y}^{k}|\Theta)]$ with respect to parameter $\mathbf{A}$, we obtain the estimate $\hat{\mathbf{A}}_{k}$ of $\mathbf{A}$ given $k$ observations as
\par\noindent\small
\begin{align}
    \hat{\mathbf{A}}_{k}=\left(\sum_{j=1}^{k}\mathbb{E}_{k}[\mathbf{x}_{j}\bm{\Phi}(\mathbf{x}_{j-1})^{T}]\right)\left(\sum_{j=1}^{k}\mathbb{E}_{k}[\bm{\Phi}(\mathbf{x}_{j-1})\bm{\Phi}(\mathbf{x}_{j-1})^{T}]\right)^{-1}.\label{eqn:RKHS A estimate}
\end{align}
\normalsize

Note that these expectations are computed using (and hence, functions of) the observations $\mathbf{Y}^{k}$ and current parameter estimate $\hat{\Theta}_{k-1}$ as described later in the E-step. However, this computation requires all $k$ observations to be processed together at time $k$, and the complexity increases as $k$ increases. To obtain approximate online estimate at low computations, we define the sum $\mathbf{S}^{x\phi}_{k}=\sum_{j=1}^{k}\mathbb{E}_{k}[\mathbf{x}_{j}\bm{\Phi}(\mathbf{x}_{j-1})^{T}]$ and $\mathbf{S}^{\phi 1}_{k}=\sum_{j=1}^{k}\mathbb{E}_{k}[\bm{\Phi}(\mathbf{x}_{j-1})\bm{\Phi}(\mathbf{x}_{j-1})^{T}]$ such that $\hat{\mathbf{A}}_{k}=\mathbf{S}^{x\phi}_{k}(\mathbf{S}^{\phi 1}_{k})^{-1}$ from \eqref{eqn:RKHS A estimate}. We approximate these sums as
\par\noindent\small
\begin{align}
    &\mathbf{S}^{x\phi}_{k}\approx\mathbf{S}^{x\phi}_{k-1}+\mathbb{E}_{k}[\mathbf{x}_{k}\bm{\Phi}(\mathbf{x}_{k-1})^{T}],\label{eqn:RKHS sum x phi}\\
    &\mathbf{S}^{\phi 1}_{k}\approx\mathbf{S}^{\phi 1}_{k-1}+\mathbb{E}_{k}[\bm{\Phi}(\mathbf{x}_{k-1})\bm{\Phi}(\mathbf{x}_{k-1})^{T}].\label{eqn:RKHS sum phi 1}
\end{align}
\normalsize
This is an approximation because we have not considered an updated parameter estimate $\hat{\Theta}_{k-1}=\{\hat{\mathbf{A}}_{k-1},\hat{\mathbf{B}}_{k-1},\hat{\mathbf{Q}}_{k-1},\hat{\mathbf{R}}_{k-1}\}$ (obtained using observations upto time `$k-1$') for computing the conditional expectations in $\mathbf{S}^{x\phi}_{k-1}$ and $\mathbf{S}^{\phi 1}_{k-1}$. The current parameter estimates are used only in computing expectations $\mathbb{E}_{k}[\mathbf{x}_{k}\bm{\Phi}(\mathbf{x}_{k-1})^{T}]$ and $\mathbb{E}_{k}[\bm{\Phi}(\mathbf{x}_{k-1})\bm{\Phi}(\mathbf{x}_{k-1})^{T}]$. Typically, in an EM algorithm, the E and M-steps are iterated a number of times considering the most recent parameter estimates in each E-step. In our RKHS-based filters, because of these approximations, the current observation is considered only once to obtain the current state estimates and update the parameter estimates.

Similarly, other approximate parameter updates (by approximating the summations as in \eqref{eqn:RKHS sum x phi} and \eqref{eqn:RKHS sum phi 1}) are obtained as
\par\noindent\small
\begin{align}
    \hat{\mathbf{Q}}_{k}&=\left(1-\frac{1}{k}\right)\hat{\mathbf{Q}}_{k-1}+\frac{1}{k}(\mathbb{E}_{k}[\mathbf{x}_{k}\mathbf{x}_{k}^{T}]-\hat{\mathbf{A}}_{k}\mathbb{E}_{k}[\bm{\Phi}(\mathbf{x}_{k-1})\mathbf{x}_{k}^{T}]-\mathbb{E}_{k}[\mathbf{x}_{k}\bm{\Phi}(\mathbf{x}_{k-1})^{T}]\hat{\mathbf{A}}_{k}^{T}\nonumber\\
    &+\hat{\mathbf{A}}_{k}\mathbb{E}_{k}[\bm{\Phi}(\mathbf{x}_{k-1})\bm{\Phi}(\mathbf{x}_{k-1})^{T}]\hat{\mathbf{A}}_{k}^{T}),\label{eqn:RKHS Q estimate}\\
    \hat{\mathbf{B}}_{k}&=\mathbf{S}^{y\phi}_{k}(\mathbf{S}^{\phi}_{k})^{-1},\label{eqn:RKHS B estimate}
\end{align}
\begin{align}
    \hat{\mathbf{R}}_{k}&=\left(1-\frac{1}{k}\right)\hat{\mathbf{R}}_{k-1}+\frac{1}{k}(\mathbb{E}_{k}[\mathbf{y}_{k}\mathbf{y}_{k}^{T}]-\hat{\mathbf{B}}_{k}\mathbb{E}_{k}[\bm{\Phi}(\mathbf{x}_{k})\mathbf{y}_{k}^{T}]-\mathbb{E}_{k}[\mathbf{y}_{k}\bm{\Phi}(\mathbf{x}_{k})^{T}]\hat{\mathbf{B}}_{k}^{T}\nonumber\\
    &+\hat{\mathbf{B}}_{k}\mathbb{E}_{k}[\bm{\Phi}(\mathbf{x}_{k})\bm{\Phi}(\mathbf{x}_{k})^{T}]\hat{\mathbf{B}}_{k}^{T}),\label{eqn:RKHS R estimate}
\end{align}
\normalsize
where the sums $\mathbf{S}^{y\phi}_{k}=\sum_{j=1}^{k}\mathbb{E}_{k}[\mathbf{y}_{j}\bm{\Phi}(\mathbf{x}_{j})^{T}]$ and $\mathbf{S}^{\phi}_{k}=\sum_{j=1}^{k}\mathbb{E}_{k}[\bm{\Phi}(\mathbf{x}_{j})\bm{\Phi}(\mathbf{x}_{j})^{T}]$ are introduced to obtain approximate online estimates and are evaluated as $\mathbf{S}^{y\phi}_{k}=\mathbf{S}^{y\phi}_{k-1}+\mathbb{E}_{k}[\mathbf{y}_{k}\bm{\Phi}(\mathbf{x}_{k})^{T}]$ and $\mathbf{S}^{\phi}_{k}=\mathbf{S}^{\phi}_{k-1}+\mathbb{E}_{k}[\bm{\Phi}(\mathbf{x}_{k})\bm{\Phi}(\mathbf{x}_{k})^{T}]$. Further, using \eqref{eqn:RKHS observation approx}, $\mathbb{E}_{k}[\mathbf{y}_{k}\mathbf{y}_{k}^{T}]=\hat{\mathbf{B}}_{k}\mathbb{E}_{k}[\bm{\Phi}(\mathbf{x}_{k})\bm{\Phi}(\mathbf{x}_{k})^{T}]\hat{\mathbf{B}}_{k}^{T}+\hat{\mathbf{R}}_{k-1}$ and $\mathbb{E}_{k}[\mathbf{y}_{k}\bm{\Phi}(\mathbf{x}_{k})^{T}]=\hat{\mathbf{B}}_{k}\mathbb{E}_{k}[\bm{\Phi}(\mathbf{x}_{k})\bm{\Phi}(\mathbf{x}_{k})^{T}]$. For the stability analysis in Section~\ref{sec:rkhs stability}, we also introduce optional projection operations $\Gamma_{a}(\cdot)$ and $\Gamma_{b}(\cdot)$, respectively, to project the coefficient matrices estimates to satisfy known bounds such that \eqref{eqn:RKHS A estimate} and \eqref{eqn:RKHS B estimate} become
\par\noindent\small
\begin{align}
    &\hat{\mathbf{A}}_{k}=\Gamma_{a}(\mathbf{S}^{x\phi}_{k}(\mathbf{S}^{\phi 1}_{k})^{-1}),\label{eqn:RKHS A estimate with projection}\\
    &\hat{\mathbf{B}}_{k}=\Gamma_{b}(\mathbf{S}^{y\phi}_{k}(\mathbf{S}^{\phi}_{k})^{-1}).\label{eqn:RKHS B estimate with projection}
\end{align}
\normalsize

\textit{Expectation computations (E-step):} As mentioned earlier, we compute the required expectations in \eqref{eqn:RKHS sum x phi}-\eqref{eqn:RKHS R estimate} using the EKF/UKF estimates. A standard EKF/UKF recursion provides a Gaussian approximation of the conditional posterior distribution of the required states given all the observations available up to the current time instant. However, the expectations here involve a non-linear transformation $\bm{\Phi}(\cdot)$. In EKF, first-order Taylor series expansion approximates these non-linear expectations. On the other hand, based on the unscented transform, RKHS-UKF generates the deterministic sigma points to approximate these expectations. We also need the statistics of $\bm{\Phi}(\mathbf{x}_{k-1})$ given $\mathbf{Y}^{k}$. Hence, we consider an augmented state $\mathbf{z}_{k}=[\mathbf{x}_{k}^{T} \;\mathbf{x}_{k-1}^{T}]^{T}$ to obtain a smoothed estimate $\hat{\mathbf{x}}_{k-1|k}$ of the previous state $\mathbf{x}_{k-1}$ given $\mathbf{Y}^{k}$. Using these approximations, we formulate the RKHS-EKF and RKHS-UKF to jointly compute estimates $\hat{\mathbf{x}}_{k}$ and $\hat{\Theta}_{k}$. The computation of the required expectations is detailed in the parameters update steps of RKHS-EKF and RKHS-UKF in Sections~\ref{subsec:RKHS-EKF} and \ref{subsec:RKHS-UKF}, respectively.

\subsection{RKHS-EKF}\label{subsec:RKHS-EKF}
In terms of the augmented state $\mathbf{z}_{k}$, the system model \eqref{eqn:RKHS state transition approx}-\eqref{eqn:RKHS observation approx} is
\par\noindent\small
\begin{align}
    &\mathbf{z}_{k}=\widetilde{f}(\mathbf{z}_{k-1})+\widetilde{\mathbf{w}}_{k-1}=\begin{bmatrix}
        \mathbf{A}\bm{\Phi}(\mathbf{x}_{k-1})\\\mathbf{x}_{k-1}
    \end{bmatrix}+\widetilde{\mathbf{w}}_{k-1},\label{eqn:RKHS-EKF state transition aug}\\
    &\mathbf{y}_{k}=\widetilde{h}(\mathbf{z}_{k})+\mathbf{v}_{k}=\mathbf{B}\bm{\Phi}(\mathbf{x}_{k})+\mathbf{v}_{k}.\label{eqn:RKHS-EKF observation aug}
\end{align}
\normalsize
The actual noise covariance matrix of $\widetilde{\mathbf{w}}_{k-1}=[\mathbf{w}_{k-1}^{T}\;\mathbf{0}_{1\times n_{x}}]^{T}$ is $\widetilde{\mathbf{Q}}=\begin{bmatrix}\mathbf{Q}&\mathbf{0}_{n_{x}\times n_{x}}\\\mathbf{0}_{n_{x}\times n_{x}}&\mathbf{0}_{n_{x}\times n_{x}}\end{bmatrix}$. At $k$-th time instant, we have from the previous recursion, an estimate $\hat{\mathbf{z}}_{k-1}=[\hat{\mathbf{x}}_{k-1|k-1}^{T}\; \hat{\mathbf{x}}_{k-2|k-1}^{T}]^{T}$ with the associated error covariance matrix $\bm{\Sigma}^{z}_{k-1}$, coefficient matrices estimates $\hat{\mathbf{A}}_{k-1}$ and $\hat{\mathbf{B}}_{k-1}$, and noise covariance matrices estimates $\hat{\mathbf{Q}}_{k-1}$ and $\hat{\mathbf{R}}_{k-1}$. We compute estimates $\hat{\mathbf{z}}_{k}$, $\hat{\mathbf{A}}_{k}$, $\hat{\mathbf{B}}_{k}$, $\hat{\mathbf{Q}}_{k}$ and $\hat{\mathbf{R}}_{k}$ based on the new observation $\mathbf{y}_{k}$ through the procedure summarized below.\\
\textit{1) Prediction:} Using $\mathbf{A}=\hat{\mathbf{A}}_{k-1}$ and $\widetilde{\mathbf{Q}}_{k-1}=\begin{bmatrix}\hat{\mathbf{Q}}_{k-1}&\mathbf{0}_{n_{x}\times n_{x}}\\\mathbf{0}_{n_{x}\times n_{x}}&\mathbf{0}_{n_{x}\times n_{x}}\end{bmatrix}$ in \eqref{eqn:RKHS-EKF state transition aug}, compute the predicted state and associated prediction error covariance matrix as
\par\noindent\small
\begin{align}
    \hat{\mathbf{z}}_{k|k-1}=\widetilde{f}(\hat{\mathbf{z}}_{k-1}),\;\;\;\bm{\Sigma}^{z}_{k|k-1}=\widetilde{\mathbf{F}}_{k-1}\bm{\Sigma}^{z}_{k-1}\widetilde{\mathbf{F}}_{k-1}^{T}+\widetilde{\mathbf{Q}}_{k-1},\label{eqn:RKHS-EKF prediction}
\end{align}
\normalsize
where $\widetilde{\mathbf{F}}_{k-1}\doteq\nabla\widetilde{f}(\mathbf{z})|_{\mathbf{z}=\hat{\mathbf{z}}_{k-1}}$.\\
\textit{2) State update:} Using $\mathbf{B}=\hat{\mathbf{B}}_{k-1}$ and $\mathbf{R}=\hat{\mathbf{R}}_{k-1}$ in \eqref{eqn:RKHS-EKF observation aug}, compute
\par\noindent\small
\begin{align}
    &\bm{\Sigma}^{y}_{k}=\widetilde{\mathbf{H}}_{k}\bm{\Sigma}^{z}_{k|k-1}\widetilde{\mathbf{H}}_{k}^{T}+\mathbf{R},\label{eqn:RKHS-EKF measurement update Sk}\\
    &\hat{\mathbf{z}}_{k}=\hat{\mathbf{z}}_{k|k-1}+\bm{\Sigma}^{z}_{k|k-1}\widetilde{\mathbf{H}}_{k}^{T}(\bm{\Sigma}^{y}_{k})^{-1}(\mathbf{y}_{k}-\widetilde{h}(\hat{\mathbf{z}}_{k|k-1})),\label{eqn:RKHS-EKF measurement update zk}\\
    &\bm{\Sigma}^{z}_{k}=\bm{\Sigma}^{z}_{k|k-1}-\bm{\Sigma}^{z}_{k|k-1}\widetilde{\mathbf{H}}_{k}^{T}(\bm{\Sigma}^{y}_{k})^{-1}\widetilde{\mathbf{H}}_{k}\bm{\Sigma}^{z}_{k|k-1},\label{eqn:RKHS-EKF measurement update sigk}
\end{align}
\normalsize
where $\widetilde{\mathbf{H}}_{k}\doteq\nabla\widetilde{h}(\mathbf{z})|_{\mathbf{z}=\hat{\mathbf{z}}_{k|k-1}}$. Here, $\hat{\mathbf{z}}_{k}=[\hat{\mathbf{x}}_{k|k}^{T}\;\hat{\mathbf{x}}_{k-1|k}^{T}]^{T}$ where $\hat{\mathbf{x}}_{k|k}$ is the RKHS-EKF's estimate of state $\mathbf{x}_{k}$. The prediction and measurement update steps follow from the standard EKF recursions to estimate the augmented state $\mathbf{z}_{k}$ with the considered system model \eqref{eqn:RKHS-EKF state transition aug} and \eqref{eqn:RKHS-EKF observation aug} given parameters $\hat{\Theta}_{k-1}$.\\
\textit{3) Parameters update:} Update the parameter estimates by approximating the required expectations in \eqref{eqn:RKHS A estimate}-\eqref{eqn:RKHS R estimate}. Consider $\mathbb{E}_{k}[\mathbf{x}_{k}\bm{\Phi}(\mathbf{x}_{k-1})^{T}]$ from \eqref{eqn:RKHS sum x phi}. Based on the standard EKF, linearize $\bm{\Phi}(\cdot)$ as
\par\noindent\small
\begin{align*}
    \bm{\Phi}(\mathbf{x}_{k-1})\approx\bm{\Phi}(\hat{\mathbf{x}}_{k-1|k})+\nabla\bm{\Phi}(\hat{\mathbf{x}}_{k-1|k})(\mathbf{x}_{k-1}-\hat{\mathbf{x}}_{k-1|k}),
\end{align*}
\normalsize
where $\nabla\bm{\Phi}(\hat{\mathbf{x}}_{k-1|k})\doteq\nabla\bm{\Phi}(\mathbf{x})|_{\mathbf{x}=\hat{\mathbf{x}}_{k-1|k}}$. Also, similar to standard EKF, we assume negligible error in computation of conditional means, i.e., $\mathbb{E}_{k}[\mathbf{x}_{k}]\approx\hat{\mathbf{x}}_{k|k}$ and $\mathbb{E}_{k}[\mathbf{x}_{k-1}]\approx\hat{\mathbf{x}}_{k-1|k}$. These yield
\par\noindent\small
\begin{align}
    &\mathbb{E}_{k}[\mathbf{x}_{k}\bm{\Phi}(\mathbf{x}_{k-1})^{T}]=\hat{\mathbf{x}}_{k|k}\bm{\Phi}(\hat{\mathbf{x}}_{k-1|k})^{T}+\textrm{Cov}(\mathbf{x}_{k}-\hat{\mathbf{x}}_{k|k},\mathbf{x}_{k-1}-\hat{\mathbf{x}}_{k-1|k})\nabla\bm{\Phi}(\hat{\mathbf{x}}_{k-1|k})^{T}.\label{eqn:Exphi}
\end{align}
\normalsize
By definition, $\textrm{Cov}(\mathbf{x}_{k}-\hat{\mathbf{x}}_{k|k},\mathbf{x}_{k-1}-\hat{\mathbf{x}}_{k-1|k})\approx[\bm{\Sigma}^{z}_{k}]_{(1:n_{x},n_{x}+1:2n_{x})}$, which was computed during the measurement update. Similarly, we have
\par\noindent\small
\begin{align}
    &\mathbb{E}_{k}[\bm{\Phi}(\mathbf{x}_{k-1})\bm{\Phi}(\mathbf{x}_{k-1})^{T}]=\bm{\Phi}(\hat{\mathbf{x}}_{k-1|k})\bm{\Phi}(\hat{\mathbf{x}}_{k-1|k})^{T}+\nabla\bm{\Phi}(\hat{\mathbf{x}}_{k-1|k})\textrm{Cov}(\mathbf{x}_{k-1}-\hat{\mathbf{x}}_{k-1|k})\nabla\bm{\Phi}(\hat{\mathbf{x}}_{k-1|k})^{T},\label{eqn:Ephi1}\\
    &\mathbb{E}_{k}[\bm{\Phi}(\mathbf{x}_{k})\bm{\Phi}(\mathbf{x}_{k})^{T}]=\bm{\Phi}(\hat{\mathbf{x}}_{k|k})\bm{\Phi}(\hat{\mathbf{x}}_{k|k})^{T}+\nabla\bm{\Phi}(\hat{\mathbf{x}}_{k|k})\textrm{Cov}(\mathbf{x}_{k}-\hat{\mathbf{x}}_{k|k})\nabla\bm{\Phi}(\hat{\mathbf{x}}_{k|k})^{T},\label{eqn:Ephi}\\
    &\mathbb{E}_{k}[\mathbf{x}_{k}\mathbf{x}_{k}^{T}]=\textrm{Cov}(\mathbf{x}_{k}-\hat{\mathbf{x}}_{k|k})+\hat{\mathbf{x}}_{k|k}\hat{\mathbf{x}}_{k|k}^{T},\label{eqn:Exx}
\end{align}
\normalsize
where $\textrm{Cov}(\mathbf{x}_{k-1}-\hat{\mathbf{x}}_{k-1|k})\approx[\bm{\Sigma}^{z}_{k}]_{(n_{x}+1:2n_{x},n_{x}+1:2n_{x})}$ and $\textrm{Cov}(\mathbf{x}_{k}-\hat{\mathbf{x}}_{k|k})\approx[\bm{\Sigma}^{z}_{k}]_{(1:n_{x},1:n_{x})}$. With these expectations, the updated parameter estimates $\hat{\Theta}_{k}$ can be computed using \eqref{eqn:RKHS A estimate}-\eqref{eqn:RKHS R estimate}.
\begin{algorithm}[t]
	\caption{RKHS-EKF initialization}
	\label{alg:RKHS-EKF initialization}
    \begin{algorithmic}[1]
    \Statex \textbf{Input:} $\hat{\mathbf{x}}_{0}$, $\bm{\Sigma}_{0}$
    \Statex \textbf{Output:} $\hat{\mathbf{z}}_{0}$, $\bm{\Sigma}^{z}_{0}$, $L$, $\{\widetilde{x}_{l}\}_{1\leq l\leq L}$, $\hat{\mathbf{A}}_{0}$, $\hat{\mathbf{B}}_{0}$, $\hat{\mathbf{Q}}_{0}$, $\hat{\mathbf{R}}_{0}$, $\mathbf{S}^{x\phi}_{0}$, $\mathbf{S}^{\phi 1}_{0}$, $\mathbf{S}^{y\phi}_{0}$ and $\mathbf{S}^{\phi}_{0}$
\State $\hat{\mathbf{z}}_{0}\gets [\hat{\mathbf{x}}_{0}^{T}\;\;\hat{\mathbf{x}}_{0}^{T}]^{T}$, and $\bm{\Sigma}^{z}_{0}\gets\begin{bmatrix}\bm{\Sigma}_{0}&\mathbf{0}_{n_{x}\times n_{x}}\\\mathbf{0}_{n_{x}\times n_{x}}&\bm{\Sigma}_{0}\end{bmatrix}$.
\State Set $L=1$ and $\widetilde{\mathbf{x}}_{1}\gets\hat{\mathbf{x}}_{0}$.
\State $\hat{\mathbf{A}}_{0}\gets\mathbf{1}_{n_{x}\times L}$ and $\hat{\mathbf{B}}_{0}\gets\mathbf{1}_{n_{y}\times L}$.
\State Initialize $\hat{\mathbf{Q}}_{0}$ and $\hat{\mathbf{R}}_{0}$ with some suitable p.d. noise covariance matrices.
\State Set $\mathbf{S}^{x\phi}_{0}=\mathbf{0}_{n_{x}\times L}$, $\mathbf{S}^{\phi 1}_{0}=\mathbf{0}_{L\times L}$, $\mathbf{S}^{y\phi}_{0}=\mathbf{0}_{n_{y}\times L}$ and $\mathbf{S}^{\phi}_{0}=\mathbf{0}_{L\times L}$.

\Statex \Return $\hat{\mathbf{z}}_{0}$, $\bm{\Sigma}^{z}_{0}$, $L$, $\{\widetilde{x}_{l}\}_{1\leq l\leq L}$, $\hat{\mathbf{A}}_{0}$, $\hat{\mathbf{B}}_{0}$, $\hat{\mathbf{Q}}_{0}$, $\hat{\mathbf{R}}_{0}$, $\mathbf{S}^{x\phi}_{0}$, $\mathbf{S}^{\phi 1}_{0}$, $\mathbf{S}^{y\phi}_{0}$ and $\mathbf{S}^{\phi}_{0}$.

    \end{algorithmic}
\end{algorithm}
\clearpage
Finally, the dictionary $\{\widetilde{\mathbf{x}}_{l}\}_{1\leq l\leq L}$ is updated using the new estimate $\hat{\mathbf{x}}_{k|k}$ based on the sliding window or ALD criterion. Algorithms~\ref{alg:RKHS-EKF initialization} and \ref{alg:RKHS-EKF recursion} summarize the RKHS-EKF's initialization and recursions, respectively. Note that the dictionary size increases when $\hat{\mathbf{x}}_{k|k}$ is added to the dictionary based on the ALD criterion or when the current size L is less than the considered window length (initial transient phase) in the sliding window criterion. For simplicity, we initialize all mixing parameters (elements of $\hat{\mathbf{A}}_{0}$ and $\hat{\mathbf{B}}_{0}$) with one in Algorithm~\ref{alg:RKHS-EKF initialization}. However, the mixing parameters can be initialized with arbitrary values.

\begin{algorithm}[t]
	\caption{RKHS-EKF recursion}
	\label{alg:RKHS-EKF recursion}
    \begin{algorithmic}[1]
    \Statex \textbf{Input:} $\hat{\mathbf{z}}_{k-1}$, $\bm{\Sigma}^{z}_{k-1}$, $\hat{\mathbf{A}}_{k-1}$, $\hat{\mathbf{B}}_{k-1}$, $\hat{\mathbf{Q}}_{k-1}$, $\hat{\mathbf{R}}_{k-1}$, $\mathbf{S}^{x\phi}_{k-1}$, $\mathbf{S}^{\phi 1}_{k-1}$, $\mathbf{S}^{y\phi}_{k-1}$, $\mathbf{S}^{\phi}_{k-1}$, and $\mathbf{y}_{k}$
    \Statex \textbf{Output:} $\hat{\mathbf{x}}_{k|k}$, $\hat{\mathbf{z}}_{k}$, $\bm{\Sigma}^{z}_{k}$, $\hat{\mathbf{A}}_{k}$, $\hat{\mathbf{B}}_{k}$, $\hat{\mathbf{Q}}_{k}$, $\hat{\mathbf{R}}_{k}$, $\mathbf{S}^{x\phi}_{k}$, $\mathbf{S}^{\phi 1}_{k}$, $\mathbf{S}^{y\phi}_{k}$, and $\mathbf{S}^{\phi}_{k}$
\State Compute $\hat{\mathbf{z}}_{k|k-1}$ and $\bm{\Sigma}^{z}_{k|k-1}$ using \eqref{eqn:RKHS-EKF prediction}.
\State Compute $\hat{\mathbf{z}}_{k}$ and $\bm{\Sigma}^{z}_{k}$ using \eqref{eqn:RKHS-EKF measurement update Sk}-\eqref{eqn:RKHS-EKF measurement update sigk}.
\State $\hat{\mathbf{x}}_{k|k}\gets[\hat{\mathbf{z}}_{k}]_{1:n_{x}}$.
\State $\textrm{Cov}(\mathbf{x}_{k}-\hat{\mathbf{x}}_{k|k},\mathbf{x}_{k-1}-\hat{\mathbf{x}}_{k-1|k})\gets[\bm{\Sigma}^{z}_{k}]_{(1:n_{x},n_{x}+1:2n_{x})}$, $\textrm{Cov}(\mathbf{x}_{k-1}-\hat{\mathbf{x}}_{k-1|k})\gets[\bm{\Sigma}^{z}_{k}]_{(n_{x}+1:2n_{x},n_{x}+1:2n_{x})}$ and $\textrm{Cov}(\mathbf{x}_{k}-\hat{\mathbf{x}}_{k|k})\gets[\bm{\Sigma}^{z}_{k}]_{(1:n_{x},1:n_{x})}$.
\State Compute $\mathbb{E}_{k}[\mathbf{x}_{k}\bm{\Phi}(\mathbf{x}_{k-1})^{T}]$, $\mathbb{E}_{k}[\bm{\Phi}(\mathbf{x}_{k-1})\bm{\Phi}(\mathbf{x}_{k-1})^{T}]$, $\mathbb{E}_{k}[\bm{\Phi}(\mathbf{x}_{k})\bm{\Phi}(\mathbf{x}_{k})^{T}]$ and $\mathbb{E}_{k}[\mathbf{x}_{k}\mathbf{x}_{k}^{T}]$ using \eqref{eqn:Exphi}-\eqref{eqn:Exx}.
\State Compute $\hat{\mathbf{A}}_{k}=\mathbf{S}^{x\phi}_{k}(\mathbf{S}^{\phi 1}_{k})^{-1}$, $\hat{\mathbf{B}}_{k}$, $\hat{\mathbf{Q}}_{k}$, and $\hat{\mathbf{R}}_{k}$ using \eqref{eqn:RKHS sum x phi}-\eqref{eqn:RKHS R estimate}.
\State Update dictionary $\{\widetilde{\mathbf{x}}_{l}\}_{1\leq l\leq L}$ using the state estimate $\hat{\mathbf{x}}_{k|k}$ based on the sliding window\cite{van2006sliding} or ALD\cite{engel2004kernel} criterion.
\If{dictionary size increases}
    \Statex Augment $\hat{\mathbf{A}}_{k}$, $\hat{\mathbf{B}}_{k}$, $\mathbf{S}^{x\phi}_{k}$, $\mathbf{S}^{\phi 1}_{k}$, $\mathbf{S}^{y\phi}_{k}$, and $\mathbf{S}^{\phi}_{k}$ with suitable initial values to take into account the updated dictionary size.
\EndIf
\Statex \Return $\hat{\mathbf{x}}_{k|k}$, $\hat{\mathbf{z}}_{k}$, $\bm{\Sigma}^{z}_{k}$, $\hat{\mathbf{A}}_{k}$, $\hat{\mathbf{B}}_{k}$, $\hat{\mathbf{Q}}_{k}$, $\hat{\mathbf{R}}_{k}$, $\mathbf{S}^{x\phi}_{k}$, $\mathbf{S}^{\phi 1}_{k}$, $\mathbf{S}^{y\phi}_{k}$, and $\mathbf{S}^{\phi}_{k}$.

    \end{algorithmic}
\end{algorithm}
\begin{remark}[Computational complexity of RKHS-EKF]\label{remark:RKHS-EKF complexity}
The computational complexity of RKHS-EKF depends on the state dimension `$n_{x}$' as well as the size of the dictionary `$L$'. In particular, the prediction and measurement update steps of RKHS-EKF follow from standard EKF recursion and hence, have a computational complexity of $\mathcal{O}(d^{3})$, where $d=2n_{x}$ is the dimension of augmented state $\mathbf{z}_{k}$. The parameters update step involves matrix multiplications of complexity $\mathcal{O}(n_{x}^{3})$ and matrix inversions of complexity $\mathcal{O}(L^{3})$. The updated parameters $\hat{\mathbf{A}}_{k}$ and $\hat{\mathbf{B}}_{k}$ require inversion of $L\times L$ matrices $\mathbf{S}^{\phi 1}_{k}$ and $\mathbf{S}^{\phi}_{k}$, respectively. Note that the dictionary size $L$ is a user-defined constant under the sliding window criterion. The conditions for a finite $L$ under the ALD criterion follow from \cite[Theorem~3.1]{engel2004kernel}.
\end{remark}

\subsection{RKHS-UKF}\label{subsec:RKHS-UKF}
RKHS-UKF's recursions closely follow that of RKHS-EKF of Section~\ref{subsec:RKHS-EKF}, and hence, we only summarize them here. The key difference is that RKHS-EKF linearizes the non-linear kernel function to approximate the statistics of a Gaussian random variable under non-linear transformation, whereas the same approximation is performed by RKHS-UKF using the unscented transform. Secondly, the prediction and measurement updates follow from a standard UKF instead of EKF. To this end, RKHS-UKF considers the scaling parameter $\kappa$ and generates a set of `$2n_{z}+1$' sigma points in the augmented state-space with dimension $n_{z}=2n_{x}$.\\
\textit{1) Prediction:} Generate the sigma points $\{\mathbf{s}_{i,k-1}\}_{0\leq i\leq 2n_{z}}=S_{gen}(\hat{\mathbf{z}}_{k-1},\bm{\Sigma}^{z}_{k-1})$ with their corresponding weights $\{\omega_{i}\}_{0\leq i\leq 2n_{z}}$ similar to \eqref{eqn:sigma points generation}. Using $\mathbf{A}=\hat{\mathbf{A}}_{k-1}$ and $\widetilde{\mathbf{Q}}_{k-1}=\begin{bsmallmatrix}\hat{\mathbf{Q}}_{k-1}&\mathbf{0}_{n_{x}\times n_{x}}\\\mathbf{0}_{n_{x}\times n_{x}}&\mathbf{0}_{n_{x}\times n_{x}}\end{bsmallmatrix}$ in \eqref{eqn:RKHS-EKF state transition aug}, propagate $\{\mathbf{s}_{i,k-1}\}$ through $\widetilde{f}(\cdot)$ to obtain the predicted state and associated prediction error covariance matrix as
\par\noindent\small
\begin{align}
    &\mathbf{s}^{*}_{i,k|k-1}=\widetilde{f}(\mathbf{s}_{i,k-1})\;\;\forall\; i=0,1,\hdots,2n_{z},\label{eqn:RKHS-UKF prediction propagate}\\  
    &\hat{\mathbf{z}}_{k|k-1}=\sum_{i=0}^{2n_{z}}\omega_{i}\mathbf{s}^{*}_{i,k|k-1},\label{eqn:RKHS-UKF predicted state}\\
    &\bm{\Sigma}^{z}_{k|k-1}=\sum_{i=0}^{2n_{z}}\omega_{i}\mathbf{s}^{*}_{i,k|k-1}(\mathbf{s}^{*}_{i,k|k-1})^{T}-\hat{\mathbf{z}}_{k|k-1}\hat{\mathbf{z}}_{k|k-1}^{T}+\widetilde{\mathbf{Q}}_{k-1}.\label{eqn:RKHS-UKF sig predict}
\end{align}
\normalsize
\textit{2) State update:} Generate the sigma points  $\{\mathbf{q}_{i,k|k-1}\}_{0\leq i\leq 2n_{z}}=S_{gen}(\hat{\mathbf{z}}_{k|k-1},\bm{\Sigma}^{z}_{k|k-1})$ and propagate through $\widetilde{h}(\cdot)$ using $\mathbf{B}=\hat{\mathbf{B}}_{k-1}$ and $\mathbf{R}=\hat{\mathbf{R}}_{k-1}$ in \eqref{eqn:RKHS-EKF observation aug} as
\par\noindent\small
\begin{align}
    &\mathbf{q}^{*}_{i,k|k-1}=\widetilde{h}(\mathbf{q}_{i,k|k-1})\;\;\forall\; i=0,1,\hdots,2n_{z},\label{eqn:RKHS-UKF state update propagate}\\
    &\hat{\mathbf{y}}_{k|k-1}=\sum_{i=0}^{2n_{z}}\omega_{i}\mathbf{q}^{*}_{i,k|k-1},\label{eqn:RKHS-UKF predicted y}\\
    &\bm{\Sigma}^{y}_{k}=\sum_{i=0}^{2n_{z}}\omega_{i}\mathbf{q}^{*}_{i,k|k-1}(\mathbf{q}^{*}_{i,k|k-1})^{T}-\hat{\mathbf{y}}_{k|k-1}\hat{\mathbf{y}}_{k|k-1}^{T}+\mathbf{R},\label{eqn:RKHS-UKF sig y predict}\\
    &\bm{\Sigma}^{zy}_{k}=\sum_{i=0}^{2n_{z}}\omega_{i}\mathbf{q}_{i,k|k-1}(\mathbf{q}^{*}_{i,k|k-1})^{T}-\hat{\mathbf{z}}_{k|k-1}\hat{\mathbf{y}}_{k|k-1}^{T},\label{eqn:RKHS-UKF sig zy predict}\\
    &\hat{\mathbf{z}}_{k}=\hat{\mathbf{z}}_{k|k-1}+\mathbf{K}_{k}(\mathbf{y}_{k}-\hat{\mathbf{y}}_{k|k-1}),\label{eqn:RKHS-UKF state update}\\
     &\bm{\Sigma}^{z}_{k}=\bm{\Sigma}^{z}_{k|k-1}-\mathbf{K}_{k}\bm{\Sigma}^{y}_{k}\mathbf{K}_{k}^{T},\label{eqn:RKHS-UKF sig update}
\end{align}
\normalsize
where gain matrix $\mathbf{K}_{k}=\bm{\Sigma}^{zy}_{k}(\bm{\Sigma}^{y}_{k})^{-1}$. Here, the prediction and state updates follow from the standard UKF recursions given parameters $\hat{\Theta}_{k-1}$.\\
\textit{3) Parameters update:} Generate sigma points $\{\mathbf{s}_{i,k}\}_{0\leq i\leq 2n_{z}}=S_{gen}(\hat{\mathbf{z}}_{k},\bm{\Sigma}^{z}_{k})$. Denote the two $n_{x}$-dimensional sub-vectors of the $i$-th sigma point, respectively, as $\mathbf{s}_{i,k}^{(1)}=[\mathbf{s}_{i,k}]_{1:n_{x}}$ and $\mathbf{s}_{i,k}^{(2)}=[\mathbf{s}_{i,k}]_{n_{x}+1:2n_{x}}$ corresponding to $\mathbf{x}_{k}$ and $\mathbf{x}_{k-1}$ parts of the augmented state $\mathbf{z}_{k}$. In order to approximate the statistics of $\bm{\Phi}(\mathbf{x}_{k-1})$ given observations $\mathbf{Y}^{k}$, we propagate the $\mathbf{x}_{k-1}$ part of the sigma points through $\bm{\Phi}(\cdot)$, i.e., $\widetilde{\mathbf{s}}_{i,k}^{(2)}=\bm{\Phi}(\mathbf{s}_{i,k}^{(2)})$. Similarly, we propagate $\mathbf{s}_{i,k}^{(1)}$ through $\bm{\Phi}(\cdot)$ to approximate the statistics of $\bm{\Phi}(\mathbf{x}_{k})$ as $\widetilde{\mathbf{s}}_{i,k}^{(1)}=\bm{\Phi}(\mathbf{s}_{i,k}^{(1)})$. Again, by definition $\textrm{Cov}(\mathbf{x}_{k}-\hat{\mathbf{x}}_{k|k})\approx[\bm{\Sigma}^{z}_{k}]_{(1:n_{x},1:n_{x})}$. With these approximations, the various expectations are computed as
\par\noindent\small
\begin{align}
    &\mathbb{E}_{k}[\mathbf{x}_{k}\mathbf{x}_{k}^{T}]=[\bm{\Sigma}^{z}_{k}]_{(1:n_{x},1:n_{x})}+\hat{\mathbf{x}}_{k|k}\hat{\mathbf{x}}_{k|k}^{T},\label{eqn:RKHS-UKF expect xx}\\
     &\mathbb{E}_{k}[\bm{\Phi}(\mathbf{x}_{k-1})\bm{\Phi}(\mathbf{x}_{k-1})^{T}]=\sum_{i=0}^{2n_{z}}\omega_{i}\widetilde{\mathbf{s}}_{i,k}^{(2)}(\widetilde{\mathbf{s}}_{i,k}^{(2)})^{T},\label{eqn:RKHS-UKF expect phi1}\\
     &\mathbb{E}_{k}[\mathbf{x}_{k}\bm{\Phi}(\mathbf{x}_{k-1})^{T}]=\sum_{i=0}^{2n_{z}}\omega_{i}\mathbf{s}_{i,k}^{(1)}(\widetilde{\mathbf{s}}_{i,k}^{(2)})^{T},\label{eqn:RKHS-UKF expect xphi}\\
     &\mathbb{E}_{k}[\bm{\Phi}(\mathbf{x}_{k})\bm{\Phi}(\mathbf{x}_{k})^{T}]=\sum_{i=0}^{2n_{z}}\omega_{i}\widetilde{\mathbf{s}}_{i,k}^{(1)}(\widetilde{\mathbf{s}}_{i,k}^{(1)})^{T}.\label{eqn:RKHS-UKF expect phi}
\end{align}
\normalsize
Using these expectations, the updated parameters estimate $\hat{\Theta}_{k}$ is obtained using \eqref{eqn:RKHS A estimate}-\eqref{eqn:RKHS R estimate} while the dictionary $\{\widetilde{\mathbf{x}}_{l}\}_{1\leq l\leq L}$ is updated using $\hat{\mathbf{x}}_{k|k}$ based on the sliding window\cite{van2006sliding} or ALD\cite{engel2004kernel} criterion.
\begin{algorithm}
	\caption{RKHS-UKF initialization}
	\label{alg:RKHS-UKF initialization}
    \begin{algorithmic}[1]
    \Statex \textbf{Input:} $\hat{\mathbf{x}}_{0}$, $\bm{\Sigma}_{0}$, $\kappa$
    \Statex \textbf{Output:} $\{\mathbf{s}_{i,0}\}_{0\leq i\leq 2n_{z}}$, $\{\omega_{i}\}_{0\leq i\leq 2n_{z}}$, $\bm{\Sigma}^{z}_{0}$, $L$, $\{\widetilde{x}_{l}\}_{1\leq l\leq L}$, $\hat{\mathbf{A}}_{0}$, $\hat{\mathbf{B}}_{0}$, $\hat{\mathbf{Q}}_{0}$, $\hat{\mathbf{R}}_{0}$, $\mathbf{S}^{x\phi}_{0}$, $\mathbf{S}^{\phi 1}_{0}$, $\mathbf{S}^{y\phi}_{0}$ and $\mathbf{S}^{\phi}_{0}$
\State $\hat{\mathbf{z}}_{0}\gets [\hat{\mathbf{x}}_{0}^{T}\;\;\hat{\mathbf{x}}_{0}^{T}]^{T}$, and $\bm{\Sigma}^{z}_{0}\gets\begin{bmatrix}\bm{\Sigma}_{0}&\mathbf{0}_{n_{x}\times n_{x}}\\\mathbf{0}_{n_{x}\times n_{x}}&\bm{\Sigma}_{0}\end{bmatrix}$.
\State $\{\mathbf{s}_{i,0},\omega_{i}\}_{0\leq i\leq 2n_{z}}\gets S_{gen}(\hat{\mathbf{z}}_{0},\bm{\Sigma}^{z}_{0})$ using \eqref{eqn:sigma points generation} and $\kappa$.
\State Set $L=1$ and $\widetilde{\mathbf{x}}_{1}\gets\hat{\mathbf{x}}_{0}$.
\State Initialize $\hat{\mathbf{A}}_{0}$ and $\hat{\mathbf{B}}_{0}$ with arbitrary $\mathbb{R}_{n_{x}\times L}$ and $\mathbb{R}_{n_{y}\times L}$ matrices, respectively.
\State Initialize $\hat{\mathbf{Q}}_{0}$ and $\hat{\mathbf{R}}_{0}$ with some suitable p.d. noise covariance matrices.
\State Set $\mathbf{S}^{x\phi}_{0}=\mathbf{0}_{n_{x}\times L}$, $\mathbf{S}^{\phi 1}_{0}=\mathbf{0}_{L\times L}$, $\mathbf{S}^{y\phi}_{0}=\mathbf{0}_{n_{y}\times L}$ and $\mathbf{S}^{\phi}_{0}=\mathbf{0}_{L\times L}$.

\Statex \Return $\{\mathbf{s}_{i,0}\}_{0\leq i\leq 2n_{z}}$, $\{\omega_{i}\}_{0\leq i\leq 2n_{z}}$, $\bm{\Sigma}^{z}_{0}$, $L$, $\{\widetilde{x}_{l}\}_{1\leq l\leq L}$, $\hat{\mathbf{A}}_{0}$, $\hat{\mathbf{B}}_{0}$, $\hat{\mathbf{Q}}_{0}$, $\hat{\mathbf{R}}_{0}$, $\mathbf{S}^{x\phi}_{0}$, $\mathbf{S}^{\phi 1}_{0}$, $\mathbf{S}^{y\phi}_{0}$ and $\mathbf{S}^{\phi}_{0}$.
    \end{algorithmic}
\end{algorithm}
\begin{algorithm}
	\caption{RKHS-UKF recursion}
	\label{alg:RKHS-UKF recursion}
    \begin{algorithmic}[1]
    \Statex \textbf{Input:} $\{\mathbf{s}_{i,k-1},\omega_{i}\}_{0\leq i\leq 2n_{z}}$, $\bm{\Sigma}^{z}_{k-1}$, $\hat{\mathbf{A}}_{k-1}$, $\hat{\mathbf{B}}_{k-1}$, $\hat{\mathbf{Q}}_{k-1}$, $\hat{\mathbf{R}}_{k-1}$, $\mathbf{S}^{x\phi}_{k-1}$, $\mathbf{S}^{\phi 1}_{k-1}$, $\mathbf{S}^{y\phi}_{k-1}$, $\mathbf{S}^{\phi}_{k-1}$, and $\mathbf{y}_{k}$
    \Statex \textbf{Output:} $\hat{\mathbf{x}}_{k|k}$, $\{\mathbf{s}_{i,k}\}_{0\leq i\leq 2n_{z}}$, $\bm{\Sigma}^{z}_{k}$, $\hat{\mathbf{A}}_{k}$, $\hat{\mathbf{B}}_{k}$, $\hat{\mathbf{Q}}_{k}$, $\hat{\mathbf{R}}_{k}$, $\mathbf{S}^{x\phi}_{k}$, $\mathbf{S}^{\phi 1}_{k}$, $\mathbf{S}^{y\phi}_{k}$, and $\mathbf{S}^{\phi}_{k}$
\State Propagate $\{\mathbf{s}_{i,k-1}\}$ through $\widetilde{f}(\cdot)$ and compute $\hat{\mathbf{z}}_{k|k-1}$ and $\bm{\Sigma}^{z}_{k|k-1}$ using \eqref{eqn:RKHS-UKF prediction propagate}-\eqref{eqn:RKHS-UKF sig predict}.
\State Generate sigma-points $\{\mathbf{q}_{i,k|k-1}\}_{0\leq i\leq 2n_{z}}=S_{gen}(\hat{\mathbf{z}}_{k|k-1},\bm{\Sigma}^{z}_{k|k-1})$ using \eqref{eqn:sigma points generation} and propagate through $\widetilde{h}(\cdot)$ using \eqref{eqn:RKHS-UKF state update propagate}.
\State Compute $\hat{\mathbf{z}}_{k}$ and $\bm{\Sigma}^{z}_{k}$ using \eqref{eqn:RKHS-UKF predicted y}-\eqref{eqn:RKHS-UKF sig update}.
\State $\hat{\mathbf{x}}_{k|k}\gets[\hat{\mathbf{z}}_{k}]_{1:n_{x}}$.
\State Generate sigma-points $\{\mathbf{s}_{i,k}\}_{0\leq i\leq 2n_{z}}=S_{gen}(\hat{\mathbf{z}}_{k},\bm{\Sigma}^{z}_{k})$ using \eqref{eqn:sigma points generation}.
\State Set $\mathbf{s}_{i,k}^{(1)}=[\mathbf{s}_{i,k}]_{1:n_{x}}$ and $\mathbf{s}_{i,k}^{(2)}=[\mathbf{s}_{i,k}]_{n_{x}+1:2n_{x}}$.
\State Compute $\widetilde{\mathbf{s}}_{i,k}^{(1)}=\bm{\Phi}(\mathbf{s}_{i,k}^{(1)})$ and $\widetilde{\mathbf{s}}_{i,k}^{(2)}=\bm{\Phi}(\mathbf{s}_{i,k}^{(2)})$.
\State Compute $\mathbb{E}_{k}[\mathbf{x}_{k}\mathbf{x}_{k}^{T}]$, $\mathbb{E}_{k}[\bm{\Phi}(\mathbf{x}_{k-1})\bm{\Phi}(\mathbf{x}_{k-1})^{T}]$, $\mathbb{E}_{k}[\mathbf{x}_{k}\bm{\Phi}(\mathbf{x}_{k-1})^{T}]$ and $\mathbb{E}_{k}[\bm{\Phi}(\mathbf{x}_{k})\bm{\Phi}(\mathbf{x}_{k})^{T}]$ using \eqref{eqn:RKHS-UKF expect xx}-\eqref{eqn:RKHS-UKF expect phi}.
\State Compute $\hat{\mathbf{A}}_{k}$, $\hat{\mathbf{B}}_{k}$, $\hat{\mathbf{Q}}_{k}$, and $\hat{\mathbf{R}}_{k}$ using \eqref{eqn:RKHS A estimate}-\eqref{eqn:RKHS R estimate}.
\State Update dictionary $\{\widetilde{\mathbf{x}}_{l}\}_{1\leq l\leq L}$ using the state estimate $\hat{\mathbf{x}}_{k|k}$ based on the sliding window\cite{van2006sliding} or ALD\cite{engel2004kernel} criterion.
\If{dictionary size increases}
    \Statex Augment $\hat{\mathbf{A}}_{k}$, $\hat{\mathbf{B}}_{k}$, $\mathbf{S}^{x\phi}_{k}$, $\mathbf{S}^{\phi 1}_{k}$, $\mathbf{S}^{y\phi}_{k}$, and $\mathbf{S}^{\phi}_{k}$ with suitable initial values to take into account the updated dictionary size.
\EndIf
\Statex \Return $\hat{\mathbf{x}}_{k|k}$, $\{\mathbf{s}_{i,k}\}_{0\leq i\leq 2n_{z}}$, $\bm{\Sigma}^{z}_{k}$, $\hat{\mathbf{A}}_{k}$, $\hat{\mathbf{B}}_{k}$, $\hat{\mathbf{Q}}_{k}$, $\hat{\mathbf{R}}_{k}$, $\mathbf{S}^{x\phi}_{k}$, $\mathbf{S}^{\phi 1}_{k}$, $\mathbf{S}^{y\phi}_{k}$, and $\mathbf{S}^{\phi}_{k}$.
    \end{algorithmic}    
\end{algorithm}

Note that in our RKHS-UKF, we need to generate only two sets of sigma points per recursion, similar to the standard UKF with a known system model. The sigma points generated in the parameters update step are the same as those obtained in the prediction step at the next time instant. Algorithm~\ref{alg:RKHS-UKF initialization} describes the initialization of the RKHS-UKF assuming initial state $\mathbf{x}_{0}\sim\mathcal{N}(\hat{\mathbf{x}}_{0},\bm{\Sigma}_{0})$, while Algorithm~\ref{alg:RKHS-UKF recursion} summarizes the RKHS-UKF recursions. The usage of the unscented transform implies that the scaling parameter $\kappa$ also impacts the parameter learning in RKHS-UKF. Alternatively, efficient numerical integration techniques can be employed and RKHS-CKF (or RKHS-QKF) can be trivially obtained from the developed RKHS-UKF.

In \cite{haykin2004kalman,zhan2006neural}, UKF has been coupled with neural networks (NNs) for non-linear system identification with the unknown weights learned by augmenting them with the state, which is computationally expensive. In \cite{kallapur2008ukf}, NN is used to model the non-linear dynamics or uncertainties with the weights updated online using previous UKF-computed state estimates. However, in general, the stability of these algorithms can not be assured\cite{haykin2004kalman}. Contrarily, our RKHS-EKF/ UKF extends the recursive Bayesian state estimation framework to unknown systems by including an additional parameter update step using the expectations from EKF/UKF recursion itself. Hence, we are able to analyze its stochastic stability using the unknown matrix approach in Section~\ref{sec:rkhs stability}.

\begin{remark}[Difference from kernel KFs]\label{remark:kernel KF}
    Our RKHS-EKF and RKHS-UKF is fundamentally different from the KKFs previously proposed in \cite{ralaivola2005time,zhu2013learning,dang2019kernel}. In \cite{ralaivola2005time}, the unknown system model is transformed to an RKHS-based feature space $\mathcal{H}$, wherein it is assumed to be linear Gaussian such that KF-based filtering recursions are applicable. The unknown parameters are represented using an orthonormal basis in $\mathcal{H}$ and learned from an exact EM algorithm. The kernel principal component analysis (PCA) provides the orthonormal basis. KKFs in \cite{zhu2013learning,dang2019kernel} consider conditional distribution embedding of the noisy but linear observation $\mathbf{y}_{k}$ to construct a new state-space model in RKHS. KF recursions are then applied to estimate the embedded observation, from which the estimate $\hat{\mathbf{x}}_{k}$ is computed. On the contrary, our RKHS-based KFs employ the kernel function approximation (with a universal Gaussian kernel) to directly represent the unknown functions in the original state space without any feature mapping. Furthermore, these KKFs involve a prior training phase to learn the orthonormal basis and the embedding operator in \cite{ralaivola2005time} and \cite{zhu2013learning,dang2019kernel}, respectively. Hence, KKFs suffer if the training data does not adequately represent the test data. On the contrary, our RKHS-based filters learn their dictionary online (from the computed estimates) based on the ALD or sliding window criterion.
\end{remark}

\section{Stability guarantees}\label{sec:rkhs stability}
We adopt the unknown matrix approach and show that our RKHS-UKF's state estimates are exponentially bounded in the mean-squared sense under some mild system conditions. As with the (forward) EKF's and UKF's stability conditions in Theorems~\ref{theorem:Forward ekf stable unknown matrix} and \ref{theorem:forward ukf stability}, respectively, RKHS-EKF's stability can also be guaranteed under RKHS-UKF's stability conditions following a similar procedure with some trivial modifications. Hence, we omit these details here.

Consider the following assumptions on the system dynamics and the RKHS-UKF of Section~\ref{subsec:RKHS-UKF} with the estimates $\hat{\mathbf{A}}_{k}$ and $\hat{\mathbf{B}}_{k}$ projected as in \eqref{eqn:RKHS A estimate with projection} and \eqref{eqn:RKHS B estimate with projection}, respectively.\\
\textbf{A6.a)} The kernel is a Gaussian kernel with width $\sigma>0$ such that $K(\mathbf{x}_{i},\mathbf{x}_{j})=\exp{\left(-\frac{\|\mathbf{x}_{i}-\mathbf{x}_{j}\|^{2}_{2}}{\sigma^2}\right)}$.\\
\textbf{A6.b)} The actual states $\mathbf{x}_{k}$ lie in a compact set $\mathcal{X}$ for all $k\geq 0$. For any state $\mathbf{x}$, $f(\mathbf{x})\in\mathcal{X}$, i.e., without any process noise, the state remains within the compact set.\\
\textbf{A6.c)} The dictionary $\{\widetilde{\mathbf{x}}_{l}\}_{1\leq l\leq L}$ is finite with fixed size.\\
\textbf{A6.d)} The true coefficient matrices $\mathbf{A}$ and $\mathbf{B}$ in \eqref{eqn:RKHS state transition approx} and \eqref{eqn:RKHS observation approx} satisfy $\|\mathbf{A}\|\leq\overline{a}$ and $\|\mathbf{B}\|\leq\overline{b}$, respectively, for some constants $\overline{a}$ and $\overline{b}$.

The conditions for a finite dictionary using the ALD criterion are discussed in \cite[Theorem~3.1]{engel2004kernel}. \textbf{A6.b} is essential for the Representer theorem to be valid\cite{scholkopf2001generalized}, the Gaussian kernel to approximate non-linear functions with arbitrarily small error \cite{steinwart2001influence} and also, for a finite dictionary under the ALD criterion\cite{engel2004kernel}. Also, the bound on coefficient matrices $\mathbf{A}$ and $\mathbf{B}$ in \textbf{A6.d} agrees with functions $f(\cdot)$ and $h(\cdot)$ being bounded, which is needed for the Gaussian kernel to approximate them with small approximation errors. Under \textbf{A6.b}, the state transition \eqref{eqn:non-linear state x} is modified to $\mathbf{x}_{k+1}=\Gamma(f(\mathbf{x}_{k})+\mathbf{w}_{k})$ where $\Gamma(\cdot)$ denotes the projection operator onto the set $\mathcal{X}$. Denote the projection error by $\bm{\eta}_{k}$ such that
\par\noindent\small
\begin{align}
    \mathbf{x}_{k+1}=f(\mathbf{x}_{k})+\mathbf{w}_{k}+\bm{\eta}_{k}.\label{eqn:state transition with projection}
\end{align}
\normalsize
Note that this projection is considered only for the actual state evolution. Intuitively, this projection represents the physical constraints on the state of the process being observed. For instance, in a radar's target localization problem, the actual target location is reasonably upper-bounded by the maximum unambiguous range and beam pattern (main lobe) of the radar. The noise $\mathbf{w}_{k}$ then represents the modeling uncertainties that are assumed to be Gaussian to obtain simplified closed-form solutions\cite{ristic2003beyond}. The RKHS-UKF's state estimates are not projected onto the set $\mathcal{X}$. However, the RKHS-UKF's coefficient matrix estimates $\hat{\mathbf{A}}_{k}$ and $\hat{\mathbf{B}}_{k}$ are projected to satisfy the bounds of \textbf{A6.d}. We denote the approximation errors in the kernel function approximation of functions $f(\cdot)$ and $h(\cdot)$, respectively, by $\bm{\delta}_{f}(\cdot)$ and $\bm{\delta}_{h}(\cdot)$ such that
\par\noindent\small
\begin{align}
    &f(\mathbf{x})=\mathbf{A}\bm{\Phi}(\mathbf{x})+\bm{\delta}_{f}(\mathbf{x}),\label{eqn:approx error in f}\\
    &h(\mathbf{x})=\mathbf{B}\bm{\Phi}(\mathbf{x})+\bm{\delta}_{h}(\mathbf{x}).\label{eqn:approx error in h}
\end{align}
\normalsize

Define the state prediction, state estimation and measurement prediction errors by $\widetilde{\mathbf{x}}_{k+1|k}\doteq\mathbf{x}_{k+1}-\hat{\mathbf{x}}_{k+1|k}$, $\widetilde{\mathbf{x}}_{k|k}\doteq\mathbf{x}_{k}-\hat{\mathbf{x}}_{k|k}$ and $\widetilde{\mathbf{y}}_{k+1}\doteq\mathbf{y}_{k+1}-\hat{\mathbf{y}}_{k+1|k}$, respectively. With $\hat{\mathbf{z}}_{k+1|k}=[\hat{\mathbf{x}}_{k+1|k}^{T}\;\hat{\mathbf{x}}_{k|k}^{T}]$, using \eqref{eqn:RKHS-EKF state transition aug}, \eqref{eqn:RKHS-UKF prediction propagate} and \eqref{eqn:RKHS-UKF predicted state}, we have $\hat{\mathbf{x}}_{k+1|k}=\sum_{i=0}^{2n_{z}}\omega_{i}\hat{\mathbf{A}}_{k}\bm{\Phi}([\mathbf{s}_{i,k}]_{1:n_{x}})$. Then, substituting \eqref{eqn:state transition with projection} and \eqref{eqn:approx error in f}, we obtain $\widetilde{\mathbf{x}}_{k+1|k}=\mathbf{A}\bm{\Phi}(\mathbf{x}_{k})+\bm{\delta}_{f}(\mathbf{x}_{k})+\mathbf{w}_{k}+\bm{\eta}_{k}-\sum_{i=0}^{2n_{z}}\omega_{i}\hat{\mathbf{A}}_{k}\bm{\Phi}([\mathbf{s}_{i,k}]_{1:n_{x}})$. Now, linearizing $\bm{\Phi}(\cdot)$ at $\hat{\mathbf{x}}_{k|k}$ and introducing unknown diagonal matrix $\mathbf{U}^{\phi 1}_{k}\in\mathbb{R}^{n_{x}\times n_{x}}$ similar to \eqref{eqn:linearized x}, we have
\par\noindent\small
\begin{align}
    \widetilde{\mathbf{x}}_{k+1|k}&=(\mathbf{A}-\hat{\mathbf{A}}_{k})\bm{\Phi}(\hat{\mathbf{x}}_{k|k})+\mathbf{U}^{\phi 1}_{k}\mathbf{A}\nabla\bm{\Phi}(\hat{\mathbf{x}}_{k|k})\widetilde{\mathbf{x}}_{k|k}+\mathbf{w}_{k}+\bm{\eta}_{k}+\bm{\delta}_{f}(\mathbf{x}_{k}),\label{eqn:RKHS-UKF state prediction error}
\end{align}
\normalsize
where $\nabla\bm{\Phi}(\hat{\mathbf{x}}_{k|k})\doteq\frac{\partial\bm{\Phi}(\mathbf{x})}{\partial\mathbf{x}}\vert_{\mathbf{x}=\hat{\mathbf{x}}_{k|k}}$. Similarly, linearizing $\bm{\Phi}(\cdot)$ at $\hat{\mathbf{x}}_{k+1|k}$, introducing unknown diagonal matrix $\mathbf{U}^{\phi 2}_{k+1}\in\mathbb{R}^{n_{y}\times n_{y}}$ and using \eqref{eqn:RKHS-EKF observation aug}, \eqref{eqn:RKHS-UKF state update propagate}, \eqref{eqn:RKHS-UKF predicted y} and \eqref{eqn:approx error in h}, we obtain
\par\noindent\small
\begin{align}
    \widetilde{\mathbf{y}}_{k+1}&=(\mathbf{B}-\hat{\mathbf{B}}_{k})\bm{\Phi}(\hat{\mathbf{x}}_{k+1|k})+\mathbf{U}^{\phi 2}_{k+1}\mathbf{B}\nabla\bm{\Phi}(\hat{\mathbf{x}}_{k+1|k})\widetilde{\mathbf{x}}_{k+1|k}+\mathbf{v}_{k+1}+\bm{\delta}_{h}(\mathbf{x}_{k+1}).\label{eqn:RKHS-UKF measurement prediction error}
\end{align}
\normalsize
Denote $\mathbf{K}^{1}_{k}$ as the sub-matrix $[\mathbf{K}_{k}]_{(1:n_{x},:)}$ such that \eqref{eqn:RKHS-UKF state update} yields $\hat{\mathbf{x}}_{k|k}=\hat{\mathbf{x}}_{k|k-1}+\mathbf{K}^{1}_{k}\widetilde{\mathbf{y}}_{k}$ and $\widetilde{\mathbf{x}}_{k|k}=\widetilde{\mathbf{x}}_{k|k-1}-\mathbf{K}^{1}_{k}\widetilde{\mathbf{y}}_{k}$. Substituting for $\widetilde{\mathbf{x}}_{k|k}$ and $\widetilde{\mathbf{y}}_{k}$ (using \eqref{eqn:RKHS-UKF measurement prediction error}) in \eqref{eqn:RKHS-UKF state prediction error}, the RKHS-UKF's prediction error dynamics becomes
\par\noindent\small
\begin{align}
    \hspace{-0.3cm}\widetilde{\mathbf{x}}_{k+1|k}&=\mathbf{U}^{\phi 1}_{k}\mathbf{A}\nabla\bm{\Phi}(\hat{\mathbf{x}}_{k|k})(\mathbf{I}-\mathbf{K}^{1}_{k}\mathbf{U}^{\phi 2}_{k}\mathbf{B}\nabla\bm{\Phi}(\hat{\mathbf{x}}_{k|k-1}))\widetilde{\mathbf{x}}_{k|k-1}+\mathbf{w}_{k}-\mathbf{U}^{\phi 1}_{k}\mathbf{A}\nabla\bm{\Phi}(\hat{\mathbf{x}}_{k|k})\mathbf{K}^{1}_{k}\mathbf{v}_{k}+(\mathbf{A}-\hat{\mathbf{A}}_{k})\bm{\Phi}(\hat{\mathbf{x}}_{k|k})\nonumber\\
    &-\mathbf{U}^{\phi 1}_{k}\mathbf{A}\nabla\bm{\Phi}(\hat{\mathbf{x}}_{k|k})\mathbf{K}^{1}_{k}(\mathbf{B}-\hat{\mathbf{B}}_{k-1})\bm{\Phi}(\hat{\mathbf{x}}_{k|k-1})+\bm{\eta}_{k}+\bm{\delta}_{f}(\mathbf{x}_{k})-\mathbf{U}^{\phi 1}_{k}\mathbf{A}\nabla\bm{\Phi}(\hat{\mathbf{x}}_{k|k})\mathbf{K}^{1}_{k}\bm{\delta}_{h}(\mathbf{x}_{k}).\label{eqn:RKHS-UKF error dynamics}
\end{align}
\normalsize

Denote $\bm{\Sigma}_{k+1|k}$ as the filter's estimate of $\mathbb{E}[\widetilde{\mathbf{x}}_{k+1|k}\widetilde{\mathbf{x}}_{k+1|k}^{T}]$ and $\bm{\Sigma}^{xy}_{k+1}$ as the estimate of $\mathbb{E}[\widetilde{\mathbf{x}}_{k+1|k}\widetilde{\mathbf{y}}_{k+1}^{T}]$. These are appropriate sub-matrices of $\bm{\Sigma}^{z}_{k+1|k}$ from \eqref{eqn:RKHS-UKF sig predict} and $\bm{\Sigma}^{zy}_{k+1}$ from \eqref{eqn:RKHS-UKF sig zy predict}, respectively. Then, following similar steps as for forward UKF's stability in Section~\ref{subsec:IUKF stability}, we obtain
\par\noindent\small
\begin{align}
    \bm{\Sigma}_{k+1|k}&=\mathbf{U}^{\phi 1}_{k}\mathbf{A}\nabla\bm{\Phi}(\hat{x}_{k|k})(\mathbf{I}-\mathbf{K}^{1}_{k}\mathbf{U}^{\phi 2}_{k}\mathbf{B}\nabla\bm{\Phi}(\hat{\mathbf{x}}_{k|k-1}))\bm{\Sigma}_{k|k-1}(\mathbf{I}-\mathbf{K}^{1}_{k}\mathbf{U}^{\phi 2}_{k}\mathbf{B}\nabla\bm{\Phi}(\hat{\mathbf{x}}_{k|k-1}))^{T}\nabla\bm{\Phi}(\hat{x}_{k|k})^{T}\mathbf{A}^{T}\mathbf{U}^{\phi 1}_{k}\nonumber\\
    &\;\;\;+\widetilde{\mathbf{Q}}_{k},\label{eqn:RKHS-UKF stability sig predict}\\
    \bm{\Sigma}^{y}_{k+1}&=\mathbf{U}^{\phi 2}_{k+1}\mathbf{B}\nabla\bm{\Phi}(\hat{\mathbf{x}}_{k+1|k})\bm{\Sigma}_{k+1|k}\nabla\bm{\Phi}(\hat{\mathbf{x}}_{k+1|k})^{T}\mathbf{B}^{T}\mathbf{U}^{\phi 2}_{k+1}+\widetilde{\mathbf{R}}_{k+1},\label{eqn:RKHS-UKF stability sig y predict}\\
    \bm{\Sigma}^{xy}_{k+1}&=\bm{\Sigma}_{k+1|k}\mathbf{U}^{xy}_{k+1}\nabla\bm{\Phi}(\hat{\mathbf{x}}_{k+1|k})^{T}\mathbf{B}^{T}\mathbf{U}^{\phi 2}_{k+1},\label{eqn:RKHS-UKF stability sig xy predict}
\end{align}
\normalsize
where $\widetilde{\mathbf{Q}}_{k}=\mathbf{Q}+\mathbf{U}^{\phi 1}_{k}\mathbf{A}\nabla\bm{\Phi}(\hat{\mathbf{x}}_{k|k})\mathbf{K}^{1}_{k}\mathbf{R}(\mathbf{K}^{1}_{k})^{T}\nabla\bm{\Phi}(\hat{\mathbf{x}}_{k|k})^{T}\mathbf{A}^{T}\mathbf{U}^{\phi 1}_{k}+\delta\mathbf{P}_{k+1|k}+\Delta\mathbf{P}_{k+1|k}$ and $\widetilde{\mathbf{R}}_{k+1}=\mathbf{R}+\delta\mathbf{P}^{y}_{k+1}+\Delta\mathbf{P}^{y}_{k+1}$ with $\delta\mathbf{P}_{k+1|k}$, $\Delta\mathbf{P}_{k+1|k}$, $\delta\mathbf{P}^{y}_{k+1}$ and $\Delta\mathbf{P}^{y}_{k+1}$ defined similarly as for the UKF stability in Section~\ref{subsec:IUKF stability}. Here, for simplicity, we have considered only the $n_{x}\geq n_{y}$ case, but the results can be trivially proved to hold for $n_{y}\geq n_{x}$ as well. Also, $\mathbf{U}^{xy}_{k+1}\in\mathbb{R}^{n_{x}\times n_{x}}$ is the unknown matrix introduced to account for errors in cross-covariance estimation. Finally, Theorem~\ref{theorem:RKHS-UKF stability} provides the stability conditions for RKHS-UKF.

\begin{theorem}[Stochastic stability of RKHS-UKF]
\label{theorem:RKHS-UKF stability}
Consider the RKHS-UKF for an unknown system model \eqref{eqn:non-linear state x} and \eqref{eqn:non-linear observation y} with the system satisfying \textbf{A6.a}-\textbf{A6.d}. The coefficient matrices estimates are projected as in \eqref{eqn:RKHS A estimate with projection} and \eqref{eqn:RKHS B estimate with projection} such that the estimates also satisfy the bounds in \textbf{A6.d}. The RKHS-UKF's estimation error $\widetilde{\mathbf{x}}_{k|k}$ is exponentially bounded in the mean-squared sense if the following hold true.\\
\textbf{C6.a)} There exist positive real numbers $\underline{\sigma}, \overline{\sigma}, \overline{\gamma}, \overline{\beta}, \widetilde{r}, \overline{\alpha}, \widetilde{q}, \overline{q}, \overline{r}, \overline{\phi}, \overline{f}$ and $\overline{h}$ such that the following bounds are fulfilled for all $k\geq 0$:
\par\noindent\small
\begin{align*}
    &\underline{\sigma}\mathbf{I}\preceq\bm{\Sigma}_{k|k-1}\preceq\overline{\sigma}\mathbf{I},\;\;\|\mathbf{U}^{\phi 1}_{k}\|\leq\overline{\alpha},\;\;\|\mathbf{U}^{\phi 2}_{k}\|\leq\overline{\beta},\;\;\|\mathbf{U}^{xy}_{k}\|\leq\overline{\gamma},\;\;\mathbf{Q}\preceq\overline{q}\mathbf{I},\;\;\mathbf{R}\leq\overline{r}\mathbf{I},\\
    &\;\;\widetilde{\mathbf{Q}}_{k}\succeq\widetilde{q}\mathbf{I},\;\;\widetilde{\mathbf{R}}_{k}\succeq\widetilde{r}\mathbf{I},\;\;\|\nabla\bm{\Phi}(\cdot)\|\leq\overline{\phi},\;\;\|\bm{\delta}_{f}(\cdot)\|_{2}\leq\overline{f},\;\;\|\bm{\delta}_{h}(\cdot)\|_{2}\leq\overline{h}.
\end{align*}
\normalsize
\textbf{C6.b)} The Jacobian $\nabla\bm{\Phi}(\mathbf{x})$ at any state $\mathbf{x}$, the actual coefficient matrix $\mathbf{A}$ and $\mathbf{U}^{\phi 1}_{k}$ are non-singular for every $k\geq 0$.\\
\textbf{C6.c)} The constants satisfy the inequality $\overline{\sigma}\overline{\gamma}\overline{\phi}^{2}\overline{\beta}^{2}\overline{b}^{2}<\widetilde{r}$.
\end{theorem}
\begin{proof}
    See Appendix~\ref{App-thm-RKHS-UKF}. An analogous method has been earlier used to prove the stability of forward EKF in Appendix~\ref{App-thm-Forward ekf stable unknown matrix}. In particular, we define $V_{k}(\widetilde{\mathbf{x}}_{k|k-1})=\widetilde{\mathbf{x}}_{k|k-1}^{T}\bm{\Sigma}_{k|k-1}^{-1}\widetilde{\mathbf{x}}_{k|k-1}$. Then, the proof involves bounding $\mathbb{E}[\mathbf{V}_{k+1}(\widetilde{\mathbf{x}}_{k+1|k})|\widetilde{\mathbf{x}}_{k|k-1}]$ such that the conditions of Lemma~\ref{lemma:exponential boundedness} are satisfied. However, unlike forward EKF, the prediction error dynamics \eqref{eqn:RKHS-UKF error dynamics} must account for additional approximation errors $\bm{\delta}_{f}(\mathbf{x}_{k})$ and $\bm{\delta}_{h}(\mathbf{x}_{k})$; projection error $\bm{\eta}_{k}$; and differences in true coefficient matrices and their estimates, i.e., $(\mathbf{A}-\hat{\mathbf{A}}_{k})$ and $(\mathbf{B}-\hat{\mathbf{B}}_{k-1})$. The corresponding additional terms in $\mathbb{E}[\mathbf{V}_{k+1}(\widetilde{\mathbf{x}}_{k+1|k})|\widetilde{\mathbf{x}}_{k|k-1}]$ expansion are bounded by suitable constants using the assumptions of Theorem~\ref{theorem:RKHS-UKF stability}.
\end{proof}

Note that for the Gaussian kernel $K(\cdot,\cdot)$, the Jacobian at state estimate $\hat{\mathbf{x}}$ is given as $\nabla\bm{\Phi}(\hat{\mathbf{x}})=\frac{2}{\sigma^{2}}\begin{bsmallmatrix}
    K(\widetilde{\mathbf{x}}_{1},\mathbf{x})(\widetilde{\mathbf{x}}_{1}-\hat{\mathbf{x}})^{T}\\
    \vdots\\
    K(\widetilde{\mathbf{x}}_{L},\mathbf{x})(\widetilde{\mathbf{x}}_{L}-\hat{\mathbf{x}})^{T}
    \end{bsmallmatrix}$. Since $K(\cdot,\cdot)$ is a bounded function,  $\|\nabla\bm{\Phi}(\cdot)\|\leq\overline{\phi}$ implies $\widetilde{\mathbf{x}}_{l}-\hat{\mathbf{x}}$ is bounded where $\hat{\mathbf{x}}$ is any state estimate and $\widetilde{\mathbf{x}}_{l}$ is a dictionary element obtained from the state estimates themselves. Hence, this assumption implies our computed state estimates (more specifically, the difference between these estimates) lie in a bounded set. However, this set need not be the same as the compact set $\mathcal{X}$ of the true states in \textbf{A6.b}.

\section{Numerical experiments}\label{sec:rkhs numericals}
We consider FM demodulator and Lorenz systems to demonstrate the estimation efficiency of our RKHS-based KFs using NCI as the performance metric. In our experiments, RKHS-EKF and RKHS-UKF are employed both as the attacker's forward filter (denoted by forward RKHS-EKF/RKHS-UKF) and the defender's inverse filter (denoted by I-RKHS-EKF/UKF) without any prior system model assumptions. Note that forward RKHS-EKF/UKF and I-RKHS-EKF/UKF are essentially the same algorithms but employed by different agents to compute their desired estimates. We further compare our RKHS-based KFs performance with forward and inverse EKFs and UKFs with the perfect system model information in Sections~\ref{subsec:FM rkhs} and \ref{subsec:lorenz rkhs}, respectively.

\subsection{FM demodulation with RKHS-EKF and RKHS-UKF}\label{subsec:FM rkhs}
Recall the FM-demodulator system of Section~\ref{subsec:FM with EKF SOEKF GSEKF} to compare RKHS-EKF and RKHS-UKF's accuracy with forward and inverse EKF (with perfect system model information). Here, the attacker knows its observation function and employs a simplified forward RKHS-EKF/UKF. On the other hand, the defender learns both its state evolution and observation model using the general RKHS-EKF/UKF as its inverse filter. The initial state estimates for both filters were drawn at random similar to Section~\ref{subsec:FM with EKF SOEKF GSEKF}. For both forward and inverse RKHS-based filters, we considered the Gaussian kernel with kernel width $\sigma$ set to $\sigma^{2}=30$ and $50$, respectively. The initial covariance estimates ($\bm{\Sigma}^{z}_{0}$) were $10\mathbf{I}_{4}$ and $5\mathbf{I}_{4}$, respectively, for forward and inverse RKHS-EKF/UKF. For RKHS-UKF, $\kappa$ was set to $5$ and $3$, respectively, for the forward and inverse filters. To construct the forward RKHS-based filters' dictionary, we used the sliding window criterion\cite{van2006sliding} with window length $2$. On the other hand, the dictionary for I-RKHS-EKF/UKF was constructed using ALD criterion \cite{engel2004kernel}. All the initial coefficient estimates (elements of $\hat{\mathbf{A}}_{0}$, and $\hat{\mathbf{B}}_{0}$) were set to $1$. The noise covariance estimates $\hat{\mathbf{Q}}_{0}$ and $\hat{\mathbf{R}}_{0}$ were $\textrm{diag}(1, 10)$ and $5$, respectively. All other system parameters were the same as in Section~\ref{subsec:FM with EKF SOEKF GSEKF}.

\begin{figure}
  \centering
  \includegraphics[width = 0.75\columnwidth]{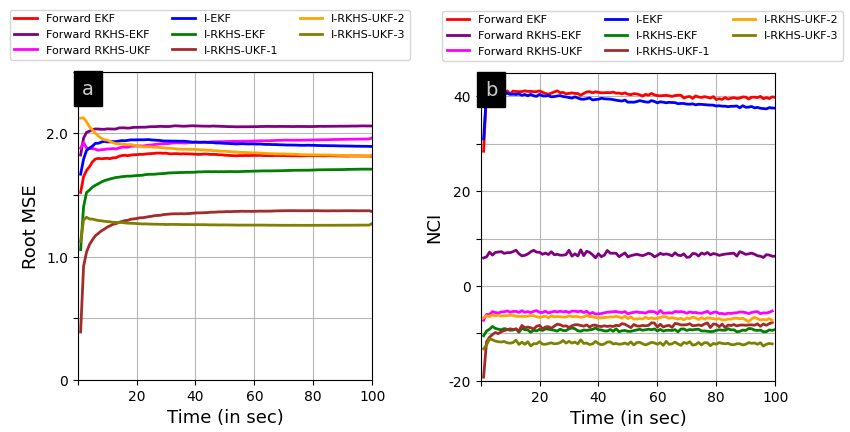}
  \caption{(a) Time-averaged RMSE, and (b) NCI for forward and inverse RKHS-EKF and UKF for FM demodulator system, compared with forward and inverse EKF.}
 \label{fig:rkhs fmdemod}
\end{figure}

Fig.~\ref{fig:rkhs fmdemod} shows the AMSE and NCI for forward and inverse EKF, RKHS-EKF, and RKHS-UKF. For I-EKF and I-RKHS-EKF, the true forward filters are EKF and RKHS-EKF, respectively. On the other hand, I-RKHS-UKF-1, I-RKHS-UKF-2, and I-RKHS-UKF-3 have RKHS-UKF, EKF, and RKHS-EKF, respectively, as true forward filters. Note that I-EKF assumes a forward EKF employed by the attacker, but I-RKHS-EKF and I-RKHS-UKFs do not make any such assumption. From Fig.~\ref{fig:rkhs fmdemod}a, we observe that both forward RKHS-EKF and RKHS-UKF have higher errors than forward EKF because they do not perfectly know the defender's state evolution function. On the other hand, both forward RKHS-EKF and RKHS-UKF are more credible than forward EKF (in Fig.~\ref{fig:rkhs fmdemod}b), but optimistic and pessimistic, respectively. However, forward RKHS-UKF is more accurate than RKHS-EKF. While all RKHS-based inverse filters perform better than I-EKF without any prior system model information, I-RKHS-UKF outperforms both I-EKF and I-RKHS-EKF when estimating the same state, i.e., I-RKHS-UKF-2 and I-RKHS-UKF-3 have lower error than I-EKF and I-RKHS-EKF, respectively. The improved performance of RKHS-based inverse filters is explained by the fact that it is a kernel function approximation and learns the unknown system dynamics completely based on its observations. On the other hand, I-EKF uses a first-order Taylor approximation for the non-linear functions. Here, the Gaussian kernel provides a better function approximation than linearization. However, RKHS-EKF linearizes the non-linear kernel function for expectation computations and thus, introduces linearization errors. RKHS-UKF, on the other hand, considers the unscented transform and hence, also outperforms RKHS-EKF. Interestingly, even though RKHS-UKF is pessimistic, it has a similar NCI for all cases, including when employed as a forward filter (forward RKHS-UKF) and hence, is more robust than all other filters.

\subsection{Lorenz system with RKHS-UKF}\label{subsec:lorenz rkhs}
Consider the same Lorenz system as in Section~\ref{subsec:lorenz QKF CQKF}. Here, we compare forward and inverse UKF (with perfect system model information) with RKHS-UKF (without any prior information). The considered system is mathematically interesting because of its three unstable equilibria\cite{ito2000gaussian}. For the attacker's state estimate, we employ a forward RKHS-UKF but unlike Section~\ref{subsec:FM rkhs}, the observation function is not known to the attacker. For both forward and inverse RKHS-UKF, we chose $\kappa=3$ and Gaussian kernel's width $\sigma^{2}=20$. The dictionaries were constructed using the sliding window criterion with a window length $15$. The initial coefficient matrix estimates ($\hat{\mathbf{A}}_{0}, \hat{\mathbf{B}}_{0}$) and noise covariance matrix estimates ($\hat{\mathbf{Q}}_{0},\hat{\mathbf{R}}_{0}$) were set to appropriate (size) all ones and identity matrices, respectively. We compare the RKHS-UKF-based forward and inverse filters with a forward UKF with $\kappa=1.5$ and I-UKF with $\overline{\kappa}=2$, respectively. All other system parameters and initial estimates were set identical to the ones in Section~\ref{subsec:lorenz QKF CQKF}.

\begin{figure}
  \centering
  \includegraphics[width = 0.7\columnwidth]{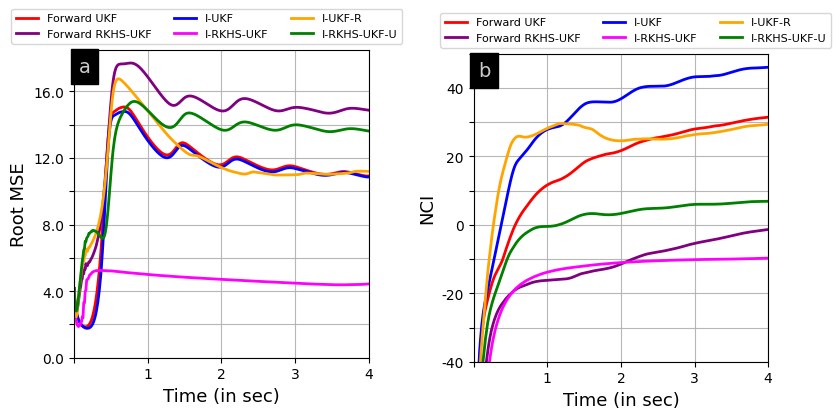}
  \caption{(a) Time-averaged RMSE, and (b) Time-averaged NCI for forward and inverse RKHS-UKF for Lorenz system, compared with forward and inverse UKF.}
 \label{fig:rkhs lorenz}
\end{figure}

Fig.~\ref{fig:rkhs lorenz} shows the AMSE and (time-averaged) NCI for forward and inverse RKHS-UKF and UKF, including the incorrect forward filter assumption case I-UKF-R which assumes a forward UKF instead of the actual forward RKHS-UKF. The true forward filters for I-RKHS-UKF and I-RKHS-UKF-U are RKHS-UKF and UKF, respectively. Similar to Fig.~\ref{fig:rkhs fmdemod}a, we observe that without perfect system information, forward RKHS-UKF has a higher estimation error than the forward UKF. While I-UKF has the same accuracy as its forward UKF, I-RKHS-UKF (I-RKHS-UKF-U case) could not achieve the same accuracy by learning the system model on its own. However, even though forward RKHS-UKF is not accurate, I-RKHS-UKF is able to learn its state estimate's evolution and has the lowest estimation error. Interestingly, I-UKF-R with incorrect forward filter assumption also shows a lower error than its forward RKHS-UKF. From Fig.~\ref{fig:rkhs lorenz}b, we observe that only I-RKHS-UKF has stable performance in terms of credibility. I-RKHS-UKF-U is slightly optimistic while I-RKHS-UKF is pessimistic.

\section{Concluding remarks}\label{sec:rkhs conclusions}
The inverse filters developed in Chapters~\ref{chap:inverse EKFs}-\ref{chap:inverse SPKFs} as well as prior works \cite{mattila2020hmm,krishnamurthy2019how}, are limited by the perfect system information assumption, including a known forward filter. In this chapter, we tackled the scenario where no prior system model information is available by developing RKHS-EKF and RKHS-UKF. These methods approximate unknown functions in an RKHS induced by a kernel, while unknown parameters are learned online using an approximate EM algorithm combined with EKF/UKF recursions. For updating parameter estimates, RKHS-EKF uses Taylor series linearization, while RKHS-UKF employs the unscented transform to approximate non-linear expectations. We provided sufficient conditions for the state estimates of the proposed filters to be exponentially bounded in the mean-squared sense. Our numerical experiments indicated that RKHS-based inverse filters can outperform even I-EKF and I-UKF with perfect system information, and RKHS-UKF was shown to provide better estimates than RKHS-EKF.

\chapter{Inverse Particle and Ensemble Kalman filters}
\label{chap:smc}
The I-EKFs and I-SPKFs developed in Chapters~\ref{chap:inverse EKFs}-\ref{chap:inverse SPKFs} assume a Gaussian posterior distribution, making them suitable only for systems with Gaussian process and measurement noises. In contrast, random sampling-based SMC approaches can approximate the arbitrary posterior distribution of the underlying state and, hence, are applicable to general non-linear and/or non-Gaussian systems. In this chapter, we develop SMC-based inverse filters under the assumption of perfect system model information. In Section~\ref{sec:IPF}, we formulate I-PF and prove its convergence to the optimal inverse filter in the $L^{4}$-sense under certain mild assumptions on the underlying system. In Section~\ref{sec:igpf and ienkf}, we explore SMC-based approaches for Gaussian systems by developing I-GPF and I-EnKF and further generalize these methods to non-Gaussian systems. Section~\ref{sec:unknown SMC} addresses the case of unknown system dynamics by proposing differentiable I-PF, differentiable I-EnKF, and RKHS-based EnKF (RKHS-EnKF). Recall that DPF and differentiable EnKF are differentiable implementations of PF and EnKF, respectively, that use NNs to learn the system dynamics. Meanwhile, RKHS-EnKF considers non-linear state-space dynamics with the state-transition and observation models unknown to the agent employing the stochastic filter. Finally, Section~\ref{sec:smc numericals} discusses numerical experiments demonstrating the performance of the proposed filters, and Section~\ref{sec:smc conclusions} provides the concluding remarks.

\section{Inverse PF}\label{sec:IPF}
In PF, the posterior distribution is approximated using an empirical distribution from a set of samples/particles with associated weights, evolving randomly in time according to the system dynamics. Such an approximation simplifies the computation of integrals to finite sums. The particles are sampled from a suitable importance density and resampled to avoid particle degeneracy \cite{ristic2003beyond,arulampalam2002tutorial}. This sequential importance sampling (SIS) algorithm is also known as bootstrap filtering \cite{gordon1993novel}, condensation algorithm \cite{maccormick2000probabilistic}, and interacting particle approximations \cite{del1998measure}. In the following, we first develop the I-PF recursions applying the SIS and resampling methods for the defender-attacker dynamics, analogous to the standard PF. For further details on the SIS-based filtering methods, we refer the readers to  \cite{arulampalam2002tutorial,chen2003bayesian}. In Section~\ref{subsec:IPF convergence}, we study the convergence of the proposed I-PF to the optimal filter \eqref{eqn:optimal filter time update}-\eqref{eqn:optimal filter measurement update} as the number of particles increases.

\subsection{Filter formulation}\label{subsec:IPF formulation}
Consider $N$ as the total number of particles. Following Section~\ref{sec:optimal filter}, in I-PF, we approximate the (joint) posterior distribution of $(\hat{\mathbf{x}}_{k},\mathbf{y}_{k})$ at $k$-th time instant as
\par\noindent\small
\begin{align*}
p(\hat{\mathbf{x}}_{k},\mathbf{y}_{k}|\mathbf{x}_{0:k},\mathbf{a}_{1:k})\approx\sum_{i=1}^{N}\omega^{i}_{k}\delta(\hat{\mathbf{x}}_{k}-\hat{\mathbf{x}}^{i}_{k},\mathbf{y}_{k}-\mathbf{y}^{i}_{k}),
\end{align*}
\normalsize
where $\{\hat{\mathbf{x}}^{i}_{k},\mathbf{y}^{i}_{k}\}_{1\leq i\leq N}$ are the current particles with associated importance weights $\{\omega^{i}_{k}\}_{1\leq i\leq N}$ and the defender knows $\mathbf{x}_{0:k}$ and $\mathbf{a}_{1:k}$. To this end, we first consider the joint conditional density $p(\hat{\mathbf{x}}_{0:k},\mathbf{y}_{1:k}|\mathbf{x}_{0:k},\mathbf{a}_{1:k})$, which we approximate using particles $\{\hat{\mathbf{x}}^{i}_{0:k},\mathbf{y}^{i}_{1:k}\}_{1\leq i\leq N}$. The particle $(\hat{\mathbf{x}}^{i}_{k},\mathbf{y}^{i}_{k})$ for approximating the marginal distribution $p(\hat{\mathbf{x}}_{k},\mathbf{y}_{k}|\mathbf{x}_{0:k},\mathbf{a}_{1:k})$ is then simply the sub-vector corresponding to $(\hat{\mathbf{x}}_{k},\mathbf{y}_{k})$ in the $i$-th particle $(\hat{\mathbf{x}}^{i}_{0:k},\mathbf{y}^{i}_{1:k})$. Furthermore, as we will show eventually, our I-PF algorithm only requires storing particles $\{\hat{\mathbf{x}}^{i}_{k}\}$, i.e., the previous particles $\{\hat{\mathbf{x}}^{i}_{0:k-1},\mathbf{y}^{i}_{1:k-1}\}$ and the current observation particles $\{\mathbf{y}^{i}_{k}\}$ are discarded. Denote $q(\cdot)$ as the chosen importance sampling density. Based on the importance sampling method \cite{ristic2003beyond}, the weights are computed as
\par\noindent\small
\begin{align}
\omega^{i}_{k}=\frac{p(\hat{\mathbf{x}}^{i}_{0:k},\mathbf{y}^{i}_{1:k}|\mathbf{x}_{0:k},\mathbf{a}_{1:k})}{q(\hat{\mathbf{x}}^{i}_{0:k},\mathbf{y}^{i}_{1:k}|\mathbf{x}_{0:k},\mathbf{a}_{1:k})},\label{eqn:IPF weight start}
\end{align}
\normalsize
for $i=1,2,\hdots,N$. In our inverse filtering problem, from \eqref{eqn:state x}-\eqref{eqn:observation a}, the joint conditional density simplifies to
\par\noindent\small
\begin{align}
p(\hat{\mathbf{x}}_{0:k},\mathbf{y}_{1:k}|\mathbf{x}_{0:k},\mathbf{a}_{1:k})&\propto p(\mathbf{a}_{k}|\hat{\mathbf{x}}_{k})p(\hat{\mathbf{x}}_{k}|\hat{\mathbf{x}}_{k-1},\mathbf{y}_{k})p(\mathbf{y}_{k}|\mathbf{x}_{k})p(\hat{\mathbf{x}}_{0:k-1},\mathbf{y}_{1:k-1}|\mathbf{x}_{0:k-1},\mathbf{a}_{1:k-1}).\label{eqn:ipf p density}
\end{align}
\normalsize
Furthermore, analogous to the standard PF \cite{ristic2003beyond}, we choose a sampling density $q(\cdot)$ that factorizes as
\par\noindent\small
\begin{align}
&q(\hat{\mathbf{x}}_{0:k},\mathbf{y}_{1:k}|\mathbf{x}_{0:k},\mathbf{a}_{1:k})=q(\hat{\mathbf{x}}_{k},\mathbf{y}_{k}|\hat{\mathbf{x}}_{k-1},\mathbf{y}_{k-1},\mathbf{x}_{k},\mathbf{a}_{k})q(\hat{\mathbf{x}}_{0:k-1},\mathbf{y}_{1:k-1}|\mathbf{x}_{0:k-1},\mathbf{a}_{1:k-1}).\label{eqn:ipf q density}
\end{align}
\normalsize
We provide the detailed steps to obtain \eqref{eqn:ipf p density} and \eqref{eqn:ipf q density} in Appendix~\ref{app:sampling}. Substituting \eqref{eqn:ipf p density} and \eqref{eqn:ipf q density} in \eqref{eqn:IPF weight start}, the optimal weights simplifies to
\par\noindent\small
\begin{align}
\omega^{i}_{k}\propto\omega^{i}_{k-1}\frac{p(\mathbf{a}_{k}|\hat{\mathbf{x}}^{i}_{k})p(\hat{\mathbf{x}}^{i}_{k}|\hat{\mathbf{x}}^{i}_{k-1},\mathbf{y}^{i}_{k})p(\mathbf{y}^{i}_{k}|\mathbf{x}_{k})}{q(\hat{\mathbf{x}}^{i}_{k},\mathbf{y}^{i}_{k}|\hat{\mathbf{x}}^{i}_{k-1},\mathbf{y}^{i}_{k-1},\mathbf{x}_{k},\mathbf{a}_{k})},\label{eqn:IPF weight 2}
\end{align}
\normalsize
However, for this choice of importance density, the variance of weights can only increase over time resulting in particle degeneracy \cite{doucet2000sequential}, i.e., with time, all but one particle will have negligible weights. To this end, resampling is employed to eliminate particles with small weights and multiply the ones with large weights. The optimal $q(\cdot)$ that minimizes the variance of weights \eqref{eqn:IPF weight 2} can be trivially obtained as \cite{doucet2000sequential}
\par\noindent\small
\begin{align}
q^{*}(\hat{\mathbf{x}}_{k},\mathbf{y}_{k}|\hat{\mathbf{x}}_{k-1},\mathbf{y}_{k-1},\mathbf{x}_{k},\mathbf{a}_{k})&=p(\hat{\mathbf{x}}_{k},\mathbf{y}_{k}|\hat{\mathbf{x}}_{k-1},\mathbf{y}_{k-1},\mathbf{x}_{k},\mathbf{a}_{k})\approx p(\hat{\mathbf{x}}_{k}|\hat{\mathbf{x}}_{k-1},\mathbf{y}_{k})p(\mathbf{y}_{k}|\mathbf{x}_{k}),\label{eqn:optimal density}
\end{align}
\normalsize
where we have ignored the correlation between attacker's observation $\mathbf{y}_{k}$ and defender's observation $\mathbf{a}_{k}$ via state estimate $\hat{\mathbf{x}}_{k}$ and hence, the approximation. The perfect knowledge of distribution $p(\mathbf{y}_{k}|\mathbf{x}_{k})$ from \eqref{eqn:observation y} can compensate for this approximation to some extent. With $q^{*}(\cdot)$ as the sampling density, the weights are computed from \eqref{eqn:IPF weight 2} as
\par\noindent\small
\begin{align}
\omega^{i}_{k}\propto\omega^{i}_{k-1}p(\mathbf{a}_{k}|\hat{\mathbf{x}}^{i}_{k}).\label{eqn:optimal weights}
\end{align}
\normalsize

\textbf{I-PF recursions:} Consider the sampling density $q^{*}(\cdot)$ from \eqref{eqn:optimal density}. We have $p(\hat{\mathbf{x}}_{k}|\hat{\mathbf{x}}_{k-1},\mathbf{y}_{k})\\=\delta(\hat{\mathbf{x}}_{k}-T(\hat{\mathbf{x}}_{k-1},\mathbf{y}_{k}))$ from \eqref{eqn:filter T} and $p(\mathbf{y}_{k}|\mathbf{x}_{k})$ is given by \eqref{eqn:observation y}. Here, we consider resampling at each time step and $p(\mathbf{a}_{k}|\hat{\mathbf{x}}_{k})$ in \eqref{eqn:optimal weights} is given by \eqref{eqn:observation a}. Note that particles $\{\mathbf{y}^{i}_{k}\}$ are sampled from \eqref{eqn:observation y} using the true state $\mathbf{x}_{k}$ only. Further, since the defender knows $\mathbf{x}_{k}$, density \eqref{eqn:optimal density} and weights \eqref{eqn:optimal weights} do not require previous particles $\{\mathbf{y}^{i}_{k-1}\}$ and hence, they need not be stored for the next recursion.

Initialize $\hat{\mathbf{x}}^{i}_{0}\sim\widetilde{\pi}^{x}_{0}(d\hat{\mathbf{x}}_{0})$ where $\widetilde{\pi}^{x}_{0}$ is the initial distribution assumed by the defender for the forward filter's initial estimate $\hat{\mathbf{x}}_{0}$. At time $(k-1)$, we have particles $\{\hat{\mathbf{x}}^{i}_{k-1}\}_{1\leq i\leq N}$ with equal weights ($1/N$) because of resampling. Finally, the I-PF recursions to compute the updated particles $\{\hat{\mathbf{x}}^{i}_{k}\}_{1\leq i\leq N}$ are as follows.
\begin{enumerate}
    \item \textit{SIS:} For $i=1,2,\hdots,N$, draw i.i.d. observation particles $\overline{\mathbf{y}}^{i}_{k}\sim\rho(\mathbf{y}_{k}|\mathbf{x}_{k})$ and then obtain state estimate particles $\hat{\overline{\mathbf{x}}}^{i}_{k}=T(\hat{\mathbf{x}}^{i}_{k-1},\overline{\mathbf{y}}^{i}_{k})$.
    \item \textit{Modification:} For a given threshold $\gamma_{k}>0$, check if $\frac{1}{N}\sum_{i=1}^{N}\beta(\mathbf{a}_{k}|\hat{\overline{\mathbf{x}}}^{i}_{k})\geq\gamma_{k}$. If the inequality is satisfied, we proceed to step 3 (similar to standard PF), otherwise, we return to the previous step and redraw particles from the sampling density.
    \item \textit{Weight computation:} Set $\hat{\widetilde{\mathbf{x}}}^{i}_{k}=\hat{\overline{\mathbf{x}}}^{i}_{k}$ and $\widetilde{\mathbf{y}}^{i}_{k}=\overline{\mathbf{y}}^{i}_{k}$ for $i=1,2,\hdots,N$. These particles estimate the prediction distribution $\pi_{k|k-1}$ as
    \par\noindent\small
    \begin{align*}
        \pi_{k|k-1}\approx\widetilde{\pi}^{N}_{k|k-1}(d\hat{\mathbf{x}}_{k},d\mathbf{y}_{k})\doteq\frac{1}{N}\sum_{i=1}^{N}\delta(\hat{\mathbf{x}}_{k}-\hat{\widetilde{\mathbf{x}}}^{i}_{k},\mathbf{y}_{k}-\widetilde{\mathbf{y}}^{i}_{k})d\hat{\mathbf{x}}_{k}d\mathbf{y}_{k}.
    \end{align*}
    \normalsize
    
    Since particles$\{\hat{\widetilde{\mathbf{x}}}^{i}_{k},\widetilde{\mathbf{y}}^{i}_{k}\}$ have equal weights due to resampling, using \eqref{eqn:optimal weights}, we compute the weights as
    \par\noindent\small
    \begin{align*}
    \widetilde{\omega}^{i}_{k}=\beta(\mathbf{a}_{k}|\hat{\widetilde{\mathbf{x}}}^{i}_{k}),\;\; i=1,2,\hdots,N,
    \end{align*}
    \normalsize
    and normalize $\omega^{i}_{k}=\widetilde{\omega}^{i}_{k}/\sum_{j=1}^{N}\widetilde{\omega}^{j}_{k}$. Using $\{\hat{\widetilde{\mathbf{x}}}^{i}_{k},\widetilde{\mathbf{y}}^{i}_{k}\}$ with weights $\{\omega^{i}_{k}\}$, we obtain the approximate posterior distribution
    \par\noindent\small
    \begin{align*}
    \pi_{k|k}\approx\widetilde{\pi}^{N}_{k|k}(d\hat{\mathbf{x}}_{k},d\mathbf{y}_{k})\doteq\sum_{i=1}^{N}\omega^{i}_{k}\delta(\hat{\mathbf{x}}_{k}-\hat{\widetilde{\mathbf{x}}}^{i}_{k},\mathbf{y}_{k}-\widetilde{\mathbf{y}}^{i}_{k})d\hat{\mathbf{x}}_{k}d\mathbf{y}_{k}.
    \end{align*}
    \normalsize
    \item \textit{Resampling:} Resample the particles by drawing $N$ independent particles as $(\hat{\mathbf{x}}^{i}_{k},\mathbf{y}^{i}_{k})\sim\widetilde{\pi}^{N}_{k|k}(d\hat{\mathbf{x}}_{k},d\mathbf{y}_{k})$. These uniformly weighted particles $\{\hat{\mathbf{x}}^{i}_{k},\mathbf{y}^{i}_{k}\}$ approximate $\pi_{k|k}$ as
    \par\noindent\small
    \begin{align*}
    \pi_{k|k}\approx\pi^{N}_{k|k}(d\hat{\mathbf{x}}_{k},d\mathbf{y}_{k})\doteq\frac{1}{N}\sum_{i=1}^{N}\delta(\hat{\mathbf{x}}_{k}-\hat{\mathbf{x}}^{i}_{k},\mathbf{y}_{k}-\mathbf{y}^{i}_{k})d\hat{\mathbf{x}}_{k}d\mathbf{y}_{k}.
    \end{align*}
    \normalsize
\end{enumerate}
Here, similar to \cite{hu2008basic}, we have introduced an optional modification (step 2) for convenience of the convergence analysis in the following section. As mentioned earlier, the resampled particles $\{\mathbf{y}^{i}_{k}\}$ in step 4 are not needed for the next recursion. If the defender estimates a function $\phi(\hat{\mathbf{x}},\mathbf{y})$ using the inverse filter's posterior distribution, the estimate $\mathbb{E}[\phi(\hat{\mathbf{x}},\mathbf{y})|\mathbf{x}_{0:k},\mathbf{a}_{1:k}]$ of $\phi(\hat{\mathbf{x}},\mathbf{y})$ is generally computed prior to resampling for better accuracy. For instance, we compute defender's state estimate $\doublehat{\mathbf{x}}_{k}$ as $\doublehat{\mathbf{x}}_{k}=\sum_{i=1}^{N}\omega^{i}_{k}\hat{\widetilde{\mathbf{x}}}^{i}_{k}$.
\begin{remark}[Threshold $\gamma_{k}$ intuition]\label{remark:threshold}
    Note that the optimal inverse filter \eqref{eqn:optimal filter time update}-\eqref{eqn:optimal filter measurement update} exists if $\langle\pi_{k|k-1},\beta\rangle>0$. In I-PF, we approximate $\pi_{k|k-1}$ by $\widetilde{\pi}^{N}_{k|k-1}$ such that
    \par\noindent\small
    \begin{align*}
    \langle\pi_{k|k-1},\beta\rangle\approx\langle\widetilde{\pi}^{N}_{k|k-1},\beta\rangle=\frac{1}{N}\sum_{i=1}^{N}\beta(\mathbf{a}_{k}|\hat{\widetilde{x}}^{i}_{k}).
    \end{align*}
    \normalsize
    In step 2, we require $\langle\widetilde{\pi}^{N}_{k|k-1},\beta\rangle\geq\gamma_{k}$. Hence, this condition is motivated by the existence of the optimal filter and has been previously used as an indicator of divergence in PFs \cite{crisan2002survey,hu2008basic}. The threshold $\gamma_{k}$ must be chosen so that the inequality is satisfied for sufficiently large $N$ and in practice, modifies the PF algorithm only for small $N$. Theorem~\ref{thm:ipf convergence} further guarantees that the algorithm will not run into an infinite loop (in steps 1 and 2) provided that $\gamma_{k}$ is chosen small enough.
\end{remark}
For the proposed I-PF, under the assumption of known forward filter, we need to be able to sample from observation distribution $\rho(\cdot)$ and compute the density $\beta(\mathbf{a}_{k}|\hat{\mathbf{x}}_{k})$ at current observation $\mathbf{a}_{k}$. I-PF can handle non-Gaussian systems if these conditions are met. Fig.~\ref{fig:ipf schematic} provides a schematic illustration of posterior distributions and their approximations.
\begin{remark}[I-PF's optimal importance density]\label{remark:IPF optimal density}
    For the considered defender-attacker dynamics, we are able to sample from the optimal density \eqref{eqn:optimal density} under the assumption of known forward filter. Another popular but suboptimal choice of $q(\cdot)$ is the transitional prior \cite{ristic2003beyond}, i.e., $q(\hat{\mathbf{x}}_{k},\mathbf{y}_{k}|\hat{\mathbf{x}}_{k-1},\mathbf{y}_{k-1},\mathbf{x}_{k},\mathbf{a}_{k})=p(\hat{\mathbf{x}}_{k},\mathbf{y}_{k}|\hat{\mathbf{x}}_{k-1},\mathbf{y}_{k-1})$ for I-PF. However, $p(\hat{\mathbf{x}}_{k},\mathbf{y}_{k}|\hat{\mathbf{x}}_{k-1},\mathbf{y}_{k-1})=p(\hat{\mathbf{x}}_{k}|\hat{\mathbf{x}}_{k-1},\mathbf{y}_{k})p(\mathbf{y}_{k}|\mathbf{y}_{k-1})$ such that to sample from the transitional prior, we need to sample from $p(\mathbf{y}_{k}|\mathbf{y}_{k-1})$. Hence, contrary to the standard PF, it is easier to sample from the I-PF's optimal density \eqref{eqn:optimal density} than the transitional prior. This is possible because of the perfect knowledge of actual state $\mathbf{x}_{k}$ available to the defender such that we can directly sample from $p(\mathbf{y}_{k}|\mathbf{x}_{k})$.
\end{remark}
\begin{figure}
  \centering
  \includegraphics[width = 0.9\columnwidth]{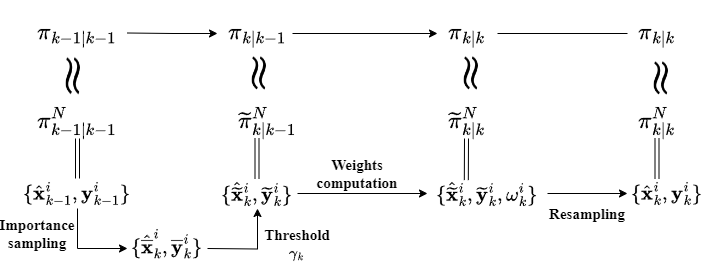}
  \caption{Graphical representation of transformation of posterior distributions in I-PF.}
 \label{fig:ipf_schematic}
\end{figure}

\textbf{I-PF variants:} As in the case of standard PF, resampling in I-PF reduces degeneracy of particles over time, but introduces sample impoverishment \cite{arulampalam2002tutorial}. Since the particles with large weights are statistically selected many times, there is a loss of diversity among particles, which may also lead to collapse to a single point in case of small system noises. Different techniques like Markov chain MC move step \cite{carlin1992monte} and regularization \cite{musso2001improving} have been proposed to address sample impoverishment. The I-PF developed here is a basic inverse filter based on the SIS and resampling techniques. Different choices of sampling density $q(\cdot)$ and/or modification of the resampling step lead to different variants of PFs. Similar modifications can also be introduced to the basic I-PF to obtain suitable variants for different applications. For instance, in auxiliary PF \cite{pitt1999filtering}, the previous particles are resampled conditioned on the current measurement before importance sampling such that the particles are most likely close to the current true state. On the other hand, regularized PF \cite{musso2001improving} considers a kernel density $K_{h}(\cdot)$ to resample from a continuous distribution instead of a discrete one, i.e., $(\hat{\mathbf{x}}^{i}_{k},\mathbf{y}^{i}_{k})\sim\sum_{i=1}^{N}\omega^{i}_{k}K_{h}(\hat{\mathbf{x}}_{k}-\hat{\widetilde{\mathbf{x}}}^{i}_{k},\widetilde{\mathbf{y}}_{k}-\mathbf{y}^{i}_{k})$ for $i=1,2,\hdots,N$.

PFs are frequently integrated into ML models for efficient state estimation and flexible parameter learning. For instance, \cite{ma2020particle} suggests PF-based recurrent NNs for robot localization and sequence prediction tasks. In \cite{gu2015neural}, a neural adaptive SMC method is shown to be effective in training a latent variable RNN. A single NN encodes the system model, and PF algorithm in \cite{karkus2018particle} to learn a model optimized for PF instead of a generic model. Other examples of PF applications include online learning of random forests \cite{ko2013human} and RL-based wireless indoor positioning system \cite{villacres2019particle}. Additionally, \cite{ma2020discriminative} builds a discriminative PF-based RL framework for decision-making in partially observable Markov decision processes (POMDPs). Other applications include variational autoencoders \cite{burda2015importance}, sequence generation \cite{le2018auto}, and variational RL \cite{igl2018deep}. Inverse filtering to infer the information learned by these systems would then resort to I-PF and, particularly, differentiable I-PF for unknown system dynamics presented in Section~\ref{subsec:dpf}.

\subsection{Convergence guarantees}\label{subsec:IPF convergence}
In PFs, the particles interact and are not statistically independent. Consequently, classical convergence results for MC methods, which rely on central limit theorems under the assumption of i.i.d. samples, are not applicable. Despite this, PFs are designed to approximate optimal filters, and their convergence to the true posterior distribution must be investigated. Several convergence results for the standard PF have been provided in the literature \cite{moral2004feynman,crisan2002survey}. In \cite{le2004stability}, a Hilbert projective metric is considered to study the optimal filter's stability, which is then used to derive uniform convergence conditions for PFs. The survey by \cite{crisan2002survey} showed almost sure convergence assuming a Feller transition kernel and a bounded, continuous, strictly positive observation likelihood. Other approaches include central limit theorems  \cite{del1999central} and large deviations \cite{crisan1999large,del1998large}. However, all these prior works assume the (estimated) function $\phi(\mathbf{x})$ of the underlying state $\mathbf{x}$ to be bounded and hence, exclude the state estimate itself, i.e., $\phi(\mathbf{x})=\mathbf{x}$. Recently, Hu et al. \cite{hu2008basic} have addressed the general case of unbounded function $\phi$ and proved PF's convergence in $L^{4}$-sense under some mild assumptions on the rate of increase of $\phi$. An extended result in the $L^{p}$-convergence sense has also been obtained using a Rosenthal-type inequality in \cite{hu2011general}.

In the following, we consider the $L^4$-approach of \cite{hu2008basic} to derive our I-PF's convergence conditions. In particular, we show that for a given time step $k$, and given observations $\mathbf{a}_{1:k}$ and known true states $\mathbf{x}_{0:k}$, the I-PF's estimate $\langle\pi^{N}_{k|k},\phi\rangle$ converges to the optimal filter's estimate $\langle\pi_{k|k},\phi\rangle$ as the number of particles $N$ increases. Note that for given $\{\mathbf{a}_{1:k},\mathbf{x}_{0:k}\}$, the optimal distribution $\pi_{k|k}$ is a deterministic function, but $\pi^{N}_{k|k}$ is random because of the randomly generated particles. Hence, all the stochastic expectations $\mathbb{E}$ and almost sure convergence are with respect to these random particles.

We assume that the system model \eqref{eqn:state x}-\eqref{eqn:observation a} and estimated function $\phi$ satisfy the following conditions.\\
\textbf{A7.a)} For given $\mathbf{x}_{0:s}$ and $\mathbf{a}_{1:s}$ for $s=1,2,\hdots,k$, $\langle\pi_{s|s-1},\beta\rangle>0$ and there exits $\{\gamma_{s}\}_{1\leq s\leq k}$ such that $0<\gamma_{s}<\langle\pi_{s|s-1},\beta\rangle$.\\
\textbf{A7.b)} The observation densities are bounded, i.e., $\beta(\mathbf{a}_{s}|\hat{\mathbf{x}}_{s})<\infty$ and $\rho(\mathbf{y}_{s}|\mathbf{x}_{s})<\infty$ for $s=1,2,\hdots,k$.\\
\textbf{A7.c)} The function $\phi$ satisfies $\textrm{sup}_{(\hat{\mathbf{x}}_{s},\mathbf{y}_{s})} |\phi(\hat{\mathbf{x}}_{s},\mathbf{y}_{s})|^{4}\beta(\mathbf{a}_{s}|\hat{\mathbf{x}}_{s})<C(\mathbf{x}_{0:s},\mathbf{a}_{1:s})$ for given $\mathbf{x}_{0:s},\mathbf{a}_{1:s}, s=1,2,\hdots,k$. Here, $C(\mathbf{x}_{0:s},\mathbf{a}_{1:s})$ is a finite constant which may depend on $\mathbf{x}_{0:s}$ and $\mathbf{a}_{1:s}$.

Recall that $\{\gamma_{s}\}$ are the thresholds introduced in the modification step (step 2) of the I-PF algorithm in Section~\ref{subsec:IPF formulation}. As discussed in Remark~\ref{remark:threshold}, assumption \textbf{A7.a} is related to the existence of the optimal filter and divergence of PFs. Intuitively, assumption \textbf{A7.c} states that the conditional observation density $\beta$ must decrease at a rate faster than the function $\phi$ increases. Furthermore, \textbf{A7.b} and \textbf{A7.c} imply that the conditional fourth moment of $\phi$ is bounded, i.e., $\langle\pi_{s|s},|\phi|^{4}\rangle<\infty$ \cite{hu2008basic}. Finally, the I-PF convergence is provided in the following theorem.
\begin{theorem}[I-PF convergence]\label{thm:ipf convergence}
Consider the I-PF developed in Section~\ref{subsec:IPF formulation} (including the modification step). If the assumptions \textbf{A7.a}-\textbf{A7.c} are satisfied, then the following hold:\\
\textbf{a)} For sufficiently large $N$, the algorithm will not run into an infinite loop in steps $1-2$.\\
\textbf{b)} For any $\phi$ satisfying \textbf{A7.c}, there exists a constant $C_{k|k}$, independent of $N$ such that
\par\noindent\small
\begin{align}
   \mathbb{E}|\langle\pi^{N}_{k|k},\phi\rangle-\langle\pi_{k|k},\phi\rangle|^{4}\leq C_{k|k}\frac{\|\phi\|^{4}_{k,4}}{N^{2}},\label{eqn:ipf converge}
\end{align}
\normalsize
where $\|\phi\|_{k,4}\doteq\textrm{max}\{1,\textrm{max}_{\;0\leq s\leq k}\langle\pi_{s|s},|\phi|^{4}\rangle^{1/4}\}$ and $\pi^{N}_{k|k}$ is generated by the I-PF algorithm.
\end{theorem}
\begin{proof}
    See Appendix~\ref{app:ipf convergence}.
\end{proof}

As a consequence of Theorem~\ref{thm:ipf convergence}, the following corollary can be obtained trivially using the Borel-Cantelli lemma as in \cite[Proposition~7.2.3(a)]{athreya2006measure}.
\begin{corollary}\label{cor:almost sure convergence}
If \textbf{A7.a}-\textbf{A7.c} holds, then for any $\phi$ satisfying \textbf{A7.c}, we have $\lim_{N\to\infty}\langle\pi^{N}_{k|k},\phi\rangle=\langle\pi_{k|k},\phi\rangle$ in the almost sure convergence sense.
\end{corollary}
\begin{remark}[Dependence of state dimension]
    From \eqref{eqn:ipf converge}, we observe that the I-PF's convergence rate does not depend on the state dimension $n_{x}$ and hence, I-PF does not suffer from the curse of dimensionality. However, for a given/desired bound on the error, the required number of particles $N$ depends on constant $C_{k|k}$, which can depend on $n_{x}$.
\end{remark}
\begin{remark}[Bound on $C_{k|k}$]
    In general, without any additional assumptions, we cannot guarantee that $C_{k|k}$ will not increase over time. In particular, if the optimal filter does not `forget' its initial condition, the approximation errors accumulate over time such that $C_{k|k}$ increases \cite{crisan2002survey}. Hence, the required number of particles $N$ also increases proportionally for the error to remain within a given bound. However, prior works \cite{del2001stability,le2004stability} provide additional conditions on the system model to ensure the optimal filter mixes quickly (forgets its initial condition) and $C_{k|k}$ does not increase with time.
\end{remark}

\section{Inverse Gaussian PF and Ensemble KF}\label{sec:igpf and ienkf}
The I-PF developed in Section~\ref{sec:IPF} considers a general inverse filtering model without any restrictive assumptions on the system model. However, in many practical applications, the process and observation noises are reasonably assumed to be Gaussian such that the Gaussian posterior assumption can provide a better tradeoff between the computational complexity and the performance of the filter. GPF and EnKF are two popular SMC-based filters that assume Gaussian posteriors. While GPF employs a PF framework to obtain the state estimates, EnKF extends the standard KF to non-linear dynamics using SMC methods. In the following, we consider similar techniques for the inverse filtering problem and develop I-GPF and I-EnKF in Section~\ref{subsec:igpf} and \ref{subsec:ienkf}, respectively. Section~\ref{subsec:non-Gaussian} discusses the modifications of these filters to handle non-Gaussian noises in the inverse filtering context. Finally, Table~\ref{tbl:differences} highlights the key differences between I-PF, I-GPF, and I-EnKF algorithms.

\subsection{Inverse GPF}\label{subsec:igpf}
Recall that GPF presents an MC-based extension of Gaussian filters. However, unlike EKF or UKF, GPF's estimate asymptotically (as number of particles $N\to\infty$) converges almost surely to the actual MMSE estimate provided that the Gaussian assumption holds true \cite{kotecha2003gaussian}. Also, contrary to PF, GPF does not suffer from particle degeneracy and hence, resampling is not required. This results in reduced computational complexity and ease of parallel implementation. As mentioned earlier, standard PF does not assume any posterior form and approximates the arbitrary posterior with randomly generated particles. Under the Gaussianity assumption, GPF propagates only the mean and covariance of the posterior, but PF propagates higher-order moments as well.

Consider the optimal filter recursions \eqref{eqn:time update} and \eqref{eqn:measurement update}. In I-GPF, we assume a marginal Gaussian distribution for the attacker's state estimate as $p(\hat{\mathbf{x}}_{k}|\mathbf{x}_{0:k},\mathbf{a}_{1:k})\approx\mathcal{N}(\hat{\mathbf{x}}_{k};\doublehat{\mathbf{x}}_{k},\overline{\bm{\Sigma}}_{k})$. This is in contrast to I-PF developed in Section~\ref{sec:IPF}, wherein we aimed to estimate the joint distribution $p(\hat{\mathbf{x}}_{k},\mathbf{y}_{k}|\mathbf{x}_{0:k},\mathbf{a}_{1:k})$. The defender employing the inverse filter has perfect knowledge of its true state $\mathbf{x}_{k}$ and hence, using \eqref{eqn:observation y}, we approximate $p(\mathbf{y}_{k}|\mathbf{x}_{0:k},\mathbf{a}_{1:k})\approx\rho(\mathbf{y}_{k}|\mathbf{x}_{k})$. Overall, we assume
\par\noindent\small
\begin{align}
    \pi_{k|k}\approx\mathcal{N}(\hat{\mathbf{x}}_{k};\doublehat{\mathbf{x}}_{k},\overline{\bm{\Sigma}}_{k})\rho(\mathbf{y}_{k}|\mathbf{x}_{k}).\label{eqn:GPF posterior}
\end{align}
\normalsize
Note that this is an approximation because $\hat{\mathbf{x}}_{k}$ and $\mathbf{y}_{k}$ are not indeed independent. Furthermore, $\mathbf{y}_{k}$ is also correlated with observation $\mathbf{a}_{k}$ via estimate $\hat{\mathbf{x}}_{k}$, but we ignore this dependence. The knowledge of true distribution $\rho(\cdot|\mathbf{x}_{k})$ can compensate for this approximation to some extent. Similarly, we assume a Gaussian prediction distribution as $p(\hat{\mathbf{x}}_{k}|\mathbf{x}_{0:k},\mathbf{a}_{1:k-1})\approx\mathcal{N}(\hat{\mathbf{x}}_{k};\doublehat{\mathbf{x}}_{k|k-1},\overline{\bm{\Sigma}}_{k|k-1})$. Finally, the means ($\doublehat{\mathbf{x}}_{k}$ and $\doublehat{\mathbf{x}}_{k|k-1}$) and covariances ($\overline{\bm{\Sigma}}_{k}$ and $\overline{\bm{\Sigma}}_{k|k-1}$) are computed as weighted MC estimates using the PF framework.

Consider $N$ as the total number of particles. We initialize I-GPF with initial distribution $\mathcal{N}(\hat{\mathbf{x}}_{0};\doublehat{\mathbf{x}}_{0},\overline{\bm{\Sigma}}_{0})$. Denote $\{(\hat{\mathbf{x}}^{i}_{k|k-1},\mathbf{y}^{i}_{k}\}_{1\leq i\leq N}$ as the particles generated for the time-update/ prediction step, while $\{\hat{\mathbf{x}}^{i}_{k}\}_{1\leq i\leq N}$ are the particles generated in the measurement update with associated weights $\{\omega^{i}_{k}\}_{1\leq i\leq N}$. Similar to I-PF, particle sampling and weight computation do not require previous particles $\{\mathbf{y}^{i}_{k-1}\}$. At the $k$-th time instant, I-GPF computes estimate $\doublehat{\mathbf{x}}_{k}$ of $\hat{\mathbf{x}}_{k}$ and the corresponding covariance estimate $\overline{\bm{\Sigma}}_{k}$ recursively as
\par\noindent\small
\begin{align}
    &\textrm{Time update:}\;\;\{\hat{\widetilde{\mathbf{x}}}^{i}_{k-1}\}_{1\leq i\leq N}\sim\mathcal{N}(\hat{\mathbf{x}}_{k-1};\doublehat{\mathbf{x}}_{k-1},\overline{\bm{\Sigma}}_{k-1}),\;\;\;\{\mathbf{y}^{i}_{k}\}_{1\leq i\leq N}\sim\rho(\mathbf{y}_{k}|\mathbf{x}_{k}),\label{eqn:IGPF y gen}\\
    &\hat{\mathbf{x}}^{i}_{k|k-1}=T(\hat{\widetilde{\mathbf{x}}}^{i}_{k-1},\mathbf{y}^{i}_{k}),\;\;\;\textrm{for}\;i=1,2,\hdots,N,\label{eqn:IGPF T}\\
    &\doublehat{\mathbf{x}}_{k|k-1}=\frac{1}{N}\sum_{i=1}^{N}\hat{\mathbf{x}}^{i}_{k|k-1},\;\;\;\overline{\bm{\Sigma}}_{k|k-1}=\frac{1}{N}\sum_{i=1}^{N}(\hat{\mathbf{x}}^{i}_{k|k-1}-\doublehat{\mathbf{x}}_{k|k-1})(\hat{\mathbf{x}}^{i}_{k|k-1}-\doublehat{\mathbf{x}}_{k|k-1})^{T},\label{eqn:IGPF Sig predict}\\
&\textrm{Measurement update:}\;\;\{\hat{\mathbf{x}}^{i}_{k}\}_{1\leq i\leq N}\sim\mathcal{N}(\hat{\mathbf{x}}_{k};\doublehat{\mathbf{x}}_{k|k-1},\overline{\bm{\Sigma}}_{k|k-1}),\nonumber\\
    &\widetilde{\omega}^{i}_{k}=\beta(\mathbf{a}_{k}|\hat{\mathbf{x}}^{i}_{k}),\;\;\;\textrm{for}\;i=1,2,\hdots,N,\;\;\;\omega^{i}_{k}=\widetilde{\omega}^{i}_{k}/\sum_{i=1}^{N}\widetilde{\omega}^{i}_{k},\;\;\;\textrm{for}\;i=1,2,\hdots,N,\nonumber\\
    &\doublehat{\mathbf{x}}_{k}=\sum_{i=1}^{N}\omega^{i}_{k}\hat{\mathbf{x}}^{i}_{k},\;\;\;\overline{\bm{\Sigma}}_{k}=\sum_{i=1}^{N}\omega^{i}_{k}(\hat{\mathbf{x}}^{i}_{k}-\doublehat{\mathbf{x}}_{k})(\hat{\mathbf{x}}^{i}_{k}-\doublehat{\mathbf{x}}_{k})^{T}.\nonumber
\end{align}
\normalsize
Here, similar to I-PF, \eqref{eqn:IGPF y gen}-\eqref{eqn:IGPF T} represent importance sampling with the optimal sampling density \eqref{eqn:optimal density} and a Gaussian posterior distribution on $\hat{\mathbf{x}}_{k-1}$. The particles are drawn identically with equal weights such that the prediction distribution $\mathcal{N}(\hat{\mathbf{x}}_{k};\doublehat{\mathbf{x}}_{k|k-1},\overline{\bm{\Sigma}}_{k|k-1})$ is obtained in \eqref{eqn:IGPF Sig predict}. Finally, in the measurement update, we sample from the prediction distribution and compute the associated weights using \eqref{eqn:optimal weights}. Overall, we need to sample twice from Gaussian densities and once from the observation distribution $\rho(\cdot)$ at each recursion. Alternatively, sampling $\{\hat{\widetilde{\mathbf{x}}}^{i}_{k-1}\}$ can be omitted and weighted samples $\{\hat{\mathbf{x}}^{i}_{k-1},\omega^{i}_{k-1}\}$ from previous time-step can be considered themselves \cite{kotecha2003gaussian}. In this case, $\hat{\widetilde{\mathbf{x}}}^{i}_{k-1}=\hat{\mathbf{x}}^{i}_{k-1}$ for all $i=1,2,\hdots,N$, and \eqref{eqn:IGPF Sig predict} becomes
\par\noindent\small
\begin{align*}
    &\doublehat{\mathbf{x}}_{k|k-1}=\sum_{i=1}^{N}\omega^{i}_{k-1}\hat{\mathbf{x}}^{i}_{k|k-1},\;\;\;\overline{\bm{\Sigma}}_{k|k-1}=\sum_{i=1}^{N}\omega^{i}_{k-1}(\hat{\mathbf{x}}^{i}_{k|k-1}-\doublehat{\mathbf{x}}_{k|k-1})(\hat{\mathbf{x}}^{i}_{k|k-1}-\doublehat{\mathbf{x}}_{k|k-1})^{T}.
\end{align*}
\normalsize

\subsection{Inverse EnKF}\label{subsec:ienkf}
Here, we develop I-EnKF for a non-linear system with additive noises given by \eqref{eqn:non-linear state x}-\eqref{eqn:non-linear observation a}. Analogous to standard EnKF, we assume $\mathbf{w}_{k}$, $\mathbf{v}_{k}$ and $\bm{\epsilon}_{k}$ to be Gaussian with distributions $\mathcal{N}(\mathbf{w}_{k};\mathbf{0},\mathbf{Q}_{k})$, $\mathcal{N}(\mathbf{v}_{k};\mathbf{0},\mathbf{R}_{k})$ and $\mathcal{N}(\bm{\epsilon}_{k};\mathbf{0},\overline{\mathbf{R}}_{k})$, respectively. Unlike I-PF and I-GPF, the attacker is assumed to employ EnKF as its forward filter instead of a general filter $T(\cdot)$.

\textit{Forward filter:} Consider $q$ as the forward EnKF's ensemble size and $\{\mathbf{x}^{i}_{k}\}_{1\leq i\leq q}$ represents the ensemble at $k$-th time step. Denote $\{\mathbf{w}^{i}_{k-1}\}_{1\leq i\leq q}$ and $\{\mathbf{v}^{i}_{k}\}_{1\leq i\leq q}$ as the i.i.d. noise samples drawn from $\mathcal{N}(\mathbf{w}_{k-1};\mathbf{0},\mathbf{Q}_{k-1})$ and $\mathcal{N}(\mathbf{v}_{k};\mathbf{0},\mathbf{R}_{k})$, respectively. In the time update step, the previous ensemble $\{\mathbf{x}^{i}_{k-1}\}$ is propagated through state evolution \eqref{eqn:non-linear state x} to obtain the predicted ensemble $\{\mathbf{x}^{i}_{k|k-1}\}$, which then yields the predicted observation ensemble $\{\mathbf{y}^{i}_{k}\}$ using \eqref{eqn:non-linear observation y}. Finally, in the measurement update, KF-like linear updates are employed to obtain the current ensemble $\{\mathbf{x}^{i}_{k}\}$ with the gain matrix approximated using the ensemble covariances. The overall forward EnKF recursions are summarized as \cite{gillijns2006ensemble}:
\par\noindent\small
\begin{align}
&\textrm{Time update:}\;\;\mathbf{x}^{i}_{k|k-1}=f(\mathbf{x}^{i}_{k-1})+\mathbf{w}^{i}_{k-1},\;\textrm{for}\;i=1,2,\hdots,q,\label{eqn:forward EnKF prediction ensemble}\\
&\hat{\mathbf{x}}_{k|k-1}=\frac{1}{q}\sum_{i=1}^{q}\mathbf{x}^{i}_{k|k-1},\;\;\;\mathbf{E}^{x}_{k|k-1}=\left[(\mathbf{x}^{1}_{k|k-1}-\hat{\mathbf{x}}_{k|k-1}),\hdots,(\mathbf{x}^{q}_{k|k-1}-\hat{\mathbf{x}}_{k|k-1})\right],\nonumber
\end{align}
\begin{align}
&\textrm{Time update (contd.):}\;\;\mathbf{y}^{i}_{k}=h(\mathbf{x}^{i}_{k|k-1})+\mathbf{v}^{i}_{k},\;\textrm{for}\;i=1,2,\hdots,q,\nonumber\\
&\hat{\mathbf{y}}_{k}=\frac{1}{q}\sum_{i=1}^{q}\mathbf{y}^{i}_{k},\;\;\mathbf{E}^{y}_{k}=\left[(\mathbf{y}^{1}_{k}-\hat{\mathbf{y}}_{k}),\hdots,(\mathbf{y}^{q}_{k}-\hat{\mathbf{y}}_{k})\right],\nonumber\\
&\bm{\Sigma}^{xy}_{k}=\frac{1}{q-1}\mathbf{E}^{x}_{k|k-1}(\mathbf{E}^{y}_{k})^{T},\;\;\bm{\Sigma}^{y}_{k}=\frac{1}{q-1}\mathbf{E}^{y}_{k}(\mathbf{E}^{y}_{k})^{T},\nonumber\\
&\textrm{Measurement update:}\;\;\mathbf{K}_{k}=\bm{\Sigma}^{xy}_{k}(\bm{\Sigma}^{y}_{k})^{-1},\nonumber\\
&\mathbf{x}^{i}_{k}=\mathbf{x}^{i}_{k|k-1}+\mathbf{K}_{k}(\mathbf{y}_{k}-\mathbf{y}^{i}_{k}),\;\textrm{for}\;i=1,2,\hdots,q,\label{eqn:forward EnKF update}
\end{align}
\normalsize
and state estimate $\hat{\mathbf{x}}_{k}=\frac{1}{q}\sum_{i=1}^{q}\mathbf{x}^{i}_{k}$. As compared to other SMC algorithms (like PF), which use reweighting and resampling, EnKF employs `linear-shifting' to update the prior ensemble to a posterior ensemble after each observation \cite{katzfuss2016understandingEnKf}. Consequently, EnKF does not suffer from any weight/particle degeneracy problem.

\textit{Inverse filter:} Unlike I-KF \cite{krishnamurthy2019how}, I-EKF, and I-UKF, it is impractical to obtain the state-transition for the forward EnKF's state estimate $\hat{\mathbf{x}}_{k}$ even under the known forward filter assumption. This is because of the random noise samples involved in the computation of the forward EnKF's gain matrix. Hence, in I-EnKF, we consider propagating the updated state ensemble further through the defender's observation model \eqref{eqn:non-linear observation a}. In particular, we start with the previous ensemble $\{\hat{\mathbf{x}}^{j}_{k-1}\}_{1\leq j\leq\overline{q}}$ of size $\overline{q}$, where $\overline{q}$ is chosen independent of the forward EnKF's size $q$. This ensemble is propagated through state evolution \eqref{eqn:non-linear state x} and attacker's observation \eqref{eqn:non-linear observation y} similar to the forward filter. An intermediate updated ensemble $\{\overline{\mathbf{x}}^{j}_{k|k-1}\}$ is then obtained from the forward EnKF's measurement update \eqref{eqn:forward EnKF update} with the actual observation $\mathbf{y}_{k}$ replaced by simulated observation generated using true state $\mathbf{x}_{k}$ according to \eqref{eqn:non-linear observation y}. Finally, the intermediate ensemble is propagated through observation model \eqref{eqn:non-linear observation a} resulting in observation ensemble $\{\mathbf{a}^{j}_{k}\}$. Linear KF-like updates then yield the I-EnKF's updated ensemble $\{\hat{\mathbf{x}}^{j}_{k}\}$. I-EnKF's recursions to infer estimate $\doublehat{\mathbf{x}}_{k}$ of $\hat{\mathbf{x}}_{k}$ are summarized as:
\par\noindent\small
\begin{align*}
&\textrm{State evolution update:}\;\;\widetilde{\mathbf{x}}^{j}_{k|k-1}=f(\hat{\mathbf{x}}^{j}_{k-1})+\widetilde{\mathbf{w}}^{j}_{k-1},\;\textrm{for}\;j=1,2,\hdots,\overline{q},\\
&\hat{\widetilde{\mathbf{x}}}_{k|k-1}=\frac{1}{\overline{q}}\sum_{j=1}^{\overline{q}}\widetilde{\mathbf{x}}^{j}_{k|k-1},\;\;\;\widetilde{\mathbf{E}}^{x}_{k|k-1}=\left[(\widetilde{\mathbf{x}}^{1}_{k|k-1}-\hat{\widetilde{\mathbf{x}}}_{k|k-1}),\hdots,(\widetilde{\mathbf{x}}^{\overline{q}}_{k|k-1}-\hat{\widetilde{\mathbf{x}}}_{k|k-1})\right],\\
&\textrm{Intermediate update:}\;\;\widetilde{\mathbf{y}}^{j}_{k}=h(\widetilde{\mathbf{x}}^{j}_{k|k-1})+\widetilde{\mathbf{v}}^{j}_{k},\;\textrm{for}\;j=1,2,\hdots,\overline{q},\\
&\hat{\widetilde{\mathbf{y}}}_{k}=\frac{1}{\overline{q}}\sum_{j=1}^{\overline{q}}\widetilde{\mathbf{y}}^{j}_{k},\;\widetilde{\mathbf{E}}^{y}_{k}=\left[(\widetilde{\mathbf{y}}^{1}_{k}-\hat{\widetilde{\mathbf{y}}}_{k}),\hdots,(\widetilde{\mathbf{y}}^{\overline{q}}_{k}-\hat{\widetilde{\mathbf{y}}}_{k})\right],\\
&\widetilde{\bm{\Sigma}}^{xy}_{k}=\frac{1}{\overline{q}-1}\widetilde{\mathbf{E}}^{x}_{k|k-1}(\widetilde{\mathbf{E}}^{y}_{k})^{T},\;\widetilde{\bm{\Sigma}}^{y}_{k}=\frac{1}{\overline{q}-1}\widetilde{\mathbf{E}}^{y}_{k}(\widetilde{\mathbf{E}}^{y}_{k})^{T},\\
&\widetilde{\mathbf{K}}_{k}=\widetilde{\bm{\Sigma}}^{xy}_{k}(\widetilde{\bm{\Sigma}}^{y}_{k})^{-1},\;\;\;\overline{\mathbf{x}}^{j}_{k|k-1}=\widetilde{\mathbf{x}}^{j}_{k|k-1}+\widetilde{\mathbf{K}}_{k}(h(\mathbf{x}_{k})+\overline{\mathbf{v}}^{j}_{k}-\widetilde{\mathbf{y}}^{j}_{k}),\;\textrm{for}\;j=1,2,\hdots,\overline{q},\\
&\doublehat{\mathbf{x}}_{k|k-1}=\frac{1}{\overline{q}}\sum_{j=1}^{\overline{q}}\overline{\mathbf{x}}^{j}_{k|k-1},\;\;\;\overline{\mathbf{E}}^{x}_{k|k-1}=\left[(\overline{\mathbf{x}}^{1}_{k|k-1}-\doublehat{\mathbf{x}}_{k|k-1}),\hdots,(\overline{\mathbf{x}}^{\overline{q}}_{k|k-1}-\doublehat{\mathbf{x}}_{k|k-1})\right],\\
&\textrm{Measurement update:}\;\;\mathbf{a}^{j}_{k}=g(\overline{\mathbf{x}}^{j}_{k|k-1})+\bm{\epsilon}^{j}_{k},\;\textrm{for}\;j=1,2,\hdots,\overline{q},\\
&\hat{\mathbf{a}}_{k}=\frac{1}{\overline{q}}\sum_{j=1}^{\overline{q}}\mathbf{a}^{j}_{k},\;\mathbf{E}^{a}_{k}=\left[(\mathbf{a}^{1}_{k}-\hat{\mathbf{a}}_{k}),\hdots,(\mathbf{a}^{\overline{q}}_{k}-\hat{\mathbf{a}}_{k})\right],
\end{align*}
\begin{align*}
&\textrm{Measurement update (contd.):}\;\;\overline{\bm{\Sigma}}^{xa}_{k}=\frac{1}{\overline{q}-1}\overline{\mathbf{E}}^{x}_{k|k-1}(\mathbf{E}^{a}_{k})^{T},\;\;\;\overline{\bm{\Sigma}}^{a}_{k}=\frac{1}{\overline{q}-1}\mathbf{E}^{a}_{k}(\mathbf{E}^{a}_{k})^{T},\\
&\overline{\mathbf{K}}_{k}=\overline{\bm{\Sigma}}^{xa}_{k}(\overline{\bm{\Sigma}}^{a}_{k})^{-1},\;\;\;\hat{\mathbf{x}}^{j}_{k}=\overline{\mathbf{x}}^{j}_{k|k-1}+\overline{\mathbf{K}}_{k}(\mathbf{a}_{k}-\mathbf{a}^{j}_{k}),\;\textrm{for}\;j=1,2,\hdots,\overline{q},
\end{align*}
\normalsize
where $\{\widetilde{\mathbf{w}}^{j}_{k-1}\}$, $\{\overline{\mathbf{v}}^{j}_{k},\widetilde{\mathbf{v}}^{j}_{k}\}$ and $\{\bm{\epsilon}^{j}_{k}\}$ are noise samples drawn from $\mathcal{N}(\mathbf{w}_{k-1};\mathbf{0},\mathbf{Q}_{k-1})$, $\mathcal{N}(\mathbf{v}_{k};\mathbf{0},\mathbf{R}_{k})$ and $\mathcal{N}(\bm{\epsilon}_{k};\mathbf{0},\overline{\mathbf{R}}_{k})$, respectively. I-EnKF's estimate $\doublehat{\mathbf{x}}_{k}=\frac{1}{\overline{q}}\sum_{j=1}^{\overline{q}}\hat{\mathbf{x}}^{j}_{k}$. Note that the intermediate gain matrix $\widetilde{\mathbf{K}}_{k}$ aims to approximate the forward EnKF's gain matrix $\mathbf{K}_{k}$. However, it differs from $\mathbf{K}_{k}$ because $\widetilde{\mathbf{K}}_{k}$ is computed from previous ensemble $\{\hat{\mathbf{x}}^{j}_{k-1}\}$, which in turn has been propagated though observation \eqref{eqn:non-linear observation a} at the previous time step. On the other hand, forward EnKF's ensemble propagates only through state evolution \eqref{eqn:non-linear state x} and observation \eqref{eqn:non-linear observation y}. But, similar to the standard EnKF, our I-EnKF is observed to be robust to this deviation in different numerical examples considered in Sections~\ref{subsec:van der pol} and \ref{subsec:heat conduction}.
\begin{remark}[Deterministic I-EnKF]
    Unlike UKF which considers deterministic sigma-points, the forward EnKF and I-EnKF considered here work with a randomly generated ensemble. While the number of sigma-points in UKF are of the same order as the state dimension $n_{x}$, the required ensemble size in EnKF is a heuristic \cite{gillijns2006ensemble}. However, deterministic update based EnKFs have also been proposed in the literature \cite{anderson2001ensemble,bishop2001adaptive,hunt2007efficient}, wherein the prior ensemble is deterministically shifted to the posterior ensemble without relying on simulated observations. In particular, the prior ensemble members are shifted and scaled such that the resulting posterior ensemble members are shifted towards the data with smaller variance than the prior. Analogous approaches can be applied to obtain deterministic variants of the proposed I-EnKF by suitably modifying the intermediate and measurement updates.
\end{remark}

\subsection{Non-Gaussian systems}\label{subsec:non-Gaussian}
In order to employ I-GPF, we only need to be able to sample from observation distribution $\rho(\cdot)$ and compute density $\beta(\cdot)$ at the generated particles. As far as these conditions are satisfied, the defender can employ I-GPF to estimate the attacker's estimate in non-Gaussian systems as well. In fact, I-GPF may outperform I-PF if the posterior state distributions are well approximated by Gaussians, as demonstrated in our numerical experiments in Section~\ref{subsec:bearing}. For the case when the posterior distribution cannot be assumed to be a single Gaussian, Gaussian-sum PFs (GSPFs) have been proposed in \cite{kotecha2003gaussian_sum} based on Gaussian mixture (GM) models using a bank of parallel EKFs or GPFs. In particular, \cite{kotecha2003gaussian_sum} considered two different approaches: (a) the prediction and posterior distributions are approximated by weighted GMs, and (b) the additive process and observation noises are modeled as a finite sum of Gaussians. In the latter case, the non-Gaussian state-space model itself is equivalent to a bank of parallel Gaussian state-space models, each of which can be approximated by (single) Gaussian or GM posteriors. However, in this case, the number of Gaussians increases exponentially with time, but only a few have significant weights. Hence, resampling is introduced to propagate a fixed number of Gaussians at each iteration. Similar approaches can be used to modify I-GPF to address severe non-Gaussianity in the inverse filtering model at a lower computational cost than I-PF. 

\noindent\textbf{Inverse GSPF:} We obtain I-GSPF by approximating the prediction and posterior densities as a mixture of $G$ Gaussians, i.e., $p(\hat{\mathbf{x}}_{k}|\mathbf{x}_{0:k},\mathbf{a}_{1:k-1})\approx\sum_{j=1}^{G}\gamma^{j}_{k|k-1}\mathcal{N}(\hat{\mathbf{x}}_{k};\bm{\mu}^{j}_{k|k-1},\bm{\Sigma}^{j}_{k|k-1})$ and $p(\hat{\mathbf{x}}_{k}|\mathbf{x}_{0:k},\mathbf{a}_{1:k})\approx\sum_{j=1}^{G}\gamma^{j}_{k}\mathcal{N}(\hat{\mathbf{x}}_{k};\bm{\mu}^{j}_{k},\bm{\Sigma}^{j}_{k})$, respectively. Here, $\gamma^{j}_{k}$ ($\gamma^{j}_{k|k-1}$), $\bm{\mu}^{j}_{k}$ ($\bm{\mu}^{j}_{k|k-1}$) and $\bm{\Sigma}^{j}_{k}$ ($\bm{\Sigma}^{j}_{k|k-1}$), respectively, denote the weight, mean and covariance of the $j$-th Gaussian of the posterior (prediction) distribution at $k$-th time step. Hence, \eqref{eqn:GPF posterior} becomes $\pi_{k|k}\approx\sum_{j=1}^{G}\gamma^{j}_{k}\mathcal{N}(\hat{\mathbf{x}}_{k};\bm{\mu}^{j}_{k},\bm{\Sigma}^{j}_{k})\\\times\rho(\mathbf{y}_{k}|\mathbf{x}_{k})$. I-GSPF generates $N$ particles per Gaussian to update these GMs using the PF framework and computes $\doublehat{\mathbf{x}}_{k}$ and $\overline{\bm{\Sigma}}_{k}$ as follows:
\par\noindent\small
\begin{align*}
&\textrm{Time update:}\;\;\{\hat{\widetilde{\mathbf{x}}}^{j,i}_{k-1}\}_{1\leq i\leq N}\sim\mathcal{N}(\hat{\mathbf{x}}_{k};\bm{\mu}^{j}_{k},\bm{\Sigma}^{j}_{k}),\;\;\{\mathbf{y}^{j,i}_{k}\}_{1\leq i\leq N}\sim\rho(\mathbf{y}_{k}|\mathbf{x}_{k})\;\forall 1\leq j\leq G,\\
&\hat{\mathbf{x}}^{j,i}_{k|k-1}=T(\hat{\widetilde{\mathbf{x}}}^{j,i}_{k-1},\mathbf{y}^{j,i}_{k})\;\forall 1\leq i\leq N, 1\leq j\leq G,\;\;\gamma^{j}_{k|k-1}=\gamma^{j}_{k-1}\;\forall 1\leq j\leq G,\\
&\bm{\mu}^{j}_{k|k-1}=\frac{1}{N}\sum_{i=1}^{N}\hat{\mathbf{x}}^{j,i}_{k|k-1},\;\;\bm{\Sigma}^{j}_{k|k-1}=\frac{1}{N}\sum_{i=1}^{N}(\hat{\mathbf{x}}^{j,i}_{k|k-1}-\bm{\mu}^{j}_{k|k-1})(\hat{\mathbf{x}}^{j,i}_{k|k-1}-\bm{\mu}^{j}_{k|k-1})^{T}\;\forall 1\leq j\leq G,\\
&\textrm{Measurement update:}\;\;\{\hat{\mathbf{x}}^{j,i}_{k}\}_{1\leq i\leq N}\sim\mathcal{N}(\hat{\mathbf{x}}_{k};\bm{\mu}^{j}_{k|k-1},\bm{\Sigma}^{j}_{k|k-1})\;\forall 1\leq j\leq G,\\
&\widetilde{\omega}^{j,i}_{k}=\beta(\mathbf{a}_{k}|\hat{\mathbf{x}}^{j,i}_{k}),\;\;\omega^{j,i}_{k}=\widetilde{\omega}^{j,i}_{k}/\sum_{i=1}^{N}\widetilde{\omega}^{j,i}_{k}\;\forall 1\leq i\leq N, 1\leq j\leq G,\\
&\bm{\mu}^{j}_{k}=\sum_{i=1}^{N}\omega^{j,i}_{k}\hat{\mathbf{x}}^{j,i}_{k},\;\;\bm{\Sigma}^{j}_{k}=\sum_{i=1}^{N}\omega^{j,i}_{k}(\hat{\mathbf{x}}^{j,i}_{k}-\bm{\mu}^{j}_{k})(\hat{\mathbf{x}}^{j,i}_{k}-\bm{\mu}^{j}_{k})^{T}\;\forall 1\leq j\leq G,\\
&\widetilde{\gamma}^{j}_{k}=\gamma^{j}_{k|k-1}\sum_{i=1}^{N}\widetilde{\omega}^{j,i}_{k}/\sum_{j=1}^{G}\sum_{i=1}^{N}\widetilde{\omega}^{j,i}_{k},\;\;\gamma^{j}_{k}=\widetilde{\gamma}^{j}_{k}/\sum_{j=1}^{G}\widetilde{\gamma}^{j}_{k}\;\forall 1\leq j\leq G,\\
&\doublehat{\mathbf{x}}_{k}=\sum_{j=1}^{G}\gamma^{j}_{k}\bm{\mu}^{j}_{k},\;\;\overline{\bm{\Sigma}}_{k}=\sum_{j=1}^{G}\gamma^{j}_{k}(\bm{\Sigma}^{j}_{k}+(\bm{\mu}^{j}_{k}-\doublehat{\mathbf{x}}_{k})(\bm{\mu}^{j}_{k}-\doublehat{\mathbf{x}}_{k})^{T}).
\end{align*}
\normalsize

\noindent\textbf{MCC-modified I-EnKF:} Although EnKF was designed for linear Gaussian systems, it is also highly efficient for non-linear and non-Gaussian systems, which is a consequence of linear Bayesian estimation. Particularly, linear Bayesian estimation seeks linear estimators given the first and second moments of the prior distributions and likelihoods \cite{goldstein2007bayes}. In the case of linear Gaussian systems, these estimates coincide with the true Bayesian posterior. However, this linear approach is inefficient in tacking highly non-Gaussian distributions \cite{katzfuss2016understandingEnKf}. As mentioned in Sections~\ref{subsec:IEKF formulation} and \ref{subsec:IUKF formulation}, MCC-modified KFs have been recently introduced to handle non-Gaussian noises \cite{izanloo2016kalman,yang2019map,tao2023maximum}. Our I-EnKF can also be similarly generalized to non-Gaussian systems. For instance, consider the predicted state $\hat{\mathbf{x}}_{k|k-1}=\frac{1}{q}\sum_{i=1}^{q}\mathbf{x}^{i}_{k|k-1}$ and its covariance matrix $\bm{\Sigma}_{k|k-1}=\frac{1}{q-1}\mathbf{E}^{x}_{k|k-1}(\mathbf{E}^{x}_{k|k-1})^{T}$ obtained from the predicted ensemble \eqref{eqn:forward EnKF prediction ensemble}. Forward MCC-EnKF in \cite{tao2023maximum} considers a scalar $l_{k}=G_{\sigma}(\|\mathbf{y}_{k}-h(\hat{\mathbf{x}}_{k|k-1})\|_{\mathbf{R}_{k}^{-1}})$ with $G_{\sigma}(\cdot)$ as the Gaussian kernel and computes the modified gain matrix $\mathbf{K}^{m}_{k}=(\bm{\Sigma}_{k|k-1}^{-1}+\mathbf{H}_{k}^{T}\widetilde{\mathbf{R}}_{k}^{-1}\mathbf{H}_{k})^{-1}\mathbf{H}^{T}_{k}\widetilde{\mathbf{R}}^{-1}_{k}$ (instead of $\mathbf{K}_{k}$), where $\widetilde{\mathbf{R}}^{-1}_{k}=\mathbf{R}_{k}/l_{k}$ and $\mathbf{H}_{k}=\nabla_{\mathbf{x}}h(\mathbf{x})|_{\mathbf{x}=\hat{\mathbf{x}}_{k|k-1}}$. All other state prediction and update steps remain same as in forward EnKF. While formulating the inverse filter, the modified gain matrix needs to be taken into account in the intermediate update step. Similarly, I-EnKF's gain matrix $\overline{\mathbf{K}}_{k}$ is also modified using scalar $\overline{l}_{k}$ which is the counterpart of $l_{k}$ for the inverse filter's dynamics.

    \begin{table}
    \footnotesize
    \caption{Summary of I-PF, I-GPF, and I-EnKF.}
    \label{tbl:differences}
    \centering
    \begin{tabular}{p{2.5cm}p{4.2cm}p{4.2cm}p{4.2cm}}
    \hline\noalign{\smallskip}
    Feature & I-PF & I-GPF & I-EnKF\\
    \noalign{\smallskip}
    \hline
    \noalign{\smallskip}
    System model & General densities \eqref{eqn:state x}, \eqref{eqn:observation y} and \eqref{eqn:observation a} & General densities \eqref{eqn:state x},\eqref{eqn:observation y} and \eqref{eqn:observation a} & Non-linear system with additive Gaussian noises, i.e., \eqref{eqn:non-linear state x}-\eqref{eqn:non-linear observation a}\\
    Forward filter & $T(\cdot)$ in \eqref{eqn:filter T} & $T(\cdot)$ in \eqref{eqn:filter T} & EnKF \\
    State posterior distribution & Empirical approximation for the optimal inverse filter's joint posterior $p(\hat{\mathbf{x}}_{k},\mathbf{y}_{k}|\mathbf{x}_{0:k},\mathbf{a}_{1:k})$ (with no assumptions) & Assumes posterior $p(\hat{\mathbf{x}}_{k}|\mathbf{x}_{0:k},\mathbf{a}_{1:k})$ as Gaussian with the mean and covariance computed as weighted MC estimates & Assumes $p(\hat{\mathbf{x}}_{k}|\mathbf{x}_{0:k},\mathbf{a}_{1:k})$ as Gaussian with the mean and covariance computed from the ensemble ones\\
    Update & SIS reweighting and resampling & SIS reweighting & KF-like updates\\
    Implementation & Sampling from $\rho(\cdot)$ and $\widetilde{\pi}^{N}_{k|k}$ (resampling), and compute $\beta(\cdot)$ at generated particles & Sampling from Gaussian densities and $\rho(\cdot)$; and compute $\beta(\cdot)$ at generated particles & Sampling from $\mathcal{N}(\mathbf{w}_{k-1};\mathbf{0},\mathbf{Q}_{k-1})$, $\mathcal{N}(\mathbf{v}_{k};\mathbf{0},\mathbf{R}_{k})$ and $\mathcal{N}(\bm{\epsilon}_{k};\mathbf{0},\overline{\mathbf{R}}_{k})$\\
    Convergence & To the optimal Bayesian filter in $L^{4}$-sense as $N\to\infty$; cf. Theorem~\ref{thm:ipf convergence} & Almost surely to the actual MMSE estimate (Gaussian posterior only). & In probability to KF's estimates as ensemble size increases (linear Gaussian only).\\
    Non-Gaussian noises & Yes, as is & No, except with GM models & No, except with MCC\\
    \noalign{\smallskip}
    \hline\noalign{\smallskip}
    \normalsize
    \end{tabular}
    \end{table}

 \section{Unknown system dynamics}\label{sec:unknown SMC}
Recall that in real-world scenarios, both the attacker and the defender may lack prior knowledge about the system model. Also, if the attacker's forward filter is unknown, assuming a simple forward EKF/UKF may sometimes be inefficient. In Chapter~\ref{chap:rkhs}, we addressed this issue for additive Gaussian systems using RKHS-EKF and RKH-UKF, but under the assumption of a Gaussian posterior. Here, we explore alternative SMC techniques to handle unknown system dynamics in general non-Gaussian systems. We introduce differentiable I-PF, differentiable I-EnKF, and RKHS-EnKF to learn the state estimates and the model parameters.

\subsection{Differentiable I-PF}\label{subsec:dpf}
DPFs construct the system dynamics and the proposal distributions using NNs and optimize them using gradient descent. However, major challenges in developing DPFs are the non-differentiable importance sampling and resampling steps. Sampling from a proposal distribution is not differentiable because of the absence of explicit dependency between the sampled particles and the distribution parameters. On the other hand, the discrete nature of multinomial resampling makes it inherently non-differentiable, i.e., a small change in input weights can lead to abrupt changes in the resampling output. Additionally, the resampled particles are equally weighted, which is a constant such that the gradients are always zero. Hence, DPFs employ reparameterization-based differentiable sampling, and various differentiable resampling techniques \cite{chen2023overview,karkus2018particle,corenflos2021differentiable,zhu2020towards}. Another important factor affecting DPFs' performance is the loss function minimized in gradient descent to optimize NNs' parameters.

Here, we discuss how this differentiable framework can be integrated into our I-PF to handle unknown dynamics case in inverse filtering. Following I-PF's formulation in Section~\ref{subsec:IPF formulation}, differentiable I-PF considers the joint density of $(\hat{\mathbf{x}}_{k},\mathbf{y}_{k})$ conditioned on the true states $\mathbf{x}_{0:k}$ and defender's observations $\mathbf{a}_{1:k}$. The formulation then closely follows from standard DPF methods, and hence, we only summarize them here. We refer the readers to \cite{chen2023overview} (and the references therein) for further details.

\noindent\textbf{Proposal distributions:} The simplest choice of proposal distribution in PFs is the system's state evolution. In DPFs, the sampled particle from this state evolution is computed as a function of the previous particle and an additional noise term. The corresponding function is parameterized by an unknown parameter $\theta$ and is differentiable with respect to both the previous particle and the noise term. Similarly, in differentiable I-PF, we can consider a differentiable model for the optimal importance sampling density \eqref{eqn:optimal density}. For instance, we can model $p(\hat{\mathbf{x}}_{k}|\hat{\mathbf{x}}_{k-1},\mathbf{y}_{k})p(\mathbf{y}_{k}|\mathbf{x}_{k})$ as a jointly Gaussian distribution $\mathcal{N}(\hat{\mathbf{x}},\mathbf{y}_{k};\mu_{\theta}(\hat{\mathbf{x}}_{k-1},\mathbf{x}_{k}),\bm{\Sigma}_{\theta})$. In this case, the sampled particles $(\hat{\widetilde{\mathbf{x}}}^{i}_{k},\widetilde{\mathbf{y}}^{i}_{k})$ are obtained by adding zero-mean Gaussian noise of covariance $\bm{\Sigma}_{\theta}$ to the mean $\mu_{\theta}(\hat{\mathbf{x}}^{i}_{k-1},\mathbf{x}_{k})$ where $\hat{\mathbf{x}}^{i}_{k-1}$ is the (state estimate) particle from previous time instant and $\mathbf{x}_{k}$ is the defender's true state known perfectly. The function $\mu_{\theta}(\cdot)$ is a differentiable function of $\hat{\mathbf{x}}_{k-1}$ while $\bm{\Sigma}_{\theta}$ can be designed manually \cite{karkus2018particle,wen2021end} or parameterized for learning \cite{kloss2021train}.
    
Alternatively, \cite{chen2021differentiable} provides a normalizing flows-based differential sampling technique, wherein samples drawn from simple distributions (like Gaussian or uniform) are transformed into arbitrary distributions through a series of invertible mappings under some mild conditions. Note that even though \eqref{eqn:optimal density} minimizes the variance of weights, the information from current observation $\mathbf{a}_{k}$ is not utilized. In general, constructing proposal distributions with observation $\mathbf{a}_{k}$ provides samples that are closer to the true posterior and more uniformly weighted. Both the Gaussian model and normalizing flow-based differential samplings can be generalized to include observation $\mathbf{a}_{k}$ in the sampling distribution \cite{chen2021differentiable,karkus2021differentiable}.

\noindent\textbf{Observation models:} In differentiable I-PF, instead of \eqref{eqn:observation a}, we consider a parameterized observation model as $p_{\theta}(\mathbf{a}_{k}|\hat{\mathbf{x}}_{k})\propto l_{\theta}(\mathbf{a}_{k},\hat{\mathbf{x}}_{k})$ where $l_{\theta}(\cdot)$ is a differentiable function with respect to $\mathbf{a}_{k}$ and $\hat{\mathbf{x}}_{k}$. To this end, we can consider known distribution with learnable parameters \cite{corenflos2021differentiable} or approximate $p_{\theta}(\mathbf{a}_{k}|\hat{\mathbf{x}}_{k})$ using a scalar function learned from NN \cite{karkus2018particle}. The observations can also be mapped to an NN-based feature space as $f_{k}=F_{\theta}(\mathbf{a}_{k})$ such that $l_{\theta}(\mathbf{a}_{k},\hat{\mathbf{x}}_{k})=h_{\theta}(f_{k},\hat{\mathbf{x}}_{k})$. In \cite{wen2021end,chen2021differentiable}, NNs are used to extract features from both the observations and states to measure similarity/discrepancy using user-defined metrics. Alternatively, conditional normalizing flows can also be employed to learn the measurement model \cite{chen2022conditional}.

\noindent\textbf{Differentiable resampling:} Soft resampling, optimal transport (OT)-based resampling, and particle transformer-based resampling are popular differentiable techniques employed in DPFs and can be readily applied to differentiable I-PF. Soft resampling \cite{karkus2018particle} aims to generate non-zero gradients by modifying the importance weights by a factor $\lambda$ as $\overline{\omega}^{i}_{k}=\lambda\omega^{i}_{k}+(1-\lambda)1/N$ where $N$ is the total number of particles. However, the particles are selected using a multinomial distribution, and hence, the outputs still change abruptly. Soft resampling can be viewed as a linear interpolation between the multinomial distribution of original weights and one with equal weights. Contrarily, resampling using entropy-regularized OT \cite{corenflos2021differentiable} is fully differentiable. In particular, in differentiable I-PF, OT provides a map between the equally weighted empirical distribution $\frac{1}{N}\sum_{i=1}^{N}\delta(\hat{\mathbf{x}}_{k}-\hat{\mathbf{x}}^{i}_{k},\mathbf{y}_{k}-\mathbf{y}^{i}_{k})$ and the target empirical distribution $\sum_{i=1}^{N}\omega^{i}_{k}\delta(\hat{\mathbf{x}}_{k}-\hat{\widetilde{\mathbf{x}}}^{i}_{k},\mathbf{y}_{k}-\widetilde{\mathbf{y}}^{i}_{k})$. Particle transformers \cite{zhu2020towards} are permutation-invariant and scale-equivalent NNs that take weighted particles $\{\omega^{i}_{k},(\hat{\widetilde{\mathbf{x}}}^{i}_{k},\widetilde{\mathbf{y}}^{i}_{k})\}_{1\leq i\leq N}$ as inputs and output resampled particles $\{(\hat{\mathbf{x}}^{i}_{k},\mathbf{y}^{i}_{k})\}_{1\leq i\leq N}$ with equal weights. Particle transformers perform differentiable resampling but require pre-training with available training data.

\noindent\textbf{Loss functions and training:} Following the standard DPFs, differentiable I-PFs can be trained using supervised losses like RMSE or negative state likelihood when the ground truth, i.e., attacker's state estimate $\hat{\mathbf{x}}_{k}$ and observation $\mathbf{y}_{k}$ are available \cite{karkus2018particle,chen2021differentiable,corenflos2021differentiable}. Semi-supervised losses like marginal observation likelihoods are helpful when unlabelled data is abundant but access to labels is limited \cite{wen2021end}. Alternatively, variational inference optimizes evidence lower bound (ELBO) instead of likelihood to learn the model and proposal distribution simultaneously \cite{maddison2017filtering,le2018auto}. With these objectives, DPFs and hence differentiable I-PFs, can be trained end-to-end \cite{karkus2018particle,wen2021end} or individually \cite{karkus2021differentiable,lee2020multimodal}. In end-to-end training, all components of the filter are jointly trained via gradient descent to minimize an overall loss function. Contrarily, in individual training, various components are first pre-trained independently and then fine-tuned jointly for a task-specific objective.

\subsection{Differentiable I-EnKF and RKHS-EnKF}\label{subsec:rkhs-en}
As in the case of PF, state augmentation-based, and MCMC-based parameter learning approaches have also been considered to generalize EnKF to unknown system dynamics \cite{anderson2001ensemble,drovandi2022ensemble}. \\
\noindent\textbf{Differentiable I-EnKF:} Recently, following DPF, auto-differentiable EnKF has been proposed in \cite{chen2022autodifferentiable} wherein EnKF's state estimation procedure is combined with ML framework for learning the unknown dynamical system. Differentiable EnKF leverages EnKF's ability to efficiently estimate high-dimensional states and automatic differentiation's ability to train high-dimensional surrogate models. In particular, EnKF algorithm provides approximate data log-likelihoods for learning the unknown dynamics. In \cite{chen2022autodifferentiable}, two cases are considered: (a) parametric system models with unknown parameters, and (b) fully unknown models approximated using NNs. Analogous approaches can be introduced in the I-EnKF framework to develop differentiable I-EnKF for inverse filtering in the unknown system case. Note that differentiable I-EnKF's (learned) state transition models the evolution of the attacker's state estimate $\hat{\mathbf{x}}_{k}$ incorporating true state $\mathbf{x}_{k}$ information while the observation model corresponds to defender's observation $\mathbf{a}_{k}$.

\noindent\textbf{RKHS-EnKF:} Our RKHS-EKF/UKF of Section~\ref{sec:RKHS formulation} coupled EKF/UKF's recursions with an approximate online EM to learn the unknown parameters. RKHS-EKF and RKHS-UKF approximate the non-linear expectation using Taylor series linearization and the unscented transform, respectively. Following EnKF, we can approximate these expectations with a randomly generated ensemble that is updated using KF-like updates. Integrating the approximate online EM and EnKF frameworks yields RKHS-EnKF. Contrary to differentiable I-EnKF, this method adopts the additive Gaussian state-space model and serves as both the attacker's forward filter and the defender's inverse filter without prior forward filter information. A key difference from standard EnKF is that the latter approximates the statistics of state $\mathbf{x}_{k}$ given $\mathbf{y}_{1:k}$. However, the EM-based parameter updates also need the statistics of $\mathbf{x}_{k-1}$ given $\mathbf{y}_{1:k}$; see Section~\ref{subsec:EM}. Therefore, RKHS-EnKF is formulated using an augmented state $\mathbf{z}_{k}=[\mathbf{x}_{k}^{T}\;\mathbf{x}_{k-1}^{T}]^{T}$, but only the ensemble corresponding to current state $\mathbf{x}_{k}$ is propagated to the next time step.

\section{Numerical experiments}\label{sec:smc numericals}
We demonstrate the estimation performance of the proposed SMC-based inverse filters using different example systems. In Section~\ref{subsec:one d non-linear}, we consider a widely used one-dimensional non-linear system model \cite{arulampalam2002tutorial,hu2008basic} and compare I-PF, I-GPF, and I-EnKF with one another as well as I-EKF. Besides the estimation error, we also consider RCRLB and NCI as performance metrics. In Section~\ref{subsec:bearing}, we consider the bearing-only tracking system \cite{bar2004estimation,lin2002comparison} for I-PF and I-GPF, while Van der Pol oscillator and one-dimensional heat conduction systems \cite{gillijns2006ensemble} are considered for I-EnKF in Sections~\ref{subsec:van der pol} and \ref{subsec:heat conduction}, respectively. We focus on estimation error and time-complexity of the proposed algorithms in Sections~\ref{subsec:bearing}-\ref{subsec:heat conduction}. The Van der Pol oscillator and heat conduction examples, respectively, represent low-dimensional non-linear and high-dimensional linear systems for which EnKF is known to be an efficient estimator.

Throughout all experiments, for simplicity, we choose EKF as the forward filter $T(\cdot)$ assumed in I-PF and I-GPF, regardless of the actual forward filter. I-EKF assumes a forward EKF while I-EnKF assumes a forward EnKF. Unless mentioned otherwise, all the forward filters are initialized with the same initial distribution. In particular, if $\mathcal{N}(\mathbf{x}_{0};\hat{\mathbf{x}}_{0},\bm{\Sigma}_{0})$ is the forward filters' initial distribution, then we initialize forward EKF/GPF with initial state $\hat{\mathbf{x}}_{0}$ and initial covariance matrix $\bm{\Sigma}_{0}$ while forward PF/EnKF consider independent samples drawn from this initial distribution. All the inverse filters are also initialized in a similar manner.

\subsection{1-D non-linear system with I-PF, I-GPF and I-EnKF}\label{subsec:one d non-linear}
Consider the non-linear system \cite{arulampalam2002tutorial}
\par\noindent\small
\begin{align*}
&x_{k+1}=\frac{x_{k}}{2}+\frac{25x_{k}}{1+x_{k}^{2}}+8\cos{(1.2k)}+w_{k},\;\; y_{k}=\frac{x_{k}^{2}}{20}+v_{k},\;\; a_{k}=\frac{\hat{x}^{2}_{k}}{10}+\epsilon_{k},
\end{align*}
\normalsize
where $w_{k}\sim\mathcal{N}(0,10)$, $v_{k}\sim\mathcal{N}(0,1)$ and $\epsilon_{k}\sim\mathcal{N}(0,5)$. We set the number of particles (or ensemble size) for forward PF, GPF and EnKF as $25$ while that for I-PF, I-GPF and I-EnKF as $50$. The initial distribution for forward and inverse filters were $\mathcal{N}(0,5)$ and $\mathcal{N}(0,10)$, respectively. Also, the function $T(\cdot)$ is initialized with distribution $\mathcal{N}(0,10)$ in I-PF and I-GPF.

\begin{figure*}
  \centering
  \includegraphics[width = 0.9\columnwidth]{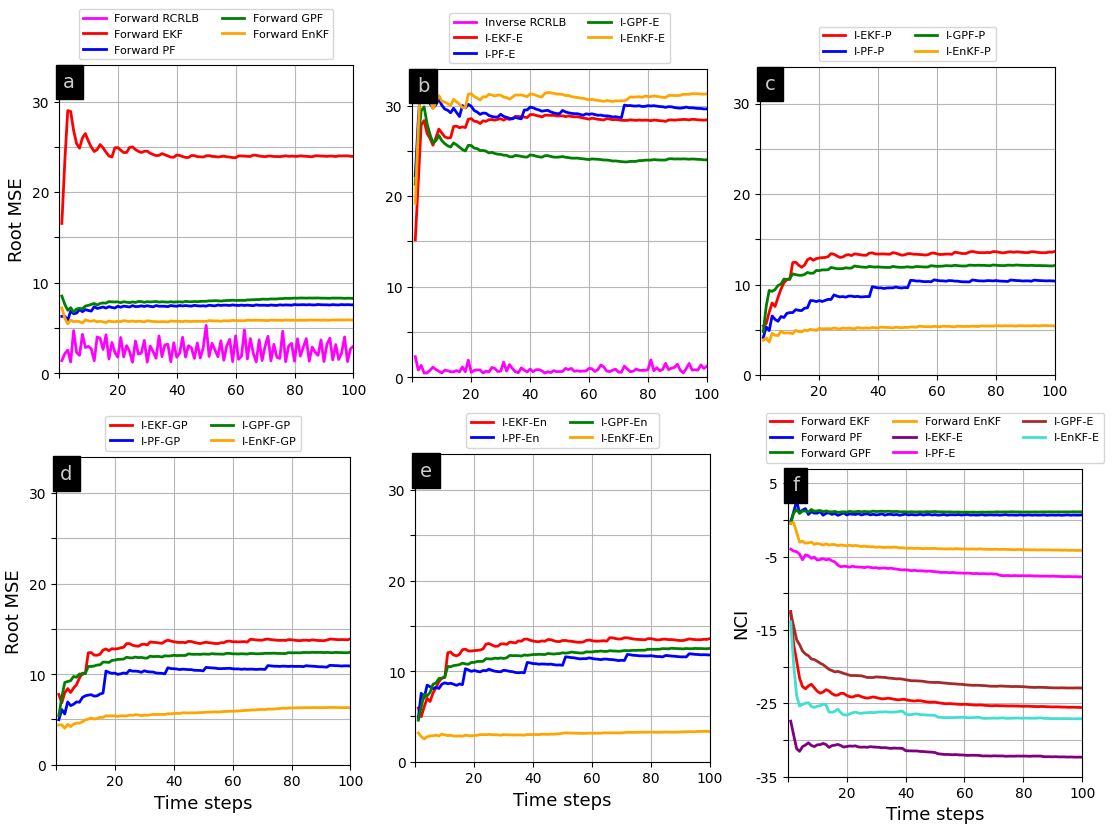}
  \caption{Time-averaged RMSE for (a) forward filters ($\mathbf{x}_{k}$ estimation), (b) forward EKF's $\hat{\mathbf{x}}_{k}$ estimation, (c) forward PF's $\hat{\mathbf{x}}_{k}$ estimation, (d) forward GPF's $\hat{\mathbf{x}}_{k}$ estimation, and (e) forward EnKF's $\hat{\mathbf{x}}_{k}$ estimation; and (f) NCI (for forward and inverse filters) for non-linear system example, compared with inverse EKF.}
 \label{fig:nonlinear}
\end{figure*}

Fig.~\ref{fig:nonlinear}a-e show the AMSE for the forward and inverse filters, averaged over $250$ runs. The I-PF's error in estimating state estimate $\hat{\mathbf{x}}_{k}$ when the attacker's actual forward filter is EKF is denoted by I-PF-E. The other notations in Fig.~\ref{fig:nonlinear} and also, in further experiments are similarly defined. In Fig.~\ref{fig:nonlinear}a \& b, we also include the corresponding RCRLBs. Note that RCRLB for forward PF/GPF/EnKF cannot be derived in closed-form because of lack of an explicit state-transition function and hence, omitted here. From Fig.~\ref{fig:nonlinear}a, we observe that forward PF, GPF and EnKF outperform forward EKF. While forward EnKF has the highest accuracy, forward PF and GPF have similar errors. This suggests that EKF's linearization does not work well for the considered non-linear system. Hence, I-EKF also has the largest error in all but I-EKF-E cases (I-EKF-P/GP/En). Similarly, I-EnKF outperforms the other inverse filters in all cases, except for I-EnKF-E in Fig.~\ref{fig:nonlinear}b. For inferring forward EKF's estimate, I-GPF turns out to be the most suitable estimator as shown in Fig.~\ref{fig:nonlinear}b. In Fig.~\ref{fig:nonlinear}c-e, I-PF and I-GPF are observed to have similar error, but slightly higher than I-EnKF. Fig.~\ref{fig:nonlinear}f shows NCI for different forward filters and inverse filters estimating forward EKF's $\hat{x}_{k}$. We observe that while forward EnKF has lowest estimation error (in Fig.~\ref{fig:nonlinear}a), both forward PF and GPF have almost perfect NCI (i.e., close to $0$) and hence, are more credible. On the other hand, in the inverse filtering case, I-PF has the smallest (in magnitude) NCI while I-GPF and I-EnKF are highly pessimistic, i.e., their estimated covariance is larger than the actual MSE matrix.

\subsection{Bearing only tracking with I-PF and I-GPF}\label{subsec:bearing}
Consider a moving sensor tracking a target moving along x-axis using bearing measurements only. Denote $p_{k}$ and $v_{k}$ as the target's position (in m) and velocity (in $\textrm{m}/\textrm{sec}$), respectively, at the $k$-th time instant. The sensor's position is $(s^{x}_{k},s^{y}_{k})$ with $s^{x}_{k}=4k+\Delta s^{x}_{k}$ and $s^{y}_{k}=20+\Delta s^{y}_{k}$ where $\Delta s^{x}_{k}$ and $\Delta s^{y}_{k}$ denote perturbations distributed as $\mathcal{N}(0,1)$. Then, the system model is \cite{lin2002comparison}
\par\noindent\small
\begin{align*}
&\mathbf{x}_{k+1}\doteq\begin{bmatrix}p_{k+1}\\v_{k+1}\end{bmatrix}=\begin{bmatrix}1&T\\0&1\end{bmatrix}\begin{bmatrix}p_{k}\\v_{k}\end{bmatrix}+\begin{bmatrix}T^{2}/2\\T\end{bmatrix}w_{k},\;y_{k}=\tan^{-1}\left(\frac{s^{y}_{k}}{p_{k}-s^{x}_{k}}\right)+v_{k},\;a_{k}=\tan^{-1}\left(\frac{s^{y}_{k}}{\hat{p}_{k}-s^{x}_{k}}\right)+\epsilon_{k},
\end{align*}
\normalsize
where $w_{k}\sim\mathcal{N}(0,0.01)$, $v_{k}\sim\mathcal{N}(0,(3^{\circ})^{2})$ and $\epsilon_{k}\sim\mathcal{N}(0,(5^{\circ})^{2})$ with $T=1$ sec. The initial state $\mathbf{x}_{0}$ was $[80,1]^{T}$. The forward filters were initialized with $\mathcal{N}(\hat{\mathbf{x}}_{0},\bm{\Sigma}_{0})$ where $\hat{\mathbf{x}}_{0}=[20/\tan^{-1}(y_{1}),0]^{T}$ and $\bm{\Sigma}_{0}=\textrm{diag}(16,1)$. The inverse filters were initialized with $\mathcal{N}(\mathbf{x}_{0},\mathbf{I}_{2})$. For both forward and inverse PFs and GPFs, the number of particles were set to $100$. The recursion function $T(\cdot)$ was also initialized with $\mathcal{N}(\mathbf{x}_{0},\mathbf{I}_{2})$. EKF's implementation for the considered system is provided in \cite{bar2004estimation}.

\begin{figure}
  \centering
  \includegraphics[width = 0.7\columnwidth]{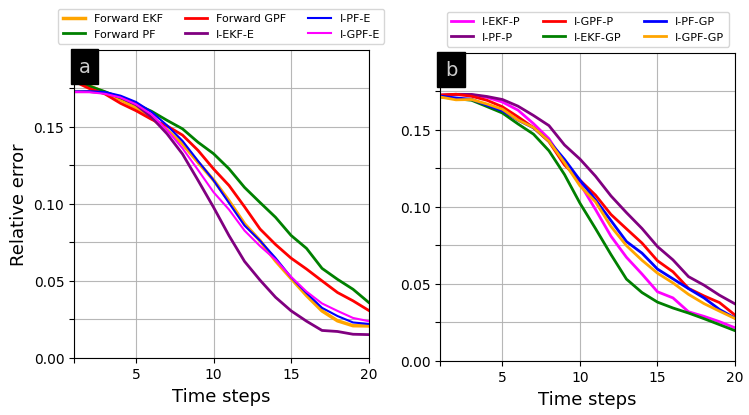}
  \caption{Relative error for (a) forward filters and forward EKF’s $\hat{\mathbf{x}}_{k}$ estimation, and (b) forward PF’s and forward GPF’s $\hat{\mathbf{x}}_{k}$ estimation for bearing-only tracking system.}
 \label{fig:bearing tracking}
\end{figure}

Fig.~\ref{fig:bearing tracking} shows the relative error in position for the forward and inverse filters, averaged over $100$ runs. For the bearing tracking example, forward PF and GPF have slightly higher error than forward EKF, but EKF requires the computation of Jacobians. The same is observed for the inverse filters, i.e., I-PF and I-GPF have higher errors than I-EKF in all cases. Interestingly, GPF and I-GPF assuming a Gaussian posterior density outperform forward PF and I-PF, respectively, with same number of particles. Note that EKF and I-EKF also consider a Gaussian posterior. This suggests that for the considered system, the Gaussian posterior assumption actually leads to better accuracy than the arbitrary distribution case of PF.

Table~\ref{tbl:bearing} lists the total run time for $20$ time steps (in one MC run) of forward and inverse filters estimating forward EKF's state estimate. As expected, forward PF and GPF have longer run times than forward EKF. While forward PF's complexity increases rapidly with the increase in number of particles, forward GPF has similar run times. The run times for the inverse filters are longer than the corresponding forward filters because an inverse filter also needs to compute function $T(\cdot)$ for each sampled particle at each recursion. Interestingly, I-PF and I-GPF have similar run times for all $N$.
    \begin{table}
    \caption{Run time (in seconds) of different filters for bearing-only tracking system.}
    \label{tbl:bearing}
    \centering
    \begin{tabular}{p{3.0cm}p{2.0cm}p{2.0cm}p{2.0cm}}
    \hline\noalign{\smallskip}
    Filter & $N=100$ & $N=250$ & $N=500$\\
    \noalign{\smallskip}
    \hline
    \noalign{\smallskip}
    Forward EKF & 0.0089 & 0.0098 & 0.0100\\
    Forward PF & 0.0191 & 0.0236 & 0.0361\\
    Forward GPF & 0.0377 & 0.0364 & 0.0486\\
    I-EKF-E & 0.0126 & 0.0103 & 0.0107\\
    I-PF-E & 0.0467 & 0.0935 & 0.1741\\
    I-GPF-E & 0.0492 & 0.0837 & 0.1429\\
    \noalign{\smallskip}
    \hline\noalign{\smallskip}
    \end{tabular}
    \end{table}
 
\subsection{Van der Pol oscillator with I-EnKF}\label{subsec:van der pol}
Consider the Van der Pol oscillator system \cite{gillijns2006ensemble}
\par\noindent\small
\begin{align*}
&\mathbf{x}_{k+1}=\begin{bmatrix}
[\mathbf{x}_{k}]_{1}+c_{1}[\mathbf{x}_{k}]_{2}\\
[\mathbf{x}_{k}]_{2}+c_{1}(c_{2}(1-[\mathbf{x}_{k}]_{1}^{2})[\mathbf{x}_{k}]_{2}-[\mathbf{x}_{k}]_{1})
\end{bmatrix}+\mathbf{w}_{k},\;y_{k}=[0\;\;\;1]\mathbf{x}_{k}+v_{k},\;\;a_{k}=[1\;\;\;0]\hat{\mathbf{x}}_{k}+\epsilon_{k},
\end{align*}
\normalsize
where $\mathbf{w}_{k}\sim\mathcal{N}(\mathbf{0},\textrm{diag}(0.0262,0.08))$, $v_{k}\sim\mathcal{N}(0,0.003)$ and $\epsilon_{k}\sim\mathcal{N}(0,0.03)$. We set parameters $c_{1}=0.1$ and $c_{2}=1$. The initial state $\mathbf{x}_{0}=[0,0]^{T}$. Considering forward EnKF's ensemble size $q=30$, the forward filters were initialized with $\mathcal{N}(\hat{\mathbf{x}}_{0},\bm{\Sigma}_{0})$ where $\hat{\mathbf{x}}_{0}=[1,-1]^{T}$ and $\bm{\Sigma}_{0}=\textrm{diag}(6.3\times 10^{-4},2.2\times 10^{-4})$. Similarly, I-EnKF's ensemble size $\overline{q}=50$ and inverse filters were initialized with $\mathcal{N}(\mathbf{x}_{0},\overline{\bm{\Sigma}}_{0})$ where $\overline{\bm{\Sigma}}_{0}=\textrm{diag}(6\times 10^{-3},2\times 10^{-3})$. I-EKF also assumes the forward EKF's $\bm{\Sigma}_{0}$ to be $\overline{\bm{\Sigma}}_{0}$.

Fig.~\ref{fig:vanderpol conduction}a shows the AMSE for the forward and inverse filters, averaged over $100$ runs. We observe that EnKF as the forward filter provides only mild improvement in estimation performance over EKF. Contrarily, I-EnKF outperforms I-EKF in estimating forward EnKF's estimate (I-EnKF-En and I-EKF-En cases) and has same error when estimating forward EKF's estimate (I-EnKF-E and I-EKF-E cases). Furthermore, incorrect assumption about the forward filter degrades I-EKF's accuracy but not I-EnKF's. Hence, we conclude that as standard EnKF is robust to system model deviations, our I-EnKF is also robust to incorrect forward filter assumptions. Table~\ref{tbl:vanderpol} lists the total run time for $500$ time steps of forward and inverse EKF and EnKF. As expected, forward EnKF and I-EnKF have longer run times than forward EKF and EnKF, respectively. However, I-EnKF's complexity is of the same order as forward EnKF which increases as the ensemble size increases.

\subsection{One-dimensional heat conduction with I-EnKF}\label{subsec:heat conduction}
\begin{figure}
  \centering
  \includegraphics[width = 0.7\columnwidth]{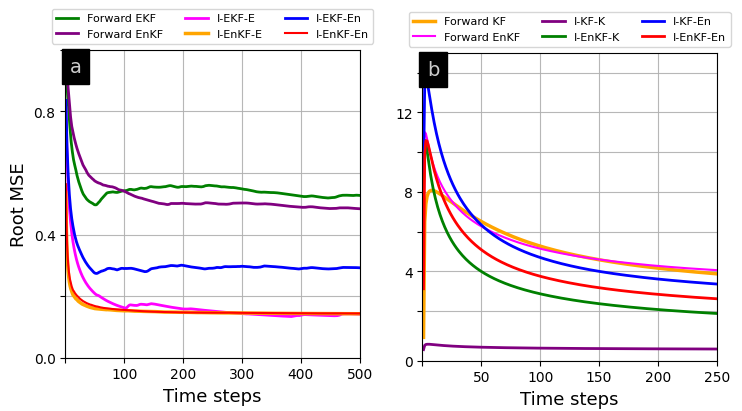}
  \caption{Time-averaged RMSE for forward and inverse EnKF for (a) Van der Pol oscillator, compared with forward and inverse EKF; and (b) One-dimensional heat conduction system, compared with forward and inverse KF.}
 \label{fig:vanderpol conduction}
\end{figure}
    \begin{table}
    \caption{Run time (in seconds) of different filters for Van der Pol oscillator system.}
    \label{tbl:vanderpol}
    \centering
    \begin{tabular}{p{3.0cm}p{2.5cm}p{2.5cm}p{2.5cm}}
    \hline\noalign{\smallskip}
    Filter & $q=\overline{q}=30$ & $q=\overline{q}=50$ & $q=\overline{q}=100$\\
    \noalign{\smallskip}
    \hline
    \noalign{\smallskip}
    Forward EKF & 0.0102 & 0.0101 & 0.0102\\
    Forward EnKF & 0.1060 & 0.1513 & 0.2679\\
    I-EKF-E & 0.0193 & 0.0186 & 0.0177\\
    I-EnKF-E & 0.1053 & 0.1591 & 0.2857\\
    \noalign{\smallskip}
    \hline\noalign{\smallskip}
    \end{tabular}
    \end{table}
Denote $\mathcal{T}(p,k)$ as the temperature of a one-dimensional bar of length $L$ at position $p$ and time $k$. We discretized the spatial grid with $100$ cells and time with time-step $0.1$ sec. Also, $\mathcal{T}(0,k)=\mathcal{T}(L,k)=300$ K for all $k$. Two external sinusoidal heat sources $u^{1}_{k}=0.1\sin{(0.1\pi k)}$ and $u^{2}_{k}=0.1\cos{(0.1\pi k)}$ act on the bar at $p=0.33L$ and $0.67L$, respectively. Defining the temperatures $\mathcal{T}(p,k)$ except $p=0$ and $p=L$ as state $\mathbf{x}_{k}$, the state evolution becomes \cite{gillijns2006ensemble}
\par\noindent\small
\begin{align*}
&\mathbf{x}_{k+1}=\mathbf{F}\mathbf{x}_{k}+\mathbf{B}\begin{bmatrix}
    u^{1}_{k}\\ u^{2}_{k}
\end{bmatrix}+\mathbf{w}_{k},
\end{align*}
\normalsize
where $\mathbf{F}$ is a tridiagonal matrix with all diagonal elements $0.8$ and off-diagonal elements $0.1$, and $\mathbf{w}_{k}\sim\mathcal{N}(0,0.5\mathbf{I}_{100})$. The observations $\mathbf{y}_{k}$ are the noisy measurements of temperatures at $p=0.1L,0.2L,\hdots,0.9L$ with $\mathbf{v}_{k}\sim\mathcal{N}(\mathbf{0},0.01\mathbf{I}_{9})$. Similarly, observations $\mathbf{a}_{k}$ are noisy measurements of the forward filter's estimates at $p=0.05L,0.15L,\hdots,0.95L$ with $\bm{\epsilon}_{k}\sim\mathcal{N}(\mathbf{0},0.1\mathbf{I}_{10})$. We set initial state and its estimates for forward and inverse KF (linear system) such that $[\mathbf{x}_{0}]_{i}=[\hat{\mathbf{x}}_{0}]_{i}=[\doublehat{\mathbf{x}}_{0}]_{i}=10$ for all $i$ and $\bm{\Sigma}_{0}=\mathbf{I}_{100}$ and $\overline{\bm{\Sigma}}_{0}=0.1\mathbf{I}_{100}$. I-KF also assumed forward KF's $\bm{\Sigma}_{0}$ to be $\overline{\bm{\Sigma}}_{0}$. On the other hand, both forward and inverse EnKF were initialized with samples drawn from $\mathcal{U}[-10,10]$ with ensemble sizes $q=100$ and $\overline{q}=500$, respectively.

Fig.~\ref{fig:vanderpol conduction}b shows the AMSE (per cell) for forward and inverse KF and EnKF, averaged over $50$ runs. While forward KF and EnKF have same accuracy, I-EnKF has higher error than I-KF when estimating forward KF's estimate (I-KF-K and I-EnKF-K cases) because of the incorrect forward filter assumption. Note that I-KF is the optimal inverse filter for the considered linear system provided that the attacker's true forward filter is also KF. On the contrary, when the forward KF assumption does not hold, I-KF's performance degrades such that I-EnKF outperforms I-KF (I-KF-En and I-EnKF-En cases). Table~\ref{tbl:conduction} lists the total run time for $250$ time-steps of forward and inverse KF and EnKF. Forward EnKF and I-EnKF have longer run times than forward KF and I-KF, respectively, which increases with the ensemble size. Because of the high-dimensionality of the system, unlike Table~\ref{tbl:vanderpol}, I-EnKF's complexity is similar to forward EnKF only for ensemble size upto $250$.
    \begin{table}
    \caption{Run time (in seconds) of different filters for heat conduction example.}
    \label{tbl:conduction}
    \centering
    \begin{tabular}{p{3.0cm}p{2.5cm}p{2.5cm}p{2.5cm}}
    \hline\noalign{\smallskip}
    Filter & $q=\overline{q}=100$ & $q=\overline{q}=250$ & $q=\overline{q}=500$\\
    \noalign{\smallskip}
    \hline
    \noalign{\smallskip}
    Forward KF & 0.0654 & 0.0517 & 0.0552\\
    Forward EnKF & 0.2176 & 0.4301 & 0.6881\\
    I-KF-K & 0.1128 & 0.1149 & 0.1164\\
    I-EnKF-K & 0.2705 & 0.5692 & 1.0258\\
    \noalign{\smallskip}
    \hline\noalign{\smallskip}
    \end{tabular}
    \end{table}
 
\section{Concluding remarks}\label{sec:smc conclusions}
In this chapter, we studied the inverse filtering problem exploiting SMC techniques and developed I-PF based on the SIS and resampling techniques. Unlike previous inverse filters, the proposed I-PF accommodates general attacker-defender dynamics, including non-Gaussian systems. We demonstrated that, under mild system model assumptions, I-PF converges to the optimal inverse filter in the $L^{4}$ sense. For practical systems wherein Gaussian posterior assumption works well, we developed I-GPF and I-EnKF, which can also be suitably generalized to non-Gaussian systems. Additionally, I-EnKF is an efficient filter for very high-dimensional systems. Differentiable I-PF, differentiable I-EnKF, and RKHS-EnKF provide the attacker’s state estimate and learn the system parameters for unknown system dynamics in the inverse
filtering context. Our numerical experiments showed that these inverse filters provide reasonable estimates even when assuming a simple forward EKF, regardless of the attacker's actual forward filter.

\chapter{Conclusions and Future scope}
\label{chap:conclusions}
In the following, we summarize the key conclusions of this dissertation in Section~\ref{sec:conclusion} while Section~\ref{sec:future scope} presents the future scope of our work.

\section{Conclusions}\label{sec:conclusion}
In this thesis, we studied the inverse filtering problem within the context of inverse cognition for counter-adversarial scenarios, where an attacker infers the defender's state and adapts accordingly. Inverse filters infer an estimate of the attacker's estimate, given the defender's noisy observations of the attacker's actions. Previous inverse filtering methods were limited to finite state-space or linear Gaussian systems with perfect information. To address these limitations, we developed inverse filters for general non-linear and non-Gaussian systems, including those with unknown dynamics. Since non-linear Bayesian filtering is inherently intractable, we explored three approximation methods: (a) Taylor series-based EKF and its variants, (b) SPKFs using deterministic sigma points, and (c) SMC methods relying on random sampling. Theoretical performance guarantees and numerical experiments validated the efficiency of these methods. The key conclusions of our work are summarized as follows.
\begin{itemize}
    \item \textbf{Generalized system models:} The I-EKF and inverse SPKFs assume additive Gaussian noises, whereas the I-PF is designed to handle non-Gaussian systems. Nonetheless, I-EKF and I-UKF can be generalized to non-Gaussian systems using MCC. Similarly, I-GPF and I-EnKF can be adapted with GM and MCC to tackle non-Gaussian noises at a lower computational cost than I-PF. I-UKF can also be modified for systems with continuous-time state evolution, and complex-valued states and observations. Notably, even with continuous-time state evolution, inverse filtering remains a discrete problem.
    \item \textbf{Unknown inputs:} Systems with and without direct feed-through unknown input necessitate conceptually different forward and inverse filters. When there is no unknown input in the attacker's observation model, it results in a one-step delay in estimating the attacker's input. Moreover, I-KFs-with-unknown-input are not merely special cases of I-EKFs-with-unknown-input. The forward filters in the former provide unbiased minimum variance estimates, whereas the latter minimize a weighted least squares criterion.
    \item \textbf{Computational complexity:} The I-EKF and I-SOEKF have the same computational complexity as the standard EKF and SOEKF, which are $\mathcal{O}(n_{x}^3)$ and $\mathcal{O}(n_{x}^{5})$, respectively. In contrast, the I-UKF estimates an  $n_{z}=n_{x}+n_{y}$ dimensional state due to the non-additive process noise, making the actual complexity dependent on the dimensions of both the defender’s state and the attacker’s observation. Similarly, I-EKFs-with-unknown-input are formulated using an augmented state, causing their complexity to depend on both the state dimension $n_{x}$ and the input dimension $n_{u}$.
    \item \textbf{Stability guarantees:} Our stability analysis, using bounded non-linearity and unknown matrix approaches, demonstrated that an inverse filter remains stable if the forward filter is stable, provided that the system satisfies certain additional assumptions. Additionally, the I-EKF and I-UKF are conservative estimators, i.e., their estimated error covariance upper bounds the true error covariance.
    \item \textbf{I-PF's optimal sampling and convergence:} At $k$-th time instant, I-PF empirically approximates the joint conditional distribution of the attacker's state estimate $\hat{\mathbf{x}}_{k}$ and observation $\mathbf{y}_{k}$ given defender's true states $\mathbf{x}_{0:k}$ and observations $\mathbf{a}_{1:k}$ up to the current instant.  Our I-PF, unlike standard PF, samples from the optimal importance sampling density and converges to the optimal inverse filter in the $L^{4}$-sense.
    \item \textbf{RKHS-based KFs:} For unknown system dynamics, we proposed RKHS-EKF and RKHS-UKF, applicable as both the attacker's forward filter and the defender's inverse filter. Unlike previously developed kernel KFs that require feature mapping or conditional embedding with prior training, our RKHS-EKF/UKF learned the dictionary online from state estimates using sliding window or ALD criteria. Integrating parameter learning into the recursive Bayesian framework allowed us to establish stability guarantees for our RKHS-based methods, which is typically not possible with other kernel-based methods.
    \item \textbf{Numerical experiments:} An inverse non-linear filter's performance depends on the system itself. Through extensive numerical experiments, we concluded the following:
    \begin{itemize}
        \item Sophisticated inverse filters, such as the I-GS-EKF for the FM demodulator system, provide better estimates even with incorrect forward filter assumptions. Unlike the I-EKF, the estimation accuracy of inverse SPKFs does not degrade with mismatched forward filters. For certain systems, the inverse filters achieve the RCRLB more efficiently than the forward filter.
        \item Incorporating second-order terms from the Taylor expansion in I-SOEKF does not necessarily improve estimation efficiency. I-DEKF can be implemented with or without prior knowledge of the attacker’s modified observation model, differing only in transient performance but converging to the same steady-state error as the I-EKF.
        \item Any change in unknown input affects the inverse filter's estimation only if it significantly impacts the forward filter.
        \item For the FM demodulator system, the Gaussian kernel-based function approximation enabled RKHS-EKF/UKF to outperform I-EKF even without any prior system model information. The RKHS-UKF provided better estimation performance than the RKHS-EKF, which suffers from linearization errors.
        \item I-GPF outperforms I-PF when a Gaussian density well approximates the posterior, as in the bearing-only tracking system. The I-PF, I-GPF, and I-EnKF showed reasonable estimation accuracy even when assuming a simple forward EKF. However, these methods have higher run-time complexity compared to the forward filters.
    \end{itemize}
\end{itemize}


\section{Future scope of work}\label{sec:future scope}
In this thesis, we developed various inverse filters to estimate the attacker's state from the perspective of a defender. A key area for future research is the practical implementation of these methods in counter-adversarial systems. Engineers often face challenges when implementing non-linear Bayesian filters and introduce techniques like modified state-space models and constraints to overcome these obstacles. Moreover, the suitability of these filters must be assessed on a case-by-case basis. The same considerations apply to inverse non-linear filters. The interaction between attacker and defender introduces further challenges in practical implementation. Therefore, further research is needed to apply these inverse filters in real-world scenarios. Some interesting applications of counter-adversarial systems for future investigation include:

\begin{itemize}
    \item A primary application of the developed inverse filters is to equip an intelligent target with inverse cognitive capabilities to evade detection by adversarial radar in systems such as bearing-only tracking, range-only tracking, and bistatic radar tracking \cite{ristic2003beyond}. Each of these systems is modeled with different state evolution and observations, presenting unique filtering challenges. The efficient implementation of the inverse filters to deduce the adversary’s information needs further exploration.
    
    \item In \cite{mattila2017inverse}, an inverse HMM is applied for real-time fault diagnosis of an automatic sleep tracking system. The inverse filters proposed in our work could be applied for fault diagnosis of state estimation methods in power systems\cite{ghahremani2011dynamic}, chemical processes\cite{kolaas2009constrained}, and structural damage identification\cite{yang2007adaptive}. This potential application requires further investigation.

    \item In cyber-physical systems, KFs are commonly utilized for secure state estimation\cite{chang2018secure} and attack detection\cite{kwon2013security} by adversaries. The application of the developed inverse filters to prevent this detection on the part of the intelligent attacker needs to be explored.

    \item Apart from practical applications of these non-linear inverse filters, the theoretical performance guarantees of I-EKFs-with-unknown-input, I-GS-EKF, I-GPF, and I-EnKF, omitted in this thesis, can also be investigated.
\end{itemize}

\clearpage
\addcontentsline{toc}{chapter}{References}
\begin{singlespace}
\small
\bibliography{references}
\appendix

\chapter{Proofs for I-EKF's and I-SOEKF's performance guarantees}
\label{chap:iekf proof}
\section{Proof of Theorem~\ref{theorem:Forward ekf stable unknown matrix}}
\label{App-thm-Forward ekf stable unknown matrix}
For simplicity, we consider the case of $n_{x}\geq n_{y}$ with $\mathbf{U}^{xy}_{k+1}\in\mathbb{R}^{n_{x}\times n_{x}}$. It is trivial to show that the proof remains valid for $n_{x}<n_{y}$ as well. Using the expressions for $\bm{\Sigma}^{xy}_{k+1}$ and $\bm{\Sigma}^{y}_{k+1}$ from \eqref{eqn:forward ekf stable sig xy} and \eqref{eqn:forward ekf stable sig y}, respectively, we have
\par\noindent\small
\begin{align*}
\mathbf{K}_{k+1}&=\bm{\Sigma}_{k+1|k}\mathbf{U}^{xy}_{k+1}\mathbf{H}_{k+1}^{T}\mathbf{U}^{y}_{k+1}\left(\mathbf{U}^{y}_{k+1}\mathbf{H}_{k+1}\bm{\Sigma}_{k+1|k}\mathbf{H}_{k+1}^{T}\mathbf{U}^{y}_{k+1}+\hat{\mathbf{R}}_{k+1}\right)^{-1},\\
\bm{\Sigma}_{k+1}&=\bm{\Sigma}_{k+1|k}-\bm{\Sigma}_{k+1|k}\mathbf{U}^{xy}_{k+1}\mathbf{H}_{k+1}^{T}\mathbf{U}^{y}_{k+1}\left(\mathbf{U}^{y}_{k+1}\mathbf{H}_{k+1}\bm{\Sigma}_{k+1|k}\mathbf{H}_{k+1}^{T}\mathbf{U}^{y}_{k+1}+\hat{\mathbf{R}}_{k+1}\right)^{-1}\mathbf{U}^{y}_{k+1}\mathbf{H}_{k+1}(\mathbf{U}^{xy}_{k+1})^{T}\bm{\Sigma}_{k+1|k}.
\end{align*}
\normalsize

Define $V_{k}(\widetilde{\mathbf{x}}_{k|k-1})=\widetilde{\mathbf{x}}_{k|k-1}^{T}\bm{\Sigma}_{k|k-1}^{-1}\widetilde{\mathbf{x}}_{k|k-1}$. Using the bounds assumed on $\bm{\Sigma}_{k|k-1}$, we have for all $k\geq 0$
\par\noindent\small
\begin{align*}
    \frac{1}{\overline{\sigma}}\|\widetilde{\mathbf{x}}_{k|k-1}\|_{2}^{2}\leq V_{k}(\widetilde{\mathbf{x}}_{k|k-1})\leq\frac{1}{\underline{\sigma}}\|\widetilde{\mathbf{x}}_{k|k-1}\|_{2}^{2}.
\end{align*}
\normalsize
Hence, the first condition of Lemma \ref{lemma:exponential boundedness} is satisfied with $v_{\textrm{min}}=1/\overline{\sigma}$ and $v_{\textrm{max}}=1/\underline{\sigma}$.

Using \eqref{eqn:forward EKF prediction error dynamics} and the independence of noise terms, we have
\par\noindent\small
\begin{align}
&\mathbb{E}\left[V_{k+1}(\widetilde{\mathbf{x}}_{k+1|k})\vert\widetilde{\mathbf{x}}_{k|k-1}\right]= \widetilde{\mathbf{x}}_{k|k-1}^{T}(\mathbf{U}^{x}_{k}\mathbf{F}_{k}(\mathbf{I}-\mathbf{K}_{k}\mathbf{U}^{y}_{k}\mathbf{H}_{k}))^{T}\bm{\Sigma}_{k+1|k}^{-1}(\mathbf{U}^{x}_{k}\mathbf{F}_{k}(\mathbf{I}-\mathbf{K}_{k}\mathbf{U}^{y}_{k}\mathbf{H}_{k}))\widetilde{\mathbf{x}}_{k|k-1}\nonumber\\
& +\mathbb{E}\left[\mathbf{v}_{k}^{T}(\mathbf{U}^{x}_{k}\mathbf{F}_{k}\mathbf{K}_{k})^{T}\bm{\Sigma}_{k+1|k}^{-1}(\mathbf{U}^{x}_{k}\mathbf{F}_{k}\mathbf{K}_{k})\mathbf{v}_{k}\vert\widetilde{\mathbf{x}}_{k|k-1}\right]+\mathbb{E}\left[\mathbf{w}_{k}^{T}\bm{\Sigma}_{k+1|k}^{-1}\mathbf{w}_{k}\vert\widetilde{\mathbf{x}}_{k|k-1}\right].\label{eqn: Vk expectation}
\end{align}
\normalsize
The difference of two matrices $\mathbf{A}-\mathbf{B}$ is invertible if maximum singular value of $\mathbf{B}$ is strictly less than the minimum singular value of $\mathbf{A}$.  Using the assumed bounds, we have $\|\mathbf{K}_{k}\|\leq\overline{k}=(\overline{\sigma}\overline{\gamma}\overline{h}\overline{\beta})/\hat{r}$. Hence, maximum singular value of $\mathbf{K}_{k}\mathbf{U}^{y}_{k}\mathbf{H}_{k}$ is upper-bounded by $(\overline{\sigma}\overline{\gamma}\overline{h}^{2}\overline{\beta}^{2})/\hat{r}$  and the inequality \textbf{C3.c} guarantees that $\mathbf{I}-\mathbf{K}_{k}\mathbf{U}^{y}_{k}\mathbf{H}_{k}$ is invertible (singular value of $\mathbf{I}$ is 1) such that
\par\noindent\small
\begin{align*}
\bm{\Sigma}_{k+1|k}&=\mathbf{U}^{x}_{k}\mathbf{F}_{k}(\mathbf{I}-\mathbf{K}_{k}\mathbf{U}^{y}_{k}\mathbf{H}_{k}) (\bm{\Sigma}_{k|k-1}+(\mathbf{U}^{x}_{k}\mathbf{F}_{k}(\mathbf{I}-\mathbf{K}_{k}\mathbf{U}^{y}_{k}\mathbf{H}_{k}))^{-1}\hat{\mathbf{Q}}_{k}((\mathbf{U}^{x}_{k}\mathbf{F}_{k}(\mathbf{I}-\mathbf{K}_{k}\mathbf{U}^{y}_{k}\mathbf{H}_{k}))^{-1})^{T})\\
&\times(\mathbf{I}-\mathbf{K}_{k}\mathbf{U}^{y}_{k}\mathbf{H}_{k})^{T}\mathbf{F}_{k}^{T}\mathbf{U}^{x}_{k},
\end{align*}
\normalsize
because $\mathbf{U}^{x}_{k}$ and $\mathbf{F}_{k}$ are also assumed to be invertible. Again with the assumed bounds, we have $\|\mathbf{U}^{x}_{k}\mathbf{F}_{k}(\mathbf{I}-\mathbf{K}_{k}\mathbf{U}^{y}_{k}\mathbf{H}_{k})\|\leq\overline{\alpha}\overline{f}(1+\overline{k}\overline{\beta}\overline{h})$ which implies
\par\noindent\small
\begin{align*}
    (\mathbf{U}^{x}_{k}\mathbf{F}_{k}(\mathbf{I}-\mathbf{K}_{k}\mathbf{U}^{y}_{k}\mathbf{H}_{k}))^{-1}\hat{\mathbf{Q}}_{k}((\mathbf{U}^{x}_{k}\mathbf{F}_{k}(\mathbf{I}-\mathbf{K}_{k}\mathbf{U}^{y}_{k}\mathbf{H}_{k}))^{-1})^{T}\succeq\frac{\hat{q}}{(\overline{\alpha}\overline{f}(1+\overline{k}\overline{\beta}\overline{h}))^{2}}\mathbf{I}.
\end{align*}
\normalsize
Using this bound in the expression of $\bm{\Sigma}_{k+1|k}$ in \eqref{eqn:forward ekf stable sig} as in \cite{li2012stochastic_ukf}, we have
\par\noindent\small
\begin{align*}
(\mathbf{U}^{x}_{k}\mathbf{F}_{k}(\mathbf{I}-\mathbf{K}_{k}\mathbf{U}^{y}_{k}\mathbf{H}_{k}))^{T}\bm{\Sigma}_{k+1|k}^{-1}(\mathbf{U}^{x}_{k}\mathbf{F}_{k}(\mathbf{I}-\mathbf{K}_{k}\mathbf{U}^{y}_{k}\mathbf{H}_{k}))\preceq(1-\lambda)\bm{\Sigma}_{k|k-1}^{-1},
\end{align*}
\normalsize
where $1-\lambda=\left(1+\frac{\hat{q}}{\overline{\sigma}(\overline{\alpha}\overline{f}(1+\overline{k}\overline{\beta}\overline{h}))^{2}}\right)^{-1}$ with $0<\lambda<1$. The last two expectation terms in \eqref{eqn: Vk expectation} can be bounded by $\mu=(\overline{r}p\overline{\alpha}^{2}\overline{f}^{2}\overline{k}^{2}/\underline{\sigma})+(\overline{q}n/\underline{\sigma})>0$ following similar steps as in \cite{li2012stochastic_ukf} such that
\par\noindent\small
\begin{align*}
\mathbb{E}\left[ V_{k+1}(\widetilde{\mathbf{x}}_{k+1|k})|\widetilde{\mathbf{x}}_{k|k-1}\right]-V_{k}(\widetilde{\mathbf{x}}_{k|k-1})\leq-\lambda V_{k}(\widetilde{\mathbf{x}}_{k|k-1})+\mu.
\end{align*}
\normalsize
Hence, the second condition of Lemma \ref{lemma:exponential boundedness} is also satisfied and the prediction error $\widetilde{\mathbf{x}}_{k|k-1}$ is exponentially bounded in mean-squared sense and bounded with probability one.

Furthermore, with the bounds assumed on various matrices, it is straightforward to show that
\par\noindent\small
\begin{align*}
\mathbb{E}\left[\|\widetilde{\mathbf{x}}_{k}\|^{2}_{2}\right]\leq(1+\overline{k}\overline{\beta}\overline{h})^{2}\mathbb{E}\left[\|\widetilde{\mathbf{x}}_{k|k-1}\|^{2}_{2}\right]+\overline{k}^{2}\overline{r}p.
\end{align*}
\normalsize
Finally, the exponential boundedness of $\widetilde{\mathbf{x}}_{k|k-1}$ leads to $\widetilde{\mathbf{x}}_{k}$ also being exponentially bounded in mean-squared sense as well as bounded with probability one.

\section{Proof of Theorem~\ref{theorem: inverse EKF stable unknown matrix}}\label{App-thm-inverse EKF stable unknown matrix}
We will show that the I-EKF's dynamics also satisfies the assumptions of Theorem \ref{theorem:Forward ekf stable unknown matrix}. For this, the following conditions \textbf{A.C1}-\textbf{A.C13} need to hold true for all $k\geq 0$ for some real positive constants $\overline{a},\overline{g},\overline{b},\overline{c},\overline{d},\hat{q},\overline{\epsilon},\hat{c},\hat{d},\underline{p},\overline{p}$.\\
\textbf{A.C1.} $\|\widetilde{\mathbf{F}}^{x}_{k}\|\leq\overline{a}$;\\
\textbf{A.C2.} $\|\overline{\mathbf{U}}^{x}_{k}\|\leq\overline{b}$;\\
\textbf{A.C3.} $\overline{\mathbf{U}}^{x}_{k}$ is non-singular;\\
\textbf{A.C4.} $\widetilde{\mathbf{F}}^{x}_{k}$ is non-singular;\\
\textbf{A.C5.} $\overline{\mathbf{Q}}_{k}\preceq\widetilde{q}\mathbf{I}$;\\
\textbf{A.C6.} $\|\mathbf{G}_{k}\|\leq\overline{g}$;\\
\textbf{A.C7.} $\|\overline{\mathbf{U}}^{a}_{k}\|\leq\overline{c}$;\\
\textbf{A.C8.} $\|\overline{\mathbf{U}}^{xa}_{k}\|\leq\overline{d}$;\\
\textbf{A.C9.} $\overline{\mathbf{R}}_{k}\preceq\overline{\epsilon}\mathbf{I}$;\\
\textbf{A.C10.} $\hat{c}\mathbf{I}\preceq\hat{\overline{\mathbf{Q}}}_{k}$;\\
\textbf{A.C11.} $\hat{d}\mathbf{I}\preceq\hat{\overline{\mathbf{R}}}_{k}$;\\
\textbf{A.C12.} $\underline{p}\mathbf{I}\preceq\overline{\bm{\Sigma}}_{k|k-1}\preceq\overline{p}\mathbf{I}$; and\\
\textbf{A.C13.} the constants satisfy the inequality $\overline{p}\overline{d}\overline{g}^{2}\overline{c}^{2}<\hat{d}$.

The conditions \textbf{A.C6-A.C13} are assumed to hold true in Theorem \ref{theorem: inverse EKF stable unknown matrix}. Next, we prove that under the assumptions of Theorem \ref{theorem: inverse EKF stable unknown matrix}, \textbf{A.C1}-\textbf{A.C5} are also satisfied for the I-EKF's error dynamics such that Theorem \ref{theorem:Forward ekf stable unknown matrix} is applicable for the I-EKF as well. From the I-EKF's state transition \eqref{eqn:iekf state transition}, the Jacobians $\widetilde{\mathbf{F}}^{x}_{k}=\mathbf{F}_{k}-\mathbf{K}_{k+1}\mathbf{H}_{k+1}\mathbf{F}_{k}$ and $\widetilde{\mathbf{F}}^{v}_{k}=\mathbf{K}_{k+1}$ such that $\overline{\mathbf{Q}}_{k}=\mathbf{K}_{k+1}\mathbf{R}_{k+1}\mathbf{K}_{k+1}^{T}$.

For \textbf{A.C1}, using $\|\mathbf{K}_{k+1}\|\leq\overline{k}$ (as proved in Theorem \ref{theorem:Forward ekf stable unknown matrix}) and the bounds on $\mathbf{F}_{k}$ and $\mathbf{H}_{k+1}$ from the assumptions of Theorem \ref{theorem:Forward ekf stable unknown matrix}, it is trivial to show that $\|\widetilde{\mathbf{F}}^{x}_{k}\|=\|\mathbf{F}_{k}-\mathbf{K}_{k+1}\mathbf{H}_{k+1}\mathbf{F}_{k}\|\leq\overline{f}+\overline{k}\overline{h}\overline{f}$. Hence, \textbf{A.C1} is satisfied with $\overline{a}=\overline{f}+\overline{k}\overline{h}\overline{f}$.

For \textbf{A.C2}-\textbf{A.C4}, consider the unknown matrix $\overline{\mathbf{U}}^{x}_{k}$ introduced to account for the residuals in linearization of $\widetilde{f}_{k}(\cdot)$. Let $\hat{\widetilde{\mathbf{x}}}_{k+1|k}$ and $\hat{\widetilde{\mathbf{x}}}_{k}$ denote the state prediction error and state estimation error of I-EKF. Similar to forward EKF with the introduction of the unknown matrix, we have
\par\noindent\small
\begin{align}
    \hat{\widetilde{\mathbf{x}}}_{k+1|k}=\overline{\mathbf{U}}^{x}_{k}(\mathbf{F}_{k}-\mathbf{K}_{k+1}\mathbf{H}_{k+1}\mathbf{F}_{k})\hat{\widetilde{\mathbf{x}}}_{k}+\mathbf{K}_{k+1}\mathbf{v}_{k+1}\label{eqn:inverse EKF error unknown alpha}.
\end{align}
\normalsize
Also, $\hat{\widetilde{\mathbf{x}}}_{k+1|k}=f(\hat{\mathbf{x}}_{k})-f(\doublehat{\mathbf{x}}_{k})-\mathbf{K}_{k+1}(h(f(\hat{\mathbf{x}}_{k}))-h(f(\doublehat{\mathbf{x}}_{k})))+\mathbf{K}_{k+1}\mathbf{v}_{k+1}$. Using the unknown matrices $\mathbf{U}^{x}_{k}$ and $\mathbf{U}^{y}_{k}$ introduced in the linearization of $f(\cdot)$ and $h(\cdot)$, respectively, we have
\par\noindent\small
\begin{align*}
    \hat{\widetilde{\mathbf{x}}}_{k+1|k}=(\mathbf{U}^{x}_{k}\mathbf{F}_{k}-\mathbf{K}_{k+1}\mathbf{U}^{y}_{k+1}\mathbf{H}_{k+1}\mathbf{U}^{x}_{k}\mathbf{F}_{k})\hat{\widetilde{\mathbf{x}}}_{k}+\mathbf{K}_{k+1}\mathbf{v}_{k+1}.
\end{align*}
\normalsize
Comparing with \eqref{eqn:inverse EKF error unknown alpha}, we have
\par\noindent\small
\begin{align}
    \overline{\mathbf{U}}^{x}_{k}(\mathbf{I}-\mathbf{K}_{k+1}\mathbf{H}_{k+1})\mathbf{F}_{k}=(\mathbf{I}-\mathbf{K}_{k+1}\mathbf{U}^{y}_{k+1}\mathbf{H}_{k+1})\mathbf{U}^{x}_{k}\mathbf{F}_{k}\label{eqn: unknown alpha for inverse EKF}.
\end{align}
\normalsize
With the additional assumption of $\underline{r}\mathbf{I}\preceq\mathbf{R}_{k}$ and using matrix inversion lemma as in proof of \cite[Lemma 3.1]{reif1999stochastic}, we have
\par\noindent\small
\begin{align*}
    (\mathbf{I}-\mathbf{K}_{k+1}\mathbf{H}_{k+1})\bm{\Sigma}_{k+1|k}=\left(\bm{\Sigma}_{k+1|k}^{-1}+\mathbf{H}_{k+1}^{T}\mathbf{R}_{k+1}^{-1}\mathbf{H}_{k+1}\right)^{-1}.
\end{align*}
\normalsize
Since $\bm{\Sigma}_{k+1|k}$ is invertible by the assumptions of Theorem \ref{theorem:Forward ekf stable unknown matrix}, $\mathbf{I}-\mathbf{K}_{k+1}\mathbf{H}_{k+1}$ is invertible for all $k\geq 0$ and
\par\noindent\small
\begin{align*}
    (\mathbf{I}-\mathbf{K}_{k+1}\mathbf{H}_{k+1})^{-1}=\mathbf{I}+\bm{\Sigma}_{k+1|k}\mathbf{H}_{k+1}^{T}\mathbf{R}_{k+1}^{-1}\mathbf{H}_{k+1}.
\end{align*}
\normalsize
With the bounds assumed on various matrices, we have $ \|(\mathbf{I}-\mathbf{K}_{k+1}\mathbf{H}_{k+1})^{-1}\|\leq 1+\overline{\sigma}\overline{h}^{2}/\underline{r}$. Furthermore, using this bound and the invertibility of $\mathbf{I}-\mathbf{K}_{k+1}\mathbf{H}_{k+1}$ in \eqref{eqn: unknown alpha for inverse EKF}, it is straightforward to show that $\overline{\mathbf{U}}^{x}_{k}=(\mathbf{I}-\mathbf{K}_{k+1}\mathbf{U}^{y}_{k+1}\mathbf{H}_{k+1})\mathbf{U}^{x}_{k}(\mathbf{I}-\mathbf{K}_{k+1}\mathbf{H}_{k+1})^{-1}$ is non-singular (both $\mathbf{U}^{x}_{k}$ and $\mathbf{I}-\mathbf{K}_{k+1}\mathbf{U}^{y}_{k+1}\mathbf{H}_{k+1}$ are invertible under the assumptions of Theorem \ref{theorem:Forward ekf stable unknown matrix}) and satisfies $\|\overline{\mathbf{U}}^{x}_{k}\|\leq\overline{\alpha}(1+\overline{k}\overline{\beta}\overline{h})(1+(\overline{\sigma}\overline{h}^{2})/\underline{r})$. Also, since both $\mathbf{I}-\mathbf{K}_{k+1}\mathbf{H}_{k+1}$ and $\mathbf{F}_{k}$ are invertible, $\widetilde{\mathbf{F}}^{x}_{k}=\mathbf{F}_{k}(\mathbf{I}-\mathbf{K}_{k+1}\mathbf{H}_{k+1})$ is non-singular. Hence, \textbf{A.C2}-\textbf{A.C4} are also satisfied with $\overline{b}=\overline{\alpha}(1+\overline{k}\overline{\beta}\overline{h})(1+(\overline{\sigma}\overline{h}^{2})/\underline{r})$.

For \textbf{A.C5}, using the upper bound on $\mathbf{R}_{k}$ from assumptions of Theorem \ref{theorem:Forward ekf stable unknown matrix}, we have $\overline{\mathbf{Q}}_{k}\preceq\overline{r}\mathbf{K}_{k+1}\mathbf{K}_{k+1}^{T}$. Since, $\|\mathbf{K}_{k+1}\|\leq\overline{k}$, the maximum eigenvalue of $\mathbf{K}_{k+1}\mathbf{K}_{k+1}^{T}$ is bounded by $\overline{k}^{2}$ such that $\overline{\mathbf{Q}}_{k}\preceq\overline{k}^{2}\overline{r}\mathbf{I}$. Hence, \textbf{A.C5} is satisfied with $\title{q}=\overline{k}^{2}\overline{r}$.

\section{Proof of Theorem~\ref{theorem: inverse ekf stable Reif}}\label{App-thm-inverse ekf stable Reif}
We will show that the error dynamics of the I-EKF given by \eqref{eqn: inverse ekf error} satisfies the following conditions for all $k\geq 0$ for some real positive constants $\underline{c},\kappa_{\bar{\phi}}, \epsilon_{\bar{\phi}}$.\\
\textbf{A.C14.} $\underline{c}\mathbf{I}\preceq\overline{\mathbf{Q}}_{k}$.\\
\textbf{A.C15.} $\widetilde{\mathbf{F}}^{x}_{k}$ is non-singular matrix for all $k\geq 0$.\\
\textbf{A.C16.} $\|\overline{\phi}_{k}(\hat{\mathbf{x}},\doublehat{\mathbf{x}})\|_{2}\leq\kappa_{\bar{\phi}}\|\hat{\mathbf{x}}-\doublehat{\mathbf{x}}\|^{2}_{2}$ for all $\|\hat{\mathbf{x}}-\doublehat{\mathbf{x}}\|_{2}\leq\epsilon_{\bar{\phi}}$ for some $\kappa_{\bar{\phi}}>0$ and $\epsilon_{\bar{\phi}}>0$.

All other conditions of Theorem \ref{theorem: ekf stable Reif} can be proved to hold true for the I-EKF's error dynamics under the assumptions of Theorem \ref{theorem: inverse ekf stable Reif} following similar approach as in proof of Theorem \ref{theorem: inverse EKF stable unknown matrix}, such that the estimation error given by \eqref{eqn: inverse ekf error} is exponentially bounded in mean-squared sense provided that the estimation error is bounded with $\overline{\epsilon}>0$ where $\overline{\epsilon}$ depends on the various bounds in the same manner as $\epsilon$ depends in the forward filter case.

For \textbf{A.C14}, using the bound on $\mathbf{R}_{k}$ from one of the assumptions of Theorem \ref{theorem: ekf stable Reif}, we have $\overline{\mathbf{Q}}_{k}=\mathbf{K}_{k}\mathbf{R}_{k}\mathbf{K}_{k}^{T}\succeq\underline{r}\mathbf{K}_{k}\mathbf{K}_{k}^{T}$.
Substituting for $\mathbf{K}_{k}$, we have
\par\noindent\small
\begin{align*}
\mathbf{K}_{k}\mathbf{K}_{k}^{T}=\mathbf{F}_{k}\bm{\Sigma}_{k}\mathbf{H}_{k}^{T}(\mathbf{H}_{k}\bm{\Sigma}_{k}\mathbf{H}_{k}^{T}+\mathbf{R}_{k})^{-2}\mathbf{H}_{k}\bm{\Sigma}_{k}\mathbf{F}_{k}^{T}.
\end{align*}
\normalsize
With the assumption that $\mathbf{H}_{k}$ is full column rank, $\mathbf{K}_{k}\mathbf{K}_{k}^{T}$ is p.d. as $\mathbf{F}_{k}$ is assumed to be non-singular in Theorem \ref{theorem: ekf stable Reif}. Hence, there exists a constant $\widetilde{q}>0$ which is the minimum eigenvalue of $\mathbf{K}_{k}\mathbf{K}_{k}^{T}$ such that $\mathbf{K}_{k}\mathbf{K}_{k}^{T}\succeq\widetilde{q}\mathbf{I}$ and $\overline{\mathbf{Q}}_{k}\succeq\underline{r}\widetilde{q}\mathbf{I}$. Hence, \textbf{A.C14} is satisfied with $\underline{c}=\underline{r}\widetilde{q}$.

For \textbf{A.C15}, $\widetilde{\mathbf{F}}^{x}_{k}=\mathbf{F}_{k}-\mathbf{K}_{k}\mathbf{H}_{k}$ is proved to be invertible for all $k\geq 0$ as an intermediate result in the proof of Theorem \ref{theorem: ekf stable Reif} in \cite[Lemma 3.1]{reif1999stochastic}.

For \textbf{A.C16}, using $\|\mathbf{K}_{k}\|\leq(\overline{f}\overline{\sigma}\overline{h}/\underline{r})$ (proved in \cite[Lemma 3.1]{reif1999stochastic}) and the bounds on functions $\phi(\cdot)$ and $\chi(\cdot)$ from the assumptions of Theorem \ref{theorem: ekf stable Reif}, we have $\|\overline{\phi}_{k}(\hat{\mathbf{x}},\doublehat{\mathbf{x}})\|_{2}\leq\|\phi(\hat{\mathbf{x}},\doublehat{\mathbf{x}})\|_{2}+\frac{\overline{f}\overline{\sigma}\overline{h}}{\underline{r}}\|\chi(\hat{\mathbf{x}},\doublehat{\mathbf{x}})\|_{2}\leq\left(\kappa_{\phi}+\frac{\overline{f}\overline{\sigma}\overline{h}}{\underline{r}}\kappa_{\chi}\right)\|\hat{\mathbf{x}}-\doublehat{\mathbf{x}}\|_{2}^{2}$, for $\|\hat{\mathbf{x}}-\doublehat{\mathbf{x}}\|_{2}\leq \textrm{min}(\epsilon_{\phi},\epsilon_{\chi})$. Hence, \textbf{A.C16} is satisfied with $\kappa_{\bar{\phi}}=\kappa_{\phi}+(\overline{f}\overline{\sigma}\overline{h}/\underline{r})\kappa_{\chi}$ and $\epsilon_{\bar{\phi}}=\textrm{min}(\epsilon_{\phi},\epsilon_{\chi})$.

\section{Proof of Theorem~\ref{thm:iekf consistency}}\label{App-thm-inverse EKF consistency}
We prove the theorem by the principle of mathematical induction. Define the prediction and estimation errors as $\hat{\widetilde{\mathbf{x}}}_{k|k-1}\doteq\hat{\mathbf{x}}_{k}-\doublehat{\mathbf{x}}_{k|k-1}$ and $\hat{\widetilde{\mathbf{x}}}_{k}\doteq\hat{\mathbf{x}}_{k}-\doublehat{\mathbf{x}}_{k}$, respectively. Assume $\mathbb{E}[\hat{\widetilde{\mathbf{x}}}_{k}\hat{\widetilde{\mathbf{x}}}_{k}^{T}]\preceq\overline{\bm{\Sigma}}_{k}$. We show that the inequality also holds for $(k+1)$-th time step. Substituting \eqref{eqn:SLT state transition} in the I-EKF's recursions, we have $\doublehat{\mathbf{x}}_{k+1|k}=\mathbf{U}^{xv}_{k}\overline{\mathbf{F}}^{x}_{k}\doublehat{\mathbf{x}}_{k}$ and $\overline{\bm{\Sigma}}_{k+1|k}=\mathbf{U}^{xv}_{k}\overline{\mathbf{F}}^{x}_{k}\overline{\bm{\Sigma}}_{k}(\overline{\mathbf{F}}^{x}_{k})^{T}\mathbf{U}^{xv}_{k}+\mathbf{U}^{xv}_{k}\overline{\mathbf{F}}^{v}_{k}\mathbf{R}_{k+1}(\overline{\mathbf{F}}^{v}_{k})^{T}\mathbf{U}^{xv}_{k}$. Hence, $\hat{\widetilde{\mathbf{x}}}_{k+1|k}=\mathbf{U}^{xv}_{k}\overline{\mathbf{F}}^{x}_{k}\hat{\widetilde{\mathbf{x}}}_{k}+\mathbf{U}^{xv}_{k}\overline{\mathbf{F}}^{v}_{k}\mathbf{v}_{k+1}$ such that $\mathbb{E}[\hat{\widetilde{\mathbf{x}}}_{k+1|k}\hat{\widetilde{\mathbf{x}}}_{k+1|k}^{T}]=\mathbf{U}^{xv}_{k}\overline{\mathbf{F}}^{x}_{k}\mathbb{E}[\hat{\widetilde{\mathbf{x}}}_{k}\hat{\widetilde{\mathbf{x}}}_{k}^{T}](\overline{\mathbf{F}}^{x}_{k})^{T}\mathbf{U}^{xv}_{k}+\mathbf{U}^{xv}_{k}\overline{\mathbf{F}}^{v}_{k}\mathbf{R}_{k+1}(\overline{\mathbf{F}}^{v}_{k})^{T}\mathbf{U}^{xv}_{k}$. Since $\mathbb{E}[\hat{\widetilde{\mathbf{x}}}_{k}\hat{\widetilde{\mathbf{x}}}_{k}^{T}]\preceq\overline{\bm{\Sigma}}_{k}$, we have $\mathbb{E}[\hat{\widetilde{\mathbf{x}}}_{k+1|k}\hat{\widetilde{\mathbf{x}}}_{k+1|k}^{T}]\preceq\overline{\bm{\Sigma}}_{k+1|k}$.

Similarly, using \eqref{eqn:SLT observation}, we predict observation $\mathbf{a}_{k+1}$ as $\hat{\mathbf{a}}_{k+1|k}=\mathbf{U}^{a}_{k+1}\overline{\mathbf{G}}_{k+1}\doublehat{\mathbf{x}}_{k+1|k}$ and $\overline{\bm{\Sigma}}^{a}_{k+1}=\mathbf{U}^{a}_{k+1}\overline{\mathbf{G}}_{k+1}\overline{\bm{\Sigma}}_{k+1|k}\overline{\mathbf{G}}_{k+1}^{T}\mathbf{U}^{a}_{k+1}+\overline{\mathbf{R}}_{k+1}$ with I-EKF's gain matrix $\overline{\mathbf{K}}_{k+1}=\overline{\bm{\Sigma}}_{k+1|k}\overline{\mathbf{G}}_{k+1}^{T}\mathbf{U}^{a}_{k+1}(\overline{\bm{\Sigma}}^{a}_{k+1})^{-1}$. Again, $\hat{\widetilde{\mathbf{x}}}_{k+1}=(\mathbf{I}-\overline{\mathbf{K}}_{k+1}\mathbf{U}^{a}_{k+1}\overline{\mathbf{G}}_{k+1})\hat{\widetilde{\mathbf{x}}}_{k+1|k}-\overline{\mathbf{K}}_{k+1}\bm{\epsilon}_{k+1}$, which implies $\mathbb{E}[\hat{\widetilde{\mathbf{x}}}_{k+1}\hat{\widetilde{\mathbf{x}}}_{k+1}^{T}]=(\mathbf{I}-\overline{\mathbf{K}}_{k+1}\mathbf{U}^{a}_{k+1}\overline{\mathbf{G}}_{k+1})\mathbb{E}[\hat{\widetilde{\mathbf{x}}}_{k+1|k}\hat{\widetilde{\mathbf{x}}}_{k+1|k}^{T}](\mathbf{I}-\overline{\mathbf{K}}_{k+1}\mathbf{U}^{a}_{k+1}\overline{\mathbf{G}}_{k+1})^{T}+\overline{\mathbf{K}}_{k+1}\overline{\mathbf{R}}_{k+1}\overline{\mathbf{K}}_{k+1}^{T}$. Finally, using\\ $\mathbb{E}[\hat{\widetilde{\mathbf{x}}}_{k+1|k}\hat{\widetilde{\mathbf{x}}}_{k+1|k}^{T}]\preceq\overline{\bm{\Sigma}}_{k+1|k}$, we have $\mathbb{E}[\hat{\widetilde{\mathbf{x}}}_{k+1}\hat{\widetilde{\mathbf{x}}}_{k+1}^{T}]\preceq\overline{\bm{\Sigma}}_{k+1}$.

\section{Proof of Theorem~\ref{theorem: forward SOEKF stability}}
\label{App-thm-forward SOEKF stability}
\subsection{Preliminaries}
\begin{lemma}
\label{lemma: SOEKF stable bounds}
Under the assumptions of Theorem \ref{theorem: forward SOEKF stability}, the following bounds hold true for all $k\geq 0$.
\begin{enumerate}
    \item $\sum_{i=1}^{n_{x}}\sum_{j=1}^{n_{x}}\widetilde{\mathbf{a}}_{i}\widetilde{\mathbf{a}}_{j}^{T}\textrm{Tr}\left(\nabla^{2}\left[f(\hat{\mathbf{x}}_{k})\right]_{i}\bm{\Sigma}_{k}\nabla^{2}\left[f(\hat{\mathbf{x}}_{k})\right]_{j}\bm{\Sigma}_{k}\right)$ is a p.s.d. matrix and satisfies the upper bound $\sum_{i=1}^{n_{x}}\sum_{j=1}^{n_{x}}\widetilde{\mathbf{a}}_{i}\widetilde{\mathbf{a}}_{j}^{T}\textrm{Tr}\left(\nabla^{2}\left[f(\hat{\mathbf{x}}_{k})\right]_{i}\bm{\Sigma}_{k}\nabla^{2}\left[f(\hat{\mathbf{x}}_{k})\right]_{j}\bm{\Sigma}_{k}\right)\preceq\overline{a}^{2}\overline{\sigma}^{2}n_{x}^{2}\mathbf{I}$.
    \item $\sum_{i=1}^{n_{y}}\sum_{j=1}^{n_{y}}\mathbf{b}_{i}\mathbf{b}_{j}^{T}\textrm{Tr}\left(\nabla^{2}\left[h(\hat{\mathbf{x}}_{k})\right]_{i}\bm{\Sigma}_{k}\nabla^{2}\left[h(\hat{\mathbf{x}}_{k})\right]_{j}\bm{\Sigma}_{k}\right)$ is a p.s.d. matrix and satisfies the upper bound $\sum_{i=1}^{n_{y}}\sum_{j=1}^{n_{y}}\mathbf{b}_{i}\mathbf{b}_{j}^{T}\textrm{Tr}\left(\nabla^{2}\left[h(\hat{\mathbf{x}}_{k})\right]_{i}\bm{\Sigma}_{k}\nabla^{2}\left[h(\hat{\mathbf{x}}_{k})\right]_{j}\bm{\Sigma}_{k}\right)\preceq\overline{b}^{2}\overline{\sigma}^{2}n_{x}n_{y}\mathbf{I}$.
    \item $\|\mathbf{M}_{k}\|\leq\beta$ with $\beta>0$.
\end{enumerate}
\end{lemma}
\begin{proof}
Using the bounds from the assumptions of Theorem \ref{theorem: forward SOEKF stability}, we have for all $i,j\in\lbrace1,2,\hdots,n_{x}\rbrace$
\par\noindent\small
\begin{align*}
\underline{a}^{2}\underline{\sigma}^{2}\mathbf{I}\preceq\nabla^{2}\left[f(\hat{\mathbf{x}}_{k})\right]_{i}\bm{\Sigma}_{k}\nabla^{2}\left[f(\hat{\mathbf{x}}_{k})\right]_{j}\bm{\Sigma}_{k}\preceq\overline{a}^{2}\overline{\sigma}^{2}\mathbf{I},
\end{align*}
which implies
\begin{align*}
\underline{a}^{2}\underline{\sigma}^{2}n_{x}\leq \textrm{Tr}\left(\nabla^{2}\left[f(\hat{\mathbf{x}}_{k})\right]_{i}\bm{\Sigma}_{k}\nabla^{2}\left[f(\hat{\mathbf{x}}_{k})\right]_{j}\bm{\Sigma}_{k}\right)\leq\overline{a}^{2}\overline{\sigma}^{2}n_{x}.
\end{align*}
\normalsize
Now, $\sum_{i=1}^{n_{x}}\sum_{j=1}^{n_{x}}\widetilde{\mathbf{a}}_{i}\widetilde{\mathbf{a}}_{j}^{T}$ is an $n_{x}\times n_{x}$ all-ones matrix with $n_{x}$ as one of its eigenvalue and all other ($n_{x}-1$) eigenvalues are zero. Hence, $\sum_{i=1}^{n_{x}}\sum_{j=1}^{n_{x}}\widetilde{\mathbf{a}}_{i}\widetilde{\mathbf{a}}_{j}^{T}\textrm{Tr}\left(\nabla^{2}\left[f(\hat{\mathbf{x}}_{k})\right]_{i}\bm{\Sigma}_{k}\nabla^{2}\left[f(\hat{\mathbf{x}}_{k})\right]_{j}\bm{\Sigma}_{k}\right)$ is a p.s.d. matrix and satisfies the first bound in the lemma. Similarly, the bound on\\ $\sum_{i=1}^{n_{y}}\sum_{j=1}^{n_{y}}\mathbf{b}_{i}\mathbf{b}_{j}^{T}\textrm{Tr}\left(\nabla^{2}\left[h(\hat{\mathbf{x}}_{k})\right]_{i}\bm{\Sigma}_{k}\nabla^{2}\left[h(\hat{\mathbf{x}}_{k})\right]_{j}\bm{\Sigma}_{k}\right)$ can be derived. Further, the maximum singular value of $\sum_{i=1}^{n_{x}}\sum_{j=1}^{n_{y}}\widetilde{\mathbf{a}}_{i}\mathbf{b}_{j}^{T}$ is $\sqrt{n_{x}n_{y}}$ and hence, using the bounds, we can show that $\|\mathbf{M}_{k}\|$ satisfies $\|\mathbf{M}_{k}\|\leq\beta$ with $\beta=\frac{1}{2}\overline{a}\overline{b}\overline{\sigma}^{2}n_{x}\sqrt{n_{x}n_{y}}$.
\end{proof}

\begin{lemma}
\label{lemma: SOEKF stable alpha term}
Under the assumptions of Theorem \ref{theorem: forward SOEKF stability}, there exists a real number $\alpha$ with $0<\alpha<1$ such that $(\mathbf{F}_{k}-\mathbf{K}_{k}\mathbf{H}_{k})^{T}\bm{\Sigma}_{k+1}^{-1}(\mathbf{F}_{k}-\mathbf{K}_{k}\mathbf{H}_{k})\preceq(1-\alpha)\bm{\Sigma}_{k}^{-1}$.
\end{lemma}
\begin{proof}
Using \eqref{eqn: one step SOEKF sig tilde} and \eqref{eqn: one step SOEKF sig update}, we have
\par\noindent\small
\begin{align*}
    \bm{\Sigma}_{k+1}&=\mathbf{F}_{k}\bm{\Sigma}_{k}\mathbf{F}_{k}^{T}+\mathbf{Q}_{k}-\mathbf{K}_{k}\bm{\Sigma}^{y}_{k}\mathbf{K}_{k}^{T}+\frac{1}{2}\sum_{i=1}^{n_{x}}\sum_{j=1}^{n_{x}}\widetilde{\mathbf{a}}_{i}\widetilde{\mathbf{a}}_{j}^{T}\textrm{Tr}\left(\nabla^{2}\left[f(\hat{\mathbf{x}}_{k})\right]_{i}\bm{\Sigma}_{k}\nabla^{2}\left[f(\hat{\mathbf{x}}_{k})\right]_{j}\bm{\Sigma}_{k}\right).
\end{align*}
\normalsize
Using the positive semi-definiteness of the last term (as proved in Lemma \ref{lemma: SOEKF stable bounds}) and substituting for $\mathbf{K}_{k}\bm{\Sigma}^{y}_{k}$ using \eqref{eqn: one step SOEKF Kk}, we have
\par\noindent\small
\begin{align*}
    \bm{\Sigma}_{k+1}\succeq\mathbf{F}_{k}\bm{\Sigma}_{k}\mathbf{F}_{k}^{T}+\mathbf{Q}_{k}-(\mathbf{F}_{k}\bm{\Sigma}_{k}\mathbf{H}_{k}^{T}+\mathbf{M}_{k})\mathbf{K}_{k}^{T}.
\end{align*}
\normalsize
Rearranging the terms,
\par\noindent\small
\begin{align*}
    \bm{\Sigma}_{k+1}\succeq&(\mathbf{F}_{k}-\mathbf{K}_{k}\mathbf{H}_{k})\bm{\Sigma}_{k}(\mathbf{F}_{k}-\mathbf{K}_{k}\mathbf{H}_{k})^{T}+\mathbf{Q}_{k}+\mathbf{K}_{k}\mathbf{H}_{k}\bm{\Sigma}_{k}(\mathbf{F}_{k}-\mathbf{K}_{k}\mathbf{H}_{k})^{T}-\mathbf{M}_{k}\mathbf{K}_{k}^{T}.
\end{align*}
\normalsize
First, we consider the last two terms. Putting $\mathbf{M}_{k}=\mathbf{K}_{k}\bm{\Sigma}^{y}_{k}-\mathbf{F}_{k}\bm{\Sigma}_{k}\mathbf{H}_{k}^{T}$, the last terms can be expressed as
\par\noindent\small
\begin{align*}
    &\mathbf{K}_{k}\mathbf{H}_{k}\bm{\Sigma}_{k}(\mathbf{F}_{k}-\mathbf{K}_{k}\mathbf{H}_{k})^{T}-\mathbf{M}_{k}\mathbf{K}_{k}^{T}=\mathbf{K}_{k}\mathbf{H}_{k}\bm{\Sigma}_{k}\mathbf{F}_{k}^{T}+\mathbf{F}_{k}\bm{\Sigma}_{k}\mathbf{H}_{k}^{T}\mathbf{K}_{k}^{T}-\mathbf{K}_{k}(\bm{\Sigma}^{y}_{k}+\mathbf{H}_{k}\bm{\Sigma}_{k}\mathbf{H}_{k}^{T})\mathbf{K}_{k}^{T}.
\end{align*}
\normalsize
Let $\mathbf{A}=\mathbf{K}_{k}\mathbf{H}_{k}\bm{\Sigma}_{k}\mathbf{F}_{k}^{T}$. Using the bounds of Theorem \ref{theorem: forward SOEKF stability} and Lemma \ref{lemma: SOEKF stable bounds}, we can show that $\|\mathbf{A}\|\leq\frac{\overline{f}\overline{\sigma}\overline{h}(\overline{f}\overline{\sigma}\overline{h}+\beta)}{\underline{r}}$. Also, $\mathbf{A}+\mathbf{A}^{T}$ is a symmetric `$n_{x}\times n_{x}$' matrix with $\|\mathbf{A}+\mathbf{A}^{T}\|\leq\frac{2\overline{f}\overline{\sigma}\overline{h}(\overline{f}\overline{\sigma}\overline{h}+\beta)}{\underline{r}}$ which implies
\par\noindent\small
\begin{align*}
    -\frac{2\overline{f}\overline{\sigma}\overline{h}(\overline{f}\overline{\sigma}\overline{h}+\beta)}{\underline{r}}\mathbf{I}\preceq\mathbf{A}+\mathbf{A}^{T}\preceq\frac{2\overline{f}\overline{\sigma}\overline{h}(\overline{f}\overline{\sigma}\overline{h}+\beta)}{\underline{r}}\mathbf{I}.
\end{align*}
\normalsize
Using this and other bounds, we have
\par\noindent\small
\begin{align*}
    \mathbf{K}_{k}\mathbf{H}_{k}\bm{\Sigma}_{k}(\mathbf{F}_{k}-\mathbf{K}_{k}\mathbf{H}_{k})^{T}-\mathbf{M}_{k}\mathbf{K}_{k}^{T}\succeq-c\mathbf{I},
\end{align*}
\normalsize
where $c=\frac{2\overline{f}\overline{\sigma}\overline{h}(\overline{f}\overline{\sigma}\overline{h}+\beta)}{\underline{r}}+\left(2\overline{\sigma}\overline{h}^{2}+\delta+\frac{1}{2}\overline{b}^{2}\overline{\sigma}^{2}n_{x}n_{y}\right)\left(\frac{\overline{f}\overline{\sigma}\overline{h}+\beta}{\underline{r}}\right)^{2}$ is a positive constant. Hence,
\par\noindent\small
\begin{align*}
    \bm{\Sigma}_{k+1}\succeq(\mathbf{F}_{k}-\mathbf{K}_{k}\mathbf{H}_{k})\bm{\Sigma}_{k}(\mathbf{F}_{k}-\mathbf{K}_{k}\mathbf{H}_{k})^{T}+\mathbf{Q}_{k}-c\mathbf{I}.
\end{align*}
\normalsize
Similar to the proof of \cite[Lemma 3.1]{reif1999stochastic}, we can prove $\mathbf{F}_{k}-\mathbf{K}_{k}\mathbf{H}_{k}$ to be invertible using matrix inversion lemma. Here,
\par\noindent\small
\begin{align*}
    \mathbf{F}_{k}^{-1}(\mathbf{F}_{k}-\mathbf{K}_{k}\mathbf{H}_{k})\bm{\Sigma}_{k}=\bm{\Sigma}_{k}-\bm{\Sigma}_{k}(\mathbf{H}_{k}^{T}+\bm{\Sigma}_{k}^{-1}\mathbf{F}_{k}^{-1}\mathbf{M}_{k})(\bm{\Sigma}^{y}_{k})^{-1}\mathbf{H}_{k}\bm{\Sigma}_{k}.
\end{align*}
\normalsize
The R.H.S. is in the form of matrix inversion lemma:
\par\noindent\small
\begin{align*}
    (\mathbf{A}+\mathbf{U}\mathbf{C}\mathbf{V})^{-1}=\mathbf{A}^{-1}-\mathbf{A}^{-1}\mathbf{U}(\mathbf{C}^{-1}+\mathbf{V}\mathbf{A}^{-1}\mathbf{U})^{-1}\mathbf{V}\mathbf{A}^{-1},
\end{align*}
\normalsize
where $\mathbf{A}$ and $\mathbf{C}$ are invertible matrices. Hence, the matrix inversion lemma is applicable if $\mathbf{R}_{k}+\frac{1}{2}\sum_{i=1}^{n_{y}}\sum_{j=1}^{n_{y}}\mathbf{b}_{i}\mathbf{b}_{j}^{T}\textrm{Tr}\left(\nabla^{2}\left[h(\hat{\mathbf{x}}_{k})\right]_{i}\bm{\Sigma}_{k}\nabla^{2}\left[h(\hat{\mathbf{x}}_{k})\right]_{j}\bm{\Sigma}_{k}\right)-\mathbf{H}_{k}\mathbf{F}_{k}^{-1}\mathbf{M}_{k}$ is invertible (because $\mathbf{C}$ needs to be invertible). Since the difference of two matrices $\mathbf{X}-\mathbf{Y}$ is invertible if maximum singular value of $\mathbf{Y}$ is strictly less than the minimum singular value of $\mathbf{X}$, the required difference is invertible if $\|\mathbf{H}_{k}\mathbf{F}_{k}^{-1}\mathbf{M}_{k}\|<\underline{r}$ because $\mathbf{R}_{k}+\frac{1}{2}\sum_{i=1}^{n_{y}}\sum_{j=1}^{n_{y}}\mathbf{b}_{i}\mathbf{b}_{j}^{T}\textrm{Tr}\left(\nabla^{2}\left[h(\hat{\mathbf{x}}_{k})\right]_{i}\bm{\Sigma}_{k}\nabla^{2}\left[h(\hat{\mathbf{x}}_{k})\right]_{j}\bm{\Sigma}_{k}\right)\succeq\underline{r}\mathbf{I}$. But $\|\mathbf{H}_{k}\mathbf{F}_{k}^{-1}\mathbf{M}_{k}\|\leq\overline{h}\beta\|\mathbf{F}_{k}^{-1}\|$. Substituting the value of $\beta$ as derived in Lemma \ref{lemma: SOEKF stable bounds}, the sufficient condition for invertibility of the required matrix is
\par\noindent\small
\begin{align*}
    \|\mathbf{F}_{k}^{-1}\|<\frac{2\underline{r}}{\overline{h}\overline{a}\overline{b}\overline{\sigma}^{2}n_{x}\sqrt{n_{x}n_{y}}}.
\end{align*}
\normalsize
The condition \eqref{eqn: SOEKF stable constraint on inverse norm} is sufficient for this to be satisfied. Note that this condition is a sufficient but not necessary condition for invertibility of $\mathbf{F}_{k}-\mathbf{K}_{k}\mathbf{H}_{k}$. With this condition, Lemma \ref{lemma: SOEKF stable alpha term} can be proved using similar approach as in \cite[Lemma 3.1]{reif1999stochastic} with
\par\noindent\small
\begin{align*}
    1-\alpha=\left(1+\frac{\underline{q}-c}{\overline{\sigma}(\overline{f}+(\overline{f}\overline{\sigma}\overline{h}^{2}+\beta\overline{h})/\underline{r})^{2}}\right)^{-1}.
\end{align*}
\normalsize
The condition \eqref{eqn: SOEKF stable constraint on q} is sufficient for $0<\alpha<1$. 

Substituting for constant $c$ and using $\underline{q}\leq\delta$ and $\underline{r}\leq\delta$, the inequality \eqref{eqn: SOEKF stable constraint on q} requires the following inequality to be satisfied
\par\noindent\small
\begin{align*}
    \frac{2\overline{f}\overline{\sigma}\overline{h}(\overline{f}\overline{\sigma}\overline{h}+\beta)}{\delta}+\left(2\overline{\sigma}\overline{h}^{2}+\delta+\frac{1}{2}\overline{b}^{2}\overline{\sigma}^{2}n_{x}n_{y}\right)\left(\frac{\overline{f}\overline{\sigma}\overline{h}+\beta}{\delta}\right)^{2}<\delta.
\end{align*}
\normalsize
Rearranging the terms, we have
\par\noindent\small
\begin{align*}
    \delta^{3}-(\overline{f}\overline{\sigma}\overline{h}+\beta)(3\overline{f}\overline{\sigma}\overline{h}+\beta)\delta-(\overline{f}\overline{\sigma}\overline{h}+\beta)^{2}\left(2\overline{\sigma}\overline{h}^{2}+\frac{1}{2}\overline{b}^{2}\overline{\sigma}^{2}n_{x}n_{y}\right)> 0.
\end{align*}
\normalsize
It can be observed that this inequality requires the noise bound $\delta$ to be large enough and other bounds on matrices like $\overline{f},\overline{h}$ etc. to be small.
\end{proof}

\begin{lemma}
\label{lemma: SOEKF stable r term}
Under the assumptions of Theorem \ref{theorem: forward SOEKF stability}, there exists positive real constants $\kappa_\textrm{nonl}$, $\epsilon'$ such that $\mathbf{r}_{k}^{T}\bm{\Sigma}_{k+1}^{-1}(2(\mathbf{F}_{k}-\mathbf{K}_{k}\mathbf{H}_{k})\mathbf{e}_{k}+\mathbf{r}_{k})\leq\kappa_\textrm{nonl}\|\mathbf{e}_{k}\|_{2}^{4}$ for $\|\mathbf{e}_{k}\|_{2}\leq\epsilon'$.
\end{lemma}
\begin{proof}
The proof follows from \cite[Lemma 3.2]{reif1999stochastic} with $\kappa_\textrm{nonl}=\frac{\kappa'}{\underline{\sigma}}\left(2\left(\overline{f}+\frac{\overline{f}\overline{\sigma}\overline{h}^{2}+\beta\overline{h}}{\underline{r}}\right)+\kappa'\epsilon'^{2}\right)$, where $\kappa'=\kappa_{\phi}+\kappa_{\chi}\frac{\overline{f}\overline{\sigma}\overline{h}+\beta}{\underline{r}}$ and $\epsilon'=\textrm{min}(\epsilon_{\phi},\epsilon_{\chi})$.
\end{proof}

\begin{lemma}
\label{lemma: SOEKF stable noise term}
Under the assumptions of Theorem \ref{theorem: forward SOEKF stability}, there exists positive real constant $\kappa_\textrm{noise}$ independent of $\delta$ such that $\mathbb{E}[\mathbf{s}_{k}^{T}\bm{\Sigma}_{k+1}^{-1}\mathbf{s}_{k}]\leq\kappa_\textrm{noise}\delta$.
\end{lemma}
\begin{proof}
The proof follows from \cite[Lemma 3.3]{reif1999stochastic} with $\kappa_\textrm{noise}=\frac{n_{x}}{\underline{\sigma}}+\frac{\overline{f}^{2}\overline{h}^{2}\overline{\sigma}^{2}n_{y}}{\underline{\sigma}\underline{r}^{2}}$.
\end{proof}

\begin{lemma}
\label{lemma: SOEKF stable q term}
Under the assumptions of Theorem \ref{theorem: forward SOEKF stability}, there exist positive real constants $\kappa_\textrm{sec}$, $c_\textrm{sec}$ such that $\mathbf{q}_{k}^{T}\bm{\Sigma}_{k+1}^{-1}(2(\mathbf{F}_{k}-\mathbf{K}_{k}\mathbf{H}_{k})\mathbf{e}_{k}+2\mathbf{r}_{k}+\mathbf{q}_{k})\leq\kappa_\textrm{sec}\|\mathbf{e}_{k}\|_{2}^{3}+c_\textrm{sec}$ for $\|\mathbf{e}_{k}\|_{2}\leq\epsilon'$ where $\epsilon'$ is same as in Lemma \ref{lemma: SOEKF stable r term}.
\end{lemma}
\begin{proof}
Using the upper bound on $\nabla^{2}\left[f(\hat{\mathbf{x}}_{k})\right]_{i}$ from the assumptions of Theorem \ref{theorem: forward SOEKF stability}, we have $\mathbf{e}_{k}^{T}\nabla^{2}\left[f(\hat{\mathbf{x}}_{k})\right]_{i}\mathbf{e}_{k}\leq\overline{a}\|\mathbf{e}_{k}\|_{2}^{2}$ which implies
\begin{align*}
&\|\sum_{i=1}^{n_{x}}\widetilde{\mathbf{a}}_{i}\mathbf{e}_{k}^{T}\nabla^{2}\left[f(\hat{\mathbf{x}}_{k})\right]_{i}\mathbf{e}_{k}\|_{2}\leq\overline{a}n_{x}\|\mathbf{e}_{k}\|_{2}^{2}.
\end{align*}
\normalsize
Similarly, using other bounds, it is trivial to show that for all $k\geq 0$, the following bounds are satisfied
\par\noindent\small
\begin{align*}
&\|\sum_{i=1}^{n_{x}}\widetilde{\mathbf{a}}_{i}\textrm{Tr}\left(\nabla^{2}\left[f(\hat{\mathbf{x}}_{k})\right]_{i}\bm{\Sigma}_{k}\right)\|_{2}\leq\overline{a}\overline{\sigma}n_{x}^{2},\\
&\|\mathbf{K}_{k}\sum_{i=1}^{n_{y}}\mathbf{b}_{i}\mathbf{e}_{k}^{T}\nabla^{2}\left[h(\hat{\mathbf{x}}_{k})\right]_{i}\mathbf{e}_{k}\|_{2}\leq\frac{\overline{b}n_{y}(\overline{f}\overline{\sigma}\overline{h}+\beta)}{\underline{r}}\|\mathbf{e}_{k}\|_{2}^{2},\\
&\|\mathbf{K}_{k}\sum_{i=1}^{n_{y}}\mathbf{b}_{i}\textrm{Tr}\left(\nabla^{2}\left[h(\hat{\mathbf{x}}_{k})\right]_{i}\bm{\Sigma}_{k}\right)\|_{2}\leq\frac{\overline{b}\overline{\sigma}n_{x}n_{y}(\overline{f}\overline{\sigma}\overline{h}+\beta)}{\underline{r}}.
\end{align*}
\normalsize
Using these bounds, we have
\par\noindent\small
\begin{align*}
    \|\mathbf{q}_{k}\|_{2}\leq\kappa_{q}\|\mathbf{e}_{k}\|_{2}^{2}+c_{q},
\end{align*}
\normalsize
where $\kappa_{q}=\frac{1}{2}\left(\overline{a}n_{x}+\frac{\overline{b}n_{y}(\overline{f}\overline{\sigma}\overline{h}+\beta)}{\underline{r}}\right)$ and $c_{q}=\frac{1}{2}\left(\overline{a}\overline{\sigma}n_{x}^{2}+\frac{\overline{b}\overline{\sigma}n_{x}n_{y}(\overline{f}\overline{\sigma}\overline{h}+\beta)}{\underline{r}}\right)$.

Using the bounds on $\|\mathbf{q}_{k}\|_{2}$ and $\|\mathbf{r}_{k}\|_{2}$ as used in \cite[Lemma 3.2]{reif1999stochastic}, the lemma can be proved with
\footnotesize
\begin{align*}
&\kappa_\textrm{sec}=\frac{\kappa_{q}}{\underline{\sigma}}\left(2\left(\overline{f}+\frac{\overline{f}\overline{\sigma}\overline{h}^{2}+\beta\overline{h}}{\underline{r}}\right)+2\kappa'\epsilon'^{2}+\kappa_{q}\epsilon'\right),\text{and}\\
&c_\textrm{sec}=\frac{c_{q}^{2}}{\underline{\sigma}}+\frac{\kappa_{q}c_{q}\epsilon'^{2}}{\underline{\sigma}}+\frac{c_{q}\epsilon'}{\underline{\sigma}}\left(2\left(\overline{f}+\frac{\overline{f}\overline{\sigma}\overline{h}^{2}+\beta\overline{h}}{\underline{r}}\right)+2\kappa'\epsilon'^{2}+\kappa_{q}\epsilon'\right).
\end{align*}
\normalsize
\end{proof}

\subsection{Proof of the theorem}
Consider $V_{k}(\mathbf{e}_{k})=\mathbf{e}_{k}^{T}\bm{\Sigma}_{k}^{-1}\mathbf{e}_{k}$. Using the bounds on $\bm{\Sigma}_{k}$ from assumptions of Theorem \ref{theorem: forward SOEKF stability}, we have
\par\noindent\small
\begin{align*}
    \frac{1}{\overline{\sigma}}\|\mathbf{e}_{k}\|_{2}^{2}\leq V_{k}(\mathbf{e}_{k})\leq\frac{1}{\underline{\sigma}}\|\mathbf{e}_{k}\|_{2}^{2}.
\end{align*}
\normalsize
Also, substituting for $\mathbf{e}_{k+1}$ using \eqref{eqn: forward SOEKF error}, we have
\par\noindent\small
\begin{align*}
V_{k+1}(\mathbf{e}_{k+1})&=\mathbf{e}_{k}^{T}(\mathbf{F}_{k}-\mathbf{K}_{k}\mathbf{H}_{k})^{T}\bm{\Sigma}_{k+1}^{-1}(\mathbf{F}_{k}-\mathbf{K}_{k}\mathbf{H}_{k})\mathbf{e}_{k}+\mathbf{r}_{k}^{T}\bm{\Sigma}_{k+1}^{-1}(2(\mathbf{F}_{k}-\mathbf{K}_{k}\mathbf{H}_{k})\mathbf{e}_{k}+\mathbf{r}_{k})\\
&+\mathbf{q}_{k}^{T}\bm{\Sigma}_{k+1}^{-1}(2(\mathbf{F}_{k}-\mathbf{K}_{k}\mathbf{H}_{k})\mathbf{e}_{k}+2\mathbf{r}_{k}+\mathbf{q}_{k})+\mathbf{s}_{k}^{T}\bm{\Sigma}_{k+1}^{-1}\mathbf{s}_{k}+2\mathbf{s}_{k}^{T}\bm{\Sigma}_{k+1}^{-1}((\mathbf{F}_{k}-\mathbf{K}_{k}\mathbf{H}_{k})\mathbf{e}_{k}+\mathbf{r}_{k}+\mathbf{q}_{k}).
\end{align*}
\normalsize
Using Lemma \ref{lemma: SOEKF stable alpha term}, \ref{lemma: SOEKF stable r term}, and \ref{lemma: SOEKF stable q term}, we have for $\|\mathbf{e}_{k}\|_{2}\leq\epsilon'$
\par\noindent\small
\begin{align*}
\hspace{-0.3cm}V_{k+1}(\mathbf{e}_{k+1})\leq&(1-\alpha)V_{k}(\mathbf{e}_{k})+\kappa_\textrm{nonl}\|\mathbf{e}_{k}\|_{2}^{4}+\kappa_\textrm{sec}\|\mathbf{e}_{k}\|_{2}^{3}+c_\textrm{sec}+\mathbf{s}_{k}^{T}\bm{\Sigma}_{k+1}^{-1}\mathbf{s}_{k}+2\mathbf{s}_{k}^{T}\bm{\Sigma}_{k+1}^{-1}((\mathbf{F}_{k}-\mathbf{K}_{k}\mathbf{H}_{k})\mathbf{e}_{k}+\mathbf{r}_{k}+\mathbf{q}_{k}).
\end{align*}
\normalsize
The last term $\mathbf{s}_{k}^{T}\bm{\Sigma}_{k+1}^{-1}((\mathbf{F}_{k}-\mathbf{K}_{k}\mathbf{H}_{k})\mathbf{e}_{k}+\mathbf{r}_{k}+\mathbf{q}_{k})$ vanishes on taking expectation conditioned on $\mathbf{e}_{k}$ and hence, for $\|\mathbf{e}_{k}\|_{2}\leq\epsilon'$,
\par\noindent\small
\begin{align*}
\mathbb{E}[V_{k+1}(\mathbf{e}_{k+1})\vert\mathbf{e}_{k}]\leq&(1-\alpha)V_{k}(\mathbf{e}_{k})+\kappa_\textrm{nonl}\|\mathbf{e}_{k}\|_{2}^{4}+\kappa_\textrm{sec}\|\mathbf{e}_{k}\|_{2}^{3}+c_\textrm{sec}+\kappa_{\textrm{noise}}\delta,
\end{align*}
\normalsize
where the bound of Lemma \ref{lemma: SOEKF stable noise term} is applied. But, for $\|\mathbf{e}_{k}\|_{2}\leq\epsilon'$,
\par\noindent\small
\begin{align*}
\kappa_\textrm{nonl}\|\mathbf{e}_{k}\|_{2}^{4}+\kappa_\textrm{sec}\|\mathbf{e}_{k}\|_{2}^{3}\leq(\kappa_\textrm{nonl}\epsilon'+\kappa_\textrm{sec})\|\mathbf{e}_{k}\|_{2}\|\mathbf{e}_{k}\|_{2}^{2}.
\end{align*}
\normalsize
Choosing $\epsilon=\textrm{min}\left(\epsilon',\frac{\alpha}{2\overline{\sigma}(\kappa_\textrm{nonl}\epsilon'+\kappa_\textrm{sec})}\right)$, we have for $\|\mathbf{e}_{k}\|_{2}\leq\epsilon$,
\par\noindent\small
\begin{align*}
\kappa_\textrm{nonl}\|\mathbf{e}_{k}\|_{2}^{4}+\kappa_\textrm{sec}\|\mathbf{e}_{k}\|_{2}^{3}\leq\frac{\alpha}{2}V_{k}(\mathbf{e}_{k}),
\end{align*}
which implies
\begin{align*}
\mathbb{E}[V_{k+1}(\mathbf{e}_{k+1})\vert\mathbf{e}_{k}]-V_{k}(\mathbf{e}_{k})\leq-\frac{\alpha}{2}V_{k}(\mathbf{e}_{k})+c_\textrm{sec}+\kappa_\textrm{noise}\delta.
\end{align*}
\normalsize
Hence, Lemma~\ref{lemma:exponential boundedness} applies here. However, to have negative mean drift, we require $\delta$ to be small enough such that there exists some $\widetilde{\epsilon}<\epsilon$ to satisfy \eqref{eqn: SOEKF stable constraint on delta}. This condition ensures that  for $\widetilde{\epsilon}\leq\|\mathbf{e}_{k}\|_{2}\leq\epsilon$, $\mathbb{E}[ V_{k+1}(\mathbf{e}_{k+1})\vert\mathbf{e}_{k}]-V_{k}(\mathbf{e}_{k})\leq 0$ is fulfilled and the estimation error $\mathbf{e}_{k}$ remains exponentially bounded in mean-squared sense if the error is within $\epsilon$ bound.

\section{Proof of Theorem~\ref{theorem: inverse SOEKF stability}}
\label{App-thm-inverse SOEKF stability}
From the state transition function of I-SOEKF \eqref{eqn: one step I-SOEKF state transition}, we have
\par\noindent\small
\begin{align*}
\nabla^{2}\left[\overline{f}_{k}(\doublehat{\mathbf{x}}_{k})\right]_{i}=\nabla^{2}\left[f(\doublehat{\mathbf{x}}_{k})\right]_{i}-\sum_{j=1}^{n_{y}}[\mathbf{K}_{k}]_{i,j}\nabla^{2}\left[h(\doublehat{\mathbf{x}}_{k})\right]_{j}.
\end{align*}
\normalsize
Using the upper bounds on Hessian matrices from the assumptions of Theorem \ref{theorem: forward SOEKF stability}, we have
\par\noindent\small
\begin{align*}
\nabla^{2}\left[\overline{f}_{k}(\doublehat{\mathbf{x}}_{k})\right]_{i}&\preceq\left(\overline{a}-\underline{b}\sum_{j=1}^{n_{y}}[\mathbf{K}_{k}]_{i,j}\right)\mathbf{I}\preceq\left(\overline{a}+|\underline{b}|\sum_{j=1}^{n_{y}}|[\mathbf{K}_{k}]_{i,j}|\right)\mathbf{I}.
\end{align*}
\normalsize
But, the row sum $\sum_{i=1}^{n_{y}}|[\mathbf{K}_{k}]_{i,j}|\leq\|\mathbf{K}_{k}\|_{\infty}$. Further, using equivalence of norms i.e. $\|\mathbf{K}_{k}\|_{\infty}\leq\sqrt{n_{y}}\|\mathbf{K}_{k}\|$, we have
\par\noindent\small
\begin{align*}
\nabla^{2}\left[\overline{f}_{k}(\doublehat{\mathbf{x}}_{k})\right]_{i}\preceq\left(\overline{a}+|\underline{b}|\sqrt{n_{y}}\left(\frac{\overline{f}\overline{\sigma}\overline{h}+\beta}{\underline{r}}\right)\right)\mathbf{I}.
\end{align*}
\normalsize
Similarly,
\par\noindent\small
\begin{align*}
\nabla^{2}\left[\overline{f}_{k}(\doublehat{\mathbf{x}}_{k})\right]_{i}&\succeq\left(\underline{a}-\overline{b}\sum_{j=1}^{n_{y}}[\mathbf{K}_{k}]_{i,j}\right)\mathbf{I}\succeq\left(\underline{a}-\frac{\overline{b}\sqrt{n_{y}}(\overline{f}\overline{\sigma}\overline{h}+\beta)}{\underline{r}}\right)\mathbf{I}.
\end{align*}
\normalsize
Hence, with $\underline{d}=\underline{a}-(\overline{b}\sqrt{n_{y}}(\overline{f}\overline{\sigma}\overline{h}+\beta)/\underline{r})$ and $\overline{d}=\overline{a}+(|\underline{b}|\sqrt{n_{y}}(\overline{f}\overline{\sigma}\overline{h}+\beta)/\underline{r})$, we have $\underline{d}\mathbf{I}\preceq\nabla^{2}\left[\overline{f}_{k}(\doublehat{x}_{k})\right]_{i}\preceq\overline{d}\mathbf{I}$ for all $i\in\lbrace 1,2,\hdots,n_{x}\rbrace$.

The proof of the remaining conditions of Theorem \ref{theorem: forward SOEKF stability} for I-SOEKF dynamics follows from the proof of Theorem~\ref{theorem: inverse ekf stable Reif} in Appendix~\ref{App-thm-inverse ekf stable Reif}.

\chapter{Proofs for forward and inverse unknown-input filters}
\label{chap:input proof}
\section{Proof of forward EKF-without-DF recursions}
\label{App-thm-forward-EKF-without-DF}
The forward EKF-without-DF is formulated based on a weighted least-squared error criterion. To this end, similar to EKF, the system model is first linearized locally at the estimates of the previous state and unknown inputs. The linearized models are then used to define a quadratic objective function of an extended state vector consisting of the current state and the unknown inputs at all time instants. Finally, recursive estimates are derived for the extended state vector and then simplified to yield forward EKF-without-DF recursions. As mentioned in Remark~\ref{remark:with and without diff}, systems without DF induce a one-step delay in input estimation.

Consider the non-linear state transition \eqref{eqn:non-linear state x with unknown input} and observation \eqref{eqn:non-linear observation y} without DF. We require estimates $\hat{\mathbf{x}}_{k|k}$ and $\hat{\mathbf{u}}_{k-1|k}$ (represented by $\hat{\mathbf{x}}_{k}$ and $\hat{\mathbf{u}}_{k-1}$ in Section~\ref{subsec:IEKF without DF}) of the state $\mathbf{x}_{k}$ and unknown input $\mathbf{u}_{k-1}$, respectively, given the observations $\{\mathbf{y}_{i}\}_{1\leq i\leq k}$. Linearize the non-linear functions $f(\cdot,\cdot)$ in \eqref{eqn:non-linear state x with unknown input} and $h(\cdot)$ in \eqref{eqn:non-linear observation y} with respect to the previous estimates as follows:
\par\noindent\small
\begin{align}
f(\mathbf{x}_{k},\mathbf{u}_{k})&=f(\hat{\mathbf{x}}_{k|k},\hat{\mathbf{u}}_{k-1|k})+\mathbf{F}_{k}(\mathbf{x}_{k}-\hat{\mathbf{x}}_{k|k})+\mathbf{B}_{k}(\mathbf{u}_{k}-\hat{\mathbf{u}}_{k-1|k}),\nonumber\\
    h(\mathbf{x}_{k+1})&=h(\hat{\mathbf{x}}_{k+1|k})+\mathbf{H}_{k+1}(\mathbf{x}_{k+1}-\hat{\mathbf{x}}_{k+1|k}),\label{eqn:supp 14}
\end{align}
\normalsize
Then, \eqref{eqn:non-linear state x with unknown input} becomes
\par\noindent\small
\begin{align}
\mathbf{x}_{k+1}=\mathbf{F}_{k}\mathbf{x}_{k}+\mathbf{B}_{k}\mathbf{u}_{k}+\overline{\mathbf{u}}_{k}+\mathbf{w}_{k},\label{eqn:supp 20}
\end{align}
\normalsize
where $\overline{\mathbf{u}}_{k}=f(\hat{\mathbf{x}}_{k|k},\hat{\mathbf{u}}_{k-1|k})-\mathbf{F}_{k}\hat{\mathbf{x}}_{k|k}-\mathbf{B}_{k}\hat{\mathbf{u}}_{k-1|k}$. Define the least-squared error objective function $J_{k+1}=\overline{\bm{\Delta}}_{k+1}^{T}\mathbf{W}_{k+1}\overline{\bm{\Delta}}_{k+1}$ where $\overline{\bm{\Delta}}_{k+1}=[\bm{\Delta}_{1}^{T},\bm{\Delta}_{2}^{T},\hdots,\bm{\Delta}_{k+1}^{T}]\in\mathbb{R}^{n_{y}(k+1)\times 1}$ with $\bm{\Delta}_{i}=\mathbf{y}_{i}-h(\mathbf{x}_{i})$. The weighting matrix $\mathbf{W}_{k+1}\in\mathbb{R}^{n_{y}(k+1)\times n_{y}(k+1)}$ is defined using the inverse of process and measurement noise covariance matrices as in \cite[Eq.~(33)]{pan2010applying}.

Define an extended state vector $\mathbf{z}_{k}=[\mathbf{x}_{k}^{T},\mathbf{u}_{1}^{T},\mathbf{u}_{2}^{T},\hdots,\mathbf{u}_{k-1}^{T}]^{T}$. We first represent the objective function $J_{k+1}$ in terms of $\mathbf{z}_{k+1}$. Rearranging \eqref{eqn:supp 20}, we obtain $\mathbf{x}_{k}=\mathbf{F}_{k}^{-1}\mathbf{x}_{k+1}-\mathbf{F}_{k}^{-1}(\mathbf{B}_{k}\mathbf{u}_{k}+\overline{\mathbf{u}}_{k}+\mathbf{w}_{k})$. Replacing $k$ by $k-1$, we have $\mathbf{x}_{k-1}=\mathbf{F}_{k-1}^{-1}\mathbf{x}_{k}-\mathbf{F}_{k-1}^{-1}(\mathbf{B}_{k-1}\mathbf{u}_{k-1}+\overline{\mathbf{u}}_{k-1}+\mathbf{w}_{k-1})$ such that $\mathbf{x}_{k-1}=\mathbf{F}_{k-1}^{-1}\mathbf{F}_{k}^{-1}\mathbf{x}_{k+1}-\mathbf{F}_{k-1}^{-1}\mathbf{F}_{k}^{-1}(\mathbf{B}_{k}\mathbf{u}_{k}+\overline{\mathbf{u}}_{k}+\mathbf{w}_{k})-\mathbf{F}_{k-1}^{-1}(\mathbf{B}_{k-1}\mathbf{u}_{k-1}+\overline{\mathbf{u}}_{k-1}+\mathbf{w}_{k-1})$. Repeating the procedure for $i=k-2,k-3,\hdots,1$, we obtain
\par\noindent\small
\begin{align}
    \mathbf{x}_{i}=\bm{\Phi}_{k+1,i}^{-1}\mathbf{x}_{k+1}-\left(\sum_{j=i}^{k}\bm{\Phi}_{j+1,i}^{-1}(\mathbf{B}_{j}\mathbf{u}_{j}+\overline{\mathbf{u}}_{j}+\mathbf{w}_{j})\right),\label{eqn:supp 23}
\end{align}
\normalsize
for $i=1,2,\hdots,k$ with $\bm{\Phi}_{q,s}^{-1}\doteq\mathbf{F}_{s}^{-1}\mathbf{F}_{s+1}^{-1}\hdots\mathbf{F}_{q-1}^{-1}$ for $q>s$ and $\bm{\Phi}_{s,s}^{-1}=\mathbf{I}$. Using \eqref{eqn:non-linear observation y}, \eqref{eqn:supp 14} and \eqref{eqn:supp 23}, we have
\par\noindent\small
\begin{align}
    &\bm{\Delta}_{i}=\mathbf{y}_{i}-\mathbf{H}_{i}\bm{\Phi}_{k+1,i}^{-1}\mathbf{x}_{k+1}+\mathbf{H}_{i}\left(\sum_{j=i}^{k}(\bm{\Phi}_{j+1,i}^{-1}\mathbf{B}_{j}\mathbf{u}_{j})\right)-\widetilde{\mathbf{u}}_{i|i-1},\;\;\textrm{where}\label{eqn:supp 28}\\
    &\widetilde{\mathbf{u}}_{i|i-1}=h(\hat{\mathbf{x}}_{i|i-1})-\mathbf{H}_{i}\left(\sum_{j=i}^{k}\bm{\Phi}_{j+1,i}^{-1}\overline{\mathbf{u}}_{j}+\hat{\mathbf{x}}_{i|i-1}\right).\label{eqn: supp 26}
\end{align}
\normalsize
for $i=1,2,\hdots,k+1$. Using \eqref{eqn:supp 28}, we express $\overline{\bm{\Delta}}_{k+1}$ as $ \overline{\bm{\Delta}}_{k+1}=\mathbf{Y}_{k+1}-\mathbf{A}_{z,k+1}\mathbf{z}_{k+1}$ where $\mathbf{Y}_{k+1}=[(\mathbf{y}_{1}-\widetilde{\mathbf{u}}_{1|0})^{T},(\mathbf{y}_{2}-\widetilde{\mathbf{u}}_{2|1})^{T},\hdots,(\mathbf{y}_{k+1}-\widetilde{\mathbf{u}}_{k+1|k})^{T}]$ and $\mathbf{A}_{z,k+1}=\begin{bmatrix}\widetilde{\mathbf{L}}_{k+1}&\widetilde{\mathbf{N}}_{k+1}\\\widetilde{\mathbf{H}}_{k+1}&\mathbf{0}_{n_{y}\times n_{u}}\end{bmatrix}$. Here,
\par\noindent\small
\begin{align}
    \widetilde{\mathbf{H}}_{k+1}=[\mathbf{H}_{k+1},\mathbf{0}_{n_{y}\times n_{u}(k-1)}],\label{eqn: supp 31}
\end{align}
\normalsize
while $\widetilde{\mathbf{N}}_{k+1}$ and $\widetilde{\mathbf{L}}_{k+1}$ are given by \cite[Eq.~32]{pan2010applying}.

Assume $n_{y}\geq n_{u}$ (condition for the existence of forward EKF-without-DF) and minimize the objective function $J_{k+1}$ with respect to the extended state vector $\mathbf{z}_{k+1}$ to yield the estimate $\hat{\mathbf{z}}_{k+1|k+1}=[\hat{\mathbf{x}}_{k+1|k+1}^{T},\hat{\mathbf{u}}_{1|k+1}^{T},\hat{\mathbf{u}}_{2|k+1}^{T},\hdots,\hat{\mathbf{u}}_{k|k+1}^{T}]$ given observations $\{\mathbf{y}_{i}\}_{1\leq i\leq k+1}$ as $\hat{\mathbf{z}}_{k+1|k+1}=\mathbf{P}_{z,k+1}(\mathbf{A}_{z,k+1}^{T}\mathbf{W}_{k+1}\mathbf{Y}_{k+1})$ where
\begin{align}
    \mathbf{P}_{z,k+1}=(\mathbf{A}_{z,k+1}^{T}\mathbf{W}_{k+1}\mathbf{A}_{z,k+1})^{-1}\label{eqn:supp 34}.
\end{align}
\normalsize

\subsection{Recursive solutions for extended state}\label{sec:extended}
In the following, we restate the updates to compute $\hat{\mathbf{z}}_{k+1|k+1}$ recursively as obtained in \cite[Appendix~A.1]{pan2010applying}. The procedure involves first expressing $\mathbf{A}_{z,k+1}$, $\mathbf{Y}_{k+1}$ and $\mathbf{W}_{k+1}$ in terms of $\mathbf{A}_{z,k}$, $\mathbf{Y}_{k}$ and $\mathbf{W}_{k}$ as follows
\par\noindent\small
\begin{align}
    &\mathbf{A}_{z,k+1}=\begin{bmatrix}
        \mathbf{A}_{z,k}\overline{\bm{\Phi}}_{k+1,k}^{-1}&-\mathbf{A}_{z,k}\overline{\bm{\Phi}}_{k+1,k}^{-1}\hat{\mathbf{B}}_{k}\\
        \widetilde{\mathbf{H}}_{k+1}&\mathbf{0}_{n_{y}\times n_{u}}
    \end{bmatrix},\label{eqn: A1 first part}\\
    &\mathbf{Y}_{k+1}=\begin{bmatrix}
        \mathbf{Y}_{k}+\mathbf{A}_{z,k}\overline{\bm{\Phi}}_{k+1,k}^{-1}\widetilde{\mathbf{U}}_{k}\\
        \mathbf{y}_{k+1}-\widetilde{\mathbf{u}}_{k+1|k}
    \end{bmatrix},\label{eqn: A2 first part}\\
    &\mathbf{W}_{k+1}=\begin{bmatrix}
        \widetilde{\mathbf{W}}_{k}&\mathbf{0}_{n_{y}k\times n_{y}}\\
        \mathbf{0}_{n_{y}\times n_{y}k}&\mathbf{R}_{k+1}^{-1}
    \end{bmatrix},\label{eqn: A3 above first part}
\end{align}
\normalsize
where $\mathbf{R}_{k}$ is the (time-varying) noise covariance matrix of measurement noise $\mathbf{v}_{k}$ in \eqref{eqn:non-linear observation y}. Also,
\par\noindent\small
\begin{align}
    &\overline{\bm{\Phi}}_{k+1,k}^{-1}=\begin{bmatrix}
        \bm{\Phi}_{k+1|k}^{-1}&\mathbf{0}_{n_{x}\times n_{u}(k-1)}\\
        \mathbf{0}_{n_{u}(k-1)\times n_{x}}&\mathbf{I}_{n_{u}(k-1)}
    \end{bmatrix},\label{eqn: A1 second part}\\
    &\widetilde{\mathbf{U}}_{k}=\begin{bmatrix}
        \overline{\mathbf{u}}_{k}\\
        \mathbf{0}_{n_{u}(k-1)\times 1}
    \end{bmatrix},\label{eqn: A2 second part}\\    &\widetilde{\mathbf{W}}_{k}=\left(\mathbf{W}_{k}^{-1}+\mathbf{A}_{z,k}\overline{\bm{\Phi}}_{k+1,k}^{-1}\widetilde{\mathbf{Q}}_{k}\overline{\bm{\Phi}}_{k+1,k}^{-T}\mathbf{A}_{z,k}^{T}\right)^{-1},\label{eqn: A3 above second part}\\
    &\widetilde{\mathbf{Q}}_{k}=\begin{bmatrix}
        \mathbf{Q}_{k}&\mathbf{0}_{n_{x}\times n_{u}(k-1)}\\
        \mathbf{0}_{n_{u}(k-1)\times n_{x}}&\mathbf{0}_{n_{u}(k-1)\times n_{u}(k-1)}
    \end{bmatrix},\label{eqn: A3 first part}\\
    &\hat{\mathbf{B}}_{k}=\begin{bmatrix}
        \mathbf{B}_{k}\\\mathbf{0}_{n_{u}(k-1)\times n_{u}}
    \end{bmatrix}.\label{eqn: A3 second part}
\end{align}
\normalsize
and $\mathbf{Q}_{k}$ is the (time-varying) noise covariance matrix of process noise $\mathbf{w}_{k}$ in \eqref{eqn:non-linear state x with unknown input}. Here, $(\cdot)^{-T}$ denotes $((\cdot)^{-1})^{T}$. Basically, the relation between $\mathbf{A}_{z,k+1}$ and $\mathbf{A}_{z,k}$ is through $\bm{\Phi}_{k+1,k}$. The relations \eqref{eqn: A1 first part}-\eqref{eqn: A3 above first part} are then obtained through comparison following a similar procedure as in \cite{yang2007adaptive} for EKF-with-DF case. Finally, consider the following matrix inversion formulas with $\mathbf{A}_{q}=\mathbf{A}_{4}-\mathbf{A}_{3}\mathbf{A}_{1}^{-1}\mathbf{A}_{2}$,
\par\noindent\small
\begin{align}
    &\begin{bmatrix}
        \mathbf{A}_{1}&\mathbf{A}_{2}\\
        \mathbf{A}_{3}&\mathbf{A}_{4}
    \end{bmatrix}^{-1}=\begin{bmatrix}
    \mathbf{A}_{1}^{-1}+\mathbf{A}_{1}^{-1}\mathbf{A}_{2}\mathbf{A}_{q}^{-1}\mathbf{A}_{3}\mathbf{A}_{1}^{-1}&-\mathbf{A}_{1}^{-1}\mathbf{A}_{2}\mathbf{A}_{q}^{-1}\\
    -\mathbf{A}_{q}^{-1}\mathbf{A}_{3}\mathbf{A}_{1}^{-1}&\mathbf{A}_{q}^{-1}
    \end{bmatrix},\label{eqn: A4}\\
    &(\mathbf{C}_{1}+\mathbf{C}_{2}\mathbf{C}_{3}\mathbf{C}_{4})^{-1}=\mathbf{C}_{1}^{-1}-\mathbf{C}_{1}^{-1}\mathbf{C}_{2}(\mathbf{C}_{3}^{-1}+\mathbf{C}_{4}\mathbf{C}_{1}^{-1}\mathbf{C}_{2})^{-1}\mathbf{C}_{4}\mathbf{C}_{1}^{-1},\label{eqn: A5}
\end{align}

Substituting \eqref{eqn: A1 first part} and \eqref{eqn: A3 above first part} in \eqref{eqn:supp 34}, we obtain $\mathbf{P}_{z,k+1}=\begin{bmatrix}
        \mathbf{P}_{1,k+1}&\mathbf{P}_{2,k+1}\\
       \mathbf{P}_{3,k+1}&\mathbf{P}_{4,k+1}        
    \end{bmatrix}^{-1}$ where $\mathbf{P}_{1,k+1}=\overline{\bm{\Phi}}_{k+1,k}^{-T}\mathbf{A}_{z,k}^{T}\widetilde{\mathbf{W}}_{k}\mathbf{A}_{z,k}\overline{\bm{\Phi}}_{k+1,k}^{-1}+\widetilde{\mathbf{H}}_{k+1}^{T}\mathbf{R}_{k+1}^{-1}\widetilde{\mathbf{H}}_{k+1}$, $\mathbf{P}_{2,k+1}=-\overline{\bm{\Phi}}_{k+1,k}^{-T}\mathbf{A}_{z,k}^{T}\widetilde{\mathbf{W}}_{k}\mathbf{A}_{z,k}\overline{\bm{\Phi}}_{k+1,k}^{-1}\hat{\mathbf{B}}_{k}$, $\mathbf{P}_{3,k+1}=\\-\hat{\mathbf{B}}_{k}^{T}\overline{\bm{\Phi}}_{k+1,k}^{-T}\mathbf{A}_{z,k}^{T}\widetilde{\mathbf{W}}_{k}\mathbf{A}_{z,k}\overline{\bm{\Phi}}_{k+1,k}^{-1}$ and $\mathbf{P}_{4,k+1}=\hat{\mathbf{B}}_{k}^{T}\overline{\bm{\Phi}}_{k+1,k}^{-T}\mathbf{A}_{z,k}^{T}\widetilde{\mathbf{W}}_{k}\mathbf{A}_{z,k}\overline{\bm{\Phi}}_{k+1,k}^{-1}\hat{\mathbf{B}}_{k}$. This is then simplified using \eqref{eqn: A4} and \eqref{eqn: A5}, similar to the procedure followed in \cite[Appendix~A]{yang2007adaptive}, to obtain the final recursive solutions for $\hat{\mathbf{z}}_{k+1|k+1}$. Define $\overline{\mathbf{P}}_{z,k+1}\doteq[\overline{\bm{\Phi}}_{k+1,k}^{-T}\mathbf{A}_{z,k}^{T}\widetilde{\mathbf{W}}_{k}\mathbf{A}_{z,k}\overline{\bm{\Phi}}_{k+1,k}^{-1}+\widetilde{\mathbf{H}}_{k+1}^{T}\mathbf{R}_{k+1}^{-1}\widetilde{\mathbf{H}}_{k+1}]^{-1}$. The recursive updates for $\hat{\mathbf{z}}_{k+1|k+1}$ are
\par\noindent\small
\begin{align}
    &\widetilde{\mathbf{P}}_{z,k+1}=\overline{\bm{\Phi}}_{k+1,k}\mathbf{P}_{z,k}\overline{\bm{\Phi}}_{k+1,k}^{T}+\widetilde{\mathbf{Q}}_{k},\label{eqn: A8 first part}\\
    &\mathbf{K}_{z,k+1}=\widetilde{\mathbf{P}}_{z,k+1}\widetilde{\mathbf{H}}_{k+1}^{T}(\mathbf{R}_{k+1}+\widetilde{\mathbf{H}}_{k+1}\widetilde{\mathbf{P}}_{z,k+1}\widetilde{\mathbf{H}}_{k+1}^{T})^{-1},\label{eqn: A8 second part}\\
    &\overline{\mathbf{P}}_{z,k+1}=(\mathbf{I}-\mathbf{K}_{z,k+1}\widetilde{\mathbf{H}}_{k+1})\widetilde{\mathbf{P}}_{z,k+1},\label{eqn: A9 first part}\\
    &\mathbf{S}_{k+1}=(\hat{\mathbf{B}}_{k}^{T}\widetilde{\mathbf{P}}_{z,k+1}^{-1}\mathbf{K}_{z,k+1}\widetilde{\mathbf{H}}_{k+1}\hat{\mathbf{B}}_{k})^{-1}\label{eqn: A9 second part}\\
    &\overline{\mathbf{z}}_{k+1}=(\overline{\bm{\Phi}}_{k+1,k}\hat{\mathbf{z}}_{k|k}+\widetilde{\mathbf{U}}_{k})+\mathbf{K}_{z,k+1}(\mathbf{y}_{k+1}-\widetilde{\mathbf{u}}_{k+1|k}-\widetilde{\mathbf{H}}_{k+1}(\overline{\bm{\Phi}}_{k+1,k}\hat{\mathbf{z}}_{k|k}+\widetilde{\mathbf{U}}_{k}))\label{eqn: A7}\\
    &\hat{\mathbf{u}}_{k|k+1}=-\mathbf{S}_{k+1}\hat{\mathbf{B}}_{k}^{T}\widetilde{\mathbf{P}}_{z,k+1}^{-1}(\overline{\bm{\Phi}}_{k+1,k}\hat{\mathbf{z}}_{k|k}+\widetilde{\mathbf{U}}_{k}-\overline{\mathbf{z}}_{k+1}),\label{eqn:A6}\\
    &\hat{\mathbf{z}}_{k+1|k+1}=\begin{bmatrix}
        \overline{\mathbf{z}}_{k+1}+\overline{\mathbf{P}}_{z,k+1}\widetilde{\mathbf{P}}_{z,k+1}^{-1}\hat{\mathbf{B}}_{k}\hat{\mathbf{u}}_{k|k+1}\\
        \hat{\mathbf{u}}_{k|k+1}
    \end{bmatrix}.\label{eqn: supp 35}
\end{align}
\normalsize

\subsection{Forward EKF-without-DF recursions}\label{sec:recursions}
From the extended state estimate $\hat{\mathbf{z}}_{k+1|k+1}$, we are only interested in $\hat{\mathbf{x}}_{k+1|k+1}$ and $\hat{\mathbf{u}}_{k|k+1}$ estimates of the current state $\mathbf{x}_{k+1}$ and unknown input $\mathbf{u}_{k}$, respectively. Hence, in this section, we simplify \eqref{eqn: A8 first part}-\eqref{eqn: supp 35} to obtain the forward EKF-without-DF recursions for computing $\hat{\mathbf{x}}_{k+1|k+1}$ and $\hat{\mathbf{u}}_{k|k+1}$. By definition, $\hat{\mathbf{z}}_{k+1|k+1}$ can be partitioned as $\hat{\mathbf{z}}_{k+1|k+1}=[\hat{\mathbf{x}}_{k+1|k+1}^{T},\hat{\mathbf{U}}_{k|k+1}^{T},\hat{\mathbf{u}}_{k|k+1}^{T}]$, where $\hat{\mathbf{U}}_{k|k+1}=[\hat{\mathbf{u}}_{1|k+1}^{T},\hat{\mathbf{u}}_{2|k+1}^{T},\hdots,\hat{\mathbf{u}}_{k-1|k+1}^{T}]$. Comparing with \eqref{eqn: supp 35}, we have
\par\noindent\small
\begin{align}
    \begin{bmatrix}
        \hat{\mathbf{x}}_{k+1|k+1}\\
        \hat{\mathbf{U}}_{k|k+1}    \end{bmatrix}=\overline{\mathbf{z}}_{k+1}+\overline{\mathbf{P}}_{z,k+1}\widetilde{\mathbf{P}}_{z,k+1}^{-1}\hat{\mathbf{B}}_{k}\hat{\mathbf{u}}_{k|k+1},\label{eqn:A11}
\end{align}
\normalsize
and $\hat{\mathbf{u}}_{k|k+1}$ is given by \eqref{eqn:A6}. Also, $\mathbf{P}_{z,k+1}$ can be partitioned as $\mathbf{P}_{z,k+1}=\begin{bmatrix}
    \mathbf{P}_{x,k+1|k+1}&\mathbf{P}_{xu,k+1|k+1}\\
    \mathbf{P}_{ux,k+1|k+1}&\mathbf{P}_{u,k+1|k+1}
\end{bmatrix}$. We can observe that the recursive solution for $\hat{\mathbf{x}}_{k+1|k+1}$ can be obtained from the top $n_x$ elements of the right side of \eqref{eqn:A11} while the recursive solution for $\hat{\mathbf{u}}_{k|k+1}$ is obtained by simplifying \eqref{eqn:A6}. Hence, we first obtain appropriate partitions for $\widetilde{\mathbf{P}}_{z,k+1}$, $\mathbf{K}_{z,k+1}$, $\overline{\mathbf{z}}_{k+1}$ and $\overline{\mathbf{P}}_{z,k+1}$ from which forward EKF-without-DF recursions are then obtained in Section~\ref{subsec: recursive estimates}-B.

\subsubsection{A. Partitions for $\widetilde{\mathbf{P}}_{z,k+1}$, $\mathbf{K}_{z,k+1}$, $\overline{\mathbf{z}}_{k+1}$ and $\overline{\mathbf{P}}_{z,k+1}$}\label{subsec:partitions}
\textit{1) Partition for $\widetilde{\mathbf{P}}_{z,k+1}$:} Applying \eqref{eqn: A4} to \eqref{eqn: A1 second part}, we obtain
\par\noindent\small
\begin{align}
    \overline{\bm{\Phi}}_{k+1,k}=\begin{bmatrix}
    \bm{\Phi}_{k+1,k}&\mathbf{0}\\ \mathbf{0}&\mathbf{I}
\end{bmatrix},\label{i}
\end{align}
\normalsize
Substituting \eqref{i}, \eqref{eqn: A3 first part} and $\mathbf{P}_{z,k}$ in \eqref{eqn: A8 first part}, we have
\par\noindent\small
\begin{align}
    &\widetilde{\mathbf{P}}_{z,k+1}=\begin{bmatrix}
        \widetilde{\mathbf{P}}_{z11,k+1}&\widetilde{\mathbf{P}}_{z12,k+1}\\
        \widetilde{\mathbf{P}}_{z21,k+1}&\widetilde{\mathbf{P}}_{z22,k+1}
    \end{bmatrix}=\begin{bmatrix}
        \bm{\Phi}_{k+1,k}\mathbf{P}_{x,k|k}\bm{\Phi}_{k+1,k}^{T}+\mathbf{Q}_{k}&\bm{\Phi}_{k+1,k}\mathbf{P}_{xu,k|k}\\
        \mathbf{P}_{ux,k|k}\bm{\Phi}_{k+1,k}^{T}&\mathbf{P}_{u,k|k}
    \end{bmatrix}.\label{A12}
\end{align}
\normalsize

\noindent\textit{2) Partition for $\mathbf{K}_{z,k+1}$:} Using \eqref{eqn: supp 31} and the partitioned form of $\widetilde{\mathbf{P}}_{z,k+1}$ from \eqref{A12}, we have
\par\noindent\small
\begin{align}
    &\widetilde{\mathbf{P}}_{z,k+1}\widetilde{\mathbf{H}}^{T}_{k+1}=\begin{bmatrix}        \widetilde{\mathbf{P}}_{z11,k+1}\mathbf{H}^{T}_{k+1}\\\widetilde{\mathbf{P}}_{z21,k+1}\mathbf{H}^{T}_{k+1}\end{bmatrix},\;\;\;\widetilde{\mathbf{H}}_{k+1}\widetilde{\mathbf{P}}_{z,k+1}\widetilde{\mathbf{H}}^{T}_{k+1}=\mathbf{H}_{k+1}\widetilde{\mathbf{P}}_{z11,k+1}\mathbf{H}^{T}_{k+1},\label{eqn:HpH}
\end{align}
\normalsize
Substituting this in \eqref{eqn: A8 second part}, we have
\par\noindent\small
\begin{align}
    &\mathbf{K}_{z,k+1}=\begin{bmatrix}
        \mathbf{K}_{x,k+1}\\
        \mathbf{K}_{u,k+1}
    \end{bmatrix}=\begin{bmatrix}
        \widetilde{\mathbf{P}}_{z11,k+1}\mathbf{H}^{T}_{k+1}(\mathbf{R}_{k+1}+\mathbf{H}_{k+1}\widetilde{\mathbf{P}}_{z11,k+1}\mathbf{H}^{T}_{k+1})^{-1}\\
        \widetilde{\mathbf{P}}_{z21,k+1}\mathbf{H}^{T}_{k+1}(\mathbf{R}_{k+1}+\mathbf{H}_{k+1}\widetilde{\mathbf{P}}_{z11,k+1}\mathbf{H}^{T}_{k+1})^{-1}
    \end{bmatrix}.\label{A13}
\end{align}
\normalsize

\noindent\textit{3) Partition for $\overline{\mathbf{z}}_{k+1}$:} By definition, $\hat{\mathbf{z}}_{k|k}=[\hat{\mathbf{x}}_{k|k}^{T},\hat{\mathbf{U}}_{k|k}^{T}]^{T}$ with $\hat{\mathbf{U}}_{k|k}=[\hat{\mathbf{u}}_{1|k}^{T},\hat{\mathbf{u}}_{2|k}^{T},\hdots,\hat{\mathbf{u}}_{k-1|k}^{T}]$. Hence, using \eqref{i} and \eqref{eqn: A2 second part}, we have $\overline{\bm{\Phi}}_{k+1,k}\hat{\mathbf{z}}_{k|k}+\widetilde{\mathbf{U}}_{k}=\begin{bmatrix}
    \bm{\Phi}_{k+1,k}\hat{\mathbf{x}}_{k|k}+\overline{\mathbf{u}}_{k}\\\hat{\mathbf{U}}_{k|k}
\end{bmatrix}$. Denote $\widetilde{\mathbf{z}}_{k}\doteq\bm{\Phi}_{k+1,k}\hat{\mathbf{x}}_{k|k}+\overline{\mathbf{u}}_{k}$. Now, using \eqref{eqn: supp 31}, we have
\par\noindent\small
\begin{align}
    \widetilde{\mathbf{H}}_{k+1}(\overline{\bm{\Phi}}_{k+1,k}\hat{\mathbf{z}}_{k|k}+\widetilde{\mathbf{U}}_{k})=\mathbf{H}_{k+1}\widetilde{\mathbf{z}}_{k}.\label{ii}
\end{align}
\normalsize
Hence, \eqref{eqn: A7} becomes $\overline{\mathbf{z}}_{k}=\begin{bmatrix}
        \widetilde{\mathbf{z}}_{k}\\\hat{\mathbf{U}}_{k|k}
    \end{bmatrix}+\mathbf{K}_{z,k+1}(\mathbf{y}_{k+1}-\widetilde{\mathbf{u}}_{k+1|k}-\mathbf{H}_{k+1}\widetilde{\mathbf{z}}_{k})$, which on using the partitioned form of $\mathbf{K}_{z,k+1}$ from \eqref{A13} yields
\par\noindent\small
\begin{align}
    \overline{\mathbf{z}}_{k}=\begin{bmatrix}
        \widetilde{\mathbf{z}}_{k}+\mathbf{K}_{x,k+1}(\mathbf{y}_{k+1}-\widetilde{\mathbf{u}}_{k+1|k}-\mathbf{H}_{k+1}\widetilde{\mathbf{z}}_{k})\\
        \hat{\mathbf{U}}_{k|k}+\mathbf{K}_{u,k+1}(\mathbf{y}_{k+1}-\widetilde{\mathbf{u}}_{k+1|k}-\mathbf{H}_{k+1}\widetilde{\mathbf{z}}_{k})
    \end{bmatrix},\label{A14}
\end{align}
\normalsize
which is the corrected \cite[Eq.~A14]{pan2010applying}.

\noindent\textit{4) Partition for $\overline{\mathbf{P}}_{z,k+1}$:} Substituting for $\mathbf{K}_{z,k+1}$ from \eqref{A13}, $\widetilde{\mathbf{H}}_{k+1}$ from \eqref{eqn: supp 31} and $\widetilde{\mathbf{P}}_{z,k+1}$ from \eqref{A12} in \eqref{eqn: A9 first part}, we have
\par\noindent\small
\begin{align}
    \overline{\mathbf{P}}_{z,k+1}=\begin{bmatrix}
       \overline{\mathbf{P}}_{1,k+1}&\overline{\mathbf{P}}_{2,k+1}\\
        \overline{\mathbf{P}}_{3,k+1}&\overline{\mathbf{P}}_{4,k+1}
    \end{bmatrix},\label{A15}
\end{align}
\normalsize
where $\overline{\mathbf{P}}_{1,k+1}=(\mathbf{I}-\mathbf{K}_{x,k+1}\mathbf{H}_{k+1})\widetilde{\mathbf{P}}_{z11,k+1}$, $\overline{\mathbf{P}}_{2,k+1}=(\mathbf{I}-\mathbf{K}_{x,k+1}\mathbf{H}_{k+1})\widetilde{\mathbf{P}}_{z12,k+1}$, $\overline{\mathbf{P}}_{3,k+1}=\widetilde{\mathbf{P}}_{e21,k+1}-\mathbf{K}_{u,k+1}\mathbf{H}_{k+1}\widetilde{\mathbf{P}}_{z11,k+1}$ and $\overline{\mathbf{P}}_{4,k+1}=\widetilde{\mathbf{P}}_{z22,k+1}-\mathbf{K}_{u,k+1}\mathbf{H}_{k+1}\widetilde{\mathbf{P}}_{z12,k+1}$.

\subsubsection{B. Recursions for $\hat{\mathbf{x}}_{k+1|k+1}$ and $\hat{\mathbf{u}}_{k|k+1}^{T}$}\label{subsec: recursive estimates}

\textit{1) $\mathbf{S}_{k+1}$ update:} From \eqref{eqn: A8 second part}, we have
\par\noindent\small
\begin{align}
    \widetilde{\mathbf{P}}_{z,k+1}^{-1}\mathbf{K}_{z,k+1}=\widetilde{\mathbf{H}}^{T}_{k+1}(\mathbf{R}_{k+1}+\widetilde{\mathbf{H}}_{k+1}\widetilde{\mathbf{P}}_{z,k+1}\widetilde{\mathbf{H}}^{T}_{k+1})^{-1},\label{v}
\end{align}
\normalsize
Substituting in \eqref{eqn: A9 second part}, we have $\mathbf{S}_{k+1}=(\hat{\mathbf{B}}^{T}_{k}\widetilde{\mathbf{H}}^{T}_{k+1}(\mathbf{R}_{k+1}+\widetilde{\mathbf{H}}_{k+1}\widetilde{\mathbf{P}}_{z,k+1}\widetilde{\mathbf{H}}^{T}_{k+1})^{-1}\widetilde{\mathbf{H}}_{k+1}\hat{\mathbf{B}}_{k})^{-1}$. Next, using \eqref{eqn: supp 31} and \eqref{eqn: A3 second part}, we have $\widetilde{\mathbf{H}}_{k+1}\hat{\mathbf{B}}_{k}=\mathbf{H}_{k+1}\mathbf{B}_{k}$. Using this and \eqref{eqn:HpH}, we have 
\par\noindent\small
\begin{align}
    \mathbf{S}_{k+1}=\left(\mathbf{B}_{k}^{T}\mathbf{H}_{k+1}^{T}(\mathbf{R}_{k+1}+\mathbf{H}_{k+1}\widetilde{\mathbf{P}}_{z11,k+1}\mathbf{H}_{k+1}^{T})^{-1}\mathbf{H}_{k+1}\mathbf{B}_{k}\right)^{-1}.\label{viS}
\end{align}
\normalsize
Now, from \eqref{A13}, $\mathbf{K}_{x,k+1}=\widetilde{\mathbf{P}}_{z11,k+1}\mathbf{H}_{k+1}^{T}(\mathbf{R}_{k+1}+\mathbf{H}_{k+1}\widetilde{\mathbf{P}}_{z11,k+1}\mathbf{H}_{k+1}^{T})^{-1}$ which implies $\mathbf{I}-\mathbf{H}_{k+1}\mathbf{K}_{x,k+1}=\mathbf{I}-\mathbf{H}_{k+1}\widetilde{\mathbf{P}}_{z11,k+1}\mathbf{H}_{k+1}^{T}(\mathbf{R}_{k+1}+\mathbf{H}_{k+1}\widetilde{\mathbf{P}}_{z11,k+1}\mathbf{H}_{k+1}^{T})^{-1}$. Comparing with \eqref{eqn: A5} with $\mathbf{C}_{1}^{-1}=\mathbf{I}$, $\mathbf{C}_{2}=\mathbf{H}_{k+1}\widetilde{\mathbf{P}}_{z11,k+1}\mathbf{H}_{k+1}^{T}$, $\mathbf{C}_{3}^{-1}=\mathbf{R}_{k+1}$ and $\mathbf{C}_{4}=\mathbf{I}$, we have $\mathbf{I}-\mathbf{H}_{k+1}\mathbf{K}_{x,k+1}=(\mathbf{I}+\mathbf{H}_{k+1}\widetilde{\mathbf{P}}_{z11,k+1}\mathbf{H}_{k+1}^{T}\mathbf{R}_{k+1}^{-1})^{-1}$ which implies $\mathbf{R}_{k+1}^{-1}(\mathbf{I}-\mathbf{H}_{k+1}\mathbf{K}_{x,k+1})=(\mathbf{R}_{k+1}+\mathbf{H}_{k+1}\widetilde{\mathbf{P}}_{z11,k+1}\mathbf{H}_{k+1}^{T})^{-1}$. Substituting in \eqref{viS}, we obtain
\par\noindent\small
\begin{align}
        \mathbf{S}_{k+1}=\left(\mathbf{B}_{k}^{T}\mathbf{H}_{k+1}^{T}\mathbf{R}_{k+1}^{-1}(\mathbf{I}-\mathbf{H}_{k+1}\mathbf{K}_{x,k+1})\mathbf{H}_{k+1}\mathbf{B}_{k}\right)^{-1}.\label{40}
\end{align}
\normalsize
Representing $\mathbf{S}_{k+1}$ by $\bm{\Sigma}^{u}_{k}$ and $\mathbf{K}_{x,k+1}$ by $\mathbf{K}^{x}_{k+1}$, \eqref{40} is the $\bm{\Sigma}^{u}_{k}$ update step of forward EKF-without-DF in Section~\ref{subsec:IEKF without DF}.

\noindent\textit{2) $\hat{\mathbf{u}}_{k|k+1}$ update:} From \eqref{eqn: A7} and \eqref{ii}, we have $\overline{\bm{\Phi}}_{k+1,k}\hat{\mathbf{z}}_{k|k}+\widetilde{\mathbf{U}}_{k}-\overline{\mathbf{z}}_{k+1}=-\mathbf{K}_{z,k+1}(\mathbf{y}_{k+1}-\widetilde{\mathbf{u}}_{k+1|k}-\mathbf{H}_{k+1|k}\widetilde{\mathbf{z}}_{k})$. Hence, from \eqref{eqn:A6}, $\hat{\mathbf{u}}_{k|k+1}=\mathbf{S}_{k+1}\hat{\mathbf{B}}_{k}^{T}\widetilde{\mathbf{P}}_{z,k+1}^{-1}\mathbf{K}_{z,k+1}(\mathbf{y}_{k+1}-\widetilde{\mathbf{u}}_{k+1|k}-\mathbf{H}_{k+1}\widetilde{\mathbf{z}}_{k})$. Again, substituting for $\widetilde{\mathbf{P}}_{z,k+1}^{-1}\mathbf{K}_{z,k+1}$ from \eqref{v}, we have $\hat{\mathbf{u}}_{k|k+1}=\mathbf{S}_{k+1}\hat{\mathbf{B}}_{k}^{T}\widetilde{\mathbf{H}}^{T}_{k+1}\times\\(\mathbf{R}_{k+1}+\widetilde{\mathbf{H}}_{k+1}\widetilde{\mathbf{P}}_{z,k+1}\widetilde{\mathbf{H}}^{T}_{k+1})^{-1}(\mathbf{y}_{k+1}-\widetilde{\mathbf{u}}_{k+1|k}-\mathbf{H}_{k+1}\widetilde{\mathbf{z}}_{k})$. Now, substituting $\widetilde{\mathbf{H}}_{k+1}\hat{\mathbf{B}}_{k}=\mathbf{H}_{k+1}\mathbf{B}_{k}$ and $(\mathbf{R}_{k+1}+\widetilde{\mathbf{H}}_{k+1}\widetilde{\mathbf{P}}_{z,k+1}\widetilde{\mathbf{H}}^{T}_{k+1})^{-1}=(\mathbf{R}_{k+1}+\mathbf{H}_{k+1}\widetilde{\mathbf{P}}_{z11,k+1}\mathbf{H}_{k+1}^{T})^{-1}=\mathbf{R}_{k+1}^{-1}(\mathbf{I}-\mathbf{H}_{k+1}\mathbf{K}_{x,k+1})$ as obtained in the previous step, we have
\par\noindent\small
\begin{align}
    \hat{\mathbf{u}}_{k|k+1}&=\mathbf{S}_{k+1}\mathbf{B}_{k}^{T}\mathbf{H}_{k+1}^{T}\mathbf{R}_{k+1}^{-1}(\mathbf{I}-\mathbf{H}_{k+1}\mathbf{K}_{x,k+1})(\mathbf{y}_{k+1}-\widetilde{\mathbf{u}}_{k+1|k}-\mathbf{H}_{k+1}\widetilde{\mathbf{z}}_{k}).\label{viu}
\end{align}
\normalsize
    
Now, we simplify $(\mathbf{y}_{k+1}-\widetilde{\mathbf{u}}_{k+1|k}-\mathbf{H}_{k+1}\widetilde{\mathbf{z}}_{k})$. From \eqref{eqn:supp 23}, using $i=k$, we have $\mathbf{x}_{k}=\bm{\Phi}^{-1}_{k+1,k}\mathbf{x}_{k+1}-\bm{\Phi}^{-1}_{k+1,k}(\mathbf{B}_{k}\mathbf{u}_{k}+\overline{\mathbf{u}}_{k}+\mathbf{w}_{k})$ which implies $\mathbf{x}_{k+1}=\bm{\Phi}_{k+1,k}\mathbf{x}_{k}+\mathbf{B}_{k}\mathbf{u}_{k}+\overline{\mathbf{u}}_{k}+\mathbf{w}_{k}$. But by definition of $\bm{\Phi}_{k+1,k}$, we have $\bm{\Phi}_{k+1,k}^{-1}=\mathbf{F}_{k}^{-1}$ such that $\bm{\Phi}_{k+1,k}=\mathbf{F}_{k}$. Hence, $\mathbf{x}_{k+1}=\mathbf{F}_{k}\mathbf{x}_{k}+\mathbf{B}_{k}\mathbf{u}_{k}+\overline{\mathbf{u}}_{k}+\mathbf{w}_{k}$. From \eqref{eqn:supp 20}, this is the linearized form of \eqref{eqn:non-linear state x with unknown input}. We obtain the predicted state $\hat{\mathbf{x}}_{k+1|k}$ by substituting $\hat{\mathbf{x}}_{k|k}$ and $\hat{\mathbf{u}}_{k-1|k}$ (previous estimates) in place of $\mathbf{x}_{k}$ and $\mathbf{u}_{k}$, respectively, with the noise $\mathbf{w}_{k}$ taken as $\mathbf{0}$. Hence,
\par\noindent\small
\begin{align}
    \hat{\mathbf{x}}_{k+1|k}=\mathbf{F}_{k}\hat{\mathbf{x}}_{k|k}+\mathbf{B}_{k}\hat{\mathbf{u}}_{k-1|k}+\overline{\mathbf{u}}_{k}.\label{vii}
\end{align}
\normalsize
But similar to EKF, instead of the linearized approximation, we use the non-linear state transition function itself for state prediction for reduced errors, i.e.,
\par\noindent\small
\begin{align}
    \hat{\mathbf{x}}_{k+1|k}=f(\hat{\mathbf{x}}_{k|k},\hat{\mathbf{u}}_{k-1|k}).\label{38}
\end{align}
\normalsize
This is the prediction step \eqref{eqn: ekfwithoutdf predict} of forward EKF-without-DF.

Now, by definition, $\widetilde{\mathbf{z}}_{k}=\bm{\Phi}_{k+1,k}\hat{\mathbf{x}}_{k|k}+\overline{\mathbf{u}}_{k}=\mathbf{F}_{k}\hat{\mathbf{x}}_{k|k}+\overline{\mathbf{u}}_{k}$. Using \eqref{vii}, we have $\widetilde{\mathbf{z}}_{k}=\hat{\mathbf{x}}_{k+1|k}-\mathbf{B}_{k}\hat{\mathbf{u}}_{k-1|k}$. Using this along with \eqref{eqn: supp 26} with $i=k+1$, we have $\mathbf{y}_{k+1}-\widetilde{\mathbf{u}}_{k+1|k}-\mathbf{H}_{k+1}\widetilde{\mathbf{z}}_{k}=\mathbf{y}_{k+1}-(h(\hat{\mathbf{x}}_{k+1|k})-\mathbf{H}_{k+1}\hat{\mathbf{x}}_{k+1|k})-\mathbf{H}_{k+1}\widetilde{\mathbf{z}}_{k}=\mathbf{y}_{k+1}-h(\hat{\mathbf{x}}_{k+1|k})+\mathbf{H}_{k+1}\mathbf{B}_{k}\hat{\mathbf{u}}_{k-1|k}$. Substituting in \eqref{viu}, we have
\par\noindent\small
\begin{align}
    \hat{\mathbf{u}}_{k|k+1}&=\mathbf{S}_{k+1}\mathbf{B}_{k}^{T}\mathbf{H}_{k+1}^{T}\mathbf{R}_{k+1}^{-1}(\mathbf{I}-\mathbf{H}_{k+1}\mathbf{K}_{x,k+1})(\mathbf{y}_{k+1}-h(\hat{\mathbf{x}}_{k+1|k})+\mathbf{H}_{k+1}\mathbf{B}_{k}\hat{\mathbf{u}}_{k-1|k}).\label{43}
\end{align}
\normalsize
Denoting $\mathbf{K}^{u}_{k}=\mathbf{S}_{k+1}\mathbf{B}_{k}^{T}\mathbf{H}_{k+1}^{T}\mathbf{R}_{k+1}^{-1}(\mathbf{I}-\mathbf{H}_{k+1}\mathbf{K}_{x,k+1})$, \eqref{43} is the update step \eqref{eqn: ekfwithoutdf update u} of forward EKF-without-DF.

\noindent\textit{3) $\hat{\mathbf{x}}_{k+1|k+1}$ update:} From \eqref{eqn: A9 first part}, \eqref{eqn: supp 31} and \eqref{A13}, we have $\overline{\mathbf{P}}_{z,k+1}\widetilde{\mathbf{P}}_{z,k+1}^{-1}=\begin{bmatrix}
    \mathbf{I}-\mathbf{K}_{x,k+1}\mathbf{H}_{k+1}&\mathbf{0}\\-\mathbf{K}_{u,k+1}\mathbf{H}_{k+1}&\mathbf{I}
    \end{bmatrix}$. Hence, using \eqref{eqn: A3 second part}, $\overline{\mathbf{P}}_{z,k+1}\widetilde{\mathbf{P}}_{z,k+1}^{-1}\hat{\mathbf{B}}_{k}=\begin{bmatrix}
    (\mathbf{I}-\mathbf{K}_{x,k+1}\mathbf{H}_{k+1})\mathbf{B}_{k}\\-\mathbf{K}_{u,k+1}\mathbf{H}_{k+1}\mathbf{B}_{k}\end{bmatrix}$. Substituting this and \eqref{A14} in \eqref{eqn:A11}, we have
\par\noindent\small
\begin{align}
    \begin{bmatrix}
        \hat{\mathbf{x}}_{k+1|k+1}\\\hat{\mathbf{U}}_{k|k+1}
    \end{bmatrix}&=\begin{bmatrix}
        \widetilde{\mathbf{z}}_{k}+\mathbf{K}_{x,k+1}(\mathbf{y}_{k+1}-\widetilde{\mathbf{u}}_{k+1|k}-\mathbf{H}_{k+1}\widetilde{\mathbf{z}}_{k})\\
        \hat{\mathbf{U}}_{k|k}+\mathbf{K}_{u,k+1}(\mathbf{y}_{k+1}-\widetilde{\mathbf{u}}_{k+1|k}-\mathbf{H}_{k+1}\widetilde{\mathbf{z}}_{k})
    \end{bmatrix}+\begin{bmatrix}
        (\mathbf{I}-\mathbf{K}_{x,k+1}\mathbf{H}_{k+1})\mathbf{B}_{k}\hat{\mathbf{u}}_{k|k+1}\\-\mathbf{K}_{u,k+1}\mathbf{H}_{k+1}\mathbf{B}_{k}\hat{\mathbf{u}}_{k|k+1}
    \end{bmatrix},\nonumber
\end{align}
\normalsize
which implies $\hat{\mathbf{x}}_{k+1|k+1}= \widetilde{\mathbf{z}}_{k}+\mathbf{K}_{x,k+1}(\mathbf{y}_{k+1}-\widetilde{\mathbf{u}}_{k+1|k}-\mathbf{H}_{k+1}\widetilde{\mathbf{z}}_{k})+(\mathbf{I}-\mathbf{K}_{x,k+1}\mathbf{H}_{k+1})\mathbf{B}_{k}\hat{\mathbf{u}}_{k|k+1}$.

Again, $\widetilde{\mathbf{z}}_{k}=\hat{\mathbf{x}}_{k+1|k}-\mathbf{B}_{k}\hat{\mathbf{u}}_{k-1|k}$ and $\mathbf{y}_{k+1}-\widetilde{\mathbf{u}}_{k+1|k}-\mathbf{H}_{k+1}\widetilde{\mathbf{z}}_{k}=\mathbf{y}_{k+1}-h(\hat{\mathbf{x}}_{k+1|k})+\mathbf{H}_{k+1}\mathbf{B}_{k}\hat{\mathbf{u}}_{k-1|k}$. Hence,
\par\noindent\small
\begin{align}
    &\hat{\mathbf{x}}_{k+1|k+1}=\hat{\mathbf{x}}_{k+1|k}-\mathbf{B}_{k}\hat{\mathbf{u}}_{k-1|k}+(\mathbf{I}-\mathbf{K}_{x,k+1}\mathbf{H}_{k+1})\mathbf{B}_{k}\hat{\mathbf{u}}_{k|k+1}+\mathbf{K}_{x,k+1}(\mathbf{y}_{k+1}-h(\hat{\mathbf{x}}_{k+1|k})+\mathbf{H}_{k+1}\mathbf{B}_{k}\hat{\mathbf{u}}_{k-1|k})\nonumber\\
    &=\hat{\mathbf{x}}_{k+1|k}+\mathbf{K}_{x,k+1}(\mathbf{y}_{k+1}-h(\hat{\mathbf{x}}_{k+1|k}))+(\mathbf{I}-\mathbf{K}_{x,k+1}\mathbf{H}_{k+1})\mathbf{B}_{k}(\hat{\mathbf{u}}_{k|k+1}-\hat{\mathbf{u}}_{k-1|k})\nonumber
\end{align}
\normalsize
Now, \cite{pan2010applying} assumed $\hat{\mathbf{u}}_{k|k+1}-\hat{\mathbf{u}}_{k-1|k}\approx\mathbf{0}$, i.e., the filter's unknown input estimate does not change abruptly (by a large value in one step) and hence,
\par\noindent\small
\begin{align}
    \hat{\mathbf{x}}_{k+1|k+1}=\hat{\mathbf{x}}_{k+1|k}+\mathbf{K}_{x,k+1}(\mathbf{y}_{k+1}-h(\hat{\mathbf{x}}_{k+1|k})).\label{42}
\end{align}
\normalsize
This is the update step \eqref{eqn: ekfwithoutdf update x} of the forward EKF-without-DF.

Also, define $\bm{\Sigma}^{x}_{k+1|k}\doteq\widetilde{\mathbf{P}}_{z11,k+1}$. From \eqref{A13}, we have
\par\noindent\small
\begin{align}
    \mathbf{K}_{x,k+1}=\bm{\Sigma}^{x}_{k+1|k}\mathbf{H}_{k+1}^{T}(\mathbf{R}_{k+1}+\mathbf{H}_{k+1}\bm{\Sigma}^{x}_{k+1|k}\mathbf{H}_{k+1}^{T})^{-1},\label{39}
\end{align}
\normalsize
which is the $\mathbf{K}^{x}_{k+1}$ (another notation for $\mathbf{K}_{x,k+1}$) update of the forward EKF-without-DF in Section~\ref{subsec:IEKF without DF}. From \eqref{A12}, $\bm{\Sigma}^{x}_{k+1|k}=\widetilde{\mathbf{P}}_{z11,k+1}=\bm{\Phi}_{k+1,k}\mathbf{P}_{x,k|k}\bm{\Phi}_{k+1,k}^{T}+\mathbf{Q}_{k}$. But $\bm{\Phi}_{k+1,k}=\mathbf{F}_{k}$. Hence,
\par\noindent\small
\begin{align}
    \bm{\Sigma}^{x}_{k+1|k}=\mathbf{F}_{k}\mathbf{P}_{x,k|k}\mathbf{F}_{k}^{T}+\mathbf{Q}_{k}.\label{41}
\end{align}
\normalsize
Representing $\bm{\Sigma}^{x}_{k}=\mathbf{P}_{x,k|k}$, \eqref{41} is the $\bm{\Sigma}^{x}_{k+1|k}$ update of the forward EKF-without-DF in Section~\ref{subsec:IEKF without DF}.

\noindent\textit{4) $\mathbf{P}_{x,k|k}$ update:} By definition, $\mathbf{P}_{z,k+1}$ can be partitioned as
\par\noindent\small
\begin{align}
    \mathbf{P}_{z,k+1}=\begin{bmatrix}
        \mathbf{P}_{x,k+1|k+1}&\mathbf{P}_{xu,k+1|k+1}\\\mathbf{P}_{ux,k+1|k+1}&\mathbf{P}_{u,k+1|k+1}.\label{viii}
    \end{bmatrix}
\end{align}
\normalsize
Substituting \eqref{eqn: A1 first part} and \eqref{eqn: A3 above first part} in \eqref{eqn:supp 34}, we have
\par\noindent\small
\begin{align}
    \mathbf{P}_{z,k+1}=\begin{bmatrix}
        \mathbf{P}^{1}_{z,k+1}&\mathbf{P}^{2}_{z,k+1}\\\mathbf{P}^{3}_{z,k+1}&\mathbf{P}^{4}_{z,k+1}
    \end{bmatrix}^{-1},\label{inv}
\end{align}
\normalsize
where
\par\noindent\small
\begin{align}
 &\mathbf{P}^{1}_{z,k+1}=\overline{\bm{\Phi}}_{k+1,k}^{-T}\mathbf{A}_{z,k}^{T}\widetilde{\mathbf{W}}_{k}\mathbf{A}_{z,k}\overline{\bm{\Phi}}_{k+1,k}^{-1}+\widetilde{\mathbf{H}}_{k+1}^{T}\mathbf{R}_{k+1}^{-1}\widetilde{\mathbf{H}}_{k+1},\;\;\;\mathbf{P}^{2}_{z,k+1}=-\overline{\bm{\Phi}}_{k+1,k}^{-T}\mathbf{A}_{z,k}^{T}\widetilde{\mathbf{W}}_{k}\mathbf{A}_{z,k}\overline{\bm{\Phi}}_{k+1,k}^{-1}\hat{\mathbf{B}}_{k},\nonumber\\
 &\mathbf{P}^{3}_{z,k+1}=-\hat{\mathbf{B}}_{k}^{T}\overline{\bm{\Phi}}_{k+1,k}^{-T}\mathbf{A}_{z,k}^{T}\widetilde{\mathbf{W}}_{k}\mathbf{A}_{z,k}\overline{\bm{\Phi}}_{k+1,k}^{-1},\;\;\;\mathbf{P}^{4}_{z,k+1}=\hat{\mathbf{B}}_{k}^{T}\overline{\bm{\Phi}}_{k+1,k}^{-T}\mathbf{A}_{z,k}^{T}\widetilde{\mathbf{W}}_{k}\mathbf{A}_{z,k}\overline{\bm{\Phi}}_{k+1,k}^{-1}\hat{\mathbf{B}}_{k}.\nonumber
\end{align}
\normalsize
Let
\par\noindent\small
\begin{align}
    \mathbf{P}_{z,k+1}=\begin{bmatrix}
        \mathbf{P}_{z11,k+1}&\mathbf{P}_{z12,k+1}\\\mathbf{P}_{z21,k+1}&\mathbf{P}_{z22,k+1}
    \end{bmatrix}.\label{ix}
\end{align}
\normalsize
In \cite[Appendix~A.1]{pan2010applying} (detailed steps in \cite[Appendix~A]{yang2007adaptive}) to obtain \eqref{eqn: A8 first part}-\eqref{eqn: supp 35}, \eqref{eqn: A4} is used to simplify \eqref{inv} after defining $\mathbf{S}_{k+1}\doteq\mathbf{P}_{z22,k+1}$. This yields
\par\noindent\small
\begin{align}
    \mathbf{P}_{z11,k+1}&=\overline{\mathbf{P}}_{z,k+1}+\overline{\mathbf{P}}_{z,k+1}\overline{\bm{\Phi}}_{k+1,k}^{-T}\mathbf{A}_{z,k}^{T}\widetilde{\mathbf{W}}_{k}\mathbf{A}_{z,k}\overline{\bm{\Phi}}_{k+1,k}^{-1}\hat{\mathbf{B}}_{k}\mathbf{S}_{k+1}\hat{\mathbf{B}}^{T}_{k}\overline{\bm{\Phi}}_{k+1,k}^{-T}\mathbf{A}_{z,k}^{T}\widetilde{\mathbf{W}}_{k}\mathbf{A}_{z,k}\overline{\bm{\Phi}}_{k+1,k}^{-1}\overline{\mathbf{P}}_{z,k+1}\nonumber
\end{align}
\normalsize

Now, \eqref{viii} and \eqref{ix} are two different partitions of $\mathbf{P}_{z,k+1}$. Comparing the dimensions, we observe that $\mathbf{P}_{x,k+1|k+1}$ is the upper-left submatrix of $\mathbf{P}_{z11,k+1}$. Denote $\mathbf{M}:=\\\overline{\mathbf{P}}_{z,k+1}\overline{\bm{\Phi}}_{k+1,k}^{-T}\mathbf{A}_{z,k}^{T}\widetilde{\mathbf{W}}_{k}\mathbf{A}_{z,k}\overline{\bm{\Phi}}_{k+1,k}^{-1}$. Then, $\mathbf{P}_{z11,k+1}=\overline{\mathbf{P}}_{z,k+1}+\mathbf{M}\hat{\mathbf{B}}_{k}\mathbf{S}_{k+1}\hat{\mathbf{B}}^{T}_{k}\mathbf{M}^{T}$. Denote the partitions of $\mathbf{M}=\begin{bmatrix}
        \mathbf{M}_{11}&\mathbf{M}_{12}\\\mathbf{M}_{21}&\mathbf{M}_{22}
\end{bmatrix}$. Hence, substituting for $\hat{\mathbf{B}}_{k}$ from \eqref{eqn: A3 second part}, we have $\mathbf{M}\hat{\mathbf{B}}_{k}\mathbf{S}_{k+1}\hat{\mathbf{B}}^{T}_{k}\mathbf{M}^{T}=\begin{bmatrix}
    \mathbf{M}_{11}\mathbf{B}_{k}\mathbf{S}_{k+1}\mathbf{B}_{k}^{T}\mathbf{M}_{11}^{T}&\mathbf{M}_{11}\mathbf{B}_{k}\mathbf{S}_{k+1}\mathbf{B}_{k}^{T}\mathbf{M}_{21}^{T}\\
    \mathbf{M}_{21}\mathbf{B}_{k}\mathbf{S}_{k+1}\mathbf{B}_{k}^{T}\mathbf{M}_{11}^{T}&\mathbf{M}_{21}\mathbf{B}_{k}\mathbf{S}_{k+1}\mathbf{B}_{k}^{T}\mathbf{M}_{21}^{T}\end{bmatrix}$, whose upper-left submatrix is $\mathbf{M}_{11}\mathbf{B}_{k}\mathbf{S}_{k+1}\mathbf{B}_{k}^{T}\mathbf{M}_{11}^{T}$. Also, from \eqref{A15}, the upper-left submatrix of $\overline{\mathbf{P}}_{z,k+1}$ is $(\mathbf{I}-\mathbf{K}_{x,k+1}\mathbf{H}_{k+1})\widetilde{\mathbf{P}}_{z11,k+1}$. Hence,
\par\noindent\small
\begin{align}
    \mathbf{P}_{x,k+1|k+1}=(\mathbf{I}-\mathbf{K}_{x,k+1}\mathbf{H}_{k+1})\widetilde{\mathbf{P}}_{z11,k+1}+\mathbf{M}_{11}\mathbf{B}_{k}\mathbf{S}_{k+1}\mathbf{B}_{k}^{T}\mathbf{M}_{11}^{T},\label{x}
\end{align}
\normalsize
where $\mathbf{M}_{11}$ is the upper-left submatrix of $\overline{\mathbf{P}}_{z,k+1}\overline{\bm{\Phi}}_{k+1,k}^{-T}\mathbf{A}_{z,k}^{T}\widetilde{\mathbf{W}}_{k}\mathbf{A}_{z,k}\overline{\bm{\Phi}}_{k+1,k}^{-1}$.

Now, we compute $\mathbf{M}_{11}$. Using \eqref{eqn: A5} in \eqref{eqn: A3 above second part}, we have $ \widetilde{\mathbf{W}}_{k}=\mathbf{W}_{k}-\mathbf{W}_{k}\mathbf{A}_{z,k}\overline{\bm{\Phi}}_{k+1,k}^{-1}\times\\\left(\widetilde{\mathbf{Q}}^{-1}_{k}+\overline{\bm{\Phi}}_{k+1,k}^{-T}\mathbf{A}_{z,k}^{T}\mathbf{W}_{k}\mathbf{A}_{z,k}\overline{\bm{\Phi}}_{k+1,k}^{-1}\right)^{-1}\overline{\bm{\Phi}}_{k+1,k}^{-T}\mathbf{A}_{z,k}^{T}\mathbf{W}_{k}$. Now, from \eqref{eqn:supp 34}, $\mathbf{A}_{z,k}^{T}\mathbf{W}_{k}\mathbf{A}_{z,k}=\mathbf{P}_{z,k}^{-1}$ and
\par\noindent\small
\begin{align}
&\widetilde{\mathbf{W}}_{k}=\mathbf{W}_{k}-\mathbf{W}_{k}\mathbf{A}_{z,k}\overline{\bm{\Phi}}_{k+1,k}^{-1}\left(\widetilde{\mathbf{Q}}^{-1}_{k}+\overline{\bm{\Phi}}_{k+1,k}^{-T}\mathbf{P}_{z,k}^{-1}\overline{\bm{\Phi}}_{k+1,k}^{-1}\right)^{-1}\overline{\bm{\Phi}}_{k+1,k}^{-T}\mathbf{A}_{z,k}^{T}\mathbf{W}_{k}\nonumber\\
    &=\mathbf{W}_{k}-\mathbf{W}_{k}\mathbf{A}_{z,k}\left[\overline{\bm{\Phi}}_{k+1,k}^{T}(\widetilde{\mathbf{Q}}^{-1}_{k}+\overline{\bm{\Phi}}_{k+1,k}^{-T}\mathbf{P}_{z,k}^{-1}\overline{\bm{\Phi}}_{k+1,k}^{-1})\overline{\bm{\Phi}}_{k+1,k}\right]^{-1}\mathbf{A}_{z,k}^{T}\mathbf{W}_{k}\nonumber\\
    &=\mathbf{W}_{k}-\mathbf{W}_{k}\mathbf{A}_{z,k}\left[\overline{\bm{\Phi}}_{k+1,k}^{T}\widetilde{\mathbf{Q}}^{-1}_{k}\overline{\bm{\Phi}}_{k+1,k}+\mathbf{P}_{z,k}^{-1}\right]^{-1}\mathbf{A}_{z,k}^{T}\mathbf{W}_{k}.\nonumber
\end{align}
\normalsize
Hence, $\mathbf{A}_{z,k}^{T}\widetilde{\mathbf{W}}_{k}\mathbf{A}_{z,k}=\mathbf{A}_{z,k}^{T}\mathbf{W}_{k}\mathbf{A}_{z,k}-\mathbf{A}_{z,k}^{T}\mathbf{W}_{k}\mathbf{A}_{z,k}\left[\overline{\bm{\Phi}}_{k+1,k}^{T}\widetilde{\mathbf{Q}}^{-1}_{k}\overline{\bm{\Phi}}_{k+1,k}+\mathbf{P}_{z,k}^{-1}\right]^{-1}\mathbf{A}_{z,k}^{T}\mathbf{W}_{k}\mathbf{A}_{z,k}$. Again, using $\mathbf{A}_{z,k}^{T}\mathbf{W}_{k}\mathbf{A}_{z,k}=\mathbf{P}_{z,k}^{-1}$, we have $\mathbf{A}_{z,k}^{T}\widetilde{\mathbf{W}}_{k}\mathbf{A}_{z,k}=\mathbf{P}_{z,k}^{-1}-\mathbf{P}_{z,k}^{-1}\left[\overline{\bm{\Phi}}_{k+1,k}^{T}\widetilde{\mathbf{Q}}^{-1}_{k}\overline{\bm{\Phi}}_{k+1,k}+\mathbf{P}_{z,k}^{-1}\right]^{-1}\mathbf{P}_{z,k}^{-1}$. Comparing with \eqref{eqn: A5} with $\mathbf{C}_{1}^{-1}=\mathbf{P}_{z,k}^{-1}$, $\mathbf{C}_{2}=\mathbf{I}$, $\mathbf{C}_{3}^{-1}=\overline{\bm{\Phi}}_{k+1,k}^{T}\widetilde{\mathbf{Q}}^{-1}_{k}\overline{\bm{\Phi}}_{k+1,k}$ and $\mathbf{C}_{4}=\mathbf{I}$, we have $\mathbf{A}_{z,k}^{T}\widetilde{\mathbf{W}}_{k}\mathbf{A}_{z,k}=\left(\mathbf{P}_{z,k}+(\overline{\bm{\Phi}}_{k+1,k}^{T}\widetilde{\mathbf{Q}}^{-1}_{k}\overline{\bm{\Phi}}_{k+1,k})^{-1}\right)^{-1}=(\mathbf{P}_{z,k}+\overline{\bm{\Phi}}_{k+1,k}^{-1}\widetilde{\mathbf{Q}}_{k}\overline{\bm{\Phi}}_{k+1,k}^{-T})^{-1}$. Now, $\overline{\bm{\Phi}}_{k+1,k}^{-T}\mathbf{A}_{z,k}^{T}\widetilde{\mathbf{W}}_{k}\mathbf{A}_{z,k}\overline{\bm{\Phi}}_{k+1,k}^{-1}=\overline{\bm{\Phi}}_{k+1,k}^{-T}(\mathbf{P}_{z,k}+\overline{\bm{\Phi}}_{k+1,k}^{-1}\widetilde{\mathbf{Q}}_{k}\overline{\bm{\Phi}}_{k+1,k}^{-T})^{-1}\overline{\bm{\Phi}}_{k+1,k}^{-1}\\=\left[\overline{\bm{\Phi}}_{k+1,k}(\mathbf{P}_{z,k}+\overline{\bm{\Phi}}_{k+1,k}^{-1}\widetilde{\mathbf{Q}}_{k}\overline{\bm{\Phi}}_{k+1,k}^{-T})\overline{\bm{\Phi}}_{k+1,k}^{T}\right]^{-1} =[\overline{\bm{\Phi}}_{k+1,k}\mathbf{P}_{z,k}\overline{\bm{\Phi}}_{k+1,k}^{T}+\widetilde{\mathbf{Q}}_{k}]^{-1}=\widetilde{\mathbf{P}}_{z,k+1}^{-1}$ using \eqref{eqn: A8 first part}. Hence, $\mathbf{M}=\overline{\mathbf{P}}_{z,k+1}\overline{\bm{\Phi}}_{k+1,k}^{-T}\mathbf{A}_{z,k}^{T}\widetilde{\mathbf{W}}_{k}\mathbf{A}_{z,k}\overline{\bm{\Phi}}_{k+1,k}^{-1}=\overline{\mathbf{P}}_{z,k+1}\widetilde{\mathbf{P}}_{z,k+1}^{-1}=\mathbf{I}-\mathbf{K}_{z,k+1}\widetilde{\mathbf{H}}_{k+1}$ using \eqref{eqn: A9 first part}. Hence, substituting for $\mathbf{K}_{z,k+1}$ from \eqref{A13} and $\widetilde{\mathbf{H}}_{k+1}$ from \eqref{eqn: supp 31}, we obtain the submatrix $\mathbf{M}_{11}=\mathbf{I}-\mathbf{K}_{x,k+1}\mathbf{H}_{k+1}$. Using this in \eqref{x}, we have $\mathbf{P}_{x,k+1|k+1}=(\mathbf{I}-\mathbf{K}_{x,k+1}\mathbf{H}_{k+1})\widetilde{\mathbf{P}}_{z11,k+1}+(\mathbf{I}-\mathbf{K}_{x,k+1}\mathbf{H}_{k+1})\mathbf{B}_{k}\mathbf{S}_{k+1}\mathbf{B}_{k}^{T}(\mathbf{I}-\mathbf{K}_{x,k+1}\mathbf{H}_{k+1})^{T}$. Using $\bm{\Sigma}^{x}_{k+1|k}=\widetilde{\mathbf{P}}_{z11,k+1}$, we obtain
\par\noindent\small
\begin{align}
    \mathbf{P}_{x,k+1|k+1}&=(\mathbf{I}-\mathbf{K}_{x,k+1}\mathbf{H}_{k+1})[\bm{\Sigma}^{x}_{k+1|k}+\mathbf{B}_{k}\mathbf{S}_{k+1}\mathbf{B}_{k}^{T}(\mathbf{I}-\mathbf{K}_{x,k+1}\mathbf{H}_{k+1}^{T}].\label{44}
\end{align}
\normalsize
With $\bm{\Sigma}^{x}_{k+1}=\mathbf{P}_{x,k+1|k+1}$, $\mathbf{K}^{x}_{k+1}=\mathbf{K}_{x,k+1}$ and $\bm{\Sigma}^{u}_{k}=\mathbf{S}_{k+1}$, \eqref{44} is the $\bm{\Sigma}^{x}_{k+1}$ update of forward EKF-without-DF in Section~\ref{subsec:IEKF without DF}. The updates \eqref{38}, \eqref{39}, \eqref{40}, \eqref{41}, \eqref{42}, \eqref{43} and \eqref{44} are the final forward EKF-without-DF recursions.

\section{Proof of Theorem~\ref{theorem: inverse kf without DF}}
\label{App-thm-kf-without-DF}
Under the stability assumption of the forward filter, $\widetilde{\mathbf{F}}_{k}$ and $\mathbf{E}_{k}$ converge to $\overline{\mathbf{F}}$ and $\overline{\mathbf{E}}$, respectively, where $\overline{\mathbf{F}}=(\mathbf{I}-\overline{\mathbf{K}}\mathbf{H})(\mathbf{I}-\mathbf{B}\overline{\mathbf{M}}\mathbf{H})\mathbf{F}$ and $\overline{\mathbf{E}}=\mathbf{B}\overline{\mathbf{M}}-\overline{\mathbf{K}}\mathbf{HB}\overline{\mathbf{M}}+\overline{\mathbf{K}}$, obtained by replacing $\mathbf{K}_{k+1}$ and $\mathbf{M}_{k+1}$ by the limiting matrices $\overline{\mathbf{K}}$ and $\overline{\mathbf{M}}$, respectively, in $\widetilde{\mathbf{F}}_{k}$ and $\mathbf{E}_{k}$. In this limiting case, the state transition equation \eqref{eqn: state for kfwithoutdf} becomes $\hat{\mathbf{x}}_{k+1}=\overline{\mathbf{F}}\hat{\mathbf{x}}_{k}+\overline{\mathbf{E}}\mathbf{Hx}_{k+1}+\overline{\mathbf{E}}\mathbf{v}_{k+1}$. From \eqref{eqn: inverse kfwithoutdf covariance predict}, \eqref{eqn: inverse kfwithoutdf gain}, and \eqref{eqn: inverse kfwithoutdf covariance update} and substituting the limiting matrices, the Riccati equation $\overline{\bm{\Sigma}}_{k+1|k}=\overline{\mathbf{F}}\left[\overline{\bm{\Sigma}}_{k|k-1}-\overline{\bm{\Sigma}}_{k|k-1}\mathbf{G}^{T}(\mathbf{G}\overline{\bm{\Sigma}}_{k|k-1}\mathbf{G}^{T}+\overline{\mathbf{R}})^{-1}\mathbf{G}\overline{\bm{\Sigma}}_{k|k-1}\right]\overline{\mathbf{F}}^{T}+\overline{\bm{Q}}$ is obtained, where $\overline{\mathbf{Q}}=\overline{\mathbf{E}}\mathbf{R}\overline{\mathbf{E}}^{T}$. For the forward filter to be stable, covariance $\mathbf{R}$ needs to be p.d.\cite{fang2012on} and hence, $\overline{\mathbf{Q}}$ is a p.s.d. matrix. With $\overline{\mathbf{R}}$ being p.d. and the observability and controllability assumptions, $\overline{\bm{\Sigma}}_{k|k-1}$ tends to a unique p.d. matrix $\overline{\bm{\Sigma}}$ satisfying $\overline{\bm{\Sigma}}=\overline{\mathbf{F}}[\overline{\bm{\Sigma}}-\overline{\bm{\Sigma}}\mathbf{G}^{T}\left(\mathbf{G}\overline{\bm{\Sigma}}\mathbf{G}^{T}+\overline{\mathbf{R}}\right)^{-1}\mathbf{G}\overline{\bm{\Sigma}}]\overline{\mathbf{F}}^{T}+\overline{\mathbf{Q}}$, and $\overline{\mathbf{F}}-\overline{\mathbf{F}}\overline{\bm{\Sigma}}\mathbf{G}^{T}(\mathbf{G}\overline{\bm{\Sigma}}\mathbf{G}^{T}+\overline{\mathbf{R}})^{-1}\mathbf{G}$ has eigenvalues strictly within the unit circle. These results follow directly from the application of \cite[Proposition 4.1, Sec. 4.1]{bertsekas1995dynamic} similar to the stability and convergence results for the standard KF for linear systems \cite[Appendix E.4]{bertsekas1995dynamic}.

In this limiting case, the inverse filter prediction and update equations take the asymptotic form $\doublehat{\mathbf{x}}_{k+1|k}=\overline{\mathbf{F}}\doublehat{\mathbf{x}}_{k}+\overline{\mathbf{E}}\mathbf{Hx}_{k+1}$ and $\doublehat{\mathbf{x}}_{k+1}=\doublehat{\mathbf{x}}_{k+1|k}+\overline{\bm{\Sigma}}\mathbf{G}^{T}(\mathbf{G}\overline{\bm{\Sigma}}\mathbf{G}^{T}+\overline{\mathbf{R}})^{-1}(\mathbf{a}_{k+1}-\mathbf{G}\doublehat{\mathbf{x}}_{k+1|k})$. Denoting the inverse filter's one-step prediction error as $\overline{\mathbf{e}}_{k+1|k}\doteq\hat{\mathbf{x}}_{k+1}-\doublehat{\mathbf{x}}_{k+1|k}$, the error dynamics for the inverse filter is obtained from this asymptotic form using \eqref{eqn: linear a} as $\overline{\mathbf{e}}_{k+1|k}=\left(\overline{\mathbf{F}}-\overline{\mathbf{F}}\overline{\bm{\Sigma}}\mathbf{G}^{T}(\mathbf{G}\overline{\bm{\Sigma}}\mathbf{G}^{T}+\overline{\mathbf{R}})^{-1}\mathbf{G}\right)\overline{\mathbf{e}}_{k|k-1}-\overline{\mathbf{F}}\overline{\bm{\Sigma}}\mathbf{G}^{T}(\mathbf{G}\overline{\bm{\Sigma}}\mathbf{G}^{T}+\overline{\mathbf{R}})^{-1}\bm{\epsilon}_{k}+\overline{\mathbf{E}}\mathbf{v}_{k+1}$. Since\\ $\overline{\mathbf{F}}-\overline{\mathbf{F}}\overline{\bm{\Sigma}}\mathbf{G}^{T}(\mathbf{G}\overline{\bm{\Sigma}}\mathbf{G}^{T}+\overline{\mathbf{R}})^{-1}\mathbf{G}$ has eigenvalues strictly within the unit circle, this error dynamics is asymptotically stable.

\chapter{Proof for I-UKF's stability}
\label{chap:iukf proof}
\section{Proof of Theorem~\ref{theorem:IUKF stability}}\label{App-thm-IUKF}
We first obtain stability results for the augmented state UKF with \emph{non-additive process noise} (from which I-UKF was formulated) in Appendix~\ref{subsec:augmented state ukf stability}. In Appendix~\ref{subsec:preliminaries}, we provide some preliminary results, including a bound on Jacobian $\widetilde{\mathbf{F}}_{k}$ of state transition \eqref{eqn: IUKF state transition detail} with respect to $\hat{\mathbf{x}}_{k}$. The I-UKF's stability then follows in Appendix~\ref{subsec:proof of IUKF theorem}. For the sake of readability, we provide the proofs of intermediate claims at last in Appendix~\ref{subsec:IUKF intermediate proofs}.

\subsection{Stochastic stability of augmented state UKF}\label{subsec:augmented state ukf stability}
Consider state-evolution \eqref{eqn:non-linear state x} with non-additive process noise as
\par\noindent\small
\begin{align}
    \mathbf{x}_{k+1}=f(\mathbf{x}_{k},\mathbf{w}_{k}).\label{eqn:augmented ukf state transition}
\end{align}
\normalsize
The non-additive noise term leads us to formulate UKF using an augmented state $\mathbf{z}_{k}=[\mathbf{x}_{k}^{T},\mathbf{w}_{k}^{T}]^{T}$ to estimate $\hat{\mathbf{x}}_{k}$ as described in \cite{wan2000unscented}. Linearizing $f(\cdot)$ at $\hat{\mathbf{z}}_{k}=[\hat{\mathbf{x}}_{k}^{T},\mathbf{0}]^{T}$, the state prediction error $\widetilde{\mathbf{x}}_{k+1|k}$ is approximated as $\widetilde{\mathbf{x}}_{k+1|k}\approx\mathbf{F}_{k}\widetilde{\mathbf{x}}_{k}+\mathbf{F}^{w}_{k}\mathbf{w}_{k}$, where $\mathbf{F}_{k}\doteq\frac{\partial f(\mathbf{x},\mathbf{0})}{\partial\mathbf{x}}\vert_{\mathbf{x}=\hat{\mathbf{x}}_{k}}$ and $\mathbf{F}^{w}_{k}\doteq\frac{\partial f(\hat{\mathbf{x}}_{k},\mathbf{w})}{\partial\mathbf{w}}\vert_{\mathbf{w}=\mathbf{0}}$. Similar to forward UKF, we introduce unknown diagonal matrices $\mathbf{U}^{x}_{k}$ and $\mathbf{U}^{w}_{k}\in\mathbb{R}^{n_{x}\times n_{x}}$ to account for the linearization errors as
\par\noindent\small
\begin{align}
    \widetilde{\mathbf{x}}_{k+1|k}=\mathbf{U}^{x}_{k}\mathbf{F}_{k}\widetilde{\mathbf{x}}_{k}+\mathbf{U}^{w}_{k}\mathbf{F}^{w}_{k}\mathbf{w}_{k}.\label{eqn:augmented state ukf linearization}
\end{align}
\normalsize
Hence, the prediction error dynamics becomes
\par\noindent\small
\begin{align}
    \widetilde{\mathbf{x}}_{k+1|k}&=\mathbf{U}^{x}_{k}\mathbf{F}_{k}(\mathbf{I}-\mathbf{K}_{k}\mathbf{U}^{y}_{k}\mathbf{H}_{k})\widetilde{\mathbf{x}}_{k|k-1}-\mathbf{U}^{x}_{k}\mathbf{F}_{k}\mathbf{K}_{k}\mathbf{v}_{k}+\mathbf{U}^{w}_{k}\mathbf{F}^{w}_{k}\mathbf{w}_{k}.\label{eqn:augmented ukf error dynamics}
\end{align}
\normalsize
The covariances $\bm{\Sigma}_{k+1|k}$, $\bm{\Sigma}^{y}_{k+1}$ and $\bm{\Sigma}^{xy}_{k+1}$ can be trivially expressed in the same forms as in forward UKF case, but with $\hat{\mathbf{Q}}_{k}=\mathbf{U}^{w}_{k}\mathbf{F}^{w}_{k}\mathbf{Q}_{k}(\mathbf{U}^{w}_{k}\mathbf{F}^{w}_{k})^{T}+\mathbf{U}^{x}_{k}\mathbf{F}_{k}\mathbf{K}_{k}\mathbf{R}_{k}\mathbf{K}_{k}^{T}\mathbf{F}_{k}^{T}\mathbf{U}^{x}_{k}+\delta\mathbf{P}_{k+1|k}+\Delta\mathbf{P}_{k+1|k}$. The following lemma extends the forward UKF's stability results to augmented state UKF.

\begin{lemma}[Stochastic stability of augmented state UKF]
\label{lemma:augmented state ukf stability}
Consider the non-linear stochastic system given by \eqref{eqn:augmented ukf state transition} and \eqref{eqn:non-linear observation y}. The augmented-state UKF's estimation error $\widetilde{\mathbf{x}}_{k}$ is exponentially bounded in mean-squared sense and bounded with probability one if all the assumptions of Theorem~\ref{theorem:forward ukf stability} hold true and additionally, there exists a constant $\overline{w}$ such that $\|\mathbf{U}^{w}_{k}\mathbf{F}^{w}_{k}\|\leq\overline{w}$ is fulfilled for all $k\geq 0$.
\end{lemma}
\begin{proof}
Define $V_{k}(\widetilde{\mathbf{x}}_{k|k-1})=\widetilde{\mathbf{x}}_{k|k-1}^{T}\bm{\Sigma}_{k|k-1}^{-1}\widetilde{\mathbf{x}}_{k|k-1}$ to apply Lemma~\ref{lemma:exponential boundedness}. Following similar steps as in Appendix~\ref{App-thm-Forward ekf stable unknown matrix}, we have
\par\noindent\small
\begin{align}
&\mathbb{E}\left[V_{k+1}(\widetilde{\mathbf{x}}_{k+1|k})\vert\widetilde{\mathbf{x}}_{k|k-1}\right]=\widetilde{\mathbf{x}}_{k|k-1}^{T}(\mathbf{U}^{x}_{k}\mathbf{F}_{k}(\mathbf{I}-\mathbf{K}_{k}\mathbf{U}^{y}_{k}\mathbf{H}_{k}))^{T}\bm{\Sigma}_{k+1|k}^{-1}(\mathbf{U}^{x}_{k}\mathbf{F}_{k}(\mathbf{I}-\mathbf{K}_{k}\mathbf{U}^{y}_{k}\mathbf{H}_{k}))\widetilde{\mathbf{x}}_{k|k-1}\nonumber\\
&+\mathbb{E}[\mathbf{w}_{k}^{T}(\mathbf{U}^{w}_{k}\mathbf{F}^{w}_{k})^{T}\bm{\Sigma}_{k+1|k}^{-1}(\mathbf{U}^{w}_{k}\mathbf{F}^{w}_{k})\mathbf{w}_{k}\vert\widetilde{\mathbf{x}}_{k|k-1}]+\mathbb{E}[\mathbf{v}_{k}^{T}(\mathbf{U}^{x}_{k}\mathbf{F}_{k}\mathbf{K}_{k})^{T}\bm{\Sigma}_{k+1|k}^{-1}(\mathbf{U}^{x}_{k}\mathbf{F}_{k}\mathbf{K}_{k})\mathbf{v}_{k}\vert\widetilde{\mathbf{x}}_{k|k-1}].\label{eqn:augmented ukf stable Vk term}
\end{align}
\normalsize

Consider the last expectation term of \eqref{eqn:augmented ukf stable Vk term}. Since, $\bm{\Sigma}_{k+1|k}\succeq\underline{\sigma}\mathbf{I}$, we have $(\mathbf{U}^{w}_{k}\mathbf{F}^{w}_{k})^{T}\bm{\Sigma}_{k+1|k}^{-1}(\mathbf{U}^{w}_{k}\mathbf{F}^{w}_{k})\\\preceq\frac{1}{\underline{\sigma}}(\mathbf{U}^{w}_{k}\mathbf{F}^{w}_{k})^{T}(\mathbf{U}^{w}_{k}\mathbf{F}^{w}_{k})$. With the bound $\|\mathbf{U}^{w}_{k}\mathbf{F}^{w}_{k}\|\leq\overline{w}$, it can be upper bounded as\\ $(\mathbf{U}^{w}_{k}\mathbf{F}^{w}_{k})^{T}\bm{\Sigma}_{k+1|k}^{-1}(\mathbf{U}^{w}_{k}\mathbf{F}^{w}_{k})\preceq\frac{\overline{w}^{2}}{\underline{\sigma}}\mathbf{I}$ such that similar to Appendix~\ref{App-thm-Forward ekf stable unknown matrix}, we have\\ $\mathbb{E}\left[\mathbf{w}_{k}^{T}(\mathbf{U}^{w}_{k}\mathbf{F}^{w}_{k})^{T}\bm{\Sigma}_{k+1|k}^{-1}(\mathbf{U}^{w}_{k}\mathbf{F}^{w}_{k})\mathbf{w}_{k}\vert\widetilde{\mathbf{x}}_{k|k-1}\right]\leq\frac{\overline{w}^{2}}{\underline{\sigma}}\mathbb{E}[\mathbf{w}_{k}^{T}\mathbf{w}_{k}]\leq\frac{\overline{w}^{2}}{\underline{\sigma}}\bar{q}n_{x}$. With this upper bound, Lemma~\ref{lemma:augmented state ukf stability} can be proved trivially following similar steps as in Appendix~\ref{App-thm-Forward ekf stable unknown matrix}.
\end{proof}
\vspace{-10pt}
\subsection{Preliminaries}\label{subsec:preliminaries}
We state Lemma~\ref{lemma:vector matrix bounds} that we employ in the sequel.
\begin{lemma}
\label{lemma:vector matrix bounds}
Bounds on a vector $\mathbf{a}$ and a `$n\times m$' matrix $\mathbf{A}$ lead to the following bounds.\\
\textbf{(a)} If $\|\mathbf{a}\|_{2}\leq\delta$, then each component satisfies $|[\mathbf{a}]_{i}|\leq\delta.$\\
\textbf{(b)} If $\|\mathbf{A}\|\leq\delta$, then the $i$-th row sum $\sum_{j=1}^{m}|[\mathbf{A}]_{i,j}|\leq\sqrt{m}\delta$.\\
\textbf{(c)} If $\|\mathbf{A}\|\leq\delta$, then the $i$-th row satisfies $\|[\mathbf{A}]_{(i,:)}\|_{2}\leq\delta$.\\
\textbf{(d)} If each component satisfies $|[\mathbf{A}]_{i,j}|\leq\delta$, then $\|\mathbf{A}\|\leq\sqrt{nm}\delta$.
\end{lemma}
\begin{proof}
For \textbf{(a)}, by the equivalence of vector norms, we have $\|\mathbf{a}\|_{\infty}\leq\|\mathbf{a}\|_{2}$. But by definition of $l_{\infty}$ norm, $\|\mathbf{a}\|_{\infty}= \textrm{max} |[\mathbf{a}]_{i}|$ and hence, $|[\mathbf{a}]_{i}|\leq\delta$.

For \textbf{(b)}, by the equivalence of matrix norms, $\|\mathbf{A}\|_{\infty}\leq\sqrt{m}\|\mathbf{A}\|\leq\sqrt{m}\delta$. But by the definition of $\|\mathbf{A}\|_{\infty}$, it is the maximum row sum such that for any i-th row of matrix $\mathbf{A}$, we have the bound $\sum_{j=1}^{m}|[\mathbf{A}]_{i,j}|\leq\sqrt{m}\delta$.

For \textbf{(c)}, by the definition of spectral norm, $\|\mathbf{A}^{T}\|=\textrm{max}_{\|\mathbf{x}\|_{2}=1}\|\mathbf{A}^{T}\mathbf{x}\|_{2}$. However, $\|\mathbf{A}^{T}\|=\|\mathbf{A}\|$ such that with $\mathbf{y}=\mathbf{x}^{T}$, we have $\|\mathbf{A}\|=\textrm{max}_{\|\mathbf{y}\|_{2}=1}\|\mathbf{yA}\|_{2}$, which implies $\|\mathbf{yA}\|_{2}\leq\|\mathbf{A}\|$ for any vector $\mathbf{y}$ with $\|\mathbf{y}\|_{2}=1$. Choosing $\mathbf{y}$ as a `$1\times n$' vector with only $i$-th component as $1$ and all other components as $0$, $\mathbf{yA}$ becomes the $i$-th row of $\mathbf{A}$ from which the result follows trivially.

For \textbf{(d)}, by the equivalence of matrix norms, we have $\|\mathbf{A}\|\leq\sqrt{nm}\|\mathbf{A}\|_{\textrm{max}}$ where $\|\mathbf{A}\|_{\textrm{max}}=\textrm{max}_{i,j} |[\mathbf{A}]_{i,j}|$. Hence, $|[\mathbf{A}]_{i,j}|\leq\delta$ leads to $\|\mathbf{A}\|\leq\sqrt{nm}\delta$.
\end{proof}

\begin{lemma}
\label{lemma: qi jacobian bound}
Under the assumptions of Theorem~\ref{theorem:IUKF stability}, the Jacobian $\frac{\partial\mathbf{q}_{i,k+1|k}}{\partial\hat{\mathbf{x}}_{k}}$ of the $i$-th sigma point generated for measurement update in the forward UKF with respect to the state estimate $\hat{\mathbf{x}}_{k}$ is bounded as $\left\|\frac{\partial\mathbf{q}_{i,k+1|k}}{\partial\hat{\mathbf{x}}_{k}}\right\|\leq c'$ for some positive real constant $c'$.
\end{lemma}

\begin{proof}
From \eqref{eqn:forward UKF update sigma points}, we observe that $\mathbf{q}_{i,k+1|k}$ is a linear function of the predicted state $\hat{\mathbf{x}}_{k+1|k}$ and the $i$-th column of $\sqrt{\bm{\Sigma}_{k+1|k}}$. In order to bound its Jacobian with respect to $\hat{\mathbf{x}}_{k}$, we need to bound the Jacobians of $\hat{\mathbf{x}}_{k+1|k}$ and $\sqrt{\bm{\Sigma}_{k+1|k}}$ with respect to $\hat{\mathbf{x}}_{k}$.

We start with $\hat{\mathbf{x}}_{k+1|k}$. Since $\mathbf{s}^{*}_{i,k+1|k}=f(\mathbf{s}_{i,k})$, the Jacobian $\frac{\partial\mathbf{s}^{*}_{i,k+1|k}}{\partial\hat{\mathbf{x}}_{k}}=\left.\frac{\partial f(\mathbf{x})}{\partial\mathbf{x}}\right\vert_{\mathbf{x}=\mathbf{s}_{i,k}}$ for all $0\leq i\leq 2n_{x}$, because $\frac{\partial\mathbf{s}_{i,k}}{\partial\hat{\mathbf{x}}_{k}}=\mathbf{I}$ from \eqref{eqn:forward ukf prediction sigma points}. Hence, differentiating \eqref{eqn:forward ukf x predict}, we have $\frac{\partial\hat{\mathbf{x}}_{k+1|k}}{\partial\hat{\mathbf{x}}_{k}}=\sum_{i=0}^{2n_{x}}\omega_{i}\left.\frac{\partial f(\mathbf{x})}{\partial\mathbf{x}}\right\vert_{\mathbf{x}=\mathbf{s}_{i,k}}$, which on using the upper-bound on Jacobian $\mathbf{F}_{k}$ from Theorem~\ref{theorem:forward ukf stability} yields
\par\noindent\small
\begin{align}
  \left\|\frac{\partial\hat{\mathbf{x}}_{k+1|k}}{\partial\hat{\mathbf{x}}_{k}}\right\|\leq\bar{f}.\label{eqn:IUKF stable bound on x predict derivative}  
\end{align}
\normalsize
The following Claim~\ref{claim:IUKF stable bound on cholesky derivative} bounds the derivative of $\sqrt{\bm{\Sigma}_{k+1|k}}$ with the detailed proof in Appendix~\ref{subsec:IUKF intermediate proofs}.
\begin{claim}
\label{claim:IUKF stable bound on cholesky derivative}
For any $i$-th column of $\sqrt{\bm{\Sigma}_{k+1|k}}$, we have the upper bound $\left\|\frac{\partial[\sqrt{\bm{\Sigma}_{k+1|k}}]_{(:,i)}}{\partial\hat{\mathbf{x}}_{k}}\right\|\leq n_{x}\delta_{\sigma}$ for some $\delta_{\sigma}>0$.
\end{claim}

Using Claim~\ref{claim:IUKF stable bound on cholesky derivative} and bound \eqref{eqn:IUKF stable bound on x predict derivative} in \eqref{eqn:forward UKF update sigma points}, we have $\left\|\frac{\partial\mathbf{q}_{i,k+1|k}}{\partial\hat{\mathbf{x}}_{k}}\right\|\leq c'$ where $c'=\bar{f}+n_{x}\delta_{\sigma}\sqrt{n_{x}+\kappa}$.
\end{proof}
\begin{lemma}
\label{lemma: jacobian bound}
Under the assumptions of Theorem~\ref{theorem:IUKF stability}, the Jacobian $\widetilde{\mathbf{F}}_{k}\doteq\left.\frac{\partial\widetilde{f}(\mathbf{x},\bm{\Sigma}_{k},\mathbf{x}_{k+1},\mathbf{0})}{\partial\mathbf{x}}\right\vert_{\mathbf{x}=\doublehat{\mathbf{x}}_{k}}$ of state transition equation \eqref{eqn: IUKF state transition detail} satisfies the bound $\|\widetilde{\mathbf{F}}_{k}\|\leq c_{f}$ for some positive real constant $c_{f}$.
\end{lemma}
\begin{proof}
Define $\mathbf{t}_{k}=h(\mathbf{x}_{k+1})+\mathbf{v}_{k+1}-\sum_{i=0}^{2n_{x}}\omega_{i}\mathbf{q}^{*}_{i,k+1|k}$ i.e. the difference in the actual observation $\mathbf{y}_{k+1}$ and its prediction. Rearranging \eqref{eqn: IUKF state transition detail} and using $\hat{\mathbf{x}}_{k+1|k}=\sum_{i=0}^{2n_{x}}\omega_{i}\mathbf{s}^{*}_{i,k+1|k}$, I-UKF's state transition becomes $\hat{\mathbf{x}}_{k+1}=\hat{\mathbf{x}}_{k+1|k}+\mathbf{K}_{k+1}\mathbf{t}_{k}$ such that its Jacobian with respect to $\hat{\mathbf{x}}_{k}$ (state to be estimated from I-UKF's state transition) is
\par\noindent\small
\begin{align}
    \widetilde{\mathbf{F}}_{k}=\left.\frac{\partial\hat{\mathbf{x}}_{k+1|k}}{\partial\hat{\mathbf{x}}_{k}}\right\vert_{\mathbf{v}_{k+1}=\mathbf{0}}+\left.\frac{\partial(\mathbf{K}_{k+1}\mathbf{t}_{k})}{\partial\hat{\mathbf{x}}_{k}}\right\vert_{\mathbf{v}_{k+1}=\mathbf{0}}.\label{eqn:IUKF stable jacobian exp}
\end{align}
\normalsize

First, consider the second derivative term. The $j$-th row of `$n_{x}\times n_{x}$' Jacobian $\frac{\partial(\mathbf{K}_{k+1}\mathbf{t}_{k})}{\partial\hat{\mathbf{x}}_{k}}$ consists of the first order partial derivatives of $j$-th element of `$n_{x}\times 1$' vector $\mathbf{K}_{k+1}\mathbf{t}_{k}$ with respect to the elements of $\hat{\mathbf{x}}_{k}$. This $j$-th element $[\mathbf{K}_{k+1}\mathbf{t}_{k}]_{j}=\sum_{m=1}^{n_{y}}[\mathbf{K}_{k+1}]_{j,m}[\mathbf{t}_{k}]_{m}$, such that the $j$-th row of the Jacobian is obtained as
\par\noindent\small
\begin{align}
    \hspace{-0.2cm}\left[\frac{\partial(\mathbf{K}_{k+1}\mathbf{t}_{k})}{\partial\hat{\mathbf{x}}_{k}}\right]_{(j,:)}=\sum_{m=1}^{n_{y}}[\mathbf{K}_{k+1}]_{j,m}\frac{\partial[\mathbf{t}_{k}]_{m}}{\partial\hat{\mathbf{x}}_{k}}+\sum_{m=1}^{n_{y}}[\mathbf{t}_{k}]_{m}\frac{\partial[\mathbf{K}_{k+1}]_{j,m}}{\partial\hat{\mathbf{x}}_{k}}.\label{eqn:IUKF stable jac K term j-th row}
\end{align}
\normalsize
The following Claim~\ref{claim:IUKF stable K j row bound} upper bounds this $j$-th row with the detailed proof provided in Appendix~\ref{subsec:IUKF intermediate proofs}.
\begin{claim}
    \label{claim:IUKF stable K j row bound}
    The $j$-th row of Jacobian of $\mathbf{K}_{k+1}\mathbf{t}_{k}$ satisfies $\left\|\left[\frac{\partial(\mathbf{K}_{k+1}\mathbf{t}_{k})}{\partial\hat{\mathbf{x}}_{k}}\right]_{(j,:)}\right\|_{2}\leq c_{t}$ for some $c_{t}>0$.
\end{claim}

With Claim~\ref{claim:IUKF stable K j row bound}, we can show that $\left\|\frac{\partial(\mathbf{K}_{k+1}\mathbf{t}_{k})}{\partial\hat{\mathbf{x}}_{k}}\right\|\leq n_{x}c_{t}$ using Lemma~\ref{lemma:vector matrix bounds}\textbf{(a)} followed by Lemma~\ref{lemma:vector matrix bounds}\textbf{(d)}. Hence, from \eqref{eqn:IUKF stable jacobian exp} along with the bound \eqref{eqn:IUKF stable bound on x predict derivative} since $\hat{\mathbf{x}}_{k+1|k}$ is independent of the noise term $\mathbf{v}_{k+1}$, we have $\|\widetilde{\mathbf{F}}_{k}\|\leq c_{f}=\bar{f}+n_{x}c_{t}$.
\end{proof}
\vspace{-5pt}
\subsection{Proof of the theorem}\label{subsec:proof of IUKF theorem}
Define $\widetilde{\mathbf{F}}^{v}_{k}\doteq\left.\frac{\partial\widetilde{f}(\doublehat{\mathbf{x}}_{k},\bm{\Sigma}_{k},\mathbf{x}_{k+1},\mathbf{v})}{\partial\mathbf{v}}\right\vert_{\mathbf{v}=\mathbf{0}}$ and $\overline{\mathbf{U}}^{v}_{k}$ as the counterpart of $\mathbf{U}^{w}_{k}$ for I-UKF dynamics. We will show that under the assumptions of Theorem~\ref{theorem:IUKF stability}, the I-UKF's dynamics satisfies the required conditions of Lemma~\ref{lemma:augmented state ukf stability}. In this regard, we show that I-UKF's dynamics satisfies the following conditions for all $k\geq 0$ for some constants $c_{f},c_{\alpha},c_{q},c_{v}$.\\
\textbf{B.C1.} $\|\widetilde{\mathbf{F}}_{k}\|\leq c_{f}$.\\
\textbf{B.C2.} $\overline{\mathbf{Q}}_{k}\preceq c_{q}\mathbf{I}$.\\
\textbf{B.C3.} $\|\overline{\mathbf{U}}^{x}_{k}\|\leq c_{\alpha}$.\\
\textbf{B.C4.} $\overline{\mathbf{U}}^{x}_{k}$ is non-singular for all $k\geq 0$.\\
\textbf{B.C5.} $\|\overline{\mathbf{U}}^{v}_{k}\widetilde{\mathbf{F}}^{v}_{k}\|\leq c_{v}$.\\
All other conditions of Lemma~\ref{lemma:augmented state ukf stability} are assumed to hold true in Theorem~\ref{theorem:IUKF stability}, such that the I-UKF's estimation error is exponentially bounded in mean-squared sense and bounded with probability one. Also, \textbf{B.C1} is satisfied by Lemma~\ref{lemma: jacobian bound}.

For \textbf{B.C2}, the noise $\mathbf{v}_{k+1}$ of state transition \eqref{eqn: IUKF state transition detail} has covariance $\overline{\mathbf{Q}}_{k}=\mathbf{R}_{k+1}$ which is upper-bounded as $\mathbf{R}_{k+1}\preceq\bar{r}\mathbf{I}$ from one of the assumptions of Theorem~\ref{theorem:forward ukf stability}. Hence, \textbf{B.C2} holds true with $c_{q}=\bar{r}$.

For \textbf{B.C3} and \textbf{B.C4}, introducing the unknown matrix $\overline{\mathbf{U}}^{x}_{k}$ and $\overline{\mathbf{U}}^{v}_{k}\widetilde{\mathbf{F}}^{v}_{k}=\mathbf{K}_{k+1}$ since the second and higher order derivatives of \eqref{eqn: IUKF state transition detail} with respect to the noise term $\mathbf{v}_{k+1}$ are zero, we have I-UKF's state prediction error $\hat{\widetilde{\mathbf{x}}}_{k+1|k}\doteq\hat{\mathbf{x}}_{k+1}-\doublehat{\mathbf{x}}_{k+1|k}$ as
\par\noindent\small
\begin{align}    \hat{\widetilde{\mathbf{x}}}_{k+1|k}=\overline{\mathbf{U}}^{x}_{k}\widetilde{\mathbf{F}}_{k}\hat{\widetilde{\mathbf{x}}}_{k}+\mathbf{K}_{k+1}\mathbf{v}_{k+1},\label{eqn:IUKF stable unknown matrix}
\end{align}
\normalsize
where $\hat{\widetilde{\mathbf{x}}}_{k}\doteq\hat{\mathbf{x}}_{k}-\doublehat{\mathbf{x}}_{k}$ is the I-UKF's state estimation error. However, substituting for $\hat{\mathbf{x}}_{k+1}$ using \eqref{eqn: IUKF state transition detail} and $\doublehat{\mathbf{x}}_{k+1|k}$ using \eqref{eqn:IUKF x predict}, we have
\par\noindent\small
\begin{align}
    \hat{\widetilde{\mathbf{x}}}_{k+1|k}&=\sum_{i=0}^{2n_{x}}\omega_{i}\mathbf{s}^{*}_{i,k+1|k}-\mathbf{K}_{k+1}\sum_{i=0}^{2n_{x}}\omega_{i}\mathbf{q}^{*}_{i,k+1|k}+\mathbf{K}_{k+1}h(\mathbf{x}_{k+1})+\mathbf{K}_{k+1}\mathbf{v}_{k+1}-\sum_{j=0}^{2n_{z}}\overline{\omega}_{j}\overline{\mathbf{s}}^{*}_{j,k+1|k}.\label{eqn:IUKF stable state prediction error}
\end{align}
\normalsize
From $\sum_{i=0}^{2n_{x}}\omega_{i}\mathbf{s}^{*}_{i,k+1|k}=\sum_{i=0}^{2n_{x}}\omega_{i}f(\mathbf{s}_{i,k})$ using the first-order Taylor series expansion, we have $\sum_{i=0}^{2n_{x}}\omega_{i}\mathbf{s}^{*}_{i,k+1|k}=f(\hat{\mathbf{x}}_{k})$. Similarly, $\sum_{i=0}^{2n_{x}}\omega_{i}\mathbf{q}^{*}_{i,k+1|k}=h(\hat{\mathbf{x}}_{k+1|k})$ and $\sum_{j=0}^{2n_{z}}\overline{\omega}_{j}\overline{\mathbf{s}}^{*}_{j,k+1|k}\\=\widetilde{f}(\doublehat{\mathbf{x}}_{k},\bm{\Sigma}_{k},\mathbf{x}_{k+1},\mathbf{0})$. Substituting \eqref{eqn: IUKF state transition detail} for $\widetilde{f}(\cdot)$ and again using the Taylor series expansion for the summation terms involving sigma points, we have $\sum_{j=0}^{2n_{z}}\overline{\omega}_{j}\overline{\mathbf{s}}^{*}_{j,k+1|k}=f(\doublehat{\mathbf{x}}_{k})-\mathbf{K}_{k+1}h(\overline{\mathbf{x}}_{k+1|k})+\mathbf{K}_{k+1}h(\mathbf{x}_{k+1})$, where $\overline{\mathbf{x}}_{k+1|k}=f(\doublehat{\mathbf{x}}_{k})$ under the assumption that the gain $\mathbf{K}_{k+1}$ computed from $\doublehat{\mathbf{x}}_{k}$ is approximately same as that computed from forward UKF's $\hat{\mathbf{x}}_{k}$. Hence, \eqref{eqn:IUKF stable state prediction error} simplifies to $\hat{\widetilde{\mathbf{x}}}_{k+1|k}=f(\hat{\mathbf{x}}_{k})-f(\doublehat{\mathbf{x}}_{k})-\mathbf{K}_{k+1}(h(\hat{\mathbf{x}}_{k+1|k})-h(\overline{\mathbf{x}}_{k+1|k}))+\mathbf{K}_{k+1}\mathbf{v}_{k+1}$. Using the unknown matrices $\mathbf{U}^{x}_{k}$ and $\mathbf{U}^{y}_{k}$ introduced in forward UKF for linearizing $f(\cdot)$ and $h(\cdot)$, respectively, at $\hat{\mathbf{x}}_{k}$ and $\hat{\mathbf{x}}_{k+1|k}$, we obtain $\hat{\widetilde{\mathbf{x}}}_{k+1|k}=\mathbf{U}^{x}_{k}\mathbf{F}_{k}\hat{\widetilde{\mathbf{x}}}_{k}-\mathbf{K}_{k+1}\mathbf{U}^{y}_{k+1}\mathbf{H}_{k+1}\mathbf{U}^{x}_{k}\mathbf{F}_{k}\hat{\widetilde{\mathbf{x}}}_{k}+\mathbf{K}_{k+1}\mathbf{v}_{k+1}$. Comparing with \eqref{eqn:IUKF stable unknown matrix}, we have $\overline{\mathbf{U}}^{x}_{k}=(\mathbf{I}-\mathbf{K}_{k+1}\mathbf{U}^{y}_{k+1}\mathbf{H}_{k+1})\mathbf{U}^{x}_{k}\mathbf{F}_{k}\widetilde{\mathbf{F}}_{k}^{-1}$ since $\widetilde{\mathbf{F}}_{k}^{-1}$ is assumed to be non-singular in Theorem~\ref{theorem:IUKF stability}. With the bounds assumed on various matrices in Theorem~\ref{theorem:forward ukf stability} and the assumption $\|\widetilde{\mathbf{F}}_{k}^{-1}\|\leq\bar{a}$ from Theorem~\ref{theorem:IUKF stability}, it is straightforward to obtain $ \|\overline{\mathbf{U}}^{x}_{k}\|\leq\bar{\alpha}\bar{f}\bar{a}(1+\bar{k}\bar{\beta}\bar{h})$ such that \textbf{B.C3} holds true with $c_{\alpha}=\bar{\alpha}\bar{f}\bar{a}(1+\bar{k}\bar{\beta}\bar{h})$. Also, \textbf{B.C4} holds true, i.e, $\overline{\mathbf{U}}^{x}_{k}$ is non-singular because $\mathbf{U}^{x}_{k}$ and $\mathbf{F}_{k}$ are non-singular from the assumptions of Theorem~\ref{theorem:forward ukf stability}, and $(\mathbf{I}-\mathbf{K}_{k+1}\mathbf{U}^{y}_{k+1}\mathbf{H}_{k+1})$ can be proved to be invertible under the forward UKF's stability conditions as proved intermediately, in Appendix~\ref{App-thm-Forward ekf stable unknown matrix}.

For \textbf{B.C5}, we have $\overline{\mathbf{U}}^{v}_{k}\widetilde{\mathbf{F}}^{v}_{k}=\mathbf{K}_{k+1}$ since the second and higher order derivatives of \eqref{eqn: IUKF state transition detail} with respect to $\mathbf{v}_{k+1}$ are zero. Under Theorem~\ref{theorem:forward ukf stability}'s assumptions, $\|\mathbf{K}_{k+1}\|\leq\bar{k}$ and \textbf{B.C5} holds with $c_{v}=\bar{k}$.

\subsection{Proofs of claims}\label{subsec:IUKF intermediate proofs}
\subsubsection{1) Proof of claim~\ref{claim:IUKF stable bound on cholesky derivative} in proof of Lemma~\ref{lemma: qi jacobian bound}}
In order to bound the derivative of $\sqrt{\bm{\Sigma}_{k+1|k}}$, we first upper-bound the derivative of $\bm{\Sigma}_{k+1|k}$.
\begin{claim}
\label{claim:IUKF stable predict sig derivative bound}
For any $(l,m)$-th element of $\bm{\Sigma}_{k+1|k}$, we have the upper bound $\left\|\frac{\partial[\bm{\Sigma}_{k+1|k}]_{l,m}}{\partial\hat{\mathbf{x}}_{k}}\right\|_{2}\leq 4\delta_{f}\bar{f}$.
\end{claim}
\begin{claimproof}
We have $\bm{\Sigma}_{k+1|k}=\sum_{i=0}^{2n_{x}}\omega_{i}\mathbf{s}^{*}_{i,k+1|k}(\mathbf{s}^{*}_{i,k+1|k})^{T}-\hat{\mathbf{x}}_{k+1|k}(\hat{\mathbf{x}}_{k+1|k})^{T}+\mathbf{Q}_{k}$. Its $(l,m)$-th element is $[\bm{\Sigma}_{k+1|k}]_{l,m}=\sum_{i=0}^{2n_{x}}\omega_{i}[\mathbf{s}^{*}_{i,k+1|k}]_{l}[\mathbf{s}^{*}_{i,k+1|k}]_{m}-[\hat{\mathbf{x}}_{k+1|k}]_{l}[\hat{\mathbf{x}}_{k+1|k}]_{m}+[\mathbf{Q}_{k}]_{l,m}$, which implies
\par\noindent\small
\begin{align*}
\frac{\partial[\bm{\Sigma}_{k+1|k}]_{l,m}}{\partial\hat{\mathbf{x}}_{k}}&=-[\hat{\mathbf{x}}_{k+1|k}]_{l}\frac{\partial[\hat{\mathbf{x}}_{k+1|k}]_{m}}{\partial\hat{\mathbf{x}}_{k}}-[\hat{\mathbf{x}}_{k+1|k}]_{m}\frac{\partial[\hat{\mathbf{x}}_{k+1|k}]_{l}}{\partial\hat{\mathbf{x}}_{k}}+\sum_{i=0}^{2n_{x}}\omega_{i}[\mathbf{s}^{*}_{i,k+1|k}]_{m}\frac{\partial[\mathbf{s}^{*}_{i,k+1|k}]_{l}}{\partial\hat{\mathbf{x}}_{k}}\\
&+\sum_{i=0}^{2n_{x}}\omega_{i}[\mathbf{s}^{*}_{i,k+1|k}]_{l}\frac{\partial[\mathbf{s}^{*}_{i,k+1|k}]_{m}}{\partial\hat{\mathbf{x}}_{k}}.
\end{align*}
\normalsize
Note that all derivatives here are gradients since $[\bm{\Sigma}_{k+1|k}]_{l,m}$ is a scalar. Theorem~\ref{theorem:IUKF stability} assumes $f(\cdot)$ has bounded outputs and hence, the magnitude of each element of $\mathbf{s}^{*}_{i,k+1|k}$ and $\hat{\mathbf{x}}_{k+1|k}$ is also bounded by $\delta_{f}$ according to Lemma~\ref{lemma:vector matrix bounds}\textbf{(a)}, which leads to
\par\noindent\small
\begin{align*}
&\left\|\frac{\partial[\bm{\Sigma}_{k+1|k}]_{l,m}}{\partial\hat{\mathbf{x}}_{k}}\right\|_{2}=\delta_{f}\left\|\frac{\partial[\hat{\mathbf{x}}_{k+1|k}]_{m}}{\partial\hat{\mathbf{x}}_{k}}\right\|_{2}+\delta_{f}\left\|\frac{\partial[\hat{\mathbf{x}}_{k+1|k}]_{l}}{\partial\hat{\mathbf{x}}_{k}}\right\|_{2}+\sum_{i=0}^{2n_{x}}\omega_{i}\delta_{f}\left\|\frac{\partial[\mathbf{s}^{*}_{i,k+1|k}]_{l}}{\partial\hat{\mathbf{x}}_{k}}\right\|_{2}+\sum_{i=0}^{2n_{x}}\omega_{i}\delta_{f}\left\|\frac{\partial[\mathbf{s}^{*}_{i,k+1|k}]_{m}}{\partial\hat{\mathbf{x}}_{k}}\right\|_{2}.
\end{align*}
\normalsize
However, $\frac{\partial[\mathbf{s}^{*}_{i,k+1|k}]_{l}}{\partial\hat{\mathbf{x}}_{k}}$ is the $l$-th row of Jacobian $\frac{\partial\mathbf{s}^{*}_{i,k+1|k}}{\partial\hat{\mathbf{x}}_{k}}$ whose spectral norm is bounded by $\bar{f}$ since $\|\mathbf{F}_{k}\|\leq\bar{f}$ (as used to obtain \eqref{eqn:IUKF stable bound on x predict derivative}). Hence, using Lemma~\ref{lemma:vector matrix bounds}\textbf{(c)} with the bounds on the spectral norms $\left\|\frac{\partial\mathbf{s}^{*}_{i,k+1|k}}{\partial\hat{\mathbf{x}}_{k}}\right\|$ and $\left\|\frac{\partial\hat{\mathbf{x}}_{k+1|k}}{\partial\hat{\mathbf{x}}_{k}}\right\|$, we have $\left\|\frac{\partial[\bm{\Sigma}_{k+1|k}]_{l,m}}{\partial\hat{\mathbf{x}}_{k}}\right\|_{2}\leq 4\delta_{f}\bar{f}$.
\end{claimproof}

From the Cholesky decomposition of $\bm{\Sigma}_{k+1|k}$, we have $[\bm{\Sigma}_{k+1|k}]_{l,m}=\sum_{j=1}^{n_{x}}[\sqrt{\bm{\Sigma}_{k+1|k}}]_{l,j}[\sqrt{\bm{\Sigma}_{k+1|k}}]_{m,j}$. However, $\bm{\Sigma}_{k+1|k}$ is a symmetric, positive definite matrix (one of the assumptions of Theorem~\ref{theorem:forward ukf stability}) and hence, $\sqrt{\bm{\Sigma}_{k+1|k}}$ is a lower triangular matrix with $[\sqrt{\bm{\Sigma}_{k+1|k}}]_{i,j}=0$ for $j\geq i$ such that $[\bm{\Sigma}_{k+1|k}]_{l,m}=\sum_{j=1}^{\textrm{min}(l,m)}[\sqrt{\bm{\Sigma}_{k+1|k}}]_{l,j}[\sqrt{\bm{\Sigma}_{k+1|k}}]_{m,j}$ where indices $l$ and $m$ ranges from $1$ to $n_{x}$. Differentiating with respect to $i$-th element of $\hat{\mathbf{x}}_{k}$, we have
\par\noindent\small
\begin{align}
&\frac{\partial[\bm{\Sigma}_{k+1|k}]_{l,m}}{\partial[\hat{\mathbf{x}}_{k}]_{i}}=\sum_{j=1}^{\textrm{min}(l,m)}[\sqrt{\bm{\Sigma}_{k+1|k}}]_{l,j}\frac{\partial[\sqrt{\bm{\Sigma}_{k+1|k}}]_{m,j}}{\partial[\hat{\mathbf{x}}_{k}]_{i}}+\sum_{j=1}^{\textrm{min}(l,m)}[\sqrt{\bm{\Sigma}_{k+1|k}}]_{m,j}\frac{\partial[\sqrt{\bm{\Sigma}_{k+1|k}}]_{l,j}}{\partial[\hat{\mathbf{x}}_{k}]_{i}}.\label{eqn:IUKF stable cholesky derivative}
\end{align}
\normalsize
Note that here, all the derivatives are scalar. We will bound each individual term of this equation. We denote $\frac{\partial[\bm{\Sigma}_{k+1|k}]_{l,m}}{\partial[\hat{\mathbf{x}}_{k}]_{i}}$ by $a_{l,m}$ which is the $i$-th element of $\frac{\partial[\bm{\Sigma}_{k+1|k}]_{l,m}}{\partial\hat{\mathbf{x}}_{k}}$. Also, denote the $(l,j)$-th element $[\sqrt{\bm{\Sigma}_{k+1|k}}]_{l,j}$ by $b_{l,j}$ and its derivative $\frac{\partial[\sqrt{\bm{\Sigma}_{k+1|k}}]_{l,j}}{\partial[\hat{\mathbf{x}}_{k}]_{i}}$ by $c_{l,j}$. We require an upper-bound on $c_{l,j}$.

\textbf{Upper-bounds on $|a_{l,m}|$ and $|b_{l,j}|$, and lower bound on $|b_{i,i}|$:} Using Claim~\ref{claim:IUKF stable predict sig derivative bound} and Lemma~\ref{lemma:vector matrix bounds}\textbf{(a)}, we have $|a_{l,m}|\leq 4\delta_{f}\bar{f}$. By the definition of spectral norm and bound $\bm{\Sigma}_{k+1|k}\preceq\bar{\sigma}\mathbf{I}$ from one of the assumptions of Theorem~\ref{theorem:forward ukf stability}, we have $\|\sqrt{\bm{\Sigma}_{k+1|k}}\|\leq\sqrt{\bar{\sigma}}$, which again using Lemma~\ref{lemma:vector matrix bounds}\textbf{(c)} followed by Lemma~\ref{lemma:vector matrix bounds}\textbf{(a)} gives $|b_{l,j}|\leq\sqrt{\bar{\sigma}}$.

Also, $\textrm{det}(\bm{\Sigma}_{k+1|k})\neq 0$ (positive definite matrix) and hence, $\textrm{det}(\sqrt{\bm{\Sigma}_{k+1|k}})\neq 0$. However, being a lower triangular matrix, $\textrm{det}(\sqrt{\bm{\Sigma}_{k+1|k}})$ is the product of its diagonal entries. Hence, no diagonal entry of $\sqrt{\bm{\Sigma}_{k+1|k}}$ is $0$ i.e. $b_{i,i}\neq 0$. Next, consider the bound $\underline{\sigma}\mathbf{I}\preceq\bm{\Sigma}_{k+1|k}$ from one of the assumptions of Theorem~\ref{theorem:IUKF stability}. Since, $\underline{\sigma}$ is a lower bound on eigenvalues of $\bm{\Sigma}_{k+1|k}$, $\textrm{det}(\bm{\Sigma}_{k+1|k})\geq\underline{\sigma}^{n_{x}}$ such that $|\textrm{det}(\sqrt{\bm{\Sigma}_{k+1|k}})|\geq\underline{\sigma}^{n_{x}/2}$. Expressing $\textrm{det}(\sqrt{\bm{\Sigma}_{k+1|k}})$ as the product of its diagonal entries $b_{i,i}$ and using the upper-bound on $|b_{i,i}|$ for all but one diagonal entry, we obtain the lower bound $|b_{i,i}|\geq c_{b}$ where $c_{b}=\underline{\sigma}^{n_{x}/2}\bar{\sigma}^{(1-n_{x})/2}$.

\textbf{Upper-bounds on $|c_{l,j}|$:} Putting different values of $l$ and $m$ in \eqref{eqn:IUKF stable cholesky derivative}, we obtain a system of $n_{x}(n_{x}+1)/2$ linear equations with same number of unknowns $c_{l,j}$ since $\bm{\Sigma}_{k+1|k}$ is symmetric. Consider $l=1$ and $m=1$ which gives $2b_{1,1}c_{1,1}=a_{1,1}$. Since, $b_{1,1}\neq 0$, $c_{1,1}=a_{1,1}/2b_{1,1}$. With the upper-bound on $|a_{1,1}|$ and lower-bound on $|b_{1,1}|$, we have $|c_{1,1}|\leq\frac{4\delta_{f}\bar{f}}{2c_{b}}$. Again, putting $l=2$ and $m=1$, we have $c_{2,1}=\frac{a_{1,2}-b_{2,1}c_{1,1}}{b_{1,1}}$ which implies $|c_{2,1}|\leq\frac{8c_{b}\delta_{f}\bar{f}+4\delta_{f}\bar{f}\sqrt{\bar{\sigma}}}{2c_{b}^{2}}$. Continuing further in the same manner for all the equations of the linear system, we can show that $|c_{l,j}|\leq\delta_{\sigma}$ for all $(l,j)$-th element of $\sqrt{\bm{\Sigma}_{k+1|k}}$ where $\delta_{\sigma}$ is the maximum of all these upper-bounds i.e. the magnitude of the partial derivative of any element of $\sqrt{\bm{\Sigma}_{k+1|k}}$ with respect to any element of $\hat{\mathbf{x}}_{k}$ is bounded. Hence, the magnitude of each element of the Jacobian $\frac{\partial[\sqrt{\bm{\Sigma}_{k+1|k}}]_{(:,i)}}{\partial\hat{\mathbf{x}}_{k}}$ of $i$-th column of $\sqrt{\bm{\Sigma}_{k+1|k}}$ is bounded such that Lemma~\ref{lemma:vector matrix bounds} \textbf{(d)} yields $\left\|\frac{\partial[\sqrt{\bm{\Sigma}_{k+1|k}}]_{(:,i)}}{\partial\hat{\mathbf{x}}_{k}}\right\|\leq n_{x}\delta_{\sigma}$.

\subsubsection{2) Proof of claim~\ref{claim:IUKF stable K j row bound} in proof of Lemma~\ref{lemma: jacobian bound}}
We will upper bound the $j$-th row in \eqref{eqn:IUKF stable jac K term j-th row} by bounding the magnitude of the terms in the R.H.S. in the following Claims~\ref{claim:IUKF stable tk term bound}-\ref{claim:IUKF stable derivative of K bound}.
\begin{claim}
\label{claim:IUKF stable tk term bound}
The $m$-the element of vector $\mathbf{t}_{k}$ satisfies $|[\mathbf{t}_{k}]_{m}|\leq 2\delta_{h}$.
\end{claim}
\begin{claimproof}
    Using $\mathbf{v}_{k+1}=\mathbf{0}$ ($\widetilde{\mathbf{F}}_{k}$ is evaluated at $\mathbf{v}_{k+1}=0$), we have $[\mathbf{t}_{k}]_{m}=[h(\mathbf{x}_{k+1})]_{m}-\sum_{i=0}^{2n_{x}}\omega_{i}[\mathbf{q}^{*}_{i,k+1|k}]_{m}$ because $h(\mathbf{x}_{k+1})$ and $\lbrace\mathbf{q}^{*}_{i,k+1|k}\rbrace_{0\leq i\leq 2n_{x}}$ do not depend on the noise term. Also, $[\mathbf{q}^{*}_{i,k+1|k}]_{m}=[h(\mathbf{q}_{i,k+1|k})]_{m}$. Since Theorem~\ref{theorem:IUKF stability} assumes $h(\cdot)$ has bounded outputs and $\sum_{i=0}^{2n_{x}}\omega_{i}=1$,\\ Lemma~\ref{lemma:vector matrix bounds}\textbf{(a)} leads to $|[\mathbf{t}_{k}]_{m}|\leq 2\delta_{h}$.
\end{claimproof}
\begin{claim}
\label{claim:IUKF stable tk derivative bound}
The derivative of $m$-th element of $\mathbf{t}_{k}$ is bounded as $\left\|\frac{\partial[\mathbf{t}_{k}]_{m}}{\partial\hat{\mathbf{x}}_{k}}\right\|_{2}\leq\bar{h}c'$ where the constant $c'$ is same as defined in Lemma~\ref{lemma: qi jacobian bound}.
\end{claim}
\begin{claimproof}
From $\mathbf{t}_{k}=h(\mathbf{x}_{k+1})+\mathbf{v}_{k+1}-\sum_{i=0}^{2n_{x}}\omega_{i}\mathbf{q}^{*}_{i,k+1|k}$, we obtain the derivative of $m$-th element as
\par\noindent\small
\begin{align}
    \frac{\partial[\mathbf{t}_{k}]_{m}}{\partial\hat{\mathbf{x}}_{k}}=-\sum_{i=0}^{2n_{x}}\omega_{i}\frac{\partial[\mathbf{q}^{*}_{i,k+1|k}]_{m}}{\partial\hat{\mathbf{x}}_{k}}.\label{eqn:IUKF stable tk term derivative}
\end{align}
\normalsize
But $\mathbf{q}^{*}_{i,k+1|k}=h(\mathbf{q}_{i,k+1|k})$ such that $\left\|\frac{\partial\mathbf{q}^{*}_{i,k+1|k}}{\partial\hat{\mathbf{x}}_{k}}\right\|=\left\|\left.\frac{\partial h(\mathbf{x})}{\partial\mathbf{x}}\right\vert_{\mathbf{x}=\mathbf{q}_{i,k+1|k}}\times\frac{\partial\mathbf{q}_{i,k+1|k}}{\partial\hat{\mathbf{x}}_{k}}\right\|$. Using the bound on Jacobian $\mathbf{H}_{k}$ from one of the assumptions of Theorem~\ref{theorem:forward ukf stability}, we have $\left\|\frac{\partial\mathbf{q}^{*}_{i,k+1|k}}{\partial\hat{\mathbf{x}}_{k}}\right\|\leq\bar{h}\left\|\frac{\partial\mathbf{q}_{i,k+1|k}}{\partial\hat{\mathbf{x}}_{k}}\right\|$. Further, with Lemma~\ref{lemma: qi jacobian bound}, we have $\left\|\frac{\partial\mathbf{q}^{*}_{i,k+1|k}}{\partial\hat{\mathbf{x}}_{k}}\right\|\leq\bar{h}c'$. Finally, using Lemma~\ref{lemma:vector matrix bounds}\textbf{(c)} ($\frac{\partial[\mathbf{q}^{*}_{i,k+1|k}]_{m}}{\partial\hat{\mathbf{x}}_{k}}$ is the $m$-th row of $\frac{\partial\mathbf{q}^{*}_{i,k+1|k}}{\partial\hat{\mathbf{x}}_{k}}$) and \eqref{eqn:IUKF stable tk term derivative}, we have $\left\|\frac{\partial[\mathbf{t}_{k}]_{m}}{\partial\hat{\mathbf{x}}_{k}}\right\|_{2}\leq\bar{h}c'$.
\end{claimproof}
\begin{claim}
\label{claim:IUKF stable derivative of K bound}
The derivative of $(j,m)$-th element of $\mathbf{K}_{k+1}$ satisfies $\left\|\frac{\partial[\mathbf{K}_{k+1}]_{j,m}}{\partial\hat{\mathbf{x}}_{k}}\right\|_{2}\leq c_{k}$ for some $c_{k}>0$.
\end{claim}
\begin{claimproof}
The $(j,m)$-th element of $\mathbf{K}_{k+1}$ is $[\mathbf{K}_{k+1}]_{j,m}=\sum_{a=1}^{n_{y}}\left[\bm{\Sigma}^{xy}_{k+1}\right]_{j,a}\left[(\bm{\Sigma}^{y}_{k+1})^{-1}\right]_{a,m}$ which implies
\par\noindent\small
\begin{align}
    &\frac{\partial[\mathbf{K}_{k+1}]_{j,m}}{\partial\hat{\mathbf{x}}_{k}}=\sum_{a=1}^{n_{y}}\left[\bm{\Sigma}^{xy}_{k+1}\right]_{j,a}\frac{\partial\left[(\bm{\Sigma}^{y}_{k+1})^{-1}\right]_{a,m}}{\partial\hat{\mathbf{x}}_{k}}+\sum_{a=1}^{n_{y}}\left[(\bm{\Sigma}^{y}_{k+1})^{-1}\right]_{a,m}\frac{\partial\left[\bm{\Sigma}^{xy}_{k+1}\right]_{j,a}}{\partial\hat{\mathbf{x}}_{k}}.\label{eqn:IUKF stable K derivative terms}
\end{align}
\normalsize
Again, we will bound the magnitude of each individual term on R.H.S.

\textbf{Bound on $\left[\bm{\Sigma}^{xy}_{k+1}\right]_{j,a}$ and its derivative:} Under the bounds on functions $f(\cdot)$ and $h(\cdot)$, we have $|[\hat{\mathbf{x}}_{k+1|k}]_{j}|\leq\delta_{f}$ and $|[\sqrt{\bm{\Sigma}_{k+1|k}}]_{j,i}|\leq\sqrt{\bar{\sigma}}$ (the bound on $|b_{j,i}|$ used to prove Claim~\ref{claim:IUKF stable bound on cholesky derivative} of Lemma~\ref{lemma: qi jacobian bound}). It is then straightforward to obtain the bound on the $a$-th element as $|[\hat{\mathbf{y}}_{k+1|k}]_{a}|\leq\delta_{h}$ and the $j$-th element as $|[\mathbf{q}_{i,k+1|k}]_{j}|\leq\delta_{f}+\sqrt{\bar{\sigma}(n_{x}+\kappa)}$ from \eqref{eqn:forward UKF update sigma points}. Since $[\bm{\Sigma}^{xy}_{k+1}]_{j,a}=\sum_{i=0}^{2n_{x}}\omega_{i}[\mathbf{q}_{i,k+1|k}]_{j}[\mathbf{q}^{*}_{i,k+1|k}]_{a}-[\hat{\mathbf{x}}_{k+1|k}]_{j}[\hat{\mathbf{y}}_{k+1|k}]_{a}$, we have $|[\bm{\Sigma}^{xy}_{k+1}]_{j,a}|\leq\delta_{f}\delta_{h}+\delta_{h}(\delta_{f}+\sqrt{\bar{\sigma}(n_{x}+\kappa)})$. The upper-bound $|[\mathbf{q}^{*}_{i,k+1|k}]_{a}|\leq\delta_{h}$ was used in proof of Claim~\ref{claim:IUKF stable tk term bound} as well.

Furthermore, following similar steps as used to bound $\left\|\frac{\partial[\bm{\Sigma}_{k+1|k}]_{l,m}}{\partial\hat{\mathbf{x}}_{k}}\right\|_{2}$ in Claim~\ref{claim:IUKF stable predict sig derivative bound} of Lemma~\ref{lemma: qi jacobian bound}, we have $\left\|\frac{\partial[\bm{\Sigma}^{xy}_{k+1}]_{l,m}}{\partial\hat{\mathbf{x}}_{k}}\right\|_{2}\leq c_{xy}$ with $c_{xy}=\bar{h}c'(\delta_{f}+\sqrt{\bar{\sigma}(n_{x}+\kappa)})+c'(\delta_{f}+\sqrt{\bar{\sigma}(n_{x}+\kappa)})+\delta_{f}\bar{h}c'+\delta_{h}\bar{f}$.

\textbf{Bound on $\left[(\bm{\Sigma}^{y}_{k+1})^{-1}\right]_{a,m}$ and its derivative:} With $\underline{y}\mathbf{I}\preceq\bm{\Sigma}^{y}_{k+1}$ as one of the assumptions of Theorem~\ref{theorem:IUKF stability}, we have $\|(\bm{\Sigma}^{y}_{k+1})^{-1}\|\leq\frac{1}{|\underline{y}|}$. Using Lemma~\ref{lemma:vector matrix bounds}\textbf{(c)} followed by Lemma~\ref{lemma:vector matrix bounds}\textbf{(a)} yields $|[(\bm{\Sigma}^{y}_{k+1})^{-1}]_{a,m}|\leq 1/|\underline{y}|$.

The derivative of $(a,m)$-th element of $(\bm{\Sigma}^{y}_{k+1})^{-1}$ can be expressed in terms of derivative of elements of $\bm{\Sigma}^{y}_{k+1}$ as $\frac{\partial[(\bm{\Sigma}^{y}_{k+1})^{-1}]_{a,m}}{\partial\hat{\mathbf{x}}_{k}}=\sum_{c,d}-[(\bm{\Sigma}^{y}_{k+1})^{-1}]_{a,c}[(\bm{\Sigma}^{y}_{k+1})^{-1}]_{d,m}\frac{\partial[\bm{\Sigma}^{y}_{k+1}]_{c,d}}{\partial\hat{\mathbf{x}}_{k}}$. Using similar steps as used to obtain an upper-bound on $\left\|\frac{\partial[\bm{\Sigma}_{k+1|k}]_{l,m}}{\partial\hat{\mathbf{x}}_{k}}\right\|_{2}$ in Claim~\ref{claim:IUKF stable predict sig derivative bound} of Lemma~\ref{lemma: qi jacobian bound}, we can show that $\left\|\frac{\partial[\bm{\Sigma}^{y}_{k+1}]_{c,d}}{\partial\hat{\mathbf{x}}_{k}}\right\|_{2}\leq 4\delta_{h}\bar{h}c'$ such that the bound on elements of $(\bm{\Sigma}^{y}_{k+1})^{-1}$ yields $\left\|\frac{\partial[(\bm{\Sigma}^{y}_{k+1})^{-1}]_{a,m}}{\partial\hat{\mathbf{x}}_{k}}\right\|_{2}\leq\frac{n_{x}^{2}4\delta_{h}\bar{h}c'}{\underline{y}^{2}}$. With these bounds on magnitudes of all $[\bm{\Sigma}^{xy}_{k+1}]_{j,a}$ and $[(\bm{\Sigma}^{y}_{k+1})^{-1}]_{a,m}$ elements along with the bounds on their derivatives, \eqref{eqn:IUKF stable K derivative terms} gives $\left\|\frac{\partial[\mathbf{K}_{k+1}]_{j,m}}{\partial\hat{\mathbf{x}}_{k}}\right\|_{2}\leq c_{k}$ where $c_{k}=\frac{c_{xy}n_{y}}{|\underline{y}|}+\frac{n_{y}n_{x}^{2}c_{xy}}{\underline{y}^{2}}(\delta_{f}\delta_{h}+\delta_{h}(\delta_{f}+\sqrt{\bar{\sigma}(n_{x}+\kappa)}))$.
\end{claimproof}

Now, under the assumptions of Theorem~\ref{theorem:forward ukf stability}, $\|\mathbf{K}_{k+1}\|\leq\bar{k}=\bar{\sigma}\bar{\gamma}\bar{h}\bar{\beta}/\hat{r}$ (as obtained intermediately in Appendix~\ref{App-thm-Forward ekf stable unknown matrix}) such that Lemma~\ref{lemma:vector matrix bounds}\textbf{(b)} yields $\sum_{m=1}^{n_{y}}|[\mathbf{K}_{k+1}]_{j,m}|\leq\sqrt{n_{y}}\bar{k}$. Finally, using this bound along with Claims~\ref{claim:IUKF stable tk term bound}-\ref{claim:IUKF stable derivative of K bound} in \eqref{eqn:IUKF stable jac K term j-th row}, we have $\left\|\left[\frac{\partial(\mathbf{K}_{k+1}\mathbf{t}_{k})}{\partial\hat{\mathbf{x}}_{k}}\right]_{(j,:)}\right\|_{2}\leq\sqrt{n_{y}}\bar{k}\bar{h}c'+2n_{y}c_{k}\delta_{h}$ such that Claim~\ref{claim:IUKF stable K j row bound} is satisfied with $c_{t}=\sqrt{n_{y}}\bar{k}\bar{h}c'+2n_{y}c_{k}\delta_{h}$.

\chapter{Proof for RKHS-UKF's stability}
\label{chap:rkhs proof}
\section{Proof of Theorem~\ref{theorem:RKHS-UKF stability}}
\label{App-thm-RKHS-UKF}
From $\mathbf{K}_{k+1}=\bm{\Sigma}^{zy}_{k+1}(\bm{\Sigma}^{y}_{k+1})^{-1}$, the sub-matrix $\mathbf{K}^{1}_{k+1}=\bm{\Sigma}^{xy}_{k+1}(\bm{\Sigma}^{y}_{k+1})^{-1}$ such that 
\par\noindent\small
\begin{align*}
    &\mathbf{K}^{1}_{k+1}=\bm{\Sigma}_{k+1|k}\mathbf{U}^{xy}_{k+1}\nabla\bm{\Phi}(\hat{\mathbf{x}}_{k+1|k})^{T}\mathbf{B}^{T}\mathbf{U}^{\phi 2}_{k+1}\left(\mathbf{U}^{\phi 2}_{k+1}\mathbf{B}\nabla\bm{\Phi}(\hat{\mathbf{x}}_{k+1|k})\bm{\Sigma}_{k+1|k}\nabla\bm{\Phi}(\hat{\mathbf{x}}_{k+1|k})^{T}\mathbf{B}^{T}\mathbf{U}^{\phi 2}_{k+1}+\widetilde{\mathbf{R}}_{k+1}\right)^{-1},
\end{align*}
\normalsize
and $\bm{\Sigma}_{k+1}=\bm{\Sigma}_{k+1|k}-\mathbf{K}^{1}_{k+1}\bm{\Sigma}^{y}_{k+1}(\mathbf{K}^{1}_{k+1})^{T}$ yields
\par\noindent\small
\begin{align*}
    &\bm{\Sigma}_{k+1}=\bm{\Sigma}_{k+1|k}-\bm{\Sigma}_{k+1|k}\mathbf{U}^{xy}_{k+1}\nabla\bm{\Phi}(\hat{\mathbf{x}}_{k+1|k})^{T}\mathbf{B}^{T}\mathbf{U}^{\phi 2}_{k+1}\\
    &\;\;\;\times\left(\mathbf{U}^{\phi 2}_{k+1}\mathbf{B}\nabla\bm{\Phi}(\hat{\mathbf{x}}_{k+1|k})\bm{\Sigma}_{k+1|k}\nabla\bm{\Phi}(\hat{\mathbf{x}}_{k+1|k})^{T}\mathbf{B}^{T}\mathbf{U}^{\phi 2}_{k+1}+\widetilde{\mathbf{R}}_{k+1}\right)^{-1}\mathbf{U}^{\phi 2}_{k+1}\mathbf{B}\nabla\bm{\Phi}(\hat{\mathbf{x}}_{k+1|k})(\mathbf{U}^{xy}_{k+1})^{T}\bm{\Sigma}_{k+1|k}.
\end{align*}
\normalsize

Define $V_{k}(\widetilde{\mathbf{x}}_{k|k-1})=\widetilde{\mathbf{x}}_{k|k-1}^{T}\bm{\Sigma}_{k|k-1}^{-1}\widetilde{\mathbf{x}}_{k|k-1}$. Then, using the independence of noise terms and zero mean, we have
\par\noindent\small
\begin{align}
    &\mathbb{E}[\mathbf{V}_{k+1}(\widetilde{\mathbf{x}}_{k+1|k})|\widetilde{\mathbf{x}}_{k|k-1}]=\mathbb{E}[\mathbf{v}_{k}^{T}(\mathbf{K}^{1}_{k})^{T}\nabla\bm{\Phi}(\hat{\mathbf{x}}_{k|k})^{T}\mathbf{A}^{T}\mathbf{U}^{\phi 1}_{k}\bm{\Sigma}_{k+1|k}^{-1}\mathbf{U}^{\phi 1}_{k}\mathbf{A}\nabla\bm{\Phi}(\hat{\mathbf{x}}_{k|k})\mathbf{K}^{1}_{k}\mathbf{v}_{k}|\widetilde{\mathbf{x}}_{k|k-1}]\nonumber\\
    &\hspace{-0.3cm}+\widetilde{\mathbf{x}}_{k|k-1}(\mathbf{I}-\mathbf{K}^{1}_{k}\mathbf{U}^{\phi 2}_{k}\mathbf{B}\nabla\bm{\Phi}(\hat{\mathbf{x}}_{k|k-1}))^{T}\nabla\bm{\Phi}(\hat{\mathbf{x}}_{k|k})^{T}\mathbf{A}^{T}\mathbf{U}^{\phi 1}_{k}\bm{\Sigma}_{k+1|k}^{-1}\mathbf{U}^{\phi 1}_{k}\mathbf{A}\nabla\bm{\Phi}(\hat{\mathbf{x}}_{k|k})(\mathbf{I}-\mathbf{K}^{1}_{k}\mathbf{U}^{\phi 2}_{k}\mathbf{B}\nabla\bm{\Phi}(\hat{\mathbf{x}}_{k|k-1}))\widetilde{\mathbf{x}}_{k|k-1}\nonumber\\
    &+\mathbb{E}[\mathbf{w}_{k}^{T}\bm{\Sigma}^{-1}_{k+1|k}\mathbf{w}_{k}|\widetilde{\mathbf{x}}_{k|k-1}]+r_{k}+s_{k}+q_{k}+u_{k}+b_{k}+\mathbb{E}[\bm{\eta}_{k}^{T}\bm{\Sigma}_{k+1|k}^{-1}\bm{\eta}_{k}|\widetilde{\mathbf{x}}_{k|k-1}]+\mathbb{E}[2\mathbf{w}_{k}^{T}\bm{\Sigma}_{k+1|k}^{-1}\bm{\eta}_{k}|\widetilde{\mathbf{x}}_{k|k-1}]\label{eqn: RKHS-UKF stability Vk expression}
\end{align}
\normalsize
where
\par\noindent\small
\begin{align*}
    r_{k}&=2\widetilde{\mathbf{x}}_{k|k-1}(\mathbf{I}-\mathbf{K}^{1}_{k}\mathbf{U}^{\phi 2}_{k}\mathbf{B}\nabla\bm{\Phi}(\hat{\mathbf{x}}_{k|k-1}))^{T}\nabla\bm{\Phi}(\hat{\mathbf{x}}_{k|k})^{T}\mathbf{A}^{T}\mathbf{U}^{\phi 1}_{k}\bm{\Sigma}_{k+1|k}^{-1}((\mathbf{A}-\hat{\mathbf{A}}_{k})\bm{\Phi}(\hat{\mathbf{x}}_{k|k})+\delta_{f}(\mathbf{x}_{k})\\
    &-\mathbf{U}^{\phi 1}_{k}\mathbf{A}\nabla\bm{\Phi}(\hat{\mathbf{x}}_{k|k})\mathbf{K}^{1}_{k}\delta_{h}(\mathbf{x}_{k}))-\mathbf{U}^{\phi 1}_{k}\mathbf{A}\nabla\bm{\Phi}(\hat{\mathbf{x}}_{k|k})\mathbf{K}^{1}_{k}(\mathbf{B}-\hat{\mathbf{B}}_{k-1})\bm{\Phi}(\hat{\mathbf{x}}_{k|k-1}),\\
    s_{k}&=\bm{\Phi}(\hat{\mathbf{x}}_{k|k})^{T}(\mathbf{A}-\hat{\mathbf{A}}_{k})^{T}\bm{\Sigma}_{k+1|k}^{-1}((\mathbf{A}-\hat{\mathbf{A}}_{k})\bm{\Phi}(\hat{\mathbf{x}}_{k|k})-2\mathbf{U}^{\phi 1}_{k}\mathbf{A}\nabla\bm{\Phi}(\hat{\mathbf{x}}_{k|k})\mathbf{K}^{1}_{k}(\mathbf{B}-\hat{\mathbf{B}}_{k-1})\bm{\Phi}(\hat{\mathbf{x}}_{k|k-1})+2\delta_{f}(\mathbf{x}_{k})\\
    &-2\mathbf{U}^{\phi 1}_{k}\mathbf{A}\nabla\bm{\Phi}(\hat{\mathbf{x}}_{k|k})\mathbf{K}^{1}_{k}\delta_{h}(\mathbf{x}_{k})),\\
    q_{k}&=\bm{\Phi}(\hat{\mathbf{x}}_{k|k-1})^{T}(\mathbf{B}-\hat{\mathbf{B}}_{k-1})^{T}(\mathbf{K}^{1}_{k})^{T}\nabla\bm{\Phi}(\hat{\mathbf{x}}_{k|k})^{T}\mathbf{A}^{T}\mathbf{U}^{\phi 1}_{k}\bm{\Sigma}_{k+1|k}^{-1}(\mathbf{U}^{\phi 1}_{k}\mathbf{A}\nabla\bm{\Phi}(\hat{\mathbf{x}}_{k|k})\mathbf{K}^{1}_{k}(\mathbf{B}-\hat{\mathbf{B}}_{k-1})\bm{\Phi}(\hat{\mathbf{x}}_{k|k-1})\\
    &-2\delta_{f}(\mathbf{x}_{k})+2\mathbf{U}^{\phi 1}_{k}\mathbf{A}\nabla\bm{\Phi}(\hat{\mathbf{x}}_{k|k})\mathbf{K}^{1}_{k}\delta_{h}(\mathbf{x}_{k})),\\
    u_{k}&=\delta_{f}(\mathbf{x}_{k})^{T}\bm{\Sigma}_{k+1|k}^{-1}(\delta_{f}(\mathbf{x}_{k})-2\mathbf{U}^{\phi 1}_{k}\mathbf{A}\nabla\bm{\Phi}(\hat{\mathbf{x}}_{k|k})\mathbf{K}^{1}_{k}\delta_{h}(\mathbf{x}_{k})),\\
    b_{k}&=\delta_{h}(\mathbf{x}_{k})^{T}(\mathbf{K}^{1}_{k})^{T}\nabla\bm{\Phi}(\hat{\mathbf{x}}_{k|k})^{T}\mathbf{A}^{T}\mathbf{U}^{\phi 1}_{k}\bm{\Sigma}_{k+1|k}^{-1}\mathbf{U}^{\phi 1}_{k}\mathbf{A}\nabla\bm{\Phi}(\hat{\mathbf{x}}_{k|k})\mathbf{K}^{1}_{k}\delta_{h}(\mathbf{x}_{k}).
\end{align*}
\normalsize

The proof involves appropriately bounding all the terms in the $\mathbb{E}[\mathbf{V}_{k+1}(\widetilde{\mathbf{x}}_{k+1|k})|\widetilde{\mathbf{x}}_{k|k-1}]$ expression such that both the conditions of Lemma~\ref{lemma:exponential boundedness} are satisfied. Following similar steps as in Appendix~\ref{App-thm-Forward ekf stable unknown matrix}, we can show that under the assumptions of Theorem~\ref{theorem:RKHS-UKF stability}, the following bounds hold.
\par\noindent\small
\begin{align}
    &(\mathbf{U}^{\phi 1}_{k}\mathbf{A}\nabla\bm{\Phi}(\hat{\mathbf{x}}_{k|k})(\mathbf{I}-\mathbf{K}^{1}_{k}\mathbf{U}^{\phi 2}_{k}\mathbf{B}\nabla\bm{\Phi}(\hat{\mathbf{x}}_{k|k-1})))^{T}\bm{\Sigma}_{k+1|k}^{-1}\mathbf{U}^{\phi 1}_{k}\mathbf{A}\nabla\bm{\Phi}(\hat{\mathbf{x}}_{k|k})(\mathbf{I}-\mathbf{K}^{1}_{k}\mathbf{U}^{\phi 2}_{k}\mathbf{B}\nabla\bm{\Phi}(\hat{\mathbf{x}}_{k|k}))\nonumber\\
    &\hspace{3cm}\preceq(1-\lambda)\bm{\Sigma}_{k|k-1}^{-1},\label{eqn:RKHS-UKF lambda bound}\\
    &\mathbb{E}[\mathbf{w}_{k}^{T}\bm{\Sigma}_{k+1|k}^{-1}\mathbf{w}_{k}|\widetilde{\mathbf{x}}_{k|k-1}]\leq c_{1},\label{eqn:RKHS-UKF wk bound}\\
    &\mathbb{E}[\mathbf{v}_{k}^{T}(\mathbf{K}^{1}_{k})^{T}\nabla\bm{\Phi}(\hat{\mathbf{x}}_{k|k})^{T}\mathbf{A}^{T}\mathbf{U}^{\phi 1}_{k}\bm{\Sigma}_{k+1|k}^{-1}\mathbf{U}^{\phi 1}_{k}\mathbf{A}\nabla\bm{\Phi}(\hat{\mathbf{x}}_{k|k})\mathbf{K}^{1}_{k}\mathbf{v}_{k}|\widetilde{\mathbf{x}}_{k|k-1}]\leq c_{2},\label{eqn:RKHS-UKF vk bound}
\end{align}
\normalsize
where $1-\lambda=\left(1+\frac{\widetilde{q}}{\overline{\sigma}(\overline{\alpha}\overline{a}\overline{\phi}(1+\overline{k\beta b\phi})^{2})}\right)^{-1}$ with $\overline{k}=\overline{\sigma\gamma\phi\beta b}/\widetilde{r}$ such that $0<\lambda<1$, $c_{1}=\overline{q}n_{x}/\underline{\sigma}$ and $c_{2}=\overline{k}^{2}\overline{\phi}^{2}\overline{\alpha}^{2}\overline{a}^{2}\overline{r}n_{y}/\underline{\sigma}$.

We will now bound the remaining terms in the right-hand side of \eqref{eqn: RKHS-UKF stability Vk expression}.
\begin{claim}\label{claim:RKHS-UKF Vk terms}
    There exist constants $c_{3}$, $c_{4}$, $c_{5}$, $c_{6}$ and $c_{7}$ satisfying $r_{k}\leq c_{3}$, $s_{k}\leq c_{4}$, $q_{k}\leq c_{5}$, $u_{k}\leq c_{6}$ and $b_{k}\leq c_{7}$ for all $k\geq 0$.
\end{claim}
\begin{claimproof}
    First, we upper-bound $r_{k}$. The Gaussian kernel function $K(\cdot,\cdot)$ has maximum value $1$ such that $\|\bm{\Phi}(\cdot)\|_{2}\leq\sqrt{L}$. Also, due to the projection of the coefficient matrix estimates, $\|\hat{\mathbf{A}}_{k}\|\leq\overline{a}$ and $\|\hat{\mathbf{B}}_{k}\|\leq\overline{b}$ for all $k\geq 0$. Then, using these bounds along with the Assumption~\textbf{A6.d} in Theorem~\ref{theorem:RKHS-UKF stability}, we have $ \|(\mathbf{A}-\hat{\mathbf{A}}_{k})\bm{\Phi}(\hat{\mathbf{x}}_{k|k})\|_{2}\leq 2\overline{a}\sqrt{L}$ and $\|(\mathbf{B}-\hat{\mathbf{B}}_{k-1})\bm{\Phi}(\hat{\mathbf{x}}_{k|k-1})\|_{2}\leq\overline{b}\sqrt{L}$.

    Next, we need an upper-bound on $\|\widetilde{\mathbf{x}}_{k|k-1}\|_{2}$. Since true state $\mathbf{x}_{k}$ lies in $\mathcal{X}$, we have $\|\mathbf{x}_{k}\|_{2}\leq\epsilon$ for some $\epsilon>0$. Also, $\hat{\mathbf{x}}_{k|k-1}=\sum_{i=0}^{2n_{z}}\omega_{i}\hat{\mathbf{A}}_{k-1}\bm{\Phi}([\mathbf{s}_{i,k-1}]_{1:n_{x}})$ as used to obtain \eqref{eqn:RKHS-UKF state prediction error} as well. Since $\hat{\mathbf{A}}_{k-1}$ and $\bm{\Phi}(\cdot)$ are bounded, we have $\|\widetilde{\mathbf{x}}_{k|k-1}\|_{2}\leq\|\mathbf{x}_{k}\|_{2}+\sum_{i=0}^{2n_{z}}\omega_{i}\|\hat{\mathbf{A}}_{k-1}\bm{\Phi}([\mathbf{s}_{i,k-1}]_{1:n_{x}})\|_{2}\leq\epsilon+\overline{a}\sqrt{L}$. With these bounds and other bounds assumed on various matrices, we obtain the required bound on $r_{k}$ with $c_{3}=2(\epsilon+\overline{a}\sqrt{L})(1+\overline{k\beta b\phi})\overline{\phi a\alpha}(2\overline{a}\sqrt{L}+\overline{\alpha a\phi k}(2\overline{b}\sqrt{L})+\overline{f}+\overline{\alpha a\phi kh})/\underline{\sigma}$.

    Following similar steps, we can obtain the other bounds of the claim with $c_{4}=4\overline{a}\sqrt{L}(\overline{a}\sqrt{L}+2\overline{\alpha a\phi kb}\sqrt{L}+\overline{f}+\overline{\alpha a\phi kh})/\underline{\sigma}$, $c_{5}=4\overline{b}\overline{k\phi a\alpha}\sqrt{L}(\overline{\alpha a\phi k}\overline{b}\sqrt{L}+\overline{f}+\overline{\alpha a\phi kh})/\underline{\sigma}$, $c_{6}=\overline{f}(\overline{f}+2\overline{\alpha a\phi kh})/\underline{\sigma}$ and $c_{7}=\overline{\alpha}^{2}\overline{a}^{2}\overline{\phi}^{2}\overline{k}^{2}\overline{h}^{2}/\underline{\sigma}$.
\end{claimproof}
\begin{claim}\label{claim:RKHS-UKF exp terms}
For the projection error $\bm{\eta}_{k}$, we have $\mathbb{E}[\bm{\eta}_{k}^{T}\bm{\Sigma}_{k+1|k}^{-1}\bm{\eta}_{k}|\widetilde{\mathbf{x}}_{k|k-1}]\leq c_{1}$ and\\ $\mathbb{E}[2\mathbf{w}_{k}^{T}\bm{\Sigma}_{k+1|k}^{-1}\bm{\eta}_{k}|\widetilde{\mathbf{x}}_{k|k-1}]\leq 2c_{1}$ where $c_{1}$ is as defined in \eqref{eqn:RKHS-UKF wk bound}.
\end{claim}
\begin{claimproof}
Using the bound on $\bm{\Sigma}_{k+1|k}$, we have $\bm{\eta}_{k}^{T}\bm{\Sigma}_{k+1|k}^{-1}\bm{\eta}_{k}\leq\frac{1}{\underline{\sigma}}\bm{\eta}_{k}^{T}\bm{\eta}_{k}$. Now, by the definition of projection $\Gamma(\cdot)$, we have $\mathbf{x}_{k+1}=\textrm{argmin}_{\mathbf{x}\in\mathcal{X}}{\|f(\mathbf{x}_{k})+\mathbf{w}_{k}-\mathbf{x}\|_{2}}$ which implies $ \|f(\mathbf{x}_{k})+\mathbf{w}_{k}-\mathbf{x}_{k+1}\|_{2}\leq\|f(\mathbf{x}_{k})+\mathbf{w}_{k}-\mathbf{x}\|_{2}$ for all $\mathbf{x}\in\mathcal{X}$. In particular, $f(\mathbf{x}_{k})\in\mathcal{X}$ and the projection error $\bm{\eta}_{k}=\mathbf{x}_{k+1}-(f(\mathbf{x}_{k})+\mathbf{w}_{k})$ such that $\|-\bm{\eta}_{k}\|_{2}\leq\|\mathbf{w}_{k}\|_{2}$ which implies $\bm{\eta}_{k}^{T}\bm{\eta}_{k}\leq\mathbf{w}_{k}^{T}\mathbf{w}_{k}$. Hence, $\bm{\eta}_{k}^{T}\bm{\Sigma}_{k+1|k}^{-1}\bm{\eta}_{k}\leq\frac{1}{\underline{\sigma}}\mathbf{w}_{k}^{T}\mathbf{w}_{k}$. Taking expectation and using the upper bound on noise covariance $\mathbf{Q}$ similar to the proof of \eqref{eqn:RKHS-UKF wk bound} (in Appendix~\ref{App-thm-Forward ekf stable unknown matrix}), we get the first bound of the claim.

Now, $\mathbb{E}[\mathbf{w}_{k}^{T}\bm{\Sigma}_{k+1|k}^{-1}\bm{\eta}_{k}|\widetilde{\mathbf{x}}_{k|k-1}]\leq\frac{1}{\underline{\sigma}}\mathbb{E}[\mathbf{w}_{k}^{T}\bm{\eta}_{k}|\widetilde{\mathbf{x}}_{k|k-1}]$ which on using Cauchy-Schwartz inequality yields $\mathbb{E}[\mathbf{w}_{k}^{T}\bm{\Sigma}_{k+1|k}^{-1}\bm{\eta}_{k}|\widetilde{\mathbf{x}}_{k|k-1}]\leq\frac{1}{\underline{\sigma}}\sqrt{\mathbb{E}[\mathbf{w}_{k}^{T}\mathbf{w}_{k}]}\sqrt{\mathbb{E}[\bm{\eta}_{k}^{T}\bm{\eta}_{k}|\widetilde{\mathbf{x}}_{k|k-1}]}$. Again, using $\bm{\eta}_{k}^{T}\bm{\eta}_{k}\leq\mathbf{w}_{k}^{T}\mathbf{w}_{k}$, we have $\mathbb{E}[\mathbf{w}_{k}^{T}\bm{\Sigma}_{k+1|k}^{-1}\bm{\eta}_{k}|\widetilde{\mathbf{x}}_{k|k-1}]\leq\frac{1}{\underline{\sigma}}\mathbb{E}[\mathbf{w}_{k}^{T}\mathbf{w}_{k}]$ such that the second bound of the claim can be obtained using $\mathbf{Q}\preceq\overline{q}\mathbf{I}$.
\end{claimproof}

Using the bounds \eqref{eqn:RKHS-UKF lambda bound}-\eqref{eqn:RKHS-UKF vk bound} and Claims~\ref{claim:RKHS-UKF Vk terms} and \ref{claim:RKHS-UKF exp terms}, Theorem~\ref{theorem:RKHS-UKF stability} can be proved using Lemma~\ref{lemma:exponential boundedness} similar to Appendix~\ref{App-thm-Forward ekf stable unknown matrix}.

\chapter{Proofs for I-PF's formulation and convergence}
\label{chap:ipf proof}
\section{Proof of eqs. (7.2) and (7.3)}\label{app:sampling}
The densities \eqref{eqn:ipf p density} and \eqref{eqn:ipf q density} follow from the conditional independence of observations and the estimation process assumed in the system model \eqref{eqn:state x}-\eqref{eqn:observation a} using the standard PF's SIS-based non-linear filtering technique \cite[Chapter~3.2]{ristic2003beyond}. Recall from Section~\ref{subsec:IPF formulation} that we consider the joint density $p(\hat{\mathbf{x}}_{0:k},\mathbf{y}_{1:k}|\mathbf{x}_{0:k},\mathbf{a}_{1:k})=p(\hat{\mathbf{x}}_{0:k},\mathbf{y}_{1:k}|\mathbf{a}_{k},\mathbf{x}_{0:k},\mathbf{a}_{1:k-1})$ to obtain I-PF's optimal sampling density and weights. Using Bayes' theorem, we have
\par\noindent\small
\begin{align}
    p(\hat{\mathbf{x}}_{0:k},\mathbf{y}_{1:k}|\mathbf{x}_{0:k},\mathbf{a}_{1:k})=\frac{p(\mathbf{a}_{k}|\hat{\mathbf{x}}_{0:k},\mathbf{y}_{1:k},\mathbf{x}_{0:k},\mathbf{a}_{1:k-1})p(\hat{\mathbf{x}}_{0:k},\mathbf{y}_{1:k}|\mathbf{x}_{0:k},\mathbf{a}_{1:k-1})}{p(\mathbf{a}_{k}|\mathbf{x}_{0:k},\mathbf{a}_{1:k-1})}.\label{eqn:ipf joint density frac}
\end{align}
\normalsize
First, consider $p(\mathbf{a}_{k}|\hat{\mathbf{x}}_{0:k},\mathbf{y}_{1:k},\mathbf{x}_{0:k},\mathbf{a}_{1:k-1})=p(\mathbf{a}_{k}|\hat{\mathbf{x}}_{k},\hat{\mathbf{x}}_{0:k-1},\mathbf{y}_{1:k},\mathbf{x}_{0:k},\mathbf{a}_{1:k-1})$. From \eqref{eqn:observation a}, the defender's observation $\mathbf{a}_{k}$ depends only on the state estimate $\hat{\mathbf{x}}_{k}$ such that $p(\mathbf{a}_{k}|\hat{\mathbf{x}}_{0:k},\mathbf{y}_{1:k},\mathbf{x}_{0:k},\mathbf{a}_{1:k-1})\\=p(\mathbf{a}_{k}|\hat{\mathbf{x}}_{k})$.

Next, we simplify $p(\hat{\mathbf{x}}_{0:k},\mathbf{y}_{1:k}|\mathbf{x}_{0:k},\mathbf{a}_{1:k-1})$ in the numerator of \eqref{eqn:ipf joint density frac}. To this end, we have $p(\hat{\mathbf{x}}_{k},\hat{\mathbf{x}}_{0:k-1},\mathbf{y}_{1:k}|\mathbf{x}_{0:k},\mathbf{a}_{1:k-1})=p(\hat{\mathbf{x}}_{k}|\hat{\mathbf{x}}_{0:k-1},\mathbf{y}_{1:k},\mathbf{x}_{0:k},\mathbf{a}_{1:k-1})p(\hat{\mathbf{x}}_{0:k-1},\mathbf{y}_{1:k}|\mathbf{x}_{0:k},\mathbf{a}_{1:k-1})$. But, from \eqref{eqn:filter T}, given $T(\cdot)$, state estimate $\hat{\mathbf{x}}_{k}$ is a function of previous estimate $\hat{\mathbf{x}}_{k-1}$ and observation $\mathbf{y}_{k}$ such that $p(\hat{\mathbf{x}}_{k}|\hat{\mathbf{x}}_{0:k-1},\mathbf{y}_{1:k},\mathbf{x}_{0:k},\mathbf{a}_{1:k-1})$ simplifies to $p(\hat{\mathbf{x}}_{k}|\hat{\mathbf{x}}_{k-1},\mathbf{y}_{k})$. Also, $p(\mathbf{y}_{k},\hat{\mathbf{x}}_{0:k-1},\mathbf{y}_{1:k-1}|\mathbf{x}_{0:k},\mathbf{a}_{1:k-1})\\=p(\mathbf{y}_{k}|\hat{\mathbf{x}}_{0:k-1},\mathbf{y}_{1:k-1},\mathbf{x}_{0:k},\mathbf{a}_{1:k-1})p(\hat{\mathbf{x}}_{0:k-1},\mathbf{y}_{1:k-1}|\mathbf{x}_{0:k},\mathbf{a}_{1:k-1})$. From \eqref{eqn:observation y}, observation $\mathbf{y}_{k}$ depends only on the true state $\mathbf{x}_{k}$ such that \small$p(\mathbf{y}_{k}|\hat{\mathbf{x}}_{0:k-1},\mathbf{y}_{1:k-1},\mathbf{x}_{0:k},\mathbf{a}_{1:k-1})=p(\mathbf{y}_{k}|\mathbf{x}_{k}).\;$\normalsize
Lastly, the attacker's estimation process and the observations upto $(k-1)$-th time instant are independent of future state $\mathbf{x}_{k}$ given the past states, i.e., $p(\hat{\mathbf{x}}_{0:k-1},\mathbf{y}_{1:k-1}|\mathbf{x}_{0:k},\mathbf{a}_{1:k-1})=p(\hat{\mathbf{x}}_{0:k-1},\mathbf{y}_{1:k-1}|\mathbf{x}_{0:k-1},\mathbf{a}_{1:k-1})$. Now, substituting $p(\mathbf{a}_{k}|\hat{\mathbf{x}}_{0:k},\mathbf{y}_{1:k},\mathbf{x}_{0:k},\mathbf{a}_{1:k-1})=p(\mathbf{a}_{k}|\hat{\mathbf{x}}_{k})$ and\\ $p(\hat{\mathbf{x}}_{0:k},\mathbf{y}_{1:k}|\mathbf{x}_{0:k},\mathbf{a}_{1:k-1})=p(\hat{\mathbf{x}}_{k}|\hat{\mathbf{x}}_{k-1},\mathbf{y}_{k})p(\mathbf{y}_{k}|\mathbf{x}_{k})p(\hat{\mathbf{x}}_{0:k-1},\mathbf{y}_{1:k-1}|\mathbf{x}_{0:k-1},\mathbf{a}_{1:k-1})$ in \eqref{eqn:ipf joint density frac} yields
\par\noindent\small
\begin{align*}
    p(\hat{\mathbf{x}}_{0:k},\mathbf{y}_{1:k}|\mathbf{x}_{0:k},\mathbf{a}_{1:k})\propto p(\mathbf{a}_{k}|\hat{\mathbf{x}}_{k})p(\hat{\mathbf{x}}_{k}|\hat{\mathbf{x}}_{k-1},\mathbf{y}_{k})p(\mathbf{y}_{k}|\mathbf{x}_{k})p(\hat{\mathbf{x}}_{0:k-1},\mathbf{y}_{1:k-1}|\mathbf{x}_{0:k-1},\mathbf{a}_{1:k-1}),
\end{align*}
\normalsize
which is \eqref{eqn:ipf p density} provided in Section~\ref{subsec:IPF formulation}.

Now, we consider the sampling density \eqref{eqn:ipf q density} of our I-PF. The standard PF estimating states $\mathbf{x}_{0:k}$ given observations $\mathbf{y}_{1:k}$ uses a sampling density $\widetilde{q}(\mathbf{x}_{0:k}|\mathbf{y}_{1:k})$ that factorizes as $\widetilde{q}(\mathbf{x}_{0:k}|\mathbf{y}_{1:k})\doteq \widetilde{q}(\mathbf{x}_{k}|\mathbf{x}_{0:k-1},\mathbf{y}_{1:k})\widetilde{q}(\mathbf{x}_{0:k-1}|\mathbf{y}_{1:k-1})$ \cite[eq.~3.12]{ristic2003beyond}. In the case of I-PF, we are estimating the joint density of $(\hat{\mathbf{x}}_{0:k},\mathbf{y}_{1:k})$ given true states $\mathbf{x}_{0:k}$ and observations $\mathbf{a}_{1:k}$. Hence, we choose a sampling density $q(\cdot)$ such that
\par\noindent\small
\begin{align*}
   q(\hat{\mathbf{x}}_{0:k},\mathbf{y}_{1:k}|\mathbf{x}_{0:k},\mathbf{a}_{1:k})\doteq q(\hat{\mathbf{x}}_{k},\mathbf{y}_{k}|\hat{\mathbf{x}}_{0:k-1},\mathbf{y}_{1:k-1},\mathbf{x}_{0:k},\mathbf{a}_{1:k})q(\hat{\mathbf{x}}_{0:k-1},\mathbf{y}_{1:k-1}|\mathbf{x}_{0:k-1},\mathbf{a}_{1:k-1}).
\end{align*}
\normalsize
Furthermore, when the standard PF requires only the filtered posterior $p(\mathbf{x}_{k}|\mathbf{y}_{1:k})$ at each time step, the sampling density $\widetilde{q}(\mathbf{x}_{k}|\mathbf{x}_{0:k-1},\mathbf{y}_{1:k})$ is commonly assumed to depend only on $\mathbf{x}_{k-1}$ and $\mathbf{y}_{k}$ \cite[Sec.~3.2]{ristic2003beyond}, i.e., $\widetilde{q}(\mathbf{x}_{k}|\mathbf{x}_{0:k-1},\mathbf{y}_{1:k})=\widetilde{q}(\mathbf{x}_{k}|\mathbf{x}_{k-1},\mathbf{y}_{k})$. Similarly, in our I-PF, we assume $q(\hat{\mathbf{x}}_{k},\mathbf{y}_{k}|\hat{\mathbf{x}}_{0:k-1},\mathbf{y}_{1:k-1},\mathbf{x}_{0:k},\mathbf{a}_{1:k})=q(\hat{\mathbf{x}}_{k},\mathbf{y}_{k}|\hat{\mathbf{x}}_{k-1},\mathbf{y}_{k-1},\mathbf{x}_{k},\mathbf{a}_{k})$. Overall, we choose I-PF's sampling density such that $q(\hat{\mathbf{x}}_{0:k},\mathbf{y}_{1:k}|\mathbf{x}_{0:k},\mathbf{a}_{1:k})=q(\hat{\mathbf{x}}_{k},\mathbf{y}_{k}|\hat{\mathbf{x}}_{k-1},\mathbf{y}_{k-1},\mathbf{x}_{k},\mathbf{a}_{k})\\\times q(\hat{\mathbf{x}}_{0:k-1},\mathbf{y}_{1:k-1}|\mathbf{x}_{0:k-1},\mathbf{a}_{1:k-1})$, which is \eqref{eqn:ipf q density} of Section~\ref{subsec:IPF formulation}.

\section{Proof of Theorem~\ref{thm:ipf convergence}}
\label{app:ipf convergence}
In the following, we first restate some useful Lemma~\ref{lemma:sum to power four inequality}-\ref{lemma:indicator inequality} from \cite{hu2008basic}. In Section~\ref{app:preliminaries}, some preliminary results are derived, from which the proof of Theorem~\ref{thm:ipf convergence} follows using a standard mathematical induction approach.
\begin{lemma}
\label{lemma:sum to power four inequality}
Let $\{\xi_{i}\}_{1\leq i\leq n}$ be conditionally independent random variables given $\sigma$-algebra $\mathcal{G}$ such that $\mathbb{E}[\xi_{i}|\mathcal{G}]=0$ and $\mathbb{E}[|\xi_{i}|^{4}|\mathcal{G}]<\infty$. Then
\par\noindent\small
\begin{align*}
   \mathbb{E}\left[\left|\sum_{i=1}^{n}\xi_{i}\right|^{4}|\mathcal{G}\right]\leq\sum_{i=1}^{n}\mathbb{E}[|\xi_{i}|^{4}|\mathcal{G}]+\left(\sum_{i=1}^{n}\mathbb{E}[|\xi_{i}|^{2}|\mathcal{G}]\right)^{2}.
\end{align*}
\normalsize
\end{lemma}
\begin{lemma}
\label{lemma:diff to power p inequality}
    If $\mathbb{E}|\xi|^{p}<\infty$, then $\mathbb{E}|\xi-\mathbb{E}[\xi]|^{p}\leq 2^{p}\mathbb{E}|\xi|^{p}$, for any $p\geq 1$.
\end{lemma}
\begin{lemma}
\label{lemma:sum to max inequality}
    Let $\{\xi_{i}\}_{1\leq i\leq n}$ be conditionally independent random variables given $\sigma$-algebra $\mathcal{G}$ such that $\mathbb{E}[\xi_{i}|\mathcal{G}]=0$ and $\mathbb{E}[|\xi_{i}|^{4}|\mathcal{G}]<\infty$. Then
    \par\noindent\small
    \begin{align*}
   \mathbb{E}\left[\left|\frac{1}{n}\sum_{i=1}^{n}\xi_{i}\right|^{4}|\mathcal{G}\right]\leq \frac{2\;\textrm{max}_{1\leq i\leq n}\mathbb{E}[|\xi_{i}|^{4}|\mathcal{G}]}{n^{2}}.
    \end{align*}
    \normalsize
\end{lemma}
\begin{lemma}
\label{lemma:indicator inequality}
    Denote $A^{c}$ as the complementary set of a given set $A$. Also, $I_{A}$ denotes the indicator function for a set $A$. Consider a random variable $\eta$ with probability density function $p(x)$ such that $\mathbb{P}(\eta\in A^{c})\leq\epsilon<1$. Define a random variable $\xi$ with probability density function as $\frac{p(x)I_{A}}{\mathbb{P}(A)}$ where $\mathbb{P}(A)=\int p(y)I_{A}dy$. If $\psi$ be a measurable function satisfying $\mathbb{E}[\psi^{2}(\eta)]<\infty$, then
    \par\noindent\small
    \begin{align*}
   |\mathbb{E}[\psi(\xi)]-\mathbb{E}[\psi(\eta)]|\leq\frac{2\sqrt{\mathbb{E}[\psi^{2}(\eta)]}}{1-\epsilon}\sqrt{\epsilon}.
    \end{align*}
    \normalsize
    In the case when $\mathbb{E}|\psi(\eta)|<\infty$, we have $\mathbb{E}|\psi(\xi)|\leq\frac{\mathbb{E}|\psi(\eta)|}{1-\epsilon}$.
\end{lemma}

\subsection{Preliminaries}\label{app:preliminaries}
 Denote $\mathcal{F}_{k-1}=\sigma\{(\hat{\mathbf{x}}^{i}_{k-1},\mathbf{y}^{i}_{k-1})\;\textrm{for}\;1\leq i\leq N\}$ as the $\sigma$-algebra generated by particles at the previous $(k-1)$-th time step. For Lemma~\ref{lemma:infinite loop}-\ref{lemma:Sampling pow 4 terms}, we assume \eqref{eqn:ipf converge} holds for the previous time instant $(k-1)$. In \eqref{eqn:initialize diff} of Appendix~\ref{app:proof}, it is shown that \eqref{eqn:ipf converge} also holds for $k=0$. Further, we assume
 \par\noindent\small
 \begin{align}
   \mathbb{E}|\langle\pi_{k-1|k-1}^{N},|\phi|^{4}\rangle|\leq M_{k-1|k-1}\|\phi\|^{4}_{k-1,4}.\label{eqn:k-1 pow 4 one}  
 \end{align}
 \normalsize
where $M_{k-1|k-1}>0$ is a constant independent of the number of particles $N$. This inequality is necessary for the proof of the theorem and also proved to hold for all $k\geq 0$ in Section~\ref{app:proof}.
 
\begin{lemma}\label{lemma:infinite loop}
Consider the particles $\{(\hat{\overline{\mathbf{x}}}^{i}_{k},\overline{\mathbf{y}}^{i}_{k})\}$ drawn in the I-PF's importance sampling step. Assume that \eqref{eqn:ipf converge} holds for $(k-1)$-th time instant. Then, for sufficiently large $N$, we have
\par\noindent\small
\begin{align}
    \mathbb{P}\left(\frac{1}{N}\sum_{i=1}^{N}\beta(\mathbf{a}_{k}|\hat{\overline{\mathbf{x}}}^{i}_{k})<\gamma_{k}|\mathcal{F}_{k-1}\right)<\epsilon_{k}\label{eqn:epsilon}
\end{align}
\normalsize
for some $0<\epsilon_{k}<1$, which implies that the I-PF algorithm will not run into an infinite loop in steps $1$ and $2$.
\end{lemma}
\begin{proof}
Note that $(\hat{\overline{\mathbf{x}}}^{i}_{k},\overline{\mathbf{y}}^{i}_{k})$ are drawn from the distribution $\langle\pi^{N}_{k-1|k-1},\delta_{T}\rho\rangle$ in the importance sampling step such that
\par\noindent\small
\begin{align}
    \mathbb{E}[\phi(\hat{\overline{\mathbf{x}}}^{i}_{k},\overline{\mathbf{y}}^{i}_{k})|\mathcal{F}_{k-1}]=\langle\pi^{N}_{k-1|k-1},\delta_{T}\rho\phi\rangle,\label{eqn:expect of predicted particles}
\end{align}
\normalsize
Further, since $(\hat{\overline{\mathbf{x}}}^{i}_{k},\overline{\mathbf{y}}^{i}_{k})$ have equal weights (because of resampling at each step), we have\\ $\frac{1}{N}\sum_{i=1}^{N}\beta(\mathbf{a}_{k}|\hat{\overline{\mathbf{x}}}^{i}_{k})=\langle\pi^{N}_{k-1|k-1},\delta_{T}\rho\beta\rangle$. Because of the modification, the distribution of $\{(\hat{\widetilde{\mathbf{x}}}^{i}_{k},\widetilde{\mathbf{y}}^{i}_{k})\}$ is obtained from the distribution of $\{(\hat{\overline{\mathbf{x}}}^{i}_{k},\overline{\mathbf{y}}^{i}_{k})\}$ conditioned on the event $\Gamma_{k}\doteq\{\langle\pi^{N}_{k-1|k-1},\delta_{T}\rho\beta\rangle\geq\gamma_{k}\}$. In order to use Lemma~\ref{lemma:indicator inequality} to handle the difference in the empirical distributions of $\{(\hat{\overline{\mathbf{x}}}^{i}_{k},\overline{\mathbf{y}}^{i}_{k})\}$ and $\{(\hat{\widetilde{\mathbf{x}}}^{i}_{k},\widetilde{\mathbf{y}}^{i}_{k})\}$, we require the probability of $\Gamma_{k}^{c}=\{\langle\pi^{N}_{k-1|k-1},\delta_{T}\rho\beta\rangle<\gamma_{k}\}$. Consider
\par\noindent\small
\begin{align*}
    \mathbb{P}(\langle\pi^{N}_{k-1|k-1},\delta_{T}\rho\beta\rangle<\gamma_{k})&=\mathbb{P}(\langle\pi^{N}_{k-1|k-1},\delta_{T}\rho\beta\rangle-\langle\pi_{k-1|k-1},\delta_{T}\rho\beta\rangle<\gamma_{k}-\langle\pi_{k-1|k-1},\delta_{T}\rho\beta\rangle)\\
    &\leq \mathbb{P}(|\langle\pi^{N}_{k-1|k-1},\delta_{T}\rho\beta\rangle-\langle\pi_{k-1|k-1},\delta_{T}\rho\beta\rangle|>|\gamma_{k}-\langle\pi_{k-1|k-1},\delta_{T}\rho\beta\rangle|),
\end{align*}
\normalsize
because $\gamma_{k}-\langle\pi_{k-1|k-1},\delta_{T}\rho\beta\rangle<0$ from assumption \textbf{A7.a}. Using Markov's inequality and then, \eqref{eqn:ipf converge} replacing $k$ by $k-1$ with the indicator function being bounded by $1$ yields
\par\noindent\small
\begin{align*}
    \mathbb{P}(\langle\pi^{N}_{k-1|k-1},\delta_{T}\rho\beta\rangle<\gamma_{k})&\leq \frac{\mathbb{E}|\langle\pi^{N}_{k-1|k-1},\delta_{T}\rho\beta\rangle-\langle\pi_{k-1|k-1},\delta_{T}\rho\beta\rangle|^{4}}{|\gamma_{k}-\langle\pi_{k-1|k-1},\delta_{T}\rho\beta\rangle|^{4}}\\
    &\leq\frac{C_{k-1|k-1}\|\rho\|_{\infty}^{4}}{|\gamma_{k}-\langle\pi_{k-1|k-1},\delta_{T}\rho\beta\rangle|^{4}}\times\frac{\|\beta\|^{4}_{k-1,4}}{N^{2}}.
\end{align*}
Hence, \eqref{eqn:epsilon} will hold for some $0<\epsilon_{k}<1$ for sufficiently large $N$. In the following, we denote $C_{\gamma_{k}}=\frac{C_{k-1|k-1}\|\rho\|_{\infty}^{4}}{|\gamma_{k}-\langle\pi_{k-1|k-1},\delta_{T}\rho\beta\rangle|^{4}}$.
\end{proof}
\begin{lemma}\label{lemma:Pi123 bounds}
Consider the optimal filter's prediction distribution $\pi_{k|k-1}$ and its approximation $\widetilde{\pi}^{N}_{k|k-1}$ obtained in I-PF. Define
\par\noindent\small
\begin{align}
    &\Pi_{1}\doteq\langle\widetilde{\pi}^{N}_{k|k-1},\phi\rangle-\frac{1}{N}\sum_{i=1}^{N}\mathbb{E}[\phi(\hat{\widetilde{\mathbf{x}}}^{i}_{k},\widetilde{\mathbf{y}}^{i}_{k})|\mathcal{F}_{k-1}],\label{eqn:def Pi1}\\
    &\Pi_{2}\doteq\frac{1}{N}\sum_{i=1}^{N}\mathbb{E}[\phi(\hat{\widetilde{\mathbf{x}}}^{i}_{k},\widetilde{\mathbf{y}}^{i}_{k})|\mathcal{F}_{k-1}]-\langle\pi^{N}_{k-1|k-1},\delta_{T}\rho\phi\rangle,\label{eqn:def Pi2}\\
    &\Pi_{3}\doteq\langle\pi^{N}_{k-1|k-1},\delta_{T}\rho\phi\rangle-\langle\pi_{k|k-1},\phi\rangle.\label{eqn:def Pi3}
\end{align}
\normalsize
Then, $\Pi_{1}$, $\Pi_{2}$ and $\Pi_{3}$ satisfy
\par\noindent\small
\begin{align}
    &\mathbb{E}[|\Pi_{1}|^{4}]\leq C_{\Pi_{1}}\frac{\|\phi\|^{4}_{k-1,4}}{N^{2}},\label{eqn:Pi1 term}\\
    &\mathbb{E}[|\Pi_{2}|^{4}]\leq C_{\Pi_{2}}\frac{\|\phi\|^{4}_{k-1,4}}{N^{2}},\label{eqn:Pi2 term}\\
    &\mathbb{E}[|\Pi_{3}|^{4}]\leq C_{\Pi_{3}}\frac{\|\phi\|^{4}_{k-1,4}}{N^{2}},\label{eqn:Pi3 term}
\end{align}
\normalsize
for suitable constants $C_{\Pi_{1}},C_{\Pi_{2}}$ and $C_{\Pi_{3}}>0$. Note that $\Pi_{1}$, $\Pi_{2}$ and $\Pi_{3}$ are defined for given $\pi_{k|k-1}$ and $\widetilde{\pi}^{N}_{k|k-1}$ distributions while constants $C_{\Pi_{1}},C_{\Pi_{2}}$ and $C_{\Pi_{3}}$ do not dependent on $N$.
\end{lemma}
\begin{proof}
\textbf{(a) $\Pi_{1}$ term:} Recall that $\widetilde{\pi}^{N}_{k|k-1}$ is the empirical distribution obtained from particles $\{(\hat{\widetilde{\mathbf{x}}}^{i}_{k},\widetilde{\mathbf{y}}^{i}_{k})\}$. Hence,
\par\noindent\small
\begin{align*}
    &\mathbb{E}[|\Pi_{1}|^{4}|\mathcal{F}_{k-1}]=\mathbb{E}\left[\left|\frac{1}{N}\sum_{i=1}^{N}(\phi(\hat{\widetilde{\mathbf{x}}}^{i}_{k},\widetilde{\mathbf{y}}^{i}_{k})-\mathbb{E}[\phi(\hat{\widetilde{\mathbf{x}}}^{i}_{k},\widetilde{\mathbf{y}}^{i}_{k})|\mathcal{F}_{k-1}])\right|^{4}|\mathcal{F}_{k-1}\right]\\
    &\leq\frac{1}{N^{4}}\sum_{i=1}^{N}\mathbb{E}[|\phi(\hat{\widetilde{\mathbf{x}}}^{i}_{k},\widetilde{\mathbf{y}}^{i}_{k})-\mathbb{E}[\phi(\hat{\widetilde{\mathbf{x}}}^{i}_{k},\widetilde{\mathbf{y}}^{i}_{k})|\mathcal{F}_{k-1}]|^{4}|\mathcal{F}_{k-1}]+\frac{1}{N^{4}}\left(\sum_{i=1}^{N}\mathbb{E}[|\phi(\hat{\widetilde{\mathbf{x}}}^{i}_{k},\widetilde{\mathbf{y}}^{i}_{k})-\mathbb{E}[\phi(\hat{\widetilde{\mathbf{x}}}^{i}_{k},\widetilde{\mathbf{y}}^{i}_{k})|\mathcal{F}_{k-1}]|^{2}|\mathcal{F}_{k-1}]\right)^{2}\\
    &\leq\frac{1}{N^{4}}\left(\sum_{i=1}^{N}2^{4}\mathbb{E}[|\phi(\hat{\widetilde{\mathbf{x}}}^{i}_{k},\widetilde{\mathbf{y}}^{i}_{k})|^{4}|\mathcal{F}_{k-1}]+\left(\sum_{i=1}^{N}2^{2}\mathbb{E}[|\phi(\hat{\widetilde{\mathbf{x}}}^{i}_{k},\widetilde{\mathbf{y}}^{i}_{k})|^{2}|\mathcal{F}_{k-1}]\right)^{2}\right),
\end{align*}
\normalsize
where the first and second inequalities are obtained using Lemma~\ref{lemma:sum to power four inequality} and \ref{lemma:diff to power p inequality}, respectively. Since $(\hat{\overline{\mathbf{x}}}^{i}_{k},\overline{\mathbf{y}}^{i}_{k})$ are obtained from $(\hat{\widetilde{\mathbf{x}}}^{i}_{k},\widetilde{\mathbf{y}}^{i}_{k})$ such that $\Gamma_{k}$ (defined in Lemma~\ref{lemma:infinite loop}) occurs and $\mathbb{P}(\Gamma_{k}^{c})<\epsilon_{k}$, using Lemma~\ref{lemma:indicator inequality} yields
\par\noindent\small
\begin{align*}
    \mathbb{E}[|\Pi_{1}|^{4}|\mathcal{F}_{k-1}]&\leq\frac{2^{4}}{N^{4}}\left(\sum_{i=1}^{N}\frac{\mathbb{E}[|\phi(\hat{\overline{\mathbf{x}}}^{i}_{k},\overline{\mathbf{y}}^{i}_{k})|^{4}|\mathcal{F}_{k-1}]}{1-\epsilon_{k}}+\left(\sum_{i=1}^{N}\frac{\mathbb{E}[|\phi(\hat{\overline{\mathbf{x}}}^{i}_{k},\overline{\mathbf{y}}^{i}_{k})|^{2}|\mathcal{F}_{k-1}]}{1-\epsilon_{k}}\right)^{2}\right)\\
    &=\frac{2^{4}}{N^{4}(1-\epsilon_{k})^{2}}\sum_{i=1}^{N}(1-\epsilon_{k})\mathbb{E}[|\phi(\hat{\overline{\mathbf{x}}}^{i}_{k},\overline{\mathbf{y}}^{i}_{k})|^{4}|\mathcal{F}_{k-1}]+\frac{2^{4}}{N^{4}(1-\epsilon_{k})^{2}}\left(\sum_{i=1}^{N}\mathbb{E}[|\phi(\hat{\overline{\mathbf{x}}}^{i}_{k},\overline{\mathbf{y}}^{i}_{k})|^{2}|\mathcal{F}_{k-1}]\right)^{2}\\
    &\leq\frac{2^{4}}{N^{4}(1-\epsilon_{k})^{2}}\left(\sum_{i=1}^{N}\mathbb{E}[|\phi(\hat{\overline{\mathbf{x}}}^{i}_{k},\overline{\mathbf{y}}^{i}_{k})|^{4}|\mathcal{F}_{k-1}]+\left(\sum_{i=1}^{N}\mathbb{E}[|\phi(\hat{\overline{\mathbf{x}}}^{i}_{k},\overline{\mathbf{y}}^{i}_{k})|^{2}|\mathcal{F}_{k-1}]\right)^{2}\right),
\end{align*}
\normalsize
where the last inequality is obtained using $1-\epsilon_{k}<1$. Expressing the expectations as in \eqref{eqn:expect of predicted particles}, we have
\par\noindent\small
\begin{align*}
    \mathbb{E}[|\Pi_{1}|^{4}|\mathcal{F}_{k-1}]&\leq\frac{2^{4}}{N^{4}(1-\epsilon_{k})^{2}}\left(\sum_{i=1}^{N}\langle\pi_{k-1|k-1}^{N},\delta_{T}\rho|\phi|^{4}\rangle+(\sum_{i=1}^{N}\langle\pi_{k-1|k-1}^{N},\delta_{T}\rho|\phi|^{2}\rangle)^{2}\right)\\
    &=\frac{2^{4}}{(1-\epsilon_{k})^{2}}\left(\frac{\langle\pi_{k-1|k-1}^{N},\delta_{T}\rho|\phi|^{4}\rangle}{N^{3}}+\frac{\langle\pi_{k-1|k-1}^{N},\delta_{T}\rho|\phi|^{2}\rangle^{2}}{N^{2}}\right)\\
    &\leq\frac{2^{4}}{(1-\epsilon_{k})^{2}}\left(\frac{\langle\pi_{k-1|k-1}^{N},\delta_{T}\rho|\phi|^{4}\rangle}{N^{3}}+\frac{\langle\pi_{k-1|k-1}^{N},\delta_{T}\rho|\phi|^{4}\rangle}{N^{2}}\right)\leq\frac{2^{5}}{(1-\epsilon_{k})^{2}}\frac{\langle\pi_{k-1|k-1}^{N},\delta_{T}\rho|\phi|^{4}\rangle}{N^{2}},
\end{align*}
\normalsize
where the second last and last inequalities result from Jensen's inequality and $N>1$, respectively. Finally, using \eqref{eqn:k-1 pow 4 one} with the indicator function being bounded by $1$, we have
\par\noindent\small
\begin{align*}
    \mathbb{E}[|\Pi_{1}|^{4}]&\leq\frac{2^{5}}{(1-\epsilon_{k})^{2}}\|\rho\|_{\infty} M_{k-1|k-1}\frac{\|\phi\|^{4}_{k-1,4}}{N^{2}},
\end{align*}
\normalsize
which gives \eqref{eqn:Pi1 term} with $C_{\Pi_{1}}\doteq\frac{2^{5}}{(1-\epsilon_{k})^{2}}\|\rho\|_{\infty} M_{k-1|k-1}$.

\textbf{(b) $\Pi_{2}$ term:} Using \eqref{eqn:expect of predicted particles}, we have
\par\noindent\small
\begin{align*}
    |\Pi_{2}|^{4}&=\left|\frac{1}{N}\sum_{i=1}^{N}(\mathbb{E}[\phi(\hat{\widetilde{\mathbf{x}}}^{i}_{k},\widetilde{\mathbf{y}}^{i}_{k})|\mathcal{F}_{k-1}]-\mathbb{E}[\phi(\hat{\overline{\mathbf{x}}}^{i}_{k},\overline{\mathbf{y}}^{i}_{k})|\mathcal{F}_{k-1}])\right|^{4}\leq\frac{1}{N}\sum_{i=1}^{N}|\mathbb{E}[\phi(\hat{\widetilde{\mathbf{x}}}^{i}_{k},\widetilde{\mathbf{y}}^{i}_{k})|\mathcal{F}_{k-1}]-\mathbb{E}[\phi(\hat{\overline{\mathbf{x}}}^{i}_{k},\overline{\mathbf{y}}^{i}_{k})|\mathcal{F}_{k-1}]|^{4},
\end{align*}
\normalsize
from the Jensen's inequality. As mentioned earlier $\{(\hat{\overline{\mathbf{x}}}^{i}_{k},\overline{\mathbf{y}}^{i}_{k})\}$ are obtained from $\{(\hat{\widetilde{\mathbf{x}}}^{i}_{k},\widetilde{\mathbf{y}}^{i}_{k})\}$ such that $\Gamma_{k}$ occurs such that using Lemma~\ref{lemma:indicator inequality} and then Jensen's inequality, we obtain
\par\noindent\small
\begin{align*}
    |\Pi_{2}|^{4}&\leq\frac{1}{N}\sum_{i=1}^{N}\left(\frac{2\sqrt{\mathbb{E}[\phi(\hat{\overline{\mathbf{x}}}^{i}_{k},\overline{\mathbf{y}}^{i}_{k})^{2}|\mathcal{F}_{k-1}]}}{1-\epsilon_{k}}\sqrt{\epsilon_{k}}\right)^{4}\leq\frac{2^{4}\epsilon_{k}^{2}}{N(1-\epsilon_{k})^{4}}\sum_{i=1}^{N}\mathbb{E}[\phi(\hat{\overline{\mathbf{x}}}^{i}_{k},\overline{\mathbf{y}}^{i}_{k})^{4}|\mathcal{F}_{k-1}].
\end{align*}
\normalsize
Substituting $\epsilon_{k}=C_{\gamma_{k}}\|\beta\|^{4}_{k-1,4}/N^{2}$ (from Lemma~\ref{lemma:infinite loop}) and \eqref{eqn:expect of predicted particles}, we have
\par\noindent\small
\begin{align*}
    |\Pi_{2}|^{4}\leq\frac{2^{4}}{(1-\epsilon_{k})^{4}}C_{\gamma_{k}}^{2}\frac{\|\beta\|^{8}_{k-1,4}}{N^{4}}\langle\pi^{N}_{k-1|k-1},\delta_{T}\rho\phi^{4}\rangle.
\end{align*}
\normalsize
Denote $C'_{\Pi_{2}}=2^{4}C_{\gamma_{k}}^{2}\|\beta\|^{8}_{k-1,4}/(1-\epsilon_{k})^{4}$. Finally, using \eqref{eqn:k-1 pow 4 one} with the indicator function being bounded by $1$, we have
\par\noindent\small
\begin{align*}
    \mathbb{E}[|\Pi_{2}|^{4}]&\leq C'_{\Pi_{2}}\frac{\mathbb{E}|\langle\pi^{N}_{k-1|k-1},\delta_{T}\rho\phi^{4}\rangle|}{N^{4}}\leq C'_{\Pi_{2}}M_{k-1|k-1}\|\rho\|_{\infty}\frac{\|\phi\|^{4}_{k-1,4}}{N^{4}},
\end{align*}
\normalsize
which yields \eqref{eqn:Pi2 term} using $N>1$ and $C_{\Pi_{2}}=C'_{\Pi_{2}}M_{k-1|k-1}\|\rho\|_{\infty}$.

\textbf{(c) $\Pi_{3}$ term:} Using \eqref{eqn:optimal filter time update} and \eqref{eqn:ipf converge} replacing $k$ by $k-1$, we have
\par\noindent\small
\begin{align*}
    \mathbb{E}[|\Pi_{3}|^{4}]&=\mathbb{E}|\langle\pi^{N}_{k-1|k-1},\delta_{T}\rho\phi\rangle-\langle\pi_{k|k-1},\phi\rangle|^{4}=\mathbb{E}|\langle\pi^{N}_{k-1|k-1},\delta_{T}\rho\phi\rangle-\langle\pi_{k-1|k-1},\delta_{T}\rho\phi\rangle|^{4}\\
    &\leq C_{k-1|k-1}\|\rho\|_{\infty}^{4}\frac{\|\phi\|^{4}_{k-1,4}}{N^{2}},
\end{align*}
\normalsize
which yields \eqref{eqn:Pi3 term} defining $C_{\Pi_{3}}\doteq C_{k-1|k-1}\|\rho\|_{\infty}^{4}$.
\end{proof}
\begin{lemma}\label{lemma:Sampling pow 4 terms}
If \eqref{eqn:k-1 pow 4 one} holds, then
\par\noindent\small
\begin{align}
    &\mathbb{E}\left|\langle\widetilde{\pi}^{N}_{k|k-1},|\phi|^{4}\rangle-\frac{1}{N}\sum_{i=1}^{N}\mathbb{E}[|\phi(\hat{\widetilde{\mathbf{x}}}^{i}_{k},\widetilde{\mathbf{y}}^{i}_{k})|^{4}|\mathcal{F}_{k-1}]\right|\leq\frac{2}{(1-\epsilon_{k})}M_{k-1|k-1}\|\rho\|_{\infty}\|\phi\|^{4}_{k-1,4},\label{eqn:T1 predict pow 4}\\
    &\mathbb{E}\left|\frac{1}{N}\sum_{i=1}^{N}\mathbb{E}[|\phi(\hat{\widetilde{\mathbf{x}}}^{i}_{k},\widetilde{\mathbf{y}}^{i}_{k})|^{4}|\mathcal{F}_{k-1}]-\langle\pi^{N}_{k-1|k-1},\delta_{T}\rho|\phi|^{4}\rangle\right|\leq\frac{2-\epsilon_{k}}{1-\epsilon_{k}}M_{k-1|k-1}\|\rho\|_{\infty}\|\phi\|^{4}_{k-1,4}\label{eqn:T2 predict pow 4}\\
    &\mathbb{E}|\langle\pi^{N}_{k-1|k-1},\delta_{T}\rho|\phi|^{4}\rangle-\langle\pi_{k|k-1},|\phi|^{4}\rangle|\leq \|\rho\|_{\infty}(M_{k-1|k-1}+1)\|\phi\|^{4}_{k-1,4}\label{eqn:T3 predict pow 4},
\end{align}
\normalsize
where $\epsilon_{k}$ is the same as defined in Lemma~\ref{lemma:infinite loop}.
\end{lemma}
\begin{proof}
Consider the first inequality \eqref{eqn:T1 predict pow 4}. Since $\widetilde{\pi}^{N}_{k|k-1}$ is the empirical distribution obtained from particles $\{(\hat{\widetilde{\mathbf{x}}}^{i}_{k},\widetilde{\mathbf{y}}^{i}_{k})\}$, we have
\par\noindent\small
\begin{align*}
    &\mathbb{E}\left[\mathbb{E}\left|\langle\widetilde{\pi}^{N}_{k|k-1},|\phi|^{4}\rangle-\frac{1}{N}\sum_{i=1}^{N}\mathbb{E}[|\phi(\hat{\widetilde{\mathbf{x}}}^{i}_{k},\widetilde{\mathbf{y}}^{i}_{k})|^{4}|\mathcal{F}_{k-1}]\right||\mathcal{F}_{k-1}\right]\\
    &=\frac{1}{N}\mathbb{E}\left[\mathbb{E}\left|\sum_{i=1}^{N}(|\phi(\hat{\widetilde{\mathbf{x}}}^{i}_{k},\widetilde{\mathbf{y}}^{i}_{k})|^{4}-\mathbb{E}[|\phi(\hat{\widetilde{\mathbf{x}}}^{i}_{k},\widetilde{\mathbf{y}}^{i}_{k})|^{4}|\mathcal{F}_{k-1}])\right||\mathcal{F}_{k-1}\right]\leq\frac{2}{N}\mathbb{E}[\sum_{i=1}^{N}\mathbb{E}[|\phi(\hat{\widetilde{\mathbf{x}}}^{i}_{k},\widetilde{\mathbf{y}}^{i}_{k})|^{4}|\mathcal{F}_{k-1}]]\nonumber\\
    &\leq\frac{2}{N(1-\epsilon_{k})}\mathbb{E}[\sum_{i=1}^{N}\mathbb{E}[|\phi(\hat{\overline{\mathbf{x}}}^{i}_{k},\overline{\mathbf{y}}^{i}_{k})|^{4}|\mathcal{F}_{k-1}]]=\frac{2}{(1-\epsilon_{k})}\mathbb{E}[\langle\pi^{N}_{k-1|k-1},\delta_{T}\rho|\phi|^{4}\rangle],
\end{align*}
\normalsize
where the first and second inequalities are obtained using Lemma~\ref{lemma:diff to power p inequality} and \ref{lemma:indicator inequality}, respectively. The last equality follows because particles $\{(\hat{\overline{\mathbf{x}}}^{i}_{k},\overline{\mathbf{y}}^{i}_{k})\}$ are drawn from distribution $\langle\pi^{N}_{k-1|k-1},\delta_{T}\rho\rangle$. Finally, using \eqref{eqn:k-1 pow 4 one}, we obtain \eqref{eqn:T1 predict pow 4}.

Next, we consider the inequality \eqref{eqn:T2 predict pow 4}. Using \eqref{eqn:expect of predicted particles} replacing $\phi$ by $|\phi|^{4}$, we have
\par\noindent\small
\begin{align*}
    &\mathbb{E}\left|\frac{1}{N}\sum_{i=1}^{N}\mathbb{E}[|\phi(\hat{\widetilde{\mathbf{x}}}^{i}_{k},\widetilde{\mathbf{y}}^{i}_{k})|^{4}|\mathcal{F}_{k-1}]-\langle\pi^{N}_{k-1|k-1},\delta_{T}\rho|\phi|^{4}\rangle\right|=\mathbb{E}\left|\frac{1}{N}\sum_{i=1}^{N}(\mathbb{E}[|\phi(\hat{\widetilde{\mathbf{x}}}^{i}_{k},\widetilde{\mathbf{y}}^{i}_{k})|^{4}|\mathcal{F}_{k-1}]-\mathbb{E}[|\phi(\hat{\overline{\mathbf{x}}}^{i}_{k},\overline{\mathbf{y}}^{i}_{k})|^{4}|\mathcal{F}_{k-1}])\right|\\
    &\leq\frac{1}{N}\sum_{i=1}^{N}\left(\mathbb{E}[\mathbb{E}[|\phi(\hat{\widetilde{\mathbf{x}}}^{i}_{k},\widetilde{\mathbf{y}}^{i}_{k})|^{4}|\mathcal{F}_{k-1}]]+\mathbb{E}[\mathbb{E}[|\phi(\hat{\overline{\mathbf{x}}}^{i}_{k},\overline{\mathbf{y}}^{i}_{k})|^{4}|\mathcal{F}_{k-1}]]\right)\nonumber.
\end{align*}
\normalsize
Now, using Lemma~\ref{lemma:indicator inequality}, we have
\par\noindent\small
\begin{align*}       
    &\mathbb{E}\left|\frac{1}{N}\sum_{i=1}^{N}\mathbb{E}[|\phi(\hat{\widetilde{\mathbf{x}}}^{i}_{k},\widetilde{\mathbf{y}}^{i}_{k})|^{4}|\mathcal{F}_{k-1}]-\langle\pi^{N}_{k-1|k-1},\delta_{T}\rho|\phi|^{4}\rangle\right|\nonumber\\
    &\leq\frac{1}{N}\sum_{i=1}^{N}\left(\frac{\mathbb{E}[\mathbb{E}[|\phi(\hat{\overline{\mathbf{x}}}^{i}_{k},\overline{\mathbf{y}}^{i}_{k})|^{4}|\mathcal{F}_{k-1}]]}{1-\epsilon_{k}}+\mathbb{E}[\mathbb{E}[|\phi(\hat{\overline{\mathbf{x}}}^{i}_{k},\overline{\mathbf{y}}^{i}_{k})|^{4}|\mathcal{F}_{k-1}]]\right)=\frac{2-\epsilon_{k}}{1-\epsilon_{k}}\mathbb{E}[\langle\pi^{N}_{k-1|k-1},\delta_{T}\rho|\phi|^{4}\rangle],\nonumber
\end{align*}
\normalsize
which yields \eqref{eqn:T2 predict pow 4} using assumption \eqref{eqn:k-1 pow 4 one}.

Finally, consider the last inequality \eqref{eqn:T3 predict pow 4}. From \eqref{eqn:optimal filter time update} and triangle inequality, we have
\par\noindent\small
\begin{align*}
    \mathbb{E}|\langle\pi^{N}_{k-1|k-1},\delta_{T}\rho|\phi|^{4}\rangle-\langle\pi_{k|k-1},|\phi|^{4}\rangle|&\leq\mathbb{E}|\langle\pi^{N}_{k-1|k-1},\delta_{T}\rho|\phi|^{4}\rangle|+|\langle\pi_{k-1|k-1},\delta_{T}\rho|\phi|^{4}\rangle|\\
    &\leq\|\rho\|_{\infty}M_{k-1|k-1}\|\phi\|^{4}_{k-1,4}+\|\rho\|_{\infty}\|\phi\|^{4}_{k-1,4},
\end{align*}
where the last inequality results from the assumption \eqref{eqn:k-1 pow 4 one} and the definition of $\|\phi\|_{k-1,4}$ (from Theorem~\ref{thm:ipf convergence}). Hence, \eqref{eqn:T3 predict pow 4} is proved.
\end{proof}

\subsection{Proof of the theorem}\label{app:proof}
As mentioned earlier, the proof employs an induction framework. To this end, we prove that \eqref{eqn:ipf converge} holds for the $k$-th time step provided that the inequality holds for $(k-1)$-th time step. Additionally, we also show that
\par\noindent\small
\begin{align}
    \mathbb{E}|\langle\pi^{N}_{k|k},|\phi|^{4}\rangle|\leq M_{k|k}\|\phi\|^{4}_{k,4},\label{eqn:M inequality k}
\end{align}
\normalsize
which is \eqref{eqn:k-1 pow 4 one} with $k-1$ replaced by $k$. The first part of Theorem~\ref{thm:ipf convergence}, i.e., the algorithm does not run into an infinite loop, follows from Lemma~\ref{lemma:infinite loop}. For the second part of the theorem, we analyze the empirical distributions obtained in different steps of the I-PF algorithm and obtain the corresponding inequalities of \eqref{eqn:ipf converge} and \eqref{eqn:M inequality k}.

\textbf{Initialization:} Following the induction framework, we first show that \eqref{eqn:ipf converge} and \eqref{eqn:M inequality k} hold for $k=0$. Consider $\{\hat{\mathbf{x}}^{i}_{0}\}_{1\leq i\leq N}$ and $\{\mathbf{y}^{i}_{0}\}_{1\leq i\leq N}$ as the i.i.d. and mutually independent (initial) samples drawn from distributions $\widetilde{\pi}^{x}_{0}(d\hat{\mathbf{x}}_{0})$ and $\rho(\mathbf{y}_{0}|\mathbf{x}_{0})$, respectively. Recall that $\widetilde{\pi}^{x}_{0}$ is the initial distribution assumed in  I-PF. Denote $\pi_{0}$ as the joint distribution for particles $\{(\hat{\mathbf{x}}^{i}_{0},\mathbf{y}^{i}_{0})\}_{1\leq i\leq N}$ such that $\langle\pi_{0},\phi\rangle=\mathbb{E}[\phi(\hat{\mathbf{x}}^{i}_{0},\mathbf{y}^{i}_{0})]$ irrespective of $i$. Also, $\phi^{N}_{0}$ is the empirical distribution obtained from particles $\{(\hat{\mathbf{x}}^{i}_{0},\mathbf{y}^{i}_{0})\}_{1\leq i\leq N}$. We have
\par\noindent\small
\begin{align*}
     \mathbb{E}|\langle\pi_{0}^{N},\phi\rangle-\langle\pi_{0},\phi\rangle|^{4}=\mathbb{E}\left|\frac{1}{N}\sum_{i=1}^{N}\left(\phi(\hat{\mathbf{x}}^{i}_{0},\mathbf{y}^{i}_{0})-\mathbb{E}[\phi(\hat{\mathbf{x}}^{i}_{0},\mathbf{y}^{i}_{0})]\right)\right|^{4}.
\end{align*}
\normalsize
Using Lemma~\ref{lemma:sum to max inequality}, we obtain
\par\noindent\small
\begin{align*}
     \mathbb{E}|\langle\pi_{0}^{N},\phi\rangle-\langle\pi_{0},\phi\rangle|^{4}\leq\frac{2}{N^{2}}\mathbb{E}\left|\phi(\hat{\mathbf{x}}^{i}_{0},\mathbf{y}^{i}_{0})-\mathbb{E}[\phi(\hat{\mathbf{x}}^{i}_{0},\mathbf{y}^{i}_{0})]\right|^{4},
\end{align*}
\normalsize
because $(\hat{\mathbf{x}}^{i}_{0},\mathbf{y}^{i}_{0})$ are identically distributed for all $i=1,2,\hdots N$. Finally, using Lemma~\ref{lemma:diff to power p inequality} with $\mathbb{E}|\phi(\hat{\mathbf{x}}^{i}_{0},\mathbf{y}^{i}_{0})|^{4}=\langle\pi_{0},|\phi|^{4}\rangle$, we have
\par\noindent\small
\begin{align}
     \mathbb{E}|\langle\pi_{0}^{N},\phi\rangle-\langle\pi_{0},\phi\rangle|^{4}\leq\frac{32}{N^{2}}\|\phi\|^{4}_{0,4}\doteq C_{0|0}\frac{\|\phi\|^{4}_{0,4}}{N^{2}},\label{eqn:initialize diff}
\end{align}
\normalsize
because $\|\phi\|_{0,4}=\textrm{max}\{1,\langle\pi_{0},|\phi|^{4}\rangle\}^{1/4}$ from Theorem~\ref{thm:ipf convergence}.

Similarly, using Lemma~\ref{lemma:diff to power p inequality}, we obtain
\par\noindent\small
\begin{align}
&\mathbb{E}|\langle\pi_{0}^{N},|\phi|^{4}\rangle-\langle\pi_{0},|\phi|^{4}\rangle|=\mathbb{E}\left|\frac{1}{N}\sum_{i=1}^{N}\left(|\phi(\hat{\mathbf{x}}^{i}_{0},\mathbf{y}^{i}_{0})|^{4}-\mathbb{E}|\phi(\hat{\mathbf{x}}^{i}_{0},\mathbf{y}^{i}_{0})|^{4}\right)\right|\leq\frac{1}{N}\sum_{i=1}^{N}2\mathbb{E}|\phi(\hat{\mathbf{x}}^{i}_{0},\mathbf{y}^{i}_{0})|^{4}=2\mathbb{E}|\phi(\hat{\mathbf{x}}^{i}_{0},\mathbf{y}^{i}_{0})|^{4}.\label{eqn:initial M inter}
\end{align}
\normalsize
From triangle inequality, we have
\par\noindent\small
\begin{align*}
    \mathbb{E}|\langle\pi_{0}^{N},|\phi|^{4}\rangle|\leq\mathbb{E}|\langle\pi_{0}^{N},|\phi|^{4}\rangle-\langle\pi_{0},|\phi|^{4}\rangle|+|\langle\pi_{0},|\phi|^{4}\rangle|.
\end{align*}
\normalsize
Note that $\pi^{N}_{0}$ is a random function obtained from the randomly generated particles, but $\pi_{0}$ is a given initial distribution. Hence, using \eqref{eqn:initial M inter} and $\langle\pi_{0},|\phi|^{4}\rangle=\mathbb{E}|\phi(\hat{\mathbf{x}}^{i}_{0},\mathbf{y}^{i}_{0})|^{4}$, we have
\par\noindent\small
\begin{align}
    \mathbb{E}|\langle\pi_{0}^{N},|\phi|^{4}\rangle|\leq 3\mathbb{E}|\phi(\hat{\mathbf{x}}^{i}_{0},\mathbf{y}^{i}_{0})|^{4}\leq\|\phi\|^{4}_{0,4}\doteq M_{0|0}\|\phi\|^{4}_{0,4}.\label{eqn:initialize pow 4}
\end{align}
\normalsize

Inequalities \eqref{eqn:initialize diff} and \eqref{eqn:initialize pow 4} show that \eqref{eqn:ipf converge} and \eqref{eqn:M inequality k} hold for $k=0$. Next, we assume that these inequalities hold at $(k-1)$-th time instant, i.e.,
\par\noindent\small
\begin{align}
    &\mathbb{E}|\langle\pi_{k-1|k-1}^{N},\phi\rangle-\langle\pi_{k-1|k-1},\phi\rangle|^{4}\leq C_{k-1|k-1}\frac{\|\phi\|^{4}_{k-1,4}}{N^{2}},\label{eqn:k-1 diff}\\
    &\mathbb{E}|\langle\pi_{k-1|k-1}^{N},|\phi|^{4}\rangle|\leq M_{k-1|k-1}\|\phi\|^{4}_{k-1,4},\label{eqn:k-1 pow 4}
\end{align}
\normalsize
where \eqref{eqn:k-1 pow 4} is same as \eqref{eqn:k-1 pow 4 one}, repeated here as a ready reference.

\textbf{Importance sampling with modification:} Recall that in the sampling step, we draw particles $\{(\hat{\overline{\mathbf{x}}}^{i}_{k},\overline{\mathbf{y}}^{i}_{k})\}$ until the inequality $\frac{1}{N}\sum_{i=1}^{N}\beta(\mathbf{a}_{k}|\hat{\overline{\mathbf{x}}}^{i}_{k})\geq\gamma_{k}$ is satisfied. Finally, $\{(\hat{\widetilde{\mathbf{x}}}^{i}_{k},\widetilde{\mathbf{y}}^{i}_{k})\}$ are the particles for which the condition holds, resulting in empirical distribution $\widetilde{\pi}^{N}_{k|k-1}$ as an approximation of $\pi_{k|k-1}$. Now, we consider bounds on $\mathbb{E}|\langle\widetilde{\pi}^{N}_{k|k-1},\phi\rangle-\langle\pi_{k|k-1},\phi\rangle|^{4}$ and $\mathbb{E}|\langle\widetilde{\pi}^{N}_{k|k-1},|\phi|^{4}\rangle-\langle\pi_{k|k-1},|\phi|^{4}\rangle|$.

Consider $\langle\widetilde{\pi}^{N}_{k|k-1},\phi\rangle-\langle\pi_{k|k-1},\phi\rangle=\Pi_{1}+\Pi_{2}+\Pi_{3}$ with $\Pi_{1}$, $\Pi_{2}$ and $\Pi_{3}$ as defined in \eqref{eqn:def Pi1}-\eqref{eqn:def Pi3} with appropriate bounds derived in Lemma~\ref{lemma:Pi123 bounds}. From Minkowski's inequality, we have
\par\noindent\small
\begin{align*}
     \mathbb{E}^{1/4}|\langle\widetilde{\pi}^{N}_{k|k-1},\phi\rangle-\langle\pi_{k|k-1},\phi\rangle|^{4}\leq\mathbb{E}^{1/4}|\Pi_{1}|^{4}+\mathbb{E}^{1/4}|\Pi_{2}|^{4}+\mathbb{E}^{1/4}|\Pi_{3}|^{4},
\end{align*}
\normalsize
such that using \eqref{eqn:Pi1 term}-\eqref{eqn:Pi3 term} yields
\par\noindent\small
\begin{align}
      \mathbb{E}|\langle\widetilde{\pi}^{N}_{k|k-1},\phi\rangle-\langle\pi_{k|k-1},\phi\rangle|^{4}\leq\widetilde{C}_{k|k-1}\frac{\|\phi\|^{4}_{k-1,4}}{N^{2}},\label{eqn:predict diff}
\end{align}
\normalsize
where constant $\widetilde{C}_{k|k-1}\doteq(C_{\Pi_{1}}^{1/4}+C_{\Pi_{2}}^{1/4}+C_{\Pi_{3}}^{1/4})^{4}$.

Next, we consider $\mathbb{E}|\langle\widetilde{\pi}^{N}_{k|k-1},|\phi|^{4}\rangle-\langle\pi_{k|k-1},|\phi|^{4}\rangle|$and employ a similar separation method. In particular, we have
\par\noindent\small
\begin{align*}
    &\langle\widetilde{\pi}^{N}_{k|k-1},|\phi|^{4}\rangle-\langle\pi_{k|k-1},|\phi|^{4}\rangle=\langle\widetilde{\pi}^{N}_{k|k-1},|\phi|^{4}\rangle-\frac{1}{N}\sum_{i=1}^{N}\mathbb{E}[|\phi(\hat{\widetilde{\mathbf{x}}}^{i}_{k},\widetilde{\mathbf{y}}^{i}_{k})|^{4}|\mathcal{F}_{k-1}]\\
    &+\frac{1}{N}\sum_{i=1}^{N}\mathbb{E}[|\phi(\hat{\widetilde{\mathbf{x}}}^{i}_{k},\widetilde{\mathbf{y}}^{i}_{k})|^{4}|\mathcal{F}_{k-1}]-\langle\pi^{N}_{k-1|k-1},\delta_{T}\rho|\phi|^{4}\rangle+\langle\pi^{N}_{k-1|k-1},\delta_{T}\rho|\phi|^{4}\rangle-\langle\pi_{k|k-1},|\phi|^{4}\rangle.
\end{align*}
\normalsize
Using bounds \eqref{eqn:T1 predict pow 4}-\eqref{eqn:T3 predict pow 4} from Lemma~\ref{lemma:Sampling pow 4 terms}, we obtain
\par\noindent\small
\begin{align}
    \mathbb{E}|\langle\widetilde{\pi}^{N}_{k|k-1},|\phi|^{4}\rangle-\langle\pi_{k|k-1},|\phi|^{4}\rangle|\leq\widetilde{M}_{k|k-1}\|\phi\|^{4}_{k-1,4},\label{eqn:predict pow 4}
\end{align}
\normalsize
where constant
\par\noindent\small
\begin{align*}
    \hspace{-0.5cm}\widetilde{M}_{k|k-1}\doteq\|\rho\|_{\infty}\left(\frac{2}{1-\epsilon_{k}}M_{k-1|k-1}+\frac{2-\epsilon_{k}}{1-\epsilon_{k}}M_{k-1|k-1}+M_{k-1|k-1}+1\right).
\end{align*}
\normalsize
The inequalities \eqref{eqn:predict diff} and \eqref{eqn:predict pow 4} are the counterparts of inequalities \eqref{eqn:ipf converge} and \eqref{eqn:M inequality k}, respectively, for the (approximate) prediction distribution $\widetilde{\pi}^{N}_{k|k-1}$ obtained from the modified importance sampling in I-PF.

\textbf{Weight computation:} The posterior distribution $\widetilde{\pi}^{N}_{k|k}$ is obtained from the prediction distribution $\widetilde{\pi}^{N}_{k|k-1}$ by associating weight $\omega^{i}_{k}$ with each particle $(\hat{\widetilde{\mathbf{x}}}^{i}_{k},\mathbf{y}^{i}_{k})$. Hence, we now analyze $\mathbb{E}|\langle\widetilde{\pi}^{N}_{k|k},\phi\rangle-\langle\pi_{k|k},\phi\rangle|^{4}$ and $\mathbb{E}|\langle\widetilde{\pi}^{N}_{k|k},|\phi|^{4}\rangle|$ based on \eqref{eqn:predict diff} and \eqref{eqn:predict pow 4}. In particular, we obtain the counterparts of inequalities \eqref{eqn:ipf converge} and \eqref{eqn:M inequality k} for $\widetilde{\pi}^{N}_{k|k}$ in following Claims~\ref{claim:update claim diff} and \ref{claim:update claim pow}, respectively.
\begin{claim}\label{claim:update claim diff}
The optimal filter's posterior distribution $\pi_{k|k}$ and its approximation $\widetilde{\pi}^{N}_{k|k}$ in I-PF satisfy
\par\noindent\small
\begin{align}
    \mathbb{E}|\langle\widetilde{\pi}^{N}_{k|k},\phi\rangle-\langle\pi_{k|k},\phi\rangle|^{4}\leq\widetilde{C}_{k|k}\frac{\|\phi\|^{4}_{k-1,4}}{N^{2}},\label{eqn:update diff}
\end{align}
\normalsize
for suitable $\widetilde{C}_{k|k}>0$.
\end{claim}
\begin{claimproof}
Using \eqref{eqn:optimal filter measurement update}, we again employ a separation method as
\par\noindent\small
\begin{align*}
    \langle\widetilde{\pi}^{N}_{k|k},\phi\rangle-\langle\pi_{k|k},\phi\rangle=\frac{\langle\widetilde{\pi}^{N}_{k|k-1},\beta\phi\rangle}{\langle\widetilde{\pi}^{N}_{k|k-1},\beta\rangle}-\frac{\langle\pi_{k|k-1},\beta\phi\rangle}{\langle\pi_{k|k-1},\beta\rangle}\doteq\widetilde{\Pi}_{1}+\widetilde{\Pi}_{2},
\end{align*}
\normalsize
where $\widetilde{\Pi}_{1}=\frac{\langle\widetilde{\pi}^{N}_{k|k-1},\beta\phi\rangle}{\langle\widetilde{\pi}^{N}_{k|k-1},\beta\rangle}-\frac{\langle\widetilde{\pi}^{N}_{k|k-1},\beta\phi\rangle}{\langle\pi_{k|k-1},\beta\rangle}$ and $\widetilde{\Pi}_{2}=\frac{\langle\widetilde{\pi}^{N}_{k|k-1},\beta\phi\rangle}{\langle\pi_{k|k-1},\beta\rangle}-\frac{\langle\pi_{k|k-1},\beta\phi\rangle}{\langle\pi_{k|k-1},\beta\rangle}$. As noted in Remark~\ref{remark:threshold}, the modification step implies that $\langle\widetilde{\pi}^{N}_{k|k},\beta\rangle\geq\gamma_{k}$. Hence, under assumptions \textbf{A7.b} and \textbf{A7.c}, we have
\par\noindent\small
\begin{align*}
    &|\widetilde{\Pi}_{1}|=\left|\frac{\langle\widetilde{\pi}^{N}_{k|k-1},\beta\phi\rangle}{\langle\widetilde{\pi}^{N}_{k|k-1},\beta\rangle}\times\frac{\langle\pi_{k|k-1},\beta\rangle-\langle\widetilde{\pi}^{N}_{k|k-1},\beta\rangle}{\langle\pi_{k|k-1},\beta\rangle}\right|\leq\frac{\|\beta\phi\|_{\infty}}{\gamma_{k}\langle\pi_{k|k-1},\beta\rangle}|\langle\pi_{k|k-1},\beta\rangle-\langle\widetilde{\pi}^{N}_{k|k-1},\beta\rangle|.
\end{align*}
\normalsize
Now, using Minkowski's inequality, we have
\par\noindent\small
\begin{align*}
    &\mathbb{E}^{1/4}|\langle\widetilde{\pi}^{N}_{k|k},\phi\rangle-\langle\pi_{k|k},\phi\rangle|^{4}\leq\mathbb{E}^{1/4}|\widetilde{\Pi}_{1}|^{4}+\mathbb{E}^{1/4}|\widetilde{\Pi}_{2}|^{4}\\
    &\leq\frac{\|\beta\phi\|_{\infty}}{\gamma_{k}\langle\pi_{k|k-1},\beta\rangle}\mathbb{E}^{1/4}|\langle\pi_{k|k-1},\beta\rangle-\langle\widetilde{\pi}^{N}_{k|k-1},\beta\rangle|^{4}+\frac{1}{\langle\pi_{k|k-1},\beta\rangle}\mathbb{E}^{1/4}|\langle\widetilde{\pi}^{N}_{k|k-1},\beta\phi\rangle-\langle\pi_{k|k-1},\beta\phi\rangle|^{4}.
\end{align*}
\normalsize
Finally, using \eqref{eqn:predict diff}, we obtain
\par\noindent\small
\begin{align*}
    \mathbb{E}^{1/4}|\langle\widetilde{\pi}^{N}_{k|k},\phi\rangle-\langle\pi_{k|k},\phi\rangle|^{4}&\leq\frac{\|\beta\phi\|_{\infty}}{\gamma_{k}\langle\pi_{k|k-1},\beta\rangle}\widetilde{C}^{1/4}_{k|k-1}\frac{\|\beta\|_{\infty}}{N^{1/2}}+\frac{1}{\langle\pi_{k|k-1},\beta\rangle}\widetilde{C}^{1/4}_{k|k-1}\|\beta\|_{\infty}\frac{\|\phi\|_{k-1,4}}{N^{1/2}}\\
    &\leq\frac{\widetilde{C}^{1/4}_{k|k-1}\|\beta\|_{\infty}}{\gamma_{k}\langle\pi_{k|k-1},\beta\rangle}(\|\beta\phi\|_{\infty}+\gamma_{k})\frac{\|\phi\|_{k-1,4}}{N^{1/2}},\\
\end{align*}
\normalsize
where the last inequality follows because by definition $\|\phi\|_{k-1,4}>1$. Defining\\ $\widetilde{C}^{1/4}_{k|k}\doteq\frac{\widetilde{C}^{1/4}_{k|k-1}\|\beta\|_{\infty}}{\gamma_{k}\langle\pi_{k|k-1},\beta\rangle}(\|\beta\phi\|_{\infty}+\gamma_{k})$ proves the claim.
\end{claimproof}
\begin{claim}\label{claim:update claim pow}
The distributions $\pi_{k|k}$ and $\widetilde{\pi}^{N}_{k|k}$ satisfy
\par\noindent\small
\begin{align}
   \mathbb{E}|\langle\widetilde{\pi}^{N}_{k|k},|\phi|^{4}\rangle|\leq\widetilde{M}_{k|k}\|\phi\|^{4}_{k,4}\label{eqn:update pow 4},
\end{align}
\normalsize
for suitable $\widetilde{M}_{k|k}>0$.
\end{claim}
\begin{claimproof}
Using a similar separation method as in proof of Claim~\ref{claim:update claim diff}, we have
\par\noindent\small
\begin{align*}
   \mathbb{E}|\langle\widetilde{\pi}^{N}_{k|k},|\phi|^{4}\rangle-\langle\pi_{k|k},|\phi|^{4}\rangle|&\leq\mathbb{E}\left|\frac{\langle\widetilde{\pi}^{N}_{k|k-1},\beta|\phi|^{4}\rangle-\langle\pi_{k|k-1},\beta|\phi|^{4}\rangle}{\langle\pi_{k|k-1},\beta\rangle}\right|\\
   &+\mathbb{E}\left|\frac{\langle\widetilde{\pi}^{N}_{k|k-1},\beta|\phi|^{4}\rangle}{\langle\widetilde{\pi}^{N}_{k|k-1},\beta\rangle}\times\frac{\langle\pi_{k|k-1},\beta\rangle-\langle\widetilde{\pi}^{N}_{k|k-1},\beta\rangle}{\langle\pi_{k|k-1},\beta\rangle}\right|.
\end{align*}
\normalsize
Again using assumption \textbf{A7.b} and \eqref{eqn:predict pow 4}, we obtain
\par\noindent\small
\begin{align*}
   \mathbb{E}|\langle\widetilde{\pi}^{N}_{k|k},|\phi|^{4}\rangle-\langle\pi_{k|k},|\phi|^{4}\rangle|&\leq\frac{\|\beta\phi^{4}\|_{\infty}}{\gamma_{k}\langle\pi_{k|k-1},\beta\rangle}\mathbb{E}|\langle\pi_{k|k-1},\beta\rangle-\langle\widetilde{\pi}^{N}_{k|k-1},\beta\rangle|+\frac{1}{\langle\pi_{k|k-1},\beta\rangle}\widetilde{M}_{k|k-1}\|\beta\|_{\infty}\|\phi\|^{4}_{k-1,4}\\
    &\leq\frac{\|\beta\phi^{4}\|_{\infty}2\|\beta\|_{\infty}}{\gamma_{k}\langle\pi_{k|k-1},\beta\rangle}\|\phi\|^{4}_{k-1,4}+\frac{\widetilde{M}_{k|k-1}\|\beta\|_{\infty}}{\langle\pi_{k|k-1},\beta\rangle}\|\phi\|^{4}_{k-1,4},
\end{align*}
\normalsize
because by definition $\|\phi\|_{k-1,4}>1$. Hence,
\par\noindent\small
\begin{align*}
    \mathbb{E}|(\widetilde{\pi}^{N}_{k|k},|\phi|^{4})|&\leq\frac{\|\beta\phi^{4}\|_{\infty}2\|\beta\|_{\infty}}{\gamma_{k}\langle\pi_{k|k-1},\beta\rangle}\|\phi\|^{4}_{k-1,4}+\frac{\widetilde{M}_{k|k-1}\|\beta\|_{\infty}}{\langle\pi_{k|k-1},\beta\rangle}\|\phi\|^{4}_{k-1,4}+\langle\pi_{k|k},|\phi|^{4}\rangle.
\end{align*}
But, $\langle\pi_{k|k},|\phi|^{4}\rangle\leq\|\phi\|^{4}_{k,4}$ and $\|\phi\|_{k,4}$ is increasing in $k$ such that
\par\noindent\small
\begin{align*}
    \mathbb{E}|(\widetilde{\pi}^{N}_{k|k},|\phi|^{4})|\leq 3\;\textrm{max}\left\{\frac{\|\beta\phi^{4}\|_{\infty}2\|\beta\|_{\infty}}{\gamma_{k}\langle\pi_{k|k-1},\beta\rangle},\frac{\widetilde{M}_{k|k-1}\|\beta\|_{\infty}}{\langle\pi_{k|k-1},\beta\rangle},1\right\}\|\phi\|^{4}_{k,4},
\end{align*}
\normalsize
which proves the claim with $\widetilde{M}_{k|k}\doteq3\;\textrm{max}\left\{\frac{\|\beta\phi^{4}\|_{\infty}2\|\beta\|_{\infty}}{\gamma_{k}\langle\pi_{k|k-1},\beta\rangle},\frac{\widetilde{M}_{k|k-1}\|\beta\|_{\infty}}{\langle\pi_{k|k-1},\beta\rangle},1\right\}$.
\end{claimproof}

\textbf{Resampling:} In this step, we draw $N$ independent particles $(\hat{\mathbf{x}}^{i}_{k},\mathbf{y}^{i}_{k})$ from the posterior distribution $\widetilde{\pi}^{N}_{k|k}$ and obtain the empirical distribution $\pi^{N}_{k|k}$ with equally weighted particles. Note that $\pi^{N}_{k|k}$ also approximates $\pi_{k|k}$ and is the I-PF's output posterior distribution. Now, we finally show that \eqref{eqn:ipf converge} and \eqref{eqn:M inequality k} hold true if we assume \eqref{eqn:k-1 diff} and \eqref{eqn:k-1 pow 4} hold at $(k-1)$-th time, which completes the induction proof. To this end, we analyze $\mathbb{E}|\langle\pi^{N}_{k|k},\phi\rangle-\langle\pi_{k|k},\phi\rangle|^{4}$ and $\mathbb{E}|\langle\pi^{N}_{k|k},|\phi|^{4}\rangle|$ based on \eqref{eqn:update diff} and \eqref{eqn:update pow 4}, and prove the following claims.
\begin{claim}\label{claim:k diff}
The distribution $\pi_{k|k}$ and its approximation $\pi^{N}_{k|k}$ satisfy
\par\noindent\small
\begin{align}
    \mathbb{E}|\langle\pi^{N}_{k|k},\phi\rangle-\langle\pi_{k|k},\phi\rangle|^{4}\leq C_{k|k}\frac{\|\phi\|^{4}_{k,4}}{N^{2}}.\label{eqn:k diff}
\end{align}
\normalsize
\end{claim}
\begin{claimproof}
Consider the separation $\langle\pi^{N}_{k|k},\phi\rangle-\langle\pi_{k|k},\phi\rangle=\overline{\Pi}_{1}+\overline{\Pi}_{2}$ where $\overline{\Pi}_{1}=\langle\pi^{N}_{k|k},\phi\rangle-\langle\widetilde{\pi}^{N}_{k|k},\phi\rangle$ and $\overline{\Pi}_{2}=\langle\widetilde{\pi}^{N}_{k|k},\phi\rangle-\langle\pi_{k|k},\phi\rangle$. Denote $\mathcal{G}_{k}$ as the $\sigma$-algebra generated by $\{(\hat{\widetilde{\mathbf{x}}}^{i}_{k},\widetilde{\mathbf{y}}^{i}_{k})\}_{i=1}^{N}$. Since $(\hat{\mathbf{x}}^{i}_{k},\mathbf{y}^{i}_{k})$ are drawn independently from $\widetilde{\pi}^{N}_{k|k}$, we have $\mathbb{E}[\phi(\hat{\mathbf{x}}^{i}_{k},\mathbf{y}^{i}_{k})|\mathcal{G}_{k}]=\langle\widetilde{\pi}^{N}_{k|k},\phi\rangle$ such that $\overline{\Pi}_{1}=\frac{1}{N}\sum_{i=1}^{N}(\phi(\hat{\mathbf{x}}^{i}_{k},\mathbf{y}^{i}_{k})-\mathbb{E}[\phi(\hat{\mathbf{x}}^{i}_{k},\mathbf{y}^{i}_{k})|\mathcal{G}_{k}])$. Using Lemma~\ref{lemma:sum to max inequality} and \ref{lemma:diff to power p inequality}, and finally \eqref{eqn:update pow 4}, we obtain
\par\noindent\small
\begin{align}
\mathbb{E}[|\overline{\Pi}_{1}|^{4}|\mathcal{G}_{k}]\leq 2^{5}\widetilde{M}_{k|k}\frac{\|\phi\|^{4}_{k,4}}{N^{2}}.\label{eqn:resample P1}
\end{align}
\normalsize
Again, using the Minkowski's inequality, and \eqref{eqn:update diff} and \eqref{eqn:resample P1}, we have
\par\noindent\small
\begin{align*}
    &\mathbb{E}^{1/4}|\langle\pi^{N}_{k|k},\phi\rangle-\langle\pi_{k|k},\phi\rangle|^{4}\leq\mathbb{E}^{1/4}|\overline{\Pi}_{1}|^{4}+\mathbb{E}^{1/4}|\overline{\Pi}_{2}|^{4}\\
    &\hspace{-0.5cm}\leq (2^{5}\widetilde{M}_{k|k})^{1/4}\frac{\|\phi\|_{k,4}}{N^{1/2}}+\widetilde{C}^{1/4}_{k|k}\frac{\|\phi\|_{k-1,4}}{N^{1/2}}\leq ((2^{5}\widetilde{M}_{k|k})^{1/4}+\widetilde{C}^{1/4}_{k|k})\frac{\|\phi\|_{k,4}}{N^{1/2}},
\end{align*}
\normalsize
because $\|\phi\|_{k,4}$ is increasing in $k$. Hence, defining $C_{k|k}^{1/4}\doteq ((2^{5}\widetilde{M}_{k|k})^{1/4}+\widetilde{C}^{1/4}_{k|k})$ yields \eqref{eqn:k diff}.
\end{claimproof}
\begin{claim}\label{claim:k pow 4}
The distributions $\pi_{k|k}$ and $\pi^{N}_{k|k}$ satisfy
\par\noindent\small
\begin{align}
    \mathbb{E}|\langle\pi^{N}_{k|k},|\phi|^{4}\rangle|\leq M_{k|k}\|\phi\|^{4}_{k,4}.\label{eqn:k pow 4}
\end{align}
\normalsize
\end{claim}
\begin{claimproof}
Since, $(\hat{\mathbf{x}}^{i}_{k},\mathbf{y}^{i}_{k})\sim\widetilde{\pi}^{N}_{k,k}$, we have $\langle\widetilde{\pi}^{N}_{k|k},|\phi|^{4}\rangle=\mathbb{E}[|\phi(\hat{\mathbf{x}}^{i}_{k},\mathbf{y}^{i}_{k})|^{4}|\mathcal{G}_{k}]$. Then, using Lemma~\ref{lemma:diff to power p inequality} and \eqref{eqn:update pow 4}, we obtain
\par\noindent\small
\begin{align*}
    &\mathbb{E}|\langle\pi^{N}_{k|k},|\phi|^{4}\rangle-\langle\pi_{k|k},|\phi|^{4}\rangle|\leq\mathbb{E}|\langle\pi^{N}_{k|k},|\phi|^{4}\rangle-\langle\widetilde{\pi}^{N}_{k|k},|\phi|^{4}\rangle|+\mathbb{E}|\langle\widetilde{\pi}^{N}_{k|k},|\phi|^{4}\rangle-\langle\pi_{k|k},|\phi|^{4}\rangle|\\
    &\leq\mathbb{E}\left|\frac{1}{N}\sum_{i=1}^{N}(|\phi(\hat{\mathbf{x}}^{i}_{k},\mathbf{y}^{i}_{k})|^{4}-\mathbb{E}[|\phi(\hat{\mathbf{x}}^{i}_{k},\mathbf{y}^{i}_{k})|^{4}|\mathcal{G}_{k}])\right|+\mathbb{E}|\langle\widetilde{\pi}_{k|k}^{N},|\phi|^{4}\rangle|+\langle\pi_{k|k},|\phi|^{4}\rangle\\
    &\leq\frac{1}{N}\sum_{i=1}^{N}2\mathbb{E}[\mathbb{E}[|\phi(\hat{\mathbf{x}}^{i}_{k},\mathbf{y}^{i}_{k})|^{4}|\mathcal{G}_{k}]]+\widetilde{M}_{k|k}\|\phi\|^{4}_{k,4}+\|\phi\|^{4}_{k,4}\\
    &=2\mathbb{E}[(\widetilde{\pi}^{N}_{k|k},|\phi|^{4})]+(\widetilde{M}_{k|k}+1)\|\phi\|^{4}_{k,4}\leq(3\widetilde{M}_{k|k}+1)\|\phi\|^{4}_{k,4},
\end{align*}
\normalsize
Hence,
\par\noindent\small
\begin{align}
    \mathbb{E}|\langle\pi^{N}_{k|k},|\phi|^{4}\rangle|\leq(3\widetilde{M}_{k|k}+2)\|\phi\|^{4}_{k,4},
\end{align}
\normalsize
which proves the claim with $M_{k|k}\leq(3\widetilde{M}_{k|k}+2)$.
\end{claimproof}
Claims~\ref{claim:k diff} and \ref{claim:k pow 4} show that inequalities \eqref{eqn:ipf converge} and \eqref{eqn:M inequality k} hold for $k$-th time instant, completing the induction proof.

\normalsize
\end{singlespace}

\clearpage
\addcontentsline{toc}{chapter}{List of Publications}
\listofpapers
\textbf{Journals}
\begin{enumerate}
\item Himali Singh, Arpan Chattopadhyay, and Kumar Vijay Mishra, \newblock
 ``\textit{Inverse Extended Kalman filter -- Part I: Fundamentals},"
 \newblock IEEE Transactions on Signal Processing, vol. 71, pp. 2936- 2951, 2023.
 \item Himali Singh, Arpan Chattopadhyay, and Kumar Vijay Mishra, \newblock
 ``\textit{Inverse Extended Kalman filter -- Part II: Highly Nonlinear and Uncertain Systems},"
 \newblock IEEE Transactions on Signal Processing, vol. 71, pp. 2952- 2967, 2023.
 \item Himali Singh, Kumar Vijay Mishra, and Arpan Chattopadhyay, \newblock
 ``\textit{Inverse Unscented Kalman filter},"
 \newblock IEEE Transactions on Signal Processing, vol. 72, pp. 2692-2709, 2024. 
\item Himali Singh, Kumar Vijay Mishra, and Arpan Chattopadhyay,  \newblock
 ``\textit{Inverse Cubature and Quadrature Kalman filters},"
  \newblock IEEE Transactions on Aerospace and Electronics Systems, vol. 60, no. 4, pp. 5431-5444, 2024.
\item Himali Singh, Arpan Chattopadhyay, and Kumar Vijay Mishra,  \newblock
 ``\textit{Inverse Particle filter}," IEEE Transactions on Signal Processing, vol. 73, pp. 1922-1938, 2025.
 \item Himali Singh, Kumar Vijay Mishra, and Arpan Chattopadhyay,  \newblock
 ``\textit{Inverse Gaussian Particle and Ensemble Kalman filters}," submitted.
\end{enumerate}
\textbf{Conferences}
\begin{enumerate}
\item Himali Singh, Arpan Chattopadhyay, and Kumar Vijay Mishra, \newblock
 ``\textit{Inverse Cognition in Nonlinear Sensing Systems},"
 \newblock in IEEE Asilomar Conference on Signals, Systems, and Computers, 2022, pp. 1116- 1120.
\item Himali Singh, Kumar Vijay Mishra, and Arpan Chattopadhyay, \newblock
 ``\textit{Counter-Adversarial Learning with Inverse Unscented Kalman Filter},"
 \newblock in IEEE Conference on Decision and Control (CDC), 2023, pp. 7661- 7666.
\end{enumerate}

\end{document}